\documentclass[RBook,envcountsame,envcountchap]{svmono}

\usepackage{newtxtext}
\usepackage{newtxmath}
\usepackage{makeidx}
\usepackage{multicol}
\usepackage{footmisc}

\usepackage{anyfontsize} 

\usepackage[LGR,T1]{fontenc}
\usepackage[american]{babel}
\usepackage{csquotes}

\DeclareFontFamilySubstitution{LGR}{ntxtlf}{Tempora-TLF}

\usepackage{amsmath}
\usepackage{slashed}
\usepackage{amscd}
\usepackage{amstext}
\usepackage{amsbsy}
\usepackage{amsopn}
\usepackage{amsthm} 
\usepackage{amsxtra}
\usepackage{upref}
\usepackage{amsfonts}
\usepackage{amssymb}
\usepackage{euscript}
\usepackage[]{latexsym}
\usepackage{graphicx}
\usepackage{color}
\usepackage[all]{xy} 
\usepackage{physics}
\usepackage{mdframed}
\usepackage{imakeidx}
\usepackage{tensor}
\usepackage{changepage} 
\usepackage{comment}
\usepackage{longtable}

\usepackage{soul}
\usepackage{graphics}
\usepackage{dynkin-diagrams}
\usepackage{textgreek}
\usepackage{tikz-cd}
\usepackage{titling}

\usepackage[style=numeric,backend=bibtex,sorting=nyt,maxbibnames=99,url=false,giveninits=true,backref=false]{biblatex}
\AtEveryBibitem{\clearlist{language}}
\renewbibmacro{in:}{}
\AtBeginBibliography{\footnotesize}
\DefineBibliographyStrings{english}{%
  backrefpage = {p.},
  backrefpages = {p.},
}

\defbibheading{bibliography}[\bibname]{%
  \chapter*{#1}
  \addcontentsline{toc}{chapter}{#1}
}

\usepackage{minitoc} 
\DeclareMathAlphabet{\pazocalalphabet}{OMS}{zplm}{m}{n}
\SetMathAlphabet\pazocalalphabet{bold}{OMS}{zplm}{bx}{n}

\ExplSyntaxOn
\NewDocumentCommand{\pazocal}{m}
 {
  \seq_set_split:Nnn \l_tmpa_seq { ~ } { #1 }
  \pazocalalphabet{\seq_use:Nn \l_tmpa_seq { \ }}
 }
\ExplSyntaxOff
\usepackage{hyperref}
\hypersetup{
	pdfauthor={C.~Kineider, G.~Kydonakis, E.~Rogozinnikov, V.~Tatitscheff and A.~Thomas},
	pdftitle={Spectral Networks},
	bookmarksnumbered=true,
	colorlinks=true,
	urlcolor=blue!50!black,
	linkcolor=black,
	citecolor=olive,
	linkbordercolor = {white}
}
\usepackage{cleveref}
\Crefname{chapter}{Chap.}{Chaps.}
\Crefname{section}{Sect.}{Sects.}
\Crefname{page}{p.}{pp.}
\Crefname{volume}{Vol.}{Vols.}
\Crefname{figure}{Fig.}{Figs.}
\Crefname{table}{Tab.}{Tabs.}
\Crefname{equation}{Eq.}{Eqs.}
\Crefname{theorem}{Thm.}{Thms.}
\Crefname{definition}{Def.}{Defs.}
\Crefname{remark}{Rmk.}{Rmks.}
\Crefname{proposition}{Prop.}{Props.}
\Crefname{lemma}{Lem.}{Lems.}
\Crefname{corollary}{Cor.}{Cors.}
\Crefname{conjecture}{Conj.}{Conjs.}
\Crefname{example}{Ex.}{Exs.}

\makeatletter
\newcommand{\labelx}[1]{
    \relax
    \ifmmode
        \label{#1} 
    \else 
        \ifnum\pdfstrcmp{\@currenvir}{document}=0
            \label{#1}
        \else
            \label[\@currenvir]{#1}
        \fi
    \fi
}
\makeatother

\usepackage{xpatch} 
\makeatletter
\xpatchcmd{\@endpart}{\vfil\newpage}{}{}{}
\xpatchcmd{\@endpart}{\newpage}{}{}{}
\makeatother

\usepackage{datatool}

\newtheorem{fact}[theorem]{Fact}

\newcommand*{\R}{\mathbb{R}}
\newcommand*{\C}{\mathbb{C}}
\newcommand*{\Z}{\mathbb{Z}}
\renewcommand*{\H}{\mathbb{H}}
\newcommand{\pslr}{\mathrm{PSL}_2(\mathbb{R})}
\newcommand{\del}{\partial}
\newcommand{\delbar}{\bar\partial}
\newcommand{\textgreek}[1]{\begingroup\fontencoding{LGR}\selectfont#1\endgroup} 

\DeclareMathOperator{\SL}{SL}
\DeclareMathOperator{\GL}{GL}
\DeclareMathOperator{\PSL}{PSL}
\DeclareMathOperator{\Hom}{Hom}

\usepackage[toc,nonumberlist]{glossaries} 
\makeglossaries 
\newglossaryentry{BPS state}{
    name={BPS state},
    description={\emph{(Physics)}: The notion of a \textit{BPS state} originates from the study of monopoles and dyons realized as classical gauge solitons in non-abelian \glspl{Yang--Mills}, as pioneered in the 1970s by 't~Hooft, Polyakov, and Julia--Zee. The squared mass of such states is bounded from below by the sum of the squares of their electric and magnetic charges, a relation known as the \textit{Bogomol'nyi bound}. States that saturate this bound are referred to as \textit{BPS states}. In four-dimensional non-abelian \glspl{gauge theory} with two \glspl{supersymmetry}, BPS states transform in specific representations of the supersymmetry algebra known as \textit{short representations}, as shown by Witten and Olive. This property grants BPS states remarkable stability against quantum corrections. These developments are reviewed in \Cref{Chap:SWtheory}. By extension, any one-particle state transforming in a short representation of a supersymmetry algebra is referred to as a BPS state. This broader definition is the one adopted in \Cref{Chapter:N=1theories}.\\[0.2cm]
    \noindent \emph{(Mathematics)}: Classically, \emph{BPS states} correspond to solutions of specific partial differential equations that arise in \glspl{Yang--Mills}. In quantum field theories with extended \glspl{supersymmetry}, however, the term \textit{BPS state} refers to distinguished irreducible representations of super-Lie algebras, specifically extended super-Poincaré algebras. General representations of these algebras satisfy the inequality $M \geq |Z|$, where $M$ is a Casimir invariant and $Z$ is a (complex) central charge. The representations corresponding to BPS states saturate this bound, $M = |Z|$. In the context of theories of class $\mathcal{S}$, spectral networks have been introduced as a powerful tool for studying the properties and spectrum of BPS states. BPS states appear in a variety of contexts within the field of enumerative geometry.}
}

\newglossaryentry{character variety}{
    name={Character variety},
    plural={character varieties},
    description={\emph{(Mathematics)}: For the fundamental group $\Gamma=\pi_1(M)$ of a manifold $M$ and a Lie group $G$ (typically a subgroup of $\mathrm{GL}_n$), the \emph{character variety} $\mathcal{R}(\Gamma,G)$ describes the isomorphism classes of representations of $\Gamma$ into $G$. It is the space of group homomorphisms $\Gamma\to G$ modulo the conjugation action of $G$ (GIT quotient). Character varieties are strongly linked to geometric structures on $M$, flat connections on $G$-principal bundles and Higgs bundles. Character varieties are discussed in \Cref{sec:character_variety}.\\[0.2cm]
    \emph{(Physics)}: In \glspl{gauge theory}, the dynamical variable is the \emph{connection one-form} (or \emph{covariant derivative}). The moduli space of connections with vanishing field strength (called flat connections) is the \emph{character variety}. This description is called the \gls{Riemann--Hilbert correspondence}, and is discussed in \Cref{sec:RH}.}
}

\newglossaryentry{class S}{
    name={Class $\mathcal S$ theory},
    plural={class $\mathcal S$ theories},
    description={\emph{(Physics)}: A four-dimensional quantum field theory with $\mathcal{N}=2$ \glspl{supersymmetry} is said to belong to \textit{class~$\mathcal{S}$} if it arises as the four-dimensional limit of an interacting six-dimensional $\mathcal{N}=(2,0)$ superconformal field theory compactified on a Riemann surface $C$. Class $\mathcal{S}$ encompasses a wide range of $4d$ $\mathcal{N}=2$ quantum field theories and is distinguished by the property that the associated \gls{Seiberg--Witten} integrable system is a \gls{Hitchin} system. \textit{Class $\mathcal{S}$ theories} are discussed in \Cref{Chap:theoriesofclassS}.},
    text={class $\mathcal S$}
}

\newglossaryentry{Teichmuller space}{
    name={Teichm\"{u}ller space},
    plural={Teichm\"{u}ller spaces},
    description={\emph{(Mathematics)}: The \textit{Teichm\"{u}ller space} is the moduli space of marked Riemannian metrics on a smooth surface $S$ with constant curvature, up to isotopy (diffeomorphism of $S$ homotopic to the identity). Equivalently, the Teichm\"uller space of $S$ is the moduli space of marked complex structures on $S$, up to isotopy. For $S$ closed and of genus $g \geq 2$, the Teichm\"uller space has the topology of a real ball of dimension $6g - 6$. Teichm\"uller spaces are endowed with various interesting structures and admit remarkable coordinate systems. The fundamental nature of Teichm\"uller spaces in the study of surfaces explains their appearance in various physical contexts. Teichm\"uller spaces are discussed in \Cref{subsec_Teich_sp}.}
}

\newglossaryentry{cluster variety}{
    name={Cluster variety},
    plural={cluster varieties},
    description={\emph{(Mathematics)}: A \textit{cluster variety} is a special kind of algebraic variety obtained by gluing copies of $(\mathbb{C}^*)^n$ along special rational transformations called mutations, which grants it remarkable properties, notably that it can be defined over semi-fields. Cluster varieties are associated to mutation classes of \glspl{quiver}. Precisely, to each mutation class of quivers correspond two cluster varieties denoted $\mathcal{A}$ and $\mathcal{X}$. The first carries a natural closed $2$-form (a presymplectic structure), and the second, a canonical Poisson structure. Cluster varieties appear in a variety of contexts in geometry and representation theory; among the main examples are the \glspl{higher-rank Teichmuller space} introduced by Fock and Goncharov. See \Cref{Sec:clustervarieties}.}
}

\newglossaryentry{component of maximal representations}{
    name={Component of maximal representations},
    plural={components of maximal representations},
    description={\emph{(Mathematics)}: Representations of surface groups into certain real Lie groups, specifically into Hermitian groups of tube type, possess a discrete invariant called the \emph{Toledo number}. This invariant is bounded, the bound being determined by the Euler characteristic of the surface and the rank of the Lie group. Representations for which the Toledo number is maximal are referred to as \emph{maximal representations}, a concept introduced by Burger, Iozzi, and Wienhard. Along with \glspl{Hitchin component}, spaces of maximal representations are fundamental examples of \glspl{higher-rank Teichmuller space}. Components of maximal representations are discussed in \Cref{sec:max-comps}.}
}

\newglossaryentry{conformal group}{
    name={Conformal group},
    plural={conformal groups},
    description={\emph{(Mathematics)}: The group of diffeomorphisms of a (pseudo-)Riemannian manifold preserving the conformal class of the metric is called the \textit{conformal group}. For example, the conformal group of the Euclidean $n$-sphere is $\mathrm{SO}(1,n+1)$, while that of the Lorentzian $n$-dimensional sphere is $\mathrm{SO}(2,n)$.\\[0.2cm]
    \emph{(Physics)}: The \textit{conformal group} is the group of spacetime symmetries for \emph{conformal field theories}, as briefly discussed in \Cref{Sec:conformal-algebras}. Conformal groups admit tightly-constrained supersymmetric extensions known as \textit{superconformal groups}.}
}

\newglossaryentry{Coulomb branch}{
    name={Coulomb branch},
    plural={Coulomb branches},
    description={\emph{(Physics)}: The \textit{Coulomb branch} is a specific branch of the \glspl{moduli space of supersymmetric vacua} in a supersymmetric quantum field theory, where the low-energy effective theory is a supersymmetric abelian gauge theory. In four-dimensional quantum field theories with $\mathcal N=2$ supersymmetries, Coulomb branches are \textit{special K\"ahler manifolds}. This is discussed in \Cref{Chap:N=2}. \gls{Seiberg--Witten} theory provides an exact description of the low-energy physics on such Coulomb branches. When the $4d$ $\mathcal{N}=2$ theory falls into the \gls{class S}, the Coulomb branch forms the base of a \gls{Hitchin} system on a Riemann surface $C$.\\[0.2cm]
    \emph{(Mathematics)}: The \textit{Coulomb branch} is an analytic variety with additional geometric structure associated with a supersymmetric classical or quantum field theory. For example, in the case of four-dimensional ($4d$) theories with $\mathcal{N}=2$ supersymmetry, which is the primary focus in \Cref{partV}, the Coulomb branch is always a special K\"ahler manifold. Notably, \gls{Seiberg--Witten} theory provides a method to compute the exact (quantum) special K\"ahler metric on Coulomb branches. In theories of class $\mathcal{S}$, which are special cases of $4d$ $\mathcal{N}=2$ theories associated to a Riemann surface $C$, the Coulomb branch corresponds to the base of a \gls{Hitchin} system on $C$, and can therefore be parametrized by tuples of holomorphic differentials on $C$.}
}

\newglossaryentry{covering space}{
    name={Covering space},
    plural={covering spaces},
    description={\emph{(Mathematics)}: A \emph{covering space} of a topological space $X$ is a pair $(\widetilde{X}, p)$, where $\widetilde{X}$ is a topological space and $p\colon \widetilde{X} \to X$ is a continuous map, such that for every $x \in X$, the preimage $p^{-1}(x)$ consists of a discrete disjoint union of copies of $X$. A natural extension of this concept is a \textit{ramified covering}, which permits the map $p$ to have ramification points. Covering spaces play a foundational role throughout this monograph and are introduced in \Cref{Sec:coveringspaces}.}
}

\newglossaryentry{electric-magnetic duality}{
    name={Electric-magnetic duality},
    plural={electric-magnetic dualities},
    description={\emph{(Physics)}: The term \textit{electric-magnetic duality} highlights the remarkable property of Maxwell's equations being invariant under the exchange of electric and magnetic fields. Quantum mechanically, this symmetry necessitates the simultaneous inversion of the electric charge, $e \mapsto e^{-1}$, and the exchange of electrically charged particles, such as electrons, with magnetic monopoles. In \glspl{Yang--Mills}, this duality extends to a broader framework known as \textit{Montonen--Olive duality}, or \textit{$S$-duality}. When $\Theta$-angles are included, $S$-duality is further enhanced to the duality group $\mathrm{SL}_2(\mathbb{Z})$. This duality is confirmed in many quantum field theories with $\mathcal{N} \geq 2$ supersymmetries, most notably in $\mathcal{N}=4$ super-Yang–Mills theories. A detailed discussion of electric-magnetic duality and $S$-duality is provided in \Cref{Sec:Sduality}.\\[0.2cm]
    \emph{(Mathematics)}: The concept of \emph{$S$-duality} is deeply intertwined with the \emph{geometric Langlands correspondence}. A key indication of this connection is that the dual of a Yang–Mills theory with gauge group $G$ corresponds to a Yang–Mills theory with gauge group $G^\vee$, where $G^\vee$ is the Langlands dual of $G$.}
}

\newglossaryentry{gauge theory}{
    name={Gauge theory},
    plural={gauge theories},
    description={\emph{(Physics)}: In classical and quantum Lagrangian field theories, a \textit{gauge symmetry} is a local redundancy in the formulation of the theory, associated with a Lie group $G$ called the \textit{gauge group}. Examples of field theories with gauge symmetries---known as \textit{gauge theories}---include Maxwell, Chern--Simons and \glspl{Yang--Mills}. The \textit{gauge field} in such theories is a massless vector field, locally described as a Lie algebra-valued one-form on spacetime. Gauge theories are ubiquitous in \Cref{partV}. Furthermore, there exist generalizations of gauge theories, known as higher-form gauge theories, where the gauge field is locally a differential form of higher degree on spacetime.\\[0.2cm]
    \emph{(Mathematics)}: \Glspl{gauge theory} involve $G$-principal bundles, their associated vector bundles, and connections on manifolds (see \Cref{Sec:bundlesandconnections}). These structures define the notion of gauge equivalence. In gauge theories, one typically aims to compute the moduli space of certain geometric objects, considered up to gauge equivalence. The notion of \textit{gerbes}, which generalizes that of principal bundles, embodies the physical concept of higher-form gauge theories.}
}

\newglossaryentry{Higgs branch}{
    name={Higgs branch},
    plural={Higgs branches},
    description={\emph{(Physics)}: The \textit{Higgs branch} is a specific branch of the \glspl{moduli space of supersymmetric vacua} in a supersymmetric quantum field theory, where the gauge group is entirely broken at low-energy. In four-dimensional quantum field theories with $\mathcal N=2$ supersymmetries, Higgs branches are \textit{hyperk\"ahler manifolds}. This is discussed in \Cref{Chap:N=2}. \\[0.2cm]
    \emph{(Mathematics)}: The \textit{Higgs branch} is an analytic variety with additional geometric structure associated with a supersymmetric classical or quantum field theory. For example, in the case of four-dimensional ($4d$) theories with $\mathcal{N}=2$ supersymmetry, which is the primary focus in \Cref{partV}, the Higgs branch is always a \emph{hyperk\"ahler manifold}.}
}

\newglossaryentry{higher-rank Teichmuller space}{
    name={Higher-rank Teichm\"uller space},
    plural={higher-rank Teichm\"uller spaces},
    description={\emph{(Mathematics)}: A \emph{$G$-Teichm\"uller space} of a surface $S$ is the union of connected components of the $G$-\glspl{character variety} $\Hom(\pi_1(S),G)/G$, where $\pi_1(S)$ is the fundamental group of $S$ and $G$ is a real Lie group, consisting exclusively of discrete and faithful morphisms. When $\mathrm{rank}(G)>1$, such a space is referred to as a higher-rank Teichm\"uller space, as opposed to \glspl{Teichmuller space} for which $G=\mathrm{PSL}_2(\mathbb{R})$. By the \gls{Riemann--Hilbert correspondence}, $G$-Teichm\"uller spaces can equivalently be described as distinguished subspaces of the moduli space of flat $G$-principal bundles on $S$. A remarkable fact is that the existence of $G$-Teichm\"uller spaces on a surface $S$ depends only on the group $G$. Typical examples of higher-Teichm\"uller spaces include the so-called \glspl{Hitchin component}, for $G$ split and adjoint, and \glspl{component of maximal representations}, for $G$ Hermitian of tube type. Higher-rank Teichm\"uller spaces are the main characters of \Cref{partIII}.}
}

\newglossaryentry{Hitchin component}{
    name={Hitchin component},
    plural={Hitchin components},
    description={\emph{(Mathematics)}: The study of moduli spaces of Higgs bundles (see \gls{Hitchin} equations) led to the discovery of distinguished connected components in the $G$-\glspl{character variety} $\Hom(\pi_1(S),G)/G$ of a smooth surface $S$, consisting solely of discrete and faithful representations when $G$ is split and real. These connected components, now known as \emph{Hitchin components}, are fundamental examples of \glspl{higher-rank Teichmuller space}. They are discussed in \Cref{sec:Hitchin_reps}.}
}

\newglossaryentry{Hitchin}{
    name={Hitchin equations},
    description={\emph{(Mathematics)}: \emph{Hitchin equations} describe self-dual $G$-\gls{Yang--Mills} connections on a Riemann surface $X$ and led to the important notion of \textit{Higgs bundles}. Through the \textit{non-abelian Hodge correspondence}, the moduli space of Higgs bundles can be identified with \glspl{character variety} of surfaces. Solving Hitchin equations centrally involves the theory of harmonic maps from the hyperbolic plane into symmetric spaces of non-compact type. \\[0.2cm]
    \emph{(Physics)}: \textit{Hitchin equations} are equations of motion for $2d$ adjoint QCD. Moduli spaces of \emph{Higgs bundles} play a prominent role in the context of \gls{class S} theories, where they form the Coulomb branch of these theories when compactified on a circle. This is briefly discussed in \Cref{Chap:theoriesofclassS}.},
    text={Hitchin}
}

\newglossaryentry{homology}{
    name={Homology},
    description={\emph{(Mathematics)}: \emph{Homology} assigns abelian groups to topological spaces in a way that is invariant under homeomorphisms, providing algebraic invariants of those spaces. Various constructions of homology exist, including simplicial, cellular, and singular homology, which are often equivalent. The dual notion, \emph{cohomology}, introduces an additional structure, turning cohomology into a ring. For smooth manifolds, the study of closed differential forms modulo exact ones leads to a cohomology theory known as de Rham cohomology.}
}

\newglossaryentry{homotopy}{
    name={Homotopy},
    description={\emph{(Mathematics)}: A \emph{homotopy} between two maps is a continuous deformation of one map into the other. This concept leads to a fundamental family of algebraic invariants of topological spaces known as \textit{homotopy groups}. The simplest case is that of the \textit{fundamental group}, discussed in \Cref{chap:fundamental_gp}, which is obtained as the group of homotopy classes of based loops in a topological space.}
}

\newglossaryentry{instanton}{
    name={Instanton},
    plural={instantons},
    description={\emph{(Physics)}: An \textit{instanton} is a classical self-dual or anti-self-dual solution to Euclidean \gls{Yang--Mills} equations. Instantons allow the efficient computation of tunneling effects between distinct classical vacua of Yang--Mills theory in Minkowski spacetime, and are typically non-perturbative effects. They play a central role in the semi-classical analysis of Yang--Mills theories. In \gls{Yang--Mills} theories, instantons are typically BPS configurations, meaning that they preserve some fraction of the supersymmetry. Instantons are discussed briefly in \Cref{Subsec:instantons}.\\[0.2cm]
    \emph{(Mathematics)}: An \textit{instanton} is a self-dual or anti-self dual solution to the \gls{Yang--Mills} equations on a four-dimensional smooth manifold. The \textit{moduli space of instantons} for a group $G$ can be described explicitly. For $G=\mathrm{SU}(2)$, this moduli space is at the heart of Donaldson's theory.}
}

\newglossaryentry{moduli space of supersymmetric vacua}{
    name={Moduli space of supersymmetric vacua},
    plural={moduli spaces of supersymmetric vacua},
    description={\emph{(Physics)}: In a quantum field theory, the set of gauge-inequivalent vacuum states defines a manifold known as the \textit{moduli space of vacua}. In general quantum field theories, this moduli space often consists of isolated points. However, in supersymmetric quantum field theories, the \textit{moduli space of supersymmetric vacua} frequently has positive dimension and is endowed with additional geometric structures, such as that of a complex K\"ahler variety.
    This space also contains distinguished subvarieties, known as \textit{branches}, which can exhibit even richer geometric properties, including structures like \textit{hyperk\"ahler manifolds} or \textit{special K\"ahler manifolds} (see \gls{Coulomb branch} and \gls{Higgs branch}). Moduli spaces of supersymmetric vacua are explored in the context of four-dimensional supersymmetric quantum field theories in \Cref{Chapter:N=1theories,Chap:N=2}.\\[0.2cm]
    \emph{(Mathematics)}: The \textit{moduli space of supersymmetric vacua} is an analytic variety associated with a supersymmetric quantum field theory, often endowed with additional geometric structure, such as that of a complex K\"ahler variety. This variety contains several physically significant subvarieties, referred to as \textit{branches}.}
}

\newglossaryentry{positivity}{
    name={Positivity},
    description={\emph{(Mathematics)}: Central to all known examples of \glspl{higher-rank Teichmuller space} is the concept of \emph{positivity}. For instance, the \glspl{Hitchin component} are characterized by \emph{totally positive} matrices and correspond to the \emph{positive} points of certain \glspl{cluster variety}. Similarly, the maximal representations valued in the symplectic group are distinguished by having \emph{positive definite} Maslov forms. These distinct notions of positivity were unified through the introduction of \emph{$\Theta$-positivity}, a broader framework that generalizes both.
    This new concept of positivity is specific to certain families of real Lie groups, including split real groups (associated with total positivity) and Hermitian groups of tube type (associated with maximality). It has revealed two additional families of higher-rank Teichm\"uller spaces: one for groups of the form $\mathrm{SO}(p,q)$ and another for certain exceptional Lie groups. Representations satisfying $\Theta$-positivity are always discrete and faithful, thereby naturally giving rise to higher-rank Teichm\"uller spaces. Positivity is discussed in \Cref{sec:positivity}.}
}

\newglossaryentry{quiver}{
    name={Quiver},
    plural={quivers},
    description={\emph{(Mathematics)}: A \textit{quiver} is a directed finite graph without 1-cycles or 2-cycles. Quivers play a central role in the definition of \glspl{cluster variety}.\\[0.2cm]
    \emph{(Physics)}: \emph{Quivers} encode the structure of certain \glspl{gauge theory}, typically the supersymmetric ones that arise on the worldvolume of branes in superstring theories. Depending on the context, notational simplifications are often made, leading to refinements such as that of $\mathcal{N}=2$ quivers, discussed in \Cref{Chap:theoriesofclassS}.}
}

\newglossaryentry{Riemann--Hilbert correspondence}{
    name={Riemann--Hilbert correspondence},
    plural={Riemann--Hilbert correspondences},
    description={\emph{(Mathematics)}: The \textit{Riemann--Hilbert correspondence}, detailed in \Cref{Riemann-Hilbert correspondence}, offers a differential geometric perspective on \glspl{character variety}. It establishes an equivalence between the representation space of a fundamental group of a manifold $M$ into a Lie group $G$, and the moduli space of flat $G$-connections on $M$, considered modulo automorphisms. In essence, the Riemann--Hilbert correspondence states that flat connections are fully characterized by their holonomies.}
}

\newglossaryentry{Seiberg--Witten}{
    name={Seiberg--Witten theory},
    description={\emph{(Physics)}: \textit{Seiberg--Witten theory} provides the framework for understanding the \glspl{Coulomb branch} of four-dimensional quantum field theories with $\mathcal{N}=2$ \glspl{supersymmetry}. Remarkably, it reveals that the low-energy dynamics on the Coulomb branch is entirely captured by an algebraic integrable system known as the \textit{Seiberg--Witten integrable system}. This theory provides a method to compute the exact special K\"ahler metric on the Coulomb branch. Seiberg--Witten theory is reviewed in \Cref{Chap:SWtheory}. Within the framework of \gls{class S} theories, the Seiberg--Witten integrable system is realized as a \gls{Hitchin} moduli space on an associated Riemann surface.},
    text={Seiberg--Witten}
}

\newglossaryentry{spectral network}{
    name={Spectral network},
    plural={spectral networks},
    description={\emph{(Mathematics)}: A \textit{spectral network} is a specific type of graph on a smooth surface $S$ that is subordinate to a ramified covering $\widetilde{S} \to S$. This structure enables the \textit{abelianization} of a non-abelian connection on $S$ into an abelian connection on $\widetilde{S}$, and vice versa (in which case the process is referred to as \textit{non-abelianization} instead). When $S$ is a Riemann surface and $\widetilde{S}$ is a spectral cover of $S$ defined by a tuple of holomorphic differentials, there exists a refined notion of an \emph{analytic spectral network}. This network is determined by the holomorphic ramified covering $T^*_{1,0}S \supset \widetilde{S} \to S$. The mathematical framework for spectral networks and their properties is elaborated in \Cref{partIV} of this monograph.\\[0.2cm]
    \emph{(Physics)}: In the context of \gls{class S} theories, \emph{spectral networks} are webs of BPS trajectories on the UV curve carrying labels known as \textit{soliton degeneracies}, and associated to an angle $\vartheta$. By studying how the topology of the spectral network changes as $\vartheta$ varies, one can extract the BPS spectrum of the theory, namely the central charge of BPS states in the theory as well as their degeneracies. This perspective on spectral networks is explored in \Cref{Sec:SpectralNetworksPhysics}.}
}

\newglossaryentry{supersymmetry}{
    name={Supersymmetry},
    plural={supersymmetries},
    description={\emph{(Physics)}: \textit{Supersymmetry} is a spacetime symmetry that links fermions to bosons. A field theory possessing such a symmetry is termed a \textit{supersymmetric field theory}. The infinitesimal symmetries of a supersymmetric field theory in flat space form a \textit{super-Poincaré algebra} or an extension thereof. Supersymmetric quantum field theories exhibit greatly enhanced behavior under renormalization compared to their non-supersymmetric counterparts, making them an ideal framework for probing deep aspects of quantum field theory that are otherwise challenging to investigate. Supersymmetric quantum field theories form a central focus of \Cref{partV} in this monograph.\\[0.2cm]
    \emph{(Mathematics)}: Lie theory has an important generalization known as \textit{super-Lie theory}. Super-Lie algebras are $\mathbb{Z}_2$-graded Lie algebras, and they can be exponentiated to form super-Lie groups. By extension, the prefix ``super'' is often used whenever a $\mathbb{Z}_2$ grading is involved.}
}

\newglossaryentry{Yang--Mills}{
    name={Yang--Mills theory},
    plural={Yang--Mills theories},
    description={\emph{(Physics)}: \textit{Yang--Mills} theories are \glspl{gauge theory} that extend Maxwell theory to non-abelian gauge groups. These theories serve as fundamental examples of non-abelian gauge theories and are crucial to the quantum field theory description of fundamental forces.\\[0.2cm]
    \emph{(Mathematics)}: \textit{Yang--Mills} theories are \glspl{gauge theory} of principal connections on manifolds. A defining feature is the \textit{Yang--Mills functional}, which associates a scalar to principal connections. The extremal points of this functional satisfy the \textit{Yang--Mills equations}, which are non-linear partial differential equations.},
    text={Yang--Mills}
}  
\setglossarystyle{altlist}

\graphicspath{{images/}}

\usepackage{authblk}

\title{Spectral Networks \\ {\large \textit{Bridging higher-rank Teichm\"uller theory and BPS states}}\\ \vspace{0cm} \begin{figure}[h!]\centering\includegraphics[scale=0.25]{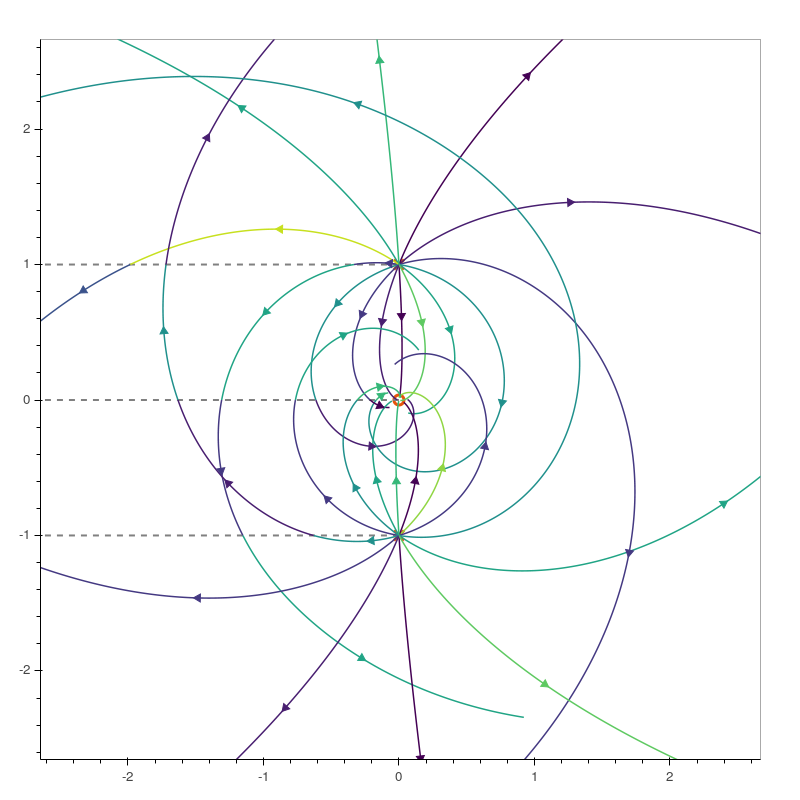} \\
{\small \textit{Spectral network of pure $4d$ $\mathcal N=2$ $\mathrm{SO}(8)$ super Yang--Mills at the origin of the Coulomb branch.}}\end{figure}\vspace{-0.5cm}}
\date{}

\author[1]{Clarence Kineider\thanks{\scriptsize clarence.kineider@gmail.com}}
\author[2]{Georgios Kydonakis\thanks{\scriptsize gkydonakis@math.upatras.gr}}
\author[3]{Eugen Rogozinnikov\thanks{\scriptsize erogozinnikov@gmail.com}}
\author[4]{Valdo Tatitscheff\thanks{\scriptsize valdo.tatitscheff@normalesup.org}}
\author[5]{Alexander Thomas\thanks{\scriptsize athomas@math.univ-lyon1.fr}}

\small{
\affil[1]{Max Planck Institute for Mathematics in the Sciences\\
Inselstraße 22, 04103 Leipzig, Germany}
\affil[2]{Department of Mathematics, University of Patras\\
University Campus, Patras 26504, Greece}
\affil[3]{School of Mathematics, Korea Institute for Advanced Study\\
85 Hoegi-ro, Dongdaemun-gu, Seoul 02455, Republic of Korea}
\affil[4]{Technical University of Munich, TUM School of Computation, Information
and Technology, Department of Mathematics\\
Boltzmannstraße 3, 85748 Garching bei München, Germany}
\affil[5]{Institut Camille Jordan, Universit\'{e} Lyon 1\\
21 Av Claude Bernard, 69100 Villeurbanne, France}}

\makeindex[intoc]

\begin{document}

\doparttoc

\maketitle
\institutename

\abstract{This book offers a comprehensive introduction to spectral networks from a unified viewpoint that bridges geometry with the physics of supersymmetric gauge theories. It provides the foundational background needed to approach the frontiers of this rapidly evolving field, treating geometric and physical aspects in parallel. After surveying fundamental topics in algebra and geometry, a detailed introduction to higher-rank Teichmüller theory is developed, including Fock--Goncharov theory for Hitchin representations, maximal representations and the more recent notion of $\Theta$-positivity. Spectral networks are subsequently introduced, emphasizing their utility in the study of character varieties via the abelianization and non-abelianization maps they define. In parallel, key aspects of four-dimensional gauge dynamics with eight supercharges are explored, including electric-magnetic duality, Seiberg--Witten theory, and class~$\mathcal S$ theories. The role of spectral networks as a framework for determining and analyzing BPS spectra in class~$\mathcal S$ theories is then examined. The final chapter outlines recent applications of spectral networks across a range of contemporary research areas. This volume is intended for researchers and advanced students in either mathematics or physics who wish to enter the field.}

\newpage
\sloppy
\hbadness=99999

\frontmatter


\preface

Spectral networks were introduced in a titular article by Davide Gaiotto, Gregory W. Moore and Andrew Neitzke published in 2013 by the journal \emph{Annales Henri Poincar\'{e}}. These are networks of trajectories on surfaces that naturally arise in the study of various four-dimensional $\mathcal{N}=2$ quantum field theories and provide a particularly efficient way to compute BPS spectra. From a purely geometric point of view, they yield a new map between flat connections over a Riemann surface and flat abelian connections on a spectral covering of the surface. At the same time, these networks of trajectories provide local coordinate systems on the moduli space of flat connections that are valuable in the study of higher-rank Teichm\"uller spaces \textit{\`a la} Vladimir V. Fock and Alexander B. Goncharov.

Since their introduction, spectral networks have been proving themselves as a profound tool in geometry and mathematical physics, with an ever-growing number of applications, generalizations and connections to other concepts. Yet, a standard first rigorous and, at the same time, expository overview of the subject is still missing from the book literature, we believe mostly due to the considerably diverse background required in order to approach their study. Motivated by our own diverse research interests, we aim to produce a book which accumulates the necessary foundation needed both by trained mathematicians and physicists for a rigorous introduction to the topic, and at the same time a book that demonstrates a substantial number of applications in geometry and mathematical physics with a view towards current research study.

The book is aiming to serve as a primary introduction to the subject, by presenting spectral networks and their many applications from the scope of geometry and mathematical physics in a unified way. We do not attempt to treat these two approaches separately but rather provide broad motivation and the necessary background to reach to the frontiers of this fast-evolving field of research, explaining simultaneously the relevant geometric and physical aspects. The targeted audience is researchers and advanced students from either a rather mathematical or physical oriented education who wish to enter the field. We have tried to make the connections between these geometric and physical aspects more transparent, by objectively linking the various topics within the book's sections. Given the largely interdisciplinary nature of spectral networks, we have endeavored to unify the various topics addressed in this book as much as possible. This approach particularly explains our decision to use both standard mathematical and physical terminology within a single volume.

Spectral networks and their applications in geometry and mathematical physics is a modern topic which currently sees a rapid development and expansion. As such, the field might undergo variations in its perception from the working mathematician or physicist. In the light of this condition, we have chosen to present in a clear way the fundamental aspects of the theory and the most solid applications up-to-date. At the same time, we have collected a series of broader topics of current research development which are presented in a less rigorous way at the end of the book. Major progress in these directions might cause relevant advanced notions to undergo certain future revisions in their guise.  

\hspace{0.3cm}

In a prelude to the main body of the book, we provide a historical overview of the key ideas from high-energy physics, geometry, and, in particular, Teichmüller theory, which led to the introduction of spectral networks as a tool for computing BPS spectra in four-dimensional gauge theories with two or eight supersymmetries, and their application in the context of higher-rank Teichmüller spaces. While not intended as a pedagogical introduction to these topics---many excellent ones already exist---we have aimed to make it accessible and enjoyable for readers unfamiliar with the field, so that everyone can absorb the key concepts and benefit from this exposition in understanding the subsequent chapters of the book. Some important mathematical notions appearing here are defined rigorously and pedagogically in later sections; we emphasize these connections when possible. \Cref{partII,partIII} are entirely independent of this introductory treatise and focus on various topics in geometry. Spectral networks and their use in geometry are introduced in \Cref{partIV}, while they are introduced from the perspective of supersymmetric gauge theories in \Cref{partV}.

Further expanding on the contents of the book, \Cref{partII} is a rigorous and pedestrian introduction to the mathematical background needed in order to discuss spectral networks. Not all notions included here are systematically and adequately explored within academic mathematics or physics curricula, however there exist already multiple standard expositions on relevant topics; this part therefore aims at setting a common ground for readers, regardless of their background. This part can be safely skipped by readers who are already comfortable with the topics addressed as well as the language in which these are introduced.

\Cref{partIII} is one of the central parts of the book and provides a rigorous description of a variety of topics from higher-rank Teichm\"{u}ller theory that are addressed in this book. In particular, we present here a unified approach to higher-rank Teichm\"{u}ller theory as a higher-rank generalization of the classical Teichm\"{u}ller theory through the central notion of $\Theta$-positivity introduced by Olivier Guichard and Anna Wienhard. Moreover, we discuss in analogy Thurston's shear coordinates over the Teichm\"{u}ller space on the one hand, and the Fock--Goncharov coordinates on higher-rank Teichm\"{u}ller spaces on the other hand. 

Spectral networks are explicitly defined in \Cref{partIV} of the book. We subsequently present and discuss the most prominent applications of spectral networks in geometry: non-abelianization and abelianization, which relate higher-rank Teichm\"{u}ller spaces of a base surface to abelian character varieties of some ramified cover, the spectral curve.
A more analytic approach to abelianization is given by the exact WKB method. From this perspective, the spectral network appears as Stokes graph for the Borel resummation. The Fock--Goncharov coordinates can be recovered (at least in low rank) from explicit  monodromy data.

\Cref{partV} is devoted to exploring the role of spectral networks in four-dimensional gauge theories with $\mathcal N=2$ supersymmetries, particularly class $\mathcal S$ theories, as a powerful framework for describing BPS spectra. We begin by reviewing the general aspects of supersymmetric gauge theories in four dimensions, and then narrow our focus on theories possessing $\mathcal{N}=2$ supersymmetry. Following this, we discuss electric-magnetic duality and provide an introduction to Seiberg–Witten theory. Lastly, we introduce theories of class $\mathcal S$ and examine the counting of BPS states in these theories, notably through the use of spectral networks. This part of the monograph is intended as a general introduction to the topic, aiming to equip readers with the foundational concepts needed to engage with the corresponding literature.

\Cref{partVI} presents more advanced applications of spectral networks both in geometry and in the context of supersymmetric quantum field theories. The discussion here is kept more informal compared to the previous chapters, and should be rather considered as an invitation to further investigation of the topics. We provide an extensive list of references, which readers are encouraged to explore for more detailed accounts.

To minimize the linguistic barrier between the mathematics and mathematical physics communities engaged with spectral networks, this monograph concludes with a glossary that aligns and clarifies the principal concepts used across both disciplines.

\hspace{0.3cm}

This book grew out of a working group between the authors on the topics addressed here, and was initiated following the conference ``Teichm\"{u}ller Theory: Classical, Higher, Super and Quantum'' hosted at CIRM, Luminy, France from Oct. 5 to Oct. 9, 2020. It rapidly became apparent to us during the works of that conference that a pedagogic and comprehensive introduction to spectral networks was absent from the book literature at the time, and that the research field was divided in two main communities between which interactions were hindered by linguistic and cultural differences, though the objects of study are essentially the same. The diversity in our own research backgrounds while interacting through our subsequent working group proved out to be especially fruitful in embracing all aspects of spectral networks, ultimately leading to the idea of distilling the working material into a hands-on volume aiming at bridging the different existing perspectives on spectral networks.

\hspace{0.3cm}

The authors are thankful to the organizers and participants of the conference at CIRM for useful discussions, as well as to CIRM itself and the peaceful inspiring nature of the Calanques of Marseille for initiating a very rewarding working group. We are grateful to the Universities of Strasbourg and Heidelberg and the Max-Planck Institute for Mathematics in Bonn for providing a fertile environment in which most of the work for this book was conducted, and the Alexander von Humboldt Foundation for its continuous support which allowed us to carry out a number of research visits between our two Institutes. V.T. is grateful to R.~Argurio, S.~Banerjee, M.~Bertolini, O.~Hulik, L.~Mol, A.~Pasternak, S.~Rota, R.~Senghaas, R.~Vandepopeliere, and J.~Walcher for stimulating discussions and feedback on preliminary versions of this work. We also extend our gratitude to all the referees for a careful review of our work and for providing a series of constructive comments and corrections. 

\hspace{0.3cm}

C.K. has received funding from the European Research Council (ERC) under the European Union’s Horizon 2020 research and innovation programme (grant agreement No 101018839).
G.K. acknowledges support from the Scientific Committee of the University
of Patras through the program ``Medicos''. E.R. was funded by the Labex IRMIA of the Universit\'e de Strasbourg, a postdoc fellowship of the German Academic Exchange Service (DAAD), funding from the European Research Council (ERC) under the European Union’s Horizon 2020 research and innovation programme (grant agreement No 101018839), and the KIAS Individual Grant (MG100901) at Korea Institute for Advanced Study. V.T. acknowledges funding from the Deutsche Forschungsgemeinschaft (DFG, German Research Foundation) under Germany’s Excellence Strategy EXC 2181/1 - 390900948 (the
Heidelberg STRUCTURES Excellence Cluster). 
The work of A.T. was partially funded by ERC-Advanced Grant 101018839, Deutsche Forschungsgemeinschaft (DFG, German
Research Foundation) - Project-ID 281071066 - TRR 191, and at the end of the project by University Claude Bernard Lyon 1.

\vspace{12mm}

\begin{flushright}
Heidelberg and Strasbourg,
\end{flushright}
\begin{flushright}
July 2023
\end{flushright}

\newpage

\tableofcontents

\newpage
\Extrachap{List of symbols}

\begin{longtable}{ll}
$G$ & Lie group   \\
$\mathfrak{g}$ & Lie algebra   \\
$G^{\vee}$ & Langlands dual group  \\
$[G,G]$ & commutator subgroup  \\
$G_{ab}$ & abelianization of a group   \\
$\mathrm{ad}$ & adjoint representation  \\
$\mathfrak{h}$ & Cartan subalgebra of complex Lie algebra  \\
$H$ & Cartan subgroup \\
$\mathfrak{g}=\mathfrak{k}\oplus\mathfrak{p}$ & Cartan decomposition of real Lie algebra\\
$\mathfrak{a}$ or $\mathfrak{t}$ & Cartan subalgebra of real Lie algebra  \\
$W$ & Weyl group  \\
$K$ & Killing form  \\
$BG$ & classifying space for a Lie group $G$  \\
$\mathfrak{g}_{\alpha}$ & eigenspace for a root  \\
$\mathrm{rad}\,{\mathfrak{g}}$ & radical of a Lie algebra  \\
$B$ and $\mathfrak{b}$ & Borel subgroup and its Lie algebra  \\
$U$ and $\mathfrak{u}$ & upper unipotent subgroup and its Lie algebra  \\
$P$ and $\mathfrak{p}$ & parabolic subgroup and its Lie algebra\\
$\Lambda^{\perp}$ & character lattice  \\
$\Lambda$ & cocharacter lattice  \\
$\mathcal{R}(G,T)$  & root datum \\
$\mathcal{R}(G,T)^{\vee}$ & dual root datum\\
$\Sigma^{+}$ & set of positive roots  \\
$\Delta$ & set of simple roots\\
$\Theta\subset \Delta$ & subset of simple roots for parabolic subgroups\\
$P_{\Theta}$ & parabolic subgroup for a set $\Theta$ of simple positive roots  \\
$\Sigma(\mathfrak{g},\alpha)$ & restricted root system  \\

\vspace{3mm} & \vspace{3mm} \\

$S$ & topological connected oriented surface \\
$S_{g,h}$ & topological surface with genus $g$ and $h$ boundary components\\
$S_{g, \vec{p}}$ & ciliated surface of genus $g$ and vector $\vec{p}$  \\
$X$ & often Riemann surface \\
$\Sigma$ & spectral curve \\
$\tilde{S}$ & universal cover of $S$\\
$\chi(S)$ & Euler characteristic of a surface  \\
$P$  & principal bundle  \\
$d_A$ & connection on a vector or principal bundle  \\
$A$ & connection matrix \\
$F(A)$ & curvature of a connection  \\
$P_{c(t)}$ & parallel transport along a curve $c(t)$ \\
$\mathcal{L}$ & local system  \\
$\mathcal{M}$ & moduli space  \\
$\mathcal{E}(S,G)$ & space of fundamental group representations modulo conjugation  \\
$\mathcal{R}(S,G)$ & character variety  \\
$\mathcal{R}^{\mathrm{fr}}(S,G)$ & character variety of framed representations \\
$\mathcal{R}^{\mathrm{fr}}_{+}(S,G)$ & character variety of positive framed representations  \\
$\mathcal{T}(S)$ & Teichm\"{u}ller space   \\
$\mathcal{T}_H(S,G)$ & $G$-Hitchin component \\
$\mathcal{X}_{S,G}$ & moduli space of framed local systems  \\
$\mathcal{A}_{S,G}$ & moduli space of decorated twisted local systems  \\
$\mathcal{M}^{\mathrm{fr}}(S,G)$ & moduli space of maximal framed representations  \\
$K_X$ & canonical line bundle of a complex manifold $X$ \\
$\omega_{MC}$ & Maurer--Cartan form  \\
$\mathbb{H}^2$ or $\mathbb{H}$ & hyperbolic plane\\
$\partial_{\infty}\mathbb{H}$ & boundary at infinity of the hyperbolic plane  \\
$\Gamma$ & graph  \\
$\gamma$ & curve on a surface  \\
$\mathcal{P}(S)$ & space of paths on a surface  \\
$\mathbb{Z}_{B'}^{\text{tw}}[\mathcal{P}(\Sigma)]$ & twisted path algebra  \\
$\mathcal{W}_{\vartheta}$ &  WKB spectral network at angle $\vartheta$ \\
$[A]$ & equivalence class of some object $A$ \\
$X//G$ & GIT quotient  \\
$\{\cdot, \cdot\}$ & Poisson bracket  \\
$\mathcal{L}[f]$ & Laplace transform of a function  \\
$S[f]$ & Borel sum of a Borel summable function  \\
$\mathrm{Res}_z\omega$ & residue of a differential at a point  \\
$\hat{\mathcal{F}}_{\omega}$ & foliation of a differential $\omega$ \\
$\Phi^{*_h}$ & hermitian conjugate  \\
$\mathbb{T}$ & standard triangle  \\
$\mathsf{T}$ & Euclidean triangle  \\
$\Delta$ & triangulation of a surface  \\
$E(\Delta)$ & edges of a triangulation  \\
$F(\Delta)$ & faces of a triangulation  \\
$E_e(\Delta)$ & external edges of a triangulation  \\
$E_i(\Delta)$ & internal edges of a triangulation  \\
$V(\Delta)$ & vertices of a triangulation  \\
$\mathrm{Lag}$ & Lagrangian Grassmannian   \\
$[L_1,M,L_2]$ & Maslov form of three pairwise transverse Lagrangians  \\
$T_{\rho}$ & Toledo number of a representation $\rho$ \\
$\mathrm{Mod}(S_{g,\vec{p}})$ & mapping class group of a ciliated surface\\
$\mathrm{PMod}(S_{g, \vec{p}})$ & pure mapping class group of a ciliated surface  \\
$\mathcal{F}$ & space of (partial) flags  \\
$Q_{\Delta}$ & quiver associated to a triangulation $\Delta$ \\
$\mathsf{S}$ & seed  \\
$\mathcal{X}_{\Lambda}$ & cluster torus of a seed  \\
$\mu_i(Q)$ & mutation of a quiver $Q$ at vertex $i$  \\
$\mathcal{A}_{\mathsf{S}}$ & $\mathcal{A}$-variety of a seed $\mathsf{S}$ \\
$\mathcal{X}_{\mathsf{S}}$ & $\mathcal{X}$-variety of a seed $\mathsf{S}$\\

\vspace{3mm} & \vspace{3mm}\\

$\mathcal{N}$ & number of supersymmetries in a quantum field theory  \\
$\mathcal{H}$ & Hilbert space   \\
$\langle .\mid. \rangle$ & Hermitian inner product on Hilbert space \\
$\ket{\phi}$ & vector in Hilbert space \\
$\bra{\phi}$ & linear functional associated to a vector\\
$\langle \mathcal{O}(x)\rangle $ & correlation function for local field $\mathcal{O}(x)$ \\
$\mathcal{Z}$ & central charge  \\
$Q$ & supercharge  \\
$Q_\alpha^I$ & generators of supersymmetries \\
$P_{\mu}$ & generators of spacetime translations  \\
$M_{\mu\nu}$ & generators of spacetime rotations\\
$\delta_{ij}$ & Kronecker function \\
$\varepsilon_{ij}$ & antisymmetric symbol (Levi--Civita symbol)\\
$\gamma^\mu$ & Dirac gamma matrices \\
$\psi$ & spinors \\
$\sigma_i$ & Pauli matrices\\
$[.,.]$ & commutator \\
$\{.,.\}$ & anticommutator\\
$a$ and $a^\dagger$ & annihilation and creation operators\\
$\star$ & Hodge star operator  \\
$S$ & action functional \\
$\mathcal{L}$ & Lagrangian \\
$E$ & electric field \\
$B$ & magnetic field \\
$\rho_{m}$ & magnetic charge density  \\
$\textbf{J}_m$ & magnetic current  \\
$q_{\mathrm{el}}$ & electric charge  \\
$q_{\mathrm{mag}}$ & magnetic charge \\
$g$ & non-abelian coupling constant \\
$C(R)$ & Casimir number for a representation $R$ \\
$I(R)$ & Dynkin index of a representation $R$ \\
$S_{\mathrm{YM}}(A)$ & Yang--Mills functional for a connection  \\
$\mathcal{T}_g(\mathcal{M}_{(d-1)})$ & topological operator or defect associated to submanifold $\mathcal{M}_{(d-1)}$\\
$\mathcal{C}$ & Coulomb branch \\
$\lambda$ & Seiberg--Witten differential (Liouville canonical 1-form)\\
$\mathcal{S}[C, \mathfrak{g}, D]$ & theory of class $\mathcal{S}$  \\
$\mu$ & BPS index  \\
$F(\wp,\vartheta)$ & generating function of framed $2d-4d$ indices  \\
$\mathcal{F}_G(X, \theta)$ & exponentiated Chern--Simons invariant  \\
$\mathrm{Vir}_c$ & Virasoro algebra  \\		
$\mathrm{Heis}$ & Heisenberg algebra  \\
$\Psi$ & conformal block  \\
$m_Z$ & energy scale of the $Z$ boson  \\
$\Lambda_{\mathrm{QCD}}$ & dynamical scale of quantum chromodynamics   \\
$L_{\wp, \vartheta}$ & supersymmetric interface  \\
$\Psi_q$ & quantum dilogarithm  \\

\vspace{3mm} & \vspace{3mm}\\

BPS & Bogomol'nyi--Prasad--Sommerfield\\
CFT & conformal field theory\\
QCD & quantum chromodynamics  \\
QED & quantum electrodynamics  \\
QFT & quantum field theory \\
SCFT & superconformal field theory\\
TQFT & topological quantum field theory\\
VEV & vacuum expectation value \\
WKB & Wentzel--Kramers--Brioullin \\

\end{longtable}



\Extrachap{Motivation and Historical Perspective}\labelx{partI}

\abstract{Our objective here is to provide a brief---hence necessarily incomplete---chronological overview of the key ideas from both high-energy physics and Teichmüller theory that contributed to the emergence of spectral networks. We will highlight how these two fields have mutually enriched one another, with concepts from one often paving the way for entirely new research directions in the other, sometimes developing independently, but more recently converging through the framework of spectral networks.
We begin by discussing significant aspects of quantum field theory, a framework for describing quantum relativistic phenomena, focusing on the crucial introduction of Yang--Mills theories and the discovery of instantons. Next, we shift to the mathematical field of gauge theories, which emerged from the study of instantons in Yang--Mills theories. We will also explore how higher-rank Teichmüller theory evolved from the study of gauge theories, significantly generalizing the classical notion of Teichmüller space.
In the third section, we examine aspects of supersymmetric quantum field theories, primarily as a fruitful working assumption that enables the exploration of deep aspects of quantum field theories that might otherwise have remained inaccessible. We will particularly emphasize electric-magnetic duality, which has motivated the investigation of BPS states in four-dimensional quantum field theories with extended supersymmetry, as well as the development of Seiberg--Witten theory.
Finally, we discuss how string-theoretic techniques have broadened the class of theories amenable to Seiberg--Witten analysis and systematized methods for studying such theories. This has led, in particular, to a large class of theories known as theories of class $\mathcal S$, in which Hitchin systems play a central role. The investigation of BPS states within these theories has revealed unexpected connections with the theory of Hitchin components, important examples of higher-rank Teichmüller spaces, and has ultimately contributed to the development of spectral networks, which have rapidly gained considerable interest in the study of higher-rank Teichmüller spaces.}

\section*{Quantum field theories and gauge theories}
\addcontentsline{toc}{section}{Quantum field theories and gauge theories}

Quantum theory is a fundamental framework that describes the behavior of nature at small scales, typically at the atomic level or smaller. It was progressively developed during the early decades of the 20th century. A central aspect of quantum theory is that the states of a quantum system are represented by rays in a separable Hilbert space, a mathematical construction known as the \textit{Hilbert space of states}\index{Hilbert space} $(\mathcal{H}, \langle \cdot \vert \cdot \rangle)$. The development of quantum theory is closely linked to the evolution of Hilbert space theory, with significant joint contributions from prominent scientific figures like John von Neumann. Wigner's 1931 theorem is an important cornerstone of quantum theory; it asserts that any symmetry of a quantum system must correspond to either a unitary or antiunitary, linear or antilinear, transformation of $\mathcal{H}$.

\textit{Special relativity}\index{special!relativity} is a theory describing the relationship between space and time, founded on two central principles: the principle of relativity and the experimentally validated postulate that the speed of light in a vacuum is constant across all inertial reference frames. Introduced by Einstein in 1905, the theory was built upon the contributions of many earlier scientists. In special relativity, space and time are unified into a single framework called \emph{spacetime}\index{spacetime}, which is mathematically modeled as a four-dimensional pseudo-Riemannian differential manifold with a metric of signature $(1,3)$. For instance, flat spacetime is represented by \emph{Minkowski spacetime}\index{spacetime!Minkowski}, i.e. $\mathbb{R}^4$ equipped with the pseudo-Riemannian metric $\mathrm{Diag}(-1,1,1,1)$. The development of special relativity was heavily influenced by 19th-century advances in geometry, particularly Riemann's groundbreaking contributions.

\subsubsection*{Quantum field theories}

In non-relativistic quantum mechanics, the states of an isolated quantum system can be  described by a \textit{wave function}\index{wave function}. The squared modulus of the wave function provides a probability density, giving a probabilistic interpretation to the system's physical quantities. The evolution of a wave function is governed by the \textit{Schrödinger equation}\index{Schrödinger equation}, which treats time and space asymmetrically, making it incompatible with the framework of special relativity. Relativistic generalizations of the Schrödinger equation exist, such as the \textit{Klein--Gordon equation}\index{Klein--Gordon equation} and the \textit{Dirac equation}\index{Dirac!equation}. However, these relativistic equations lack a straightforward probabilistic interpretation of their solutions. Ultimately, this is because the number of particles is not conserved in special relativity (this is, for example, discussed in \cite{Alvarez-Gaume:2005ojg}).

This realization prompted the development of quantum field theory (QFT), providing a framework for quantum relativistic systems. In particular, all quantum field theories in Minkowski spacetime respect the symmetries of spacetime, specifically the Poincaré group\index{Poincaré group} $\mathbb{R}^4 \rtimes \mathrm{O}(3,1)$, as a fundamental subgroup of their symmetries.

The development of quantum field theory was significantly shaped by the study of the quantum interactions between light and matter, specifically through \textit{quantum electrodynamics}\index{quantum!electrodynamics} (QED) developed during the 1940s. QED is the quantum field theory extension of classical Maxwell theory, incorporating charged matter fields, whose excitations describe electrons and positrons, into the framework. It stands as the simplest example of a gauge theory, where the gauge group is $\mathrm{U}(1)$. The success of QED, especially in accurately predicting experimental results, made it a cornerstone in the broader development of quantum field theories, serving as a prototype for more complex gauge theories in particle physics.

\subsubsection*{Renormalization}

In quantum electrodynamics (QED), many calculations yield infinities, calling for regularization methods extensively developed in the late 1940s. However, it gradually became apparent that these infinities are not problems to be eliminated, but rather intrinsic features of quantum field theories. This realization led to the concept of \textit{renormalization}\index{renormalization}: in quantum field theories, correlation functions depend on the energy scale at which experiments are performed. In the Wilsonian paradigm, this induces a dependence of the parameters of the theory, such as masses, charges, and coupling constants, on the energy. For instance, the fine structure constant $\alpha$, which is dimensionless and related to the elementary electric charge $e$ by the relation:
\begin{equation*}
	4\pi\epsilon_0\hbar c\alpha(\mu) = e^2(\mu)\; ,
\end{equation*}
exhibits energy scale dependence as $\mu$ changes. It is measured to be approximately $1/137$ at low energies, but increases to about $1/127$ at the energy scale $m_Z$ of the $Z$ boson (around 90 GeV). This variation arises because each electron interacts with the scale-dependent electromagnetic field it itself creates. 

The theory of renormalization was significantly enhanced by concepts from condensed matter physics, leading in particular to the development of Wilson's \textit{renormalization group}\index{renormalization!group} in the 1960s and 70s. This framework describes how the parameters $g_1,\dots,g_k$ of a theory evolve with the energy scale $\mu$ in a manner that is local in $\mu$. Specifically, one can express this variation as:
\begin{equation*}
	\mu\frac{\mathrm{d}g_i}{\mathrm{d}\mu}(\mu_0)=:\beta_{g_i}\left(g_1(\mu_0),\dots,g_k(\mu_0)\right)\; , \quad \text{for all } i=1,\dots,k\; ,
\end{equation*}
where $\beta$ is a theory-dependent function of the parameters evaluated at the energy scale $\mu_0$. This relationship induces a \textit{flow}\index{renormalization!group!flow} within the parameter space of the theory, known as the \textit{renormalization group flow}. Within this framework, certain points in the parameter space may be fixed under this flow; the corresponding theories at these points exhibit scale invariance. A prominent example of this phenomenon occurs in the context of \textit{conformal field theories}\index{conformal!field theory}, for which the group of conformal symmetries of spacetime is a subgroup of their symmetries, implying in particular scale-invariance.

\subsubsection*{Yang--Mills theories} 

A cornerstone of the framework of quantum field theory is the introduction of Yang–Mills theories in 1954 \cite{Yang:1954}. These theories generalize Maxwell's theory; while they are both gauge theories, Yang–Mills theories can be associated with any reductive compact simple Lie group $G$ rather than just the abelian group $\mathrm{U}(1)$ used in electromagnetism. In this context, the \textit{gauge field}\index{gauge!field} serves as the generalization of the four-potential $A_\mu$ from Maxwell's theory and is represented as a connection on a $G$-principal bundle $P$ over spacetime. Alongside the gauge field, additional fields may also be included, such as sections of vector bundles associated with the principal bundle $P$. The equations of motion for the gauge field, known as the \textit{Yang–Mills equations}\index{Yang–Mills equations}, form a system of partial differential equations that extend Maxwell's equations. These theories provide a successful description of the weak and strong interactions within atomic nuclei. Furthermore, Yang–Mills equations have found widespread applications in various branches of mathematics, which will be discussed shortly.

\subsubsection*{Anomalies} 

\emph{Chiral anomalies}\index{anomaly!chiral} were first studied by Adler and Bell–Jackiw in 1969, motivated by the anomalous rate of the decay $\pi^0\rightarrow \gamma+\gamma$ of the neutral pion $\pi^0$ into two photons. In general, a classical symmetry in field theory is said to be \textit{anomalous}\index{anomaly} if it fails to persist as a quantum symmetry. Chiral anomalies are deeply connected to the \textit{Atiyah--Singer index theorem}\index{Atiyah--Singer index theorem} (1963), a relationship which notably led to Fujikawa’s analysis of chiral anomalies in 1984. Since then, the concept has broadened significantly. A related and equally fundamental notion is that of \textit{'t~Hooft anomalies}\index{'t~Hooft!anomaly}\index{anomaly!'t~Hooft} of global symmetries (1980), which are intrinsic renormalization group invariant quantities. 't~Hooft anomalies impose strict constraints on the dynamics of quantum field theories, even when standard computational tools are lacking.

\subsubsection*{Asymptotic freedom, QCD, and confinement}

The study of the weak nuclear force, particularly through the Wu experiment (1956), eventually led to the formulation of the \textit{Weinberg–Salam electroweak theory}\index{electroweak model} in 1967, unifying electromagnetism and the weak force. This theory, a specific case of Yang–Mills theory with gauge group $\mathrm{SU}(2) \times \mathrm{U}(1)$, received its full experimental confirmation with the discovery of the Higgs boson at CERN in 2012.

As for the strong nuclear force, the \textit{quark model}\index{quark} was introduced in the early 1960s, with quarks being observed experimentally in 1968. The realization that strong interactions are best described by an $\mathrm{SU}(3)$ Yang–Mills theory was formulated in 1973, leading to what is now known as \textit{Quantum Chromodynamics}\index{quantum!chromodynamics} (QCD). Both the electroweak theory and QCD form key components of the Standard Model of particle physics, which was fully developed by the late 1970s. The advancement of QCD was notably driven by the discovery of \textit{asymptotic freedom}\index{asymptotic freedom} in 1973.

QCD is asymptotically free, meaning that as the energy scale increases, the theory increasingly resembles a free quantum field theory, i.e., a theory without interactions that can be solved exactly. At energy scales much higher than $\Lambda_{\mathrm{QCD}} \sim 330$ MeV, QCD can be approximated as a free theory of quarks and gluons. The strength of the quark-gluon interactions is determined by the gauge coupling constant $\alpha_s$, which diminishes as the energy scale rises, approaching zero at extremely high energies. The scale $\Lambda_{\mathrm{QCD}}$ is known as the \textit{dynamical scale}\index{dynamical scale} of QCD and emerges due to \textit{dimensional transmutation}\index{dimensional transmutation}.

In \textit{weakly-coupled}\index{weakly-coupled} quantum field theories, that is, when interactions are nearly negligible---as in QCD at very high energies---a variety of \textit{perturbative techniques}\index{perturbation theory} can be used for analysis, such as \textit{Feynman diagrams}\index{Feynman diagrams}. These techniques express physical quantities as series expansions in the coupling constants, allowing the first few terms to be computed, providing increasingly accurate approximations as the coupling becomes weaker. The closer the theory is to being free, the more precise this perturbative approximation becomes.

In contrast, when quantum field theories are \textit{strongly-coupled}, there are very few established techniques to analyze them directly. This is particularly true for QCD at energy scales below $\Lambda_{\mathrm{QCD}}$. Unfortunately, it is within this strongly coupled regime that key phenomena occur, such as the \textit{confinement}\index{confinement} of quarks and gluons into protons and neutrons, which form the nuclei of atoms. Understanding confinement rigorously remains, as of today, one of the Clay Institute's Millennium Prize Problems.

\subsubsection*{Electric-magnetic duality} 

Maxwell's equations read:
\begin{align*}
    \nabla \cdot \mathbf{E} &= \frac{\rho}{\epsilon_0}\; , \\
    \nabla \cdot \mathbf{B} &= 0\; , \\
    \nabla \times \mathbf{E} &= -\frac{\partial\mathbf{B}}{\partial t}\; ,\\
    \nabla \times \mathbf{B} &= \mu_0\left(\mathbf{J}+\epsilon_0\frac{\partial\mathbf{E}}{\partial t}\right)\; .
\end{align*}
In the vacuum, where there are no charges or currents (i.e., $\rho = 0$ and $\mathbf{J} = \mathbf{0}$), the equations remain invariant under the transformation $(\mathbf{E}, \mathbf{B}) \rightarrow (\mathbf{B}, -\mathbf{E})$ (assuming natural units with $\mu_0\epsilon_0 = c^2 = 1$). This transformation is known as \textit{electric-magnetic duality}\index{electric-magnetic duality}. To restore this duality in the presence of sources, one would need to introduce both a magnetic charge density $\rho_m$ and a magnetic current $\mathbf{J}_m$, which would be produced by \textit{magnetic monopoles}\index{magnetic monopole}. In 1931, Dirac famously demonstrated what is now known as the \textit{Dirac quantization condition}\index{Dirac!quantization condition}, showing that the existence of magnetic monopoles would require the quantization of electric charge, a phenomenon for which there is significant experimental evidence.

Further theoretical support for magnetic monopoles was provided by 't Hooft~\cite{tHooft:1974kcl} and Polyakov~\cite{Polyakov:1974ek} in 1974, who demonstrated that magnetic monopoles arise naturally as classical solitonic solutions in certain non-abelian Yang–Mills theories. Building on their work, Julia and Zee showed in 1975 that particles carrying both electric and magnetic charge, known as \textit{dyons}\index{dyon}, also emerge as solitonic configurations in this context \cite{Julia:1975ff}. Moreover, the analyses of 't Hooft, Polyakov, and Julia–Zee extend to non-abelian Yang–Mills theories, replacing Maxwell theory. This generalization led Goddard, Nyuts, and Olive to show in 1975 that magnetic monopoles in a Yang–Mills theory with gauge group $G$ belong to representations of the so-called \textit{GNO dual gauge group}\index{GNO dual group} $G^\vee$ \cite{Goddard:1976qe}, which is also the \textit{Langlands dual}\index{Langlands!dual group} of $G$ (the Langlands program\index{Langlands!program} was initiated by Langlands' 1967 letter to Weil). This foundational work led Montonen and Olive to formulate a non-abelian version of electric-magnetic duality for (classical) Yang–Mills theories in 1977 \cite{Montonen:1977sn}.

\subsubsection*{Instantons}

\textit{Instantons}\index{instanton} in Yang–Mills theories were discovered in 1975 by Belavin, Polyakov, Schwarz, and Tyupkin \cite{Belavin:1975fg}. Formally, an instanton is a classical solution to the Euclidean Yang–Mills equations that has a finite, non-zero action. This implies that the solution is a dual or self-dual principal Yang--Mills connection. Physically, instantons capture significant non-perturbative effects in Yang–Mills theories, enabling the computation of tunneling amplitudes between different classical vacua. This, in turn, aids in understanding the structure of the quantum vacuum in these theories. Because of their non-perturbative nature, instanton effects in Yang--Mills theory are tiny when the theory is weakly-coupled, explaining why these robust predictions of the quantum field theory formalism have not yet been directly observed.


\section*{Self-duality equations and geometry}
\addcontentsline{toc}{section}{Self-duality equations and geometry}

\subsubsection*{Gauge theories in mathematics}

The question of finding solutions to the self-dual Yang--Mills equations became of profound relevance for physicists working in quantum field theory, with some solutions having been constructed already in the years following the work of Yang and Mills. The problem of solving the equations had nevertheless a very clear differential geometric meaning, and was brought by Singer to the Oxford school of geometers in 1977. A connection $A$ on a principal $\mathrm{SU}(2)$-bundle $P$ over $\mathbb{R}^4$ satisfies the self-dual Yang--Mills equations if its curvature $F(A)$ is invariant under the Hodge star operator on $\Omega^2(\mathbb{R}^4; \mathrm{ad}(P))$:
\[\star F(A) =F(A)\; .\]

The Atiyah--Singer index theorem\index{Atiyah--Singer index theorem} was providing more solutions to the equations than the ones produced already by the physicists. However, the goal was now to find \textit{all} solutions in concrete form. In pursuit of this endeavor, Atiyah and Ward \cite{Atiyah-Ward} first adapted the work of Penrose and Ward \cite{Penrose}, \cite{Ward} on the twistor geometry and kinematics for flat spacetime, bringing ideas from complex geometry into mathematical physics. A more precise understanding came with the appearance of the Atiyah--Drinfeld--Hitchin--Manin (ADHM) construction \cite{Atiyah:1978a}, which allowed for a description of all finite-energy solutions of the self-dual Yang--Mills equations over the 4-sphere $\mathbb{S}^4$. This treatment was building on the twistorial construction that was translating the problem into one about holomorphic vector bundles over $\mathbb{CP}^3$, thus relating the study of the equations with bundle theory in algebraic geometry. 

The next fundamental step was made by Atiyah and Bott in 1983 \cite{Atiyah-Bott} who studied the Yang--Mills equations over a Riemann surface of arbitrary genus. In this situation, the Yang--Mills functional has critical points of arbitrarily high Morse index, and there exists a perfect Morse theory providing information about the space of Yang--Mills minima. The Narasimhan--Seshadri approach \cite{NS} to irreducible unitary connections by stable holomorphic vector bundles seemed to fit naturally into this picture, because the bundles arising from fundamental group representations were giving the critical points. Thus, a broader picture started to emerge concerning the relationship between stable principal bundles and the \emph{existence of Hermite--Einstein connections}\index{Hermite--Einstein connection}. It was Donaldson \cite{Don} who reformulated this problem to an infinite-dimensional moment map question. Around the same time, Hitchin--Karlhede--Lindstr\"{o}m--Ro\u{c}ek \cite{Hitchin:1986ea} gave two constructions of hyperk\"{a}hler manifolds\index{hyperkähler!manifold} which were making more transparent the connection between Supersymmetry and modern Differential Geometry. 

These works were pointing towards the existence of a space of solutions with a very rich geometric interpretation. Indeed, Hitchin in 1987 \cite{Hitchin87} has finally demonstrated this clearly by constructing analytically the moduli space $\mathcal{M}$ of solutions to the $\mathrm{SU}(2)$ self-dual Yang--Mills equations over a Riemann surface modulo gauge transformations, as a complete hyperk\"{a}hler manifold. In this particular context, the equations may be written as  
\begin{align*}
    F(A)+[\Phi, \Phi^*] & =0\; , \\
    d''_A \Phi & =0\; ,
\end{align*}
for a connection $A$ on a principal bundle $P$ over the Riemann surface and a holomorphic (1,0)-form $\Phi$ on the Riemann surface valued in the complex Lie algebra bundle of $P$. 

The solutions to these equations also have an intimate relevance to the internal structure of the Riemann surface. Namely, the fixed point set of the involution on $\mathcal{M}$ induced by the map $(A, \Phi) \mapsto (A, -\Phi)$ has a component corresponding to flat bundles with Euler class equal to $2g-2$, where $g$ denotes the genus of the underlying surface. Solutions to the self-duality equations then lead to metrics on the surface of constant negative curvature -4, thus giving a natural diffeomorphism from a solution set to the \emph{Teichm\"{u}ller space} of the surface.

\subsubsection*{Classical Teichm\"{u}ller theory}

For a closed connected oriented topological surface $S$ of genus $g \geq 2$, the \emph{Teichm\"{u}ller space}\index{Teichm\"uller!space} $\mathcal{T}(S)$ of $S$ is defined as the space of all marked hyperbolic structures on $S$ modulo isotopy, that is, diffeomorphisms which can be continuously deformed to the identity. This space enjoys a canonical complex manifold structure and admits an abundance of metrics. The study of geometric features of these various structures has a long tradition starting with Riemann,  Poincar\'{e} and Klein, among others. The Uniformization Theorem\index{uniformization theorem} of Poincar\'{e} and Koebe established in 1907 also describes $\mathcal{T}(S)$ as the moduli space of complex structures on $S$ modulo diffeomorphisms isotopic to the identity.

An in-depth study of this moduli space was undertaken by Teichm\"{u}ller in the late 1930s, thus bearing his name ever since. Teichm\"{u}ller recognized the usefulness of considering not only conformal mappings, but also quasiconformal mappings for the study of this space. Quasiconformal mappings were already introduced in complex analysis by Gr\"{o}tzsch in 1928; these are homeomorphisms between plane domains which to first order take small circles to small ellipses of bounded eccentricity. Among various results, Teichm\"{u}ller introduced a metric on $\mathcal{T}(S)$, now called the \emph{Teichm\"{u}ller metric}, and discovered an intimate relation between extremal quasiconformal mappings\index{quasiconformal mapping} and holomorphic quadratic differentials, thus asserting that $\mathcal{T}(S)$ is homeomorphic to $\mathbb{R}^{6g-6}$. This fundamental approach was later further extended by Ahlfors and Bers in the 1950s and 1960s using more analytic methods, providing also rigorous proofs to Teichm\"{u}ller's results. 

It was then Thurston who introduced an even more geometric flavor to the subject, particularly through his studies of the action of the mapping class group on $\mathcal{T}(S)$. This group is the group of all diffeomorphisms of $S$ modulo isotopy and is a fundamental tool used for the study of $\mathcal{T}(S)$. The Teichm\"{u}ller space also admits global real analytic coordinates, called the \emph{Fenchel--Nielsen coordinates}\index{Fenchel--Nielsen!coordinates}. These coordinates depend on a choice of $3g-3$ distinct, non-trivial homotopy classes of simple closed curves on the surface with disjoint representatives. The coordinates are given by the  hyperbolic lengths of geodesics in each homotopy class, together with twists around these geodesics. 

Thurston, moreover, introduced a new Riemannian metric on the Teichm\"{u}ller space where the scalar product of two tangent vectors at a point represented by a hyperbolic surface is the second derivative of the length of a uniformly distributed sequence of closed geodesics on the surface. This metric uses only the hyperbolic geometry of the surface, in contrast to the existing Teichm\"{u}ller metric\index{Teichm\"uller!metric}, which is based on quasiconformal theory, and also in contrast to the Weil--Petersson metric\index{Weil--Petersson metric}, whose definition used the Petersson hermitian product, defined in the context of modular forms and used by number theorists. In fact, it was shown later by Wolpert that Thurston’s metric coincides with the Weil--Petersson metric. 

For a very thorough and pedagogical overview of this extremely rich area of modern geometry, we refer the reader to the book of Imayoshi and Taniguchi \cite{ImaTan}, as well as the book series of Hubbard \cite{Hubbard} and the seven-volume \emph{Handbook of Teichm\"{u}ller Theory} edited by Papadopoulos \cite{Papadopoulos-Handbook}. Various aspects of the Teichm\"{u}ller space will be studied in the later chapters of the present work in detail.

\subsubsection*{Hitchin components}

The Teichm\"{u}ller space of a genus $g \geq 2$ surface $S$ enjoys a more algebraic interpretation as the space of irreducible, discrete, and faithful representations of the fundamental group $\pi_1(S)$ into the group $\mathrm{PSL}_2(\mathbb{R})$, the group of orientation preserving isometries of the hyperbolic plane. This alternative, yet far-reaching scope is due to Fricke and can be seen via the so-called \emph{holonomy representation}\index{holonomy!representation}\index{representation!holonomy}. A hyperbolic structure $(M,f)$ over $S$ gives rise to a homomorphism $\rho\colon \pi_1(S) \to \mathrm{PSL}_2(\mathbb{R})=\mathrm{Isom}^{+}(\mathbb{H}^2)$, whereas representations induced by equivalent hyperbolic structures are conjugate by an element in $\mathrm{PSL}_2(\mathbb{R})$, and conversely. Thus, 
\[\mathcal{T}(S) \subset \mathcal{R}(S, \mathrm{PSL}_2(\mathbb{R})):=\mathrm{Hom}^{+}(\pi_1(S), \mathrm{PSL}_2(\mathbb{R}))/\mathrm{PSL}_2(\mathbb{R})\; ,\]
the moduli space of reductive fundamental group representations into the group $\mathrm{PSL}_2(\mathbb{R})$. The latter is called the \emph{character variety} for the group $\mathrm{PSL}_2(\mathbb{R})$. Actually, the space $\mathcal{T}(S)$ is a connected component inside the character variety $\mathcal{R}(S, \mathrm{PSL}_2(\mathbb{R}))$, as shown by Goldman in 1980. It is one of the two connected components consisting entirely of discrete and faithful representations; the other such component is $\mathcal{T}(\bar{S})$, where $\bar{S}$ denotes the surface with the opposite orientation. 

A natural question that emerged from this algebraic realization of the Teichm\"{u}ller space as a connected component of the $\mathrm{PSL}_2(\mathbb{R})$-character variety was whether one could identify connected components inside character varieties $\mathcal{R}(S,G)$ of special geometric significance on the surface $S$, for groups with higher rank than $\mathrm{SL}_2(\mathbb{R})$. The pluralistic nature of the classical Teichm\"{u}ller space as investigated throughout the 20th century made this a promising new direction of study. 

The new insights brought by Hitchin in relating the Teichm\"{u}ller space with certain solutions of the self-dual Yang--Mills equations over a Riemann surface soon gave the first examples of connected components that were realizing this particular endeavor in higher rank. Namely, for an adjoint split real semisimple Lie group $G$, there exists a unique embedding $\sigma\colon \mathrm{SL}_2(\mathbb{R} )\to G$, which is the associated Lie group homomorphism to a principal 3-dimensional subalgebra of $\mathfrak{g}$, called the \emph{Kostant principal subalgebra}\index{Kostant principal subalgebra} $\mathfrak{sl}_2(\mathbb{R})\subset \mathfrak{g}$.
Upon fixing a discrete embedding $\iota\colon {{\pi }_{1}}(S)\to \mathrm{SL}_2(\mathbb{R})$, then Hitchin \cite{Hit92} showed in 1992 using analytic tools that the connected component containing $\sigma \circ \iota \colon {{\pi }_{1}}(S)\to G$ is topologically a ball of real dimension $(2g-2) \mathrm{dim}G$. In the special case when the group is $G=\mathrm{PSL}_2(\mathbb{R})$, this component is exactly the Teichm\"{u}ller space. These connected components are nowadays called the \emph{Hitchin components}\index{Hitchin!component}. 

Yet, as was already pointed out in the original work of Hitchin \cite{Hit92}, the analytic tools used for the realization of these connected components give no indication about their geometric significance. This started to become apparent in 2006 with the works of Fock--Goncharov \cite{FG}, Guichard \cite{Guichard} and Labourie \cite{Labourie}. In particular, fundamental group representations lying inside the Hitchin components are all discrete, faithful, and Anosov. The latter is a notion introduced by Labourie and is referring to representations encoding a type of dynamical structure that is analogous to that of an \emph{Anosov flow}\index{Anosov flow}. Moreover, an important characterization was obtained for the representations in the Hitchin components which pertains to the existence of a continuous equivariant map sending positive triples in $\mathbb{RP}^1$ to positive triples in certain flag varieties associated with the Lie group $G$. This characterization proved out to be essentially what opened the way for a more holistic approach to the problem of identifying the generalizations of the classical Teichm\"{u}ller space for higher rank Lie groups.  

\subsubsection*{Higher-rank Teichm\"{u}ller spaces}

\emph{Higher-rank Teichm\"{u}ller Theory}\index{higher-rank Teichm\"uller!theory} is concerned with the study of the properties of fundamental group representations lying in distinguished subsets of the character variety $\mathcal{R}(S, G)$, for simple real groups $G$. An abundance of methods from geometry, gauge theory, algebraic geometry and dynamics is used in order to understand the features of these subsets, many methods of which having been provided by the non-abelian Hodge theory for the moduli space $\mathcal{R}(S, G)$. The term ``higher Teichm\"{u}ller space''\index{higher-rank Teichm\"uller!space} originates in the work of Fock and Goncharov \cite{FG}, who developed a more algebro-geometric approach to the notion of total positivity\index{total positivity} by Lusztig \cite{Lu} in the context of general split real semisimple reductive Lie groups,  and defined \emph{positive representations}\index{representation!positive} of the fundamental group of $S$ into these groups. Among establishing significant geometric results, Fock and Goncharov constructed all positive representations and showed that they are faithful, discrete and positive hyperbolic. Today, the term ``higher-rank Teichm\"{u}ller space'' refers more generally to a connected component of a character variety $\mathcal{R}(S, G)$, for a reductive Lie group $G$, that entirely consists of faithful representations with discrete image.

Several essential features of higher-rank  Teichm\"{u}ller spaces can be traced back to the ideas and work of Thurston, and a very active area of modern research pertains to establishing in an appropriate way the various geometric features that the classical Teichm\"{u}ller space enjoys into this higher-rank setting. For instance, Thurston’s shear coordinates have been extended in this setting by Fock and Goncharov \cite{FG}, nowadays called the \emph{Fock--Goncharov coordinates}\index{Fock--Goncharov coordinates}. Another example concerns the use of the Thermodynamic Formalism in producing mapping class group invariant Riemannian metrics on the higher-rank Teichm\"{u}ller spaces \cite{BCS}. 

Nonetheless, the characterization of fundamental group representations lying in the Hitchin components relevant to the existence of a continuous equivariant map sending positive triples in $\mathbb{RP}^1$ to positive triples in certain flag varieties associated with the Lie group $G$, provided the motivation to propose a unified approach in terms of positivity that could, in fact, distinguish all higher-rank Teichm\"{u}ller spaces. In search of such a general characterization, Guichard and Wienhard introduced in \cite{GW} a notion of positivity for surface group representations which unified previously introduced notions of positivity for particular families of Lie groups $G$. This characterization accounts to the existence of very special boundary maps sending positive triples to positive triples in the flag variety $G/P_{\Theta}$, where $P_{\Theta} <G$ is a parabolic subgroup determined by a set of simple positive roots $\Theta$. As a matter of fact, Guichard and Wienhard showed in \cite{GW} that there is a full classification of the Lie groups where such a positivity structure emerges. Moreover, subsets of \emph{$\Theta$-positive representations}\index{representation!$\Theta$-positive} in the character varieties for these groups are open and closed, thus connected components of $\mathcal{R}(S, G)$ \cite{BGLPW}. 

\emph{Spectral networks} were introduced by Gaiotto, Moore and Neitzke in 2013 \cite{Gaiotto:2012rg} as certain networks of trajectories on a Riemann surface as a means to calculate the BPS spectra for various four-dimensional $\mathcal{N}=2$ quantum field theories, as will be discussed shortly. However, these networks of trajectories can be used in order to define local coordinate systems on moduli spaces of flat $G$-connections and thus study particularly higher-rank Teichm\"{u}ller spaces. Indeed, one of the primary clear relationships between spectral networks and higher-rank Teichm\"{u}ller theory was established by Gaiotto, Moore and Neitzke in \cite{Gaiotto:2012db} who showed that the Darboux coordinates on moduli spaces of flat connections which come from special spectral networks coincide with the Fock--Goncharov coordinates. Moreover, in \cite{HoNe} Hollands and Neitzke showed that there are two special types of spectral networks, related to ideal triangulations and pants decompositions of the Riemann surface, that lead to Fock--Goncharov and complexified Fenchel--Nielsen coordinates respectively, on the moduli space of flat $\mathrm{SL}_2(\mathbb{C})$-connections over the surface. The effectiveness of spectral networks in studying moduli spaces of flat connections and related objects in geometry is rapidly increasing ever since.

\section*{Supersymmetric quantum field theories}
\addcontentsline{toc}{section}{Supersymmetric quantum field theories}

\subsubsection*{Supersymmetry}

The 1967 \textit{Coleman--Mandula theorem}\index{Coleman--Mandula theorem} \cite{Coleman:1967ad} imposes stringent constraints on the symmetries of physical quantum field theories that can be described by an \textit{$\mathcal S$-matrix}\index{$\mathcal S$-matrix}, where $\mathcal S$ stands for ``scattering''. Here, ``physical'' refers to a set of precise axioms, and the $\mathcal S$-matrix is a unitary matrix that maps the set of so-called \textit{in-states} to the set of \textit{out-states} involved in scattering processes. The theorem asserts that if the symmetries of such a quantum field theory form a Lie group, then it must be a direct product of the \textit{Poincaré group}\index{Poincaré group} of spacetime symmetries and a group of \textit{internal symmetries}\index{internal symmetries}. However, \textit{supersymmetric quantum field theories}\index{supersymmetric!quantum field theory} extend the symmetry structure without violating the Coleman–Mandula theorem because their symmetries form a \textit{super-Lie group}\index{super-Lie!group} rather than a conventional Lie group. This larger symmetry group allows SUSY theories to overcome the theorem's restrictions while providing a richer framework for unifying internal and spacetime symmetries. The most general symmetry super-groups of physical supersymmetric QFTs are constrained by the \textit{Haag--Lopuszanski--Sohnius theorem}\index{Haag--Lopuszanski--Sohnius theorem} \cite{Haag:1974qh}.

Supersymmetries are spacetime symmetries that relate \textit{bosons}\index{boson} to \textit{fermions}\index{fermion} and vice-versa. In a supersymmetric quantum field theory (QFT), the infinitesimal symmetries form a \textit{super-Lie algebra}\index{super-Lie!algebra}, a $\mathbb{Z}_2$-graded Lie algebra. In addition to the usual Poincaré generators, the 4-dimensional \textit{super-Poincaré algebra}\index{super-Poincaré!algebra} includes odd generators $Q_{\alpha}$ and $\overline{Q}_{\dot\alpha}$, where~$\alpha=1,2$ and $\dot\alpha=\dot1,\dot2$, are the \textit{supercharges}\index{supercharge}. Crucially, the supercharges do not commute with the Lorentz generators, meaning they change the spin of particles and thus map fermions to bosons and vice-versa. Furthermore, they satisfy the following anticommutation relation: 
\begin{equation*}
    \left\{Q_{\alpha},\overline{Q}_{\dot\beta}\right\} \propto \sigma^\mu_{\alpha\dot\beta} P_\mu\; ,
\end{equation*}
where $P_\mu$ are the generators of spacetime translations. In this way, the supercharges can be interpreted as ``odd square roots'' of spacetime translations. There are also extended versions of the 4-dimensional super-Poincaré algebras. In the $\mathcal N$-extended super-Poincaré algebra, there are $\mathcal N$ copies of the odd generators $Q_{\alpha}^I$ and $\overline{Q}_{\dot\alpha}^I$, where $I=1,\dots,\mathcal N$. If one is interested in genuine quantum field theories---specifically, theories that do not include elementary particles with spin greater than 1---then the number of supersymmetries, $\mathcal N$, must be less than or equal to 4.  

In 1974, Wess and Zumino observed that \textit{supersymmetry}\index{supersymmetry} (SUSY) endows quantum field theories with remarkable renormalization properties \cite{Wess:1974tw}, making them highly appealing from a phenomenological perspective. SUSY soon emerged as a promising solution to several shortcomings of the Standard Model of particle physics, particularly in addressing the \textit{hierarchy problem}\index{hierarchy problem}, the potential for \textit{grand unification}\index{grand unification} of gauge interactions, and providing candidates for the constituents of \textit{dark matter}\index{dark matter}. The supersymmetric extension of the Standard Model known as the \textit{Minimal Supersymmetric Standard Model} (MSSM), was proposed in 1981 \cite{Dimopoulos:1981zb}. The 1980s saw many developments towards the phenomenological realization of supersymmetry, notably the discovery and study of dynamical supersymmetry breaking \cite{Witten:1981nf,Affleck:1984xz}.

The exceptional renormalization properties of supersymmetric quantum field theories are encapsulated in what are known as \textit{SUSY non-renormalization theorems}\index{SUSY non-renormalization theorems} \cite{Grisaru:1979wc}. These theorems played a key role in Seiberg's development of the concept of \textit{holomorphy}\index{holomorphy} in the early 1990s \cite{Seiberg:1988ur,Seiberg:1993vc,Seiberg:1994bz,Seiberg:1994bp}. Holomorphy asserts that certain renormalized quantities in SUSY theories are analytic functions of the (bare) fields and couplings, as opposed to more general smooth functions. In this way, the distinction between supersymmetric quantum field theories and ordinary ones mirrors the difference between analytic and smooth geometry. This profound structure imposes strong constraints on quantum corrections in super-symmetric theories, making them more tractable compared to their non-supersymmetric counterparts. One of the early successful applications of this holomorphic principle was the detailed study of $\mathcal N=1$ super-QCD (SQCD) \cite{Seiberg:1994bz,Seiberg:1994bp}, which led to the discovery of \textit{Seiberg duality}\index{Seiberg duality} \cite{Seiberg:1994pq}. Seiberg duality is a significant result that notably reveals that $\mathcal N=1$ SQCD with gauge group $\mathrm{SU}(N)$ and $N+2\leq F\leq 3N$ flavors of fundamental quarks, and a modified version of $\mathcal N=1$ SQCD with gauge group $\mathrm{SU}(F-N)$ and $F$ quarks, are different microscopic descriptions of the same macroscopic physics. This duality provided deep insights into the non-perturbative behavior of supersymmetric gauge theories.

\subsubsection*{Electric-magnetic duality and BPS states}

Building on the work of Polyakov, 't Hooft, and Julia–Zee, Bogomol'nyi derived in 1975 a classical inequality for the mass of a \textit{dyon}\index{dyon} obtained as a soliton in the \textit{Georgi--Glashow model}\index{Georgi--Glashow model}, in \cite{Bogomolny:1975de}:
\begin{equation*}
    M_{q_\mathrm{el},q_\mathrm{mag}}\geq \alpha\sqrt{q_\mathrm{el}^2+q_\mathrm{mag}^2}\; ,
\end{equation*}
where $q_\mathrm{el}$ represents the dyon's electric charge, $q_\mathrm{mag}$ its magnetic charge, and $\alpha$ is a constant. Equality holds only in the so-called \textit{Bogomol'nyi–Prasad–Sommerfield (BPS) limit}\index{Bogomol'nyi–Prasad–Sommerfield limit} of the underlying non-abelian gauge theory \cite{Prasad:1975kr}, where dyons saturating the Bogomol'nyi bound are referred to as \textit{BPS states}\index{BPS!state}. All of these results are purely classical, and there is no fundamental reason why they should remain valid once quantum effects are introduced. Specifically, the mass of dyons could, in principle, be altered by quantum corrections.

In quantum field theories with extended supersymmetry, specifically for $\mathcal N\geq 2$, the super-Poincaré algebra incorporates central generators known as \textit{central charges}\index{central charge}. For $\mathcal N=2$ the central charge takes the form of a complex number. Super Yang–Mills theories with $\mathcal N=2$ supersymmetries represent particular generalizations of the Georgi–Glashow model in the Bogomol'nyi–Prasad–Sommerfield (BPS) limit, which includes fermions. In 1978, Witten and Olive demonstrated that the central charge of a dyon in such theories receives contributions from its low-energy electric and magnetic charges \cite{Witten:1978mh}. They also established that BPS states correspond to specific representations of the $\mathcal N=2$ extended super-Poincaré algebra, referred to as \textit{short representations}\index{representation!short}. More precisely, in theories with $\mathcal N=2$ supersymmetries, the mass of any supermultiplet is bounded from below by the absolute value of its central charge $\mathcal Z$:
\begin{equation*}
    M \geq \vert\mathcal Z\vert\; .
\end{equation*}
When the inequality is strict, the supermultiplet is classified as a \textit{long representation}\index{representation!long} of the supersymmetry algebra, wherein none of the supercharges $Q_{\alpha}^I$ and $\overline{Q}_{\dot\alpha}^I$ are trivially realized. Conversely, when the inequality saturates, half of the supercharges act as the zero operator on the representation, implying that the dimension of a short supermultiplet is half that of a long one. Since quantum corrections cannot account for this difference, in $\mathcal N=2$ quantum field theories, BPS states are protected from quantum corrections by representation theory.

Unlike pure Maxwell theory, which is free, the gauge coupling in quantum Yang–Mills theories renormalize, complicating the understanding and formulation of Montonen–Olive duality for non-abelian quantum Yang–Mills theories. However, in super-Yang–Mills (SYM) theories with $\mathcal N=2$ supersymmetry, the situation improves significantly due to the insights provided by Witten and Olive. In these theories, BPS states emerge as quantum states that can also be analyzed classically. While most $\mathcal N=2$ SYM theories are subject to renormalization, $\mathcal N=4$ SYM theories---special cases of $\mathcal N=2$ theories---are not. This makes $\mathcal N=4$ SYM theories an ideal framework for stating and investigating Montonen–Olive duality, as initially conjectured in 1979 \cite{Osborn:1979tq}. Subsequent developments have provided substantial evidence supporting the realization of Montonen–Olive duality, commonly referred to as \textit{S-duality}\index{$S$-duality} in this context. It is now widely accepted that S-duality is indeed realized in $\mathcal N=4$ SYM theories, establishing a remarkable connection between the realm of quantum field theories and the geometric Langlands program \cite{Kapustin:2006pk}.

\subsubsection*{Seiberg--Witten theory}

Building on the discovery of \textit{holomorphy}\index{holomorphy} in the early 1990s and the subsequent insights into the properties of $\mathcal N=1$ SQCD theories, Seiberg and Witten demonstrated in 1994 that 4-dimensional $\mathcal N=2$ supersymmetry imposes such strong constraints that it allows for an exact solution of its low-energy Coulomb dynamics. This was first addressed in pure $\mathcal N=2$ $\mathrm{SU}(2)$ SYM \cite{Seiberg:1994rs}, and in $\mathcal N=2$ $\mathrm{SU}(2)$ SQCD with $1\leq F\leq 4$ flavors \cite{Seiberg:1994aj}, and subsequently extended to more general $4d$ $\mathcal N=2$ SYM and SQCD theories with various gauge groups.

Supersymmetric quantum field theories often exhibit a continuum of inequivalent quantum vacua, known as the \textit{moduli space of supersymmetric vacua}\index{moduli space!of supersymmetric vacua}. This is particularly true for $\mathcal N=2$ SYM theories. In these theories, the moduli space contains a notable subspace referred to as the \textit{Coulomb branch}\index{branch of moduli space!Coulomb}, which consists of vacua where the low-energy physics is an $\mathcal N=2$ supersymmetric generalization of Maxwell theory. Mathematically, the Coulomb branch is a \textit{special Kähler manifold}\index{special!Kähler manifold} diffeomorphic to~$\mathbb{C}^r$, implying the existence of a set of local holomorphic functions, denoted as $a_i$ and $a_D^i$ for $i=1,\dots,r$, called \textit{special coordinates}\index{special!coordinates}. These functions encapsulate both the low-energy effective abelian gauge theory and the central charge of low-energy dyons, hence the mass of BPS states. The findings of Seiberg and Witten provide exact expressions for these special coordinates, thereby completely solving the low-energy dynamics of such theories.

Specifically, the special coordinates $a_i$ and $a_D^i$ are derived as period integrals of a meromorphic 1-form defined on an auxiliary Riemann surface $\Sigma$ known as the \textit{Seiberg--Witten curve}\index{Seiberg--Witten!curve}. This curve is dependent on the vacuum within the Coulomb branch. The Seiberg--Witten curves are organized into a holomorphic fibration over the Coulomb branch, which is entirely characterized by the singular behavior of $a_i$ and $a_D^i$, determined through physical considerations. The techniques developed by Seiberg and Witten swiftly led to the identification of remarkable new $4d$ $\mathcal N=2$ superconformal field theories (SCFTs), among which the \textit{Argyres--Douglas theories}\index{Argyres--Douglas theory} \cite{Argyres:1995jj} (1995) and the \textit{Minahan--Nemeschansky theories}\index{Minahan--Nemeschansky theory} \cite{Minahan:1996fg,Minahan:1996cj} (1996).

A significant outcome of Seiberg--Witten theory is the understanding that a notion of $S$-duality exists in $4d$ $\mathcal N=2$ SYM theories, even in cases where the gauge coupling renormalizes, though this realization is more intricate compared to that in $4d$ $\mathcal N=4$ SYM theories. Unlike in $4d$ $\mathcal N=4$ SYM theories, the spectrum of BPS states in general $4d$ $\mathcal N=2$ theories is not invariant under the monodromies caused by metric singularities on the Coulomb branch. However, a phenomenon known as \textit{wall-crossing}\index{wall-crossing!phenomenon} ensures that this non-invariance does not contradict the realization of $S$-duality in these theories. There are codimension-1 \textit{walls of marginal stability}\index{wall of marginal stability} on the Coulomb branch, across which BPS states can decay. These walls pass through the singularities on the Coulomb branch, indicating that as one encircles such singularities, a BPS state may sometimes transform into two states upon crossing the wall.

In 1995, Donagi and Witten demonstrated that the low-energy physics structure on the Coulomb branch of general $4d$ $\mathcal N=2$ theories defines a complex completely integrable system \cite{Donagi:1995cf}. This framework encompasses all previously studied theories using the methods of Seiberg and Witten, as well as new theories, namely $4d$ $\mathcal N=2^*$ theories. The complex completely integrable systems associated with these $4d$ $\mathcal N=2$ quantum field theories are special cases of those introduced by Hitchin in \cite{Hitchin87,Hitchin87b}, along with their generalizations to non-closed Riemann surfaces.

\hspace{0.3cm}

Supersymmetry has yet to be observed in particle collider experiments, casting doubt on its specific realization in models like the MSSM and leaving open the question of how the unresolved issues of the Standard Model are addressed in nature. However, even if SUSY is not realized exactly as in the MSSM, it remains a highly compelling framework for addressing many unanswered questions in high-energy physics. In the string theory community, SUSY is often considered a natural prediction of string theory, though this does not necessarily mean it should appear at energy scales currently reachable by colliders. This uncertainty raises important scientific and philosophical questions. Nevertheless, whatever the ultimate resolution may be, supersymmetry has undeniably played a pivotal role in advancing the study of the quantum field theory formalism, even if this pursuit leans more toward the mathematical realm. The results highlighted above demonstrate its profound impact on our understanding of quantum field theory.

\section*{Spectral networks}
\addcontentsline{toc}{section}{Spectral networks}

\subsubsection*{String theory and branes}

\textit{String theory}\index{string!theory} was first developed in the early 1970s under the name \textit{dual resonance model}\index{dual resonance model}, originally as a theory for describing strong nuclear interactions. An important outcome from this period is the \textit{Ramond–Neveu–Schwarz formalism}\index{Ramond--Neuveu--Schwarz formalism} for superstrings, introduced in 1971. However, as quantum chromodynamics became recognized as a more accurate description of strong interactions, the dual resonance model was set aside.

String theory was revived in 1974 when it was discovered that one of the vibrational modes of a closed string had the required properties to potentially describe \textit{gravitons}\index{graviton}, positioning string theory as a candidate for a unified \textit{theory of everything}\index{theory of everything}. In its perturbative framework, string theory describes strings as elementary objects, rather than point-like particles. At scales larger than the typical string length, strings effectively appear to be point-like particles, with their mass and spin determined by the specific vibrational modes of the string.

Significant milestones in the development of string theory in the late 1970s and 1980s include the discovery of the \textit{Gliozzi–Scherck–Olive projection}\index{Gliozzi--Scherck--Olive projection} in 1977, the cancellation of anomalies via the \textit{Green–Schwarz mechanism}\index{Green--Schwarz mechanism} in 1984, and the formulation of \textit{heterotic string theory}\index{heterotic string theory} by Gross, Harvey, Martinec, and Rohm in 1985, which brought to five the number of known superstring theories in ten dimensions: type I, types IIA and IIB, and types HO and HE. Also in 1985, the importance of \textit{Calabi–Yau manifolds}\index{Calabi--Yau manifold} for compactifying extra dimensions was recognized (Candelas, Horowitz, Strominger, and Witten, 1985). This proved instrumental to the discovery of \textit{mirror symmetry}\index{mirror symmetry} at the turn of the 1980s and 1990s (Dixon, 1988; Lerche, Vafa, and Warner, 1989; Candelas, de la Ossa, and Parks, 1991). The study of mirror symmetry greatly benefited from the introduction of \textit{topological string theory}\index{topological!string theory}\index{string!theory!topological} by Witten in 1990, which notably led to Kontsevich's \textit{homological mirror symmetry}\index{homological mirror symmetry} conjecture (1994).

The low-energy limits of all 10-dimensional superstring theories are the so-called 10-dimensional \textit{supergravity theories}\index{supergravity theory}, also referred to as type I, IIA, IIB, HO, and HE. In the early 1990s, it became increasingly clear that all five versions of supergravity and superstring theories are related by dualities, particularly \textit{$S$-duality}\index{$S$-duality} and \textit{$T$-duality}\index{$T$-duality}, suggesting that they should be viewed as different descriptions of the same underlying theory, rather than distinct theories. In 1995, Witten proposed \textit{M-theory}\index{M-theory} as the strong-coupling limit of type IIA superstrings. Unlike the 10-dimensional string theories, M-theory is an eleven-dimensional theory of membranes. Its low-energy limit is 11-dimensional supergravity, which is in some sense universal among supergravity theories.

Another breakthrough occurred in 1995 when Polchinski demonstrated that superstring theories require the inclusion of additional extended objects beyond strings: \textit{D-branes}\index{brane!D-} \cite{Polchinski:1995mt}. For example, type IIA superstrings admit D0-, D2-, D4-, D6-, and D8-branes, which correspond to objects that are respectively 1-, 3-, 5-, 7-, and 9-dimensional in spacetime. D-branes are loci where open strings can end. At low energies, the modes of open strings ending on a D-brane induce a \textit{super-Yang--Mills theory}\index{super-Yang--Mills theory} on it, preserving half of the total supercharges in type IIA superstrings—specifically, 16 supercharges. In other words, supersymmetric QFTs are ubiquitous in superstring theories (and M-theory). Besides D-branes and fundamental strings (often referred to as NS1-branes), type IIA superstrings also admit so-called \textit{NS5-branes}\index{brane!NS5-}.

Shortly after Polchinski's breakthrough, it was realized that the ability of fundamental strings to end on D-branes implies, through sequences of dualities, that branes can often end on other branes as well \cite{Strominger:1995ac,Townsend:1995af}. This opened the door to studying the dynamics of \textit{brane configurations}\index{brane!configuration}, which host supersymmetric QFTs on their \textit{worldvolumes}\index{worldvolume}---the volumes they sweep out in spacetime. Much of the non-trivial dynamics of supersymmetric Yang–Mills theories can be efficiently explored through such brane configurations, as demonstrated, for instance, by the pioneering work of Hanany and Witten \cite{Hanany:1996ie}.

D-branes have also played a crucial role in the discovery of \textit{holographic dualities}\index{holographic duality}, under which $d$-dimensional quantum field theories are equivalent to $(d+1)$-dimensional quantum theories of gravity (or more broadly, generally covariant theories). The most prominent example of a holographic duality is the AdS/CFT correspondence, discovered by Maldacena in 1997 \cite{Maldacena:1997re}, which relates 4-dimensional $\mathcal N=4$ $\mathrm{SU}(N)$ SYM theory on $S^4$ to type IIB superstring theory on $\mathrm{AdS}_5\times S^5$ with $N$ units of Ramond--Ramond flux through the 5-sphere $S^5$. An early review covering many aspects of AdS/CFT correspondences can be found in \cite{Aharony:1999ti}. Since then, holographic dualities have been extended significantly to include less supersymmetric and non-conformal quantum field theories. At a fundamental level, holographic dualities provide non-perturbative definitions of string theory on specific backgrounds.

Mirroring the last paragraph of the previous section, the precise way string theory must be implemented to describe the physics of our universe---if it can do so at all---remains elusive to this day, raising significant scientific and philosophical questions. Nonetheless, substantial evidence suggests that string theory is a consistent quantum theory of gravity, justifying its study as a valuable tool for addressing deep questions in theoretical high-energy physics, particularly the nature of quantum gravity. Furthermore---and this is the primary aspect of string theory that will concern us in this monograph---string theory has proven to be a powerful tool for studying supersymmetric quantum field theories, enhancing our understanding of the general quantum field theory framework and illuminating the rich interplay between fundamental mathematics and physics.

\subsubsection*{Seiberg--Witten curves from M-theory}

A large class of 4-dimensional $\mathcal N=2$ super-Yang–Mills (SYM) theories can be studied through configurations of D4-branes stretched between NS5-branes in type IIA superstring theory, where the SYM theory of interest arises as the worldvolume theory on the D4-branes. A significant challenge in studying the low-energy physics of these SYM theories via brane setups is that D4-branes end on NS5-branes in the configurations being considered, and the nature of the D4-brane endpoints on NS5-branes has proven difficult to fully understand. In 1997, Witten showed that this issue disappears when these brane configurations are \textit{lifted}\index{M-theory!uplift} to M-theory \cite{Witten:1997sc}. By simultaneously increasing the type IIA string coupling and rescaling certain directions in spacetime, one can reach the strong coupling regime, where M-theory offers a more appropriate description without altering the $4d$ $\mathcal N=2$ theory under study. In M-theory, the system generically corresponds to a single M5-brane wrapping a smooth Riemann surface $\Sigma$, which is the Seiberg–Witten curve of the $4d$ $\mathcal N=2$ theory. Thus, this brane-based approach provides a unified framework to construct Seiberg–Witten curves for an infinite family of $4d$ $\mathcal N=2$ SYM theories. These can be described by so-called $\mathcal N=2$ \textit{linear quivers}\index{quiver!$\mathcal N=2$!linear} of special unitary gauge groups, and flavor groups. The algebraic completely integrable system that encodes the low-energy physics of any such $4d$ $\mathcal N=2$ SYM theory is a \textit{Hitchin integrable system}\index{Hitchin!system} on a punctured sphere or torus. 

\subsubsection*{Class $\mathcal S$ theories}

One can analyze the $S$-duality properties of general $\mathcal{N}=2$ conformal linear quivers involving special unitary gauge and flavor groups by assuming that all but one of the gauge groups are very weakly coupled, and then performing the dualization on one gauge factor at a time. This approach naturally leads to the study of $4d$ $\mathcal{N}=2$ theories described by \textit{generalized quivers}\index{quiver!$\mathcal N=2$!generalized} \cite{Gaiotto:2009we}, which not only cover previously studied cases \cite{Witten:1997sc}, but also include many more general ones. These generalized quivers correspond to specific descriptions of \textit{$4d$ $\mathcal{N}=2$ theories in class $\mathcal{S}$}. A class $\mathcal{S}$ theory is specified by the data of a punctured Riemann surface $C$, a compact simple Lie algebra $\mathfrak{g}$ of type $A$, $D$, or $E$, and certain constraints $D$ at the punctures of $C$ (where $D$ stands for "defects"). Theories in class $\mathcal{S}$ are typically denoted by $\mathcal{S}[C, \mathfrak{g}, D]$.

There exist remarkable $6d$ superconformal field theories (SCFTs) with $\mathcal{N}=(2,0)$ supersymmetry, which are classified by compact simple Lie algebras $\mathfrak{g}$ of type $A$, $D$, or $E$. These theories admit a specific set of codimension-2 \textit{defects}\index{defect}. Class $\mathcal{S}$ theories can be uniformly described in terms of these $6d$ SCFTs: the theory $\mathcal{S}[C,\mathfrak{g},D]$ corresponds to the $4d$ $\mathcal{N}=2$ theory obtained by compactifying the $6d$ SCFT of type $\mathfrak{g}$ on the surface $C$ (with a partial topological twist), with codimension-2 defects inserted at the punctures of $C$. The moduli space of $6d$ $\mathcal{N}=(2,0)$ theories admits a \textit{Coulomb branch}\index{branch of moduli space!Coulomb}, where the theory reduces to an abelian $6d$ $\mathcal{N}=(2,0)$ theory at low energies. This Coulomb branch, in turn, allows the analysis of the Coulomb branch of the $4d$ $\mathcal{N}=2$ theories of class $\mathcal{S}$ that are derived from it \cite{Gaiotto:2009we, Gaiotto:2009hg}.

For $\mathfrak{g} = \mathrm{su}(N)$, the $6d$ $\mathcal{N}=(2,0)$ theory describes the worldvolume theory on a stack of $N$ M5-branes. The Coulomb branch corresponds to separating the M5-branes in directions transverse to the original stack. Therefore, the theory $\mathcal{S}[C, \mathfrak{su}(N), D]$ is defined by wrapping a stack of $N$ M5-branes on $C$, with boundary conditions determined by the defects $D$. The Coulomb branch of this $4d$ theory corresponds to normalizable deformations of the stack of M5-branes in the holomorphic cotangent bundle of $C$. These deformations typically define a holomorphic ramified covering of $C$ of degree $N$, which is identified with the Seiberg–Witten curve of $\mathcal{S}[C, \mathfrak{su}(N), D]$ at a particular point on its Coulomb branch.

The integrable systems governing the low-energy physics on the Coulomb branch of class $\mathcal{S}$ theories are Hitchin systems. Unlike in four-dimensions where the Coulomb branch corresponds only to the base of the Hitchin fibration, when class $\mathcal{S}$ theories are compactified on a circle $S^1$, the entire Hitchin system---which is hyperkähler---emerges as the Coulomb branch of the compactified theory \cite{Seiberg:1996nz}. To derive this from $6d$, one can first compactify the $6d$ $\mathcal{N}=(2,0)$ theory on $S^1$, which gives a $5d$ SYM theory. Further compactification on a Riemann surface $C$ (with a partial topological twist) leads to the class $\mathcal{S}$ theory on $\mathbb{R}^3\times S^1$. The BPS equations\index{BPS!equations} for the $5d$ theory compactified on $C$ are precisely Hitchin's equations \cite{Gaiotto:2009hg}.

\subsubsection*{Hyperkähler metrics and wall-crossing phenomenon}

In general, $4d$ $\mathcal{N}=2$ theories compactified on a circle $S^1$ of radius $R$ have a hyperkähler Coulomb branch with a smooth hyperkähler metric $g_R$. As $R \to \infty$, this metric $g_R$ approaches a specific hyperkähler "semi-flat" metric $g^\mathrm{sf}$. While $g^\mathrm{sf}$ can be explicitly written down, it exhibits singularities along codimension-2 manifolds. Quantum corrections arise at finite $R$ and smooth out these singularities, yielding the non-singular metric $g_R$. These quantum corrections depend solely on the BPS states of the $4d$ $\mathcal{N}=2$ theory. Crucially, the requirement that $g_R$ remains smooth imposes constraints on the wall-crossing behavior of BPS spectra \cite{Gaiotto:2010okc}: they must satisfy the \textit{Kontsevich--Soibelman wall-crossing formula}\index{Kontsevich--Soibelman wall-crossing formula}\index{wall-crossing!formula}, derived in \cite{Kontsevich-Soibelman} in a different context. The (twistorial) construction of the hyperkähler metric $g_R$ involves a crucial set of so-called \textit{Darboux coordinates}\index{Darboux coordinates} $\mathcal X^\mathrm{RH}_\gamma$.

In the specific case of $4d$ $\mathcal{N}=2$ theories of class $\mathcal{S}$, the $3d$ Coulomb branch corresponds to a Hitchin system, i.e., the moduli space of flat connections on a Riemann surface $C$. In this context, the Darboux coordinates $\mathcal{X}^\mathrm{RH}_\gamma$ can be derived from the coordinates on moduli spaces of flat connections introduced by Fock and Goncharov \cite{FG}. The seminal work of Gaiotto, Moore, and Neitzke \cite{Gaiotto:2009hg} marks a major step in reconnecting higher-rank Teichmüller theory---specifically the Hitchin components---and the study of $4d$ $\mathcal{N}=2$ theories of class $\mathcal{S}$.

\subsubsection*{WKB networks}

In theories of class $\mathcal{S}$ of type $A$, which arise from compactifying $N$ M5-branes on a Riemann surface $C$, BPS states at a generic point on the Coulomb branch, corresponding to a ramified covering $\Sigma \rightarrow C$ of degree $N$, are represented by M2-branes stretching between the sheets of the covering. The projection of such an M2-brane onto $C$ forms a curve. Locally, the sheets of the covering can be labeled by $1, \dots, N$, allowing one to label the curves by $\gamma^{ij}$. For each $i = 1, \dots, N$, let $\lambda_i$ denote the restriction of the Seiberg--Witten differential $\lambda$ to the $i$-th sheet. The central charge $Z_{\gamma^{ij}}$ and mass $M_{\gamma^{ij}}$ associated with $\gamma^{ij}$ are given by:
\begin{equation*}
	Z_{\gamma^{ij}} = \frac{1}{\pi}\int_{\gamma^{ij}} \lambda_i -\lambda_j\; , \quad \text{and}    \quad M_{\gamma^{ij}} = \frac{1}{\pi}\int_{\gamma^{ij}} \vert\lambda_i -\lambda_j\vert\; .
\end{equation*}

The BPS inequality $M_{\gamma^{ij}}\geq \vert Z_{\gamma^{ij}}\vert$ automatically holds. The inequality is saturated if and only if the argument of $\lambda_i - \lambda_j$ remains constant along $\gamma^{ij}$. Therefore, BPS states in the $4d$ theory are described by curves on $C$ that satisfy:
\begin{equation*}
	\lambda_{ij}\cdot\partial_s\in \E^{\I\vartheta}\mathbb{R}_+\; ,
\end{equation*}
where $s$ is a coordinate on $\gamma^{ij}$, and $\vartheta \in [0, 2\pi)$. For any value of $\vartheta$, this differential equation defines a singular foliation of $C$, for $N=2$. When $N \geq 2$, there can be junctions of $ij$-curves on $C$. This method of representing BPS states as curves on the surface $C$ was first investigated in the context of geometrically engineered $4d$ $\mathcal{N}=2$ theories \cite{Klemm:1996bj,Shapere:1999xr}.

The $4d$ BPS states associated with a web of curves on $C$ that satisfy this differential equation have finite mass if and only if the total length of the web is finite. This naturally leads to considering the specific webs on $C$ that originate from the branch points of the covering $\Sigma \rightarrow C$, where three integral curves meet. These webs define the \textit{WKB network}\index{WKB!spectral network}\index{spectral network!WKB} $\mathcal W_\vartheta$ on $C$ for a given angle $\vartheta$ \cite{Gaiotto:2009hg} (so named due to the \textit{WKB approximation}\index{WKB!approximation}). For generic values of $\vartheta$, no finite-length webs exist. However, as $\vartheta$ varies, the WKB network evolves and undergoes sudden changes at particular values of $\vartheta$, where finite webs, or equivalently, genuine BPS states, emerge. By continuously varying $\vartheta$ from $0$ to $\pi$, one can describe all the BPS states of the theory at a given point on its Coulomb branch.  Examples of WKB networks for pure $4d$ $\mathcal{N}=2$ $\mathrm{SU}(2)$ super-Yang--Mills (where $C = \mathbb{CP}^1$) at the origin of the Coulomb branch, for different values of $\vartheta$, are illustrated in \Cref{Fig:SNpureSU2}. At $\vartheta = 0$ and $\vartheta = \pi/2$, finite BPS curves emerge on $C$, corresponding to the two BPS states at the origin of the Coulomb branch for this theory—namely, the monopole and dyon with charges $(2,1)$.

\begin{figure}[!ht]
	\centering
	\includegraphics[width=\textwidth]{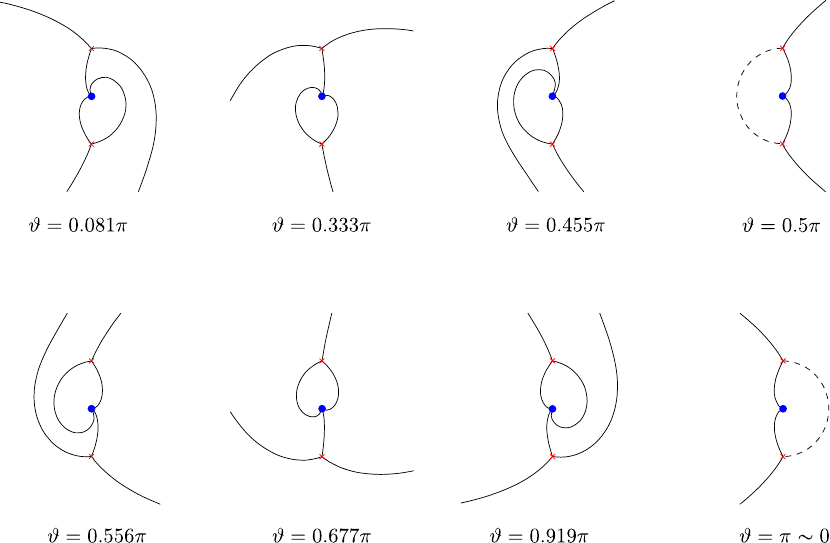}
	\caption{BPS curves for certain values of $\vartheta$ at the origin of the Coulomb branch $\mathcal{C}$. The orange crosses represent the branch points of the ramified covering $\Sigma \rightarrow C$, while the blue dot indicates a puncture on $C=\mathbb{CP}^1$, depicted here as a plane using stereographic projection (there is another puncture at infinity).}\labelx{Fig:SNpureSU2}
\end{figure}

\subsubsection*{Spectral networks} 

Spectral networks, which are the central focus of this monograph, were formally introduced in \cite{Gaiotto:2012rg}. They build upon the concept of \textit{framed BPS states}\index{BPS!framed state}, introduced in the context of four-dimensional theories in \cite{Gaiotto:2010be}, and later extended to coupled $2d$-$4d$ systems in \cite{Gaiotto:2011tf}.

Theories of class $\mathcal{S}$ admit canonical half-BPS surface defects $S_z$ parameterized by points $z \in C$ \cite{Gaiotto:2012,Gaiotto:2011tf,Alday:2009fs}. When $z \in C$ is a generic point of the covering $\Sigma \rightarrow C$ corresponding to a fixed point on the Coulomb branch of the theory, the defect $S_z$ admits $N$ massive vacua, which correspond to the $N$ preimages of $z$ in $\Sigma$. In the presence of a surface defect $S_z$, the theory supports BPS particles known as \textit{solitons}\index{soliton}, which are bound to the defect and interpolate between different vacua of $S_z$. Consequently, the charge $a$ of a soliton is (roughly) a homology class of paths between two distinct lifts of $z$ in $\Sigma$, relative to its boundaries. A soliton with charge $a$ is associated with a central charge denoted $Z_a(z)$. Geometrically, solitons are represented as finite webs of strings on $C$, similar to those discussed previously, except that one of the strings must terminate at $z$. The degeneracies of solitons are captured by a BPS index $\mu(a)$. BPS solitons from WKB networks are studied in detail in the lecture notes \cite{BHM} and \cite{Neitzke_BPS-lec}.

The \textit{spectral network}\index{spectral network} on $C$ for a parameter $\vartheta\in\mathbb{R}$ is defined as the set of points $z\in C$ such that there exists a homology class $a$ between two distinct lifts of $z$ in $\Sigma$ (relative to its boundaries), for which $Z_a(z)/\E^{\I\vartheta}\in\mathbb{R}_-$ and $\mu(a)\neq 0$.

Given two points $z_1, z_2 \in C$, along with a path $\wp$ from $z_1$ to $z_2$ and a real parameter $\vartheta$, one can define a supersymmetric interface $L_{\wp,\vartheta}$ between the surface defects $S_{z_1}$ and $S_{z_2}$, preserving two supercharges. The $2d$-$4d$ framed BPS states correspond to the different vacua of the interface $L_{\wp,\vartheta}$ for generic values of $\vartheta$, which interpolate between a lift of $z_1$ and a lift of $z_2$ in $\Sigma$. As with solitons, the degeneracies of these $2d$-$4d$ framed BPS states are encoded in an index denoted $\overline{\underline{\Omega}}'(L_{\wp,\vartheta}, a)$, where $a$ is (roughly) a homology class of paths from a lift of $z_1$ to a lift of $z_2$. Framed $2d$-$4d$ indices are gathered in a generating function denoted $F(\wp,\vartheta)$, which is a linear combination of all homology classes of paths from lifts of $z_1$ to lifts of $z_2$, weighted by the framed $2d$-$4d$ indices. For any paths $\wp_1$ from $z_1$ to $z_2$ and $\wp_2$ from $z_2$ to $z_3$, the generating functions $F(\wp_1,\vartheta)$ and $F(\wp_2,\vartheta)$ can be composed, with the composition descending from the concatenation product in the path algebra on $\Sigma$. The function $F(\wp,\vartheta)$ depends in a crucial way on how the path $\wp$ intersects the spectral network $W_\vartheta\subset C$. 

Physically, the supersymmetric interface $L_{\wp,\vartheta}$ is expected to depend only on the homotopy class of the path $\wp$. This homotopy-invariance property imposes strict constraints on both $F(\wp,\vartheta)$ and the soliton degeneracies $\mu(a)$. Furthermore, while $F(\wp,\vartheta)$ remains locally constant for generic values of $\vartheta$, there are specific values of $\vartheta$ where $F(\wp,\vartheta)$ exhibits jumps. These jumps occur precisely at the values of $\vartheta$ where the $4d$ theory has BPS states whose central charge has an argument of $e^{i\vartheta}$. Additionally, the degeneracies of the $4d$ BPS states can be inferred from analyzing the behavior of the functions $F(\wp,\vartheta)$. Therefore, spectral networks provide a framework to compute the $4d$ BPS spectra of theories of class $\mathcal{S}$ for cases where the Lie algebra $\mathfrak{g}$ has rank greater than $1$, greatly extending the results obtained using WKB networks introduced in \cite{Gaiotto:2009hg}.

The fact that the generating functions $F(\wp,\vartheta)$ associate a combination of paths on $\Sigma$ with a path $\wp$ on $C$ in a homotopy-invariant manner is a crucial feature. This property allows these functions to be utilized as tools for constructing non-abelian flat connections on $C$ from abelian flat connections on $\Sigma$. This process is referred to as \textit{non-abelianization}\index{non-abelianization}, while the reverse process is called \textit{abelianization}\index{abelianization} \cite{HoNe}. This property serves as a guiding principle in \cite{Gaiotto:2012rg} for defining a topological notion of a spectral network, which can be viewed as graphs on $C$ associated with a not-necessarily holomorphic ramified covering $\Sigma \rightarrow C$. It is also the primary reason behind the increasing interest in spectral networks as a tool for studying higher-rank Teichmüller spaces. 
While spectral networks were originally introduced in \cite{Gaiotto:2012rg} as a way to generalize Fock--Goncharov coordinates (of type $\mathcal{A}$, and also of type $\mathcal{X}$ with a slight modification), these have now proven to encompass many more subjects of geometric and physical relevance.

We close this brief overview introducing spectral networks by mentioning that a very effective tool for drawing spectral networks is the program ``Loom'' developed by P. Longhi, C. Park, and A. Neitzke, which is freely available online on GitHub \cite{LonghiLOOM2016}. The front page picture of the book as well as \Cref{Fig:SNpureSU2,fig:KwallsSU2} were obtained using Loom.

\mainmatter

\newpage
\part{Mathematical Foundation}\labelx{partII}

\vspace{2cm}

The objective of this first part is to assemble the mathematical framework required for the study of spectral networks from a geometric perspective. We begin with foundational notions in algebraic topology, including homotopy equivalence and the fundamental group. We then turn to the theory of covering spaces, bundles, and connections, as well as various forms of the Riemann--Hilbert correspondence, which relates flat bundles on a manifold to representations of its fundamental group. This groundwork will serve as a basis for the analysis of character varieties and higher-rank Teichm\"uller spaces in \Cref{partIII}. In addition, we introduce essential concepts from hyperbolic geometry and provide an overview of cluster varieties, which will play a complementary role in the subsequent developments.

\vspace{2cm}

\parttoc


\chapter{The Fundamental Group}\labelx{chap:fundamental_gp}



\abstract{In Algebraic Topology one is interested in studying properties of topological spaces using algebraic methods. Two topological spaces are considered to be ``the same'' if they are homeomorphic, and appropriate topological invariants may be used in order to distinguish spaces that are non-homeomorphic. The main such invariants are the \emph{homology groups} and the \emph{homotopy groups}. Homology is a tool that counts holes and boundary components of a topological manifold; it appears in a very central way in many branches of mathematics, but also in physics, for example in supersymmetric string theory compactifications as pioneered in \cite{CHSW85}. Homotopy, on the other hand, is also categorizing topological spaces, but is considering \emph{paths} on these spaces rather than their boundaries. The simplest among the homotopy groups is the first homotopy group, also called the \emph{fundamental group} of a topological space. This first chapter is a basic, yet detailed introduction to some standard topological notions that will be important for the understanding of the construction of spectral networks.
After some basic notions, we introduce the fundamental group and examine some of its basic properties, as well as a number of applications in distinguishing basic topological spaces up to homeomorphism. Wonderful introductory resources for further information on the topics outlined in this chapter is the textbook by Hatcher \cite{Hatcher} (which is freely available online), as well as \cite{Dieck}, \cite{Massey} and \cite{Munkres}.}

\section{Quotient topology}

Topology is the study of shapes under continuous maps with continuous inverse. One important operation changing the topology is the gluing operation. This is made precise with the notion of \emph{quotient topology}\index{quotient topology}.

Consider a topological space $X$. To define a quotient, we have to gather the data of the points to be glued together. This is described by an equivalence relation $\sim$. Recall that an \emph{equivalence relation}\index{equivalence relation} $\sim$ is a binary relation which is reflexive, symmetric and transitive. We denote by $[x]$ the equivalence class of $x\in X$.

\begin{definition}
The \emph{quotient space}\index{space!quotient} of $X$ modulo $\sim$ is the set of all equivalence classes of points in $X$, that is, 
\[{X}/{\sim }\;=\left\{ \left[ x \right] \mid x\in X \right\}\; .\]
\end{definition}

The \emph{projection map}\index{projection map}, $pr\colon X\to {X}/{\sim }\;$ with $x\mapsto \left[ x \right]$, can be used to define a topology on the quotient space ${X}/{\sim }\;$, namely, a subset $U\subseteq {X}/{\sim }\;$ is open if and only if its inverse image, ${{pr}^{-1}}(U)$, is an open subset of the topological space $X$. 
Note that this topology is the biggest topology, meaning that it includes the most open subsets with respect to which the map $pr$ is continuous.

\begin{example}
Let $X=[0,1]$ and $\sim$ an equivalence relation on $X$, such that $0 \sim 1$ and $\left[ x \right]=\left\{ x \right\}$, for every $x\in X\setminus \left\{ 0,1 \right\}$. Then the quotient space ${X}/{\sim }\;$ is homeomorphic to the unit circle ${{S}^{1}}$. Indeed, a homeomorphism is given by the map $f\colon {X}/{\sim }\;\,\,\to \,\,{{S}^{1}}$ defined by 
$x\mapsto \left( \mathrm{cos}(2\pi x),\mathrm{sin}(2\pi x) \right)$; this map is well-defined since $f(0)= f(1)$, injective, surjective and its inverse is continuous. 
\end{example}

Another example makes precise the idea of obtaining a torus from gluing opposite sides of a square.
\begin{example}
Let $X=[0,1]^2$ be the unit square and $\sim$ be the equivalence relation on $X$
defined by $(0,x)\sim (1,x)$ and $(x,0)\sim (x,1)$, for all $x\in[0,1]$. Note that the point $(0,0)$ gets identified with $(1,0), (0,1)$ and $(1,1)$.
Then the quotient space ${X}/{\sim }\;$ is homeomorphic to a torus.
\end{example}

\section{Path homotopy}

The idea of homotopy theory is to understand maps between spaces up to deformations. One of the simplest examples is to understand the space of paths in a topological space $X$ up to deformations.

A \emph{path}\index{path} on a topological space $X$ is a map $p\colon [0,1] \to X$. We call the point $p(0)$ the \emph{initial point} and respectively $p(1)$ the \emph{terminal point} of the path.
A topological space $X$ is called \emph{path-connected}\index{space!path-connected} if for every pair of points $x,y \in X$ there is a path connecting them. In the sequel, unless stated otherwise, we restrict attention to path-connected spaces.

\begin{example}
Note that not all connected spaces are also path-connected. For instance, the union of the graph of the function $f(x)=\mathrm{sin}(\frac{1}{x})$, for $x\in \left( 0,\infty  \right)$, with the interval $\left[ -1,1 \right]$ on the $y$-axis is connected but not a path-connected space (see \Cref{graph_sin}).
\end{example}

\begin{figure}[!ht]
\centering
\begin{tikzpicture}[x=5cm]
    \draw[xstep=.2,ystep=.5,lightgray,ultra thin] (-0.1,-1.5) grid (1.1,1.5);
    \draw[->] (0,0) -- (1.1,0) node[right] {$x$};
    \draw[->] (0,-1) -- (0,1.1) node[above] {$\sin (1/x)$};
  \draw[black,domain=0.01:1,samples=5000] plot (\x, {sin((1/\x)r)});
\end{tikzpicture}
 \caption{The graph of the function $f(x)=\mathrm{sin}(\frac{1}{x})$, for $x\in \left( 0,\infty  \right)$.}
 \labelx{graph_sin}
\end{figure}

Now we introduce the fundamental concept of a homotopy between two maps. Roughly speaking, a homotopy deforms the first map continuously into the second. In other words, it includes both maps as extremal points of a 1-parameter family of maps.

\begin{definition}
Let $X,Y$ be topological spaces and $f_0, f_1$ be continuous maps from $X$ to $Y$. The maps $f_0, f_1$ will be called \emph{homotopic}\index{homotopic!map} if there exists a continuous map $F\colon X\times [0,1]\to Y$, such that for every $x\in X$ it holds 
\begin{align*}
F\left( x,0 \right) & = {{f}_{0}}\left( x \right), \text{ and }\\
F\left( x,1 \right)& = {{f}_{1}}\left( x \right)\; .
\end{align*}
We say that the map $F$ is a \emph{homotopy}\index{homotopy} that connects the maps $f_0, f_1$ and we write ${{f}_{0}}\simeq {{f}_{1}}$; notice that the binary relation $\simeq $ defined above is an equivalence relation. 
 \end{definition}

The notion of homotopic maps leads to the notion of homotopic spaces. 

\begin{definition}
Let $X,Y$ be topological spaces. A map $f\colon X \to Y$ will be called a \emph{homotopy equivalence}\index{homotopy!equivalence} if there exists a map $g\colon Y\to X$ such that $f \circ g \simeq \mathrm{id}_Y$ and $g \circ f \simeq \mathrm{id}_X$, where $\mathrm{id}_X$, $\mathrm{id}_Y$ are the identity maps on $X$ and $Y$ respectively.\\
Whenever there exists a homotopy equivalence $f\colon X \to Y$, we then say that the spaces $X$ and $Y$ are \emph{homotopicaly equivalent}\index{homotopic!spaces}; alternatively, we say that $X$ and $Y$ have the \emph{same homotopy type} or are \emph{homotopic}. 
\end{definition}

The binary relation of homotopy equivalence is an equivalence relation. It is weaker than being homeomorphic. For instance, the annulus and the circle have the same homotopy type without being homeomorphic. On the contrary, two homeomorphic spaces are always homotopic (using a homeomorphism and its inverse).

As a nice exercise, we invite the reader to classify the letters of the alphabet by homotopy type. There are two versions of the exercise: first consider thick letters, second consider thin letters, i.e. which are 1-dimensional\footnote{Here is the answer: thick letters are classified by the number of holes. For thin letters we have the groups (C,G,I,J,L,M,N,S,U,V,W,Z), (D,O), (E,F,T,Y), (K,X) and (Q,R). All other letters are alone in their class.}.

\begin{definition}
A topological space $X$ is called \emph{contractible}\index{space!contractible} if it is homotopic to a point.
\end{definition}

We will be particularly interested in the case when paths on a topological space $X$ are connected by a continuous map $F$ as above. More precisely, we introduce:
\begin{definition}
Let $X$ be a topological space and $I:=[0,1]$. A \emph{path homotopy}\index{homotopy!path} on $X$ is a family of maps $f_t\colon  I \to X$, for $t \in [0,1]$, such that
\begin{enumerate}
    \item The initial and terminal points, ${{f}_{t}}\left( 0 \right)={{x}_{0}}$ and ${{f}_{t}}\left( 1 \right)={{x}_{1}}$ respectively, are independent of $t$. 
    \item The map $F\colon I\times I\to X$ where $F\left( s,t \right)={{f}_{t}}\left( s \right)$ is continuous. 
\end{enumerate}
Whenever the paths ${{f}_{0}}$, ${{f}_{1}}$ are connected via a homotopy as introduced above, we say that they are \emph{homotopic}\index{homotopic!paths} and write ${{f}_{0}}\simeq {{f}_{1}}$. 
\end{definition}

\begin{figure}[!ht]
	\centering
	\includegraphics{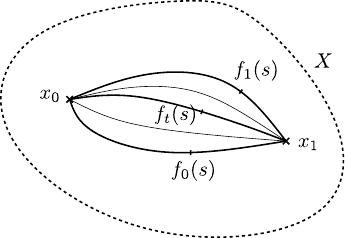}
	\caption{A path homotopy in $X$.}
\end{figure}

\begin{proposition}
The homotopy of paths with fixed endpoints as defined above is an equivalence relation.
\end{proposition}
 \begin{proof}
\begin{enumerate}
\item The binary relation $\simeq $ is obviously reflexive considering ${{f}_{t}}=f$, for every $t\in \left[ 0,1 \right]$.
\item Let ${{f}_{0}}\simeq {{f}_{1}}$ under the homotopy ${{f}_{t}}$; then, ${{f}_{1}}\simeq {{f}_{0}}$ under the homotopy ${{f}_{1-t}}$, thus $\simeq$ is symmetric.
\item Let $f\simeq g$ under the homotopy defined by a map $F$ and let $g\simeq h$ under the homotopy defined by a map $G$; then $f\simeq h$ via the map 
	\[H\left( s,t \right)=\left\{ \begin{matrix}
   F\left( s,2t \right),\,\,\,\,\,\text{for  }t\in \left[ 0,\frac{1}{2} \right]  \\
   G\left( s,2t-1 \right),\,\,\,\,\,\text{for  }t\in \left[ \frac{1}{2},1 \right]\; ,  \\
\end{matrix} \right.\]
thus implying transitivity.
\end{enumerate}
\end{proof}
\noindent We shall denote the equivalence class of a path $f$ by $\left[ f \right]$. It is referred to as the \emph{homotopy class}\index{homotopy!class} of $f$. There is an intuitive notion of combining two paths when the endpoint of the first coincides with the starting point of the second. This is the notion of concatenation:

\begin{definition}
For a pair of paths $f,g\colon [0,1] \to X $ on a topological space $X$ with $f\left( 1 \right)=g\left( 0 \right)$, we define the \emph{concatenation}\index{concatenation} of $f$ and $g$ as the path
\[f\cdot g\left( s \right)=\left\{ \begin{matrix}
   f\left( 2s \right),\,\,\,\,\text{if  }s\in \left[ 0,\frac{1}{2} \right]  \\
   g\left( 2s-1 \right),\,\,\,\,\text{if  }s\in \left[ \frac{1}{2},1 \right]\; .  \\
\end{matrix} \right. \]
\end{definition}

This operation is well-defined on equivalence classes of paths; indeed, if ${{f}_{0}}\simeq {{f}_{1}}$ and ${{g}_{0}}\simeq {{g}_{1}}$ under homotopies ${{f}_{t}}$ and ${{g}_{t}}$ respectively, then ${{f}_{0}}\cdot {{g}_{0}}\simeq {{f}_{1}}\cdot {{g}_{1}}$ via the homotopy ${{f}_{t}}\cdot {{g}_{t}}$.

\section{The Fundamental Group}

A special case of paths involves those paths $f\colon I \to X$ with the same initial and terminal point, in other words, for which $f(0)=f(1)=x_0$, for a point $x_0 \in X$; we call these paths \emph{loops}\index{loop} on the topological space $X$ and the point $x_0$ is then called a \emph{base point}\index{base point}.

A \emph{reparameterization}\index{reparameterization} of $f$ (or change of coordinates) is the map $f\circ \varphi $ where $\varphi\colon I \to I$ is a continuous map on the unit interval, such that $\varphi(0)=0$ and $\varphi(1)=1$. 

The reparameterization of a loop is homotopic to the loop, in other words, we claim that $f\circ \varphi \simeq f$, and so $\left[ f\circ \varphi  \right]\simeq \left[ f \right]$.  Indeed, the map 
	\[{{\varphi }_{t}}\left( s \right)=\left( 1-t \right)\varphi \left( s \right)+ts\; ,\]
for $t,s \in I$ defines a homotopy $f\circ {{\varphi }_{t}}$ from $f\circ \varphi $ to $f$.

\medskip

The definition of the fundamental group is now the following:

\begin{definition}
For a topological space $X$ and a fixed point $x_0 \in X$, consider the set of homotopy classes of loops on $X$ with base point $x_0$. This set is, in fact, a group with product defined by the concatenation of loops; it is called the \emph{fundamental group}\index{fundamental group} of $X$ at $x_0$ and is denoted by $\pi_{1} \left(X, x_0 \right)$.   
\end{definition}

\noindent We owe a proof that the set of equivalence classes of loops on $X$ is indeed a group:

\begin{proposition}
The set $\pi_{1} \left(X, x_0 \right)$ is a group with product defined by the concatenation $\left[ f \right]\left[ g \right]=\left[ f\cdot g \right]$ of two loops $f,g$ on $X$ with base point $x_0$.
\end{proposition}

\begin{proof}
We need to show that the product is an associative operation, there exists an identity element and there exists an inverse for every point in $\pi_{1} \left(X, x_0 \right)$. Indeed, for three loops $f,g,h \in \pi_{1} \left(X, x_0 \right)$, the map $\left( f\cdot g \right)\cdot h$ is a reparameterization of $f\cdot \left( g\cdot h \right)$, thus 
$\left[ \left( f\cdot g \right)\cdot h \right]=\left[ f\cdot \left( g\cdot h \right) \right]$, which implies associativity. \\
For every loop $f \in \pi_{1} \left(X, x_0 \right)$, the constant loop $c$ defined by $c(t)=x_0$, for all $t \in I$, defines an identity element, since then $f$ is a reparameterization of $f\cdot c$ and $c\cdot f$, in other words, $\left[ f \right]=\left[ f\cdot c \right]=\left[ c\cdot f \right]$.\\
For the property of existence of an inverse, consider the map $${{f}_{t}}\left( s \right)=f\left( \left( 1-t \right)s \right)\; ,$$
for which it is ${{f}_{0}}=f$ and ${{f}_{1}}=c=f\left( 0 \right)$. Moreover, consider 
$${{\bar{f}}_{t}}\left( s \right)=\bar{f}\left( t+\left( 1-t \right)s \right)\; ,$$ 
for which ${{\bar{f}}_{0}}=\bar{f}$ and ${{\bar{f}}_{1}}=c=\bar{f}\left( 1 \right)=f\left( 0 \right)$. Then, ${{f}_{t}}\cdot {{\bar{f}}_{t}}$ defines a homotopy in ${{\pi }_{1}}\left( X,{{x}_{0}} \right)$ from $f\cdot \bar{f}$ to the constant loop $c$ based at the point  $x_0$. The proof of the proposition is complete. 
\end{proof}

A natural question is how does the fundamental group depend on the base point $x_0$. For a path-connected topological space $X$, let $x_0$ and $x_1$ be two points on $X$, and consider the fundamental groups $\pi_{1} \left(X, x_0 \right)$ and $\pi_{1} \left(X, x_1 \right)$. Since $X$ is path-connected, there is a path on $X$, say $h$, from $x_0$ to $x_1$. Then, for any loop $f$ based at the point $x_1$, the path  $h\cdot f\cdot \bar{h}$ is a loop based at $x_0$, where $\bar{h}$ is the path from  $x_1$ to $x_0$ defined by $\bar{h}\left( s \right)=h\left( 1-s \right)$, for $s\in I$.  One then has the following proposition, the proof of which is left to the reader: 

\begin{proposition}
For a path-connected topological space $X$ and for $h$ a path between two points  $x_0$ and $x_1$ on $X$, the map $\left[ f \right]\mapsto \left[ h\cdot f\cdot \bar{h} \right]$ defines an isomorphism of fundamental groups ${{\pi }_{1}}\left( X,{{x}_{1}} \right)\cong {{\pi }_{1}}\left( X,{{x}_{0}} \right)$.
\end{proposition}

\begin{figure}[!ht]
	\centering
        	\includegraphics{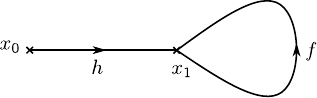}
	\caption{Changing the base point.}
\end{figure}

\noindent In view of the proposition above, when the topological space $X$ is path-connected, we shall omit reference to the base point of a fundamental group and simply write ${{\pi }_{1}}( X)$. 

\begin{definition}
Let $X$ be a path-connected topological space. We say that $X$ is \emph{simply connected}\index{space!simply connected} if   ${{\pi }_{1}}\left( X \right) = \{e\}$ is trivial. Equivalently, for any two points on $X$, there exists a unique homotopy class of paths connecting them. Four examples are shown in \Cref{Fig:four_simply_con}.
\end{definition}

\begin{figure}[!ht]
	\centering
	\includegraphics{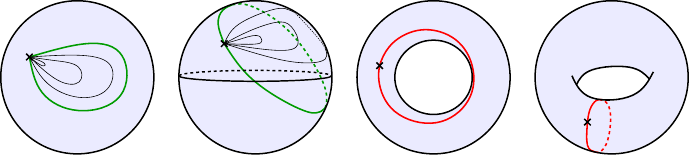}
	\caption{The disk and the sphere are simply connected; the annulus and the torus are not.}\labelx{Fig:four_simply_con}
\end{figure}

\begin{proposition}\labelx{Prop:product-pi-1}
Let $X,Y$ be two path-connected topological spaces. Then, the fundamental group of the Cartesian product ${{\pi }_{1}}(X\times Y)$ is isomorphic to the product of fundamental groups ${{\pi }_{1}}(X)\times {{\pi }_{1}}(Y)$.
\end{proposition}

\begin{proof}
For a loop $f(s)$ on the space $X\times Y$ based at the point $({{x}_{0}},{{y}_{0}})$, one may write $f(s)=(x(s),y(s))$ where $x(s)$, $y(s)$ are loops on $X,Y$ with base points $x_0$, $y_0$ respectively. We consider the map 
\[\varphi\colon {{\pi }_{1}}(X\times Y)\to {{\pi }_{1}}(X)\times {{\pi }_{1}}(Y)\; ,\]
with $\varphi \left( \left[ f\left( s \right) \right] \right)=\left( \left[ x(s) \right],\left[ y(s) \right] \right)$. It is then easy to check that $\varphi$ is a well-defined homomorphism, which is 1-1 and surjective; the full proof is left to the reader.
 \end{proof}
 
The fundamental group can be seen as a functor between the category of topological spaces and the category of groups. In more basic terms, this means that any continuous map $\varphi \colon X \to Y$ between topological spaces $X$ and $Y$ induces a homomorphism between the fundamental groups:
\[{{\varphi }_{*}}\colon {{\pi }_{1}}\left( X,{{x}_{0}} \right)\to {{\pi }_{1}}\left( Y,{{y}_{0}} \right)\; .\]
This map is defined by ${{\varphi }_{*}}\left( \left[ f \right] \right)=\left[ \varphi \circ f \right]$, for a loop $f$ on $X$. The map $\varphi$ is called \emph{induced homomorphism}\index{homomorphism!induced}.
The proof that the map ${{\varphi }_{*}}$ is indeed a well-defined group homomorphism is left to the reader.

In addition, if $\mathrm{id}\colon (X,{{x}_{0}})\to (X,{{x}_{0}})$ denotes the identity map on a topological space $X$, then the induced homomorphism $i{{d}_{*}}$ is the identity morphism on the fundamental group ${{\pi }_{1}}\left( X,{{x}_{0}} \right)$. Moreover, for a pair of continuous maps between three topological spaces with base points, $\varphi\colon (X,{{x}_{0}})\to (Y,{{y}_{0}})$ and $\psi\colon (Y,{{y}_{0}})\to (Z,{{z}_{0}})$, the composition of their induced homomorphisms satisfies 
\[{{\left( \psi \circ \varphi  \right)}_{*}}={{\psi }_{*}}\circ {{\varphi }_{*}}\; .\]

\begin{proposition}
For a pair of topological spaces with base points, 	
$(X,{{x}_{0}})$ and $(Y,{{y}_{0}})$, if $\varphi\colon (X,{{x}_{0}})\to (Y,{{y}_{0}})$ is a homeomorphism, then the induced map ${{\varphi }_{*}}\colon {{\pi }_{1}}\left( X,{{x}_{0}} \right)\to {{\pi }_{1}}\left( Y,{{y}_{0}} \right)$ is an isomorphism of fundamental groups. 
\end{proposition}

\begin{proof}
Let $\psi\colon (Y,{{y}_{0}})\to (X,{{x}_{0}})$ be the inverse of $\varphi$. Then, one can check that ${{\psi }_{*}}\circ {{\varphi }_{*}}={{\left( \psi \circ \varphi  \right)}_{*}}=\mathrm{id}_*=\mathrm{id}_{{{\pi }_{1}}\left( X,{{x}_{0}} \right)}$, and similarly ${{\varphi }_{*}}\circ {{\psi }_{*}}=\mathrm{id}_{{{\pi }_{1}}\left( Y,{{y}_{0}} \right)}$, namely, the homomorphism ${{\varphi }_{*}}$ is both injective and surjective.
\end{proof}

Note that the previous proposition implies that the fundamental group is, in fact, \emph{invariant under homeomorphisms} of topological spaces.

\section{Applications and Computations}

The following seminal computation of the fundamental group of the $n$-spheres $S^n$, for $n \ge 1$, leads to a series of highly significant corollaries. For a complete proof the reader is directed for instance to \cite[Sect. 1.1]{Hatcher} or \cite[Sect. 2.7]{Dieck}.

\begin{theorem}
\begin{enumerate}
\item The fundamental group of the circle is isomorphic to the group of the integers, namely, $\pi_{1} (S^1 ) \cong \mathbb{Z}$.
\item For $n \ge 2$, the fundamental group of the $n$-sphere is trivial, ${{\pi }_{1}}( {{S}^{n}})=\{e\}$.
\end{enumerate}
\end{theorem}

Using \Cref{Prop:product-pi-1} we can immediately deduce:

\begin{corollary}
The fundamental group of the 2-torus $T^2 = S^1 \times S^1$ is  ${{\pi }_{1}}({{T}^{2}})={{\pi }_{1}}({{S}^{1}})\times {{\pi }_{1}}({{S}^{1}})\cong \mathbb{Z}\times \mathbb{Z}$; more generally, for the $n$-dimensional torus ${{T}^{n}}={{S}^{1}}\times {{S}^{1}}\times \cdots \times {{S}^{1}}$, one has ${{\pi }_{1}}({{T}^{n}})\cong {{\mathbb{Z}}^{n}}$, and so, for $n \ge 2$, the $n$-torus is not homeomorphic to the $n$-sphere.
\end{corollary}

\begin{corollary}
The fundamental group of the spaces $\mathbb{R}^n \setminus \{0\}$, for $n \ge 2$, is given as follows
\[{{\pi }_{1}}\left( {{\mathbb{R}}^{n}}\setminus \{0\} \right)=\left\{ \begin{matrix}
   \mathbb{Z},\text{   for }n=2  \\
   0,\text{   for }n>2\; .  \\
\end{matrix} \right.\]
\end{corollary}

\begin{proof}
Indeed, using polar coordinates one sees that, for $n \ge 2$,  ${{\mathbb{R}}^{n}}\setminus \{0\}$ is homeomorphic to ${{S}^{n-1}}\times \mathbb{R}$. Then, since for every pair of distinct points in $\mathbb{R}$ there is a unique equivalence class of paths connecting them, the space $\mathbb{R}$ is simply connected, in other words, ${{\pi }_{1}}(\mathbb{R})=\{e\}$. We thus see that ${{\pi }_{1}}({{\mathbb{R}}^{n}}\setminus \{0\}) \cong {{\pi }_{1}}( {{S}^{n-1}}) \times {{\pi }_{1}}(\mathbb{R}) \cong {{\pi }_{1}}( {{S}^{n-1}})$, and so the asserted computation follows.
\end{proof}
 
\begin{corollary}
The real plane ${{\mathbb{R}}^{2}}$ is not homeomorphic to the space ${{\mathbb{R}}^{n}}$, for any $n\ne 2$. 
\end{corollary}

\begin{proof}
Assume that there were a homeomorphism $f\colon {{\mathbb{R}}^{2}}\to {{\mathbb{R}}^{n}}$. Then for $n=1$, one has that  ${{\mathbb{R}}^{2}}\setminus \{0\}$ is a connected space, however, $\mathbb{R}\setminus \left\{ f(0) \right\}$ is not, a contradiction. Moreover, for  $n>2$, there is also a contradiction, since we saw that 
\[{{\pi }_{1}}( {{\mathbb{R}}^{2}}\setminus \{0\})\ne {{\pi }_{1}}( {{\mathbb{R}}^{n}}\setminus \{0\})\; .\]
\end{proof}

\section{The fundamental group and the first homology group}

The fundamental group provides extremely useful information about the topology of a space $X$ especially if $X$ is path-connected. One can define \emph{higher homotopy groups}\index{homotopy!group} in a similar way -- for $n\in\mathbb{Z}_{\geq 2}$, the $n$th homotopy group $\pi_n(X)$ has the set of homotopy classes of maps $[0,1]^n\rightarrow X$ as elements. Another important family of topological invariants is the one of \emph{homology groups}\index{homology group} $H_n(X)$. A connection between these two families of topological invariants is provided by the \emph{Hurewicz Theorem}\index{Hurewicz Theorem}, which describes a homomorphism from the $n$th homotopy group into the $n$th homology group. In practice, homology groups are easier to compute than homotopy groups, though they contain in general less sharp information than the latter, as we will see shortly. We refer to \cite{Dieck} for an introduction to higher homotopy groups, as well as a detailed exposition and proof of the Hurewicz Theorem. 

For the particular case $n=1$ involving the fundamental group, more can be said here; we first introduce some terminology.

\begin{definition}
For a group $G$, the \emph{commutator subgroup}\index{commutator!subgroup} of $G$, denoted by $[G,G]$, is defined as the subgroup generated by all elements of the form $aba^{-1}b^{-1}$, for any $a,b \in G$; the elements $[a,b]:=aba^{-1}b^{-1}$ are called the \emph{commutators}\index{commutator} of $G$. Namely,
$[G,G ]= \left\langle { ab{{a}^{-1}}{{b}^{-1}}, \text{ for all } a,b\in G} \right\rangle$.
\end{definition}
Note that, in a sense, the commutator subgroup measures how far is the group $G$ from being an abelian group, since if $[G,G]=\{e\}$, this is equivalent to having $ab=ba$, for every $a,b \in G$. Moreover, the commutator subgroup is a normal subgroup of $G$, in other words, $[G,G]$ is closed under conjugation; indeed, one checks that for every $a,b,c \in G$: 
\begin{align*}
c\left[ a,b \right]{{c}^{-1}} & = c\left( ab{{a}^{-1}}{{b}^{-1}} \right){{c}^{-1}}\\
& = \left( ca{{c}^{-1}} \right)\left( cb{{c}^{-1}} \right){{\left( ca{{c}^{-1}} \right)}^{-1}}{{\left( cb{{c}^{-1}} \right)}^{-1}}\\
& = \left[ ca{{c}^{-1}},cb{{c}^{-1}} \right]\; .
\end{align*}
This fact provides that the quotient 
\[{{G}_{ab}}:={G}/{\left[ G,G \right]}\;\]
has a well-defined group structure; we call ${{G}_{ab}}$ the \emph{abelianization}\index{group!abelianization} of $G$ and it is the largest quotient of $G$ which is an abelian group. The Hurewicz Theorem\index{Hurewicz Theorem}, or more properly the \emph{Hurewicz--Poincar\'{e} Theorem} for the case of fundamental groups, states the following:
\begin{theorem}[Hurewicz--Poincar\'{e} Theorem]\labelx{th:Hurewicz-Poincare}
If $X$ is a path-connected topological space and $x_0 \in X$ is any point of $X$, then the abelianization of ${{\pi }_{1}}\left( X,{{x}_{0}} \right)$ is isomorphic to the first homology group $H_1(X)$. 
\end{theorem}
\noindent An obvious corollary is that when $X$ is path-connected with abelian fundamental group, then for any base point $x_0 \in X$, ${{\pi }_{1}}\left( X,{{x}_{0}} \right)\cong {{H}_{1}}\left( X \right)$. This implies for example that $H_1(T^n)=\mathbb{Z}^n$, and $H_1(S^n) = 0$ when $n\geq2$. Note that when $X$ is not path-connected, then the statement of the theorem fails; for instance, for $X={{S}^{1}}\coprod{{{S}^{1}}}$ the disjoint union of two circles, then ${{\pi }_{1}}\left( X,{{x}_{0}} \right)\cong \mathbb{Z}$, whereas ${{H}_{1}}\left( X \right)\cong \mathbb{Z}\oplus \mathbb{Z}$.
\chapter{Covering spaces}\labelx{Sec:coveringspaces}


\abstract{The theory of coverings allows to link properties between different manifolds whenever a special map exists between them. This permits the study of complicated manifolds from simpler ones. In this chapter, we explore unramified and ramified spaces which involve fundamental notions of covering spaces over manifolds. Roughly speaking, an unramified covering is a surjective map between manifolds which makes them locally homeomorphic. Fixing the base manifold, there is a beautiful relationship between all covering spaces and the fundamental group of the base manifold, similar to the Galois theory of field extensions. Allowing for special points at which the manifolds are not locally homeomorphic leads to ramified coverings, a much more flexible notion.}

\section{Unramified coverings}\labelx{sec:deck_tran}

We make the idea of a covering space described above precise:

\begin{definition}\labelx{def_cov_space_unramified}
Let $X, \tilde{X}$ be two topological spaces and $p\colon \tilde{X} \to X$  continuous. The map $p$ is called a \emph{covering space projection} if for every $x \in X$ there is a neighborhood $U$ of $x$ such that $p^{-1}(U)$ is a disjoint union of open sets $\{ U_a\}_{a \in A}$, and for each index $a$, the map $p\rvert_{U_{a}}\colon U_a \to U$ is a homeomorphism. We say in this case that the space $\tilde{X}$ is a \emph{covering space}\index{covering!space}\index{space!covering} of $X$ and sometimes denote it as a pair $(\tilde{X}, p)$. Moreover, the open set $U$ as above is called a \emph{covering neighborhood} of $X$. 
\end{definition}

\begin{example}
The real line $\mathbb{R}$ is a covering space for the unit circle $S^1$ via the covering space projection $p\colon \mathbb{R} \to S^1$ given by $p(x)=(\mathrm{cos}(2\pi x), \mathrm{sin}(2\pi x))$.
\end{example}

\begin{figure}[!ht]
	\centering
	\includegraphics{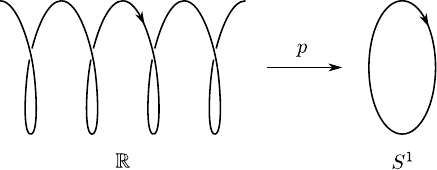}
	\caption{The real line as a covering space of the circle.}\labelx{Fig:RcoversS1}
\end{figure}

Fix a covering $p\colon \tilde{X}\to X$. Since the spaces $\tilde{X}$ and $X$ are locally the same, they can be distinguished only by their global properties. 

\begin{definition}
Let $p\colon \tilde{X} \to X$ be a covering space projection and let $f\colon Y \to X$ be a continuous map from another topological space $Y$ to $X$. We call \emph{lift}\index{lift} of $f$ a map $\tilde{f}\colon Y \to \tilde{X}$ such that $f=p \circ \tilde{f}$, i.e. such that the following diagram commutes:
\begin{equation*}
	\xymatrix{
	 & \tilde X \ar[d]^p \\
 	Y \ar[r]_f \ar@{-->}[ur]^{\tilde f} & X}
\end{equation*}  
\end{definition}

Particularly interesting is the existence of lifts for paths. This leads to a link between the fundamental groups of $X$ and $\tilde{X}$.

\begin{proposition}\labelx{uniq_path_lift}
Let $f\colon  I \to X$ be a path with endpoints $f(0)=x$ and $f(1)=y$. For each  $\tilde{x} \in p^{-1}(x)$, there exists a unique lift $\tilde{f} \colon  I \to \tilde{X}$ such that  $\tilde{f}(0) = \tilde{x}$ and $p \circ \tilde{f} = f$.
\end{proposition}

\begin{proof}
Take an open covering $\{ U_{a}\}$ of $X$ such that each open set $U_a$ is a covering neighborhood. For each $s \in I$ there exists a neighborhood $V_s$ of $s$ such that $f(V_{s}) \subset U_a$, for some $a$. Thus, $\{ V_s \mid s \in I \}$ is an open covering of $I$. But since $I$ is compact, from Lebesgue' s lemma we have that there is an $\epsilon   > 0$ such that if $\lvert s-t \rvert \le \varepsilon $, then $f([s,t])\subset {{U}_{a}}$ for some $a$. Now, let $n\in \mathbb{N}$ such that $\frac{1}{n}<\varepsilon $ and ${{s}_{k}}=\frac{k}{n}$. Assume without loss of generality that $f(0)\in {{U}_{a}}$. \\
Define $\tilde{f}(0)=\tilde{x}\in U_{a}^{i}\subset {{p}^{-1}}({{U}_{a}})$, where $U_{a}^{i}$ are open disjoint with $p \rvert_{U_{a}^{i}} $ homeomorphism. Then, for each $q\in [{{s}_{0}},{{s}_{1}}]$, we have $f(q)\in {{U}_{a}}$ and also $p\colon U_{a}^{i}\to {{U}_{a}}$ is a homeomorphism. We may now define  
\[\tilde{f}(q):={{p}^{-1}}(f(q))\; ,\]
for each $q\in [{{s}_{0}},{{s}_{1}}]$. Since $[{{s}_{0}},{{s}_{1}}]$ is connected, we imply that $\tilde{f}([{{s}_{0}},{{s}_{1}}])\subset U_{a}^{i}$ and so 	$\tilde{f}\rvert _{[{{s}_{0}},{{s}_{1}}]}$ is uniquely defined. We may now extend $\tilde{f}$ over the entire $I$ by iterating the process.
\end{proof}

Consider base points $x_0 \in X$ and $\tilde{x}_0 \in p^{-1}(x_0)\subset \tilde{X}$.
\begin{proposition}
The induced homomorphism on fundamental groups ${{p}_{*}}\colon {{\pi }_{1}}( \tilde{X},{{{\tilde{x}}}_{0}})\to {{\pi }_{1}}\left( X,{{x}_{0}} \right)$ is injective. Moreover, the subgroup ${{p}_{*}}\,{{\pi }_{1}}(\tilde{X},{{{\tilde{x}}}_{0}})$ is consisting of loops based at $x_0$ which lift to loops based at ${{\tilde{x}}_{0}}$.
\end{proposition}

\begin{proof}
For a point $[{{\tilde{f}}_{0}}]\in \mathrm{ker}({{p}_{*}})$, we have that ${{\tilde{f}}_{0}}\colon I\to \tilde{X}$ is a loop based at ${{\tilde{x}}_{0}}$ and the path ${{f}_{0}}=p\circ {{\tilde{f}}_{0}}$ is homotopic to the constant path ${{f}_{1}}(t)={{x}_{0}}$, for every $t \in I$. For the homotopy ${{f}_{t}}$ between the paths $f_0$ and $f_1$, one can show that this lifts to a unique homotopy ${{\tilde{f}}_{t}}\colon I\to \tilde{X}$ from 
${{\tilde{f}}_{0}}$ to ${{\tilde{f}}_{1}}$, with $p\circ {{\tilde{f}}_{t}}={{f}_{t}}$. We thus conclude that $[{{\tilde{f}}_{0}}]=1$. 
\end{proof}

\begin{proposition}
 The number of sheets of the covering $p$ is equal to the index of the subgroup  ${{p}_{*}}\bigl( {{\pi }_{1}}(\tilde{X},{{{\tilde{x}}}_{0}}) \bigr)$ in $\pi_1 (X,x_0)$. 
\end{proposition}

\begin{proof}
Let us first denote $H:={{p}_{*}}\bigl( {{\pi }_{1}}(\tilde{X},{{{\tilde{x}}}_{0}}) \bigr)$. For a loop $f$ on $x$ based at $x_0$, we consider a map from the right co-sets $\varphi\colon \left\{ H[f] \right\}\to \left\{ {{p}^{-1}}({{x}_{0}}) \right\}$ with $\varphi (H[f])=\tilde{f}(1)$, where $\tilde{f}$ is a lift of $f$ from the point $\tilde{x}_0$. One can then check that this map $\varphi$ is well-defined, injective and surjective. 
\end{proof}

The next theorem is fundamental for the problem of classifying the covering spaces of a topological space; we first introduce the following technical terminology:

\begin{definition}
A topological space $X$ is called \emph{semi-locally simply connected}\index{space!semi-locally simply connected} if for every $x\in X$ there exists a neighborhood $U$ of $x$ such that the homomorphism $i_{*}\colon \pi_1(U,x) \to \pi_1 (X,x)$ is trivial. We call $X$ \emph{locally simply connected}\index{space!locally simply connected} if for every $x\in X$ and every neighborhood $V$ of $x$, there is a neighborhood $U \subset V$ of $x$ such that $\pi_1 (U,x)=\{ e\}$ is trivial. 
\end{definition}

\begin{figure}[!ht]
	\centering
	\includegraphics{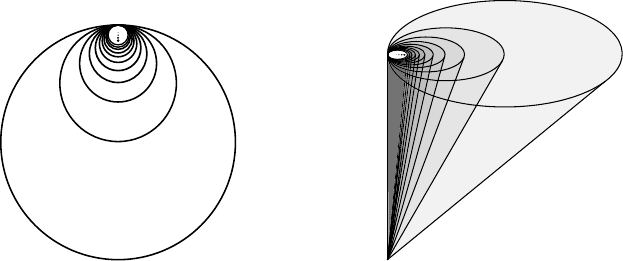}
	\caption{The Hawaiian earring (left) and the cone over it (right).}\labelx{Fig:Hawaian}
\end{figure}

The topological space defined as
\begin{equation*}
	H = \bigcup_{n=1}^\infty \left\{(x,y)\in\R^2~\left\vert~\left(x-\frac{1}{n}\right)^2+y^2 = \frac{1}{n^2}\right.\right\}
\end{equation*}
is dubbed the \emph{Hawaiian earring}\index{Hawaiian earring}; it is displayed on the left of \Cref{Fig:Hawaian}. It is a standard example of a not semi-locally simply-connected topological space: any neighborhood of $(0,0)$ contains infinitely many circles, and hence non-contractible loops in it which are still non-contractible in $H$. The cone over the Hawaiian earring is displayed on the right of the same figure; it is an example of a semi-locally simply connected but not locally simply connected topological space. 

\begin{theorem}\labelx{them:exist_universal_cov}
Let $X$ be a path-connected, locally path-connected and semi-locally simply connected topological space. Then there exists a covering space $\tilde{X}$ of $X$ with $\pi_1(\tilde{X})=1$. The space $\tilde{X}$ is called the \emph{universal covering space}\index{covering!universal} of $X$.  
\end{theorem}

\begin{remark}
In the sequel of the book, we will restrict our attention to topological manifolds which satisfy the conditions of the theorem above. Thus, the reader can think of the statement and proof of this theorem in this context.   
\end{remark}

\begin{proof}[Proof of \Cref{them:exist_universal_cov}]
We only sketch the main ideas here; for a complete proof we refer for example to \cite[Sect. 1.3]{Hatcher}. We may view the points of $\tilde{X}$ as homotopy classes of points of $X$, namely, we define 	
\[\tilde{X}=\left\{ [\gamma ]:\gamma \text{ a path on }X\text{ with }\gamma (0)={{x}_{0}} \right\}\] and consider the map $p\colon \tilde{X}\to X$ with $p([\gamma ])=\gamma (1)$. Since $X$ is assumed to be path-connected, one implies then that the map $p$ is surjective. \\
For an open path-connected neighborhood $U$ of $x$, such that the map ${{\pi }_{1}}(U,x)\hookrightarrow {{\pi }_{1}}(X,x)$ is trivial, and for a path $\gamma$ such that $\gamma (1) \in U$, we may now consider the collection of subsets
\[{{U}_{[\gamma ]}}:=\left\{ [\gamma \cdot \eta ] \mid \eta \text{ a path on }U\text{ with }\eta (0)=\gamma (1) \right\}\; .\]
Then one shows that this collection ${{U}_{[\gamma ]}}$ defines a basis of open sets for a topology on $\tilde{X}$. With respect to this topology the covering map  $p\colon \tilde{X}\to X$ is a covering space projection and one lastly verifies that $\tilde{X}$ is path-connected and simply connected.
\end{proof}

\begin{definition}
Let $X$ be a path-connected topological space and let ${{p}_{1}}\colon {{\tilde{X}}_{1}}\to X$, ${{p}_{2}}\colon {{\tilde{X}}_{2}}\to X$ be covering spaces of $X$. A map $f\colon {{\tilde{X}}_{1}}\to {{\tilde{X}}_{2}}$ is an \emph{isomorphism of covering spaces}\index{covering!space!isomorphism} if $f$ is a homeomorphism and ${{p}_{1}}={{p}_{2}}\circ f$; if only the latter condition holds, then we call $f$ a \emph{homomorphism of covering spaces}\index{covering!space!homomorphism}. Moreover, for a covering space $p\colon \tilde{X}\to X$ of $X$, we say that $f\colon \tilde{X}\to \tilde{X}$ is an \emph{automorphism}\index{covering!space!automorphism} if $f$ is an isomorphism of covering spaces.  
\end{definition}

The automorphisms of a covering space form a group, which is of central importance for what we are going to see in the sequel.

\begin{definition}
Let $X$ be a path-connected topological space and let $p\colon \tilde{X} \to X$ be a covering space. The automorphisms $f\colon \tilde{X}\to \tilde{X}$ are called \emph{deck transformations}\index{deck transformation}. Under the operation of concatenation of loops, these form a group which shall be denoted by $G(\tilde{X})$. 
\end{definition}

\begin{example}
\begin{enumerate}
\item Let $p\colon \mathbb{R}\to S^1$ be the covering space of the unit circle for $p(x)=(\mathrm{cos}(2\pi x), \mathrm{sin}(2 \pi x))$ as in \Cref{Fig:RcoversS1}. The deck transformations for this covering are the maps $f_n\colon  \mathbb{R} \to \mathbb{R}$ given by $x \mapsto x+n$, for $n \in \mathbb{Z}$. Therefore, we have $G(\tilde{\mathbb{R}}) \cong \mathbb{Z}$. 
\item Let $p\colon S^1\to S^1$ be another covering space of the unit circle for $p(z)=z^n$, for $n \in \mathbb{Z}$. In this case, we have $G(\tilde{{S^1}}) \cong \mathbb{Z}_n$.
\end{enumerate}
\end{example}

Note that the action $G(\tilde{X})\curvearrowright X$ on the connected space $\tilde{X}$ is a \emph{covering space action}\index{covering!space!action}, namely, for any point $\tilde{x} \in \tilde{X}$, a nontrivial $g \in G(\tilde{X})$, and a covering neighborhood $\tilde{U}\ni\tilde x$, one has $g(\tilde{U})\cap \tilde{U}=\varnothing $. In particular,  if $\tilde{X}$ is the universal covering space, then since $G(\tilde{X})={{\pi }_{1}}(X,{{x}_{0}})$, we have that ${{\pi }_{1}}(X,{{x}_{0}})\curvearrowright \tilde{X}$ is a covering space action.

\section{Ramified coverings}\labelx{Sec:ramified-coverings}

We have seen in  \Cref{def_cov_space_unramified} that a surjective map $p\colon \tilde{X} \to X$ between two topological spaces is a covering if they are locally homeomorphic via $p$. This type of covering is, in fact, called an \emph{unramified covering}. It imposes many restrictions on $\tilde{X}$. There is a much more flexible notion of a ramified covering which originates in the study of multi-valued functions in complex analysis. In these cases, over isolated points of the base space, sheets of the covering can merge, thus the group of covering maps can have fixed points. 

\begin{example}
Consider the map $f\colon \mathbb{C}\to\mathbb{C}$ given by $f(z)=z^2$. Restricted to $\mathbb{C}^*$, it defines a covering with 2 sheets. At the origin, the two sheets merge. In the language defined below, we will say that the origin is a branch point.

The inverse function $z\mapsto \sqrt{z}$ is not well-defined on all of $\mathbb{C}$. There are several ways to understand this function. The square root is a well-defined function if one removes a ray originating from the origin, for example the negative reals. One can also consider $\sqrt{z}$ as a multi-valued function, with two values for every non-zero complex number. Yet another way is to say that $\sqrt{z}$ is a well-defined function not on $\mathbb{C}$, but on the ramified double cover defined by $f$. The two possible values of the square root are assigned to the two sheets, and when you go along a loop around the origin, you permute the two sheets.
\end{example}

We make the notion of ramified covering more precise now: 

\begin{definition}
Let $X, \tilde{X}$ be two topological surfaces. A continuous surjective map $p\colon \tilde{X} \to X$ is called a \emph{(topological) ramified covering}\index{covering!ramified} if there exists a nonempty discrete set $Y\subset X$ such that $p\vert_{p^{-1}(X\setminus Y)}\colon {p^{-1}(X\setminus Y)} \to X\setminus Y$ is a covering in the sense of \Cref{def_cov_space_unramified}. The subset $Y$ above is called the \emph{branch locus}\index{branch!locus} of $p$ and we will denote it by $B(p)$. A point $b\in Y$ is called a \emph{branch point}\index{branch!point}.
\end{definition}

We further introduce the following terms:

\begin{definition}
Let $X, \tilde{X}$ and $p\colon \tilde{X} \to X$ be as above. Consider a point $\tilde{x} \in \tilde{X}$ and its image through $p$, $x:=p(\tilde{x})$. The \emph{ramification index}\index{ramification!index} of the map $p$ at the point $\tilde{x}$ is defined as the positive integer $m_{\tilde{x}}(p)$ such that there is an open neighborhood $U$ of $\tilde{x}$ so that the point $x\in X$ has only one pre-image in $U$, in other words, $p^{-1}(x) \cap U =\tilde{x}$, and for all other points $z\in p(U)$ it holds that $\vert p^{-1}(z) \cap U \vert = m_{\tilde{x}}(p)$. A point $\tilde{x} \in \tilde{X}$ is called a \emph{ramification point}\index{ramification!point} of $\tilde{X}$ if it is $m_{\tilde{x}}(p)>1$. In this case, $p(\tilde{x})$ is a \emph{branch point}. Moreover, the \emph{ramification locus}\index{ramification!locus} $R(p)$ of the map $p$ is the subset of $\tilde{X}$ consisting of all the ramification points; see also  \Cref{Fig:branch_ramif_points} for an example.
\end{definition}

\begin{figure}[!ht]
	\centering
	\includegraphics[width=0.9\textwidth]{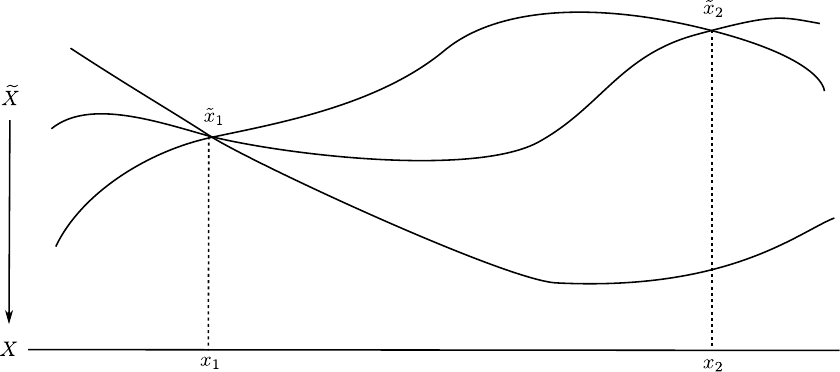}
	\caption{The map $p\colon \tilde X\to X$ has two branch points $x_1, x_2\in X$ and two ramification points $\tilde x_1,\tilde x_2\in\tilde X$. Note that here $m_{\tilde{x}_1}(p)=3$ and $m_{\tilde{x}_2}(p)=2$.}\labelx{Fig:branch_ramif_points}
\end{figure}

In view of the definitions above, it is always the case that $R(p) \subseteq p^{-1}(B(p))$. Moreover, whenever $R(p)=\varnothing $, the map $p$ is an unramified covering. When all branch points are of index 2, we speak about a \emph{simply ramified covering}\index{covering!simply ramified}. For each point $x\in X$ we now consider the sum of ramification indices of the map $p$ at the pre-images:
\[\mathrm{d}_{x}(p):= \sum\limits_{\tilde{x}\in {{p}^{-1}}(x)}{{{m}_{{\tilde{x}}}}(p)}\; .\]
The number $\mathrm{d}_{x}(p)$ is, in fact, constant for all points of $x\in X$ (see \cite[Chap. II, Prop. 4.8]{Miranda} for a proof). One is thus led to the following:

\begin{definition}
Let $X, \tilde{X}$ and $p\colon \tilde{X} \to X$ be as above. The \emph{degree} of the map $p$ is the integer $\mathrm{d}_{x}(p)$, for any point $x\in X$, and will thus be simply denoted by $\mathrm{deg}(p)$. 
\end{definition}

The following example gives a local model for ramified coverings on Riemann surfaces, notably \emph{spectral curves} which we will discuss further in \Cref{Sec:Spectral-curve}.

\begin{example}\labelx{Ex:spectral-curve-loc}
Consider an open set $U\subset \mathbb{C}$ and the trivial bundle $E=U\times \mathbb{C}$. Denote by $z$ a coordinate on $U$ and $p$ a coordinate on the fiber. For holomorphic functions $a_2(z), a_3(z), ..., a_n(z)$, define the function $P(p,z)=p^n+a_2(z)p^{n-2}+a_3(z)p^{n-3}+...+a_n(z)$, which is a polynomial in $p$. Finally, consider
$$\tilde{U}=\{(z,p)\in E\mid P(p,z)=0\}\; .$$
Restricting the canonical projection $E\to U$ to $\tilde{U}$, we get a covering $\pi\colon \tilde{U}\to U$, ramified over the multiple zeros of the polynomial $P$. The degree of $\pi$ is $n$, the degree of $P$ as polynomial in $p$. The covering is simply ramified if $P$ has at most double zeros. 
\end{example}

We are now ready to state the fundamental \emph{Riemann--Hurwitz formula}\index{Riemann--Hurwitz formula} in the context of a pair of topological surfaces: 

\begin{theorem}[Riemann--Hurwitz formula]\labelx{RiHu_formula}    
Let  $S_{g_1,h_1}$, $S_{g_2,h_2}$ be a pair of compact connected orientable topological surfaces of respective genera $g_1$ and $g_2$ and with respective number of boundary components $h_1$ and $h_2$. Let $p\colon S_{g_1,h_1} \to S_{g_2,h_2}$ be a (possibly ramified) covering of topological surfaces. Then,
\[2{{g}_{1}}+h_1-2=\deg (p)(2{{g}_{2}}+h_2-2)+\sum\limits_{x\in {{S}_{{{g}_{1},h_1}}}}{\left( {{m}_{x}}(p)-1 \right)}\; .\]
\end{theorem}

The proof idea is to consider a triangulation of $S_{g_2,h_2}$ adapted to the branch points (i.e. all branch points appear as vertices of the triangulation), and to take its pre-image under $p$ which gives a triangulation on $S_{g_1,h_1}$. Then one computes the Euler characteristic of $S_{g_1,h_1}$ by Euler's formula $\chi(S_{g_1,h_1})=V-E+F$, where $V, E, F$ are the number of vertices, edges and faces of the triangulation on $S_{g_1,h_1}$. We refer to \cite[Chap. II, Prop. 4.16]{Miranda} for a full proof.

The Riemann--Hurwitz formula is a powerful tool with an immense number of applications. Its origins are traced in the study of Riemann surfaces, which we will more carefully introduce in  \Cref{sec:smooth_holo_v_bundles}. A Riemann surface is a connected 1-dimensional complex manifold and assuming that a covering map 
$p\colon S_{g_1,h_1} \to S_{g_2,h_2}$ between a pair of Riemann surfaces is complex analytic, the map $p$ is ramified at a point $x\in S_{g_1,h_1}$ if the map $p$ is locally described as $p(z)=z^n$ in local coordinates $z$ around the point $x$, for some integer $n>1$. This number $n$ is then the ramification index at the point $x\in S_{g_1,h_1}$.

Besides their role in \Cref{partIII} and \Cref{partIV} of this monograph, ramified coverings and Riemann surfaces will reappear prominently in \Cref{Chap:theoriesofclassS,Chap:BPSstatesclassS}, where class~$\mathcal S$ theories and their BPS states will be discussed.
\chapter{Bundles and connections}\labelx{Sec:bundlesandconnections}



\abstract{The notion of a connection on a vector bundle is of tremendous importance both in differential geometry as well as in mathematical physics. Using connections one can understand the local geometry at points of a vector bundle and differentiate sections along vector fields. Moreover, connections give rise to fundamental geometric invariants such as the curvature. Physically, connections are called gauge fields or vector fields, and their curvature is the corresponding field strength. Gauge theories provide the appropriate mathematical framework to describe forces in Physics. For example, the relativistic vector potential in Maxwell's theory is a connection, of which the electromagnetic tensor describing the electric and magnetic fields is the curvature. The Riemann--Hilbert correspondence relates the geometry of flat vector bundles to the topological structure of the base manifold. In this chapter, we briefly review the theory of vector bundles with connections over a Riemann surface; an abundance of texts can be found in the literature for a proper introduction in differential geometry of complex vector bundles, for instance \cite{KoNo} or \cite{Wells}. For readers with more physical background, we warmly recommend \cite{Baez}.}

\section{Smooth and holomorphic vector bundles}\labelx{sec:smooth_holo_v_bundles}

The theory of manifolds grew out of the search for a generalized notion of Euclidean space $\mathbb{R}^n$, a concept which only locally looks like $\mathbb{R}^n$. Very similarly, the theory of fiber bundles originates in the search for a notion generalizing functions $f\colon M\to F$ on a manifold $M$ into some space $F$. It turns out that the appropriate global object is a \emph{section of a bundle with fiber $F$}\index{vector bundle!section}. Here we restrict attention to the case where $F$ is a vector space. The precise definition of a smooth complex vector bundle is more technical.

\begin{definition} Let $E,M$ be two differentiable manifolds and $p\colon E\to M$ be a smooth surjective mapping. By a \emph{vector bundle chart}\index{vector bundle!chart} on the triple $\left( E,p,M \right)$ we mean a pair $( U,\varphi )$, where $U$ is an open subset in $M$ and $\varphi $ a diffeomorphism such that the following diagram is commutative

\end{definition}

\begin{equation*}
	\xymatrix{
		p^{-1}(U) \ar[rr]^{\varphi } \ar[rd]_p & & U\times \mathbb{C}^{k} \ar[ld]^{\mathrm{pr}_1} \\
	& U & 
	}
\end{equation*}

\begin{definition}Two vector bundle charts $( U,\varphi  )$ and $( V,\psi )$ will be called \emph{compatible}\index{vector bundle!chart!compatible}, if the mapping $\varphi \circ {{\psi }^{-1}}$ is a linear isomorphism, namely  $\varphi \circ {{\psi }^{-1}}( x,\lambda )=( x,{{g}_{UV}}( x )\cdot \lambda  )$, for some mapping ${{g}_{UV}}\colon U\cap V\to \mathrm{GL}( {{\mathbb{C}}^{k}})$ and for $( x,\lambda )\in V\times {{\mathbb{C}}^{k}}$. The mapping ${{g}_{UV}}$ is then unique and smooth, and it is called the \emph{transition function} between the two vector bundle charts. 
\end{definition}

\begin{definition}
A \emph{vector bundle atlas}\index{atlas} ${{( {{U}_{\alpha }},{{\varphi }_{\alpha }} )}_{\alpha \in A}}$ for $( E,p,M )$ is defined to be a set of pairwise compatible vector bundle charts $( {{U}_{\alpha }},{{\varphi }_{\alpha }} )$, such that  ${{\left\{ {{U}_{\alpha }} \right\}}_{\alpha \in A}}$ is an open cover of $M$. Moreover, two vector bundle atlases will be called \emph{equivalent} if their union is again a vector bundle atlas. 
 \end{definition}

\begin{definition}
A \emph{smooth vector bundle}\index{vector bundle!smooth}\index{bundle!vector} of rank $k$ is a triple $( E,p,M)$, which consists of two smooth manifolds $E$ and $M$, and a smooth mapping $p\colon E\to M$, together with an equivalence class of vector bundle atlases. The manifold $E$ is called the \emph{total space}\index{vector bundle!total space},  $M$ the \emph{base space}\index{vector bundle!base space} and  $p$ the \emph{projection mapping}. 
\end{definition}

The complex vector spaces ${{\left\{ {{E}_{x}} \right\}}_{x\in M}}$, where $E=\cup_{x\in M}{{{E}_{x}}}$ are called the \emph{fibers}\index{vector bundle!fiber} of the bundle. The diffeomorphism ${{\varphi }_{U}}\colon {{p}^{-1}}( U )\to U\times {{\mathbb{C}}^{k}}$, which maps the vector space ${{E}_{x}}$ isomorphically into $\left\{ x \right\}\times {{\mathbb{C}}^{k}}$ for each $x\in U$ is called a \emph{local trivialization} of the bundle $E$ over $U$.
The group $G$ into which the transition functions of a vector bundle are taking their values is called the \emph{structure group}\index{vector bundle!structure group} of the bundle. A vector bundle of rank 1 is called in particular a \emph{line bundle}\index{vector bundle!line}.

If $( U,\varphi )$ and $( V,\psi  )$ are vector bundle charts then for the transition functions ${{g}_{UV}}\colon U\cap V\to \mathrm{GL}( {{\mathbb{C}}^{k}} )$ with $g_{UV}( x )=( \varphi \circ {\psi }^{-1}) \vert_{\left\{ x \right\}\times {{\mathbb{C}}^{k}}}$ the following identities are satisfied 
\begin{align}
   {{g}_{UV}}( x )\cdot {{g}_{VU}}( x ) & = I, \text{ for all }  x\in U\cap V  \labelx{cocycle_cond1}\\
   {{g}_{UV}}( x )\cdot {{g}_{VW}}( x )\cdot {{g}_{WU}}( x ) & =I, \text{ for all } x\in U\cap V\cap W  \labelx{cocycle_cond2}
\end{align}

\noindent The above two identities are called the \emph{cocycle condition}\index{cocycle condition} and we will call the family $\left\{ {{g}_{UV}} \right\}$, the cocycle of \emph{transition functions} for the vector bundle atlas ${{( {{U}_{\alpha }},{{\varphi }_{\alpha }} )}_{\alpha \in A}}$.  Conversely, given an open cover ${{\left\{ {{U}_{\alpha }} \right\}}_{\alpha \in A}}$ of $M$ and smooth mappings
${{g}_{\alpha \beta }}\colon {{U}_{\alpha }}\cap {{U}_{\beta }}\to \mathrm{GL}( {{\mathbb{C}}^{k}} )$, which satisfy the above identities, there is a unique smooth complex vector bundle $( E,p,M )$ with transition functions $\left\{ {{g}_{\alpha \beta }} \right\}$.

\begin{definition}A \emph{subbundle}\index{vector bundle!subbundle} $F\subset E$ of a vector bundle $E$ is a collection ${{\left\{ {{F}_{x}}\subset {{E}_{x}} \right\}}_{x\in M}}$ of subspaces of the fibers ${{E}_{x}}$ of $E$, such that $F=\cup {{F}_{x}}$ is a submanifold of $E$. 
\end{definition}

\begin{definition}
A \emph{vector bundle homomorphism}\index{vector bundle!homomorphism}\index{homomorphism!vector bundle} between two smooth vector bundles $E$ and $F$ is a smooth mapping $f\colon E\to F$, such that $f\left( {{E}_{x}} \right)\subset {{F}_{x}}$ and ${{f}_{x}}=f \vert_{{{E}_{x}}}\colon {{E}_{x}}\to {{F}_{x}}$ is a linear mapping of complex vector spaces. Additionally, two smooth vector bundles $E,F$ are called \emph{isomorphic}\index{isomorphic vector bundles}\index{vector bundle!isomorphism} if there is a vector bundle homomorphism $f\colon E\to F$ with ${{f}_{x}}\colon {{E}_{x}}\to {{F}_{x}}$ an isomorphism of complex vector spaces, for all $ x\in M$.
\end{definition}

A smooth vector bundle of rank $k$ will be called \emph{trivial}\index{vector bundle!trivial} if it is isomorphic to $M\times {{\mathbb{C}}^{k}}$.

An important particular case in the above definition is $E=F$ for which we get the notion of a \emph{bundle automorphism}, also called a \emph{gauge transformation}.

\begin{definition} For a smooth complex vector bundle $V$, a \emph{gauge transformation}\index{gauge!transformation} is defined as a smooth automorphism of the bundle $V$. The set of all gauge transformations of $V$ is called the \emph{gauge group}\index{gauge!group} of $V$.
\end{definition}

Now that we have introduced vector bundles, the notion of a section, which generalizes a vector-valued function, comes naturally next:

\begin{definition}
A \emph{section}\index{vector bundle!section} of a smooth vector bundle $(E,p,M)$ is a smooth mapping $\sigma\colon M\to E$, such that $\sigma \left( x \right)\in {{E}_{x}}$ for every $x\in M$. The space of sections is denoted by $\mathcal{A}^0(M;E)$. Similarly, we denote by $\mathcal{A}^k(M;E)$ the space of smooth differential $k$-forms on $E$ over $M$. 
\end{definition}

\begin{definition}
A smooth \emph{local frame}\index{local frame} for the bundle $E$ over $U\subset M$ is a collection ${{\sigma }_{1}},\ldots ,{{\sigma }_{k}}$ of smooth sections such that $\left\{ {{\sigma }_{1}}\left( x \right),\ldots ,{{\sigma }_{k}}\left( x \right) \right\}$ is a basis for the complex vector space ${{E}_{x}}$ for every $x\in U$.
\end{definition} 

A smooth local frame for a smooth vector bundle $E$ of rank $k$ on $U$ is actually a trivialization of $E$ over $U$. Indeed, given a trivialization ${{\varphi }_{U}}\colon {{E}_{U}}={{p}^{-1}}\left( U \right)\to U\times {{\mathbb{C}}^{k}}$, then the sections ${{\sigma }_{i}}\left( x \right)=\varphi _{U}^{-1}\left( x,{{e}_{i}} \right)$ constitute a smooth frame. Conversely, if ${{\sigma }_{1}},\ldots ,{{\sigma }_{k}}$ is a smooth frame then we can define a trivialization ${{\varphi }_{U}}$ by 
	\[{{\varphi }_{U}}\left( \lambda  \right)=\left( x,\left( {{\lambda }_{1}},\ldots ,{{\lambda }_{k}} \right) \right)\; ,\]
where $\lambda =\sum{{{\lambda }_{i}}}{{\sigma }_{i}}\left( x \right)\in {{E}_{x}}$.

\begin{remark}
Locally, having chosen a frame for the bundle $V$, a gauge transformation is a ${{C}^{\infty }}$-mapping with its values in $\mathrm{GL}_n(\mathbb{C})$.         
\end{remark}

\medskip
We next define holomorphic vector bundles in a way similar to smooth vector bundles. To define them, we need a base manifold which is complex.

\begin{definition} A \emph{complex manifold}\index{complex!manifold} $M$ of (complex) dimension $n$ is a differentiable manifold, which admits an open covering ${{\left\{ {{U}_{\alpha }} \right\}}_{\alpha \in A}}$ and transition functions ${{\varphi }_{\alpha }}\colon {{U}_{\alpha }}\to {{\mathbb{C}}^{n}}$, such that ${{\varphi }_{\alpha }}\circ \varphi _{\beta }^{-1}$ is holomorphic on ${{\varphi }_{\beta }}\left( {{U}_{\alpha }}\cap {{U}_{\beta }} \right)\subset {{\mathbb{C}}^{n}}$ for all $\alpha $ and $\beta $. A complex manifold of dimension 1 is called a \emph{Riemann surface}\index{Riemann!surface}.
\end{definition}

\begin{definition}
Let $M$ be a complex manifold. A \emph{holomorphic vector bundle}\index{vector bundle!holomorphic} $\left( E,p,M \right)$ of rank $k$ is a smooth complex vector bundle with the extra property that $E$ is endowed with a complex manifold structure, such that for all $ x\in M$ there exists an open $U\subset M$ with $x\in U$ and trivialization
${{\varphi }_{U}}\colon {{E}_{U}}\to U\times {{\mathbb{C}}^{k}}$, which is a biholomorphic mapping between complex manifolds. 
\end{definition}

Such a trivialization is called a \emph{holomorphic trivialization}\index{holomorphic!trivialization}. Furthermore, similarly to the case of smooth complex vector bundles, given a family
$\left\{ {{\varphi }_{\alpha }}\colon {{E}_{{{U}_{\alpha }}}}\to {{U}_{\alpha }}\times {{\mathbb{C}}^{k}} \right\}$ of trivializations for the holomorphic vector bundle, the transition functions with respect to this family are then holomorphic. Conversely, given holomorphic mappings ${{g}_{\alpha \beta }}\colon {{U}_{\alpha }}\cap {{U}_{\beta }}\to \mathrm{GL}\left( {{\mathbb{C}}^{k}} \right)$ which satisfy identities (\ref{cocycle_cond1}), (\ref{cocycle_cond2}), we can construct a holomorphic vector bundle with transition functions ${{g}_{\alpha \beta }}$. 

All aspects we discussed on smooth complex vector bundles apply analogously for the category of holomorphic vector bundles, with the difference that here the mappings in the definitions are taken to be holomorphic.

As an example, let us analyze holomorphic line bundles over the Riemann sphere.
\begin{example}
The Riemann sphere $\mathbb{CP}^1$ is given by two charts $U_0 = \mathbb{C}$ and $U_1=\mathbb{C}$ with transition map $U_0\cap U_1 = \mathbb{C}^* \to \mathbb{C}^*$ given by $z\mapsto 1/z$. Define the line bundle $\mathcal{O}(k)$ in the following way: it is made of two pieces, $U_0\times \mathbb{C}$ and $U_1\times \mathbb{C}$ which are glued together via the map
\begin{align*}
 (U_0\cap U_1)\times \mathbb{C} &\to (U_0\cap U_1)\times \mathbb{C}\\
(z,v) &\mapsto (1/z, z^kv)\; .
\end{align*}
It turns out that up to isomorphism, the line bundles $\mathcal{O}(k)$ describe all holomorphic line bundles on $\mathbb{CP}^1$.
\end{example}

To any vector bundle $E\to M$, we can associate a line bundle in the following way. There is an open covering of $M$ by open sets $U_\alpha$ where $E$ is trivial and with transition maps $f_{\alpha \beta}$. Define the \emph{determinant line bundle}\index{vector bundle!determinant} $\det(E)$ to be the line bundle which is trivial over the $U_\alpha$ with transition functions $\det(f_{\alpha \beta})$.

For a general complex manifold, an important holomorphic bundle is the canonical bundle:
\begin{definition}
    The \emph{canonical bundle}\index{vector bundle!line!canonical} $K_X$ of a complex manifold $X$ is the determinant line bundle associated to the holomorphic cotangent bundle $T^*X$. Equivalently, it is the line bundle of holomorphic $n$-forms where $n$ is the dimension of $X$.
\end{definition}

Over a compact topological surface $S$, complex vector bundles can be easily classified. The isomorphism class of a line bundle is uniquely determined by its \emph{degree}\index{vector bundle!line!degree} or \emph{first Chern class} which is an element in $H^2(S,\mathbb{Z})\cong \mathbb{Z}$. Complex vector bundles of higher rank are classified by their determinant line bundle:
\begin{proposition}
    A complex bundle $E\to S$ of rank $k$ over a surface $S$ is isomorphic to $\det(E)\oplus \mathbb{C}^{k-1}$ (as a complex bundle).
\end{proposition}

Holomorphic bundles are much more abundant. On a compact Riemann surface $X$ of genus at least 2, there are smooth families of holomorphic line bundles, described by the \emph{Picard variety}\index{Picard variety}. Line bundles of degree zero are parameterized by the \emph{Jacobian variety}\index{Jacobian variety}. Moduli spaces of holomorphic bundles of higher rank over $X$ can be constructed using the notion of \emph{slope stability}\index{slope stability} by Mumford \cite{Mumford}.

To finish this section, we give an example illustrating the difference between complex and holomorphic bundles:
\begin{example}
On $\mathbb{CP}^1$, consider the holomorphic rank-two bundle $\mathcal{O}(1)\oplus \mathcal{O}(1)$. Its underlying complex vector bundle is $\mathbb{C}\oplus T\mathbb{CP}^1$, since it can be easily checked that $\det(\mathcal{O}(1)\oplus\mathcal{O}(1))=\mathcal{O}(2)$ which is the tangent bundle of $\mathbb{CP}^1$. Note that $\mathbb{C}\oplus T\mathbb{CP}^1$ is not trivial, since there is no non-vanishing section of $T\mathbb{CP}^1$ (the ``hairy ball theorem''\index{hairy ball theorem}).

Note that the underlying real bundle is \emph{trivial}. One can check that $T(\mathbb{S}^2) \oplus \mathbb{R}$ is a trivial bundle. Indeed, one has to take the unit sphere $\mathbb{S}^2\subset \mathbb{R}^3$ and consider its normal bundle, which is trivial (there is a constant section). Hence its direct sum with the tangent bundle gives the tangent bundle of $\R^3$, which is trivial, restricted to the sphere. Therefore, $T(\mathbb{S}^2)\oplus \mathbb{R}^2 \cong \mathbb{R}^4$.
\end{example}

\section{Connections}\labelx{Sec:connections}

Connections allow one to define directional derivatives of sections. They permit us to connect different fibers of the same fiber bundle via the notion of parallel transport.

\begin{definition} Let $V$ be a smooth complex vector bundle of rank $n$ over a complex manifold $M$.  We define a \emph{connection}\index{connection} on $V$ to be a smooth linear mapping \[{{d}_{A}}\colon {{\mathcal{A}}^{0}}( M;V )\to {{\mathcal{A}}^{1}}( M;V )\cong {{\mathcal{A}}^{1}}( M ){\otimes } {{\mathcal{A}}^{0}}( M,V)\; ,\]
which satisfies the Leibniz rule
	\begin{equation}{{d}_{A}}( fs )=\D f\otimes s+f{{d}_{A}}s\; ,\end{equation}
for every smooth function $f$ and for every section $s\in {{\mathcal{A}}^{0}}( M;V )$.
\end{definition}

\noindent Let $e=( {{e}_{1}},\ldots ,{{e}_{n}} )$ be a local frame for the bundle $V$. We may then write ${{d}_{A}}{{e}_{i}}$ as a linear combination of the elements of this frame as:
	\[{{d}_{A}}{{e}_{i}}=\sum A_{i}^{j}{{e}_{j}}\; .\]
The matrix $A=\left( {{A}_{ij}} \right)$ of 1-forms is called the \emph{connection matrix}\index{connection!matrix} of ${{d}_{A}}$ with respect to the frame $e$. 
\footnote{In gauge theories, connections are considered as fields on the base manifold $M$ which is often interpreted as the spacetime manifold. They are called
\emph{gauge fields}\index{gauge!field}, and their quanta, \emph{gauge bosons}\index{gauge!boson}. One simple example is the (relativistic) vector potential of
electromagnetism, which is a $1\times 1$ matrix of 1-forms which represents locally a connection on line bundles over spacetime.}

One sees that the frame and connection matrix fully determine the connection; indeed, writing any section  $s\in {{\mathcal{A}}^{0}}( M;V )$ in the form  $s=\sum{{{s}^{i}}}{{e}_{i}}$,  for ${{s}^{i}}\in {{C}^{\infty }}\left( M \right)$, we then have 
\begin{equation*}
{{d}_{A}}s =\sum{\D{{s}^{i}}}\otimes {{e}_{i}}+\sum{{{s}^{i}}}\cdot {{d}_{A}}{{e}_{i}}=\sum\limits_{j}{( \D{{s}^{j}}+\sum\limits_{i}{{{s}^{i}}{{A}_{i}^{j}}})}\,{{e}_{j}}\; .
\end{equation*} 
Thus, considering a local frame $e=( {{e}_{1}},\ldots ,{{e}_{n}})$ for the bundle $V$, the connection can be locally written as
\[{{d}_{A}}=\D +A=\D +\sum_i A_i dx_i \; ,\]
where we fixed local coordinates $(x_1,...,x_n)$ on $M$ giving the 1-forms $\D x^i$, $A$ is the connection matrix and $A_i$ are matrices whose entries are ${{C}^{\infty }}$-functions on $M$.

However, the connection matrix $A$ for a point $z\in M$ does depend on the choice of the frame in a neighborhood of $z$. Indeed, choosing another frame $e'$ on a neighborhood of $z$, then the connection matrix of ${{d}_{A}}$ with respect to ${e}'$ is written as
\begin{equation}
A_{e'}={{g}^{-1}}\cdot ( \D g )+{{g}^{-1}}\cdot {{A}_{e}}\cdot g\; ;
\end{equation}
the above describes the action of the group of gauge transformations on the space of connections.

The local expression of a connection helps us realize that connections provide a means of transporting vectors between fibers of a vector bundle along a curve on the base manifold in a parallel manner. We describe this picture next. 
\begin{definition}
Let $(E,p,M)$ be a smooth complex vector bundle equipped with a connection $d_A$. A local section $s$ of $E$ will be called \emph{parallel}\index{vector bundle!section!parallel}\index{parallel!section} with respect to the connection $d_A$, if $d_A s=0$.
\end{definition}

Consider a smooth curve $c\colon [0,1] \to M$ on the manifold $M$. With respect to a local frame, the condition $\D_{c'(t)} s=0$, for all $t\in [0,1]$, can be expressed as a system of first order linear ODEs:
\begin{equation}\labelx{sys_ODEs_paral}
\frac{\D {{s}^{i}}}{\D t}+\sum\limits_{i,j}{A_{i}^{j}({c}'(t)){{s}^{j}}}=0\; ,
\end{equation}
where $(A_{i}^{j})$ are the entries in the connection matrix of $d_A$ with respect to the chosen local frame. If the initial condition is given by a point $v \in E_{c(0)}$, then a solution to the system (\ref{sys_ODEs_paral}) is a time-dependent section $s$ such that $s(c(0))=v$ and $\D_{c'(t)} s=0$. Standard techniques from the theory of ODEs provide existence and uniqueness of solutions to the system (\ref{sys_ODEs_paral}); moreover, these solutions are defined for all $t \in [0,1]$, since the 1-forms $A_{i}^{j}$ that appear in the connection matrix are smooth and our system of ODEs is of first order and linear. Therefore, this time-dependent unique parallel section $s \in \mathcal{A}^0(E)$ with $s(c(0))=v$ and $\D_{c'(t)} s=0$ is well-defined, and is called the \emph{parallel transport of $v$ along the curve $c$}, alternatively the \emph{parallel section along $c$}\index{vector bundle!section!parallel}. We make the following definition:
\begin{definition}\labelx{defn_parallel transport}
Let $(E,p,M)$ be a smooth complex vector bundle and let  $c\colon [0,1] \to M$ be a smooth curve on the base manifold $M$. For a fixed $t \in [0,1]$, the \emph{parallel transport}\index{parallel!transport} from $c(0)$ to $c(t)$ along the curve $c$ is a linear map 
\[P_{c(t)}\colon E_{c(0)}\to E_{c(t)}\; ,\]
which maps a point $v \in E_{c(0)}$ to the vector $P_{c(t)}(v)=s(c(t))\in E_{c(t)}$, for $s$ the unique parallel section along the curve $c$ and for $s(c(0))=v$.
\end{definition}
Thus, a connection prescribes a way of lifting curves from the base manifold to the bundle, and the parallel transport map is a linear isomorphism between the fibers of a vector bundle; see  \Cref{fig:Parallel_transport}.

\begin{figure}[!ht]
\centering
\includegraphics{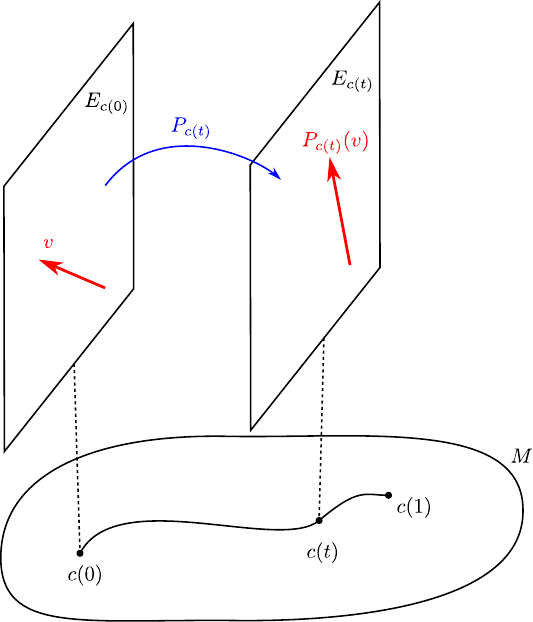}
\caption{Parallel transport along a path $c$ on $M$.}
\labelx{fig:Parallel_transport}
\end{figure}

\section{Curvature and flatness}

We have described two main perspectives in the study of connections; one is global and involves viewing connections as differential operators, while the other considers their local expression as a matrix of 1-forms using frames and provides a means of parallel transport along a curve on the base manifold. The \emph{curvature} of any given connection on a smooth complex vector bundle can be analogously defined globally as a linear differential operator of order two by uniquely extending the operator defined by a connection, and it can be described locally by matrices of 2-forms.   

\begin{definition}
Let $(V,p,M)$ be a smooth complex vector bundle equipped with a connection $d_A$. We define the \emph{curvature}\index{curvature} of the connection ${{d}_{A}}$ on $V$ to be the operator
\begin{equation*}
    F_A=d_{A}^{2}\colon \mathcal{A}^0( M; V)\to \mathcal{A}^2( M; V )\; ,
\end{equation*}
which is a smooth linear mapping and can be considered as an element in $\mathcal{A}^2( M;\mathrm{End}V) =\mathcal{A}^2(\Hom(V,V))$.
\end{definition}

In a similar way to connections, we can represent the curvature operator locally with respect to a frame for the bundle. Indeed, let $e=( e_1,\ldots , e_n)$ be a local frame for the bundle $V$. We then write $F_{A}e_i$ as a linear combination of the frame’s elements
\[{{F}_{A}}( e_i)=\sum{{\Theta _{i}^{j}}}{{e}_{j}}\; .\]
This matrix of 2-forms, ${{\Theta }_{e}}=( {{\Theta }_{i}^{j}})$, is called the \emph{curvature matrix}\index{curvature!matrix} for the connection. For a different choice of a local frame $e'=(e'_1,\ldots ,e'_n)$, one checks that 
\[\Theta_{e'}=g^{-1}\cdot \Theta_e \cdot g\; .\]
We may also express the curvature matrix with respect to the connection matrix. Indeed, the following formula can be easily established
\begin{equation}\labelx{eq_Cartan structure equation}
\Theta_e=\D A_e+A_e\wedge A_e    
\end{equation}  
and this is usually called the \emph{Cartan structure equation}\index{Cartan structure equation}.
\footnote{Physically, the curvature operator is called \emph{field strength}\index{field strength} of the corresponding gauge field. In electromagnetism, the field strength of the vector potential is the electromagnetic tensor, which describes both the electric and magnetic fields.}

\begin{definition}
If the curvature $F_A$ of a connection on a smooth complex vector bundle vanishes everywhere, then we say that the connection is \emph{flat}\index{connection!flat}. 
\end{definition}

\begin{proposition}If $d_A$ is a flat connection, then there exist $n$ linearly independent local solutions of the equation $d_A s=0$.
\end{proposition}

\begin{proof} It suffices to show that there is a local frame $e=( s_1,\ldots, s_n)$ of parallel sections. The matrix $A$ for the connection $d_A$ with respect to such local frame is then the zero matrix.
Let $e'=(e'_1,\ldots,e'_n)$ be a local frame and $A'$ the connection matrix of $d_A$ with respect to $e'$. As was previously mentioned, the transition from the frame $e$ to $e'$ is determined by the relationship $A=g^{-1}(\D g )+g^{-1}A'g$ between the respective matrices for the frames. 
Then, the condition $A=0$, is equivalent to the system of linear differential equations 
\begin{equation}\labelx{sys_diff-eq_flat}
g^{-1}( \D g)+g^{-1}A'g=0\; .
\end{equation}
Therefore, finding a solution $g$ for this system of linear differential equations is equivalent to proving that there exists a frame of parallel sections. Now, taking the image of \Cref{sys_diff-eq_flat} via $g$ we have
$A'g+\D g=0$ and differentiating this we get the integral condition
\[0=( \D A')g-A'\wedge \D g=( \D A' )g+( A'\wedge A' )g=\Theta'g\; ,\]
using that $\D g=-A'g$ and the Cartan structure equation (\ref{eq_Cartan structure equation}) above.

\noindent We thus see that the integral condition for this system is precisely the vanishing of the matrix $\Theta'$, a condition which is independent of the choice of the local frame.  Therefore, if the connection is flat then obviously the curvature matrix is the zero matrix, so from the existence of a solution for a system of linear differential equations, there exists the solution $g$ we seek.
\end{proof}

\begin{remark}\labelx{const_tran_funct} In the case when there are two such bases of solutions, $(e_1,\ldots, e_n)$ and $( \tilde{e}_{1},\ldots,\tilde{e}_{n})$ with $\tilde{e}_{i}=\sum{a_{i}^{j}s_j}$, we then have
\[0=d_A \tilde{e}_{i}=d_A ( \textstyle\sum{a_{i}^{j}e_j}) = \sum{\D a_{i}^{j} \otimes e_j}\; ,\]
thus $\D a_{i}^{j}=0$ and so $a_{i}^{j}$ are \emph{constant functions}.
\end{remark}

We conclude that a flat connection induces a family of \emph{locally constant transition functions} for the bundle $V$.
The converse is also true: If $V$ has locally constant transition functions, then it admits a \emph{flat connection}.
Indeed, if $\left\{ U,{{e}_{U}} \right\}$ is a family of local frames for the bundle with constant transition functions, we define the connection $d_A$ by $d_A e_U = 0$.  Then, since the transition functions are constant, it follows that the condition $d_A e_U = 0$ is compatible with $d_A e_W = 0$ on $U\cap W\subset V$, and so the connection is well defined.  Now it is obvious that this connection is flat. A bundle endowed with a flat connection is called a \emph{local system}\index{local system}.

\section{The Riemann--Hilbert correspondence}\labelx{sec:RH}

We will now see a profound equivalence between representations
 $\rho\colon \pi_{1}( M )\to \mathrm{GL}_n(\mathbb{C})$ of the fundamental group of the manifold $M$ and flat ${{C}^{\infty }}$-complex vector bundles $E$ of rank $n$ over $M$.
The image $H$ of $\rho $ is called the \emph{holonomy group}\index{holonomy!group}. First we construct a flat bundle out of a representation.

Let $\rho\colon \pi_{1}( M )\to \mathrm{GL}_n(\mathbb{C})$ be a fundamental group representation. We view $\pi_1 ( M )$ as the group of deck transformations which acts on the universal covering space $\tilde{M}$ of $M$, and consider the action
\[\delta \colon \pi_1( M )\to \mathrm{Homeo}( \tilde{M}\times {{\mathbb{C}}^{n}})\]
given by $\delta ( \gamma )( x,v )=( \gamma ( x ),\rho ( \gamma  )^{-1}\cdot v )$.
This action induces the commutative diagram

\begin{equation*}
	\xymatrix{ \tilde{M}\times\mathbb C^n \ar[r]^{\mathrm{pr}} \ar[d]^{\pi} & \tilde{M} \ar[d]^q \\ E = (\tilde{M}\times\mathbb{C}^n)/\delta \ar[r]^-p & M\; ,}
\end{equation*}
where $\mathrm{pr}$ is the natural projection, which is obviously equivariant, $\pi $ is the quotient map, $q$ is the covering map and $p$ is induced uniquely by the projection $\mathrm{pr}$. We may then prove the following:
 
\begin{proposition}For the above diagram, it holds that
\begin{enumerate} 
\item The mapping $\pi\colon \tilde{M}\times \mathbb{C}^n\to E$ is a covering map.
\item If the holonomy group $H=\mathrm{Im}\rho $ is equipped with the induced topology from $\mathrm{GL}_n(\mathbb{C} )$, then $\xi_H = ( E,p,M )$ is a flat $C^{\infty}$ complex vector bundle with discrete structure group, the group $H$.
\end{enumerate}
\end{proposition}

 \begin{proof}
 \begin{enumerate}  
\item Since the action $\delta $ is a free and properly discontinuous action, it follows that the mapping $\pi $ is a covering map.
\item We choose an open covering $\left\{ U_i \right\}$ of $M$ of contractible sets. 
Moreover, for every $i$, we choose a point $b_i\in U_i$ and a path $c_i$ from $b_0$ to $b_i$, where $b_0\in M$ is a base point, in other words, $\pi_1 ( M )=\pi_1 ( M,b_0)$.\\
For each $U_i$ we ask for a trivialization $\varphi _i\colon p^{-1}( U_i) \to U_i\times \mathbb{C}^n$. Let $\tilde{c}_i$ be the lifting of the path $c_i$ from the base point $\tilde{b}_0\in q^{-1}( b_0)\subset \tilde{M}$ and let $\tilde{U}_i$ be the subset of $q^{-1}(U_i)$ which contains the final point of the path $\tilde{c}_i$. Since every non trivial element of $\pi _1( M,b_0)$ commutes the subsets of $pr^{-1}( q^{-1}( U_i) )$, it follows that 
\[\pi_i:=\pi \vert_{\tilde{U}_i\times \mathbb{C}^n}\colon \tilde{U}_i\times \mathbb{C}^n\to p^{-1}(U_i)\]
is a homeomorphism.\\
Then, \[\varphi_i=q\vert_{\tilde{U}_i\times \mathbb{C}^n} \circ \pi_{i}^{-1}\colon p^{-1}(U_i)\to U_i\times \mathbb{C}^n\]
is a local trivialization of the bundle $\xi_H$ over $U_i$, as follows from the commutativity of the diagram.\\
Now, let $U_i,U_j$ be two elements of the covering of $M$ with $U_i \cap U_j \ne \varnothing $.  For every $b\in U_i\cap U_j$, we consider paths $w_{ib}$ on $U_i$ from $b_i$ to $b$, and $w_{jb}$ from $b_j$ to $b$.
\begin{figure}[!ht]
\centering
\includegraphics{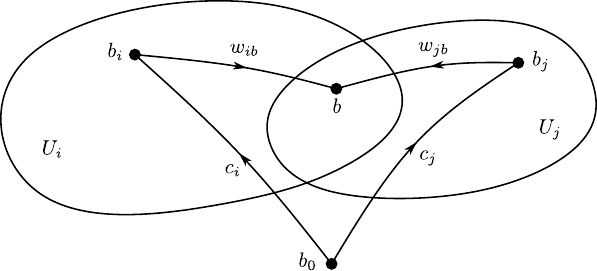}
\caption{Construction of a loop based at $b_0$.}\labelx{Fig:constr_loop_based_b0}
\end{figure}
Then the loop $\gamma_{ij}=\left[ c_i \circ w_{ib} \circ w_{jb}^{-1} \circ c_{j}^{-1} \right]\in \pi_1 ( M,b_0)$ (where by $\circ$ here we denote the concatenation of paths) does not depend on the choice of $w_{ib}$ and $w_{jb}$ (\Cref{Fig:constr_loop_based_b0}). Thus, it follows that the mapping
\[ g_{ij}\colon U_i\cap U_j \to \mathrm{Im}\rho  \] 
with $b\mapsto \rho ( \gamma_{ij})$ 
is locally constant. Therefore, we have constructed a vector bundle with locally constant transition functions, and so by  \Cref{const_tran_funct}, this bundle is flat.   
\end{enumerate} 
\end{proof}                                                        
We have seen that a representation $\rho\colon {{\pi }_{1}}( M )\to \mathrm{GL}_n(\mathbb{C})$ determines a smooth complex vector bundle $E$ of rank $n$ over $M$ with locally constant transition functions. We will next see that all smooth complex vector bundles with locally constant transition functions are actually constructed in this way.

\begin{definition}\labelx{def:Holonomy_representation}
Let $( E,p,M )$ be a ${{C}^{\infty }}$-complex vector bundle with locally constant transition functions which take values on a structure group $G\subseteq \mathrm{GL}_n(\mathbb{C} )$ and let $F_0 , F_1$ be the fibers over the points $b_0 , b_1 \in M$ respectively. If $c\colon [ 0,1 ]\to M$ is a path from $b_0$ to $b_1$, then for every $y\in F_0$, there exists a unique lifting $\tilde{c}_{y}$ of the path $c$ from $y$. We may thus define a homeomorphism
\[T_c\colon F_0\to F_1\; ,\]
which is called the \emph{translation}\index{translation} of $F_0$ to $F_1$ along $c$ and it only depends on its homotopy class (with constant endpoints).
\end{definition}
If we now take $b_0 = b_1$ and fix a trivialization $( U_0,\varphi_0)$ with $b_0\in U_0$, then a homeomorphism $T_{\gamma }\in G$ is constructed for every $\gamma \in \pi_1 ( M)$. 
It is easy to see that $T_{\gamma {\gamma }'}=T_{{\gamma }'}\circ T_{\gamma }$. Thus, if we define $H_{\xi }( \gamma  )=T_{\gamma }^{-1}$ for a bundle $\xi$, then the mapping 
\[H_{\xi }\colon \pi_1 ( M )\to G\]
is a homomorphism and is called the \emph{holonomy representation}\index{holonomy!representation} (or \emph{monodromy representation}\index{monodromy!representation}) of the bundle $\xi $. 

\begin{definition}
Let $p\colon \tilde{M} \to M$ be a covering space projection and let $\gamma$ be a loop on $M$ based at a point $x_0 \in M$ which lifts to a unique path $\tilde{\gamma}$ in $\tilde{M}$ based at a point $y_0 \in p^{-1}(x_0)$. For the point $x_0$, define the \emph{monodromy action}\index{monodromy!action} as the right group action on $p^{-1}(x_0)$ by the group $\pi_1 (M, x_0)$ given by 
\begin{align*}
p^{-1}(x_0) \times \pi_1 (M, x_0) & \to  p^{-1}(x_0) \\
(y_0, [\gamma]) & \mapsto \tilde{\gamma}(1)\; .
\end{align*}
The representation $\rho\colon \pi_1 (M, x_0) \to \mathrm{GL}(p^{-1}(x_0))$ associated to the monodromy action is called the \emph{monodromy representation}\index{monodromy!representation}. Two monodromy representations $\rho$ and $\rho'$ shall be called \emph{equivalent}\index{monodromy!representation!equivalence} if for every homotopy class $[\gamma] \in \pi_1 (M, x_0)$, there exists a linear automorphism of the fiber $p^{-1}(x_0)$, $A\in \mathrm{GL}(p^{-1}(x_0))$, such that $\rho'([\gamma])=A\rho([\gamma])A^{-1}$.  
\end{definition}

\begin{proposition}
A ${{C}^{\infty }}$-vector bundle $\xi =( E,p,M)$ with fiber $\mathbb{C}^n$ and locally constant transition functions into structure group $G\le \mathrm{GL}_n(\mathbb{C})$ is trivial if and only if its holonomy representation is trivial. 
\end{proposition}

\begin{proof}
The straight direction is obvious. 
For the converse, let us suppose that \[H_{\xi }\colon \pi_1 ( M )\to G\] is trivial. Then, for every path $c$ in $M$ with initial point $b$ and final point $b_0$, the translation $T_c\colon F_b\to F_{b_{0}}$ is independent of the path $c$ and shall be denoted from now on as $T_b$.  We can identify the fiber $F_{b_{0}}$ with $\mathbb{C}^n$ via the corresponding chart, and then the mapping 
$ h\colon E\to M\times \mathbb{C}^n$ with $x\mapsto ( p( x ),T_{p( x )}( x ) )$ 
is an isomorphism.
\end{proof}

\begin{theorem}Every vector bundle $\xi =( E,p,M )$ with locally constant transition functions into the group $G\le \mathrm{GL}_n(\mathbb{C} )$ is isomorphic to the vector bundle induced by a representation $H_{\xi }\colon \pi_1 ( M )\to G$.
\end{theorem}

\begin{proof} Let $\tilde{\xi }=( \tilde{E},\tilde{p},\tilde{M} )$ be the vector bundle induced by the universal covering $q\colon \tilde{M}\to M$.  We thus have the commutative diagram
\begin{equation*}
\xymatrix{ \tilde{E} \ar[r]^{\tilde{p}} \ar[d]_{\tilde\pi} & \tilde{M} \ar[d]^q \\ E \ar[r]^-p & M\; .}
\end{equation*}

\noindent Since the universal covering space $\tilde{M}$ is simply connected, we have that $H_{\tilde{\xi }}\equiv id$ is the identity, and so by the preceding lemma, the bundle $\tilde{\xi }$ is isomorphic to the trivial bundle $( \tilde{M}\times \mathbb{C}^n,pr,\tilde{M} )$.  Using this isomorphism the above diagram becomes
\begin{equation*}
\xymatrix{ \tilde{M}\times\C^n \ar[r]^{\mathrm{pr}} \ar[d]_{\pi} & \tilde{M} \ar[d]^q \\ E \ar[r]^-p & M\; .}
\end{equation*}

\noindent We deduce that $\pi $ is a regular covering with the group of deck transformations given by $\pi_1 ( M )$, with its action on $\tilde{M}\times \mathbb{C}^n$ given by 
\begin{align*}
\delta\colon \pi_1 ( M )\times ( \tilde{M}\times \mathbb{C}^n ) & \to \tilde{M}\times \mathbb{C}^n \\ 
 ( \gamma , ( \tilde{m},y ) ) & \mapsto ( \gamma ( \tilde{m}),H_{\xi } ( \gamma ) ( y ) )\; . 
\end{align*}
\end{proof}

\noindent Summing up the preceding results we have the following:

\begin{theorem}\labelx{Riemann-Hilbert1} If $E$ is a vector bundle of rank $n$ over a manifold $M$ then the following statements are equivalent
\begin{enumerate}
\item There exists a family of locally constant transition functions which define the vector bundle $E$.  
\item The vector bundle $E$ is flat, that is, it admits a flat connection.
\item The vector bundle $E$ is defined by a representation $\rho \colon \pi_1 ( M )\to \mathrm{GL}_n(\mathbb{C} )$.
\end{enumerate}
\end{theorem}

The holonomy of a flat vector bundle\index{vector bundle!flat} can be directly described geometrically by the connection. Indeed, in view of the definition of parallel transport of a vector along a curve on the base manifold (\Cref{defn_parallel transport}), the parallel transport in any direction over a grid on the manifold does not change the direction of the vector. Thus, since every homotopy class with constant endpoints determines a grid between the loops on the manifold, the parallel transport determines a unique element for each homotopy class.  Therefore, we get a well-defined homomorphism 
$\rho\colon \pi_1 ( M )\to \mathrm{GL}_n(\mathbb{C})$, which is the  holonomy representation of the bundle in terms of \Cref{def:Holonomy_representation}.

Given \Cref{Riemann-Hilbert1} one can now see that gauge equivalent flat connections give rise to equivalent holonomy representations (cf. \cite{Vassiliou}); we then have the following theorem:

\begin{theorem}[Riemann--Hilbert Correspondence]\labelx{Riemann-Hilbert correspondence}\index{Riemann--Hilbert correspondence}
There exists an isomorphism between the space of gauge equivalence classes of flat rank $n$ vector bundles $E$ over a manifold $M$ and equivalence classes of fundamental group representations $\rho\colon \pi_1 (M, x) \to \mathrm{GL}_n(\mathbb{C})$.
\end{theorem}

\begin{remark}\labelx{rk:line_bundles}
    When $n=1$, $\GL_n(\mathbb{C})=\mathbb{C}^\times$ is abelian, hence any homomorphism $\rho\colon \pi_1(M,x)\to \mathbb{C}^\times$ factors through the first homology group $\tilde{\rho}\colon H_1(M)\to \mathbb{C}^\times$ by \Cref{th:Hurewicz-Poincare}. The space of gauge equivalent classes of flat \emph{line} bundles over $M$ is isomorphic to the space of homomorphisms $\tilde{\rho}\colon  H_1(M)\to \mathbb{C}^\times$.
\end{remark}

\section{Principal bundles}

Principal fiber bundles are of fundamental importance in differential geometry and global analysis as they provide the right formal language for several geometric and analytic questions. In mathematical physics they play a central role in setting the  necessary framework for physical gauge theories and the standard model in particle physics. 

In this subsection we introduce the basic notions and features of principal fiber bundles, and review the construction of vector bundles associated to principal bundles. For a reference which is also close to the gauge theoretic occurrence of principal bundles, we refer to \cite{Morgan}.

\begin{definition}
Let $G$ be a Lie group, $E$, $M$ be two differentiable manifolds and $p\colon E\to M$ be a smooth surjective mapping. The tuple $(E,p,M,G)$ is called a \emph{$G$-principal (fiber) bundle}\index{principal bundle}\index{bundle!principal} if the following conditions are satisfied:
\begin{enumerate}
    \item There is a continuous free right action $E\times G \to E$ with respect to which the map $p$ is equivariant, that is, $p(eg)=p(e)$, for every $e\in E$ and $g\in G$.
    \item There is an open covering $U_{\alpha}$ of the base manifold $M$ and local trivializations $\varphi_{\alpha}$ such that the next diagram is commutative
\begin{equation*}
	\xymatrix{
		p^{-1}(U_{\alpha}) \ar[rr]^{\varphi _{\alpha}} \ar[rd]_p & & U_{\alpha}\times G \ar[ld]^{\mathrm{pr}_1} \\
	& U_\alpha & 
	}
\end{equation*}

In the diagram above, $pr_1$ denotes the projection onto the first factor and the maps $\varphi_{\alpha}$ are $G$-equivariant with respect to the right $G$-action.
\end{enumerate}
\end{definition}

Note that the base space $M$ can be viewed as the quotient manifold of $E$ by the group action.

An \emph{isomorphism}\index{bundle!principal!isomorphism} between two principal $G$-bundles over the same base manifold is a homeomorphism of the total spaces which is $G$-equivariant and commutes with the projections to the base space.

One of the most important questions regarding any fiber bundle is whether or not it is trivial, that is, isomorphic to a product bundle. For principal bundles there is a convenient characterization of triviality, which we now describe. Any local trivialization $\varphi_{\alpha}$ -- which is a diffeomorphism -- gives rise to a local section $\sigma \colon U_{\alpha}\to E$, since $\varphi _{\alpha }^{-1}(x,g)\in {{p}^{-1}}({{U}_{\alpha }})$, for any $g\in G$ and $x \in U_{\alpha}$. Conversely, any local section of a principal $G$-bundle gives a local trivialization since the map $p$ is $G$-equivariant. This gives the following characterization:   

\begin{proposition} 
A smooth $G$-principal bundle $(E,p,M,G)$ is trivial, that is, $E\cong M\times G$, if and only if it admits a smooth global section $\sigma\colon M \to E$.
\end{proposition}

This characterization is not true in general for other fiber bundles. For instance, vector bundles always have a zero section whether they are trivial or not and sphere bundles may admit many global sections without being trivial. 

We now focus on the construction of certain classes of bundles starting from $G$-principal bundles. For a principal $G$-bundle $(E,p,M,G)$ and a continuous map $f\colon N \to M$, we can pull-back to get another principal $G$-bundle $(f^{*}E,p',N,G)$. The total space $f^{*}E$ in this case is the fibered product
\[f^{*}E=\{(y,e) \mid f(y)=p(e)\} \subset N \times E\; .\]
The following two examples are of central importance and we will be making frequent use of them within the next chapters:

\begin{example}
Let $(E,p,M,G)$ be a principal $G$-bundle and let $\rho \colon G \to \mathrm{GL}(V)$ be a linear representation of the Lie group $G$ on some vector space $V$. The representation gives a natural left action of the group $G$ on the space $V$. Thus, we may consider the total space $E\times_{G}V$, which is the quotient manifold of $E\times V$ under the equivalence $( e\cdot g,v)\sim ( e,g\cdot v )$, which is a vector bundle over the base space $M$. This will be called the \emph{vector bundle associated to} the principal $G$-bundle $E$ via the representation $\rho$.
\end{example}

An important special case of associated vector bundle is then the following: 
\begin{example}
For a complex semisimple Lie algebra $\mathfrak{g}$, consider the \emph{adjoint representation}\index{representation!adjoint} of the associated Lie group $G$ on $\mathfrak{g}$: 
\begin{align*}
G \times \mathfrak{g} & \to \mathfrak{g} \\
(g,X) & \mapsto gXg^{-1}\; . 
\end{align*}
The vector bundle associated to the principal $G$-bundle $(E,p,M,G)$ via this representation, is called the \emph{adjoint bundle}\index{bundle!adjoint} of $E$ and is denoted by $\mathrm{ad}E$.
\end{example}

Analogously to the case of vector bundles, we may introduce the notion of connection and curvature for principal $G$-bundles. For a Lie group $G$ with corresponding Lie algebra $\mathfrak{g}$, there is a unique 1-form ${{\omega }_{MC}}\in {{\mathcal{A} }^{1}}\left( G,\mathfrak{g} \right)$ which is invariant under left multiplication by $G$ and for the identity element $\mathrm{id} \in G$ it is the linear identification map ${{T}_{\mathrm{id}}}G \to \mathfrak{g}$. This is called the \emph{Maurer--Cartan form}\index{Maurer--Cartan form} and for a point $g \in G$ and a point $v\in {{T}_{g}}G$, it is defined by $${{\omega }_{MC}}(v)={{g}^{-1}}\cdot v\in {{T}_{\mathrm{id}}}G=\mathfrak{g}\; .$$

We may now state the following definition for a connection on a smooth principal $G$-bundle in terms of a differential 1-form:

\begin{definition}
Let $(E,p,M,G)$ be a principal $G$-bundle for a Lie group $G$. A \emph{connection} on $E$ is a 1-form $\omega \in {{\mathcal{A} }^{1}}\left( E,\mathfrak{g} \right)$ on $E$, satisfying the following properties:
\begin{enumerate}
\item Under right multiplication by $G$, $\omega$ transforms via the adjoint representation of $G$ on $\mathfrak{g}$ as 
\[{{\omega }_{eg}}(v\cdot g)=\mathrm{ad}_{g}(\omega_e(v))\; ,\]
for any $e \in E$, any $v \in T_{e}E$ and any point $g\in G$.
\item For the embedding $\iota_e \colon G \to E$ defined at any point $e \in E$ by the rule $g \mapsto eg$, the pullback $\iota_{e}^{*} \omega$ agrees with the Maurer--Cartan form $\omega_{MC}$.
\end{enumerate}
\end{definition}

As in the case of vector bundles, the existence of a connection as above amounts to the existence of solutions of a particular system of ODEs.

For a connection 1-form $\omega \in {{\mathcal{A} }^{1}}\left( E,\mathfrak{g} \right)$ on a principal $G$-bundle $(E,p,M,G)$, we can also define its \emph{curvature}\index{curvature} $F_{\omega}$ as an appropriate 2-form $F_{\omega} \in {{\mathcal{A} }^{2}}\left( M, \mathrm{ad}(E) \right)$. In particular, for $\eta \in {{\mathcal{A} }^{1}}\left( M,\mathrm{ad}(E) \right)$, then $\eta \wedge \eta$ is defined by  $(v,w) \mapsto \frac{1}{2}\left[ \eta (v),\eta (w) \right]$, where $[,\,]\colon \mathrm{ad}(E)\otimes \mathrm{ad}(E)\to \mathrm{ad}(E)$ is the Lie bracket. Then, one has that ${{p}^{*}}{{F}_{\omega }}=d\omega +\omega \wedge \omega $. A principal $G$-bundle endowed with a flat connection is called a \emph{$G$-local system}\index{local system!$G$-}.

The Riemann--Hilbert correspondence directly generalizes to $G$-principal bundles:
\begin{theorem}\index{Riemann--Hilbert correspondence}
There exists an isomorphism between the space of gauge equivalence classes of flat $G$-principal bundles $P$ over a manifold $M$ and conjugacy classes of fundamental group representations $\rho\colon  \pi_1 (M, x) \to G$.
\end{theorem}
\chapter{Hyperbolic geometry}\labelx{Subsec:hypgeom}



\abstract{Every compact surface admits a constant curvature metric, endowing it with a geometry. For surfaces of genus at least 2, this leads to hyperbolic geometry. The model space is the hyperbolic plane. We give a brief introduction to the beautiful world of hyperbolic geometry in dimension 2 starting from the hyperbolic plane and its isometries and then discussing hyperbolic surfaces. References for an extended treatment include, for example, \cite{thurston1979geometry} or \cite{BePe}, \cite{Marden}, \cite{RaRi}.}

\section{The hyperbolic plane}
In dimension 2, there are three types of geometry: Euclidean geometry which is taught at school, spherical and hyperbolic geometry. They are distinguished by the curvature of the underlying space: zero for the Euclidean plane, 1 for the unit sphere and $-1$ for the hyperbolic plane.

To be more precise, the hyperbolic plane $\mathbb{H}$ is the unique complete, simply connected surface with Riemannian metric $\mathrm{d}_{\mathbb{H}}$ of constant Gauss curvature $-1$. One model for the hyperbolic plane is the \emph{Poincar\'{e} upper half-plane}\index{model of $\mathbb{H}$!half-plane} $\left\{z=x+iy\in \mathbb{C} \mid  y>0\right\}$, endowed with the metric:
\begin{equation*}\labelx{Eq:metrichalfplane}
\mathrm{d}s^2 = \frac{\mathrm{d}x^2+\mathrm{d}y^2}{y^2}~\; .
\end{equation*}
The geodesics in the upper half-plane model are vertical lines and half-circles perpendicular to the real axis, and hence any two distinct points in $\mathbb{R}\cup\ \{\infty\}$ are the endpoints of a unique geodesic (\Cref{Fig: hyp_models_lines} on the right). Another model is the \emph{Poincar\'{e} disk}\index{model of $\mathbb{H}$!disk} $\left\{z=x+iy\in \mathbb{C} \mid \vert z \vert < 1 \right\}$, with the metric
\begin{equation*}\labelx{Eq:metricdisk}
\mathrm{d}s^2 = 4\displaystyle\frac{\lvert \mathrm{d}z \rvert^2}{(1-\vert z \vert^2)^2}~\; ,
\end{equation*}
in which geodesics are arcs and lines in the disk perpendicular to the unit circle in $\mathbb{R}^2$, and hence any two distinct points on the unit circle $S^1$ are the endpoints of a unique geodesic (\Cref{Fig: hyp_models_lines} on the left). The map
\begin{equation}\labelx{Eq:mapHPtodisk}
z\longmapsto \frac{z-i}{z+i}
\end{equation}
is an isometry from the half-plane to the disk model of $\mathbb{H}$.

\begin{figure}[!ht]
\centering
\includegraphics[scale=0.83]{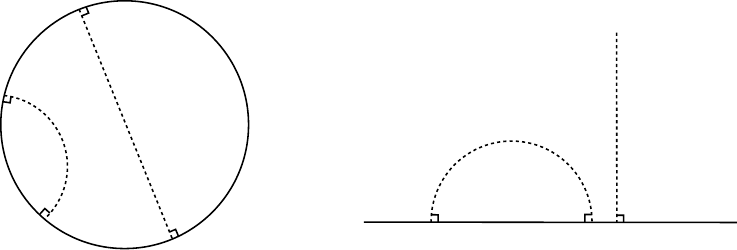}
\caption{On the left, the Poincar\'e disk model of $\mathbb H$. On the right, the upper-half plane model of $\mathbb H$. In both models, some geodesics are drawn in dotted lines.}\labelx{Fig: hyp_models_lines}
\end{figure}

The pictures of the models of $\mathbb{H}$ suggest the natural notion of the boundary at infinity of $\mathbb{H}$.
Let us define an equivalence relation on the space of unit-speed geodesic rays $\gamma\colon [0,\infty)\rightarrow \mathbb{H}$ by imposing that $\gamma$ and $\gamma'$ are equivalent if there exists $D\in\mathbb{R}_{>0}$ such that for all $t\in[0,\infty)$:
\begin{equation*}
\mathrm{d}_{\mathbb{H}}(\gamma(t),\gamma'(t))\leq D~\; .
\end{equation*}

\begin{definition}\labelx{Def:boundaryatinfinity}
	The \emph{boundary at infinity}\index{boundary!at infinity} $\partial_\infty\mathbb{H}$ of $\mathbb{H}$ is the set of equivalence classes of unit-speed geodesic rays in $\mathbb{H}$ for the above-defined equivalence relation. One denotes $\overline{\mathbb{H}}:=\mathbb{H}\cup\partial_\infty \mathbb{H}$.
\end{definition}

The boundary $\partial_\infty\mathbb{H}$ is naturally identified with $\mathbb{R}\cup\{\infty\}=\mathbb{P}^1(\mathbb{R})$ in the half-plane model, and with the unit circle in the disk model. Moreover, the map of \Cref{Eq:mapHPtodisk} extends to the boundary: $\mathbb{P}^1(\mathbb{R})\rightarrow S^1$.

\vspace{0.3cm}

Isometries of $\mathbb{H}$ are naturally represented by elements of $\mathrm{PSL}_2 (\mathbb{R})$ acting on $\mathbb{H}$ by homographies:
\begin{equation}\labelx{Eq:actionpslr}
\left[\begin{array}{cc}
a & b \\ c & d 
\end{array}\right]\cdot z = \frac{az+b}{cz+d}\; ,
\end{equation}
where $z\in \mathbb{C}$ is such that $\mathrm{Im}(z)>0$, and where $ad-bc=1$. This action extends to $\partial_\infty\mathbb{H}$, and hence every isometry of $\mathbb{H}$ is a homeomorphism from the closed unit disk in $\mathbb{C}$ to itself. The Brouwer fixed point theorem\index{Brouwer fixed point theorem} then implies that it has at least one fixed point in $\overline{\mathbb{H}}$. At such a fixed point $z_0$, \Cref{Eq:actionpslr} becomes
\begin{equation}\labelx{Eq:actionpslr2}
P(z):= cz_0^2+(d-a)z_0-b = 0~\; ,
\end{equation}
and three cases can be distinguished:
\begin{itemize}
	\item $P$ has two conjugated roots in $\mathbb{C}$, with exactly one in $\mathbb{H}$. Equivalently, $(d-a)^2+4bc<0$ from which one deduces that $(d+a)^2<4$ since $ad-bc=1$, and hence $\vert d+a \vert<2$. The corresponding elements of $\mathrm{PSL}_2(\mathbb{R})$ are said to be \emph{elliptic}\index{isometry of $\mathbb{H}$!elliptic}. They are rotations about their fixed point.
	\item $P$ has a double real root. This is equivalent to having the absolute value of the trace equal to $2$; the corresponding elements in $\mathrm{PSL}_2(\mathbb{R})$ are said to be \emph{parabolic}\index{isometry of $\mathbb{H}$!parabolic}.
	\item $P$ has two distinct real roots. This is equivalent to having the absolute value of the trace strictly bigger than $2$; the corresponding elements in $\mathrm{PSL}_2(\mathbb{R})$ are said to be \emph{hyperbolic}\index{isometry of $\mathbb{H}$!hyperbolic}. Hyperbolic isometries of $\mathbb{H}$ have a unique geodesic orbit, shown as dotted red lines in \Cref{Fig:typelepslr}.
\end{itemize}

It follows from \Cref{Eq:actionpslr2} that the only element in $\mathrm{PSL}_2(\mathbb{R})$ which fixes three distinct points or more in $\overline{\mathbb{H}}$ is the identity. The orbits of typical elliptic, parabolic and hyperbolic elements in $\mathrm{PSL}_2(\mathbb{R})$ acting on the half-plane or disk models of $\mathbb{H}$ are shown in \Cref{Fig:typelepslr}, where the boundary $\partial_\infty \mathbb{H}$ is shown as a dashed line.

\begin{figure}[!ht]
	\centering
	\includegraphics[scale=0.7]{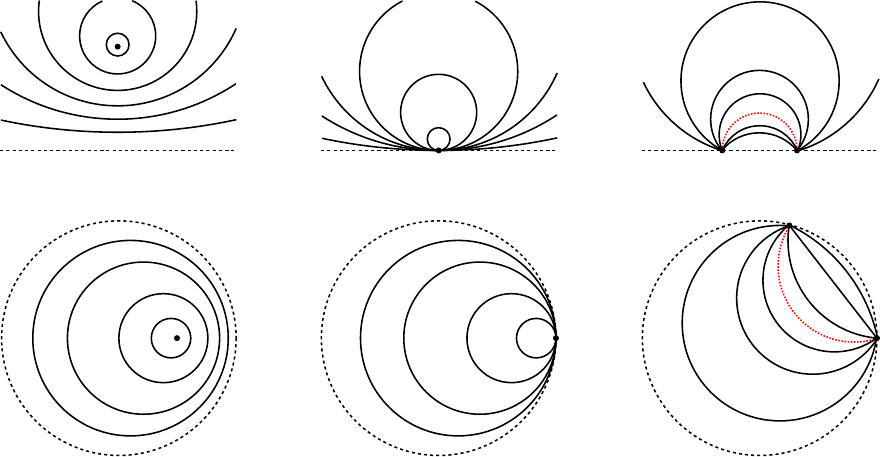}
	\caption{Orbits of an elliptic (left), parabolic (middle) and hyperbolic (right) element in $\mathrm{PSL}_2(\mathbb{R})$.}\labelx{Fig:typelepslr}
\end{figure}

Another tool to study the boundary at infinity of subspaces of the hyperbolic plane are the so-called quasi-geodesics. Roughly speaking these are curves in $\mathbb H$ that approximate the distance good enough. We give a precise definition next:

\begin{definition}
    A map $\gamma\colon \mathbb R\to \mathbb H$ is called a \emph{$(C,D)$-quasi-geodesic}\index{quasi-geodesic} if there exist positive numbers $C,D>0$ such that for all $t_1,t_2\in\mathbb R$, 
    $$C^{-1}\vert t_2-t_1\vert-D\leq \mathrm{d}_{\mathbb H}(\gamma(t_1),\gamma(t_2))\leq C\vert t_2-t_1\vert+D\; .$$ 
\end{definition}

In the same way, a \emph{$(C,D)$-quasi-geodesic ray} can be defined as a map $\gamma\colon [0,+\infty)\to \mathbb H$ satisfying the condition above.

The following Morse lemma shows that every quasi-geodesic stays within bounded distance away from a true geodesic (the proof specializes from more general results in \cite{KLP18}):
\begin{lemma}
    Let $C,D>0$. There exists $M>0$ such that for every $(C,D)$-quasi-geodesic ray $\gamma$ there exists a geodesic ray $\alpha\colon [0,\infty) \to \mathbb H$ such that $\mathrm{d}_{\mathbb H}(\gamma(t),\alpha([0,\infty)))\leq M$ for all $t\in [0,\infty)$.
\end{lemma}

Therefore, we can say that every quasi-geodesic ray represents a point of $\partial_{\infty}\mathbb H$, namely, the one corresponding to a geodesic which stays within bounded distance. If $X\subseteq \mathbb H$ is any subset of $\mathbb H$, we can define $\partial_\infty X\subseteq\H$ as the set of equivalence classes of all quasi-geodesics in $X$ that stay within bounded distance from each other.  If $X$ is a convex subset of $\H$, that is, if for every $p,q\in X$, $X$ contains the geodesic segment connecting $p$ and $q$, then in every equivalence class of quasi-geodesic rays described above there is a geodesic ray.

\section{Compact hyperbolic surfaces}\labelx{sec:hyp.surfaces}

Consider a compact, connected, orientable surface possibly with boundary components. Topologically it is classified by its genus $g$ and the number of boundary components $n$, so we will write $\bar S_{g,n}$ for such a topological surface of genus $g$ with $n$ boundary components.

A surface $\bar S_{g,n}$ is said to be of \emph{hyperbolic type}\index{hyperbolic!type} if its Euler characteristic $\chi(\bar S_{g,n})=2-2g-n$ is negative. In this case $\bar S_{g,n}$ admits a \emph{hyperbolic structure with geodesic boundary}\index{hyperbolic!structure!with geodesic boundary}, that is, a Riemannian metric of curvature $-1$ such that all the boundary components are closed geodesics. This hyperbolic structure is in general not unique. A surface equipped with such a hyperbolic structure is called a \emph{hyperbolic surface}\index{hyperbolic!surface}. 

A closed surface $S:=\bar S_{g,0}$ is of hyperbolic type if and only if $g\geq 2$. This can be seen as follows: similar to the torus which can be obtained by the gluing of a square identifying opposite sides, one can obtain the surface $S$ by gluing the sides of a $4g$-gon. The universal cover of $S$ is then a space tiled by $4g$-gons.  For $g=1$ this gives the plane $\mathbb{R}^2$ tiled by squares, and for $g\geq 2$ one can identify the universal cover of $S$ with the hyperbolic plane which is tiled by $4g$-gons glued by isometries. 

\begin{figure}[!ht]
	\centering
	\includegraphics[scale=1]{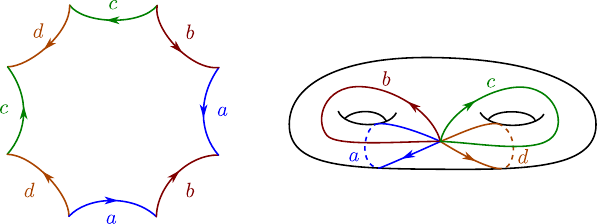}
	\caption{An octagon glued to a genus $2$ surface.}\labelx{Fig:Octogon_gluing}
\end{figure}

The hyperbolic surface $S$ can be recovered from its universal cover $\tilde S$ which is a closed submanifold (possibly with geodesic boundary) of $\mathbb{H}$ by taking the quotient of $\tilde S$ by the deck transformations. Since $\tilde S\subseteq \mathbb{H}$ comes with a natural metric of constant curvature $-1$, and the deck transformations act by isometries, this provides a metric of constant curvature $-1$ on the quotient $S$.

One important property of a hyperbolic surface is that in each free homotopy class of a closed curve, there is a unique geodesic:
\begin{proposition}[Prop. 5.3.1 in \cite{thurston1979geometry}]\labelx{Prop:uniquegeodesic}
	Let $\alpha\in\pi_1(S)$ be a free homotopy class of closed curves in $S$. There exists a unique geodesic in $\alpha$. In the case when $S$ has a non-empty boundary, let $\beta\in\pi_1(S,\{p_0,p_1\})$ be a homotopy class of curves in $S$ from $p_0\in \partial S$ to $p_1\in \partial S$ such that the pre-images of $p_0$ and $p_1$ under the covering map $\mathbb{H}\rightarrow S$ lie on $\partial_\infty \mathbb{H}$, and are considered relatively to $\{p_0,p_1\}$. There exists a unique geodesic in $\beta$.
\end{proposition}

\begin{proof}
	Let $M_\alpha\in \mathrm{PSL}_2(\mathbb{R})$ be the covering automorphism of $\mathbb{H}$ corresponding to $\alpha$. It cannot be elliptic since $S$ is smooth. It cannot be parabolic, for otherwise there would be simple closed curves in $\alpha$ of arbitrarily small length, which goes against the fact that $S$ is compact. Hence $M_\alpha$ is a hyperbolic element of $\mathrm{PSL}_2(\mathbb{R})$, and it preserves a unique geodesic which projects to a closed geodesic on $S$ in the homotopy class $\alpha$. The geodesics $\gamma$ and $\gamma'$ corresponding to two hyperbolic elements $M_\alpha$ and $M_{\alpha'}$ in $\mathrm{PSL}_2(\mathbb{R})$ project to the same geodesic in $S$ if and only if there is a covering transformation which maps the first to the second, namely, $\alpha'=g\alpha g^{-1}$ for $g\in \pi_1(S)$, which is equivalent to saying that  $\alpha'$ and $\alpha$ lie in the same free homotopy class. 
	
	Let $\tilde p_0$ (respectively $\tilde p_1$) be a pre-image of $p_0$ (respectively $p_1$) under the covering map $\mathbb{H}\rightarrow S$. The points $\tilde p_0$ and $\tilde p_1$ are distinct points on $\partial_\infty\mathbb{H}$. There exists a one-parameter family of hyperbolic elements in $\mathrm{PSL}_2(\mathbb{R})$ which preserves $\tilde p_0$ and $\tilde p_1$ and fixes a unique geodesic from $\tilde p_0$ to $\tilde p_1$ (again shown as dotted red lines on the right of \Cref{Fig:typelepslr}). This geodesic projects to a geodesic on $S$ in the class $\beta$. If $\tilde p_0'$ and $\tilde p_1'$ are two other lifts of $p_0$ and $p_1$ in $\partial_\infty\mathbb{H}$, the unique geodesic from $\tilde p_0'$ to $\tilde p_1'$ projects to a geodesic in $\beta$ if and only if it is related to the unique one from $\tilde p_0$ to $\tilde p_1$ by a covering transformation, which implies that both geodesics in $\mathbb{H}$ project to the same one in $S$.
\end{proof}

\begin{remark}
    There is also an argument using the Gauss--Bonnet formula\index{Gauss--Bonnet formula} to prove that there cannot exist two geodesics in the same homotopy class. Assume the contrary, and let $\gamma_1$ and $\gamma_2$ be two distinct geodesics and consider a homotopy $H$ between them. On the one hand we know that the image of $H$ is topologically a cylinder which has Euler characteristic 0. On the other hand, the Gauss--Bonnet formula provides that the Euler characteristic is strictly negative, since the boundary of $H$ is geodesic (so there is no contribution from the extrinsic curvature) and the interior has constant negative curvature. This gives the contradiction.
\end{remark}

The fundamental group of a surface $\bar S_{g,n}$ (not necessarily of hyperbolic type) is always finitely presented. Namely,
\begin{equation}\labelx{fund_gp_presentation}
{{\pi }_{1}}(\bar S_{g,n})=\left\langle {{a}_{1}},{{b}_{1}},\ldots ,{{a}_{g}},{{b}_{g}},c_1,\ldots,c_n \left\vert~ \displaystyle\prod_{i=1}^g{[ {{a}_{i}},{{b}_{i}]}\prod_{i=1}^n c_i=1} \right.\right\rangle\; .
\end{equation}
In particular, the fundamental group is always free when $n>0$. For a hyperbolic surface $S$, the fundamental group $\pi_1(S)$ can be embedded into the universal covering $\tilde{S}\subseteq\mathbb H$ as follows: we choose a closed connected fundamental domain $S_0\subset \tilde S$ and a point $p\in S_0$. Since $S$ is compact, $S_0$ is also compact. The group  $\pi_1(S)$ acts on $\tilde S$ by deck transformations (see  \Cref{sec:deck_tran}). Let $P$ be the orbit of $p$ under this action. Further, for every $p_1,p_2\in P$ we connect them by a geodesic segment if $p_2=h(p_1)$ where $h$ is one of the elements $a_1,\dots a_g$, $b_1,\dots b_g$, $c_1,\dots c_n$ presenting the fundamental group. The resulting subspace in $\mathbb H$ (which is, in fact, topologically the Cayley graph\index{Cayley graph} of $\pi_1(S)$ with respect to the generating set $a_1,\dots a_g$, $b_1,\dots b_g$, $c_1,\dots c_n$) will be denoted by $\Gamma$. Notice that for every $h\in\pi_1(S)$ there exists a shortest word in the variables $a_1,\dots a_g$, $b_1,\dots b_g$, $c_1,\dots c_n$ presenting $h$. Every such word induces a path in $\Gamma$ which is a quasi-geodesic ray because the fundamental domain $S_0$ is compact and hence bounded. 

We may therefore define the \emph{boundary}\index{boundary!at infinity}  $\partial_\infty\Gamma$, which, in fact, agrees with $\partial_\infty \tilde S$. In particular, if $S$ is closed, then $\partial_\infty \Gamma=\partial_\infty\tilde S=\partial_\infty \mathbb H\cong S^1$. If $S$ has boundary components, then $\tilde S\subset \mathbb H$ and $\partial_\infty \Gamma=\partial_\infty\tilde S\subset\partial_\infty \mathbb H$ is a Cantor set embedded into $S^1$ and hence equipped with a cyclic order that defines the topology on $\partial_\infty \Gamma$. The group $\pi_1(S)$ acts on $\partial_\infty \Gamma$ preserving the cyclic order.

Actually, since any fundamental domain $S_0$ of $S$ is bounded, this construction does not depend on the chosen base point $p$ and on the choice of the fundamental domain. Moreover, the topology and the cyclic order of $\partial_\infty \Gamma$ do not depend on the choice of the hyperbolic structure on $S$. The only thing that depends on the hyperbolic structure is the embedding of $S^1$ in the case of closed surfaces, or the embedding of the Cantor set in the case of surfaces with boundary into $\partial_\infty \mathbb H$.

So we can define the topological space $\partial\pi_1(S)$ with a cyclic order as $\partial_\infty \Gamma$ for some choice of a hyperbolic structure on $S$. The topology on $\partial\pi_1(S)$ is defined using the cyclic order. Note, finally, that the group $\pi_1(S)$ acts on $\partial\pi_1(S)$ preserving the cyclic order. The space $\partial\pi_1(S)$ is called the \emph{Gromov boundary}\index{boundary!Gromov} of the group $\pi_1(S)$.

A more detailed survey on the definition and known results regarding the boundary of word-hyperbolic groups may be found in \cite{KaBe}.

\section{Ciliated surfaces}\labelx{Sec:ciliatedsurfaces}

The notion of \emph{ciliated surfaces}\index{surface!ciliated} refers to hyperbolic surfaces which are not compact but have  additional data on the boundary of the surface. This extra data is called \emph{cilia}\index{cilia} and can be seen as points at infinity. Moreover, the boundary segments connecting cilia are in this case infinite geodesics. In this Section we introduce rigorously marked and ciliated surfaces; the former are natural compactifications of ciliated surfaces. We also discuss here the ideal triangulations of these surfaces.

\begin{definition}
	A \emph{marked surface}\index{surface!marked} is a compact oriented smooth surface obtained from a closed oriented smooth surface by removing a finite number of disjoint open disks and by marking a finite set of points so that on each boundary component at least one point is marked. These points are called \emph{marked points}\index{marked point}. A marked point is called \emph{internal} if it lies in the interior of the surface. Otherwise it is called \emph{external}.
\end{definition}

Let $k\in\Z_{\geq0}$ be the number of disjoint open disks one removes from the closed oriented smooth surface $\bar S_g$ of genus $g$, and let us label the boundary components of the resulting surface by $1,\dots,k$. Let $p_i\in\Z_{>0}$ be the number of marked points on the $i$-th boundary circle, for $i\in\{1,\dots,k\}$. We also denote by $p_0$ the number of internal marked points. The resulting ciliated surface is completely determined by $g$ and the vector\footnote{We emphasize the difference in the nature of $p_0$ compared to the $p_i$'s for $i \geq 1$ by separating $p_0$ and $p_1$ with a semi-colon.} $\vec p:=(p_0;p_{1},\dots,p_k)\in\mathbb Z_{\geq 0}\times\mathbb Z_{>0}^{k}$ as a smooth (equivalently, topological) surface, and hence we will denote it by $\bar S_{g,\vec p}$ (and $\bar S_{g,p_0}$ for short when there is no external marked points). Three examples of marked surfaces are shown in \Cref{Fig:exciliatedsurf}.

\begin{figure}[!ht]
	\centering
	\includegraphics[width=\textwidth]{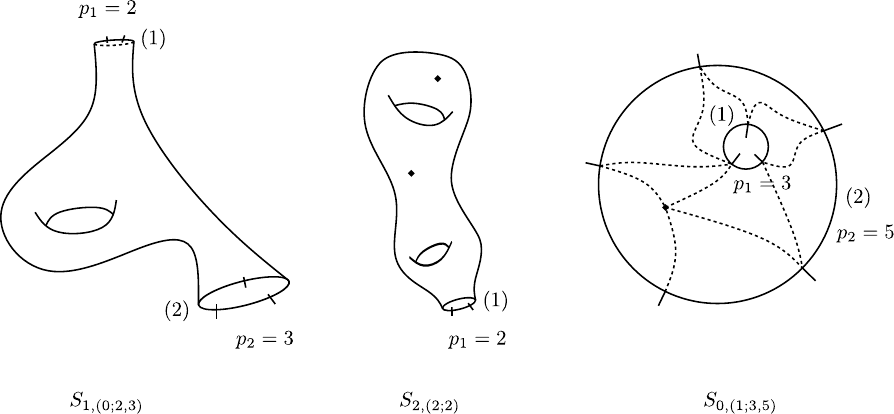}
	\caption{Three ciliated surfaces (the labels of boundary components are shown in parentheses).}\labelx{Fig:exciliatedsurf}
\end{figure}

\begin{definition}
    A \emph{ciliated surface}\index{surface!ciliated} $S_{g,\vec p}$ is a surface obtained from a marked surface $\bar S_{g,\vec p}$ by removing all the marked points. Formally, if $P$  denotes the set of all marked points of $\bar S_{g,\vec p}$, then $S_{g,\vec p}:=\bar S_{g,\vec p}\setminus P$. Points of $P$ are called punctures of $S_{g,\vec p}$. Internal marked points of $\bar S_{g,\vec p}$ are called \emph{internal punctures}\index{puncture!internal} of $S_{g,\vec p}$. External marked points of $\bar S_{g,\vec p}$ are called \emph{external punctures} or \emph{cilia}\index{puncture!external} of $S_{g,\vec p}$.
\end{definition}

In the sequel, unless explicitly specified, we will assume every ciliated surface to be non-closed, i.e. $\#P\geq 1$, and hyperbolic, i.e. $\chi(S_{g,\vec p})<0$. These conditions are equivalent to saying that either $g=0$ and $\#P\geq 3$, or $g\geq 1$ and $\#P\geq 1$. 

\begin{remark}\labelx{Def:boundaryciliated}
	The boundary of a ciliated surface\index{boundary!of a ciliated surface} is the disjoint union of the boundary segments connecting two adjacent cilia.
\end{remark}

A ciliated surface can be equipped with a marked, possibly incomplete, convex, finite area hyperbolic structure possibly with boundary such that the boundary segments between cilia are infinite geodesics and cilia are points at infinity, in other words, neighborhoods of these points are of infinite diameter. We introduce the following definition.

\begin{definition}\labelx{def:length_puncture}
For a ciliated surface equipped with a hyperbolic structure as above, we define the \emph{length of an internal puncture}\index{puncture!internal!length} to be the infimum of the lengths of all simple loops surrounding the puncture.     
\end{definition}

For every puncture, this length is either positive or zero. If a puncture has length zero, then there exists a neighborhood of this puncture homeomorphic to a punctured disk which is of infinite diameter and complete as a metric space. The puncture in this case is a point at infinity and will be called a \emph{cusp}\index{cusp}. If a puncture has positive length $l>0$, then every neighborhood of this puncture which is homeomorphic to an annulus is incomplete as a metric space and has finite diameter. The metric completion of such a neighborhood adds a closed boundary geodesic of length $l$ to the annulus making it compact; see \Cref{Fig:ciliated_truncated} for an example.

\begin{figure}[!ht]
	\centering
	\includegraphics{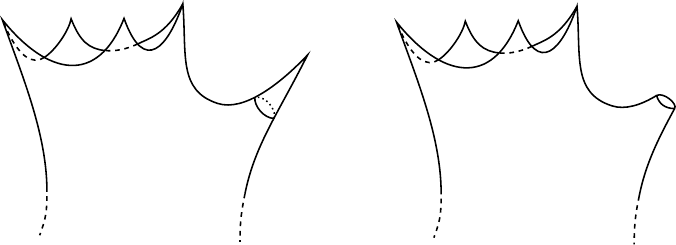}
	\caption{Example of two hyperbolic ciliated surfaces: with a cusp (left) and with a boundary geodesic (right).}\labelx{Fig:ciliated_truncated}
\end{figure} 

As a non-compact manifold,  a ciliated surface $S_{g,\vec p}$ can be equipped with a hyperbolic metric in many different non-quasi-isometric ways. This implies that the universal covering of $S_{g,\vec p}$, seen as a subspace of the hyperbolic plane, will have different boundaries at infinity depending on the chosen hyperbolic structure. This distinguishes ciliated surfaces from compact surfaces and makes it impossible to define the boundary at infinity of the universal covering independently of the hyperbolic structure chosen. We will discuss two different ways of doing this. A more rigorous study of hyperbolic structures on ciliated surfaces is provided in \Cref{Subec:Shearcoord} where the Teichm\"uller space for ciliated surfaces is introduced. 

\begin{remark}\labelx{complex_str_ciliated}
    When such a hyperbolic structure on the ciliated surface $S_{g,\vec p}$ has been chosen, the universal cover $\tilde{S}_{g,\vec p}$ can be identified with a convex subset of $\mathbb{H}$ with geodesic boundary. We endow $S_{g,\vec p}$ with the natural complex structure induced by the one on $\mathbb{H}$.
\end{remark}

Every ciliated surface $S_{g,\vec p}$ can be equipped with a convex hyperbolic structure of finite volume with totally geodesic boundary such that every internal puncture admits a neighborhood of finite diameter. In particular, every such structure is incomplete as a metric space in neighborhoods of internal punctures. With respect to such a structure, all boundary curves of $S_{g,\vec p}$ are bi-infinite geodesics connecting cilia. Once equipped with a hyperbolic structure as above, the universal covering $\tilde S_{g,\vec p}$ of $S_{g,\vec p}$ can be seen as a convex subset of the hyperbolic plane $\mathbb H^2$, which we denote by $\Sigma$. Its closure $\bar\Sigma$ in $\mathbb H^2$ has totally geodesic boundary, which is invariant under the natural action of $\pi_1(S_{g,\vec p})$ on $\mathbb H^2$ by the holonomy representation. The space $\bar\Sigma\setminus\Sigma$ is a collection of infinite geodesics on the boundary of $\bar\Sigma$. The ends of the boundary geodesics of $\bar \Sigma$ are points in the ideal boundary of $\mathbb H^2$ which are called \emph{cilia}\index{cilia} of $\Sigma$. The set of all cilia will be denoted by 
\[\tilde P\subseteq \partial_\infty \Sigma=\partial_\infty \bar \Sigma\subseteq \partial_\infty\mathbb H^2\; .\] 
For any two hyperbolic structures as above the spaces $\partial_\infty \Sigma$ are isomorphic as cyclically ordered topological spaces. Thus, we denote this set by $\partial_\infty \tilde S_{g,\vec p}$ and call it the \emph{boundary at infinity of the universal covering of $S_{g,\vec p}$}\index{boundary!at infinity}. Notice that $\tilde P = \partial_\infty \tilde S_{g,\vec p}$ if and only if $g=0$ and $\vec p=(0,n)$ for some $n>0$, i.e. the corresponding marked surface $\bar S_{g,\vec p}$ is a polygon. This happens if and only if $\tilde P$ is finite. Otherwise, $\tilde P$ is infinite countable and $\partial_\infty \tilde S_{g,\vec p}$ is a Cantor set in $\partial_\infty\mathbb H^2$.

Similarly, every ciliated surface $S_{g,\vec p}$ can be equipped with a complete hyperbolic structure of finite  volume with totally geodesic boundary. Every such hyperbolic structure turns $S_{g,\vec p}$ into a complete metric space and every neighborhood of each puncture has infinite diameter. All boundary curves of $S_{g,\vec p}$ are bi-infinite geodesics connecting cilia. Once equipped with a hyperbolic structure as above, the universal covering $\tilde S_{g,\vec p}$ of $S_{g,\vec p}$ can be seen as a closed convex subset of the hyperbolic plane $\mathbb H^2$, which we denote by $\Sigma'$. The space $\Sigma'$ has totally geodesic boundary and is invariant under the natural action of $\pi_1(S_{g,\vec p})$ on $\mathbb H^2$ by the holonomy representation. The ends of the boundary geodesics of $\tilde S_{g,\vec p}$ are points of the ideal boundary of $\mathbb H^2$ which are called \emph{cilia} of $\Sigma'$. The collection of all cilia of $\Sigma'$ will be denoted by 
\[\tilde P\subseteq \partial_\infty \Sigma'\subseteq \partial_\infty\mathbb H^2\; .\]
For any two hyperbolic structures as above, the spaces $\partial_\infty \Sigma'$ are isomorphic as cyclically ordered topological spaces. Thus, we denote this space by $\partial_\infty^{\mathrm{red}} \tilde S_{g,\vec p}$ and call the \emph{reduced boundary at infinity of the universal covering of $S_{g,\vec p}$}\index{boundary!at infinity!reduced}. If $S_{g,\vec p}$ has no internal punctures, i.e. $p_0=0$, then $\partial_\infty^{\mathrm{red}} \tilde S_{g,\vec p}=\partial_\infty \tilde S_{g,\vec p}$. However, if there are internal punctures, then the boundary $\partial_\infty^{\mathrm{red}} \tilde S_{g,\vec p}$ can be seen as a quotient of $\partial_\infty \tilde S_{g,\vec p}$ by the following equivalence relation: $p_1\sim p_2$ if and only if $p_1$ and $p_2$ are two ends of a geodesic of $\bar\Sigma\setminus \Sigma$. Since between $p_1$ and $p_2$ there are no further points of  $\partial_\infty \tilde S_{g,\vec p}$, this equivalence relation preserves the cyclic order, i.e. the quotient map $\partial_\infty \tilde S_{g,\vec p}\to \partial_\infty^{\mathrm{red}} \tilde S_{g,\vec p}$ is order preserving, continuous and surjective.

Lastly, we give a more detailed description of the complex structure induced on the compactification of a surface with punctures $S$ equipped with a complex structure in the sense of \Cref{complex_str_ciliated}. Namely, 
for a set of distinct points $P=\{s_1,...,s_k\}$ with $k \geq 1$ on $S$, there exists an open subset $U_P :=U_{s_1} \cup \cdots \cup U_{s_k} \subset S$, such that $S\setminus U_P$ is compact and each neighborhood $U_{s_i}$ of $s_i$ is homeomorphic to a punctured disk $z_i\colon U_i \to \mathbb{D}_2 \setminus \{0\} = \{z \in \mathbb{C} \mid 0< \vert z \vert < 1\}$, for $i = 1,...,k$, where the homeomorphisms $z_i$ are holomorphically compatible with the complex structure on $S$. The maps $z_i$ can be extended to 
\begin{equation}\labelx{ext_hol_coords}
	z_i\colon \hat{U}_{s_i}:=U_{s_i} \cup \{s_i\} \to \mathbb{D}_2=\{z \in \mathbb{C} \mid \vert z\vert <1\}\; ,
\end{equation}
by setting $z_i(s_i)=0$, for each $i=1,...,k$. Then, a complex atlas for the new Riemann surface $\bar{S}:=S \cup P$, the compactification of the Riemann surface $S$ with punctures, is given by the union of a complex atlas $\mathcal{A}$ of $S$ with the coordinate charts (\ref{ext_hol_coords})
compatible with $\mathcal{A}$.

\section{Hyperbolic triangles, hexagons and pairs of pants}\labelx{sec: triangles_hexagons}

Ideal triangles are of fundamental importance in hyperbolic geometry and are widely employed into the study of hyperbolic surfaces and the geometry of Teichm\"{u}ller space. In \Cref{sec: Fenchel--Nielsen}, we will see how global coordinates on the Teichm\"{u}ller space can be obtained from the decomposition of a hyperbolic surface into simpler geometric pieces. These geometric pieces are often called \emph{three-circle domains}\index{three-circle domain} or \emph{$Y$-pieces}\index{$Y$-piece} or \emph{pairs of pants}\index{pair of pants} in the literature. Among these terms, we adopt  the latter in the text and review in this Section a gluing construction of pairs of pants out of right-angled hexagons. A central result is then that the moduli space of hyperbolic structures on a pair of pants is, in fact, trivial; we refer to \cite{Papadopoulos} for a survey on relevant results and methods.

\begin{definition}\labelx{def:ideal_triangle}
Let $x,y,z$ be distinct points on the boundary at infinity $\partial_\infty\mathbb{H}$ of the hyperbolic plane $\mathbb{H}$. An \emph{ideal triangle}\index{ideal!triangle} with vertices at the points $x,y,z$ is the closed subset of $\mathbb{H}$ bounded by the three geodesics that join pairwise the points $x,y,z$ (see \Cref{Fig:ideal_triangle}). Note that any two ideal triangles are isometric. 
\end{definition}

\begin{figure}[!ht]
	\centering
	\includegraphics[scale=1]{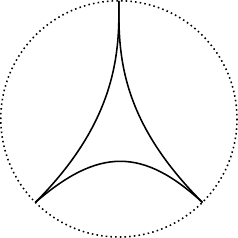}
	\caption{An ideal triangle in the Poincar\'{e} disk model as a fundamental example of a ciliated surface. }\labelx{Fig:ideal_triangle}
\end{figure}

\begin{definition}
A \emph{right-angled hexagon} $H \subset \mathbb{H}^2$ is a compact simply connected subset of the hyperbolic plane whose boundary consists of six geodesic segments $\gamma_i$, $i=1,...,6$, that meet each other orthogonally (see \Cref{Fig:right-angled hexagon}).
\end{definition}

\begin{figure}[!ht]
	\centering
	\includegraphics[width=\textwidth]{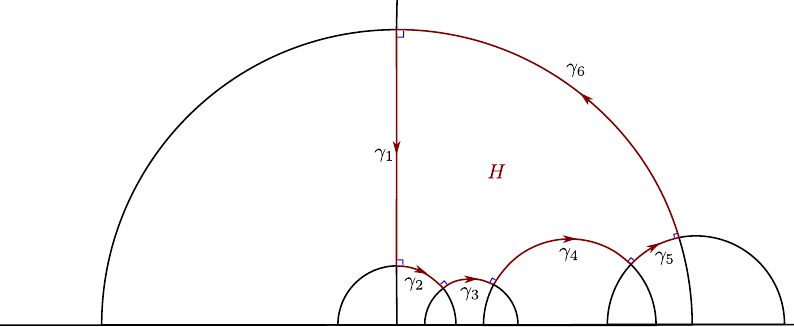}
	\caption{A right-angled hexagon $H \subset \mathbb{H}^2$ in the Poincar\'{e} half plane model.}\labelx{Fig:right-angled hexagon}
\end{figure}

It is not hard to show that any right-angled hexagon is uniquely determined up to isometry by the lengths of three of its non-consecutive sides (see \cite[Lem. 4.3.2]{Jost}).

\begin{definition}
Let $l_1, l_2, l_3 \in [0, \infty)$. A \emph{hyperbolic pair of pants}\index{hyperbolic!pair of pants} of type $(l_1,l_2,l_3)$ is a ciliated surface of genus 0 with three internal punctures equipped with a hyperbolic structure such that the punctures have length equal to $l_1, l_2, l_3$ in the sense of \Cref{def:length_puncture}    (see \Cref{Fig:pair_of_pants}). 
\end{definition}

\begin{figure}[!ht]
	\centering
	\includegraphics[scale=1]{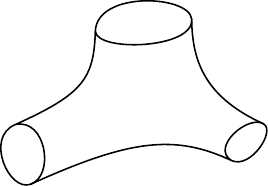}
	\caption{A pair of pants.}\labelx{Fig:pair_of_pants}
\end{figure}

Note that in the definition of a hyperbolic pair of pants cuspidal boundaries are allowed, that is when the puncture  lengths $l_i =0$.
There are two combinatorially distinct ways to glue two ideal triangles along their boundaries to get a pair of pants as a topological space; these two ways are described in 
\Cref{Fig:gluing_ideal_triangles}. From a geometric point of view, these two combinatorial gluings may give four different types of hyperbolic pairs of pants, depending on whether these are complete or have 1, 2 or 3 cusps (see \Cref{Fig:four_geom_pop}).

\begin{figure}[!ht]
	\centering
	\includegraphics[width=\textwidth]{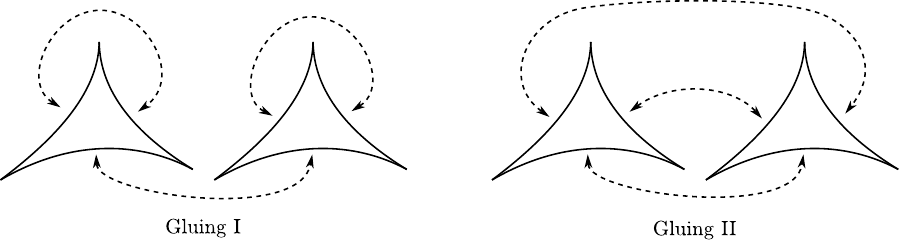}
	\caption{Two ways to get a pair of pants from gluing two ideal triangles.}\labelx{Fig:gluing_ideal_triangles}
\end{figure}

\vspace{5mm}

\begin{figure}[!ht]
	\centering
	\includegraphics[width=\textwidth]{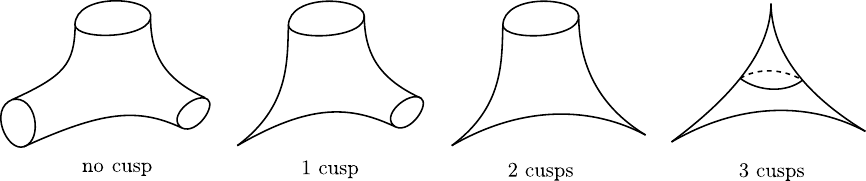}
	\caption{Four different hyperbolic pairs of pants obtained by gluing two ideal triangles.}\labelx{Fig:four_geom_pop}
\end{figure}

\begin{theorem}\labelx{thm:existence_hyp_pair_pants}
For any triple $(l_1, l_2, l_3)$ of non-negative real numbers, there exists a unique hyperbolic pair of pants $P$ with punctures of lengths $l_i$, for $i=1,2,3$ respectively.
\end{theorem}

In the case when there are no cusps, that is, whenever $l_i \neq 0$ for each $i=1,2,3$, \Cref{thm:existence_hyp_pair_pants} above can be obtained from the existence and uniqueness of hexagons for given lengths of three of its consecutive sides. In particular, we first consider a pair of right-angled hexagons with lengths $\frac{l_1}{2}$, $\frac{l_2}{2}$, $\frac{l_3}{2}$
of alternating sides. These hexagons are uniquely determined up to isometry and we glue them together along their remaining sides. This gives a compact hyperbolic surface of genus 0 with three boundary components of length $l_1$, $l_2$, $l_3$ respectively. Removing these boundary components turns this compact surface into a ciliated surface with punctures of length $l_1$, $l_2$, $l_3$ respectively. For a full proof the reader is referred to \cite[Thm. 4.3.1]{Jost}.

Now consider the moduli space of all hyperbolic pairs of pants $P$ with labeled punctures of fixed length $l_1$, $l_2$, $l_3$ modulo isometry. We denote the isometry class of $P$ by $[P]$. The fact that every label-preserving self-homeomorphism on a pair of pants is homotopy equivalent to the identity map, implies the following:

\begin{corollary}
The moduli space of hyperbolic structures on a hyperbolic pair of pants $P$ with fixed length of punctures is trivial $\{[P]\}$.
\end{corollary}

In the case when all punctures have positive length, one can take the metric completion of the pair of pants $P$ and obtain a compact hyperbolic surface $\bar{P}$ with three geodesic boundaries. In fact, the only simple closed geodesics on the compactified pair of pants  $\bar{P}$ are the three geodesics $\gamma_1, \gamma_2, \gamma_3$ in the boundary of $\bar{P}$. Moreover, there is a unique simple geodesic arc denoted by $c_{ij}=c_{ji}$, called an orthogeodesic, joining the boundaries $\gamma_i, \gamma_j$ (see \Cref{Fig:pop_orthogeodesics}). Therefore, the hyperbolic structure of the compactified pair of pants is uniquely determined by the lengths of its three boundary curves. 

\begin{figure}[!ht]
	\centering
	\includegraphics[scale=0.83]{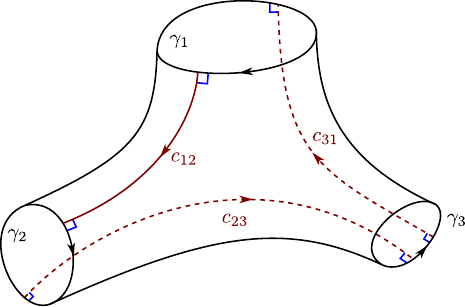}
	\caption{A pair of pants with its orthogeodesics.}\labelx{Fig:pop_orthogeodesics}
\end{figure}

Furthermore, we can cut the compactified pair of pants along the geodesic arcs $c_{12}$, $c_{23}$, $c_{31}$ and obtain two hyperbolic right-angled hexagons (see  \Cref{Fig:cutting_pop}).

\begin{figure}[!ht]
	\centering
	\includegraphics[width=\textwidth]{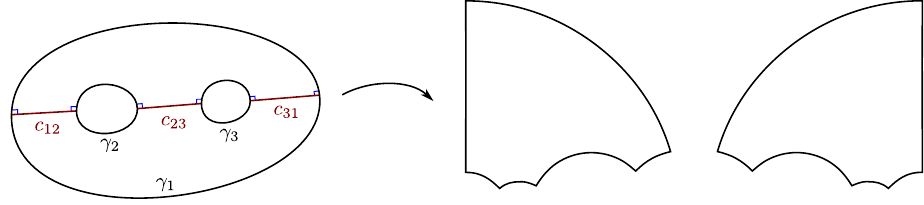}
	\caption{Cutting a pair of pants along geodesic arcs.}\labelx{Fig:cutting_pop}
\end{figure}

\section{Triangulations}\labelx{sec:hyptriang}

Ciliated surfaces can be always decomposed into simpler pieces, namely, into ideal triangles\index{ideal!triangle} as introduced in \Cref{def:ideal_triangle}. This decomposition will become particularly useful for a number of topics and applications discussed in the sequel of the book. 

\begin{definition}\labelx{Def:triang}
	A \emph{triangulation}\index{triangulation} $\Delta$ of a marked surface $\bar S_{g,\vec p}$ such that $\#P\geq 1$ and $\chi(S_{g,\vec p})<0$ is a maximal isotopy class of non self-intersecting in the interior, pairwise non-intersecting and non-isotopic curves on $\bar S_{g,\vec p}$ called \emph{edges}\index{triangulation!edge} whose endpoints are marked points. A triangulation of $\bar S_{g,\vec p}$ decomposes the marked surface into topological triangles such that each vertex is a marked point.
\end{definition}

Every triangulation $\Delta$ of a marked surface $\bar S_{g,\vec p}$ can be turned into a geodesic \emph{ideal triangulation}\index{ideal!triangulation}\index{triangulation!ideal} of the ciliated surface $S_{g,\vec p}$ whenever a hyperbolic structure on $S_{g,\vec p}$ as above is chosen. This means that every edge of $\Delta$ is represented by a unique infinite geodesic connecting two punctures of $S_{g,\vec p}$. In particular, all the boundary geodesics are contained in $\Delta$. A triangulation of $S_{g,\vec p}$ decomposes it into ideal geodesic triangles.

A triangulation of $S_{0,(1;3,5)}$ is shown on the right of \Cref{Fig:exciliatedsurf}.
If $\Delta$ is a triangulation of a marked or a ciliated surface, the endpoints of its edges are called \emph{vertices}\index{triangulation!vertex}, while the connected components of $S_{g,\vec p}\setminus \Delta$ are called \emph{faces}\index{triangulation!face}. One can distinguish the {\it external edges}\index{triangulation!edge!external}, which belong to the boundary of $S_{g,\vec p}$, from the ones belonging in the interior of $S_{g,\vec p}$, dubbed {\it internal edges}\index{triangulation!edge!internal}.

Let $V(\Delta)$, $E(\Delta)$, $F(\Delta)$, $E_e(\Delta)$, $E_i(\Delta)$ be respectively the set of vertices, edges, faces, external edges and internal edges of $\Delta$. 
Let also $c:=\sum_{i=1}^k p_i$ be the total number of cilia of $S_{g,\vec p}$. One has:
\begin{align*}
\#E_e(\Delta) &= c\; , \\
\#V(\Delta) &= p_0+c=\sum_{i=0}^k p_i\; .
\end{align*}
The Euler characteristic of $S_{g,\vec p}$ is
\begin{equation*}
\#F(\Delta)-\#E(\Delta)+\#V(\Delta) = 2-2g-p_0-k\; ,
\end{equation*}
and since $\Delta$ decomposes $S_{g,\vec p}$ in triangles
\begin{equation*}
3\#F(\Delta) = 2\#E(\Delta)-\#E_e(\Delta) = 2\#E(\Delta)-c\; ,
\end{equation*}
then from these two last equations one deduces that:
\begin{align*}
\#E(\Delta) &= 6g-6+2c+3k+3p_0\; , \\
\#E_i(\Delta) &= 6g-6+c+3k+3p_0\; , \\
\#F(\Delta) &= 4g-4+c+2k+2p_0\; .
\end{align*}
Hence $\#E(\Delta)$, $\#E_e(\Delta)$, $\#E_i(\Delta)$, $\#V(\Delta)$, $\#F(\Delta)$ are entirely determined by the topology of $S_{g,\vec p}$.

\vspace{0.3cm}

\begin{definition}\labelx{def:pure_Mod_marked}
The \emph{pure mapping class group}\index{mapping class group} $\mathrm{PMod}(\bar{S}_{g,\vec{p}})$ of a marked surface $\bar{S}_{g,\vec{p}}$, that is, the group of all orientation-preserving diffeomorphisms of $\bar{S}_{g,\vec{p}}$ fixing all marked points of the surface modulo the diffeomorphisms which are isotopic relative to boundary to the identity, acts naturally on ideal triangulations.
\end{definition}

\begin{remark}\labelx{total_num_triang}
Note that the total number of ideal triangulations of a marked surface is in general infinite, however, up to the action of $\mathrm{PMod}(\bar{S}_{g,\vec{p}})$ on the surface their number is finite. 
\end{remark}

\begin{remark}
    In the framework of class~$\mathcal S$ theories, as presented in \Cref{Chap:theoriesofclassS}, the mapping class group of the \emph{UV curve} naturally realizes the theory’s \emph{$S$-duality group}.
\end{remark}

The following classical result will be of prime importance to us in order to construct later on in \Cref{Subec:Shearcoord} coordinates on certain spaces that are realized over a ciliated surface $S_{g,\vec p}$. These coordinates will be associated to the edges of an ideal triangulation $\Delta$ of $S_{g,\vec p}$, and so in order to compute the transition functions between any two charts, i.e. for any two ideal triangulations of $S_{g,\vec p}$, one only needs to know what the transition functions are when the two triangulations are related by a mere flip of an internal edge.

\begin{theorem}[\cite{mosher1988tiling,hatcher1991triangulations,burman1999triangulations}]\labelx{Thm:finiteseqflips}
	Any two triangulations of a ciliated surface can be related through a finite sequence of flips\index{flip} (or Whitehead moves\index{Whitehead move}) displayed in \Cref{Fig:flip}.
\end{theorem}

\begin{figure}[!ht]
	\centering
	\includegraphics[width=\textwidth]{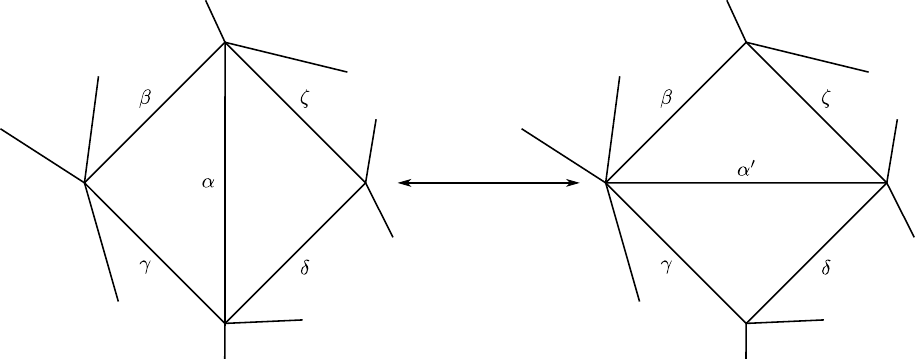}
	\caption{A flip of a triangulation at the inner edge $\alpha$.}\labelx{Fig:flip}
\end{figure}

\section{The Riemann--Hilbert correspondence for framed bundles}

An important generalization of the classical Riemann--Hilbert correspondence, which will be used later in ~\Cref{sec:non-abel}, is the Riemann--Hilbert correspondence for so-called framed vector bundles over ciliated surfaces.

Let us now denote a ciliated surface\index{surface!ciliated} $S_{g,\vec p}$ simply by $S$. Let $\bar S$ be the corresponding compact marked surface, such that $S=\bar S\setminus P$ for a finite set $P$ of punctures of $S$. For every $p\in P$ we fix a neighborhood $\bar S_p\subset \bar S$ of $p$, which is diffeomorphic to a punctured disk with $p$ in the center of it if $p$ is a puncture, and to a half-disk with $p$ as the midpoint of its diameter if $p$ is a cilium. We denote $S_p:=\bar S_p\setminus \{p\}$. 

Let $E\to S$ be a smooth flat vector bundle of rank $n$ over $S$. Let $p\in P$, and $\mathcal E_p=(E_0,\dots,E_{k_p})$ be a sequence of vector subbundles of $E|_{S_p}$ such that $0\leq\mathrm{rk}(E_{i-1})<\mathrm{rk}(E_{i})\leq n$ for all $i\in\{1,\dots,k_p\}$. We say that $\mathcal E_p$ is a \emph{framing of $E$ at $p$}\index{framing} if each $E_i$ is parallel. A family $\mathcal E=\{\mathcal E_p\}_{p\in P}$ is called a \emph{framing of $E$} if for every $p\in P$, $\mathcal E_p$ is a framing of $E$ at $p$. A pair $(E,\mathcal E)$ is called a \emph{framed vector bundle}\index{framed!vector bundle}\index{vector bundle!framed} if $\mathcal E$ is a framing of $E$.

As we have seen, the classical Riemann--Hilbert correspondence provides a 1-1 correspondence between gauge equivalence classes of rank $n$ vector bundles over a manifold $M$ and conjugacy classes of the representations of the fundamental group of $M$ into $\mathrm{GL}_n(\mathbb C)$. To provide a similar correspondence for framed vector bundles, we need to equip fundamental group representations with additional information also called \emph{framing}.

Let $\rho\colon \pi_1(S)\to\mathrm{GL}_n(\mathbb C)$ be a homomorphism of the fundamental group of a ciliated surface $S$ as above. Let $\mathcal F$ be the space of all partial flags of $\mathbb C^n$, that is, 
$$\mathcal F:=\{(V_0,\dots,V_k)\mid 0\leq k\leq n,\;V_k\leq\mathbb C^n,\;\dim(V_{i-1})<\dim(V_{i}),\; \forall\; 1\leq i\leq k\}\; .$$
Let $\tilde S$ be the universal covering of $S$, let $\partial_\infty^{\mathrm{red}}\tilde S$ be its reduced boundary at infinity and $\tilde P$ be the set of cilia of $\tilde S$ as it was defined in \Cref{Sec:ciliatedsurfaces}. A \emph{framing of $\rho$}\index{framing} is a map $F\colon\tilde P\to \mathcal F$ which is $\rho$-equivariant, i.e. for all $x\in\tilde P$ and $\gamma\in\pi_1(S)$, $\rho(\gamma)F(x)=F(\gamma(x))$. A pair $(\rho,F)$ is called a \emph{framed homomorphism}\index{framed!homomorphism}\index{homomorphism!framed} if $F$ is a framing of $\rho$. The group $\mathrm{GL}_n(\mathbb C)$ acts on the space of all framed homomorphisms by conjugation in the first component and by the standard linear action in the second one. The quotient space is called the \emph{space of framed representations} of $\pi_1(S)$ into $\mathrm{GL}_n(\mathbb C)$.

Similarly to \Cref{Riemann-Hilbert correspondence}, we now formulate the framed version of the Riemann--Hilbert correspondence:

\begin{theorem}[Riemann--Hilbert Correspondence for framed bundles]\labelx{framed Riemann-Hilbert correspondence}\index{Riemann--Hilbert correspondence!for framed bundles}
There exists an isomorphism between the space of gauge equivalence classes of flat rank $n$ framed vector bundles $E$ over a ciliated surface $S$ and the space of framed representations of $\pi_1(S)$ into $\mathrm{GL}_n(\mathbb C)$.
\end{theorem}
\chapter{Cluster varieties}\labelx{sec:Clusters}





\abstract{In this chapter, we introduce cluster varieties and cluster algebras. Cluster varieties allow a special atlas in which the transition functions are certain birational transformations called mutations. Many moduli spaces associated to ciliated surfaces allow this structure, for example the moduli space of $n$ points in the projective line. On certain spaces associated to ciliated surfaces there exist natural coordinate systems, one for each triangulation of the surface. In this sense, a triangulation provides a chart on such a space, where a chart is to be understood rather in the sense of algebraic geometry than differential geometry since each chart covers an open dense subset of the space. Cluster algebras provide a profound algebraic structure which helps us understand the transition functions between two such charts or, more explicitly, how to relate the coordinate systems corresponding to two distinct triangulations of the same underlying ciliated surface. Cluster algebras are rather recent algebraic structures introduced by Fomin, Zelevinsky and Berenstein in a series of papers \cite{FominZelev1:2002,FominZelev2:2003,FominZelev3:2005,FominZelev4:2007}.}

\section{Starting example}\labelx{Sec:basic-ex-cluster}

Before introducing the main concepts for cluster varieties, let us go through a concrete example which illustrates nicely the global structure of cluster varieties.

We consider the space of $n\geq 3$ vectorial lines in $\mathbb{C}^2$ modulo the natural action by $\mathrm{PGL}_2(\mathbb{C})$, i.e. the space of $n$ points on $\mathbb{CP}^1$ modulo projective transformations. Let us call this moduli space $\mathcal{M}_n$.

\medskip
Consider first the case $n=3$. Since the action of $\mathrm{PGL}_2(\mathbb{C})$ is 3-transitive on generic triples of lines, there is only one orbit for $(\ell_1,\ell_2,\ell_3)$ if $\ell_i\neq\ell_j$ for all $i\neq j$. There are 4 other orbits when some lines coincide: three orbits where exactly two lines coincide and $(\ell_1,\ell_1,\ell_1)$. Using the quotient topology, we notice that any neighborhood of these non-generic orbits intersects the generic orbit. This means that the naive topological quotient gives a space which is not Hausdorff (in our case here: 5 points where 4 of the 5 points are infinitely close to the generic point). To solve this issue, we use the so-called \emph{GIT quotient} in order to get a moduli space $\mathcal{M}_n$ which is Hausdorff. 

The \emph{GIT quotient}\index{GIT quotient} of a variety $X$ by a group $G$ acting on $X$ is a Hausdorff space $X//G$ which has the universal property that all $G$-equivariant maps from $X$ to $Y$ (a variety with a $G$-action) factor through $X//G$. In practice, the GIT quotient is the naive topological quotient where one merges non-Hausdorff points. In the case of $\mathcal{M}_3$, we merge the two non-generic points with the generic point. Hence, $\mathcal{M}_3$ is just one point.

\medskip
Consider now $n=4$. For a 4-tuple of points $(P_1,P_2,P_3,P_4)$, there is a projective invariant, which is the \emph{cross ratio}\index{cross ratio} $(P_1,P_2,P_3,P_4):=\frac{P_1-P_3}{P_2-P_3}\frac{P_2-P_4}{P_1-P_4} \in \mathbb{C}\cup \{\infty\}$ (we identify $\mathbb{CP}^1$ with $\mathbb{C}\cup \{\infty\}$). For a generic 4-tuple, i.e. when $P_i\neq P_j$ for all $i\neq j$, the cross ratio takes values in $\mathbb{CP}^1\setminus\{0,1,\infty\}$. The remaining values 0, 1 and $\infty$ are taken when two of the 4 lines coincide, for which we have $\binom{4}{2}=6$ possible ways. Each of the special values is taken by exactly two pairs. For example $P_1=P_2$ gives the cross ratio 1, which is also taken for $P_3=P_4$. The two pairs $(P_1,P_1,P_3,P_4)$ and $(P_1,P_2,P_3,P_3)$ are infinitely close (any two neighborhoods of the two orbits intersect), so they are identified via the GIT quotient. Finally, whenever more than 2 points coincide, the orbit is infinitely close to all other points, so they are absorbed by the others.

The final result is that $\mathcal{M}_4\cong \mathbb{CP}^1$ which has two charts, given by the cross ratio $x=(P_1,P_2,P_3,P_4)\in\mathbb{C}$
and $y=(P_2,P_3,P_4,P_1)\in\mathbb{C}$, which are linked by the transformation $y=x^{-1}$. The cross ratio gives the usual charts of the sphere $\mathbb{CP}^1\cong \mathbb{S}^2$ given by the north and south pole.
This is a typical feature of a cluster variety: every chart is open and dense, and the transition functions are special rational maps.

\medskip
In order to see the appearance of ciliated surfaces, let us finally consider $n=5$. So we are interested in the moduli space of 5 vectorial lines in $\mathbb{C}^2$.

Consider a pentagon and associate to each of its vertices one of the lines, say line $\ell_i$ for vertex $i$. Note that the pentagon is a very special case of a ciliated surface. Then fix a triangulation of the pentagon. To each of the two diagonals, associate a complex number in the following way: a diagonal sits inside a unique quadrilateral. The number associated to the diagonal is the cross ratio of the four lines of this quadrilateral (in clockwise order and starting from a point on the diagonal). In  \Cref{fig:pentagon-start-example} for example, we have $x=(\ell_1,\ell_3,\ell_4,\ell_5)$ and $y=(\ell_1,\ell_2,\ell_3,\ell_4)$. The invariant property of the cross ratio shows that this number is independent of the starting point.

\begin{figure}[!ht]
	\centering
	\includegraphics{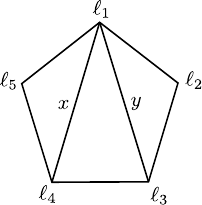}
	\caption{A fixed triangulation of the pentagon gives coordinates on $\mathcal{M}_5$.}\labelx{fig:pentagon-start-example}
\end{figure}

We claim that these two numbers form a coordinate system of $\mathcal{M}_5$. Indeed, we can reconstruct the five lines from the two numbers: Since $\mathrm{PGL}_2(\mathbb{C})$ acts 3-transitively on the set of lines in $\mathbb{C}^2$, we can fix three of the four lines in the first quadrilateral (say $\ell_1,\ell_2,\ell_3$). The fourth line is then determined by the cross ratio written on the diagonal. The fifth line is then determined by the other cross ratio.

We see that this reconstruction does only give generic 5-tuples of lines. So one chart, given by a fixed triangulation of the pentagon, describes an open dense subset of $\mathcal{M}_5$. 

When we change the triangulation by a flip, one can use the properties of the cross ratio to determine how the coordinates change. After flipping the diagonal associated to $x$, the new coordinates are given by $(x^{-1}, y(1+x))$. If we flip the diagonal associated to $y$, we obtain $(x(1+y^{-1})^{-1}, y^{-1})$. Note that these transformations are only birational (for instance, the expression $x(1+y^{-1})^{-1}$ is not well-defined for $y=-1$).

\medskip
The above picture generalizes to all $n$: the moduli space $\mathcal{M}_n$ of $n$ points in $\mathbb{CP}^1$ is a variety, where charts are indexed by triangulations of an $n$-gon, and the transition functions are special birational transformations.
The ubiquitous appearance of such kind of transformation laws lies at the origin of the notion of cluster varieties.

\section{Triangulations, quivers and lattices}

We start from the initial data of an ideal triangulation $\Delta$ on an oriented ciliated surface $S$. The core combinatorial data provided by this triangulation is a finite set (the set of arcs of $\Delta$), together with the relation of two edges belonging to the same triangle.

We can represent all this information given by $\Delta$ by a special class of  graph called a \emph{quiver}.

\begin{definition}\labelx{def:quiver}
    A \emph{quiver}\index{quiver} is an oriented graph without 1- and 2-cycles. To be more precise, it is a triple $Q = (Q_0,Q_0^f,Q_1)$ where $Q_0$ is a finite set, $Q_0^f\subset Q_0$, and $Q_1$ is a finite set of pairs $(v,v')\in Q_0^2$ such that if $(v,v')\in Q_1$, then $v\neq v'$ and $(v',v)\notin Q_1$. Each pair $(v,v')\in Q_1$ has a multiplicity $m_{(v,v')}>0$. If $(v',v)\in Q_1$, we set $m_{(v,v')} = - m_{(v',v)}$ and if neither $(v,v')$ nor $(v',v)$ belong to $Q_1$, then we set $m_{(v,v')} =0$.
    The elements of $Q_0$ are called \emph{vertices}\index{quiver!vertex}, the vertices in $Q_0^f$ are called \emph{frozen}\index{quiver!vertex!frozen}, and a pair $(v,v')\in Q_1$ is called an \emph{arrow}\index{quiver!arrow} from $v$ to $v'$.
\end{definition}

The quiver $Q_\Delta = (Q_0,Q_0^f,Q_1)$ associated to a triangulation $\Delta$ of a ciliated surface is obtained as follows: the set $Q_0$ is the set of edges of $\Delta$, the set of frozen vertices $Q_0^f$ is the set of external edges, and for any oriented triangle $\gamma,\gamma',\gamma''$ of $\Delta$ we have arrows from $\gamma$ to $\gamma'$, from $\gamma'$ to $\gamma''$ and from $\gamma''$ to $\gamma$ (see \Cref{fig:triangulation_quiver}).

\begin{figure}[!ht]
	\centering
	\includegraphics{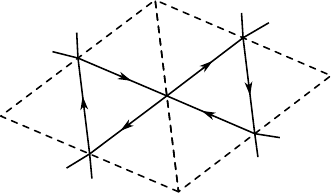}
	\caption{The quiver $Q_\Delta$ associated to a triangulation $\Delta$.}\labelx{fig:triangulation_quiver}
\end{figure}

Given a quiver $Q = (Q_0,Q_0^f,Q_1)$, consider the $\mathbb{Z}$-module $\Lambda$ freely generated by the vertices $(e_v)_{v\in Q_0}$. We call $\Lambda$ the \emph{lattice}\index{quiver!associated lattice} associated to $Q$. This lattice is naturally endowed with a skew-symmetric bilinear form $(\cdot,\cdot)$ defined by
$$(e_v,e_{v'}) = m_{(v,v')}\; .$$
The lattice $\Lambda$ is also endowed with a canonical basis $(e_v)_{v\in Q_0}$ by construction, and the subset of frozen vertices $Q_0^f$ induces a subfamily of this basis given by $(e_v)_{v\in Q_0^f}$.

\begin{definition}\labelx{def:seed}
    A \emph{seed}\index{seed} is a quadruple $\mathsf{S}=(\Lambda, (\cdot,\cdot), (e_i)_{i\in I}, J)$ where $\Lambda$ is a lattice (i.e. a finitely generated free $\mathbb{Z}$-module), $(\cdot,\cdot)$ is a skew-symmetric bilinear form on $\Lambda^2$, $(e_i)_{i\in I}$ is a basis of $\Lambda$ and $J$ is a (possibly empty) subset of $I$.
\end{definition}

\begin{remark}
    There is a slightly more general notion of seed obtained by adding the data of a positive integer (called a multiplier) on each vertex of the quiver, but we will not need this level of generality for the subjects discussed here. More details about this construction can be found in \cite{FG:2009,Marsh:2014}.
\end{remark}

Notice that one can retrieve the data of the quiver from a seed $(\Lambda, (\cdot,\cdot), (e_i)_{i\in I}, J)$ in the following way: we have $Q_0 = I$, $Q_0^f = J$ and there is an arrow of multiplicity $(e_v,e_{v'})$ from $v$ to $v'$, for all $v,v'\in I$.

\section{Cluster seed tori and coordinates}

Every seed defines two split algebraic tori $(\mathbb{C}^\times)^n$ of interest. In fact, every lattice $\Lambda$ of rank $n$ defines a complex split torus 
\begin{equation*}
    \mathcal X_\Lambda := \mathrm{Hom}(\Lambda,\mathbb{C}^\times)\cong (\mathbb{C}^\times)^n\; .
\end{equation*}
The assignment $\Lambda \rightarrow \mathcal X_\Lambda$ induces a contravariant equivalence of categories from the category of (finite-rank) lattices to the category of split algebraic tori. 

Any seed $(\Lambda, (\cdot,\cdot), (e_i)_{i\in I}, J)$ defines the \emph{cluster torus}\index{cluster!torus} $\mathcal X_\Lambda$ as well as $\mathcal A_\Lambda := \mathcal X_{\Lambda^*}$, where $\Lambda^*$ is the lattice dual to $\Lambda$. We will see that the antisymmetric bilinear form $(\cdot,\cdot)$ itself induces interesting structures on $\mathcal X_\Lambda$ and $\mathcal A_\Lambda$: the former is endowed with a Poisson bracket (a Lie bracket which is also a derivation), and the latter, with a (possibly degenerate) closed 2-form. Moreover, $(\cdot,\cdot)$ defines a map $\mathcal A_\Lambda\rightarrow\mathcal X_\Lambda$.

Note first that any $\lambda\in\Lambda$ defines an element $x_\lambda\in \Hom(\mathcal X_\Lambda,\mathbb{C}^\times)$, by
\begin{equation*}
    \begin{array}{rcl}
        x_\lambda\colon \mathcal X_\Lambda & \longmapsto & \mathbb{C}^\times \\
        x & \longmapsto & x(\lambda)\; .
    \end{array}
\end{equation*}
For all $\lambda,\mu \in \Lambda$, one has $x_\lambda x_\mu = x_{\lambda+\mu}$.
The basis $\{e_1,\dots,e_n\}$ of $\Lambda$ given by the seed defines $\mathbb{C}^\times$-valued coordinates  on $\mathcal X_\Lambda$ that we denote by $x_1,\dots,x_n$. Let us moreover denote $\epsilon_{ij}:=(e_i,e_j)$, for all $i,j=1,\dots,n$. 

\begin{proposition}
    The form $(\cdot,\cdot)$ induces a Poisson bracket on $\mathcal X_\Lambda$, defined by
\begin{equation*}
    \{x_\lambda,x_\mu\} = (\lambda,\mu)\cdot x_\lambda x_\mu\; .
\end{equation*}  
\end{proposition}

\begin{proof}
    One can easily verify that $\{x_\lambda x_{\lambda'},x_\mu\} = x_\lambda \{x_{\lambda'},x_\mu\}+x_{\lambda'}\{x_\lambda,x_\mu\}$, and that $\{\cdot,\cdot\}$ satisfies all other properties necessary to make it a Poisson bracket.
    In coordinates, $\{\cdot,\cdot\}$ reads
    \begin{equation*}
      \{x_i,x_j\} = \epsilon_{ij}~x_i x_j\; .
    \end{equation*}
\end{proof}

Let us now consider the dual lattice $\Lambda^*=\Hom(\Lambda,\mathbb{Z})$ and let $\{e_1^*,\dots,e_n^*\}$ be the basis of $\Lambda^*$ dual to $\{e_1,\dots,e_n\}$. Similarly to the above, this basis defines $\mathbb{C}^\times$-valued coordinates $a_1,\dots,a_n$ on 
\begin{equation*}
    \mathcal A_\Lambda := \mathrm{Hom}(\Lambda^*,\mathbb{C}^\times)\cong (\mathbb{C}^\times)^n\; .
\end{equation*}

\begin{proposition}
    The form $(\cdot,\cdot)$ on $\Lambda$ induces a closed 2-form on $\mathcal A_\Lambda$.
\end{proposition}

\begin{proof}
    In coordinates, the form $(\cdot,\cdot)$ reads
    \begin{equation*}
        (\cdot,\cdot) = \sum_{i<j}\epsilon_{ij}~e_i^*\wedge e_j^* \in \bigwedge^2 \Lambda^*\; .
    \end{equation*}
    Dually, it defines the function 
    \begin{equation*}
        \sum_{i<j} \epsilon_{ij} ~a_i\wedge a_j
    \end{equation*}
    on $\mathcal A_\Lambda$. Note that the coordinates $a_i$ never vanish on $\mathcal A_\Lambda$ since they are $\mathbb{C}^\times$-valued. Whereas the logarithm $\log(a_i)$ of each $a_i$ is only defined up to $2\pi$ shifts, the 1-form $\mathrm{d}\log(a_i)$ is unambiguously defined on $\mathcal A_\Lambda$. Hence,
    \begin{equation*}
        \sum_{i<j} \epsilon_{ij}~ \mathrm{d}\log (a_i)\wedge \mathrm{d}\log(a_j)
    \end{equation*}
    is a 2-form on $\mathcal A_\Lambda$, and it is clearly closed. 
    
    In fact, the 2-form can be defined independently of the coordinates. The bilinear antisymmetric form $(\cdot,\cdot)$ induces an element $W\in\mathcal O(\mathcal A_\Lambda)^*\wedge \mathcal O(\mathcal A_\Lambda)^*$, where $\mathcal O(\mathcal A_\Lambda)^*$ is the group of invertible (regular) functions on $\mathcal A_\Lambda$. Applying $\mathrm{d}\log\wedge \mathrm{d}\log$ to any element of $\mathcal O(\mathcal A_\Lambda)^*\wedge \mathcal O(\mathcal A_\Lambda)^*$ yields a 2-form on $\mathcal A_\Lambda$; in particular it is the case for $W$. More details can be found in \cite{FG:2009}.
\end{proof}

\begin{proposition}\labelx{prop:projection_map_cluster}
    There exists a natural map $p\colon \mathcal A_\Lambda \rightarrow \mathcal X_\Lambda$.
\end{proposition}

\begin{proof}
    The bilinear form $(\cdot,\cdot)$ on $\Lambda$ induces a map $\Lambda\rightarrow\Lambda^*$ by
    \begin{equation*}
        \lambda  \longmapsto  \sum_i (\lambda,e_i)e_i^*\; .
    \end{equation*}
    Since the functor $\Hom(\cdot,\mathbb{C}^\times)$ is contravariant, we obtain $p\colon \mathcal A_\Lambda \rightarrow \mathcal X_\Lambda$ as required. In coordinates, the map $p$ reads
    \begin{equation*}
        p(a_1,\dots,a_n) = \left(\prod_j a_j^{\epsilon_{1j}},\dots,\prod_j a_j^{\epsilon_{nj}} \right)\; .
    \end{equation*}
\end{proof}

\section{Mutations}

In this Section we are interested in studying the dependency on the chosen triangulation of the previously introduced objects. Indeed, the aim of the cluster varieties construction is to build objects that only depend on the surface, and not on the choice of an ideal triangulation.

First of all, recall \Cref{Thm:finiteseqflips} stating that any two ideal triangulations on a ciliated surface $S$ can be related by a finite sequence of elementary transformations called \emph{flips}. This theorem implies that we only need to understand the changes induced by a flip of the triangulation to understand any change of triangulation.

The first step is to write the flip operation in the language of quivers. We thus now define the notion of \emph{mutation}\index{quiver!mutation}\index{mutation} of a quiver. Let $Q=(Q_0,Q_0^f,Q_1)$ be a quiver, and let $i\in Q_0\setminus Q_0^f$ be a non-frozen vertex. The quiver $Q' = \mu_i(Q)$ obtained by \emph{mutating $Q$ at the vertex $i$} is constructed as follows: 

\begin{enumerate}
    \item Reverse all those arrows which have $i$ as a starting point or an endpoint (keeping their multiplicity).
    \item For all pairs of arrows of the form $j\to i\to k$ with $j,k\in Q_0$, add a new arrow $k\to j$ with multiplicity $m_{(k,j)} =m_{(j,i)}m_{(i,k)}$. 
    \item If after these modifications there are pairs of vertices $(j,k)$ with both an arrow $j\to k$ of multiplicity $m_{(j,k)}$ and an arrow $k\to j$ of multiplicity $m_{(k,j)}$ with $m_{(j,k)}< m_{(k,j)}$, then remove the arrow $j\to k$ and set the multiplicity of $k\to j$ to $m_{(k,j)}-m_{(j,k)}$. If $m_{(j,k)}= m_{(k,j)}$, then both arrows are removed. This last step is equivalent to counting the algebraic multiplicity of the arrows.
\end{enumerate}

\begin{example}
    Let us look at a mutation of the quiver $Q_\Delta$ associated to an ideal triangulation $\Delta$ of a quadrilateral. The mutation at any non-frozen vertex is described step by step in \Cref{fig:triangulation_quiver_flip}. Notice that after the mutation, the quiver obtained is the one associated to the triangulation obtained by flipping the edge of the vertex mutated. The same phenomenon stays true for any ideal triangulation of any ciliated surface. Note that while there is in general an infinite number of different ideal triangulations of a given ciliated surface $S$, the number of different quivers that one can obtain from triangulations of $S$ is finite. This is because there are only finitely many different orbits of ideal triangulations for the action of the pure mapping class group as mentioned in \Cref{total_num_triang}, and all the quivers in a given orbit are isomorphic.

    \begin{figure}[!ht]
	\centering
	\includegraphics{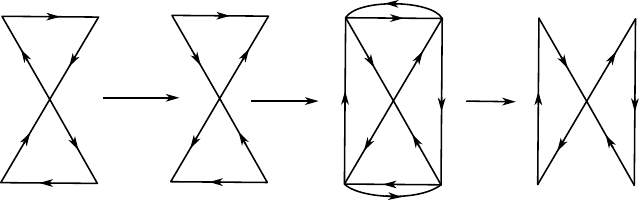}
    
	\caption{A mutation of the quiver associated to a triangulation of a quadrilateral. The mutation results in the quiver associated to the other triangulation of the quadrilateral.}\labelx{fig:triangulation_quiver_flip}
\end{figure}
\end{example}

For the purpose of computations it will be convenient to have the effect of a mutation on the seed associated to the quiver. The only changes induced by a mutation on a seed $\mathsf{S} = (\Lambda, (\cdot,\cdot), (e_i)_{i\in I}, J)$ will be on the basis $e = (e_i)_{i\in I}$. Let $i\in I$. The new basis obtained by the mutation $\mu_i$ is $\mu_i(e)$ given by
$$(\mu_i(e))_{k} = \left\lbrace\begin{array}{ll}
     e_k + \mathrm{max}(\epsilon_{ki},0)e_i, &\text{ if } k\neq i  \\
    -e_i, & \text{ if } k= i\; .
\end{array}\right.$$
We denote by $\mu_i(\mathsf{S})$ the new seed $(\Lambda, (\cdot,\cdot), \mu_i(e), J)$. Note that although the skew-symmetric bilinear form does not change under mutation, its matrix $\epsilon$ in the basis given by the seed does change. The matrix $\mu_i(\epsilon)$ of $(\cdot,\cdot)$ in the basis $\mu_i(e)$ is given by
$$(\mu_i(\epsilon))_{k\ell} = \left\lbrace\begin{array}{ll}
    -\epsilon_{k\ell}, & \text{ if }i \in\lbrace k,\ell\rbrace\\
     \epsilon_{k\ell}, &\text{ if } \epsilon_{ki}\epsilon_{i\ell}\leq 0, i \notin\lbrace k,\ell\rbrace  \\
    \epsilon_{k\ell} +\abs{\epsilon_{k,i}}\epsilon_{i\ell}, &\text{ if } \epsilon_{ki}\epsilon_{i\ell}>0, i \notin\lbrace k,\ell\rbrace\; . 
\end{array}\right.$$

Since the mutation of a seed does not change the lattice or the skew-symmetric bilinear form, the two tori $\mathcal{X}_\Lambda$ and $\mathcal{A}_\Lambda$ are invariant under mutation, together with the structures they carry:

\begin{proposition}
    Let $\mathsf{S}$ be a seed. The Poisson bracket on $\mathcal{X}_\Lambda$ and the closed 2-form on $\mathcal{A}_\Lambda$ are invariant under mutation of the seed $\mathsf{S}$.
\end{proposition}

However, the $\mathbb{C}^\times$-valued coordinates $(x_i)_{i\in I}$ on $\mathcal{X}_\Lambda$ and $(a_i)_{i\in I}$ on $\mathcal{A}_\Lambda$ are modified by the mutation in the following way
\begin{equation}\labelx{Eq:cluster-X-mutation} \mu_i(x_j) = \left\lbrace\begin{array}{ll}
    x_i^{-1}, & \text{ if }j = i\\
    x_j(1+x_i^{\mathrm{sgn}(\epsilon_{ij})})^{\epsilon_{ij}}, &\text{ if } j\neq i\; ,
\end{array}\right.\end{equation}
and
\begin{equation}\labelx{Eq:cluster-A-mutation} \mu_i(a_j) = \left\lbrace\begin{array}{ll}    a_i^{-1}\left(\displaystyle\prod_{k\vert\epsilon_{ki}>0}a_k^{\epsilon_{ki}}+\prod_{k\vert\epsilon_{ki}<0}a_k^{-\epsilon_{ki}}\right), & \text{ if }j = i\\
    a_j, &\text{ if } j\neq i\; .
\end{array}\right.\end{equation}
These formulas are called \emph{mutation formulas}\index{mutation!formulas}\index{cluster!mutation} of $\mathcal{X}$-type and of $\mathcal{A}$-type respectively. 

\begin{remark}
    In particular, if $x_1,\dots,x_n$ (resp. $a_1,\dots,a_n$) are the coordinates of a point $x\in\mathcal{X}_\Lambda$ (resp. $a\in\mathcal{A}_\Lambda$) in a given seed $\mathsf{S}$, then for any seed $\mathsf{S}'$ obtained from $\mathsf{S}$ by successive mutations, the coordinates $x'_1,\dots,x'_n$ of $x$ (resp $a'_1,\dots,a'_n$ of $a$) are rational functions of $x_1,\dots,x_n$ (resp. $a_1,\dots,a_n$).
\end{remark}

As seen in \Cref{Sec:ciliatedsurfaces}, any two sequences of flips transforming an ideal triangulation $\Delta$ into another $\Delta'$ differ by a finite number of commutativity relations and pentagon relations. It follows from the formulas that for any given triangulation $\Delta$ with associated seed $\mathsf{S}$ and for any $i,j$ arcs of $\Delta$ not belonging to the same triangle, $\mu_i\circ\mu_j(x_k) = \mu_j\circ\mu_i(x_k)$ and $\mu_i\circ\mu_j(a_k) = \mu_j\circ\mu_i(a_k)$, for all $k\in \Delta$.

\begin{proposition}[Pentagon relation]
    Let $\Delta$ be an ideal triangulation of $S$ and let $\mathsf{S}$ be the associated seed. Let $i,j\in \Delta$ such that $m_{i,j} = 1$. Then for all $k\in \Delta$ we have
    $$\mu_i\circ\mu_j\circ\mu_i\circ\mu_j\circ\mu_i(x_k) = \left\lbrace\begin{array}{cc}
        x_k, & \text{ if } k\neq i,j \\
        x_j, & \text{ if } k = i \\
        x_i, & \text{ if } k = j\; ,
    \end{array}\right.$$ and $$\mu_i\circ\mu_j\circ\mu_i\circ\mu_j\circ\mu_i(a_k) = \left\lbrace\begin{array}{cc}
        a_k, & \text{ if } k\neq i,j \\
        a_j, & \text{ if } k = i \\
        a_i, & \text{ if } k = j\; . 
    \end{array}\right.$$
\end{proposition}

\begin{proof}
    The proof is straightforward but involves a quite long computation. We present here the computation for $\mathcal{A}$-type coordinates on a pentagon with all external edge coordinates frozen and equal to $1$ (see \Cref{fig:pentagon_relation_cluster}). Then, the two internal $\mathcal{A}$-coordinates are changed as follows:
    $$\begin{array}{c}
         (a_i,a_j) \overset{\mu_i}{\longrightarrow} \left(\dfrac{a_j+1}{a_i}, a_j\right) \overset{\mu_j}{\longrightarrow} \left(\dfrac{a_j+1}{a_i}, \dfrac{1+a_i+a_j}{a_ia_j}\right) \overset{\mu_i}{\longrightarrow} \left(\dfrac{a_i+1}{a_j}, \dfrac{1+a_i+a_j}{a_ia_j}\right) \\ \overset{\mu_j}{\longrightarrow} \left(\dfrac{a_i+1}{a_j}, a_i\right) \overset{\mu_i}{\longrightarrow} (a_j,a_i)\; .
    \end{array}$$
\end{proof}

\begin{figure}[!ht]
	\centering
	\includegraphics{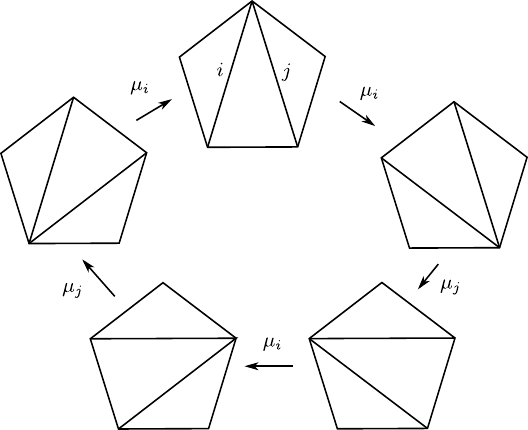}
    
	\caption{The sequence of mutations associated to the pentagon relation. Notice that the two internal edges are swapped by the sequence of mutation.}\labelx{fig:pentagon_relation_cluster}
\end{figure}

One of the fundamental properties of the mutation formulas is that since they do not involve any subtraction, they preserve the set of real positive coordinates: if a point in  $\mathcal{X}_\Lambda$ or $\mathcal{A}_\Lambda$ has all its coordinates real positive in a given seed $\mathsf{S}$, then it will have real positive coordinates in any seed obtained by mutations from $\mathsf{S}$. This property will be fundamental in relating cluster varieties to higher-rank Teichmüller theory in \Cref{Sec:FockGoncharov}.

Another important property of $\mathcal{A}$-type coordinates is the so-called \emph{Laurent phenomenon}:

\begin{theorem}[\cite{FominZelev1:2002}]
    Let $\mathsf{S}$ be a seed and let $a_1,\dots,a_n$ be the coordinates of a point $a\in\mathcal{A}_\Lambda$ in $\mathsf{S}$. Let $\mathsf{S}'$ be a seed obtained from $\mathsf{S}$ by successive mutations. Then the coordinates  $a'_1,\dots,a'_n$ of the point $a$ are Laurent polynomials in the variables $a_1,\dots,a_n$.
\end{theorem}

This property of $\mathcal{A}$-coordinates also has consequences on the $\mathcal{X}$-coordinates, although they are less straightforward, see \cite{FG:2009} for more details.

\section{Cluster varieties}\labelx{Sec:clustervarieties}

\begin{definition}
    Let $\mathsf{S}$ be a seed and let $|\mathsf{S}|$ denote the set of all seeds that can be obtained from $\mathsf{S}$ by a sequence of mutations. We define the \emph{$\mathcal{A}$-variety}\index{cluster!$\mathcal{A}$-variety} of $\mathsf{S}$ to be the space $$\mathcal{A}_\mathsf{S} :=\left(\bigsqcup_{\mathsf{S}'\in|\mathsf{S}|}\mathcal{A}_{\Lambda_{\mathsf{S}'}}\right)/\sim \; ,$$
    where $\Lambda_{\mathsf{S}'}$ denotes the lattice of the seed $\mathsf{S}$ and $a'\in\mathcal{A}_{\Lambda_{\mathsf{S}'}}\sim a''\in\mathcal{A}_{\Lambda_{\mathsf{S}''}}$ if for some $i$ we have $\mu_i(\mathsf{S}')=\mathsf{S}''$ and $\mu_i(a')=a''$. 
    Similarly, we define the \emph{$\mathcal{X}$-variety}\index{cluster!$\mathcal{X}$-variety} of $\mathsf{S}$ by $$ \mathcal{X}_\mathsf{S}:=\left(\bigsqcup_{\mathsf{S}'\in|\mathsf{S}|}\mathcal{X}_{\Lambda_{\mathsf{S}'}}\right)/\sim \; , $$
    where $x'\in\mathcal{X}_{\Lambda_{\mathsf{S}'}}\sim x''\in\mathcal{X}_{\Lambda_{\mathsf{S}''}}$ if for some $i$ we have $\mu_i(\mathsf{S}')=\mathsf{S}''$ and $\mu_i(x')=x''$.
\end{definition}

A seed $\mathsf{S}'\in|\mathsf{S}|$ corresponds to an open dense chart with values in $(\mathbb{C}^\times)^n$ on both $\mathcal{A}$- and $\mathcal{X}$-varieties, given respectively by the $\mathcal{A}$- and $\mathcal{X}$-coordinates associated to the seed $\mathsf{S}'$. The changes of charts are given by the mutation functions, hence are birational. Since the mutation functions preserve the closed 2-form $(\cdot,\cdot)$ on the $\mathcal{A}$-tori and the Poisson bracket $\lbrace\cdot,\cdot\rbrace$ on the $\mathcal{X}$-tori, these structures extend to a closed 2-form $(\cdot,\cdot)$ on $\mathcal{A}_\mathsf{S}$ and a Poisson bracket $\lbrace\cdot,\cdot\rbrace$ on $\mathcal{X}_\mathsf{S}$. The map $p$ defined in \Cref{prop:projection_map_cluster} also extends to a map 
$$p\colon \mathcal{A}_\mathsf{S}\to \mathcal{X}_\mathsf{S}\; .$$

\begin{example}
    Let us come back to the basic example of the moduli space of 5 lines in $\mathbb{C}^2$ given in \Cref{Sec:basic-ex-cluster} and let us describe its structure as a cluster $\mathcal{X}$-variety. 

    The quiver associated to $\mathcal{M}_5$ is the quiver of a triangulated pentagon. There are 7 vertices, 5 of which are frozen (the ones corresponding to outer edges of the pentagon). This quiver gives also the initial seed $(\Lambda, (\cdot,\cdot), (e_i)_{i\in Q_0}, Q_0^f)$.

    The crucial point is that the cross ratio transformation which gives the coordinate change $(x,y)\mapsto (x^{-1},y(1+x))$ or $(x,y)\mapsto (x^{-1},y(1+x^{-1})^{-1})$ is exactly the $\mathcal{X}$-mutation from Equation \eqref{Eq:cluster-X-mutation}. Hence, gluing the $\mathcal{X}$-torus associated to the initial seed to another seed obtained by a flip represents exactly the transition between the coordinate charts.

    The same method can be used to describe the moduli space of $n$ lines in $\mathbb{C}^2$, using triangulations of an $n$-gon. Cluster theory shows that the moduli space is a Poisson variety.
\end{example}

\begin{example}\labelx{Ex:n-vectors}
    Very similar to the moduli space of $n$ points in $\mathbb{CP}^1$, there is the moduli space of $n$ vectors in $\mathbb{C}^2$, which is a cluster $\mathcal{A}$-variety.

    Equip $\mathbb{C}^2$ with a complex volume form $\omega\in\Omega^2(\mathbb{C}^2)$. We consider the space $\mathcal{V}_n$ of $n$ vectors $(v_1,...,v_n)$ in $\mathbb{C}^2$ modulo the natural action of $\mathrm{SL}_n(\mathbb{C})$, using the GIT quotient.

    We consider again an $n$-gon where to vertex $i$ we associate the vector $v_i$. Fix a triangulation of the polygon. To each edge (also the external sides of the polygon), we associate $\omega(v_i,v_j)\in\mathbb{C}$, where $v_i$ and $v_j$ are the vectors at the vertices of the edge.
    There is an obvious problem though: since $\omega$ is antisymmetric, $\omega(v_j,v_i)=-\omega(v_i,v_j)$, so the label is only defined up to sign. To resolve this issue, we actually have to associate to the vertex $i$ of the polygon the two vectors $v_i$ and $v_{n+i}:=-v_i$. The rule is then to take one vector $v_k\in\{v_i,v_{n+i},v_j,v_{n+j}\}$ of one endpoint of the chosen edge, to take the distance $d$ in clockwise direction to the other endpoint, and to associate $\omega(v_k,v_{k+d})$, where we compute $k+d$ modulo $2n$. In this way, the coordinate on the edge is well-defined.

    One can check that the $n$ vectors can be reconstructed from the coordinates, and that a flip will induce a cluster $\mathcal{A}$-mutation of the coordinates, as in Equation \eqref{Eq:cluster-A-mutation} where the quiver is the same as for the moduli space of $n$ lines in $\mathbb{C}^2$. Hence, the moduli space $\mathcal{V}_n$ is a cluster $\mathcal{A}$-variety. Note that there is a canonical map $\pi \colon  \mathcal{V}_n\to \mathcal{M}_n$ obtained by taking the lines spanned by each vector. The canonical map $p$ between the corresponding $\mathcal{A}$- and $\mathcal{X}$-varieties is such that if $a\in\mathcal{A}$ is the coordinate of an $n$-tuple of vectors $v = (v_1,\dots,v_n)\in\mathcal{V}_n$, then $x=p(a)\in\mathcal{X}$ is the coordinate of $\pi(v)\in\mathcal{M}_n$.
\end{example}

As mentioned before, since the mutation formulas do not involve subtractions, all definitions above make sense if we replace the field $\mathbb{C}$ by a semifield, for instance $\mathbb{R}_{>0}$. This particular semifield will be of great importance in the study of higher-rank Teichmüller spaces; see \Cref{Sec:FockGoncharov}. 

Other semifields of geometric significance are the \emph{tropical} semifields, e.g. $\mathbb{Z}^{\mathrm{t}}=(\mathbb{Z},\max,+)$. In the specific case of cluster classical Teichmüller spaces, the tropical points of these cluster varieties embody the notion of \emph{measured laminations}\index{measured laminations} on surfaces \cite{FG_book}. These objects appear at the boundary of the \emph{Thurston compactification} of Teichm\"uller space. It is a common feature to use tropical objects to compactify moduli spaces. For higher-rank Teichm\"uller spaces, the tropical versions are algebraically well-defined, although their geometric interpretation is only partly understood \cite{Xie, Le:2016, Hecke-TQFT}.

\begin{remark}
    Tropical mutation formulas admit a natural physical realization in supersymmetric quantum field theories, for instance as Seiberg duality in $4d$ $\mathcal N=1$ quiver gauge theories, a topic we will discuss in \Cref{Chapter:N=1theories}.
\end{remark}

The cluster varieties $\mathcal{A}_\mathsf{S}$ and $\mathcal{X}_\mathsf{S}$ corresponding to a mutation class $\vert \mathsf{S}\vert$ are related by a profound notion of duality called \emph{cluster duality}\index{cluster!duality}, conjectured in \cite[Conj. 12.3]{FG} and (partially) proved in \cite{Gross-Kontsevich}. The specific case of Teichm\"uller spaces is treated in \cite{FG_book}. Every cluster variety admits a distinguished space of functions: the regular functions which are Laurent polynomials with positive coefficients in each cluster chart. This space is a cone. In other words, cluster dualities state that the extremal elements of this cone for the cluster $\mathcal{A}_\mathsf{S}$ (resp. $\mathcal{X}_\mathsf{S}$) variety are in one-to-one correspondence with the $\mathbb{Z}^t$-tropical points of the the cluster $\mathcal{X}_\mathsf{S}$ (resp. $\mathcal{A}_\mathsf{S}$) variety.

The special basis given by the duality conjecture has a remarkable property: if $(f_\mu)$ denotes such a basis, then the structure constants $c_{\mu,\nu}^\rho$, obtained by
$$f_\mu f_\nu =\sum_{\rho} c_{\mu,\nu}^\rho f_\rho\; ,$$
are positive integers. 
The question about the combinatorial interpretation of these structure constants is open in general. Similar phenomena appear for the Lusztig canonical basis of quantum groups, for the Kazhdan--Lusztig basis of the Hecke algebra, and for the fusion coefficients in representation theory of Lie groups.

Lastly, we would like to highlight that the fundamental properties of cluster varieties—namely, that cluster $\mathcal A$ (resp. $\mathcal X$) varieties admit a closed 2-form (resp. a Poisson bracket)—are shadows of deeper structures, for a general discussion of which we refer to \cite{FG}. Specifically, the closed 2-form on cluster $\mathcal A$-varieties actually descends from a so-called \emph{motivic avatar}\index{motivic avatar}, while the Poisson structure on cluster $\mathcal{X}$-varieties admits a canonical \emph{quantization}. In the context of Teichmüller spaces, this quantization was explicitly carried out in \cite{Kashaev:1998,FG:2009}. A central role is played by the \emph{quantum dilogarithm}\index{quantum!dilogarithm} 
$$\Psi_q(z)=\prod_{k=0}^\infty \frac{1}{1+q^{2k+1}z}\; ,$$ 
which appears both in the quantum mutation rule and in the action of the mapping class group on the deformed moduli space. These developments led to the notion of a \emph{quantum $\mathcal X$-variety}, where the mutation rule is replaced by conjugation by the quantum dilogarithm. The limit as $q \to 1$ recovers the classical mutation.

\newpage
\part{Higher-rank Teichm\"uller theory}\labelx{partIII}

\vspace{2cm}

We introduce Teichmüller spaces and their higher-rank generalizations, taking the character variety as the central object of study. The character variety, defined as the space of group homomorphisms of the fundamental group into a Lie group modulo conjugation, admits several equivalent formulations: as the moduli space of geometric structures, as the moduli of flat connections, or, in the holomorphic setting, as the moduli of Higgs bundles. Geometric structures are associated with discrete and faithful representations, and connected components of the character variety consisting entirely of such representations are referred to as higher-rank Teichmüller spaces. These spaces extend the classical Teichmüller space of hyperbolic structures and form the focus of ongoing and active research. More recently, the notion of $\Theta$-positivity has been introduced as a unifying framework that encompasses all known examples of higher-rank Teichmüller spaces.

\vspace{2cm}

\parttoc


\chapter{The Teichm\"{u}ller space}\labelx{subsec_Teich_sp}



\abstract{The Teichmüller space of a smooth surface $S$ is the space of hyperbolic metrics on $S$, modulo diffeomorphisms isotopic to the identity. Equivalently, it parameterizes conformal (or complex) structures on $S$ under the same equivalence. Via the holonomy representation, Teichmüller space can also be identified with a connected component of the $\mathrm{PSL}_2(\mathbb{R})$ character variety. In this chapter, we set the main definitions and exhibit the holonomy representation explicitly.}

\vspace{0.5cm}

Since Teichmüller spaces encode metrics on surfaces, they naturally play an important role in various areas of physics, most prominently in two- and three-dimensional gravity. In addition, they assume a particularly distinguished role in the study of $4d$ $\mathcal N=2$ supersymmetric quantum field theories of class~$\mathcal S$, where they are most naturally interpreted as parameter spaces of complex structures on smooth surfaces. The appearance of Teichmüller spaces in this context will be discussed in greater detail in \Cref{Chap:theoriesofclassS}.

\section{Definition}

A big theme in mathematics is the classification of objects up to some equivalence relation. A simple example is given by the classification of orientable compact surfaces up to homeomorphism. Such a surface is characterized by the genus $g\in\mathbb{N}$ and the number of boundary components $h$. Teichm\"{u}ller spaces refine this topological classification by adding geometry. More specifically, they classify surfaces endowed with constant curvature metrics.
We give here a very short glimpse on Teichm\"uller theory. For more details, we recommend the books by Hubbard \cite{Hubbard}, Farb--Margalit \cite{Farb-Margalit}, and the series of books \cite{Papadopoulos-Handbook} edited by Papadopoulos.

Any surface can be equipped with a metric with constant curvature (coming for example from its universal cover, which is either $\mathbb{S}^2, \mathbb{R}^2$ or $\mathbb{H}^2$) and geodesic boundary. The Gauss--Bonnet theorem allows to link the curvature constant $K$ to the topology: $K>0$ for positive Euler characteristic, $K=0$ for zero Euler characteristic and $K<0$ for negative Euler characteristic.

We are interested in the last case, so we consider compact connected and oriented topological surfaces $\bar S_{g,h}$ of genus $g$ with $h$ boundary components such that $\chi(\bar S_{g,h}) = 2-2g-h<0$ (also called of \emph{hyperbolic type}). Then any surface $\bar S_{g,h}$ admits a hyperbolic structure, as defined next:

\begin{definition}
A \emph{hyperbolic structure}\index{hyperbolic!structure} on a topological surface $\bar S_{g,h}$ as above is a pair $(M, f)$, where $M$ is a hyperbolic surface, that is, a topological surface with a metric of constant negative curvature $K=-1$ and geodesic boundary, and $f\colon \bar S_{g,h} \to M$ an orientation preserving homeomorphism. 
\end{definition}

The map $f\colon \bar S_{g,h}\to M$ is called \emph{marking}\index{hyperbolic!marked structure}. It is responsible for equipping the surface $\bar S_{g,h}$ with a Riemannian metric of constant negative curvature -1. The \emph{Teichm\"{u}ller space} is then the quotient space of these pairs subject to an equivalence relation:

\begin{definition}
The \emph{Teichm\"{u}ller space}\index{Teichm\"uller!space}\index{space!Teichm\"uller} $\mathcal{T}(\bar S_{g,h})$ of $\bar S_{g,h}$ is the space
\[\mathcal{T}(\bar S_{g,h}):=\left\{(M,f) \right\}/{\sim }\; ,\]
where
\begin{itemize}
    \item $(M,f)$ is a hyperbolic structure on $\bar S_{g,h}$,
    \item $(M,f) \sim (M',f')$ if and only if there is an isometry $\iota: M \to M'$, such that $\iota  \circ f$ is homotopic to $f'$; this equivalence relation is called an \emph{isotopy}.
\end{itemize}
\end{definition}

Surprisingly, besides hyperbolic structures there are several other geometric structures on the surface, e.g. conformal and complex structures, which are parameterized by the Teichm\"{u}ller space. This comes from the famous
\begin{theorem}[Poincar\'{e} uniformization]\index{uniformization theorem}
    Any simply connected Riemann surface is biholomorphic to either $\mathbb{CP}^1, \mathbb{C}$ or the hyperbolic plane $\mathbb{H}^2$.
\end{theorem}

From the uniformization theorem using a doubling argument one can extend the result to Riemann surfaces with non-empty boundary:

\begin{theorem}
Let $\bar S_{g,h}$ be a Riemann surface with non-empty boundary. Then the universal cover of $\bar S_{g,h}$ is biholomorphic to a simply connected closed proper subset of $\mathbb{C}$ with non-empty interior.
\end{theorem}

It follows that $\bar S_{g,h}$ is of hyperbolic type if and only if its universal cover is biholomorphic to a closed subset of $\mathbb H^2$.

Consider a complex structure on $\bar S_{g,h}$, i.e. an atlas with charts in $\mathbb{C}$ and holomorphic transition functions. Then its universal cover $\tilde{S}_{g,h}$ gets also equipped with a complex structure since locally $\bar  S_{g,h}$ and $\tilde{S}_{g,h}$ are homeomorphic. We recover $\bar S_{g,h}$ from $\tilde{S}_{g,h}\subset\mathbb{H}^2$ by quotienting by the deck transformations acting by isometries. Therefore, to any complex structure on $\bar S_{g,h}$ we can associate a hyperbolic structure on $\bar S_{g,h}$ and conversely, all hyperbolic structures arise in this way. 
\begin{theorem}
    The Teichm\"{u}ller space $\mathcal{T}(\bar S_{g,h})$ is the space of complex structures modulo diffeomorphisms of $\bar S_{g,h}$ isotopic to the identity.
\end{theorem}

The \emph{pure mapping class group}\index{mapping class group} $\mathrm{PMod}(\bar S_{g,h})$, that is, the group of all orientation-preserving diffeomorphisms of $\bar S_{g,h}$ fixing pointwise the boundary components modulo the ones which are isotopic relative to boundary to the identity, acts naturally on $\mathcal{T}(\bar S_{g,h})$ by changing the marking; this action is properly discontinuous and the quotient is called the \emph{moduli space $\mathcal{M}(\bar S_{g,h})$ of Riemann surfaces}\index{moduli space!of Riemann surfaces} with underlying topological surface $\bar S_{g,h}$. Here, by \emph{properly discontinuous}\index{group!action!properly discontinuous} we mean that for every compact subset $K\subset \mathcal{T}(\bar S_{g,h})$, the set $\left\{g \in \mathrm{PMod}(\bar S_{g,h}): gK \cap K \neq \varnothing  \right\}$ is finite. 

A well-known result about the Teichm\"{u}ller space is the following:

\begin{theorem}
The Teichm\"{u}ller space $\mathcal{T}(\bar S_{g,h})$ of a compact connected oriented surface $\bar S_{g,h}$ is homeomorphic to an open ball of real dimension $6g-6+3h$. 
\end{theorem}

One direct way to prove this theorem is by parameterizing $\mathcal{T}(\bar S_{g,h})$ by Fenchel--Nielsen coordinates which shall be studied more closely in \Cref{Sec:Teichcoordinates}; a complete proof may be found in \cite[Thm. 9.7.4]{Ratcliffe}.

\section{The holonomy representation}\labelx{alg_realiz_Teich} 
An alternative, algebraic realization of the Teichm\"{u}ller space $\mathcal{T}(\bar S_{g,h})$ can be conceived through the \emph{holonomy representation}\index{holonomy!representation}\index{representation!holonomy}, which will be described next. This allows one to interpret $\mathcal{T}(\bar S_{g,h})$ purely in terms of the fundamental group $\pi_1(\bar S_{g,h})$ and its representations into the group $\mathrm{PSL}_2(\mathbb{R})$ of the orientation-preserving isometries of the upper-half plane $\mathbb H^2$.

Let $(M,f)$ be a hyperbolic structure on the surface $\bar S_{g,h}$. Upon fixing a base point, the homeomorphism $f: \bar S_{g,h} \to M$ induces an isomorphism of fundamental groups 
\[ f_{*}\colon \pi_1(\bar S_{g,h}) \to \pi_{1}(M)\; .\]
Moreover, the group $\pi_{1}(M)$ acts as the group of deck transformations on the universal cover $\tilde{M}$ with orientation preserving isometries. But since $M$ is a hyperbolic surface, by the Poincar\'{e} uniformization theorem, we know that $\tilde{M}$ is biholomorphic to a closed subset of $\mathbb{H}^2$ and the group $\mathrm{PSL}_2(\mathbb{R})$ is precisely the group $\mathrm{Isom}^{+}(\mathbb{H}^2)$ of orientation-preserving isometries. Therefore, we deduce that the group $\pi_{1}(M)$ embeds in the group $\mathrm{PSL}_2(\mathbb{R})$ as a discrete, cocompact subgroup; a group action on a topological space is called \emph{cocompact}\index{group!action!cocompact} if the associated quotient space is a compact space. 

In summary, any hyperbolic structure $(M,f)$ defines an injective homomorphism $\rho\colon \pi_1(\bar S_{g,h}) \to  \mathrm{PSL}_2(\mathbb{R})$ with discrete, cocompact image. Moreover, it turns out that such representations induced by equivalent hyperbolic structures are conjugate by an element in $\mathrm{PSL}_2(\mathbb{R})$, and conversely. Thus, we have realized:
\[\mathcal{T}(\bar S_{g,h}) \subset \mathrm{Hom}\left( \pi_1(\bar S_{g,h}), \mathrm{PSL}_2(\mathbb{R}) \right)/{\mathrm{PSL}_2(\mathbb{R}) }\; .\]
When $h=0$, i.e. $\bar S_{g,h} = S_g$ is closed, the next theorem implies that actually more can be said (see \cite{Goldman88}, \cite{Hitchin87}):

\begin{theorem}
The Teichm\"{u}ller space $\mathcal{T}(S_{g})$ of a closed surface $S_g$ is a connected component of the space of representations $\mathrm{Hom}\left( \pi_1(S_{g}), \mathrm{PSL}_2(\mathbb{R}) \right)/{\mathrm{PSL}_2(\mathbb{R}) }\; $.
\end{theorem}

\begin{remark}
In fact, when $\bar S_{g,h} = S_g$ is closed, then $\mathcal{T}(S_g)$ is one of the two connected components of the space $\mathrm{Hom}\left( \pi_1(S_g), \mathrm{PSL}_2(\mathbb{R}) \right)/{\mathrm{PSL}_2(\mathbb{R}) }\; $ consisting entirely of discrete and faithful representations. The other such component is $\mathcal{T}(S^\mathrm{op}_g)$, where $S^\mathrm{op}_{g}$ denotes the surface with the opposite orientation. From this point of view, as a set of discrete and faithful representations, the Teichm\"{u}ller space first appeared in the work of Fricke and Klein \cite{FrKl} defined in terms of Fuchsian groups. This approach will motivate the introduction of the character variety in \Cref{sec:character_variety}.
\end{remark}

The study of Teichmüller spaces is facilitated by the existence of coordinates which we shall describe next.
\chapter{Coordinates on Teichm\"uller space}\labelx{Sec:Teichcoordinates}




\abstract{The Teichmüller space can be coordinatized via the Fenchel--Nielsen length and twist parameters associated with a pants decomposition. For ciliated surfaces, one can alternatively use Thurston’s shear coordinates, whose transformation rules are particularly well-structured and naturally lead to cluster algebra structures.}

\section{Fenchel--Nielsen coordinates}\labelx{sec: Fenchel--Nielsen}

For a pair of pants $P$ and positive real numbers $a,b,c\in\R_{>0}$, there is a unique hyperbolic structure on $P$ such that the boundary components of $P$ are geodesics of respective hyperbolic length $a$, $b$ and $c$ (see \Cref{sec: triangles_hexagons}). Given two hyperbolic pairs of pants $P$ and $P'$ such that both have a boundary component (denoted by $C$ and $C'$ respectively) of hyperbolic length $a\in\R_{>0}$, one can glue $P$ and $P'$ together by identifying $C$ and $C'$. There is a 1-parameter family of inequivalent ways to do this such that the hyperbolic metrics on the two original surfaces glue together in a smooth way. This real parameter is called the \emph{twist} corresponding to the simple closed curve $C=C'$ in the resulting surface. One can also glue together different boundary components of the same hyperbolic pair of pants, as shown in \Cref{Fig:gluingpairpants}.

\begin{figure}[!ht]	
	\centering
	\includegraphics[scale=0.7]{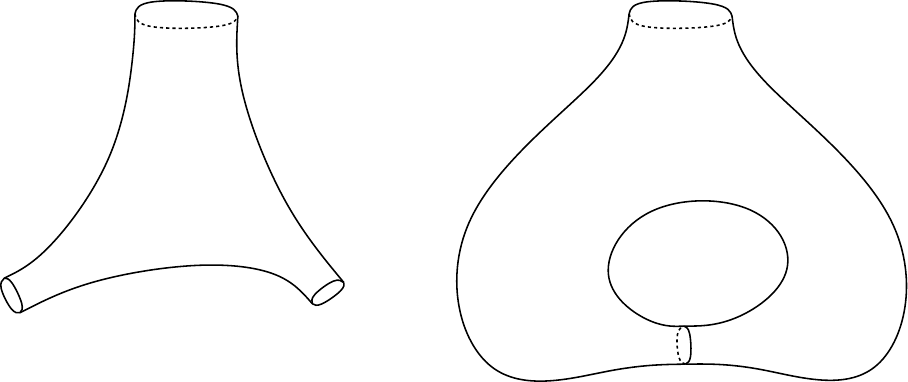}
	\caption{Gluing two boundary components of a hyperbolic pair of pants.}\labelx{Fig:gluingpairpants}
\end{figure}

Let $S_g$ be a closed connected oriented topological surface of genus $g$. One needs exactly $3g-3$ simple closed curves $\gamma_1,\dots,\gamma_{3g-3}$ to define a pants decomposition, and the latter contains $2g-2$ pairs of pants. Given a hyperbolic structure $\kappa\in \mathcal{T}(S_g)$, each of these simple closed curves $\gamma_i$ is assigned the hyperbolic length\index{Fenchel--Nielsen!coordinates!length parameter} $l_i$ of the unique geodesic representative in its free homotopy class. Further, one chooses a collection of simple closed curves $\nu_1,\dots,\nu_r$ called a \emph{marking} which decompose every pair of pants into two hexagons. Notice that the number $r$ is not fixed a priori and depends on the choice of $\nu_1,\dots,\nu_r$. This additional choice provides a twist parameter $\theta_i\in\mathbb R$ along every $\gamma_i$ for $\kappa\in\mathcal{T}(S_g)$, hence the \emph{Fenchel--Nielsen coordinates}\index{Fenchel--Nielsen!coordinates} are defined by a map
\begin{equation}\labelx{def:FN_map}
\begin{array}{crcl}
    FN\colon & \mathcal{T}(S_g) & \longrightarrow & \mathbb{R}_{>0}^{3g-3}\times \mathbb{R}^{3g-3}\\
    & \kappa & \mapsto & (l_1,\dots,l_{3g-3},\theta_1,\dots,\theta_{3g-3})\; .
\end{array}
\end{equation}
An example of a pants decomposition of the closed oriented smooth surface of genus $g=3$ is shown on the left of \Cref{Fig:FenchelNielsen}.

\begin{figure}[!ht]
	\centering
	\includegraphics[width=\textwidth]{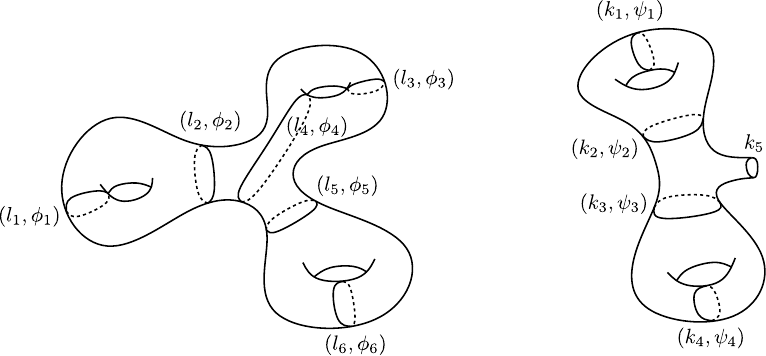}
	\caption{Hyperbolic lengths are denoted by $l_i$ and $k_i$ while the twists by $\phi_i$ and $\psi_i$.}\labelx{Fig:FenchelNielsen}
\end{figure}

Without loss of generality, we take $i=1$ and define the \emph{twist parameter}\index{Fenchel--Nielsen!coordinates!twist parameter} $\theta_1$ along $\gamma_1$ (we are following here the discussion of \cite[Sect. 7.3.1]{Martelli}). We take a hyperbolic structure $\kappa\in \mathcal{T}(S_g)$ and assume that all $\gamma_i$ are geodesics with respect to $\kappa$. The loop $\gamma_1$ separates two (not necessarily distinct) geodesic pairs of pants (see \Cref{Fig:FenchelNielsen2}). We denote by $S'$ the subsurface of $S_g$ which is the union of these two pairs of pants glued along $\gamma_1$. Let $\lambda$ and $\lambda'$ be the two curves that are restrictions of curves of the family $(\nu_1,\dots,\nu_k)$ to $S'$ that intersect $\gamma_1$. Without loss of generality, we pick $\lambda$. Let $p:=\gamma_1\cap\nu_1$. We fix a lift $\tilde p\in \tilde S=\mathbb H^2$ and lift all the curves incident to $p$: the geodesic loop $\gamma_1$ lifts to the geodesic $\tilde\gamma_1$, the (non-geodesic) path $\lambda$ lifts to a curve $\tilde\lambda$ that connects two lifts $\tilde\gamma_2$ and $\tilde\gamma_3$ of the closed geodesics $\gamma_2$ and $\gamma_3$ respectively (see \Cref{Fig:FenchelNielsen1}).

\begin{figure}[!ht]
	\centering
	\includegraphics[scale=1]{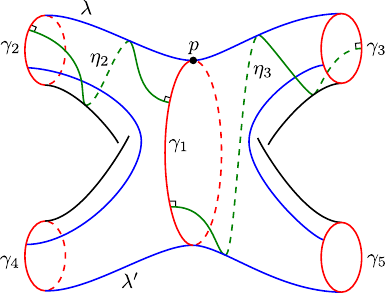}
	\caption{The subsurface $S'$, which is a union of two hyperbolic pairs of pants glued along $\gamma_1$.}\labelx{Fig:FenchelNielsen2}
\end{figure}

\begin{figure}[!ht]
	\centering
	\includegraphics[scale=1]{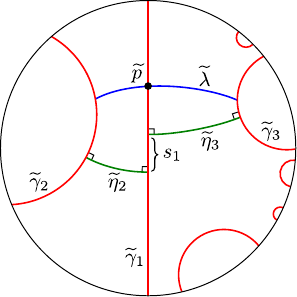}
	\caption{A lift of the subsurface $S'$ to the universal covering.}\labelx{Fig:FenchelNielsen1}
\end{figure}

Now we take two unique orthogonal geodesics $\tilde\eta_2$ and $\tilde\eta_3$ that connect $\tilde\gamma_1$ with $\tilde\gamma_2$ and $\tilde\gamma_3$ respectively.

\begin{definition}\labelx{def:shear}
    The \emph{shear}\index{shear} $s_1\in\mathbb R$ between $\tilde\eta_2$ and $\tilde\eta_3$ is the signed length of the geodesic segment of $\tilde\gamma_1$ between the intersection points $\tilde\gamma_1\cap\tilde\eta_2$ and $\tilde\gamma_1\cap\tilde\eta_3$. The sign of $s_1$ is positive if an observer walking along $\gamma_1$ first meets another geodesic on its left (as in \Cref{Fig:FenchelNielsen1}).
\end{definition}

Finally, we define the twist parameter $\theta_1$ along $\gamma_1$ as $\theta_1:=\frac{2\pi s_1}{l_1}.$

The following theorem shows that the map $FN$ from \eqref{def:FN_map} is well-defined and provides a coordinatization of $\mathcal T(S_g)$:

\begin{theorem}[Fenchel--Nielsen coordinates, {\cite[Thm. 7.3.1]{Martelli}}]
    The map $FN$ is well-defined, bijective and depends only on the choice of pants curves $\gamma_1\dots,\gamma_{3g-3}$ and of a marking $\nu_1,\dots,\nu_k$.
\end{theorem}

There are in general many different pants decompositions with a marking of the same surface, and each one yields a different set of Fenchel--Nielsen coordinates on the Teichmüller space $\mathcal{T}(S_g)$ of the surface $S_g$, which can be interpreted as different global charts on $\mathcal{T}(S_g)$. One can compute the transition functions from one chart to another, but these \emph{do not enjoy nice algebraic properties}. 

In the case of a finite type compact surface $\bar S_{g,h}$ obtained by removing $h>0$ many open disks from a genus $g\geq 0$ oriented closed surface where $g$ and $h$ are such that $\chi(\bar S_{g,h})=2-2g-h<0$, one needs (in addition to the $k$ boundary loops) $3g-3+h$ simple closed curves to define a pants decomposition, and the latter involves $2g-2+h$ pairs of pants. Given a hyperbolic structure $\kappa\in\mathcal T(\bar S_{g,h})$, the boundary components of $\bar S_{g,h}$ are hyperbolic geodesics entirely described by their length, and hence the Fenchel--Nielsen coordinates provide a map
\begin{equation}\labelx{eq:FN-compact}
FN\colon\mathcal{T}(\bar S_{g,h})\longrightarrow \R_{>0}^{3g-3+2h}\times \R^{3g-3+h}\; .
\end{equation}
An example corresponding to $g=2$ and $h=1$ is shown on the right of \Cref{Fig:FenchelNielsen}.

We can enlarge the domain of definition of the map $FN$ allowing boundary components of $\bar S_{g,h}$ to degenerate into cusps. To make the statement precise, instead of considering a compact surface $\bar S_{g,h}$, we take a ciliated surface $S_{g,h}$ with $h$ internal punctures\index{puncture!internal} and without boundary. We can equip such a surface with a (possibly incomplete) convex hyperbolic metric of finite area. We denote the space of all such metrics modulo isotopy by $\mathcal T(S_{g,h})$. Then for every puncture there is a well-defined length function\index{puncture!internal!length} on $\mathcal T(S_{g,h})$, which is the infimum of lengths of all simple loops surrounding the puncture. These length functions take values in the closed half-ray $\mathbb R_{\geq 0}$. Note that a puncture is a cusp\index{cusp} if its length is equal to zero. Then the Fenchel--Nielsen map can also parameterize the space $\mathcal T(S_{g,h})$:
\begin{equation*}
FN\colon\mathcal{T}(S_{g,h})\longrightarrow (\R_{>0}\times\R)^{3g-3+h}\times\R_{\geq0}^{h}\; .
\end{equation*}

Notice that if a hyperbolic structure from $\mathcal{T}(S_{g,h})$ is chosen, then for every puncture with positive length $l>0$, a neighborhood of this puncture, which is homeomorphic to an open annulus, can be compactified by adding a closed geodesic of length $l$, which corresponds to the metric completion of this annulus. If all punctures have positive lengths, this provides a compactification of the hyperbolic ciliated surface $S_{g,h}$ which is the compact hyperbolic surface $\bar S_{g,h}$ in the sense of \Cref{sec:hyp.surfaces} parameterized by~\eqref{eq:FN-compact}.

\section{Shear coordinates}\labelx{Subec:Shearcoord}

In this Section, we will introduce another set of coordinates on a slightly different version of the Teichm\"uller space for a ciliated surface $S:=S_{g,\vec p}$ .

Remember that a ciliated surface\index{surface!ciliated} $S$ can be equipped with a marked, possibly incomplete, convex, finite area hyperbolic structure possibly with boundary  such that the boundary segments between cilia are infinite geodesics and cilia\index{cilia} are points at infinity. Moreover, to every internal puncture there is an associated length (see \Cref{def:length_puncture}). The metric completion of each puncture of positive length adds a closed boundary geodesic to the surface. Note that there are two different orientations for each of these boundary geodesics. 

\begin{definition}\labelx{def:framed_Teichm}
We denote by $\mathcal T^x(S)$ the space of all marked hyperbolic structures on $S$ as above up to isotopy together with a choice of an orientation of the boundary geodesics corresponding to the metric completion of every puncture with positive length. We will call the space $\mathcal T^x(S)$, the \emph{framed Teichm\"{u}ller space} of the ciliated surface $S$.
\end{definition}

\begin{remark}
Note that the notion of a ``Teichm\"{u}ller space of surfaces with holes'' was first introduced in \cite{Fock97} (and further studied in \cite{FG}, \cite{FG_book}) in order to describe the space of all hyperbolic structures on surfaces with compact geodesic boundary components together with a choice of an orientation of the boundary geodesics. Our context here is slightly more general in a sense that we consider surfaces with punctures and cilia. This is a convention that will prove out to be useful in order to study later on in \Cref{Sec:FockGoncharov} framed fundamental group representations and the notion of positivity.    
\end{remark}

The global coordinate systems on $\mathcal T^x(S)$ that we will define in this Section are associated with triangulations of $S$ (see \Cref{Sec:ciliatedsurfaces}). Given a triangulation $\Delta$ of $S$ and a hyperbolic structure $\kappa\in \mathcal T^x(S)$, one can associate a real number $s_\gamma$ to every internal edge of $\Delta$; this defines a map called \emph{shear coordinates}\index{shear!coordinates}
\begin{equation*}
    \begin{array}{crcl}
	Sh\colon & \mathcal T^x(S) & \rightarrow & \mathbb{R}^{E_i(\Delta)}\\
            & \kappa & \mapsto & (s_\gamma)_{\gamma\in E_i(\Delta)}\; ,
    \end{array}
\end{equation*}
where $E_i(\Delta)$ is the set of internal edges of $\Delta$.

Given a hyperbolic structure $\kappa\in\mathcal T^x(S)$, every ideal triangulation can be turned into a geodesic ideal triangulation\index{ideal!triangulation}\index{triangulation!ideal}, i.e. a triangulation with geodesic edges either connecting points at infinity (cusps and cilia) or spiraling around boundary geodesics in the direction prescribed by $\kappa$ for punctures with positive length. Every face of an ideal triangulation is an ideal geodesic hyperbolic triangle. 

The lift of $\Delta$ to the universal covering $\tilde S\subseteq \mathbb H^2$ of $S$ is an infinite $\pi_1(S)$-invariant ideal geodesic triangulation $\tilde\Delta$ of the convex domain $\tilde S$. We consider an ideal quadrilateral of $\tilde\Delta$ with vertices $x_1,x_2,x_3,x_4\in\partial_\infty\tilde S$ that is triangulated by an edge $\gamma$ connecting $x_1$ and $x_3$  (see \Cref{Fig:gluingtriangles}).

\begin{figure}[!ht]
	\centering
	\includegraphics[width=\textwidth]{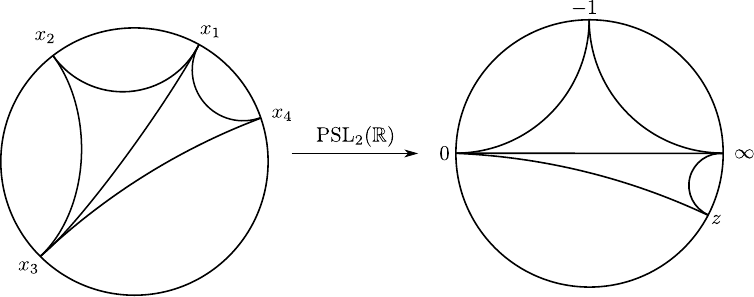}
	\caption{A normal form for a triangulated quadrilateral in $\H$.}\labelx{Fig:gluingtriangles}
\end{figure}

Now we take two unique orthogonal geodesics $\tilde\eta_2$ and $\tilde\eta_4$ that connect $\tilde\gamma$ with $x_2$ and $x_4$ respectively. We associate in this way the shear $s_\gamma$ of $\tilde\eta_2$ and $\tilde\eta_4$, defined as in \Cref{def:shear}, to the edge $\gamma$. Since the construction is $\pi_1(S)$-invariant, the coordinates on $\tilde\Delta$ descend to coordinates on $\Delta$.

\begin{theorem}[Shear coordinates, {\cite[Prop. 7.4.9]{Martelli}}]
    The map $Sh$ is well-defined, bijective and depends only on the choice of a triangulation $\Delta$.
\end{theorem}

Given a hyperbolic structure $\kappa\in\mathcal T^x(S)$, it is easy to recognize in terms of shear coordinates if a puncture on $S$ is a cusp or admits a boundary geodesic, and in the latter case, how it is oriented.

\begin{proposition}[{\cite[Sects. 7.4.4-7.4.5]{Martelli}}]
    Let $\gamma_1,\dots,\gamma_k$ be all the edges of $\Delta$ that are incident to a puncture $p$ of $S$, let $\kappa\in\mathcal T^x(S)$, and let $c_i$ be the shear coordinate associated to $\gamma_i$, for $i\in\{1,\dots,k\}$. The number $C=\sum_{i=1}^kc_i$ is zero if and only if $p$ is a cusp. The number $C>0$ if and only if $p$ has length $C$ and the corresponding boundary geodesic is oriented accordingly to the orientation of $S$. The number $C<0$ if and only if $p$ has length $-C$ and the corresponding boundary geodesic is oriented oppositely to the orientation of $S$.
\end{proposition}

The shear coordinates admit an interpretation as cross ratios. Since the points $x_1,\dots,x_4\in\partial_\infty\tilde S\subseteq\partial_\infty \mathbb H^2\cong \mathbb RP^1$,
we can compute the cross ratio\index{cross ratio}:
\begin{equation}\labelx{Eq:crossratio}
z_\gamma:=z:=[x_1,x_2,x_3,x_4]=- \frac{(x_1-x_2)(x_3-x_4)}{(x_2-x_3)(x_4-x_1)}\; .    
\end{equation}
The direct computation shows that $z_\gamma=\E^{s_\gamma}$.

The number $z$ can be also interpreted as follows: Since the element of $\pslr$ which maps $(x_1,x_2,x_3)$ to $(\infty, -1, 0)$ is unique, $z$ is uniquely defined by the cyclically ordered quadruple $(x_1,x_2,x_3,x_4)$ and the choice of a first element $x_1$. The only other possible choice of the first element for the cyclic set $(x_1,x_2,x_3,x_4)$ which respects the triangulation in the sense that the first three elements of the quadruple are the vertices of a triangle is $x_3$. If instead of considering the quadruple $(x_1,x_2,x_3,x_4)$, one takes $(x_3,x_4,x_1,x_2)$, it is the triangle with vertices $(x_3,x_4,x_1)$ which is mapped to $\infty,-1,0$ while the second one has vertices at $x_1,x_2,x_3$. It is a very happy fact (which can be checked by an explicit computation and which is also a direct consequence of \Cref{Eq:crossratio}) that the shear parameters in both cases actually coincide. Hence the shear parameter we have defined is $\pslr$-invariant and depends only on the vertices $x_1,x_2,x_3,x_4$ of the quadrilateral and on how it is triangulated. 

\begin{remark}
Note that shear coordinates as discussed above provide the simplest example of a cluster algebra when the structure group is the group $\mathrm{SL}_2(\mathbb{R})$ studied in detail in \Cref{sec:Clusters}.
\end{remark}
\chapter{Differentials on a Riemann surface}\labelx{Sec:Differentials}

\abstract{We review basic facts about meromorphic differentials that will be needed later on for the analytic construction of spectral networks. In particular, quadratic differentials induce natural decompositions of surfaces that lead to triangulations, thereby providing insight into some of its differential geometric aspects. Standard references for the role of quadratic differentials in Teichmüller theory include \cite{Gardiner} or \cite{Strebel}.}

\section{Abelian differentials and their trajectories}\labelx{Sec:holo_abel_diff_main_defs}

Let $X$ be a compact Riemann surface with underlying topological surface $S$. Since the complex dimension of $X$ is by definition one, there are up to degree-2 differential forms defined on $X$. Differential forms are invariant under coordinate changes, thus are defined over the entire space $X$. We may thus define:

\begin{definition}
Let $X$ be a Riemann surface and choose a local coordinate $z$ on $X$. An \emph{abelian differential}\index{abelian differential} on $X$ is locally described as a $1$-form $\omega=f(z)\D z$, where $f$ is a smooth function on $X$. In other words, it is a smooth section of $K$ where $K=\Omega^{1,0}(S)$ denotes the canonical bundle. If every function defining locally the differential $\omega$ is holomorphic (resp. meromorphic), then $\omega$ is called a \emph{holomorphic abelian differential} (resp. meromorphic abelian differential)\index{holomorphic!abelian differential}\index{abelian differential!holomorphic}\index{abelian differential!meromorphic}.
\end{definition}

In the sequel unless stated otherwise, we consider meromorphic abelian differentials. 

An application of the Riemann--Roch theorem shows that the dimension of the vector space $H^0(X, \Omega)$ of holomorphic abelian differentials on $X$ is equal to the genus $g$ of the Riemann surface. This can be proved using analytic methods (see for instance \cite{FaKr}); explicit descriptions of a basis $\omega_1,..., \omega_g$ of holomorphic abelian differentials for certain cases of Riemann surfaces can be found in \cite{Les}.

Zeroes and poles of the function $f(z)$ are defined independent of the choice of local coordinates, thus respectively one  defines the zeroes and poles of a corresponding meromorphic differential. For the Laurent series expansion 
\[f(z)=\sum_{n=-{\infty} }^{\infty} \alpha_{n}(z-z_{0})^{n}\]
around a singular point $z_{0} \in X$, the \emph{residue}\index{residue}\index{abelian differential!residue} $\mathrm{Res}_{z_{0}}\omega$ of an abelian differential $\omega$ at the point $z_{0}$ is defined as the coefficient $\alpha_{-1}$ in this expansion.  The residue is also independent of the choice of a local coordinate since it can be written as
\[\mathrm{Res}_{z_{0}}\omega= \frac{1}{2\pi\I} \int_{\del B_{z_{0}}} \omega(z)\; ,\]
for $B_{z_{0}}$ a disk around the point $z_{0}$ such that the differential form $\omega$ has no other singular point within the closure of $B_{z_{0}}$. Note that for a finite collection of singular points $z_{0}, ..., z_{n}$ of a differential $\omega$ on $X$, one has that $\sum_{j=0}^{n} \mathrm{Res}_{z_{j}}\omega =0$. Indeed, for a disk $B_j$ around each singularity $z_j$ containing no other singularity within its closure, one sees that
\[\sum_{j=0}^{n} \mathrm{Res}_{z_{j}}\omega = \frac{1}{2\pi \I} \sum_{j=0}^{n} \int_{\del B_{z_{j}}} \omega(z)= -\frac{1}{2\pi \I}\int_{X\setminus \cup B_{z_{j}}} \D\omega (z)=0\; ,\]
since the differential $\omega$ is holomorphic on $X\setminus \cup B_{z_{j}}$, and so $\D\omega =0$ there.

\begin{definition}
 Let $X$ be a Riemann surface and let $\omega$ be an abelian differential on $X$. The \emph{critical points}\index{critical!point}\index{abelian differential!critical point} of $\omega$ are its zeroes and poles, whilst all other points on $X$ are called \emph{regular}\index{abelian differential!regular point}. Note that there is a finite number of critical points for any abelian differential $\omega$. 
\end{definition}

For a generically non-zero abelian differential $\omega$ let now $P$ denote the set of poles of $\omega$ and by $Z$ the set of zeroes of $\omega$.  The imaginary part $\mathrm{Im}(\omega)$ of $\omega$ is a field of real 1-forms on $X$. Since the tangent bundle of $X$ has real rank two, the form $\mathrm{Im}(\omega)$ has generically 1-dimensional kernel $\mathrm{Ker}(\mathrm{Im}(\omega_z))\subset T_zX$, for every $z\in X$. Such an object is usually called a \emph{distribution}\index{distribution}. More precisely, a 1-dimensional distribution in the tangent bundle $TX$ of $X$ is a 1-dimensional subbundle of $TX$. In our case, the form $\omega$ can only have isolated zeroes, so the distribution $\mathrm{Ker}(\mathrm{Im}(\omega))$ is well-defined over $X':=X\setminus(P\cup Z)$.

A distribution $\mathrm{Ker}(\mathrm{Im}(\omega))$ gives rise to a \emph{foliation}\index{foliation}\index{abelian differential!foliation} $\hat{\mathcal F}_\omega$ on $X'$, i.e. a family of smooth unoriented curves (called \emph{leaves of the foliation}\index{abelian differential!foliation!leaf})\index{foliation!leaf} that are tangent to $\mathrm{Ker}(\mathrm{Im}(\omega))$ at every point and one leaf of $\hat{\mathcal F}_\omega$ for every point of $X'$. We can assign an orientation to $\hat{\mathcal F}_\omega$ choosing smoothly at every $z\in X\setminus Z$ a direction $\mathbb R_+ v_z\subset \mathrm{Ker}(\mathrm{Im}(\omega))$ such that $\mathrm{Re}(\omega(v_z))>0$. The oriented foliation will be denoted by $\mathcal F_\omega$ (\Cref{fig:regular_horizontal_foliation}).

Notice, that $\mathcal F_{-\omega}$ can be obtained from $\mathcal F_{\omega}$ by inverting orientation.

\begin{example}
Let $X=\mathbb{C}$ and consider two abelian differentials, $\omega_1=\D z$ and $\omega_2=(z-z_0)^2\D z$, for $z_0 \in X$. Then, the foliations $\mathcal F_{\omega_1}$ and $\mathcal F_{\omega_2}$ in a small neighborhood of $z_0$ look as in \Cref{fig:regular_horizontal_foliation}.
\begin{figure}[!ht]
\centering
\includegraphics[]{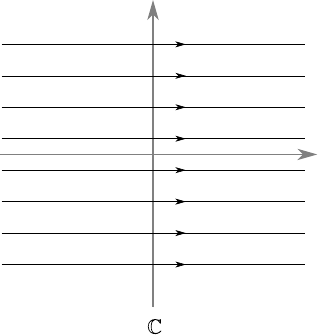}\qquad
\includegraphics[]{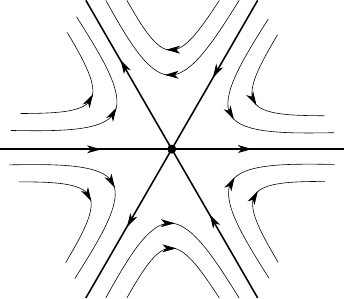}
\caption{Foliations $\mathcal F_{\omega_1}$ (top) and $\mathcal F_{\omega_2}$ (bottom) around $z_0$.}
\labelx{fig:regular_horizontal_foliation}
\end{figure}
\end{example}

\section{Quadratic differentials and their trajectories}\labelx{Sec:Quaddifftrajectories}

Instead of considering abelian differentials which are 1-forms, we can consider differentials of higher order $k\in\mathbb{N}_{\geq 2}$. These are symmetric tensors locally of the form $f(z)\D z^k$ where $f$ is a local smooth function. Particularly interesting is the case of \emph{quadratic differentials} which we study next.

\begin{definition}
Let $X$ be a Riemann surface with underlying topological surface $S$ and choose a local coordinate $z$ on $X$. A \emph{quadratic differential}\index{quadratic differential} on $X$ is locally described as $\omega=f(z)\D z^2$, where $f$ is a smooth function on $X$. In other words, it is a smooth section of $K\otimes K$ where $K=\Omega^{1,0}(S)$ denotes the canonical bundle. If every function defining locally the differential $\omega$ is holomorphic (resp. meromorphic), then $\omega$ is called a \emph{holomorphic quadratic differential} (resp. meromorphic quadratic differential)\index{quadratic differential}\index{quadratic differential!holomorphic}\index{quadratic differential!meromorphic}.
\end{definition}

In the sequel unless stated otherwise, we consider meromorphic quadratic differentials. 

\begin{remark}\labelx{non-can_orient}
In contrast to the case of abelian differentials, there is no canonical choice of orientation for trajectories of quadratic differentials. Indeed, one can consider similarly for a meromorphic quadratic differential $\omega$ a distribution\index{quadratic differential!distribution} $\mathrm{Ker}(\mathrm{Im}(\omega))$ giving rise to a foliation\index{quadratic differential!foliation} $\hat{\mathcal F}_\omega$ on $X'$. However, a canonical orientation cannot be assigned to $\hat{\mathcal F_\omega}$. This is because  the sign of $\mathrm{Re}(\omega_z)(v)$ at every $z$ and every $0\neq v\in \mathrm{Ker}(\mathrm{Im}(\omega))$ does not depend on $v$ and is given by the sign of $\mathrm{Re}(f(z))$. We will see below that in fact there are quadratic differentials such that the corresponding non-oriented foliation $\hat{\mathcal F_\omega}$ is non-orientable.
\end{remark}

A classical application of the Riemann--Roch theorem provides that on a closed Riemann surface of genus at least 2, the space $Q(X)$ of all quadratic holomorphic differentials on $X$ has complex dimension $3g-3$; we refer, for instance, to \cite[Cor. 5.4.2]{Jost} for a proof. In fact, the Teichm\"{u}ller space  $\mathcal{T}(S_g)$ is isomorphic to the vector space  $Q(X)$ as seen in \cite{Wolf} using Hopf differentials of appropriate harmonic maps.

\begin{definition}
Let $\omega$ be a quadratic differential over a Riemann surface $X$ locally described as $\omega=f(z)\D z^2$. For $\theta \in [0,2\pi)$, a \emph{straight $\theta$-arc}\index{quadratic differential!straight $\theta$-arc} with respect to $\omega$ is a smooth curve $\gamma$ along which $\mathrm{arg}(\omega)=\theta \mod 2\pi$. The curve $\gamma$ is called a \emph{$\theta$-trajectory} if it is maximal, that is, $\gamma$ is not properly contained in any other straight arc. 
\end{definition}

Note that a straight arc contains only regular points of a quadratic differential $\omega$. We will be interested in the so-called \emph{horizontal trajectories}\index{quadratic differential!horizontal trajectory}, that is, the 0-trajectories, as well as the \emph{vertical trajectories}\index{quadratic differential!vertical trajectory}, that is, the $\frac{\pi}{2}$-trajectories. 
The next theorem provides the existence and uniqueness of maximal trajectories:

\begin{theorem}\cite[Thm. 5.5]{Strebel}
Let $\omega$ be a quadratic differential on a Riemann surface $X$. Then through every regular point of $\omega$, there exists a uniquely determined maximal trajectory.  
\end{theorem}

From this Theorem, one gets that a maximal trajectory is parameterized by an open interval. Then one can consider the two limits in the two limit points of the parameterizing interval. In the case when such a limit exists, we say that this defines an \emph{end} of the trajectory. Note that the ends of a maximal trajectory cannot be regular points, otherwise this would violate maximality.

In the light of \Cref{non-can_orient}, note that it is also common to refer to the trajectories of quadratic differentials as the \emph{leaves}\index{quadratic differential!foliation!leaf}\index{foliation!leaf} of the foliation $\hat{\mathcal F_\omega}$. \Cref{fig:foliation_zero} illustrates the behavior of horizontal and vertical leaves of the foliation around a simple zero and \Cref{fig:poles} analogously around poles of various orders of a meromorphic quadratic differential.

\begin{figure}[!ht]
\centering
\includegraphics[scale=0.7]{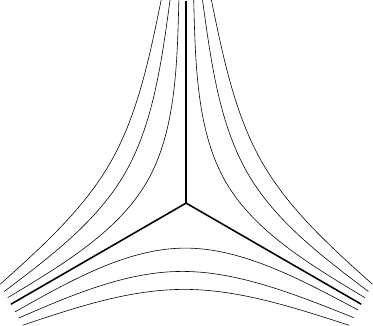}
\caption{Foliation $\hat{\mathcal F_\omega}$ around a simple zero.}
\labelx{fig:foliation_zero}
\end{figure}

\begin{figure}[!ht]
\centering
\includegraphics[width=\textwidth]{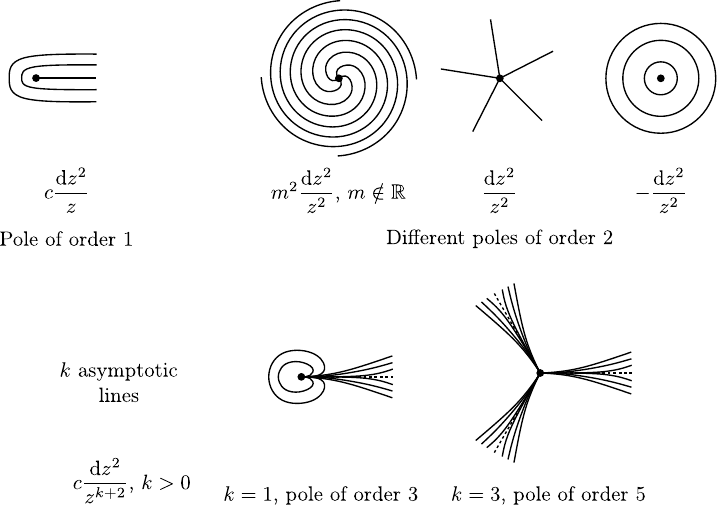}
\caption{Foliation $\hat{\mathcal F_\omega}$ around poles of different orders.}
\labelx{fig:poles}
\end{figure}

\section{Critical graphs of abelian and quadratic differentials}\labelx{sec:Crit_graph_quadratic}

The information from the leaves of the non-oriented foliation $\hat{\mathcal F_\omega}$ in the case of an abelian or quadratic differential $\omega$ can be collected to the notion of a critical graph as defined below.

\begin{definition}\labelx{defn:critical_leaf-generic_qd} Let $\omega$ be an abelian or quadratic meromorphic differential on $X$. A leaf of $\hat{\mathcal F_\omega}$ is called \emph{critical}\index{quadratic differential!foliation!leaf!critical} if one of its ends is a zero of $\omega$. A critical leaf is called \emph{regular}\index{quadratic differential!foliation!leaf!critical!regular} if the other end is a pole of $\omega$. A critical leaf is called a \emph{saddle connection}\index{saddle!connection}\index{quadratic differential!saddle connection} if both its ends are zeroes of $\omega$. A leaf is moreover called \emph{recurrent}\index{quadratic differential!foliation!leaf!recurrent} if its closure as a subset of $X$ has positive measure. The \emph{critical graph}\index{critical!graph}\index{quadratic differential!critical graph} of $\omega$ is the collection of all critical leaves of $\hat{\mathcal F_\omega}$. The quadratic differential is called \emph{generic}\index{quadratic differential!generic} if all its critical leaves are regular.
\end{definition}

Let $\omega$ be a generic meromorphic quadratic differential with at least one pole of order at least two and only simple zeroes. Then $\omega$ defines a triangulation. More precisely, for every zero of $\omega$ we have exactly three leaves that end at poles. For every two such poles, there are leaves of the foliation connecting them. We fix one of them for every pair of poles as above, thus obtaining a triangulation of $X$ (see \Cref{fig:triangulation}).
\begin{figure}[!ht]
\centering
\includegraphics[scale=0.8]{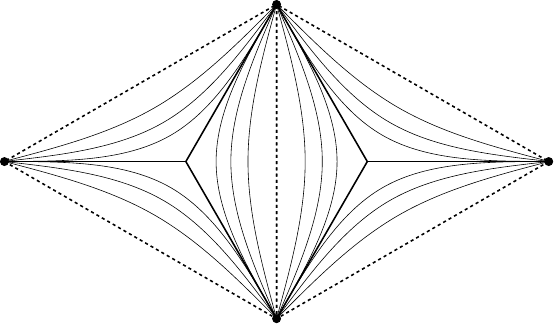}
\caption{Ideal triangulation given by a foliation of a meromorphic quadratic differential.}
\labelx{fig:triangulation}
\end{figure}

A fundamental theorem of Strebel \cite[Thm. 23.5]{Strebel} now shows the following; for the present form below see \cite[Thm. 2.1]{SCLX}: 
\begin{theorem}
Given a compact Riemann surface $X$ of genus $g\ge 0$ with $n \ge 1$ many marked points $p_i$ and  $n \ge 1$ many positive integers $a_i$, for $i=1,...,n$, with $2g-2+n>0$, there exists a unique quadratic meromorphic differential $\omega$ on $X$ satisfying the following conditions:
\begin{enumerate}
    \item The differential $\omega$ has a double pole at each marked point $p_i$ with residue $a_i$ and it has no other poles.
    \item Every closed leaf of $\omega$ is circling around exactly one of the marked points. 
    \item The quadratic differential $\omega$ has no recurrent leaf.
\end{enumerate}
\end{theorem}

A generic quadratic differential on a Riemann surface has horizontal and vertical leaves which naturally describe a way to cut up the surface into Euclidean rectangles. The horizontal and vertical trajectories together with the quadratic differential determine two measured foliations which are transverse. Two foliations are transverse at a singular point if there are $C^1$-topological structures equivalent to the horizontal and vertical leaves of $z^r \D z^2$ for some integer $r \ge 1$ in a local coordinate $z$. 

\begin{example}\labelx{example_traj_diff_proj}This example appears in  \cite[Ex. 1.2.4]{Ionita}. Consider
$X={{\mathbb{CP}}^{1}}$, the complex projective line. The meromorphic quadratic differential $\omega (x)=x\D x\otimes \D x$, determines a real projective vector field $V_{\omega}$ on $X\setminus \mathrm{Crit}(\omega)$, where $\mathrm{Crit}(\omega)$ is the set of critical points of $\omega$ defined by $ \pm \sqrt{\omega} (V_{\omega}) \in \mathbb{R}$. The integral curves of this vector field satisfy: 
\[\int_{{{x}_{0}}}^{{{x}_{1}}}{\pm \sqrt{x}\D x}=\pm \frac{2}{3}\left( x_{1}^{3/2}-x_{0}^{3/2} \right)\in \mathbb{R}\; ,\]
for all $x_0 < x_1 \in \mathbb{R}$.

\noindent When ${{x}_{0}}\ne 0$, then there exists a neighborhood $U$ of $x_0$ such that the map ${{x}_{1}}\mapsto x_{1}^{3/2}$ restricted to $U$ is injective, for either choice of branch of the square root function. Thus, $U$ contains a unique trajectory of $\omega$ passing through the point $x_0$  parameterized by 
	\[{{x}_{1}}(t)={{\left( x_{0}^{3/2}+t \right)}^{2/3}}, \text{ for }
t\in \mathbb{R}\; .\]
When ${{x}_{0}}=0$, a neighborhood $U$ of $x_0$ contains three trajectories starting at 0. These are parameterized by  
	\[{{x}_{1,k}}(t)=t{{e}^{2\pi \I k/3}}, \text{ for }
t\in {{\mathbb{R}}_{+}} \text{ and } k\in \{0,1,2\}\; .\] 
Finally, in a neighborhood of $x=\infty $, using the local coordinate  $y={{x}^{-1}}$, the quadratic differential $\omega$ has a pole of order five:
	\[\omega (y)={{y}^{-1}}\D({{y}^{-1}})\otimes \D({{y}^{-1}})={{y}^{-5}}\D y\otimes \D y\; .\]
\Cref{trajectories_omega_proj} describes the trajectory structure around the points $x=0$ and $x=\infty $.

\begin{figure}[!ht]
\centering
\includegraphics[scale=0.8]{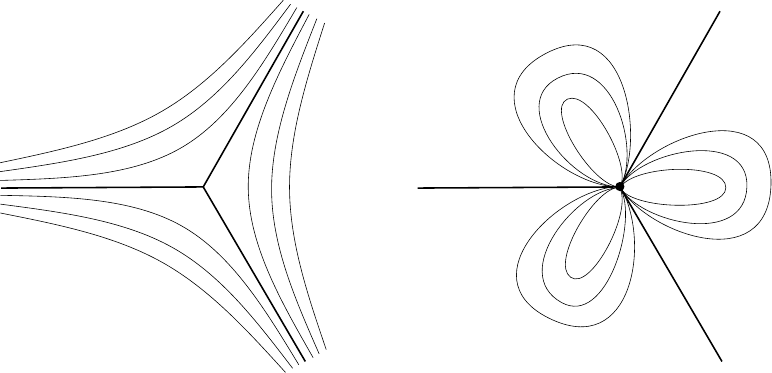}
\caption{Trajectories of the differential $\omega$ in  \Cref{example_traj_diff_proj} in a neighborhood of $x=0$ (left) and of $x=\infty$ (right).}
\labelx{trajectories_omega_proj}
\end{figure}

\end{example}

\section{Spectral curve}\labelx{Sec:Spectral-curve}

To a Riemann surface together with a collection of holomorphic differentials, we associate a certain ramified covering, the so-called \emph{spectral curve}. Spectral curves were introduced in Hitchin's papers \cite{Hitchin87b,Hitchin87} in the process of abelianization of the moduli spaces of stable vector bundles.  \Cref{Ex:spectral-curve-loc} in \Cref{Sec:ramified-coverings} about ramified coverings gives the local model of a spectral curve.
For our purposes, we adopt the following version of the notion:

\begin{definition}\labelx{def:spectral_curve}
Let $X$ be a smooth complex algebraic curve of genus $g \ge 2$, and $K=\Omega_X^1$ the canonical line bundle over $X$. A \emph{spectral curve}\index{spectral!curve} or \emph{spectral cover}\index{spectral!cover} of degree $r$ is a complete smooth algebraic curve $\Sigma$ embedded in the cotangent bundle $T^*X$ such that the projection $\pi:T^*X \to X$ is a ramified covering of degree $r$:
\begin{equation}\labelx{spectral_curve_diag}
	\xymatrix{ f: \Sigma \ar[r] & T^*X \ar[d]^{\pi} \\  & X\; .}
\end{equation}
\end{definition}

Let $\eta \in \mathrm{H}^0(T^*X, \pi^{*}K)$ be the tautological 1-form on $T^*X$ such that $-\D \eta$ is the canonical symplectic form on $T^*X$ ($\eta$ is also called the \emph{Liouville form}\index{Liouville form}). 

\begin{example}
For a holomorphic quadratic differential $q_2\in\mathrm{H}^0(X,K^{\otimes 2})$, consider the solutions to the equation
$$\eta^{\otimes 2}+q_2=0\; .$$
The set of solutions forms a spectral curve $\Sigma\subset T^*X$, with ramification points at the zeros of $q_2$. The order of a zero of $q_2$ equals the ramification index minus 1. Hence if all zeros of $q_2$ are simple, $\Sigma$ is a simple ramified covering.
\end{example}

We can generalize this example. Consider the vector space of tuples of holomorphic differential $k$-forms, where $k$ varies from 1 to $r$:
\[\mathcal{B}:= \oplus_{i=1}^{r} \mathrm{H}^0(X, K^{\otimes i})\; .\]
Using the Riemann--Roch formula, one can show that the dimension of $\mathcal{B}$ equals $r^2(g-1)+1$. A point $q=(q_1,...,q_r) \in \mathcal{B}$ gives a spectral curve via the characteristic equation  
\[\eta^{\otimes r} + \sum_{i=1}^r q_i \, \eta^{\otimes (r-i)} =0\; .\]
Alternatively, the curve $\Sigma$ can be viewed as the 0-locus of the global section 
\[\eta^{\otimes r} + \sum_{i=1}^r \pi^{*}q_i \, \eta^{\otimes (r-i)} \in \mathrm{H}^0(T^*X,\pi^{*}K^{\otimes r})\; .\]
A point $q=(q_1,...,q_r) \in \mathcal{B}$ is called \emph{generic} if the spectral curve $\Sigma$ has only simple ramification points.

\begin{remark}\labelx{rem:ram_path_lifting}
Let $X=\Sigma=\mathbb{C}$ and $\pi\colon \Sigma\to X$, $\pi(z)=z^2$. This is a ramified covering with the only ramification point $z=0$. We consider two homotopic loops $\gamma_1$ and $\gamma_2$ in $X$ as in \Cref{fig:ex2.1_1}. Both lifts of $\gamma_1$ are closed, but none of the lifts of $\gamma_2$ are closed, so they cannot be homotopic.

\begin{figure}[ht]
\centering
\includegraphics[]{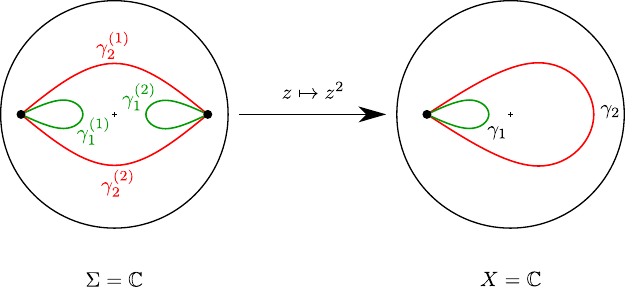}
\caption{Non-homotopy invariant path lifting for ramified coverings.}
\labelx{fig:ex2.1_1}
\end{figure}

Thus the path-lifting property of unramified coverings (\Cref{uniq_path_lift}) does not hold for ramified coverings. Later on, we will see that spectral networks can provide a way to fix this problem (see \Cref{sec:path_lifting}).

\end{remark}

\section{Strebel differentials}\labelx{sec:Strebel_difs}

In \Cref{defn:critical_leaf-generic_qd}, we called a meromorphic quadratic differential $\omega$ over a Riemann surface \emph{generic}, if all critical leaves of the foliation $\hat{\mathcal{F}}_{\omega}$ are regular, that is, each leaf of $\hat{\mathcal{F}}_{\omega}$ has one of its end at a zero of $\omega$ and its other end at a pole of $\omega$. This generic situation enabled the definition of an ideal triangulation of the Riemann surface (see \Cref{fig:triangulation}). We now consider a \emph{non-generic} case; in particular, we are interested in the case when $\omega$ is a so-called Strebel differential in the sense described below:

\begin{definition}\labelx{defn:Strebel_dif}
Let $\bar{X}$ be a closed Riemann surface of genus $g \geq 2$ and fix a finite collection of marked points $P=\{s_1,...,s_n\}$ of $\bar{X}$, for $n \in \mathbb{Z}_{>0}$, and let $X=\bar{X}\backslash P$ be the corresponding open Riemann surface. Let also $\omega$ be a meromorphic quadratic differential on $X$ with double poles at the points of $P$ and let $\hat{\mathcal{F}}_{\omega}$ be the non-oriented foliation of $\omega$. If the foliation $\hat{\mathcal{F}}_{\omega}$ has no leaves terminating at punctures (poles of $\omega$) and, moreover, all leaves are compact, then we say that $\omega$ is a \emph{Strebel (quadratic) differential}\index{quadratic differential!Strebel}.     
\end{definition}
From this definition, one sees that for a Strebel differential the leaves of the foliation $\hat{\mathcal{F}}_{\omega}$ are all either closed trajectories or saddle connections. That Strebel differentials indeed exist is a fundamental theorem by Strebel (cf. \cite[Chap. VI]{Strebel}).

\begin{example}\labelx{ex:Strebel}
This example is taken from \cite[Sect. 3.3]{HoNe}
Let $X$ be the sphere minus 3 points, and consider the meromorphic quadratic differential $\omega$ locally described by 
\[\omega = -\frac{m^2_{\infty}z^2-(m_{\infty}^2+m_0^2-m_1^2)z+m_0^2}{z^2(z-1)^2}\D z^2\; ,\]
where $m_i \in \mathbb{R}$, for $i=0,1,\infty$. Notice that $\omega$ is a Strebel differential with double poles at the punctures $\{0,1,\infty\}$ and residues $-m_0^2, -m_1^2, -m_{\infty}^2$, respectively. 

\begin{figure}[ht]
\centering
\includegraphics[]{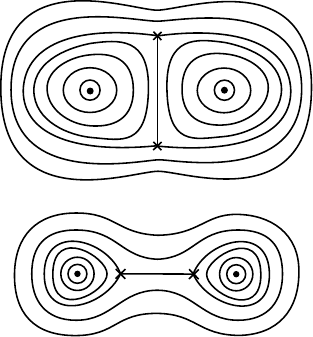}
\caption{The two different kinds of foliation $\hat{\mathcal{F}}_{\omega}$ obtained from \Cref{ex:Strebel}. The top one is called "molecule I" and is obtained when $m_\infty<m_0+m_1$. The bottom one is called "molecule II" and is obtained when $m_\infty>m_0+m_1$.}
\labelx{fig:Strebel_fol}
\end{figure}
\end{example}

\chapter{Higher-rank Teichm\"{u}ller spaces}\labelx{chap_higher Teich th}


\abstract{The algebraic approach to Teichm\"{u}ller spaces hints towards the more general consideration of representations of $\pi_1(S_g)$ to Lie groups other than $\mathrm{PSL}_2(\mathbb{R})$. This leads naturally to the consideration of character varieties and the notion of higher-rank Teichm\"{u}ller spaces, which are connected components of the character variety entirely consisting of discrete and faithful representations. Two main examples of these spaces are Hitchin components and maximal components studied in the following chapters.}

\section{Character Varieties}\labelx{sec:character_variety}

Character varieties lie at the intersection of Representation Theory, Geometry and Mathematical Physics. They deal with the representation theory of fundamental groups of manifolds, encoding in a algebraic way several geometric objects like Kleinian geometries and flat connections. We give an introduction to this rich theory, restricting attention to surfaces.

For a closed connected and oriented topological surface $S_g$ of genus $g \ge 2$, we are interested in studying the set of fundamental group representations, that is, group homomorphisms $\rho : \pi_{1}(S_g) \to G$ into a Lie group $G$. In fact, one can more generally take the set of representations $\mathrm{Hom}(\Gamma, G)$ for any finitely generated group $\Gamma$, but we will primarily focus on the case of fundamental group representations (in order to keep the link to geometry). As usual in representation theory, we are only interested in isomorphism classes of representations. Two representations are identified whenever one is the conjugate of the other by some group element. Hence we are interested in the quotient
$$\mathcal{E}(S_g,G):=\mathrm{Hom}(\pi_1(S_g),G)/G\; ,$$ where $G$ acts by conjugation.

Under suitable conditions discussed below, the space $\mathrm{Hom}(\pi_1(S_g), G)/G$ has a subspace with the structure of a variety, usually called the \emph{character variety}\index{character!variety}.
One of the most famous links to geometry arises when $G=\mathrm{PSL}_2(\mathbb{R})$, the group of the orientation-preserving isometries of the hyperbolic plane. We will see that in this case, one can single out a connected component of the character variety which can be identified with the Teichm\"{u}ller space of the surface $S_g$.

To do so, we need the presentation of the fundamental group by generators and relations as introduced in \Cref{fund_gp_presentation}. However, for a closed surface $S_g$ the presentation is simply
\begin{equation}\labelx{Eq:pi1-presentation}
{{\pi }_{1}}(S_g)=\left\langle {{a}_{1}},{{b}_{1}},\ldots ,{{a}_{g}},{{b}_{g}}  \left\vert~ \displaystyle\prod_{i=1}^g{[ {{a}_{i}},{{b}_{i}} ]=1} \right.\right\rangle\; ,
\end{equation}
where $\left[ {{a}_{i}},{{b}_{i}} \right]={{a}_{i}}{{b}_{i}}a_{i}^{-1}b_{i}^{-1}$ is the commutator. The set $\mathrm{Hom}\left( {{\pi }_{1}}(S_g),G \right)$ of all representations of ${{\pi }_{1}}(S_g)$ into a connected reductive real Lie group $G$ can be naturally identified with the subset of ${{G}^{2g}}$ consisting of $2g$-tuples $\left( {{a}_{1}},{{b}_{1}},\ldots ,{{a}_{g}},{{b}_{g}} \right)$ satisfying the algebraic equation $\prod_i{\left[ {{a}_{i}},{{b}_{i}} \right]}=1$.
The group $G$ acts on $\mathrm{Hom}\left( {{\pi }_{1}}(S_g),G \right)$ by conjugation
	\[\left( g\cdot \rho  \right)=g\rho \left( \gamma  \right){{g}^{-1}}\; ,\]
where $g\in G$, $\rho \in \mathrm{Hom}\left( {{\pi }_{1}}(S_g),G \right)$ and $\gamma \in {{\pi }_{1}}(S_g)$. The restriction of this action to the subspace $\mathrm{Hom}^{\mathrm{red}}\left( {{\pi }_{1}}(S_g),G \right)$ of reductive representations provides that the orbit space is Hausdorff. Here, by a reductive representation\index{representation!reductive} we mean one that composed with the adjoint representation in the Lie algebra of $G$ decomposes as a sum of irreducible representations. When $G$ is algebraic, this is equivalent to the Zariski closure of the image of ${{\pi }_{1}}(S_g)$ in $G$ being a reductive group. Define the \emph{moduli space of reductive representations of ${{\pi }_{1}}(S_g)$ into $G$}\index{moduli space!of reductive representations} to be the orbit space
\[\mathcal{R}(S_g, G) := {\mathrm{Hom}^{\mathrm{red}}\left( {{\pi }_{1}}(S_g),G \right)}/{G}\; .\]
It was shown in \cite{Goldman84} that this space has a stratification by real analytic varieties indexed by the stabilizers of representations, thus $\mathcal{R}\left(S_g, G \right)$ is usually called the \emph{character variety}\index{character!variety} or the \emph{representation variety}\index{representation!variety}.

\begin{theorem}
The moduli space $\mathcal{R}\left(S_g, G \right)$ has the structure of a real analytic variety, which is algebraic if $G$ is algebraic and is a complex variety if $G$ is complex.
\end{theorem}

\begin{remark}
Character varieties are considered and studied also for representations $\rho\colon \Gamma \to G$, for any finitely generated group $\Gamma$ and Lie group $G$. The $G$-character varieties are Hausdorff topological spaces. In general, they are not manifolds, since they can have singularities, however, they are always locally contractible. 
\end{remark}

\begin{remark}
The condition on the representations to be reductive can be seen as a \emph{stability condition} in the context of a GIT quotient. See \cite[Sect. 7]{Sikora} for more details.
\end{remark}

\begin{example}
The simplest example of a character variety is for the abelian group $G=\mathrm{U}(1)=\mathbb{S}^1$. In that case, the character variety can be identified with the Jacobian of the surface. Indeed, the conjugation action is trivial and all representations are reductive, and so $\mathcal{R}(S_g, \mathrm{U}(1)) = \mathcal{E}(S_g, \mathrm{U}(1)) = \mathrm{Hom}(\pi_1(S_g), \mathrm{U}(1))$.
\end{example}

\begin{example}
For $G=\mathrm{SU}(2)$, the character variety $\mathcal{R}(S_g,\mathrm{SU}(2))$ is a symplectic manifold with singularities. Its quantization of level $k$ can be used to define a topological quantum field theory (TQFT). The dimension of the associated vector space $V_k(S_g)$ is given by the so-called \emph{Verlinde formula}. In addition, $V_k(S_g)$ gives finite dimensional representations of the mapping class group $\mathrm{Mod}(S_g)$  (see \cite{ChMa, MaNa}).
\end{example}

The \emph{Riemann--Hilbert correspondence}\index{Riemann--Hilbert correspondence}, discussed in \Cref{sec:RH}, can be restated in the case of surfaces $S_g$ as follows:
\begin{theorem}
The space $\mathcal{E}(S_g, G)$ is isomorphic to the moduli space of flat principal $G$-bundles over $S_g$ modulo gauge transformations.
\end{theorem}
This gives a differential geometric description of the character variety.

\section{Connected components}\labelx{Sec:connectedComponents}

Character varieties have an intriguing topology, since even the simplest topological question about the number of connected components is a highly non-trivial one for general Lie groups.
\emph{Higher-rank Teichm\"uller theory}\index{higher-rank Teichm\"uller!theory} studies connected components where all representations are discrete and faithful, hinting towards a geometric interpretation of these representations.

The \emph{primary topological invariant}\index{primary topological invariant} associates to each connected component of $\mathcal{R}(S_{g},G)$ an element in $\pi_1(G)$ in the following way:
consider $\rho\colon \pi_1(S_g)\to G$ and choose a set of generators $(a_1,b_1,...,a_g,b_g)$ of $\pi_1(S_g)$. For each generator, choose a lift $\tilde{\rho}(a_i)$ and $\tilde{\rho}(b_i)$ in the universal cover $\tilde{G}$. Since $\prod_i [a_i,b_i]=1$ (see Equation \eqref{Eq:pi1-presentation}), we see that 
$$e(\rho):=\prod_{i=1}^g [\tilde{\rho}(a_i),\tilde{\rho}(b_i)] \in \tilde{G}$$
projects to $1\in G$ under the covering map $\pi\colon \tilde{G}\to G$. Hence we can identify $e(\rho)$ with an element of $\pi_1(G)=\pi^{-1}(\{1\})$. It is easy to check that $e(\rho)$ does not depend on the choice of generators or the choice of lifts. It is a discrete invariant associated to the representation $\rho$, and hence is constant on the connected component containing $\rho$.

Equivalently, the primary topological invariant $e(\rho)$ is the \emph{Euler class} of the associated flat bundle to $\rho$, given by the Riemann--Hilbert correspondence (see \Cref{Sec:bundlesandconnections}). Note that this Euler class measures, in fact, the obstruction to lift $\rho$ to a representation of $\pi_1(S_g)$ in the universal cover $\tilde{G}$. The subvarieties $\mathcal{R}_d(S_{g},G) \subset \mathcal{R}(S_{g},G)$ which are consisted of those representations with a fixed value $d$ of the invariant $e(\rho)$, form disjoint closed subspaces but not necessarily connected components. The description and exact count of all connected components is a problem that has received considerable attention in the recent past.

For connected compact semisimple Lie groups $G$, the subvarieties $\mathcal{R}_d(S_{g},G)$ are non-empty and connected, for every $d \in \pi_1(G)$. This can be seen upon fixing a complex structure on the surface $S_g$, which allows one to introduce holomorphic bundle methods over the resulting Riemann surface $X$ with underlying curve $S_g$. Indeed, Ramanathan \cite{Ramanathan} identified the spaces  $\mathcal{R}_d(S_{g},G)$ with the moduli spaces $\mathcal{M}_d(X,G)$ of polystable holomorphic principal $G^{\mathbb{C}}$-bundles over $X$ with topological class $d$, where $G^{\mathbb{C}}$ denotes the complexification of the group $G$, and showed that the latter space is, in fact, connected.

For complex semisimple Lie groups $G$, holomorphic methods can still provide the connectedness of the spaces $\mathcal{R}_d(S_{g},G)$, for every $d \in \pi_1(G)$, via their identification to moduli spaces $\mathcal{M}_d(X,G)$ of polystable \emph{$G$-Higgs bundles}, that is, pairs consisting of a holomorphic principal $G$-bundle and a holomorphic section of the adjoint bundle twisted with the canonical line bundle $T^*X$ of $X$. The identification of the spaces $\mathcal{R}_d(S_{g},G)$ with the spaces $\mathcal{M}_d(X,G)$ in this complex group setting is the product of combined work by Hitchin \cite{Hitchin87} and Donaldson \cite{Don87} for $G=\mathrm{SL}_2(\mathbb{C})$, and Corlette \cite{Corlette} and Simpson \cite{Simpson} for general $G$. The connectedness of the spaces $\mathcal{M}_d(X,G)$ was obtained using Morse-theoretic methods introduced by Hitchin in \cite{Hitchin87, Hit92}. Alternative techniques for showing that the spaces $\mathcal{M}_d(X,G)$ are connected, were also provided by Goldman \cite{Goldman88}, for $G=\mathrm{SL}_2(\mathbb{C})$ 
and $G=\mathrm{PSL}_2(\mathbb{C})$, and by Li \cite{Li}, for an arbitrary semisimple complex Lie group $G$.

We summarize this discussion into the following: 
\begin{theorem}
Let $G$ be a connected semisimple complex Lie group or a compact Lie group. Then the connected components of the character variety $\mathcal{R}(S,G)$ are exactly given by the primary topological invariants. 
\end{theorem}

Note that $\pi_1(G)\cong \pi_1(K)$, where $K$ denotes a maximal compact subgroup of $G$, and that $\pi_1(K)$ is finite as a consequence of Myers theorem. Therefore, we get a finite number of connected components. 

For general real Lie groups, we still have a finite number of connected components but the count is much harder to obtain. Several components can have the same primary topological invariant and not all values in $\pi_1(G)$ are reached, as we will see in the case of Hermitian Lie groups in \Cref{sec:Maximal_reps}. 

In the Teichm\"uller case, we have $$\pi_1(\mathrm{PSL}_2(\mathbb{R}))\cong \pi_1(\mathrm{SO}(2))\cong \mathbb{Z}\; ,$$
but only values with absolute value smaller or equal to $\lvert \chi(S_g)\rvert = 2g-2$ are reached (this follows from the Milnor--Wood inequality that will be discussed in \Cref{sec:Maximal_reps}). In that case, it turns out that each value is reached exactly once, so that there are $4g-3$ connected components. The Teichm\"uller space\index{Teichm\"uller!space} is the connected component with Euler number $2g-2$, the maximal value. The trivial representation lies in the connected component of invariant 0. A fundamental counting result using techniques from Morse theory was obtained by Hitchin in \cite{Hit92}:
\begin{theorem}
The $\mathrm{PSL}_n(\mathbb{R})$-character variety has 3 connected components, for $n$ odd, and 6 connected components, for $n>2$ even. 
\end{theorem}

For the case of real Lie groups, the exact count of connected components was achieved in the realm of \emph{non-abelian Hodge theory} using an abundance of techniques allowing for a complex analytic description of the $G$-character varieties in terms of moduli spaces of polystable $G$-Higgs bundles. 

For $G=\mathrm{GL_n(\mathbb{R})}$, a \emph{Higgs bundle}\index{Higgs bundle} over a Riemann surface $X$ is a holomorphic bundle $E$ together with a holomorphic 1-form $\Phi\in\mathrm{H}^0(X,\mathrm{End}(E)\otimes K)$, where $K$ denotes the canonical bundle of $X$. Non-abelian Hodge theory associates to a (polystable) Higgs bundle a flat connection (i.e. a point in the character variety) in the following way. For a given hermitian metric $h$ on $E$, we can consider the connection 
$$\Phi+\D_A+\Phi^{*_h}\; ,$$ 
where $\D_A$ is the Chern connection on $E$ determined by $h$, and $\Phi^{*_h}$ is the hermitian conjugate of $\Phi$. 

The completely non-trivial fact is that there is a choice for $h$ such that this connection is \emph{flat}. The equation expressing the vanishing curvature is called the \emph{Hitchin equation}\index{Hitchin!equation}: 
$$F(\D_A)+[\Phi\wedge \Phi^{*_h}]=0\; ,$$
where $F(\D_A)$ denotes the curvature of the Chern connection.

This beautiful theory lies outside the scope of the present book and we recommend \cite{Guichard_notes} and \cite{Thomas:gentle} for an introduction, alongside with the original works \cite{Corlette}, \cite{Don87}, \cite{Hitchin87}, \cite{Simpson}. For a most recent and detailed survey of results concerning the counting of these components in cases where the Lie group is a real form of a complex simple Lie group, we recommend \cite{Bradlow}.

\section{Higher-rank Teichm\"uller spaces}

Starting from the work of Labourie \cite{Labourie}, a fundamental idea emerged for describing connected components of character varieties using the dynamical properties of the representations in the component. In this perspective, the following generalization of Teichm\"uller space is defined:
\begin{definition}
    A \emph{higher-rank Teichm\"uller space}\index{higher-rank Teichm\"uller!space} is a union of connected components of a character variety $\mathcal{R}(S_g,G)$ consisting entirely of discrete and faithful representations.
\end{definition}

The term ``higher Teichm\"{u}ller space'' (which later turned into ``higher-rank Teichm\"uller space'') originates in the work of Fock and Goncharov \cite{FG}, who developed a more algebro-geometric approach to
Lusztig’s notion of total positivity from \cite{Lu}, in the context of general split real semisimple reductive Lie groups and defined positive fundamental representations into these groups. 

In \cite{GW, GW22}, Guichard and Wienhard introduced the concept of a \emph{$\Theta$-positive structure}\index{$\Theta$-positive!structure} on a Lie group $G$ and showed that this leads to a \emph{secondary topological invariant} of connected components of the $G$-character variety. Very recently, Beyrer--Guichard--Labourie--Pozzetti--Wienhard \cite{BGLPW} showed that the primary and secondary topological invariants give a complete invariant for higher-rank Teichm\"uller spaces, thus culminating several decades of research on the subject. There are four known main families of higher-rank  Teichm\"uller spaces identified using this central notion of $\Theta$-positivity:
\begin{itemize}
    \item \emph{Hitchin components}, in the case of split real groups $G$,
    \item \emph{Maximal components}, in the case of Hermitian Lie groups $G$ of tube type,
    \item Connected components of the $G$-character variety, for groups $G$  locally isomorphic to $\mathrm{SO}(p,q)$,
    \item Connected components of exceptional type, that is, for the case when the group $G$ is of type $F_4, E_6, E_7, E_8$ with restricted root system of type $F_4$.
\end{itemize}
In particular, that the Hitchin components are entirely consisted of discrete and faithful representations was shown by Labourie who introduced in \cite{Labourie} the notion of an \emph{Anosov representation}\index{representation!Anosov} and used techniques from dynamical systems. This was also established by Fock and Goncharov in \cite{FG} who constructed all positive representations for split real Lie groups and  showed that they are faithful, discrete and positive hyperbolic. The assertion for maximal components in the case when the group $G$ is a Hermitian Lie group of tube type was established by Burger, Iozzi, Labourie and Wienhard in \cite{BILW} and \cite{BIW}, while for the case of Lie groups locally isomorphic to $\mathrm{SO}(p,q)$, it was obtained by Beyrer and Pozzetti \cite{BP}. 

\vspace{2mm}

In the sequel, we discuss two main families of higher-rank Teichm\"uller spaces in detail: \emph{Hitchin components} in \Cref{sec:hitchin-comps} and \emph{maximal components} in \Cref{sec:max-comps}. The general notion of $\Theta$-positivity is then studied in \Cref{sec:positivity}.
\chapter{Hitchin representations}\labelx{sec:hitchin-comps}


\abstract{The composition of a Fuchsian representation with the Kostant principal embedding of $\mathrm{PSL}_2(\mathbb{R})$ into an adjoint split real semisimple Lie group $G$ provides a class of fundamental group representations of profound importance. All these representations lie in one connected component of the character variety, the so-called Hitchin component, and all representations in this component are discrete and faithful.
    Framed versions of character varieties allow for a cluster structure, i.e. coordinate systems with special birational transition maps. These are in particular subtraction-free giving the well-defined notion of points with positive coordinates. The positive part of the character variety is the Hitchin component.}

\section{Hitchin representations}\labelx{sec:Hitchin_reps}
The first examples of higher-rank Teichm\"{u}ller spaces have appeared long before the term was introduced by Fock and Goncharov in \cite{FG}. This family of examples concerns connected components of the character variety for an adjoint split real semisimple Lie group $G$ and were introduced by Hitchin in \cite{Hit92}. They emerge by using an appropriate embedding $\mathrm{SL}_2(\mathbb{R}) \hookrightarrow G$ that allows to extend many of the fundamental topological and dynamical properties of the classical Teichm\"{u}ller space to these higher-rank analogues. The components were originally defined as \emph{Teichm\"{u}ller components}, but nowadays they are more commonly referred to as \emph{Hitchin components} and the fundamental group representations they contain as \emph{Hitchin representations}\index{representation!Hitchin}\index{Hitchin!representation}. This chapter will make extensive use of the theory of real Lie groups and algebras. The essentials on real Lie groups are covered in \Cref{sec:real_Lie_groups}.

\subsection{Configurations of flags}
\labelx{sec:conf-flags}

We now focus on the split real Lie group $\mathrm{SL}_n(\mathbb{R})$ and describe the notion of a \emph{flag variety}\index{flag!variety}. This notion is central for the characterization of the Hitchin components for any split real Lie group; another example of this notion will appear in \Cref{sec:maslov-index}. 

Consider the full flag variety, which is defined as $$\mathsf{\mathcal{F}}({{\mathbb{R}}^{n}}):=\{F=({{F}_{1}},{{F}_{2}},...,{{F}_{n-1}}) \mid F_i\subset {{\mathbb{R}}^{n}},\mathrm{dim}({{F}_{i}})=i,{{F}_{i}}\subset {{F}_{i+1}}\}\; .$$ Two flags $F, F' \in \mathcal{F}$ are said to be \emph{transverse}\index{flag!transversality} whenever ${{F}_{i}}\cap {{F}'_{n-i}}=\{0\}$. A triple of flags $(F,G,H)$ is said to be \emph{transverse} if for all integers $1\leq i,j,k\leq n$ such that $i+j+k = 2n$, we have $F_i\cap G_j\cap H_k = \left\lbrace 0\right\rbrace $. In particular, this implies that for integers $1\leq i,j,k\leq n$ such that $i+j+k = 2n+1$, the space $F_i\cap G_j\cap H_k$ is a line in $\mathbb{R}^n$. 

For every integer~$d\geq 2$, let us denote by $\mathrm{Conf}^d(\mathsf{\mathcal{F}}({{\mathbb{R}}^{n}}))$ the moduli space of $d$-tuples of flags, i.e. the quotient of $(\mathsf{\mathcal{F}}({{\mathbb{R}}^{n}}))^{d}$ by the diagonal action of~$\mathrm{SL}_{n}(\mathbb{R})$. The natural action of the symmetric group~$\mathfrak{S}_d$ on~$(\mathsf{\mathcal{F}}({{\mathbb{R}}^{n}}))^{d}$ descends to $\mathrm{Conf}^d(\mathsf{\mathcal{F}}({{\mathbb{R}}^{n}}))$.

We will be particularly interested in the spaces $\mathrm{Conf}^2(\mathsf{\mathcal{F}}({{\mathbb{R}}^{n}})),\mathrm{Conf}^3(\mathsf{\mathcal{F}}({{\mathbb{R}}^{n}}))$ and certain of their subspaces. We will denote by
$\mathrm{Conf}^{d\ast}(\mathsf{\mathcal{F}}({{\mathbb{R}}^{n}}))$ the configuration space of $d$-tuples of pairwise transverse flags\index{flag!configuration space of}. The subspace $\mathrm{Conf}^{d\ast}(\mathsf{\mathcal{F}}({{\mathbb{R}}^{n}})) \subset \mathrm{Conf}^d(\mathsf{\mathcal{F}}({{\mathbb{R}}^{n}}))$ is invariant under all permutations. 

An easy linear algebra exercise shows that the group $\mathrm{SL}_n(\mathbb R)$ acts on $\mathrm{Conf}^{2\ast}(\mathsf{\mathcal{F}}({{\mathbb{R}}^{n}}))$ transitively and the stabilizer of any element in $\mathrm{Conf}^{2\ast}(\mathsf{\mathcal{F}}({{\mathbb{R}}^{n}}))$ is conjugate in $\mathrm{SL}_{n}(\mathbb R)$ to the subgroup of diagonal matrices.

\subsection{Positive flags}\labelx{sec:positive_flags}

Now for a fixed standard basis $({{e}_{1}},...,{{e}_{n}})$ of $\mathbb{R}^{n}$, let  $F\in \mathsf{\mathcal{F}}$ be the flag with  ${{F}_{i}}=\mathrm{Span}({{e}_{1}},...,{{e}_{i}})$ and let $E\in \mathsf{\mathcal{F}}$ be the flag with ${{E}_{i}}=\mathrm{Span}({{e}_{n}},...,{{e}_{n-i+1}})$. Then any flag $T\in \mathsf{\mathcal{F}}$ transverse to $F$ is the image of $E$ under a unique unipotent matrix $u_T$. A unipotent matrix is called \emph{totally positive}\index{totally positive!unipotent matrix} if and only if all its minors are positive, except those that have to be zero by the condition that the matrix is unipotent. A more general account on the properties of such matrices can be found in \Cref{sec:total_positivity}. The triple of flags $(E,T,F)$ is said to be \emph{positive}\index{totally positive!triple of flags} if and only if the matrix $u_T$ is a totally positive unipotent matrix. Note that any two transverse flags $(F_1, F_2)$ can be mapped to the pair of flags $(E,F)$ by an element in $\mathrm{SL}_n(\mathbb{R})$. So the above notion of positivity can be extended to the notion of positivity for any triple in $\mathcal{F}$ of pairwise transverse flags. The moduli space of positive triples of flags is denoted by $\mathrm{Conf}^{3+}(\mathcal F(\mathbb R))$. Since any positive triple of flags is automatically transverse, in particular its flags are pairwise transverse and we have $\mathrm{Conf}^{3+}(\mathcal F(\mathbb R))\subset \mathrm{Conf}^{3*}(\mathcal F(\mathbb R))$. A $k$-tuple of flags $(F_1,\dots,F_k)$ is called \emph{positive} if there exist $u_1,\dots,u_{k-2}$ totally positive unipotent matrices such that $(F_1,\dots,F_k)$ can be mapped to $(F, u_1F, u_1u_2F,\dots, u_1\dots u_{k-2}F,E)$.

Positivity of tuples of flags in the flag variety $\mathcal{F}_G$ can be defined for any split real Lie group $G$ in an analogous manner to the $\mathrm{SL}_n(\mathbb{R})$-case involving the Borel subgroup of $G$ by the work of Lusztig \cite{Lu}.

\subsection{Hitchin representations into adjoint split real Lie groups}
Let $S_g$ be a closed oriented surface. For an adjoint split real semisimple Lie group $G$, there exists a unique embedding $\pi\colon\mathrm{PSL}_2(\mathbb R)\to G$, which is the associated Lie group homomorphism to a principal 3-dimensional subalgebra of $\mathfrak{g}$, called \emph{Kostant's principal subalgebra}\index{Kostant principal subalgebra} $\mathfrak{sl}_2(\mathbb{R})\subset \mathfrak{g}$ and introduced in \cite[Lem. 5.2]{Kos} (refer to \Cref{principal_sl2} for the complete description).

For a fixed discrete embedding $\iota \colon{{\pi}_{1}}(S_g)\to \mathrm{PSL}_2(\mathbb R)$, Hitchin in~\cite{Hit92} showed that the subspace containing the \emph{principal Fuchsian representation} $\pi \circ \iota \colon\pi_{1}(S_g)\to G$ is a connected component and, in fact, topologically trivial of dimension $\left( 2g-2 \right)\mathrm{dim}_{\mathbb{R}} G$. In the special case when the group is $G=\mathrm{PSL}_2(\mathbb R)$, this component is the Teichm\"{u}ller space\index{Teichm\"uller!space}. Namely, we have: 

\begin{theorem}\cite[Thm. A]{Hit92}
Let $S_g$ be a compact closed connected topological surface of genus $g \geq 2$ and $G$ be an adjoint split real semisimple Lie group. Then, the character variety $\mathcal{R}(S_g, G)$ has a connected component homeomorphic to a Euclidean ball of dimension $(2g-2)\text{dim}_{\mathbb{R}}G$.
\end{theorem}

We thus define the \emph{Hitchin component} as follows:

\begin{definition} For an adjoint split real semisimple Lie group $G$, and for a compact closed connected topological surface $S_g$ of genus $g\ge2$, the \emph{Hitchin component}\index{Hitchin!component} $\mathcal{T}_{H}(S_g,G)$ is the connected component of the character variety $\mathcal{R}(S_g, G)$ containing the principal Fuchsian representation $\pi \circ \iota :{{\pi }_{1}}\left( S_g \right)\to G$.
\end{definition}

The fundamental group representations contained in the Hitchin component were shown to be discrete and faithful in~\cite{FG, Labourie}. A very useful characterization of the Hitchin components was also obtained in terms of the positive flags discussed above in the special case of $\mathrm{PSL}_n(\mathbb R)$:

\begin{theorem}\cite{FG, Guichard, Labourie}
Let $\rho: \pi_1(S_g)\to G$ be a representation and let $\mathcal{F}_G$ denote the space of flags for any split real Lie group $G$. Then $\rho \in \mathcal{T}_{H}(S_g, G)$ if and only if there exists a continuous $\rho$-equivariant map $\xi\colon\partial\pi_1(S_g) \to \mathcal{F}_G$ sending positive tuples in $\partial\pi_1(S_g)$ to positive tuples in $\mathcal{F}_G$.   
\end{theorem}

\begin{remark}
    If $\bar{S}_{g,h}$ is a compact surface of genus $g$  with $n$-many boundary components, a similar construction can be provided. We need to fix a discrete embedding $\iota \colon{{\pi}_{1}}(\bar{S}_{g,h})\to \mathrm{PSL}_2(\mathbb R)$ coming from a hyperbolic structure on $\bar{S}_{g,h}$ with geodesic boundary. However, then $\mathcal{R}(\bar{S}_{g,h}, G)$ is a connected space if $G$ is connected, and we cannot directly apply the construction above. In this case, we can consider the following subspace of $\mathcal{R}(\bar{S}_{g,h}, G)$: Let $\mathcal{R}^p(\bar{S}_{g,h},G)$ be the space of all representations $\rho$ such that for every $\gamma\in\pi_1(\bar{S}_{g,h})$ which is homotopic to a boundary component, $\rho(\gamma)$ fixes an element of $\mathcal F_G$. Such representations are called \emph{peripherally parabolic}. Then let $\mathcal{T}_{H}(\bar{S}_{g,h},G)$ be the connected component in $\mathcal{R}^p(\bar{S}_{g,h},G)$ containing $\pi \circ \iota :{{\pi }_{1}}\left(\bar{S}_{g,h}\right)\to G$. Then $\mathcal{T}_{H}(\bar{S}_{g,h},G)$ consists of representations that are discrete and faithful~\cite{FG_book}.
\end{remark}

\begin{remark}
    The map $\xi$ sends positive tuples in $\partial\pi_1(S_g)$ to positive tuples in $\mathcal{F}_G$ if and only if it sends positive quadruples in $\partial\pi_1(S_g)$ to positive quadruples in $\mathcal{F}_G$.
\end{remark}

\section{Fock--Goncharov coordinates}\labelx{Sec:FockGoncharov}

\subsection{Transverse and positive framed representations}

A general principle in the theory of moduli spaces is that it might be easier to describe the moduli space of the structures we are interested in if we add to this structure some additional data. This principle comes into form in this Section as we will enrich our representations with an additional piece of data called \emph{framing}. While the moduli space of framed representations might seem more complicated than the character variety, it is easier to describe the former using a special coordinate system introduced by Fock and Goncharov in \cite{FG}. From this point on, information about the original character variety can be obtained by forgetting the framing.

Let $S$ be a non-compact ciliated surface equipped with a complete hyperbolic structure of finite volume with geodesic boundary. Let $\tilde S\subseteq \mathbb H^2$ be the universal covering of $S$, and let $\tilde P\subseteq \partial^{red}_{\infty}\tilde S$ be the lift on the boundary of $\tilde{S}$ of all punctures and cilia of $S$. Let $G$ be a split real Lie group with a Borel subgroup $B$ and let $\mathcal F_G=G/B$ be the full flag variety. 

\begin{definition}
A \emph{(flag) framing}\index{framing} is a map $F\colon \tilde P\to \mathcal F_G$. Let $\rho\colon\pi_1(S)\to G$ be a homomorphism. A \emph{framing of $\rho$} is a $\pi_1(S)$-equivariant framing $F\colon \tilde P\to \mathcal F_G$, i.e. for every $\gamma\in\pi_1(S)$, $F(\gamma(\tilde p))=\rho(\gamma)F(\tilde p)$, for all $\tilde p\in \tilde P$. A \emph{framed homomorphism}\index{framed!homomorphism} is a pair $(\rho,F)$ where $F$ is a framing of $\rho$.

The space of all framed homomorphisms is denoted by $\mathrm{Hom}^{\mathrm{fr}}(\pi_1(S),G)$. The space
$$\mathcal E^{\mathrm{fr}}(S,G):= \mathrm{Hom}^{\mathrm{fr}}(\pi_1(S),G)/G$$
is called the \emph{moduli space of framed representations}\index{moduli space!of framed representations}. A \emph{framed representation}\index{framed!representation} is an element of $\mathcal E^{\mathrm{fr}}(S,G)$.   
The \emph{framed character variety}\index{character!variety!framed}\index{framed!character variety} is the Hausdorff quotient
$$\mathcal R^{\mathrm{fr}}(S,G):= \mathrm{Hom}^{\mathrm{red},fr}(\pi_1(S),G)/G $$
of the space of framed reductive representations.
\end{definition}

\begin{remark}
Notice that not every homomorphism $\rho\colon\pi_1(S)\to G$ admits a framing when $G$ is a real Lie group. Homomorphisms that admit a framing are called \emph{peripherally parabolic}\index{homomorphism!peripherally parabolic}.
\end{remark}

Let $\Delta$ be an ideal triangulation of $S$ and $\tilde{\Delta}$ denote the lift of $\Delta$ to the universal covering $\tilde S$.

\begin{definition}\labelx{df:transverse_rep_FG}
A framed homomorphism $(\rho,F)$ is called \emph{$\Delta$-transverse}\index{framing!transverse} if for every triangle of $\tilde{\Delta}$ with vertices $\tilde p_1,\tilde p_2,\tilde{p}_3 \in\tilde P$, the triple of flags $(F(\tilde p_1), F(\tilde p_2), F(\tilde p_3))$ is transverse.

The space of all $\Delta$-transverse framed homomorphisms is denoted by $\mathrm{Hom}^{\mathrm{fr}}_\Delta(\pi_1(S),G)$. The space
$$\mathcal E^{\mathrm{fr}}_\Delta(S,G) :=  \mathrm{Hom}^{\mathrm{fr}}_\Delta(\pi_1(S),G)/G$$
is called the \emph{moduli space of $\Delta$-transverse framed representations}\index{moduli space!of transverse framed representations}.
\end{definition}

\begin{definition}\labelx{Def:modspaceposrep} A framing $F$ is called \emph{positive}\index{framing!positive} if for every cyclically oriented tuple $(\tilde p_1,\dots,\tilde p_k)\in\tilde P^k$, the tuple of flags $(F(\tilde p_1),\dots,F(\tilde p_k))$ is positive (in the sense of \Cref{sec:positive_flags}).
A homomorphism $\rho\in\mathrm{Hom}(\pi_1(S),G)$ is called \emph{positive} if it admits a positive framing $F$. The pair $(\rho, F)$ is called a \emph{positive framed homomorphism}\index{framed!positive homomorphism}.

The space of all positive framed homomorphisms is denoted by $\mathrm{Hom}^{\mathrm{fr}}_+(\pi_1(S),G)$. The space
$$\mathcal R_+^{\mathrm{fr}}(S,G) :=  \mathrm{Hom}^{\mathrm{fr}}_{+}(\pi_1(S),G)/G$$
is called the \emph{moduli space of positive framed representations}\index{moduli space!of positive framed representations}.
\end{definition}

\begin{remark}
From this definition follows immediately that positive framed representations are transverse with respect to any ideal triangulation of $S$ (cf. \Cref{sec:positive_flags}).
\end{remark}

\begin{remark}\labelx{pos.Hausdorff}
Notice, that in general not all orbits of the $G$-actions on $\mathrm{Hom}(\pi_1(S),G)$ and on $\mathrm{Hom}^{\mathrm{fr}}(\pi_1(S),G)$ are closed. So the quotient spaces $\mathcal{E}(S,G)$ and $\mathcal{E}^{\mathrm{fr}}(S,G)$ are not Hausdorff. However the space of positive framed representations $\mathcal{R}_+^{\mathrm{fr}}(S,G)$ inside $\mathcal{E}^{\mathrm{fr}}(S,G)$ form a Hausdorff subspace.

Moreover, the natural forgetful map $\mathcal{E}^{\mathrm{fr}}(S,G)\to \mathcal{E}(S,G)$ maps the space of positive framed representations $\mathcal{R}_+^{\mathrm{fr}}(S,G)$ onto the Hitchin component $\mathcal{T}_H(S,G)$ discussed in \Cref{sec:Hitchin_reps} surjectively with finite fibers. This means that every Hitchin representation admits a framing; in fact, it admits at most finitely many framings. Note that this is false in general: There are representations that do not admit framings as well as representations that admit infinitely many framings (for more details we address to~\cite{FG_book}).
\end{remark}

In fact, in order to check if a framing is positive, it is enough to check this only for all positively oriented quadrilaterals of a triangulation and not for all possible cyclically oriented tuples in $\tilde P$:

\begin{proposition}[\cite{FG_book}]
Let $\Delta$ be a triangulation of $S$ and let $F$ be a framing such that for every positively oriented quadrilateral $(\tilde p_1,\tilde p_2,\tilde p_3,\tilde p_4)$ of $\tilde{\Delta}$,  the quadruple of flags $(F(\tilde p_1),F(\tilde p_2),F(\tilde p_3),F(\tilde p_4))$ is positive. Then the framing $F$ is positive.
\end{proposition}

\begin{remark}
For a more general discussion about positive framings, we refer to \cite{GRW}. 
\end{remark}

\begin{definition}
    The \emph{$G$-higher framed Teichm\"uller space} (or Teichm\"uller $\mathcal X$-space)\index{framed!higher-rank Teichm\"uller space}\index{higher-rank Teichm\"uller!space!framed} of $S$ is the moduli space of positive framed representations $\mathcal R^{\mathrm{fr}}_+(S,G)$ defined in \Cref{Def:modspaceposrep}.
\end{definition}
Naming such spaces higher-rank Teichm\"uller spaces is justified by the following proposition.
    
\begin{proposition}[Thms 1.9 and 1.10 in \cite{FG}]\labelx{Prop:teichareteich}
    Any positive representation ${\rho\in\mathcal{R}^{\mathrm{fr}}_+(G,S)}$ is discrete and faithful.
\end{proposition}

The image of $\mathcal{R}^{\mathrm{fr}}_+(S,G)$ under the natural forgetful map of \Cref{pos.Hausdorff} is a space of representations of $\pi_1(S)$ into $G$ consisting only of discrete and faithful representations, i.e. this is also a higher-rank Teichm\"uller space which describes representations of the fundamental group only. However, looking at $\mathcal{R}^{\mathrm{fr}}_+(S,G)$ instead of the latter space is justified by the fact that $\mathcal{R}^{\mathrm{fr}}_+(S,G)$ admits natural coordinates, which are both easily defined and interesting \emph{per se}.

\begin{remark}
    When $G=\mathrm{PSL}_2(\mathbb{R})$, the image of $\mathcal{R}^{\mathrm{fr}}_+(S,\mathrm{PSL}_2(\mathbb{R}))$ under the forgetful map is what is most naturally called Teichm\"uller space of $S$. When $S$ is non-closed however, there exist other (slightly different) versions of Teichm\"uller spaces for $S$, such as $\mathcal{R}^{\mathrm{fr}}_+(S,\mathrm{PSL}_2(\mathbb{R}))$. The latter is the \emph{framed Teichm\"{u}ller space}\index{framed!Teichm\"uller space}\index{Teichm\"uller!space!framed} $\mathcal T^x(S)$ introduced in \Cref{def:framed_Teichm} and is naturally parameterized by shear coordinates. Another instance of Teichm\"uller space when $G=\mathrm{SL}_2(\mathbb{R})$ is the so-called \emph{decorated Teichm\"uller space}\index{Teichm\"uller!space!decorated}, introduced by Penner \cite{P1, Penner} and carries natural coordinates which endow this space with the structure of an $\mathcal{A}$-type cluster variety (cf. \Cref{Sec:clustervarieties}). In this book we only consider the higher-rank generalizations of framed Teichm\"uller spaces, i.e. the spaces $\mathcal{R}^{\mathrm{fr}}_+(S,G)$, and refer to \cite{FG} for the definition of the higher-rank generalizations of decorated Teichm\"uller spaces.
\end{remark}

\subsection{Parameterization of the space of framed representations}\labelx{sec:FG_coord}

The cornerstone of the construction of the natural coordinates on $\mathcal{R}^{\mathrm{fr}}(S,G)$ (or rather an open dense subset of it) in which we are interested, is the following decomposition theorem. It shows that when $G$ has a trivial center, given a triangulation $\Delta$ of $S$ and on an open dense subset of $\mathcal{R}^{\mathrm{fr}}_+(S,G)$, one can describe a positive framed representation entirely as the data of its restriction to every face -- triangle $T$ -- of $\Delta$  as well as gluing data: an element of the Cartan subgroup of $G$ on each edge of $\Delta$ along which two faces are glued.

\begin{theorem}[Decomposition theorem (\cite{FG}, Sect. 6)]
    Let $\Delta$ be an ideal triangulation of $S$, and let $F(\Delta)$ (resp. $E(\Delta)$) be the set of faces (resp. edges) of $\Delta$. If the center of $G$ is trivial, the moduli space of framed representations $\mathcal{R}^{\mathrm{fr}}(S,G)$ is birationally equivalent to 
    \begin{equation}
        \prod_{T\in F(\Delta)}     \mathcal{R}^{\mathrm{fr}}(T,G) \times    \prod_{e\in E(\Delta)} H\; ,
    \end{equation}
    where $H$ is the Cartan subgroup corresponding to the Borel $B$. 
\end{theorem}

\begin{remark}
    The notion of rational maps and birational equivalences is of an algebraic nature. Informally, saying that two spaces are birationally equivalent means that they are isomorphic up to subvarieties of positive codimension. Here, the latter consist of non-generic framed representations, e.g. those for which the image of two distinct punctures in $\tilde P$ in $\mathcal F_G$ coincide. We refer to \cite{hartshorne2013algebraic} for more details on algebraic geometry and rational maps.
\end{remark}

The moduli space $\mathcal{R}^{\mathrm{fr}}(T,G) =\mathcal{E}^{\mathrm{fr}}(T,G)$, where $T$ is an ideal triangle, is identified with the space of configurations of triples of flags. We thus need to define coordinates on the space of triples of flags, and the coordinates associated to the edges of the triangulation. We will not describe the construction in its full generality, we refer the reader to \cite{FG}. We will detail the construction of the coordinates when $G=\mathrm{PGL}_n(\mathbb R)$ in \Cref{sec:FG_coord_PGLn}, and will see how to retrieve the framed representation from them in \Cref{sec:snakes}. We focus here on giving the general properties of the coordinates on $\mathcal{R}^{\mathrm{fr}}(S,G)$. These coordinates endow the space of framed representations with a structure of an $\mathcal{X}$-\emph{cluster variety}\index{cluster!variety} (see \Cref{sec:Clusters}):

\begin{theorem}[\cite{FG_book}]
    For any ciliated surface\index{surface!ciliated}  $S$ and for any centerless split real Lie group $G$, there exists a seed $\mathsf{S}$ (in the sense of \Cref{def:seed}) and a map
	\begin{equation}\labelx{Eq:paramXspace}
		  \mathcal{R}^{\mathrm{fr}}(S,G) \rightarrow \mathcal{X}_{\mathsf{S}}
	\end{equation}
	which is a birational isomorphism.
\end{theorem}

The seed $\mathsf{S}$ depends on $G$ and on the choice of an ideal triangulation $\Delta$ of $S$, however the seed obtained from the triangulation after a flip of $\Delta$ can be obtained from a sequence of mutations from $\mathsf{S}$, thus providing the same cluster variety. Therefore, we can denote this cluster variety by $\mathcal{X}_{\mathsf{S}} = \mathcal X_{S,G}$. Note that in most cases, after a sequence of mutations one does not obtain a seed corresponding to a triangulation of $S$. Since $\mathcal{R}^{\mathrm{fr}}(S,G)$ is a cluster $\mathcal{X}$-variety, it is endowed with a canonical Poisson structure. 

\vspace{0.3cm}

Remember from \Cref{sec:Clusters} that in a real cluster variety the locus of positive points is well defined. In the character variety, this locus is exactly the space of positive framed representations:

\begin{theorem}[\cite{FG_book}]
    The space $\mathcal{R}^{\mathrm{fr}}_+(S,G)$ is diffeomorphic to the set of real positive points of $\mathcal{X}_{S,G}$.
\end{theorem}

Subsequently, after taking logarithms of the coordinates one sees that the $G$-higher Teichm\"uller spaces $\mathcal{X}_{S,G}^+$ are homeomorphic to open balls $\mathbb{R}^N$, for some $N\in\mathbb{R}$.

\begin{remark}\labelx{rem:Amodulispace}
    For a simply-connected Lie group $G$, there is a similar construction of a moduli space $\mathcal{A}_{S,G}$ with a structure of cluster $\mathcal{A}$-variety \index{cluster!$\mathcal{A}$-variety}. It is called the moduli space of \emph{twisted decorated representations}\index{representation!twisted decorated}. 
    
    \noindent There is one important technical complication: as we have seen in the basic \Cref{Ex:n-vectors} of the moduli space of $n$ vectors, there is a \emph{twisted} shift map $\tau\colon(v_1,v_2,...,v_n)\mapsto (v_2,...,v_n,s_G v_1)$, where $s_G$ is a very special element in $G$ of order 2. For $G=\mathrm{SL}_m(\mathbb{C})$, it is $-\mathrm{id}$ for $m$ even, and $\mathrm{id}$ for $m$ odd. 
    
    \noindent In order to realize the twisted shift in a local system, we consider the punctured tangent bundle $T'S:= TS\setminus S$. Its fundamental group fits into the short exact sequence
    $$1\to \mathbb{Z}\to \pi_1(T'S)\to \pi_1(S)\to 1\; .$$
    A \emph{twisted} $G$-local system on $S$ is a $G$-local system on $T'S$ with monodromy $s_G$ around the generator in any tangent space. Finally, we decorate a twisted $G$-local system $L$ with a choice of a \emph{decorated flag}\index{flag!decorated}, i.e. a complete flag with a (complex) volume form on each subspace, on each boundary component of the ciliated surface $S$. A decorated flag can be seen as a tuple $(v_1,v_1\wedge v_2,v_1\wedge v_2\wedge v_3,...,\omega)$ where $\omega$ is a fixed volume form (which gives the reduction of structure group to $\mathrm{SL}_m(\mathbb{C})$). More generally, it is an element of the coset space $G/U$, where $U$ is a maximal unipotent subgroup.
    
    \noindent The cluster structure on the moduli space of decorated twisted $G$-local systems $\mathcal{A}_{S,G}$ for $G=\mathrm{SL}_m(\mathbb{C})$ is obtained in a similar manner as the $\mathcal{X}$-coordinates: we start from an ideal triangulation and consider its $m$-th subdivision. Inside a triangle with decorated flags $(f_1,f_1\wedge f_2,...), (g_1,g_1\wedge g_2, ...)$ and $(h_1,h_1\wedge h_2,...)$ at its outer vertices, we associate a coordinate to each inner vertex of barycentric coordinates $(a,b,c)$ (with $a+b+c=m$) by 
    $$\Delta_{(a,b,c)}:=\omega(f_1\wedge f_2\wedge...\wedge f_a\wedge g_1\wedge ...\wedge g_b\wedge h_1\wedge...\wedge h_c)\; .$$
    The collection of the $\Delta_{(a,b,c)}$ is a positive regular atlas on $\mathcal{A}_{S,\mathrm{SL}_m(\mathbb{C})}$. For more details, we refer the reader to \cite[Chap. 8]{FG}.
\end{remark}

\begin{remark}
    The moduli spaces $\mathcal X_{S,G}$ and $\mathcal A_{S,G}$ form the pair of cluster varieties that is associated to the same mutation class of seeds; such a pair is called a \emph{cluster ensemble}\index{cluster!ensemble}. In this context, the \emph{cluster dualities}\index{cluster!duality} discussed in \Cref{Sec:clustervarieties} assert that the integral tropical points $\mathcal A_{S,G}(\mathbb{Z}^t)$ parameterize a basis of a family of important functions on $\mathcal X_{S,G}$, and vice-versa.
\end{remark}

\subsection{Special coordinates when \texorpdfstring{$G=\mathrm{PGL}_n$}{G=PGL(n)}.}\labelx{sec:FG_coord_PGLn} 

In all this part, $\mathbb K$ will denote either $\mathbb R$ or $\mathbb C$.

In this part, we will detail the construction of the cluster coordinates on $\mathcal{R}(S,\mathrm{PGL}_n(\mathbb K))$. When $\mathbb K = \mathbb R$, this generalizes the construction of coordinates on $\mathcal T^x(S)$ presented in \Cref{Sec:Teichcoordinates}. This construction will be a prime example of a cluster $\mathcal{X}$-variety, as defined in \Cref{sec:Clusters}. Let $\Delta$ be a triangulation of a ciliated surface $S$ as defined in \Cref{Sec:ciliatedsurfaces}, such that $\Delta$ does not have any self-folded triangle nor any bigon (see \Cref{fig:monogon_bigon}). 

\begin{figure}[ht]
		\centering
			\includegraphics{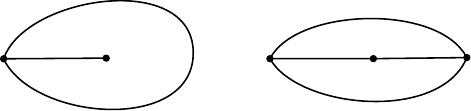}
			\caption{On the left a self-folded triangle, and on the right a bigon.}
            \labelx{fig:monogon_bigon}
\end{figure}

Let $\mathsf{T}$ be the Euclidean triangle in $\mathbb{R}^3$ that is the convex hull of the points of coordinates $A=(1,n,n)$, $B=(n,1,n)$ and $C=(n,n,1)$. The set of points of $\mathsf{T}$ of the form $(a,b,n)$ with $a+b = n+1$ (resp. $(n,b,c)$ with $b+c = n+1$, $(a,n,c)$ with $a+c = n+1$) is called the \emph{side} $AB$ (resp, $BC$, $AC$), or the side opposed to $C$ (resp. $B$, $A$). This euclidean triangle $\mathsf{T}$ can be decomposed into smaller triangles as in \Cref{fig:triangle_subdivision}.
	
	\begin{figure}[ht]
		\centering
			\includegraphics[scale=0.5]{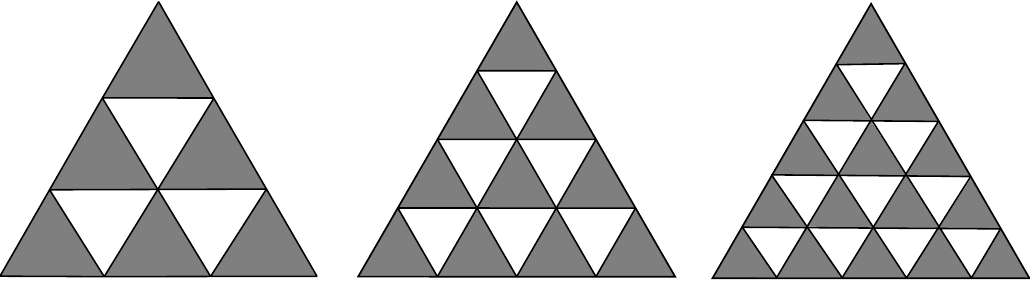}
			\caption{From left to right, a 4-subdivision, a 5-subdivision and a 6-subdivision of $\mathsf{T}$.}
			\labelx{fig:triangle_subdivision}
	\end{figure}
	
	The vertices of those smaller triangles are the points of coordinates $(a,b,c)$ with $a+b+c = 2n+1$ for integers $a,b,c\geq 1$, and each smaller triangle is one of two types: \emph{upward} triangles have vertices $(a-1,b,c),(a,b-1,c),(a,b,c-1)$ with $a+b+c=2n+2$, and \emph{downward} triangles have vertices $(a+1,b,c),(a,b+1,c),(a,b,c+1)$ with $a+b+c=2n$. We call this subdivision into smaller triangles an \emph{$n$-subdivision} of $\mathsf{T}$. On each side of $\mathsf{T}$ there are $n-1$ smaller upward triangles (grayed out on \Cref{fig:triangle_subdivision}). For each triangle $T$ of $\Delta$, we fix a diffeomorphism from $\mathsf{T}$ to $T$ such that for all edges $\gamma$ of $\Delta$, the $n$-subdivision of the two triangles adjacent to $\gamma$ agree on $\gamma$.

We now define a quiver $Q_{n,\Delta}$ embedded in $S$ as follows: the vertices of $Q_{n,\Delta}$ are all the vertices of the $(n+1)$-subdivision of each triangle of $\Delta$ except for the ones that are punctures of $S$ (i.e. the vertices of the triangles of $\Delta$ itself). For each downward triangle of the $(n+1)$-subdivision, there is an arrow of multiplicity 1 between each vertex such that the boundary of the small triangle is oriented clockwise, see \Cref{fig:An_quivers}. The vertices of $Q_{n,\Delta}$ that lie on an edge of $\Delta$ homotopic to a boundary component of $S$ are frozen.

\begin{figure}[!ht]
	\centering
	\includegraphics[width=\textwidth]{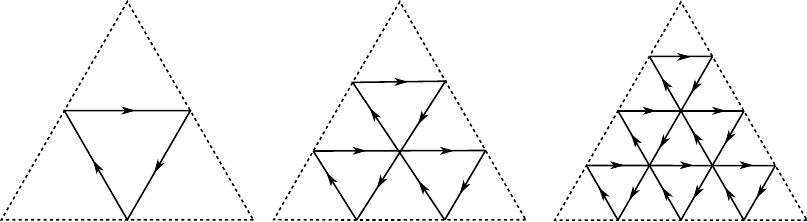}
	\caption{The quivers $Q_{n,\Delta}$ inside a triangle of $\Delta$ for $n =2,3,4$ from left to right.} \labelx{fig:An_quivers}
\end{figure}

We have seen in \Cref{sec:Clusters} that this quiver gives rise to a seed $\mathsf{S}_\Delta$, which defines an $\mathcal{X}$-torus $\mathcal{X}_\Delta$ endowed with coordinate functions. We will now show that a $\Delta$-transverse framed representation $(\rho, F)$ will correspond to a point of $\mathcal{X}_\Delta$. For this, we will construct the $\mathcal{X}$-coordinates corresponding to $(\rho, F)$.  We will denote by $\mathcal{X}_{S,\mathrm{PGL}_n(\mathbb{K})}$, the $\mathcal{X}$-variety\index{cluster!variety} associated to the initial seed\index{seed} $\mathsf{S}_\Delta$. A framed representation will correspond to a point in the $\mathcal{X}$-variety $\mathcal{X}_{S,G}$ only if it is $\Delta$-transverse for some triangulation $\Delta$ of $S$.

Let $(\rho, F)$ be a $\Delta$-transverse framed representation\index{representation!transverse framed}. We start with the coordinates associated to the vertices of $Q_{n,\Delta}$ lying on the interior of a triangle $T$ of $\Delta$ of vertices $p_1,p_2,p_3$. We start by defining a notion called \emph{triple ratio}\index{triple ratio} for triples of flags in $\mathbb K^3$. Let $(A,B,C)$ be a transverse triple of flags in $\mathbb{K}^3$, and let $\omega$ be a volume form on $\mathbb{K}^3$. Let $a_1\in A_1$, $a_2\in A_2/A_1$, $b_1\in B_1$, $b_2\in B_2/B_1$, $c_1\in C_1$ and $c_2\in C_2/C_1$ all nonzero. We define the \emph{triple ratio} of the triple $(A,B,C)$ as 
$$r(A,B,C) = \frac{\omega(a_1,a_2,b_1)\omega(b_1,b_2,c_1)\omega(c_1,c_2,a_1)}{\omega(a_1,a_2,c_1)\omega(b_1,b_2,a_1)\omega(c_1,c_2,b_1)}\; .$$
Note that the triple ratio does not depend on the choice of $a_1,a_2,b_1,b_2,c_1,c_2$ and does not depend on the choice of the volume form $\omega$.
	
Let $(A,B,C)$ be a transverse triple of flags in $\mathbb{K}^n$, and let $1\leq a,b,c\leq n$ integers such that $a+b+c = 2n+3$. The space $P_{a,b,c} = A_a\cap B_b\cap C_c$ is of dimension 3 and the flags $A,B$ and $C$ induce a transverse triple of flags on $P_{a,b,c}$. We denote the associated cross ratio\index{cross ratio} by $r_{a,b,c}(A,B,C)$. Since the non-frozen vertices of $Q_{n,\Delta}$ in a triangle $T$ of $\Delta$ are labeled by points of the $(n+1)$-subdivision of coordinates $(a,b,c)$ such that $a+b+c=2n+3$, we can associate to such a vertex the triple ratio $r_{a,b,c}(F(p_1),F(p_2),F(p_3))$.

Now let us consider the vertices of $Q_{n,T}$ that lie on an internal edge $\gamma$ of $\Delta$, and let $p_1$ and $p_2$ be the endpoints of $\gamma$. Let $p_3$ and $p_4$ be the remaining vertices associated to the two triangles $T_1$ and $T_2$ surrounding $\gamma$ as in \Cref{fig:Quadrilateral_triangulated}. A vertex $v$ of $Q_{n,\Delta}$ on $\gamma$ will have coordinates $(a,b,n+1)$ (resp. $(a,n+1,b)$) in the triangle $T_1$ (resp. $T_2$) with $a+b = n+2$. The space $P_{a,b} = F_a(p_1)\cap F_b(p_2)$ is a plane and the four flags $F(p_1), F(p_2),F(p_3),F(p_4)$ induce in $P_{a,b}$ a quadruple of lines. We denote by $r_{a,b}(F(p_1), F(p_2),F(p_3),F(p_4))$ the cross ratio associated to this quadruple, and associate it to the vertex $v$. Coordinates on the external edges of $\Delta$ are frozen and set equal to 1.

\begin{figure}[!ht]
	\centering
	\includegraphics[scale=1]{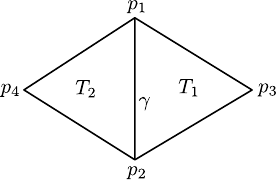}
	\caption{An edge $\gamma$ of the triangulation contained in two triangles $T_1$ and $T_2$ with surrounding punctures $p_1,p_2,p_3,p_4$.} 
 \labelx{fig:Quadrilateral_triangulated}
\end{figure}

One can show that the triple and cross ratios follow the $\mathcal{X}$-mutation formulas given in \Cref{sec:Clusters}, and we will show in \Cref{sec:snakes} that one can reconstruct a framed representation from its $\mathcal{X}$-coordinates, thus giving us the following theorem:

\begin{theorem}[\cite{FG_book}]
    The map
	\begin{equation}\labelx{Eq:paramXspaceGLn}
		  \mathcal{R}^{\mathrm{fr}}(S,\mathrm{PGL}_n(\mathbb K)) \rightarrow \mathcal{X}_{S,\mathrm{PGL}_n(\mathbb{K})}\; ,
	\end{equation}
	sending a framed representation to its coordinates constructed above, is a birational isomorphism.
\end{theorem}

Note, however, that in most cases, after a sequence of mutations one does not obtain a subdivision of a triangulation of $S$, and the coordinates cannot be expressed as triple ratios and cross ratios of the flags at the punctures and cilia of $S$. However, these charts do exist: they are defined by sequences of mutations starting at a subdivision of a triangulation of $S$.

\subsection{Snakes}\labelx{sec:snakes}

In all this section, $\mathbb K$ will denote either $\mathbb R$ or $\mathbb C$.

\emph{Snakes} are combinatorial tools introduced by Fock and Goncharov in \cite{FG_book} to describe \emph{projective bases} associated to a triple of flags, used to reconstruct a framed $\mathrm{PGL}_n(\mathbb{K})$-representation of the fundamental group of a surface $S$ from the data of the $\mathcal{X}$-coordinates. There is a notion of \emph{transformation} of snakes, which induces easily computable changes of the projective bases associated.
	
Let us first introduce snakes and their transformations. Let $n\in\mathbb{N}$. Recall the definition of an $n$-subdivision of a Euclidean triangle $\mathsf{T}$ given in \Cref{sec:FG_coord_PGLn}. We see this $n$-subdivision as a (non-oriented) graph $S_{n}$ on $\mathsf{T}$. 
	
	\begin{definition}
		A \emph{snake}\index{snake} on $S_{n}$ based at $A$ (resp. $B$, $C$) is an oriented path $A=v_1,\dots,v_{n}$ of length $n$ from $A$ (resp. $B$, $C$) to a vertex $v_{n}$ on the side $BC$ (resp. $AC$, $AB$) (\Cref{exam: snake on S5}).
	\end{definition}

    \begin{figure}[ht]
		\centering
			\includegraphics[scale=0.8]{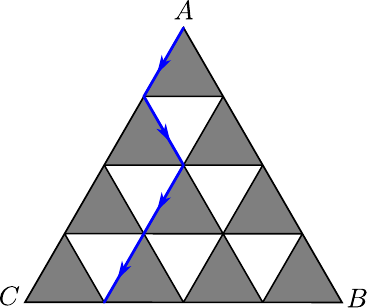}
			\caption{A snake on $S_5$ based at $A$.}\labelx{exam: snake on S5}
	\end{figure}
	
	\begin{remark}
		In particular, the sequence of the first coordinates of the vertices of a snake based at $A$ is increasing.
	\end{remark}

Let $(P,Q)\in\left\lbrace A,B,C\right\rbrace^2$, $P\neq Q$ and let $s_{PQ}$ denote the unique snake based at $P$ and ending in $Q$. These snakes stay within the side $PQ$.
	
	Let $(F^A,F^B,F^C)\in \mathrm{Conf}^{3}(\mathsf{\mathcal{F}}({{\mathbb{K}}^{n}}))$ be a transverse triple of flags in $\mathbb{K}^n$. Recall that $\mathrm{PGL}_n(\mathbb{K})$ acts on triples of flags of $\mathbb{K}^n$.
	We call \emph{projective basis}\index{projective basis} a linear basis of $\mathbb{K}^n$ up to multiplication of all vectors by the same scalar. Projective bases are in one-to-one correspondence with elements of $\mathrm{PGL}_n(\mathbb{K})$.
	
	To a vertex of $S_{n}$ of coordinates $(a,b,c)$ we associate the space $F_{a,b,c} = F^A_a\cap F^B_b\cap F^C_c$ which is a line since the triple of flags is transverse and $a+b+c=2n+1$. 
	
	Let $e$ be an oriented edge of $S_{n}$. There is exactly one upward triangle $t$ adjacent to $e$. Let $V_1$ (resp. $V_2$) be the line associated to the source (resp. the target) of $e$ and let $V_3$ be the line associated to the vertex of $t$ not on $e$. Then given a non-zero vector $v_1\in V_1$, there exists a unique non-zero vector $v_2\in V_2$ such that either $v_1+v_2\in V_3$ if $t$ is on the right of $e$, or such that $v_2-v_1\in V_3$ if $t$ is on the left of $e$, see \Cref{fig:snake_basis_construction}. Iterating this process for all edges of a snake, allows us to associate to a snake and a family of $n$ vectors depending only on the choice of the first vector, thus it is unique up to multiplication of all vectors by the same scalar. 

    \begin{figure}[ht]
		\centering
			\includegraphics[scale=1]{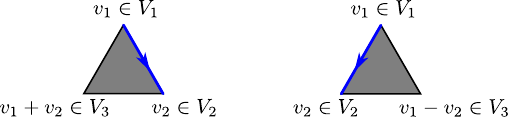}
			\caption{The vector $v_2$ is uniquely determined by $v_1$ and the triple of lines $(V_1,V_2,V_3)$.}
			\labelx{fig:snake_basis_construction}
	\end{figure}
	
	\begin{proposition}
		Let $(F^A,F^B,F^C)\in \mathrm{Conf}^{3}(\mathsf{\mathcal{F}}({{\mathbb{K}}^{n}}))$ be a transverse triple of flags. Let $s$ be a snake on $S_n$, and let $(v_1,\dots,v_n)$ be the family of vectors constructed above. Then $(v_1,\dots,v_n)$ is a projective basis.
	\end{proposition}
	
	\begin{proof}
		Without loss of generality, assume $s$ is based at $A$. Then $v_1$ is a basis of $F^A_1$, and if $(v_1,\dots,v_k)$ is a basis of $F^A_k$, since $v_{k+1}\in F^A_{k+1}\setminus F^A_k$ then $(v_1,\dots,v_{k+1})$ is a basis of $F^A_{k+1}$. We get the result by induction.
	\end{proof}

	\begin{example}
		Let $s$ be the snake on $S_4$ based at $A$ pictured in \Cref{fig:snake_basis_example}.
		Let $v_1\in A_1$ non-zero. The vector $v_2\in A_2\cap B_3$ is uniquely determined by the relation $v_2 - v_1\in A_2\cap C_3$, the vector  $v_3\in A_3\cap B_3\cap C_3$ is then determined by $v_2+v_3\in A_3\cap B_2$, and finally $v_4\in B_2\cap C_3$ is determined by $v_4-v_3\in B_3\cap C_2$.
	\end{example}

 \begin{figure}[ht]
		\centering
			\includegraphics[scale=0.8]{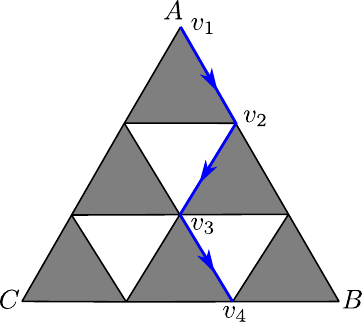}
			\caption{An example of projective basis associated to a snake.}
			\labelx{fig:snake_basis_example}
	\end{figure}

	We now describe the two types of elementary snakes \emph{moves}\index{snake!elementary move}, together with the induced transformations of their projective bases. The first type of move, which we call \emph{type I}\index{snake!elementary move!of type I} is pictured in \Cref{fig:snake_moves}.
	This move is only possible at the end of the snake as it changes the final vertex. The second type of move, called \emph{type II}\index{snake!elementary move!of type II}, is also pictured in \Cref{fig:snake_moves}.

 \begin{figure}[ht]
		\centering
			\includegraphics[scale=1]{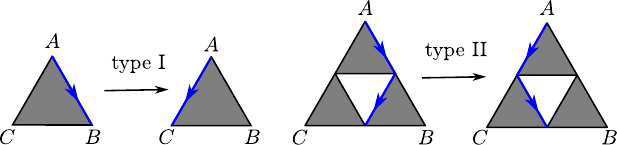}
			\caption{On the left, the move of type I. On the right, the move of type II.}
			\labelx{fig:snake_moves}
	\end{figure}
	
	It is straightforward to see that any two snakes based at the same vertex of $S_n$ are related by a finite sequence of moves of type I and II; see the proof of \Cref{th:snake_transformations}
	for an example of a sequence of moves from $s_{AB}$ to $s_{AC}$.
	
	We now describe the effect of the transformations on the associated projective bases. Consider the snake on $S_2$ pictured on the left of \Cref{fig:snake_moves}.
	The projective basis $(v_1,v_2)$ associated to $s_{AB}$ is determined by a choice of $v_1\in A_1$ and by the relation $v_1+v_2\in C_1$. The projective basis associated to $s_{AC}$ is then $(v_1,v_1+v_2)$, and the matrix associated to this move is $$M_I = \begin{pmatrix}
		1&1\\
		0&1
	\end{pmatrix}\; .$$

	Now consider the move of type II on the right of \Cref{fig:snake_moves}.
	Let $(v_1,v_2,v_3)$ be the projective basis associated to the first snake. Then the projective basis associated to the second snake is $(v_1,v'_2,v'_3)$ with $v'_2 =v_1+v_2$, and since $v_3$ and $v'_3$ belong to the same line, $v'_3 = \lambda v_3$ for some $\lambda\neq 0$. In fact, $\lambda$ is the inverse of the triple ratio of the triple of flags. Indeed, let us choose the following vectors: $v_1\in A_1$, $v_2\in A_2/A_1$, $b_1 = v_3-v_2\in B_1$, $v_3\in B_2/B_1$, $c_1 = v'_2+v'_3 \in C_1$ and $v'_2\in C_2/C_1$. Then we have
	\begin{align*}
		r(A,B,C) &= \frac{\omega(v_1,v_2,b_1)\omega(b_1,v_3,c_1)\omega(c_1,v'_2,v_1)}{\omega(v_1,v_2,c_1)\omega(b_1,v_3,v_1)\omega(c_1,v'_2,b_1)}\\
		& = \frac{\omega(v_1,v_2,-v_3)\omega(v_2,v_3,c_1)\omega(c_1,v_2,v_1)}{\omega(v_1,v_2,c_1)\omega(v_2,v_3,v_1)\omega(v'_3,v'_2,b_1)}\\
		& = \frac{\omega(v_2,v_3,c_1)\omega(c_1,v_2,v_1)}{\omega(v_1,v_2,c_1)\omega(v'_3,v'_2,b_1)}\\
		& = \frac{-\omega(v_2,v_3,c_1)}{\omega(v'_3,v'_2,b_1)}\\
		& = \frac{-\omega(v_2,v_3,c_1)}{\omega(v'_3,v'_2,v_2)}\\
		& = \frac{-\omega(v_2,v_3,c_1)}{\omega(v'_3,c_1,v_2)}\\
		& = \frac{1}{\lambda}\; .
	\end{align*}
	
	The matrix associated to this transformation is then
	\begin{align*}M_{II} &= \begin{pmatrix}
		1&1&0\\
		0&1&0\\
		0&0&r(A,B,C)^{-1}
	\end{pmatrix}\\ &= \begin{pmatrix}
	1&1&0\\
	0&1&0\\
	0&0&1
\end{pmatrix}\begin{pmatrix}
1&0&0\\
0&1&0\\
0&0&r(A,B,C)^{-1}
\end{pmatrix}\\& = \begin{pmatrix}
1&0&0\\
0&1&0\\
0&0&r(A,B,C)^{-1}
\end{pmatrix}\begin{pmatrix}
1&1&0\\
0&1&0\\
0&0&1
\end{pmatrix}\; .\end{align*}

	We now deal with the general case: for $i\leq n$, let $\varphi_i : \mathrm{SL}_2(\mathbb{K})\to \mathrm{GL}_n(\mathbb{K})$ associated to the $i$-th simple root. Let $$E_i := \varphi_i\begin{pmatrix}
		1&1\\
		0&1
	\end{pmatrix}$$
and $$H_i(x) := \mathrm{diag}(\underbrace{x,\dots,x}_{i \mathrm{-times}},1,\dots,1)\; .$$
Notice that $H_i(x)$ commutes with $E_j$ if $i\neq j$.	

\begin{theorem}\labelx{th:snake_transformations}
	Let $(A,B,C)$ be a transverse triple of flags in $\mathbb{K}^n$, and for $i,j,k\geq 1$ such that $i+j+k = 2n+3$, let $r_{i,j,k}$ be the triple ratio of the triple of flags induced by $(A,B,C)$ on $A_i\cap B_j\cap C_k$. Let $M_{AB\to AC}$ be the matrix associated to the transformation of $s_{AB}$ into $s_{AC}$, and let $M_{AB\to BA}$ be the matrix associated to the transformation of $s_{AB}$ into $s_{BA}$. Then
	$$ M_{AB\to AC} = \prod_{j=n-1}^{1}\left[\left[ \prod_{i=j}^{n-2} H_{i+1}(r_{n-i+1,n-j+1,i+j+1})E_i\right] E_{n-1} \right] $$
	and $M_{AB\to BA}$ is given by $M_{AB\to BA}(e_i) = (-1)^{i-1}e_{n+1-i}$.
\end{theorem}

\begin{proof}
	For the first point, the innermost product results from the sequence of moves described in \Cref{fig:snake_big_move1},
	and the outermost product corresponds to the sequence of moves of \Cref{fig:snake_big_move2}.

  \begin{figure}[ht]
		\centering
			\includegraphics[width=\textwidth]{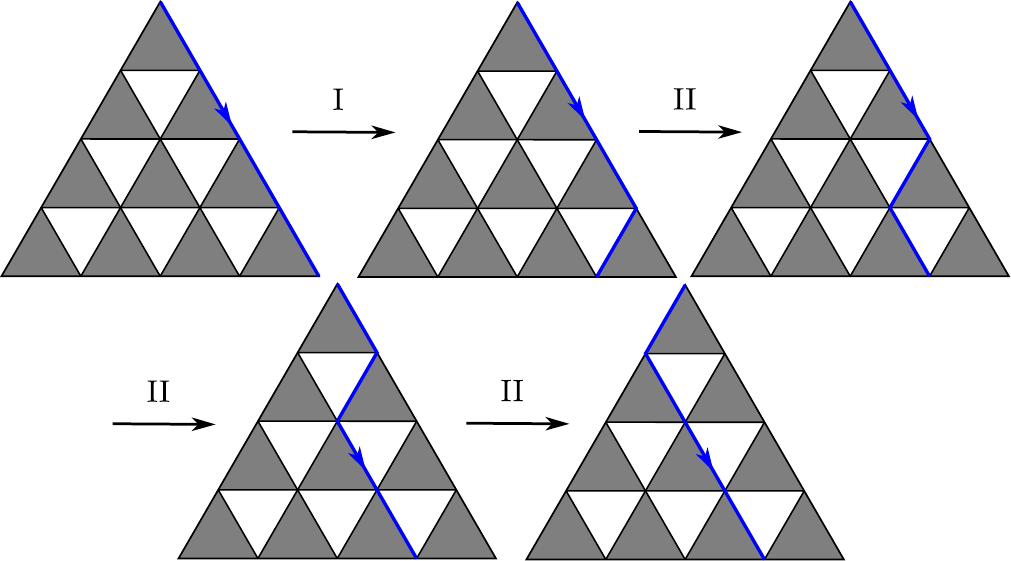}
			\caption{A sequence of moves inducing the transformation given by the innermost product of \Cref{th:snake_transformations}.}
			\labelx{fig:snake_big_move1}
	\end{figure}

  \begin{figure}[ht]
		\centering
			\includegraphics[width=\textwidth]{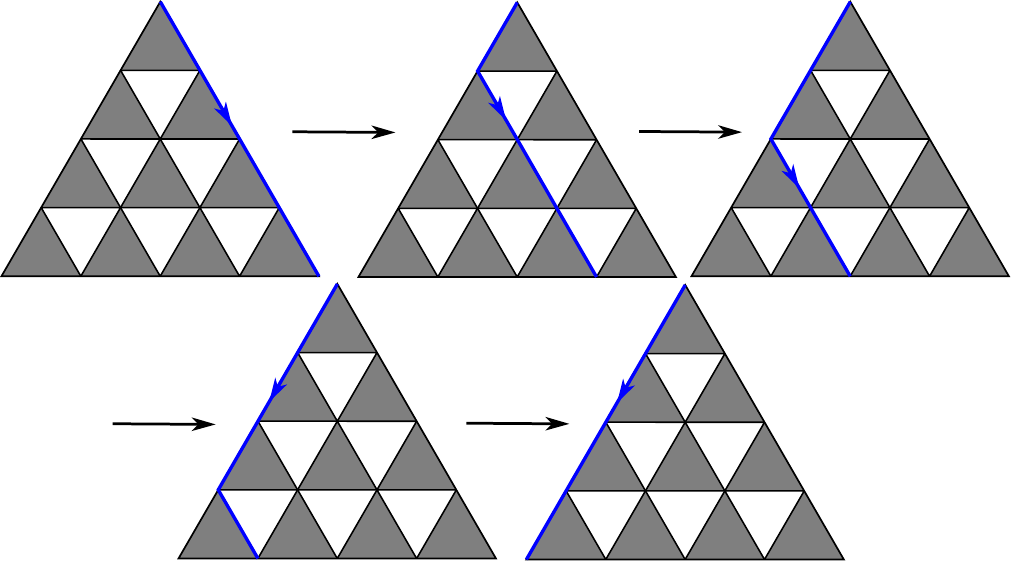}
			\caption{A sequence of moves inducing the transformation given by the outermost product of \Cref{th:snake_transformations}.}
			\labelx{fig:snake_big_move2}
	\end{figure}
	
	For the second point, let $(v_1,\dots,v_n)$ be the projective basis associated to $s_{AB}$ and let $(v'_1,\dots,v'_n)$ be the projective basis associated to $s_{BA}$. Up to scalar multiplication, we can assume $v'_1 = v_n$. Then we have $v'_2-v'_1 = v_{n-1}+v_n$ hence $v_{n-1} = -v'_2$. We get the result by induction.
\end{proof}

Let $(V_1,V_2,V_3,V_4)$ be four transverse lines in $\mathbb{K}^2$. Let $s$ be the snake from $V_1$ to $V_2$ on $S_2$ in the triangle $(V_1,V_2,V_3)$ and let $s'$ be the snake from $V_1$ to $V_2$ on $S_2$ in the triangle $(V_1,V_4,V_2)$. Let $(v_1,v_2)$ be the projective basis associated to $s$ and let $(v_1, v'_2)$ be the projective basis associated to $s'$, with $v'_2=\lambda v_2$. By definition, $v_1+v_2\in V_3$ and $v'_2-v_1\in V_4$. Let $\omega$ be a volume form on $\mathbb{K}^2$. The cross ratio of the quadruple of lines is \begin{align*}
    x & = \frac{\omega(v_1,v_1+v_2)\omega(v_2,v'_2-v_1)}{\omega(v_1,v'_2-v_1)\omega(v_1+v_2,v_2)}\\
    & = \frac{\omega(v_1,v_2)\omega(v_2,-v_1)}{\omega(v_1,v'_2)\omega(v_1,v_2)}\\
    & = \frac{1}{\lambda}\; .
\end{align*}

\begin{figure}[ht]
		\centering
			\includegraphics[scale=1]{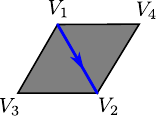}
			\caption{An edge of a snake belonging to two different triangles.}
			\labelx{fig:snake_triangle_change}
	\end{figure}

By repeating this argument for each edge of a snake belonging to two different triangles (\Cref{fig:snake_triangle_change}), we get the following result:

\begin{theorem}
    Let $(A,B,C)$ and $(A,D,B)$ be two transverse triples of flags. Let $s_{T^+(AB)}$ be the snake from $A$ to $B$ in the triangle $(A,B,C)$ and let $s_{T^-(AB)}$ be the snake from $A$ to $B$ in the triangle $(A,D,B)$. Let $(v_1,\dots,v_n)$ be the projective basis associated to $s_{T^+(AB)}$ and let $(v'_1,\dots,v'_n)$ be the projective basis associated to $s_{T^-(AB)}$. Let $x_1,\dots,x_{n-1}$ be the cross ratios of the quadruple of lines surrounding $s_{AB}$ in order. Then we have 
    $$M_{T^+(AB)\to T^-(AB)} = \mathrm{diag}(1,x_1^{-1},\dots,x_1^{-1}\dots x_{n-1}^{-1})\; .$$
\end{theorem}

We can now reconstruct a framed representation $\rho\colon \pi_1(S)\to\mathrm{PGL}_n(\mathbb{K})$ from the data of the $\mathcal{X}$-coordinates associated to a triangulation $\Delta$ of $S$. Notice that all the matrices associated to transformations of snakes depend only on the $\mathcal{X}$-coordinates associated to the chosen triangulation. We first construct a graph $\Gamma$ by gluing together elementary pieces on each triangle of $\Delta$, see \Cref{fig:local_system_reconstruction}. We associate to each oriented edge of $\Gamma$ an invertible matrix corresponding to a transformation of snakes as in \Cref{fig:local_system_reconstruction}. Now any loop $\gamma$ on $S$ can be retracted on the graph $\Gamma$ and its image by $\rho$ will be the product of the matrices associated to the edges of $\Gamma$ taken by $\gamma$. The matrix obtained does not depend on which retraction of $\gamma$ to $\Gamma$ is chosen because the holonomies around the hexagonal and quadrilateral faces of $\Gamma$ are trivial. The projective basis associated to any snake starting from a puncture induces a flag, hence the constructed representation admits a natural framing.

\begin{figure}[!ht]
    \centering
        \includegraphics{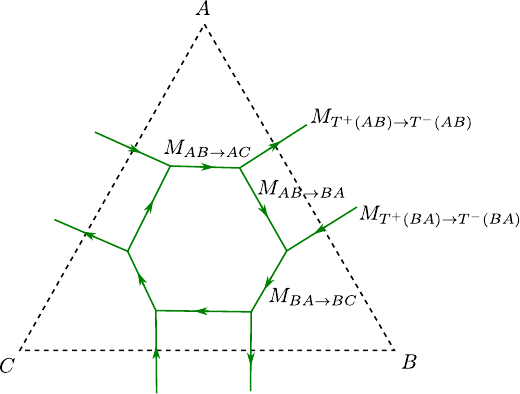}
	\caption{The local system on the graph $\Gamma$ used to reconstruct a framed local system on $S$.}
	\labelx{fig:local_system_reconstruction}
\end{figure}
\chapter{Maximal representations}\labelx{sec:max-comps}



\abstract{Connected components of the $G$-character variety are distinguished by a primary topological invariant taking values in $\pi_1(G)$. For Hermitian Lie groups, this is the Toledo invariant, which is bounded in absolute value by a Milnor–Wood type inequality. In the case of Hermitian groups of tube type, the components where the Toledo invariant attains its maximal value are known as maximal components. These form another class of higher-rank Teichmüller spaces. In this chapter we discuss coordinate systems that allow their explicit study.}

\section{Maximal representations}\labelx{sec:Maximal_reps}

Hermitian Lie groups of tube type form a very important class of semisimple Lie groups for higher-rank Teichm\"uller theory. Similarly to split real Lie groups, Hermitian Lie groups of tube type admit a certain notion of positivity that we will discuss in this Section as well as in \Cref{sec:HLG_positivity}. Moreover, roughly speaking, Hermitian Lie groups of tube type can bee seen as ``big'' $\mathrm{SL}_2(\mathbb R)$ --- or more precisely  ``big'' $\mathrm{Sp}_2(\mathbb R)$ --- Lie groups. This means, they can be treated as symplectic Lie groups over noncommutative algebras with some positive structure. For more details about this point of view on Hermitian Lie groups of tube type, we address the reader to~\cite{ABRRW}.

Spaces of maximal representations of a surface group into a Hermitian Lie group of tube type provide another example of higher-rank Teichm\"uller spaces. In this Section, we define Hermitian Lie groups and maximal representations and discuss some of their properties. 

\subsection{Hermitian symmetric spaces and Hermitian Lie groups}\labelx{sec:herm_lie}

Let $M$ be a connected complex manifold and $g$ be a Hermitian metric on $M$.
\begin{definition}
$(M,g)$ is called a \emph{Hermitian symmetric space}\index{Hermitian symmetric space} if every $p\in M$ is an isolated fixed point of a holomorphic involutive isometry $s_p\colon M\to M$.
\end{definition}

\begin{theorem}[{Harish--Chandra embedding theorem~\cite[Thm. 7.1]{Helgason}}]
Every non-compact Hermitian symmetric space is biholomorphic to an open bounded domain in a complex vector space.
\end{theorem}

\begin{definition}
A Hermitian symmetric space is called \emph{of tube type}\index{Hermitian symmetric space!of tube type} if it is biholomorphic to a domain of the form $V + \I\Omega$, where $V$ is a real vector space
and $\Omega\subset V$ is an open proper convex cone.
\end{definition}

Let $G$ be a connected non-compact semisimple Lie group with finite (or more generally with compact) center, and $K$ be a maximal compact subgroup of $G$.
\begin{definition}
The group $G$ is called \emph{Hermitian Lie group} if the symmetric space $G/K$ is Hermitian. $G$ is called \emph{Hermitian Lie group of tube type} if the symmetric space $G/K$ is Hermitian of tube type.
\end{definition}

\begin{fact}
Hermitian Lie groups are classified. Up to coverings, these are:
$$\mathrm{Sp}_{2n}(\mathbb R),\;\;\mathrm{SU}(n,m),\;\; \mathrm{SO}^*(2n),\;\; \mathrm{SO}_0(2,n+2),\;\; E_{7(-25)},\;\;E_{6(-14)}$$
for $n,m\in \mathbb N$. The first four groups are called \emph{classical} whereas the last two are called \emph{exceptional}. The groups $$\mathrm{Sp}_{2n}(\mathbb R),\;\;\mathrm{SU}(n,n),\;\; \mathrm{SO}^*(4n),\;\; \mathrm{SO}_0(2,n+2),\;\; E_{7(-25)}$$
are Hermitian of tube type.
\end{fact}

Our main example in this Section will be the real symplectic group $\mathrm{Sp}_{2n}(\mathbb R)$. Its maximal compact subgroup $K=\left\{\begin{pmatrix}u_1 & u_2 \\ -u_2 & u_1 \end{pmatrix} \mid u_1+\I u_2\in \mathrm U(n) \right\}$ is isomorphic to $\mathrm{U}(n)$. The symmetric space $\mathrm{Sp}_{2n}(\mathbb R)/K$ can be identified with the Siegel upper-half space\index{Siegel upper-half space} 
$$\mathcal S:=\{a+\I b\in \mathrm{Sym}_n(\mathbb C)\mid a\in \mathrm{Sym}_n(\mathbb R),\;b\in\mathrm{Sym}^+_n(\mathbb R) \}\; .$$
The group $G$ acts on $\mathcal S$ by generalized M\"obius transformations:
$$\begin{pmatrix}a & b\\ c & d\end{pmatrix}.z=(az+b)(cz+d)^{-1},\;\text{for $z\in\mathcal S$ and $\begin{pmatrix}a & b\\ c & d\end{pmatrix}\in \mathrm{Sp}_{2n}(\mathbb R)$\; .}$$

\subsection{The Lagrangian Grassmannian}\labelx{sec:lagr}

We consider the symplectic vector space $(\mathbb{R}^{2n},\omega)$ where~$\omega$ is
the standard symplectic form on~$\mathbb{R}^{2n}$, i.e.\
\begin{equation} \labelx{form:standard symplectic form}
  \omega(x,y)= x^t \begin{pmatrix}
    0 & \mathrm{Id} \\ -\mathrm{Id} & 0
  \end{pmatrix} y\; ,
\end{equation}
for~$x$ and~$y$ in~$\mathbb{R}^{2n}$.

Every basis of $\mathbb{R}^{2n}$ such that $\omega$, expressed in that basis, has the
form~\eqref{form:standard symplectic form} is called a \emph{symplectic basis}
(hence the standard basis is a symplectic basis). We will usually write a
symplectic basis as a pair $(\mathbf{e},\mathbf{f})$, where
$\mathbf{e}=(e_1,\dots,e_n), \mathbf{f}=(f_1,\dots,f_n)$; thus, one has, for
all~$i, j$, that $\omega(e_i,f_j)=\delta_{ij}$. More generally, given
two families $\mathbf{v}=(v_1, \dots, v_n)$ and $\mathbf{w}=(w_1,\dots, w_m)$
we will write $\omega( \mathbf{v}, \mathbf{w})$ for the $m\times n$-matrix
whose coefficients are $\omega(v_i,w_j)$; one has $\omega(\mathbf{w},
\mathbf{v}) = -\omega(\mathbf{v}, \mathbf{w})^t$.

The real symplectic group\index{group!real symplectic} can be seen as the group of isometries of the form $\omega$:
\[ \mathrm{Sp}_{2n}(\mathbb R)= \{g\in\mathrm{GL}_{2n}(\mathbb{R}) \mid g^t \omega g=\omega\}\; .\] 
Its adjoint form $\mathrm{PSp}_{2n}(\mathbb R) := \mathrm{Sp}_{2n}(\mathbb R)/\{\pm \mathrm{Id}\}$ is called the \emph{projective real symplectic group}\index{group!real symplectic!projective}.

\begin{definition}
  A subspace $L$ of $\mathbb{R}^{2n}$ is called \emph{Lagrangian}\index{Lagrangian!subspace} if $\dim(L)=n$ and
  $\omega(u,v)=0$ for all $u,v\in L$. The space of all Lagrangian subspaces of
  $(\mathbb{R}^{2n},\omega)$ is called the \emph{Lagrangian Grassmannian}\index{Lagrangian!Grassmannian}, and is denoted by
  $\mathrm{Lag}_{n}$.   Two Lagrangians $L_1, L_2 $ are called \emph{transverse} if their   intersection is trivial.
\end{definition}

\begin{remark}
  \labelx{rem:action-on-basis}
  \begin{enumerate}
  \item Recall that for any $n$-dimensional real vector space~$L$, the
    group~$\mathrm{GL}_n(\mathbb{R})$ acts on the right on the space of bases of $L$: if
    $\mathbf{v}=(v_1,\dots, v_n)$ is a basis of~$L$ and
    $g=(g_{i,j})_{i,j\in\{1,\dots, n\}}$ is in~$\mathrm{GL}_n(\mathbb{R})$ then
    $\mathbf{w}:= \mathbf{v}g$ is the family $(w_1,\dots, w_n)$ defined for all $j\in\{ 1,\dots, n\}$ by $w_j= \sum_{i=1}^{n} v_i
    g_{i,j}$. This is the simply transitive action behind the principal bundle structure mentioned above.

  \item In the case of a symplectic vector space of dimension~$2n$, we get in this way a simply transitive action of $\mathrm{Sp}_{2n}(\mathbb R)$ on the space of
    symplectic bases.

  \item We will use also the notation $\mathbf{v}\cdot g$ (or sometimes $\mathbf{v}g$) when $\mathbf{v}$ is a family (not necessarily free, nor
    generating) of~$n$ elements in a vector space~$V$ and when $g$ is an $m\times n$-matrix.

  \item\labelx{item:4:rem:action-on-basis} A morphism $\psi\colon L\to L'$ and
    its matrix~$g$ with respect to bases~$\mathbf{v}$ of~$L$ and~$\mathbf{v}'$
    of~$L'$ are related by the well-known formula
    $\psi(\mathbf{v}) =\mathbf{v}'g$.

  \item For two families $\mathbf{v}=(v_1,\dots,v_n)$ and $\mathbf{w}=(w_1,
    \dots, w_n)$, the family $(v_1+w_1, \dots, v_n+w_n)$ is denoted by
    $\mathbf{v}+ \mathbf{w}$.
  \end{enumerate}
\end{remark}

The group $\mathrm{Sp}_{2n}(\mathbb R)$ has a natural left action on~$\mathrm{Lag}_{n}$ given by
\begin{align*}
  g\cdot L&:= \{g(x) \}_{  x\in L}\; .
\end{align*}

This action is transitive, hence the space $\mathrm{Lag}_{n}$ is a homogeneous space under the symplectic group.

Let $(\mathbf{e}_0, \mathbf{f}_0)$ be the standard symplectic basis
of~$\mathbb{R}^{2n}$. The Lagrangians $L_0=\mathrm{Span}(\mathbf{e}_0)$ and $L_0^{\mathrm{opp}}=\mathrm{Span}(\mathbf{f}_0)$ are called \emph{standard}\index{Lagrangian!subspace!standard}. Notice that they are transverse. We consider the stabilizers
$$P = \mathrm{Stab}_{\mathrm{Sp}_{2n}(\mathbb R)}(L_0)=\left\{ \begin{pmatrix}A & AB \\ 0 & (A^t)^{-1}\end{pmatrix} \mid A\in\mathrm{GL}_n(\mathbb R), B\in\mathrm{Sym}_n(\mathbb R) \right\}\; ,$$
$$U = \mathrm{Stab}_{\mathrm{Sp}_{2n}(\mathbb R)}(\mathbf{e}_0)=\left\{ \begin{pmatrix}I_n & B \\ 0 & I_n\end{pmatrix} \mid B\in\mathrm{Sym}_n(\mathbb R) \right\}\; ,$$
$$P^{\mathrm{opp}} = \mathrm{Stab}_{\mathrm{Sp}_{2n}(\mathbb R)}(L_0^{\mathrm{opp}})=\left\{ \begin{pmatrix}A & 0 \\ BA & (A^t)^{-1}\end{pmatrix} \mid A\in\mathrm{GL}_n(\mathbb R), B\in\mathrm{Sym}_n(\mathbb R) \right\}\; ,$$
$$U^{\mathrm{opp}} = \mathrm{Stab}_{\mathrm{Sp}_{2n}(\mathbb R)}(\mathbf{f}_0)=\left\{ 
\begin{pmatrix}I_n & 0 \\ B & I_n\end{pmatrix} \mid B\in\mathrm{Sym}_n(\mathbb R) \right\}\; ,$$
where $I_n$ is the identity $n\times n$-matrix.

The groups $P$ and $P^{\mathrm{opp}}$ are maximal parabolic subgroups\index{subgroup!maximal parabolic} of $\mathrm{Sp}_{2n}(\mathbb R)$ opposite to each other, and $U \subset P$ and $U^{\mathrm{opp}}\subset P^{\mathrm{opp}}$ are their unipotent radicals. Further, we have the Lie algebras of the unipotent radicals:
$$\mathfrak u=\mathrm{Lie}(U)=\left\{ \begin{pmatrix}0 & B \\ 0 & 0\end{pmatrix} \mid B\in\mathrm{Sym}_n(\mathbb R) \right\}\; ,$$
$$\mathfrak u^{\mathrm{opp}}=\mathrm{Lie}(U^{\mathrm{opp}})=\left\{ \begin{pmatrix}0 & 0 \\ B & 0\end{pmatrix} \mid B\in\mathrm{Sym}_n(\mathbb R) \right\}\; ,$$
and the intersection of $P$ and $P^{\mathrm{opp}}$ is called the \emph{Levi subgroup}\index{subgroup!Levi} of $\mathrm{Sp}_{2n}(\mathbb R)$:
$$L:=P\cap P^{\mathrm{opp}}=\left\{\begin{pmatrix} A & 0 \\ 0 & (A^t)^{-1}\end{pmatrix}\mid A\in \mathrm{GL}_n(\mathbb R)\right\}\; .$$
The Levi subgroup acts on $\mathfrak u$ by the adjoint action (which is in our case just the conjugation) preserving the following sharp open convex cone
$$\mathfrak u^+=\left\{ \begin{pmatrix}0 & B \\ 0 & 0\end{pmatrix} \mid B\in\mathrm{Sym}^+_n(\mathbb R) \right\}\; .$$

Since the action of $\mathrm{Sp}_{2n}(\mathbb R)$ on $\mathrm{Lag}_{n}$ is transitive, the Lagrangian Grassmannian is a smooth manifold diffeomorphic to the homogeneous space $\mathrm{Sp}_{2n}(\mathbb R)/P$. The action of $\mathrm{Sp}_{2n}(\mathbb R)$ on $\mathrm{Lag}_{n}$ is not effective, its kernel is $\{\pm \mathrm{Id}\}$. The group of symmetries of $\mathrm{Lag}_{n}$ is the projective symplectic group $\mathrm{PSp}_{2n}(\mathbb R)$.

\subsection{Configurations of Lagrangians}
\labelx{sec:conf-lagr}

For every integer~$d\geq 2$, let us denote by $\mathrm{Conf}^d(\mathrm{Lag}_{n})$ the moduli space of $d$-tuples of Lagrangians, i.e.\ the quotient of $(\mathrm{Lag}_{n})^{d}$ by the diagonal action of~$\mathrm{Sp}_{2n}(\mathbb{R})$. The natural action of the symmetric group~$\mathfrak{S}_d$ on~$(\mathrm{Lag}_{n})^{d}$
descends to $\mathrm{Conf}^d(\mathrm{Lag}_{n})$.

We will be particularly interested in $\mathrm{Conf}^2(\mathrm{Lag}_{n}),\mathrm{Conf}^3(\mathrm{Lag}_{n})$  and certain of their subspaces. We will denote by
$\mathrm{Conf}^{d\ast}( \mathrm{Lag}_{n})$ the configuration space of $d$-tuples of pairwise
transverse Lagrangians. The subspace
$\mathrm{Conf}^{d\ast}( \mathrm{Lag}_{n}) \subset \mathrm{Conf}^d( \mathrm{Lag}_{n})$ is invariant under all permutations. An easy linear algebra exercise shows that the real symplectic group acts on $\mathrm{Conf}^{2\ast}( \mathrm{Lag}_{n})$ transitively and the stabilizer of any element of  $\mathrm{Conf}^{2\ast}( \mathrm{Lag}_{n})$ is conjugated in $\mathrm{Sp}_{2n}(\mathbb R)$ to the Levi subgroup.

\subsection{Maslov index}
\labelx{sec:maslov-index}

In this Section we review properties of the Maslov index.

Let $L_1, M, L_2$ be three pairwise transverse Lagrangians. There is a
unique linear map~$M_{L_1 \to L_2}$ from~$L_1$ to~$L_2$ such that
\[M = \{ v\in \mathbb{R}^{2n} \mid \text{ there exists  } e\in L_1,\, \text{ such that }v=e+M_{L_1\to L_2}(e)\}\; .\] When this does not cause confusion, it will be denoted just by $M$. Using the symplectic form $\omega$, we can define a bilinear form $\beta$ on
$L_1$ in the following way: for $v_1,v_2\in L_1$,
\[\beta(v_1,v_2):= \omega(v_1,M(v_2))\; .\]

\begin{definition}\labelx{def:maslov-form}
  The bilinear form $\beta$ is called the \emph{Maslov form}\index{Maslov!form} and is denoted
  by~$[L_1,M,L_2]$.
\end{definition}
The following Proposition is an easy linear algebra exercise:

\begin{proposition}
  The Maslov form $[L_1,M,L_2]$ is symmetric and non-degenerate.
\end{proposition}

\begin{remark}
  \labelx{rem:maslov-index-matrix}
  Let $\mathbf{e}$ be a basis of $L_1$ and let~$\mathbf{f}$ be the basis of
  $L_2$ such that $(\mathbf{e}, \mathbf{f})$ is a symplectic basis. Then the
  matrix $[M]_{\mathbf{e}, \mathbf{f}}$ of $M$ in these bases and the matrix
  $[\beta]_{\mathbf{e}}$ are equal:
  $[M]_{\mathbf{e}, \mathbf{f}} = [\beta]_{\mathbf{e}}$.
\end{remark}

We will denote the signature of $\beta$ by
\[\mathrm{sgn}(\beta):=p-q\; ,\]
where $p$ is the dimension of a maximal subspace of $L_1$ on which $\beta$
is positive definite and $q$ is the dimension of a maximal subspace of $L_1$
on which $\beta$ is negative definite. They satisfy $p+q=n$ so that
$\mathrm{sgn}(\beta) = n \mod 2$.

\begin{definition}
The \emph{Maslov index}\index{Maslov!index} of the triple of Lagrangians $(L_1,M,L_2)$ is the signature $\mathrm{sgn}([L_1,M,L_2])$ and is denoted by $\mu_n(L_1,M,L_2)$. The triple $(L_1,M,L_2)$ is called \emph{maximal} if its Maslov index is equal to $n$. The triple $(L_0,L^+,L_0^{\mathrm{opp}})$ where $L^+$ is the Lagrangian generated by the basis $\mathbf e_0+\mathbf f_0$ is called the \emph{standard maximal triple}.
\end{definition}

For $n=1$, the three Lagrangians $(L_1,M,L_2)$ are pairwise
distinct points in the circle $\mathbb{RP}^1$. The Maslov index is~$1$ if the three
points are cyclically ordered, and~$-1$ otherwise.

\begin{remark}
There is a slightly more general definition of the Maslov index that works for any triple of Lagrangians, not just for the pairwise transverse triples. It can be defined as the signature $\mathrm{sgn}(\gamma)$ of the quadratic form
\begin{align*}
  \gamma\colon  L_1 \oplus M \oplus L_2 &\longrightarrow \mathbb{R}\\
  (v,w,x) &\longmapsto \omega(v,w) +\omega(w,x) +\omega(x,v)\; .
\end{align*}
When the triple is pairwise transverse, the two definitions agree.
\end{remark}

\begin{proposition}[Properties of Maslov index]\labelx{maslov_prop}
The Maslov index
\begin{itemize}
    \item has range $\{-n, -n+2, \dots, n\}$;
    \item is invariant under the action of $\mathrm{Sp}_{2n}(\mathbb R)$ on $\mathrm{Lag}_{n}^{3}$;
    \item is antisymmetric and, as a result, is cyclically invariant;
    \item satisfies the cocycle relation, namely, for all $L_1,L_2,L_3,L_4\in \mathrm{Lag}_{n}$, it holds that
    \[\mu_n(L_1,L_2,L_3) - \mu_n(L_1,L_2,L_4) + \mu_n(L_1,L_3,L_4) - \mu_n(L_2,L_3,L_4)=0\; .\]
\end{itemize}
The group $\mathrm{Sp}_{2n}(\mathbb R)$ acts transitively on the set of triples of pairwise transverse Lagrangians with the same Maslov index. 
The triple $(L_1,L_2,L_3)$ is maximal\index{Lagrangian!maximal triple of} if and only if there exists an element $g\in\mathrm{Sp}_{2n}(\mathbb R)$ and an element $u\in\mathfrak u^+$ such that $g(L_1,L_2,L_3)=(L_0,\exp(u)L_0^{\mathrm{opp}},L_0^{\mathrm{opp}})$.
\end{proposition}

\begin{proposition}[Action on maximal triples of Lagrangians]\labelx{act_pos_triples}
The group $\mathrm{Sp}_{2n}(\mathbb R)$ acts transitively on maximal triples of Lagrangians and
$$\mathcal K=\left\{
\begin{pmatrix}
u & 0 \\
0 & u
\end{pmatrix}
\mid u\in \mathrm{O}(n)\right\}\cong \mathrm{O}(n)$$
is the stabilizer of the standard maximal triple. 
\end{proposition}

\begin{definition}
A quadruple $(L_1,L_2,L_3,L_4)$ of Lagrangians is called \emph{positive}\index{Lagrangian!positive quadruple of} if the triples $(L_1,L_2,L_3)$ and $(L_1,L_3,L_4)$ are maximal.
\end{definition}

\begin{proposition}[Action on positive quadruples of Lagrangians]\labelx{act_pos_quadr}
Let $(L_1,L_2,L_3,L_4)$ be a positive quadruple. There exists an element $g\in\mathrm{Sp}_{2n}(\mathbb R)$ and $b\in \mathrm{Sym}^+_n(\mathbb R)$ such that
$g(L_1,L_2,L_3,L_4)=(L_0,L^+,L_0^{\mathrm{opp}},L(b))$, where $L(b)$ is the Lagrangian generated by the basis $\mathbf f_0-\mathbf e_0b$.
The element $g$ is unique up to the left multiplication by an element of $\mathcal K$. The element $b$ is determined by the quadruple $(L_1,L_2,L_3,L_4)$ uniquely up to conjugation by an element of $\mathrm{O}(n)$.
\end{proposition}

\begin{remark}
    In the notation of the previous proposition, the eigenvalues of $b$ are well-defined by the quadruple $(L_1,L_2,L_3,L_4)$ and they are all positive. The spectrum of $b$ is sometimes called the \emph{cross ratio} of the positive quadruple $(L_1,L_2,L_3,L_4)$.
\end{remark}

\subsection{Maximal representations into Hermitian Lie groups of tube type}\labelx{sec:maximal_reps}

Let $S$ be an oriented surface of finite type. An important invariant of a homomorphism $\rho\colon \pi_1(S)\to \mathrm{Sp}_{2n}(\mathbb R)$ is the Toledo invariant\index{Toledo invariant} $T_\rho$, which was introduced in~\cite{BIW} using bounded cohomology for surfaces with punctures or boundary, and in~\cite{BILW} for closed surfaces. It is important to emphasize that in the case of surfaces with punctures or boundary, the Toledo number depends on the topological surface $S$ and not only on its fundamental group. It is a real number which satisfies an inequality of Milnor--Wood type:
$$-n\lvert\chi(S)\rvert\leq T_\rho\leq  n\lvert\chi(S)\rvert\; .$$
In the case of a closed surface, this number is moreover an integer. Further, $T_\rho$ is invariant under the action of $\mathrm{Sp}_{2n}(\mathbb R)$ on $\mathrm{Hom}(\pi_1(S),\mathrm{Sp}_{2n}(\mathbb R))$ by conjugation, and so it is well defined for representations in $\mathcal{E}(S_{g,n},\mathrm{Sp}_{2n}(\mathbb R))$.

\begin{definition}
A representation $\rho\in\mathcal{E}(S,\mathrm{Sp}_{2n}(\mathbb R))$ is called \emph{maximal}\index{representation!maximal} if $T_\rho=n\lvert \chi(S)\rvert$. The subspace of all maximal representations in $\mathcal{E}(S,\mathrm{Sp}_{2n}(\mathbb R))$ is denoted by $\mathcal{M}(S,\mathrm{Sp}_{2n}(\mathbb R))$.
\end{definition}

Maximal representations have important geometric properties that we state in the next theorem. They were proved in~\cite{BIW,BILW,AGRW}.

\begin{theorem}\labelx{max.prop} Let $S$ be an oriented surface of finite type.
\begin{itemize}
    \item Every maximal representation $\rho$ is  reductive i.e. $\mathcal{M}(S,\mathrm{Sp}_{2n}(\mathbb R)) \subset \mathcal{R}(S,\mathrm{Sp}_{2n}(\mathbb R))$.
    \item Every maximal representation $\rho$ is discrete and faithful, i.e. $\rho$ has no kernel and the image of $\rho$ is discrete in $\mathrm{Sp}_{2n}(\mathbb R)$.
    \item The space $\mathcal{M}(S,\mathrm{Sp}_{2n}(\mathbb R))$ is Hausdorff.
    \item If $S$ is closed, then $\mathcal{M}(S,\mathrm{Sp}_{2n}(\mathbb R))$ is an open and closed subset of $\mathcal{R}(S,\mathrm{Sp}_{2n}(\mathbb R))$, i.e. it is a union of connected components of $\mathcal{R}(S,\mathrm{Sp}_{2n}(\mathbb R))$.
    \item A homomorphism $\rho$ is maximal if and only if it admits a continuous maximal boundary map, i.e. there exists a map $\xi\colon\partial\pi_1(S) \to \mathrm{Lag}_n$ which is $\rho$-equivariant and maximal in the sense that for every positive triple $(x_1,x_2,x_3)\in \partial\pi_1(S)^3$, the image $(\xi(x_1),\xi(x_2),\xi(x_3))\in \mathrm{Lag}_n^3$ is maximal.
\end{itemize} 
\end{theorem}

\begin{remark}
    If $S_{g,n}$ is a compact surface of genus $g$  with $n$-many boundary components, then $\mathcal{R}(S_{g,n},\mathrm{Sp}_{2n}(\mathbb R))$ is a connected space, so $\mathcal{M}(S_{g,n},\mathrm{Sp}_{2n}(\mathbb R))$ has no chance to be open and closed. However, we can consider the following subspace of $\mathcal{R}(S_{g,n},\mathrm{Sp}_{2n}(\mathbb R))$: Let $\mathcal{R}^p(S_{g,n},\mathrm{Sp}_{2n}(\mathbb R))$ be the space of all representations $\rho$ such that for every $\gamma\in\pi_1(S_{g,n})$ which is homotopic to a boundary component, $\rho(\gamma)$ fixes an element of $\mathrm{Lag}_n$. Such representations are called \emph{peripherally parabolic}. Then $\mathcal{M}(S_{g,n},\mathrm{Sp}_{2n}(\mathbb R))$ is an open and closed subset of $\mathcal{R}^p(S_{g,n},\mathrm{Sp}_{2n}(\mathbb R))$.
\end{remark}

\section{Coordinates on the space of framed maximal representations}

In this Section, we introduce coordinates on spaces of maximal framed representations similarly to the Fock--Goncharov coordinates\index{Fock--Goncharov coordinates} for positive framed representations into a split real Lie group. 

\subsection{Transverse and maximal framed representations}
Let $S$ be a ciliated surface\index{surface!ciliated} equipped with a complete hyperbolic structure of finite volume with geodesic boundary. Let $\tilde S\subseteq \mathbb H^2$ be the universal covering of $S$, and let $\tilde P\subseteq \partial^{red}_{\infty}\tilde S$ be the lift on the boundary of $\tilde{S}$ of all punctures and cilia of $S$. 

\begin{definition}
A \emph{(Lagrangian) framing}\index{framing} is a map $F\colon \tilde P\to \mathrm{Lag}_n$. Let $\rho\colon\pi_1(S)\to \mathrm{Sp}_{2n}(\mathbb R)$ be a homomorphism. A \emph{framing of $\rho$} is a $\pi_1(S)$-equivariant framing $F\colon \tilde P\to \mathrm{Lag}_n$, i.e. for every $\gamma\in\pi_1(S)$, $F(\gamma(\tilde p))=\rho(\gamma)F(\tilde p)$, for all $\tilde p\in \tilde P$. A \emph{framed homomorphism}\index{framed!homomorphism}\index{homomorphism!framed} is a pair $(\rho,F)$, where $F$ is a framing of $\rho$.

The space of all framed homomorphisms is denoted by $\mathrm{Hom}^{\mathrm{fr}}(\pi_1(S),\mathrm{Sp}_{2n}(\mathbb R))$. The space
$$\mathcal E^{\mathrm{fr}}(S,\mathrm{Sp}_{2n}(\mathbb R)):= \mathrm{Hom}^{\mathrm{fr}}(\pi_1(S),\mathrm{Sp}_{2n}(\mathbb R))/\mathrm{Sp}_{2n}(\mathbb R)$$
is called the \emph{moduli space of framed representations}\index{moduli space!of framed representations}. A \emph{framed representation}\index{representation!framed} is an element of $\mathcal E^{\mathrm{fr}}(S,\mathrm{Sp}_{2n}(\mathbb R))$.    
\end{definition}

\begin{remark}
Notice that not every homomorphism $\rho\colon\pi_1(S)\to \mathrm{Sp}_{2n}(\mathbb R)$ admits a framing. Homomorphisms that admit a framing are called \emph{peripherally parabolic}\index{homomorphism!peripherally parabolic}.
\end{remark}

Let $\Delta$ be an ideal triangulation of $S$ and $\tilde{\Delta}$ denote the lift of $\Delta$ to the universal covering $\tilde S$.

\begin{definition}\labelx{df:transverse_rep}
A framed homomorphism $(\rho,F)$ is called \emph{$\Delta$-transverse}\index{framed!homomorphism!transverse} if for every two punctures $\tilde p_1,\tilde p_2\in\tilde P$ that are connected by an edge of $\tilde{\Delta}$, the Lagrangians $F(\tilde p_1)$ and $F(\tilde p_2)$ are transverse.

The space of all $\Delta$-transverse framed homomorphisms is denoted by $\mathrm{Hom}^{\mathrm{fr}}_\Delta(\pi_1(S),\mathrm{Sp}_{2n}(\mathbb R))$. The space
$$\mathcal E^{\mathrm{fr}}_\Delta(S,\mathrm{Sp}_{2n}(\mathbb R)) :=  \mathrm{Hom}^{\mathrm{fr}}_\Delta(\pi_1(S),\mathrm{Sp}_{2n}(\mathbb R))/\mathrm{Sp}_{2n}(\mathbb R)$$
is called the \emph{moduli space of $\Delta$-transverse framed representations}\index{moduli space!of transverse framed representations}.
\end{definition}

\begin{definition} A framing $F$ is called \emph{maximal}\index{framing!maximal} if for every cyclically oriented triple $(\tilde p_1,\tilde p_2,\tilde p_3)\in\tilde P^3$, the triple of Lagrangians $(F(\tilde p_1),F(\tilde p_2),F(\tilde p_3))$ is maximal.
A homomorphism $\rho\in\mathrm{Hom}(\pi_1(S),\mathrm{Sp}_{2n}(\mathbb R))$ is called \emph{maximal}\index{homomorphism!maximal} if it admits a maximal framing $F$. The pair $(\rho, F)$ is called a \emph{maximal framed homomorphism}\index{homomorphism!framed!maximal}.

The space of all maximal framed homomorphisms is denoted by $\mathrm{Hom}^{\mathrm{fr}}_{max}(\pi_1(S),\mathrm{Sp}_{2n}(\mathbb R))$. The space
$$\mathcal M^{\mathrm{fr}}(S,\mathrm{Sp}_{2n}(\mathbb R)) :=  \mathrm{Hom}^{\mathrm{fr}}_{max}(\pi_1(S),\mathrm{Sp}_{2n}(\mathbb R))/\mathrm{Sp}_{2n}(\mathbb R)$$
is called the \emph{moduli space of maximal framed representations}\index{moduli space!of maximal framed representations}.
\end{definition}

\begin{remark}
From this definition it follows immediately that maximal framed representations are transverse with respect to any ideal triangulation of $S$.
\end{remark}

\begin{remark}\labelx{max.Hausdorff}
Notice, that in general not all orbits of the $\mathrm{Sp}_{2n}(\mathbb R)$-actions on $\mathrm{Hom}(\pi_1(S),\mathrm{Sp}_{2n}(\mathbb R))$ and on $\mathrm{Hom}^{\mathrm{fr}}(\pi_1(S),\mathrm{Sp}_{2n}(\mathbb R))$ are closed. So the quotient spaces  $\mathcal{E}(S,\mathrm{Sp}_{2n}(\mathbb R))$ and $\mathcal{E}^{\mathrm{fr}}(S,\mathrm{Sp}_{2n}(\mathbb R))$ are not Hausdorff. However the space of maximal framed representations $\mathcal{M}^{\mathrm{fr}}(S,\mathrm{Sp}_{2n}(\mathbb R))$ inside $\mathcal{E}^{\mathrm{fr}}(S,\mathrm{Sp}_{2n}(\mathbb R))$ forms a Hausdorff subspace.

Moreover, the natural map $\mathcal{E}^{\mathrm{fr}}(S,\mathrm{Sp}_{2n}(\mathbb R))\to \mathcal{E}(S,\mathrm{Sp}_{2n}(\mathbb R))$ forgetting the framing maps the space of maximal framed representations $\mathcal{M}^{\mathrm{fr}}(S,\mathrm{Sp}_{2n}(\mathbb R))$ onto the space of maximal representations $\mathcal{M}(S,\mathrm{Sp}_{2n}(\mathbb R))$ discussed in \Cref{sec:maximal_reps} surjectively with finite fiber. This means that every maximal representation admits a framing and has at most finitely many framings. Notice, that this is false in general. There are representations that do not admit framings as well as representations that admit infinitely many framings (for more details we refer to~\cite{AGRW}).
\end{remark}

In fact, to check if the framing is maximal, it is enough to check it only for all positively oriented triangles of one triangulation and not for all possible cyclically oriented triples in $\tilde P$.

\begin{proposition}[Sect. 4.12 in \cite{AGRW}]
Let $\Delta$ be a triangulation of $S$ and let $F$ be a framing such that for every positively oriented triangle $(\tilde p_1,\tilde p_2,\tilde p_3)$ of $\tilde{\Delta}$  the triple of Lagrangians $(F(\tilde p_1),F(\tilde p_2),F(\tilde p_3))$ is maximal. Then the framing $F$ is maximal.
\end{proposition}

\subsection{Framed trivial local systems on graphs}\labelx{sec:graph_loc_systems}

As we mentioned at the beginning of \Cref{sec:Maximal_reps}, Hermitian Lie groups of tube type have a structure of a symplectic group over a noncommutative algebra. So the reader can expect that the coordinates for maximal representations will be similar to coordinates for positive representations in the $\mathrm{PGL}_2(\mathbb R)$-case but with values in some appropriate noncommutative algebra. However, this is only to some extent true because in noncommutative algebras there is an additional complication related to the fact that the group of invertible elements of an algebra acts nontrivially on the algebra by conjugation. This makes impossible to define coordinates for maximal representations locally as we did for the Fock--Goncharov coordinates for every internal edge of the triangulation and for every ideal triangle. To respect this nontrivial action by conjugation, we need to  make some choice in one triangle and then propagate this choice over the~entire surface. Moreover, there are some additional parameters that are not visible in the $\mathrm{PGL}_2(\mathbb R)$-case, that can be seen as global twists around curves on the surface and that are also not of a local nature.

To parameterize spaces of maximal framed representations, we will apply the following strategy: we will construct a certain graph $\Gamma$ on the surface $S$ adapted to a given ideal triangulation $\Delta$; then for a given maximal framed representation we will construct a framed local system on this graph and, finally, provide parameters using this local system.

\subsubsection{The Graph \texorpdfstring{$\Gamma$}{Gamma}}\labelx{sec:Graph_Gamma}

We construct the following graph $\Gamma$ on $S$. The set of vertices of $\Gamma$ is $V_\Gamma := \{(\tau,r)\in T\times E\mid r\subset\bar \tau\}$. A vertex $(\tau,r)\in V_\Gamma$ can be seen as a point in the triangle $\tau\in T$ lying close to the internal edge $r\in E$.

\begin{figure}[ht]
\centering
\includegraphics[scale=1]{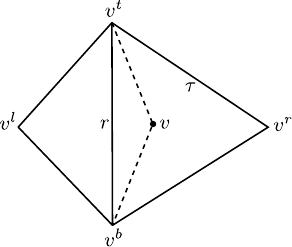}
\caption{Definition of $v^t$, $v^b$, $v^r$ and $v^l$ for $v\in V_\Gamma$.}\labelx{vt-vb}
\end{figure}

Let $v=(\tau,r)\in V_\Gamma$ and let the edge $r$ connect two punctures $p,p'\in P$. If we connect $v$ with $p$ and $p'$ by two simple non-intersecting segments in $\tau$, we obtain a triangle with vertices $v,p,p'$ in $\tau$. We denote $v^t :=  p$, $v^b :=  p'$ if the orientation of the triangle $(v,p,p')$ agrees with the orientation of the surface (see \Cref{vt-vb}). Further, we define $v^r$ as the unique puncture of $P$ such that $(v^t,v^b,v^r)=\tau$. If $r$ is an internal edge of $\Delta$, we also define $v^l$ as the unique puncture of $P$ such that $(v^b,v^t,v^l)$ is a triangle of $\Delta$. This triangle is adjacent to $\tau$ along $r$.

\begin{figure}[ht]
\centering
\includegraphics[scale=1]{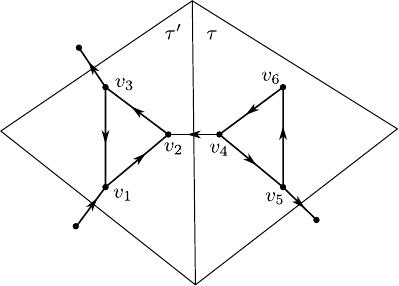}
\caption{Example of a graph $\Gamma$. Here, the upper right edge of the triangulation is external, all other edges are internal and $\tau<\tau'$.}\labelx{Graph_Gamma}
\end{figure}

We now describe the set of oriented edges $E_\Gamma^+$ of $\Gamma$. First we fix the following notation: the oriented edge from the vertex $v$ to the vertex $v'$ is denoted by $(v\to v')$ and $(v'\gets v)$.
\begin{itemize}
    \item Let $v=(\tau_1,r)$, $v'=(\tau_2,r)$ and $\tau_1<\tau_2$, then $(v \to v')$ is an edge of $\Gamma$. It crosses the internal edge $r$ of the triangulation $\Delta$.
    \item Let $v_1$ and $v_2$ be two vertices lying in one triangle such that $v_1^b=v_2^t$, then $(v_1\to v_2)$ is an edge of $\Gamma$ (see \Cref{Graph_Gamma}).
\end{itemize}

\begin{remark}
The graph $\Gamma$ has no multiple edges and no two-cycles.
\end{remark}

We also denote $E^-_\Gamma :=  \{(v'\gets v)\mid (v\gets v')\in E^+_\Gamma\}$. Since $\Gamma$ has no two-cycles, $E^+_\Gamma\cap E^-_\Gamma=\emptyset$. We denote $E_\Gamma :=  E^+_\Gamma\cup E^-_\Gamma$.

We denote by $\tilde\Gamma$ the lift of $\Gamma$ to the universal covering $\tilde S\subset \mathbb{H}^2$ of $S$. We also choose a point $b\in V_\Gamma$ and fix one of its lifts $\tilde b\in V_{\tilde\Gamma}$. Assume $b\in \tau_0$ and $\tilde b\in \tilde \tau_0$ for $\tau_0\in T$ and $\tilde \tau_0\in \tilde T$.

\begin{remark}
Every path on $S$ that starts and ends in $V_\Gamma$ can be deformed to a non-oriented path in $\Gamma$.
\end{remark}

To simplify notation, until the end of this chapter, we shall denote $G=\mathrm{Sp}_{2n}(\mathbb R)$.

\begin{definition}
A \emph{transverse framing} of $\tilde\Gamma$ is a pair of maps $F^t,F^b\colon V_{\tilde\Gamma}\to \mathrm{Lag}_n$ such that $F^t(v)$ and $F^b(v)$ are transverse for every $v\in V_{\tilde\Gamma}$.
\end{definition}

\begin{definition}\labelx{df:triv_loc_sys} A \emph{trivial $G$-local system on $\tilde\Gamma$} is a map $T\colon V_{\tilde\Gamma}\cup E_{\tilde\Gamma}\to G$ such that for any edge $(v'\gets v)\in E_{\tilde\Gamma}$, $v,v'\in V_{\tilde\Gamma}$, $T(v'\gets v)=T(v\gets v')^{-1}$ and  $T(v')=T(v'\gets v)T(v)$. A trivial $G$-local system on $\tilde\Gamma$ defined by the map $T$ is denoted by $(\tilde\Gamma,T)$.
\end{definition}

\begin{remark}
From \Cref{df:triv_loc_sys} it follows that $T(e_k)...T(e_1)=1$, for every cycle $(e_1,\dots e_k)$, where all $e_i\in E_{\tilde\Gamma}$, i.e. holonomies around cycles are trivial. This is the reason to call such a local system trivial.

Let $v\in V_{\tilde\Gamma}$ be a fixed vertex of ${\tilde\Gamma}$. The map $T$ is determined by $T\vert_{E^+_{\tilde\Gamma}}$ and $T(v)$. The elements $T(v)$ and $T(e)$, for all $e\in E^+_{{\tilde\Gamma}}$, can be arbitrary elements of $G$ satisfying the following condition: $T(e_k)...T(e_1)=1$, for every cycle $(e_1,\dots e_k)$, where all $e_i\in E_{\tilde\Gamma}$.
\end{remark}

Let $(F^t,F^b)$ be a transverse framing, let $({\tilde\Gamma},T)$ be a trivial $G$-local system on ${\tilde\Gamma}$. We say that the transverse framing $(F^t,F^b)$ is \emph{adapted} to the trivial local system $({\tilde\Gamma},T)$ if for every $v\in V_{\tilde\Gamma}$ holds $T(v)F^t(v)=L_0^{\mathrm{opp}}$ and $T(v)F^b(v)=L_0$. A quadruple $\mathfrak L:=({\tilde\Gamma},T,F^t,F^b)$ is called a \emph{transverse framed trivial $G$-local system} if $(F^t,F^b)$ is a transverse framing adapted to $({\tilde\Gamma},T)$.

Since ${\tilde\Gamma}$ has no multiple edges, any path $\gamma$ in ${\tilde\Gamma}$ can be written in a unique way as a sequence of vertices $\gamma=(v_k\gets v_{k-1}\gets\dots\gets v_1)$, where $v_1$ is the starting vertex of $\gamma$ and $v_k$ is the end-vertex of $\gamma$ and $(v_{i+1}\gets v_{i})\in E_{\tilde\Gamma}$, for all $i\in\{1,\dots,k-1\}$. 
We also extend the map $T$ to the space of paths as follows: if $\gamma=(v_k\gets v_{k-1}\gets\dots\gets v_1)$, then $T(\gamma) :=  T(v_k\gets v_{k-1})\dots T(v_2\gets v_1)$.

\begin{remark}
Let $\gamma$ be a path in ${\tilde\Gamma}$ connecting two vertices $v$ and $w$. Then, in fact, $T(\gamma)$ depends only on $v$ and $w$ and not on the entire path $\gamma$. So sometimes in order to emphasize this fact, we will just write $T(w\gets v)$ instead of $T(\gamma)$.
\end{remark}

\subsubsection{From local systems to representations}\labelx{Loc_to_Rep}

Let $\mathfrak L=({\tilde\Gamma},T,F^t,F^b)$ be a transverse framed trivial $G$-local system on ${\tilde\Gamma}$. The action of the fundamental group $\pi_1(S)$ on $\tilde S$ restricts to the action on $\tilde\Gamma$. 

\begin{definition}
The local system $\mathfrak L=({\tilde\Gamma},T,F^t,F^b)$ is called \emph{$\pi_1(S)$-invariant} if, for all $\gamma\in \pi_1(S)$ and for all edges $(w\gets v)\in E_{\tilde\Gamma}$, it is $v,w\in V_{\tilde\Gamma}$, $T(w\gets v)=T(\gamma w\gets \gamma v)$.
\end{definition}

A $\pi_1(S)$-invariant $G$-local system always gives rise to a homomorphism $\rho\colon \pi_1(S)\to G$ in the following way: let $v\in V_{\tilde\Gamma}$, we define for every $\gamma \in \pi_1(S)$:
\begin{equation}\labelx{Graph_representation}
\rho_\mathfrak L(\gamma) :=  T(\gamma v)^{-1}T(v)\; .
\end{equation}

\begin{remark}
If $\mathfrak L=({\tilde\Gamma},T,F^t,F^b)$ is a $\pi_1(S)$-invariant local system, then the framing $(F^t,F^b)$ is $\rho_\mathfrak L$-equivariant, i.e. for all $v\in V_{\tilde\Gamma}$, $F_t(\gamma v)=\rho_\mathfrak L(\gamma)(F_t(v))$ and $F_b(\gamma v)=\rho_\mathfrak L(\gamma)(F_b(v))$.
\end{remark}

\begin{proposition}
The map $\rho_\mathfrak L$ is a group homomorphism that does not depend on the choice of $v\in V_{\tilde\Gamma}$.
\end{proposition}

\begin{proof}
Let $w\in V_{\tilde\Gamma}$ be another vertex. Then
\begin{align*}
T(\gamma w)^{-1}T(w)
& =(T(\gamma w\gets \gamma v)T(\gamma v))^{-1}(T(w\gets v)T(v))\\
&=T(\gamma v)^{-1}T(\gamma w\gets \gamma v)^{-1}T(w\gets v)T(v)\\
&=T(\gamma v)^{-1}T(v)\; ,
\end{align*}
since $T(\gamma w\gets \gamma v)=T(w\gets v)$ by $H$-invariance. So $\rho_\mathfrak L$ does not depend on the choice of $v\in V_{\tilde\Gamma}$.

Let now $\gamma_1,\gamma_2\in \pi_1(S)$. We obtain $\rho_\mathfrak L(\gamma_2)=T(\gamma_2 (\gamma_1 v))^{-1}T(\gamma_1 v)$ and $\rho_\mathfrak L(\gamma_1)=T(\gamma_1 v)^{-1}T(v)$. Therefore,
\begin{equation*}
   \begin{split}
       \rho_\mathfrak L(\gamma_2)\rho_\mathfrak L(\gamma_1) &=T(\gamma_2 (\gamma_1 v))^{-1}T(\gamma_1 v)T(\gamma_1 v)^{-1}T(v) \\
       &=T(\gamma_2 (\gamma_1 v))^{-1}T(v)=\rho_\mathfrak L(\gamma_2\gamma_1)\; .
   \end{split}
\end{equation*}
So $\rho_\mathfrak L$ is a group homomorphism.
\end{proof}

Every framing $F\colon \tilde P\to\mathrm{Lag}_n$ defines the unique framing $(F^t,F^b)$ on $\tilde\Gamma$ as follows: for every vertex $v$ of $\tilde\Gamma$ define $F^t(v):=F(v^t)$ and $F^b(v):=F(v^b)$ (see \Cref{vt-vb}). In this case we say that $(F^t,F^b)$ is \emph{induced} by $F$. We also denote $F^r(v):=F(v^r)$ and $F^l(v):=F(v^l)$.

Let $r$ be an internal edge of $\tilde\Gamma$ that separates two triangles $\tau$ and $\tau'$, and $v:=(\tau,r)$, $v':=(\tau',r)$. Then $F^t(v')=F^b(v)$, $F^b(v')=F^t(v)$, $F^r(v')=F^l(v)$, and $F^l(v')=F^r(v)$. Further, let $v$ and $v'$ be two vertices of $\tilde\Gamma$ that lie in the same triangle and there is the edge $(v\to v')$ in $\tilde\Gamma$. Then $F^t(v')=F^b(v)$, $F^b(v')=F^r(v)$, $F^r(v')=F^t(v)$.

The following proposition is immediate:

\begin{proposition}
Let $(F^t,F^b)$ be a transverse framing on $\tilde\Gamma$ such that
\begin{itemize}
    \item for every internal edge $r$ of $\tilde\Gamma$ that separates two triangles $\tau$ and $\tau'$, and $v:=(\tau,r)$, $v':=(\tau',r)$ holds $F^t(v')=F^b(v)$;
    \item for every two vertices $v$ and $v'$ of $\tilde\Gamma$ that lie in the same triangle and there is the edge $(v\to v')$ in $\tilde\Gamma$ holds $F^t(v')=F^b(v)$.
\end{itemize}
Then there exists a unique framing $F\colon \tilde P\to\mathrm{Lag}_n$ that induces $(F^t,F^b)$.
\end{proposition}

\begin{remark} Let $(\rho, F)$ be a framed homomorphism. If $(F^t,F^b)$ is the framing induced by $F$, then $(F^t,F^b)$ is $\rho$-equivariant as well.
If $\mathfrak L=(\tilde\Gamma,T,F^t,F^b)$ is a $\pi_1(S)$-invariant transverse framed trivial $G$-local system on $\tilde\Gamma$, then $F$ is a framing of $\rho_\mathfrak L$, but in general, $\rho\neq \rho_\mathfrak L$.
\end{remark}

\begin{definition}
We say that a transverse framed trivial $G$-local system $(\tilde\Gamma,T,F^t,F^b)$ is \emph{maximal} if the framing $(F^t,F^b)$ is induced by a maximal framing $F\colon \tilde P\to\mathrm{Lag}_n$.
\end{definition}

\subsubsection{From representations to local systems}\labelx{Moves}

In this Section, we show how to construct a maximal framed trivial $G$-local system on $\Gamma$ out of a maximal framed representation. 
Let $\Delta$ be an ideal triangulation of $S$ and $\tilde\Gamma$ be a graph on $\tilde S$ as defined in \Cref{sec:Graph_Gamma}. Given a maximal framed homomorphism $(\rho,F)\in\mathrm{Hom}^{\mathrm{fr}}_{max}(\pi_1(S,b),G)$, then the framing $F$ gives rise to the $\rho$-equivariant transverse framing $(F^t,F^b)$ of $\tilde\Gamma$. We are going to define a local system $\mathfrak L=(\tilde\Gamma,T)$ such that $(F^t,F^b)$ will be its transverse framing and  $\rho_\mathfrak L=\rho$.

We will start with the two simplest cases when $S$ is a triangle and a triangulated quadrilateral, and then we describe a procedure for a general surface $S$.

\subsubsection*{Triangle}\labelx{Turn}

Let $S=\tilde S$ be an ideal triangle with vertices $p_1,p_2,p_3$. The only triangulation $\Delta$ of $S$ consists of all external edges of $S$. Let $(\rho,F)$ be a maximal framed homomorphism. Since $\pi_1(S)$ is trivial, $\rho$ is trivial. Let $F(p_i)=F_i$ for all $i\in\{1,2,3\}$. Then $(F_1,F_2,F_3)$ is a maximal triple. Let the graph $\Gamma=\tilde\Gamma$ consist of three vertices $v$, $v'$ and $v''$ and three edges $(v'\gets v)$, $(v''\gets v')$ and $(v\gets v'')$.

Because of maximality, there exists an element $g\in G$ such that $g(F_1,F_2,F_3)=(L_0^{\mathrm{opp}},L_0,L^+)$
(see \Cref{Turn_figure}).

\begin{figure}[ht]
\centering
\includegraphics{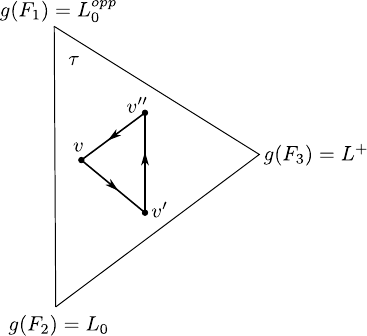}
\caption{Graph $\Gamma$ in a triangle.}\labelx{Turn_figure}
\end{figure}

We define $T(v) :=  g$ and $T(v'\gets v) :=  T(v''\gets v') :=  T(v\gets v'') :=  \begin{pmatrix} -1 & 1 \\ -1 & 0 \end{pmatrix}$, then $T(v')=T(v'\gets v)T(v)$, $T(v'')=T(v''\gets v')T(v')$ and
$$T(v')F^t(v')=T(v')F_2=T(v'\gets v)L_0=L_0^{\mathrm{opp}}\; ,$$
$$T(v')F^b(v')=T(v')F_3=T(v'\gets v)L^+=L_0\; ,$$
$$T(v')F^r(v')=T(v')F_1=T(v'\gets v)L_0^{\mathrm{opp}}=L^+\; .$$
Similarly, one sees that $T(v'')F^t(v'')=L_0^{\mathrm{opp}}$, $T(v'')F^b(v'')=L_0$ and $T(v'')F^r(v'')=L^+$. Moreover, $T(v\gets v'')T(v''\gets v')T(v'\gets v)=\mathrm{Id}$. So we have constructed a well defined maximal framed trivial $G$-local system in $\Gamma$. Notice, that the local system is not unique, it depends on the choice of $g\in G$. This $g$ is uniquely defined up to left multiplication by an element of $\mathrm{O}(n)$ (see \Cref{act_pos_triples}).

\subsubsection*{Quadrilateral. Crossing an edge of triangulation}\labelx{Quadruple}

Let $S=\tilde S$ be an ideal quadrilateral with cyclically ordered vertices $p_1,p_3,p_2,p_4$. Let $\Delta$ be the triangulation that consists of all external edges of $S$ and one internal edge connecting $p_1$ and $p_2$. Let $(\rho,F)$ be a maximal framed homomorphism. Since $\pi_1(S)$ is trivial, $\rho$ is trivial. Let $F(p_i)=F_i$ for all $i\in\{1,\dots,4\}$. Then $(F_1,F_4,F_2,F_3)$ is a positive quadruple. The graph $\Gamma=\tilde\Gamma$ consists of six vertices, four of them lie close to external edges and two, $v$ and $v'$, lie close to the internal edge. So there exists an edge $(v'\gets v)$ or $(v\gets v')$. Without loss of generality, we assume the first case.

\begin{figure}[ht]
\centering
\includegraphics[width=\textwidth]{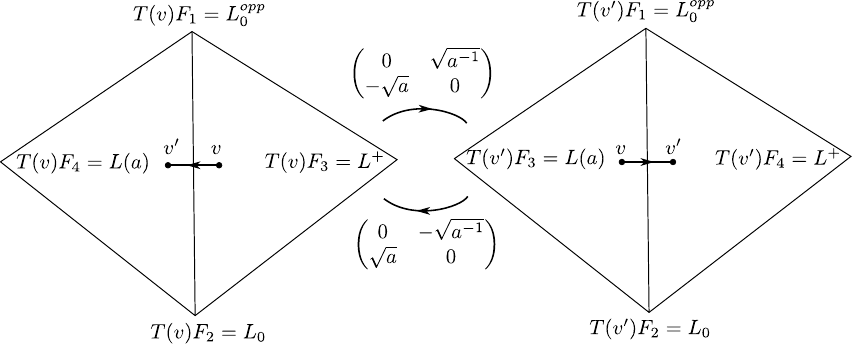}
\caption{Crossing an edge of triangulation.}\labelx{Quadr_figure}
\end{figure}

Because of transversality by \Cref{act_pos_quadr}, there exists an element $g\in G$ such that
$$g(F_2,F_3,F_1,F_4)=(L_0,L^+,L_0^{\mathrm{opp}},L(a))\; ,$$
where $L(a)$ is generated by the basis $\mathbf f_0-\mathbf e_0a$ with $a\in \mathrm{Sym}^+_n(\mathbb R)$ (see \Cref{Quadr_figure}). For every matrix $a\in \mathrm{Sym}^+_n(\mathbb R)$, there exists the unique square root $q\in \mathrm{Sym}^+_n(\mathbb R)$ such that $q^2=a$. We define $T(v) :=  g$, $T(v'\gets v) := \begin{pmatrix}0 & \sqrt{a^{-1}} \\ -\sqrt{a} & 0\end{pmatrix}$, then $T(v')=T(v'\gets v)T(v)$. All other edges of $\Gamma$ lie inside one of two triangles of $S$, so we construct the map $T$ for them as in the previous subsection. We have therefore constructed a maximal framed trivial $G$-local system on $\Gamma$. The element $g$ is unique up to the left multiplication by an element of $\mathrm{O}(n)$.

Let now $S$ be any ciliated surface, $\Delta$ be an ideal triangulation of $S$ and $(\rho,F)\in\mathrm{Hom}^{\mathrm{fr}}_+(\pi_1(S),G)$. Let $\tilde\Gamma$ be the graph as in \Cref{sec:Graph_Gamma} with the framing $(F^t,F^b)$ induced by $F$. We can assume that the base point of the fundamental group agrees with some point $b\in V_\Gamma$ that lies in a triangle $\tau$. We take $\tilde b\in V_{\tilde\Gamma}$ a lift of $b$.

Starting at $\tilde b$ and applying inductively the procedure described in \Cref{Turn} for every triangle and in \Cref{Quadruple} for every pair of adjacent triangles, we obtain a maximal framed local system on $\tilde\Gamma$. This local system has the property that the representation described in (\ref{Graph_representation}) agrees with $\rho$. However, this local system is in general not $\pi_1(S)$-invariant, although the framing is $\rho$-equivariant. In the proof of \Cref{main_thm} in \Cref{section:parametrization}, we will modify this local system to make it $\pi_1(S)$-invariant.

\subsection{Parametrization of the space of maximal representations}\labelx{section:parametrization}

Now we are ready to provide a parameterization of the space of maximal framed representations. As before, let $S$ be a ciliated surface with Euler characteristic $\chi(S)$ and with $p_e$ cilia.

\begin{theorem}\labelx{main_thm}
The space $\mathcal{M}^{\mathrm{fr}}(S,G)$ is homeomorphic to the following space:
$$(\mathrm{Sym}^+_n(\mathbb R)^{p_e-3\chi(S)}\times \mathrm{O}(n)^{1-\chi(S)})/\mathrm{O}(n)\; ,$$
where $\mathrm{O}(n)$ acts by conjugation in every factor.
\end{theorem}

\begin{remark}
    As we announced at the beginning of \Cref{sec:graph_loc_systems}, some parameters are indeed elements of the noncommutative algebra $\mathrm{Mat}_n(\mathbb R)$, namely of its ``positive part'' $\mathrm{Sym}^+_n(\mathbb R)$. As in the $\mathrm{PGL}_2(\mathbb R)$-case, they are associated to internal edges of the triangulation. However, we also see additional $\mathrm{O}(n)$-parameters as well as the quotient by the $\mathrm{O}(n)$-action by conjugation which we did not see in the Fock--Goncharov case. To clarify their meaning, we provide the proof of this theorem although it is quite technical.
\end{remark}

\begin{proof} Let $(\rho,F)$ be a maximal framed homomorphism. We pick some ideal triangulation $\Delta$ of $S$ with the set of internal edges $E_{in}$. The framing $F$ induces a framing $(F^t,F^b)$ of $\tilde\Gamma$.

\noindent Let $b\in S$ be the base point of $\pi_1(S,b)$ and $\tilde b\in\tilde S$ be a lift of $b$. As before, we assume $\tilde b\in V_{\tilde\Gamma}$. Let $g\in G$ be such that $g(F^t(\tilde b),F^b(\tilde b),F^r(\tilde b))=(L_0^{\mathrm{opp}},L_0,L^+)$. We now define
\begin{align*}
X := \{(\rho,F,g)\mid & (\rho,F)\in \mathrm{Hom}^{\mathrm{fr}}_+(\pi_1(S,b),G),  \,\, g\in G, \text{ such that:} \\ & g(F^t(\tilde b),F^b(\tilde b),F^r(\tilde b))=(L_0^{\mathrm{opp}},L_0,L^+)\}\; ,
\end{align*}
and the equivalence relation $\sim$ on $X$ as follows: \[(\rho,F,g)\sim (\rho',F',g'), \text{ if and only if }(g\rho g^{-1},gF)=(g'\rho' g'^{-1},g'F')\; .\]

\noindent Step 1: Assume $S=\tilde S$ is a triangulated polygon and $\Gamma$ is defined as in \Cref{sec:Graph_Gamma}. Let $(\rho,F,g)\in X$. We define $T(b)=g$ and then apply the procedure defined in \Cref{Moves}. We obtain in this way a framed transverse $G$-local system on $\Gamma$. Let $v=(\tau,r)\in V_\Gamma$ for $r\in E_{in}$, then $v$ lies in the triangle $\tau$ with vertices $v^t,v^b,v^r\in P$. Let $\tau'$ be the triangle adjacent to $\tau$ along the edge $r=(v^t,v^b)$, and $v^b$, $v^t$ and $v^l$ be the vertices of $\tau'$. Let $b_v\in \mathrm{Sym}^+_n(\mathbb R)$ be defined as $T(v)F(v^l)=L(b_v)$, where $L(b_v)$ is the Lagrangian generated by the basis $\mathbf f_0-\mathbf e_0b_v$. The elements $b_v$ are uniquely defined for all $v\in V_\Gamma$. Moreover, by the construction in \Cref{Moves}, $b_v=b_{v'}$ where $v'=(\tau',r)$. Therefore, we can assign the element $b_v$ not to the vertex of $\Gamma$ but to the internal edge $r$ of $\Delta$. So we will write $b_r$ instead.

\noindent Therefore, we obtain the following map:
$$\begin{matrix}
\mathcal X\colon & X & \to & \mathrm{Sym}^+_n(\mathbb R)^{E_{in}}\\
 & (\rho,F,g) & \mapsto & (b_r)_{r\in E_{in}}\; .
\end{matrix}$$
Notice that $\mathcal X(\rho,F,g)=\mathcal X(\rho',F',g')$ if and only if  $(\rho,F,g)\sim(\rho',F',g')$. This map is surjective because any choice of $(b_e)_{e\in E_{in}}$ defines a transverse framed $G$-local system on $\Gamma$ that by \Cref{Loc_to_Rep} defines a framed homomorphism $(\rho,F)\in\mathrm{Hom}^{\mathrm{fr}}_\Delta(\pi_1(S),G)$.

\noindent We also have the $\mathrm{O}(n)$-action on $X$ by the left multiplication in the last component and by componentwise conjugation on $\mathrm{Sym}^+_n(\mathbb R)^{E_{in}}$ that is compatible with $\mathcal X$, i.e. let $\mathrm{diag}(u,u)=k\in\mathcal K$, for $u\in \mathrm{O}(n)$, then $\mathcal X(\rho,D,kg)=(ub_ru^{-1})_{r\in E_{in}}$.
We obtain the quotient map:
$$\bar{\mathcal X}\colon \mathcal{M}^{\mathrm{fr}}(S,G)\to \mathrm{Sym}^+_n(\mathbb R)^{E_{in}}/\mathrm{O}(n)\; ,$$
which is a homeomorphism.

\noindent Step 2: Now let us turn to the general case. We choose a connected fundamental domain $S_0$ for $S$ in $\tilde S$ that completely consists of ideal triangles and $\tilde b$ lies in one of them. The fundamental domain $S_0$ is an ideal polygon, so we can apply the procedure above and obtain a transverse framed $G$-local system over $\Gamma_0$ which is a subgraph of $\tilde \Gamma$ contained in $S_0$. We want to extend this local system to the entire graph $\tilde\Gamma$ in a $\pi_1(S)$-invariant way.

\noindent To obtain a $\pi_1(S)$-invariant local system, according to \Cref{Loc_to_Rep}, we need to define $T(\tilde v)$ for every $\tilde v\in V_{\tilde\Gamma}$ as follows: let $\gamma\in\pi_1(S)$ be the unique element such that $\gamma v\in S_0$. We define $T(\tilde v) :=  T(\gamma \tilde v)\rho(\gamma)$. From this follows immediately that for every $\tilde v\in V_{\tilde\Gamma}$ and for every $\gamma\in\pi_1(S)$, $T(\gamma\tilde v)=T(\tilde v)\rho(\gamma)^{-1}$.

\noindent The local system $(\tilde\Gamma,T)$ is indeed $\pi_1(S)$-invariant because for $(w\gets v)\in E_{\tilde\Gamma}$ and $\gamma\in\pi_1(S)$, then 
\begin{align*}
T(\gamma\tilde w\gets\gamma\tilde v)&=T(\gamma\tilde w)T(\gamma\tilde v)^{-1}\\
&=T(\tilde w)\rho(\gamma)^{-1}\rho(\gamma)T(\tilde v)^{-1}\\
&=T(\tilde w)T(\tilde v)^{-1}\\
&=T(\tilde w\gets\tilde v)\; .
\end{align*}

\noindent Moreover, the framing $(F^t,F^b)$ of $(\tilde\Gamma,T)$ is adapted to $(\tilde\Gamma, T)$. Indeed, if $\tilde v$ is contained in $S_0$, then by construction $T(\tilde v)(F^t(\tilde v),F^b(\tilde v))=(L_0^{\mathrm{opp}},L_0)$. If $\tilde v\in V_{\tilde\Gamma}$ is any vertex, then there exists the unique $\gamma\in\pi_1(S)$ with $\gamma\tilde v$ contained in $S_0$. Then,
\begin{align*}
T(\tilde v)(F^t(\tilde v),F^b(\tilde v)) & = T(\gamma \tilde v)\rho(\gamma)\rho(\gamma)^{-1}(F^t(\gamma\tilde v),F^b(\gamma\tilde v))\\
& = T(\gamma \tilde v)(F^t(\gamma\tilde v),F^b(\gamma\tilde v))\\ & = (L_0^{\mathrm{opp}},L_0)\; .
\end{align*}

\noindent The framed local system $(\tilde \Gamma,T, F^t,F^b)$ is determined by its restriction to $\Gamma_0$ and the maps $T(\tilde w\gets \tilde v)$ for $\tilde v\in S_0$ and $\tilde w\notin S_0$. We thus need to understand the maps $T(\tilde w\gets \tilde v)$.

\noindent Let $e,e'$ be edges of the triangulation $\tilde{\Delta}$ that are in the boundary of $S_0$ but not for $\tilde S$, and let $\gamma\in\pi_1(S)$ be the unique element such that $\gamma(e)=e'$. Without loss of generality assume $(\tilde w\gets \tilde v)\in E^+_{\tilde\Gamma}$ that crosses $e$ and $\tilde v\in S_0$, $\tilde w\notin S_0$. Then the edge $(\gamma_i\tilde w\gets \gamma_i\tilde v)\in E^+_{\tilde\Gamma}$ crosses $e'$ and $\gamma_i\tilde v\notin S_0$, $\gamma_i\tilde w\in S_0$. Moreover,
$$T(\tilde w\gets \tilde v)= T(\tilde w)T(\tilde v)^{-1}=T(\gamma \tilde w)\rho(\gamma)T(\tilde v)^{-1}\; ,$$
and so
\begin{align*}
T(\tilde w\gets \tilde v)(L_0^{\mathrm{opp}},L_0)
& = T(\gamma \tilde w)\rho(\gamma)T(\tilde v)^{-1}(L_0^{\mathrm{opp}},L_0)\\
&=T(\gamma \tilde w)\rho(\gamma)(F^t(\tilde v),F^b(\tilde v))\\
& = T(\gamma \tilde w)(F^t(\gamma\tilde v),F^b(\gamma\tilde v))\\
&=T(\gamma \tilde w)(F^b(\gamma\tilde w),F^t(\gamma\tilde w))\\
&=(L_0,L_0^{\mathrm{opp}})\; .
\end{align*}
This means that $T(\tilde w\gets \tilde v)=l_e\omega$, where $l_e\in L$ is the uniquely defined element and $\omega=\begin{pmatrix}0 & \mathrm{Id} \\ -\mathrm{Id} & 0 \end{pmatrix}$. So for every pair $\{e,e'\}$ of external edges that are related by a nontrivial element of $\pi_1(S)$, we get an element of $L$. By the polar decomposition, the element $l_e$ can be written in the unique way as $l_e=\mathrm{diag}(u_eb_e,u_eb_e^{-1})$, where $u_e\in \mathrm{O}(n)$, $b_e\in \mathrm{Sym}^+_n(\mathbb R)$. We denote the set of all $\{e,e'\}$ pairs as above by $E_0$. Its cardinality is $1-\chi(S)$.

\noindent As in Step 1, we obtain a map
$$\begin{matrix}
\mathcal X\colon & X & \to & \mathrm{Sym}^+_n(\mathbb R)^{E_{in}}\times \mathrm{O}(n)^{E_0}\\
 & (\rho,F,g) & \mapsto & (b_r,u_e)_{r\in E_{in},\;e\in E_0}\; .
\end{matrix}$$
Notice that $\mathcal X(\rho,F,g)=\mathcal X(\rho',F',g')$ if and only if  $(\rho,F,g)\sim(\rho',F',g')$. Further, this map is surjective because any choice of $(b_r,u_e)_{r\in E_{in},\;e\in E_0}$ defines a $\pi_1(S)$-invariant transverse framed $G$-local system on $\tilde\Gamma$, which in turn defines by \Cref{Loc_to_Rep} a framed homomorphism $(\rho,F)\in\mathrm{Hom}^{\mathrm{fr}}_{max}(\pi_1(S),G)$.

\noindent We also have the $\mathrm{O}(n)$-action on $X$ by the left multiplication in the last component and by componentwise conjugation on $\mathrm{Sym}^+_n(\mathbb R)^{E_{in}}\times \mathrm{O}(n)^{E_0}$ that is compatible with $\mathcal X$, i.e. 
\[\mathcal X(\rho,D,kg)=(ub_ru^{-1},uu_eu^{-1})_{r\in E_{in},\;e\in E_0}\; ,\] 
where $k=\mathrm{diag}(u,u)$. We thus obtain the quotient map
$$\bar{\mathcal X}\colon \mathcal{M}^{\mathrm{fr}}(S,G)\to (\mathrm{Sym}^+_n(\mathbb R)^{E_{in}}\times \mathrm{O}(n)^{E_0})/\mathrm{O}(n)\; ,$$
which is a homeomorphism.
\end{proof}

\begin{corollary}
The space $\mathcal{M}^{\mathrm{fr}}(S,G)$ is homotopy equivalent to $\mathrm{O}(n)^{1-\chi(S)}/\mathrm{O}(n)$ and has $2^{1-\chi(S)}$ connected components.
\end{corollary}

For the proof of this result we refer to~\cite[Sect. 7]{AGRW}.
\chapter{Positivity}\labelx{sec:positivity}


\abstract{Fock--Goncharov coordinates describe Hitchin components as the positive part of the character variety. We give an introduction to the underlying notion of \emph{positivity} for split real Lie groups, as well as a generalized notion of positivity, called $\Theta$-positivity, for real semisimple Lie groups introduced by Guichard--Wienhard in \cite{GW,GW22} that characterizes higher-rank Teichm\"{u}ller spaces. The notion emerges from a choice of a certain subset of simple positive roots $\Theta$ among all the roots of the corresponding Lie algebra $\mathfrak{g}$ of the group $G$. Accidentally, the Greek letter $\Theta$ is also the initial of the word \textgreek{Θετικότητα} in Greek which translates to ``positivity''. A detailed analysis of the $\Theta$-positive structure for the groups $\mathrm{SL}_n(\mathbb{R})$ and $\mathrm{Sp}_{2n}(\mathbb{R})$ is included.}

\section{Total positivity}\labelx{sec:total_positivity}

The notion of total positivity is apparent in a surprisingly broad spectrum of mathematical fields. In this Section we briefly recall the definition of a \emph{totally positive} matrix given in \Cref{sec:positive_flags}, end expand on its properties. An invertible $n \times n$ matrix with real entries is called \emph{totally positive}, if the determinants of all square sub-matrices are positive. We exhibit important features of the notion next and refer to \cite{Ando,GasMic,Karlin} for a general treatise.

\begin{definition}
A matrix $m\in\GL_n(\R)$ is said to be \emph{totally non-negative}\index{matrix!totally non-negative} (resp. \emph{totally positive}\index{totally positive!matrix}\index{matrix!totally positive}\labelx{matrix!totally positive}), if for all $I,J\subset [\vert1,n\vert]$ such that $\vert I\vert=\vert J\vert$, one has $\Delta_{I,J}(m)\in\R_{\geq0}$ (resp. $\Delta_{I,J}(m)\in\R_{>0}$), where $\Delta_{I,J}(m)$ is the minor of $m$ containing rows of $m$ indexed by $I$ and columns of $m$ indexed by $J$.
\end{definition}

\begin{example}\labelx{Ex:tot-pos-matrix}
    The matrix $M=\begin{pmatrix} 22 & 9 & 6\\ 8 & 4 & 3 \\ 1 & 1 & 1\end{pmatrix}$ is totally positive, since all entries, all $2\times 2$-determinants and $\mathrm{det}(M)$ are strictly positive.
\end{example}

\begin{proposition}
	Totally non-negative (resp. positive) matrices form a multiplicative sub-semigroup of $\mathrm{GL}_n(\mathbb{R})$ denoted by $\mathrm{GL}_{\geq 0}$ (resp. $\mathrm{GL}_{>0}$).
\end{proposition}

\begin{proof}
This is a direct consequence of the property of the determinant of the product of a pair of square matrices being equal to the product of their determinants.
\end{proof}

In order to understand better the structure of totally positive matrices in $\mathrm{GL}_n(\mathbb{R})$, we first introduce several subgroups of it.
    Let $B$ be the subgroup of $\mathrm{GL}_n(\mathbb R)$ of all upper triangular matrices called the \emph{Borel subgroup}\index{subgroup!Borel} (or \emph{maximal parabolic subgroup}\index{subgroup!maximal parabolic}, and let $B^{\mathrm{opp}}$ be the subgroup of $\mathrm{GL}_n(\mathbb R)$ of all lower triangular matrices called the \emph{opposite Borel subgroup}.  Let also $U$ be the subgroup of $B$ of all matrices with 1s along the diagonal called the \emph{upper unipotent subgroup}\index{subgroup!unipotent} (or just the \emph{unipotent subgroup}), and let $U^{\mathrm{opp}}$ be the subgroup of $B^{\mathrm{opp}}$ of all matrices with 1s along the diagonal called the \emph{lower unipotent subgroup}.

The notion of total positivity can be extended for matrices in $U$, resp. $U^{\mathrm{opp}}$ as follows:

\begin{definition}
A matrix $m\in U\cup U^{\mathrm{opp}}$ is said to be \emph{totally non-negative} (resp. \emph{totally positive}), if for all $I,J\subset [\vert1,n\vert]$ such that $\vert I\vert=\vert J\vert$, one has $\Delta_{I,J}(m)\in\R_{\geq0}$ (resp. $\Delta_{I,J}(m)\in\R_{>0}$), except those that have to be zero by the condition that $m\in U$ (resp. $m\in U^{\mathrm{opp}}$). The set of all totally non-negative (resp. totally positive) of $U$ (resp. $U^{\mathrm{opp}}$) is denoted by $U_{\geq 0}$ (resp. $U_{>0}$), and $U^{\mathrm{opp}}_{\geq 0}$ (resp. $U^{\mathrm{opp}}_{>0}$).
\end{definition}

\begin{example}\labelx{Ex:tot-pos-matrix-U}
    The matrix $U_1=\begin{pmatrix} 1 & 3 & 6\\ 0 & 1 & 3 \\ 0 & 0 & 1\end{pmatrix}$ is unipotent and it is easy to check that it is a totally positive element of $U$.
\end{example}

The following property is a key fact for the study of totally positive matrices:

\begin{proposition}[Cryer's splitting lemma\labelx{Cryer's splitting lemma} \cite{cryer1973lu,cryer1976some}]
An invertible matrix $m\in\GL_n(\R)$ is totally non-negative if and only if it has a Gaussian decomposition $M=U_1 DU_2$ where $U_1\in U_{\geq 0}$, $D$ is diagonal with non-negative entries and $U_2\in U^{\mathrm{opp}}_{\geq 0}$. 
\end{proposition}

\begin{example}
    Coming back to \Cref{Ex:tot-pos-matrix}, the Gauss decomposition of the matrix $M$ there is given by
    $$M=\begin{pmatrix} 22 & 9 & 6\\ 8 & 4 & 3 \\ 1 & 1 & 1\end{pmatrix}=\begin{pmatrix} 1 & 3 & 6\\ 0 & 1 & 3 \\ 0 & 0 & 1\end{pmatrix}\begin{pmatrix} 1 & 0 & 0\\ 5 & 1 & 0 \\ 1 & 1 & 1\end{pmatrix}\; .$$ 
    The diagonal part $D$ is trivial, $U_1$ is totally positive by \Cref{Ex:tot-pos-matrix-U}, and it is easy to check that $U_2$ is also totally positive.
\end{example}

Subsequently, the study of the semigroup $\mathrm{GL}_{\geq 0}$ reduces to the study of its sub-semigroup $U_{\geq 0} \subset \mathrm{GL}_{\geq 0}$ of upper triangular unipotent totally non-negative matrices. The next fundamental reduction theorem for totally non-negative matrices allows one to identify the infinitesimal generators of $U_{\geq 0}$ as the Chevalley generators of the corresponding Lie algebra: 

\begin{theorem}[Loewner--Whitney\index{Loewner-Whitney theorem} \cite{Loewner,Whitney}] 
Any invertible totally non-negative matrix is a product of elementary matrices $x_i(t)$ (see (\ref{eq:elementay_matrix})) with $t>0$.
\end{theorem}

The theorem implies that every non-negative unipotent upper-triangular matrix $u\in U_{\geq 0}$ can be written as a product of totally non-negative matrices $x_i(t)$, where $t\in\R_{\geq0}$ and where $x_i(t)$ is the matrix in $\GL_n(\R)$ with $1$'s on the diagonal, $t$ at the entry $(i,i+1)$ and $0$ elsewhere:
\begin{equation}\labelx{eq:elementay_matrix}
x_i(t)=\left[\begin{array}{cccccc}
1 & \cdots & 0 & 0 & \cdots & 0 \\
\vdots & \ddots & \vdots & \vdots & \ddots & \vdots \\
0 & \cdots & 1 & t & \cdots & 0 \\
0 & \dots & 0 & 1 & \cdots & 0 \\
\vdots & \ddots & \vdots & \vdots & \ddots & \vdots \\
0 & \cdots & 0 & 0 & \cdots & 1
\end{array}\right]\; .
\end{equation}

\begin{example}
    Coming back to \Cref{Ex:tot-pos-matrix-U}, we can find easily the decomposition of the totally positive matrix $U_1$:
    $$U_1=\begin{pmatrix} 1 & 3 & 6\\ 0 & 1 & 3 \\ 0 & 0 & 1\end{pmatrix}=\begin{pmatrix} 1 & 2 & 0\\ 0 & 1 & 0 \\ 0 & 0 & 1\end{pmatrix}\begin{pmatrix} 1 & 0 & 0\\ 0 & 1 & 3 \\ 0 & 0 & 1\end{pmatrix}\begin{pmatrix} 1 & 1 & 0\\ 0 & 1 & 0 \\ 0 & 0 & 1\end{pmatrix}=x_1(2)x_2(3)x_1(1)\; .$$
\end{example}

This property provided the idea for extending the notion of total positivity beyond the group $\mathrm{GL}_n(\mathbb{R})$. Namely, Lusztig in \cite{Lu} defined the set $G_{\geq0}$ of totally non-negative elements in a split real reductive Lie group $G$ as the semigroup generated by the associated Chevalley generators. The main idea involves considering one parameter subgroups obtained by exponentiating the simple root spaces of the Lie algebra and the changes of coordinates are then given by positive rational maps.
The notions of Borel and unipotent subgroups can be generalized accordingly for any split real Lie group, see \Cref{Sec:Levi-flags}. 

The subsemigroup of totally positive elements in a split real group $G$ is also closely related to the notion of positivity for triples in the generalized flag variety $G/B$, where $B$ is the Borel subgroup of $G$. Elements of $G/B$ can be naturally identified with subgroups of $G$ that are conjugated in $G$ to the standard Borel subgroup $B$. The pair of flags $(B,B^{\mathrm{opp}})$, where $B^{\mathrm{opp}}$ is the opposite Borel subgroup of $G$, is called the \emph{standard transverse pair}. Two flags $E$ and $F$ are called \emph{transverse}\index{flag!transverse} if there exists $g\in G$ such that $(gE,gF)=(B,B^{\mathrm{opp}})$.

\begin{definition}
Let $G$ be a split real Lie group. The triple of flags $(E,T,F)$ is said to be \emph{positive}\index{flag!positive triple of} if and only if there exist $g\in G$ and $u\in U_{>0}$ such that $(gE,gT,gF)=(B,uB^{\mathrm{opp}},B^{\mathrm{opp}})$.
\end{definition}

\begin{example}
    For $G=\mathrm{GL}_3(\mathbb{R})$, a flag can be represented as a straight line in $\mathbb{RP}^2$ with a marked point on it. A triple of flags is positive if the triple ratio is.
\end{example}

\section{\texorpdfstring{$\Theta$}{Theta}-positive representations}

The notion of a $\Theta$-positive representation is a generalization of Lusztig's total positivity condition in \cite{Lu} and applies for real semisimple Lie groups which are not necessarily split. At the same time, it generalizes important properties of maximality for Hermitian Lie groups as those were illustrated in \cite{BIW}, while its significance in higher-rank Teichm\"{u}ller theory is already apparent from the results in \cite{GW} or \cite{GLW}. We present the notion here along with detailed examples in the following subsections; for further reference on $\Theta$-positive representations we direct the reader to the original resources  \cite{GW}, \cite{GW22}, \cite{GLW}. Explicit computations are presented in the next sections which can provide a great help to familiarize with the following objects.

Background on Lie theory, including the Cartan subalgebra, the root space, Levi and parabolic subalgebras, can be found in \Cref{app:Lietheory}. The definition of a $\Theta$-positive structure for a real semisimple Lie group $G$ is given pertaining to properties of the Lie algebra of parabolic subgroups $P_{\Theta}<G$ defined by a subset of simple positive roots $\Theta \subset \Delta $. In these terms, let us start by considering a real semisimple Lie group $G$ with finite center and let $\mathfrak{g}$ denote the Lie algebra of $G$. Let $K$ be a maximal compact subgroup of $G$, $\mathfrak{k}$ be its Lie algebra and ${\mathfrak{k}}^\perp$ be the orthogonal complement of $\mathfrak{k}$ with respect to the Killing form. Furthermore, let $\mathfrak{a}$ be a maximal abelian Cartan subspace of ${\mathfrak{k}}^\perp$ and denote by  
$\Sigma = \Sigma(\mathfrak{g},\mathfrak{a})$ the system of restricted roots. Lastly, choose $\Delta \subset \Sigma$ a set of simple roots and let $\Sigma^{+}$ denote the set of positive roots with respect to $\Delta$.  

Now, let $\Theta \subset \Delta $ be any subset of simple roots ($\Theta$ can be empty). Let $\Sigma _{\Theta }^{+}={{\Sigma }^{+}}\setminus \mathrm{Span}\left( \Delta \setminus \Theta  \right)$ and let 
\[ \mathfrak{u}_\Theta = \bigoplus_{\alpha \in \Sigma_\Theta^+} \mathfrak{g}_\alpha,~~~ \mathfrak{u}_\Theta^{\mathrm{opp}} = \bigoplus_{\alpha\in\Sigma_\Theta^+}\mathfrak{g}_{-\alpha} \]
 and 
\[ \mathfrak{l}_\Theta = \mathfrak{g}_0 \oplus \bigoplus_{\alpha\in\Sigma^+\cap\mathrm{Span}(\Delta\setminus\Theta)}\mathfrak{g}_\alpha \oplus\mathfrak{g}_{-\alpha}\; .\]
 As a vector space, $\mathfrak{g}$ splits into Lie subalgebras $\mathfrak{g} = \mathfrak{u}_\Theta^{\mathrm{opp}}\oplus \mathfrak{l}_\Theta \oplus \mathfrak{u}_\Theta$. Moreover, $\mathfrak{l}_\Theta$ acts on $\mathfrak{u}_\Theta$ by the Lie bracket, and we now want to understand this action and its properties. 
 
 The group $G$ acts on $\mathfrak{g}$ by the adjoint representation; let $P_{\Theta}$ be the normalizer in $G$ of $\mathfrak{u}_{\Theta}$. The Lie algebra of $P_\Theta$ is $\mathfrak{p}_\Theta = \mathfrak{l}_\Theta \oplus \mathfrak{u}_\Theta$. Similarly, denote by $P_{\Theta}^{\mathrm{opp}}$ the normalizer of $\mathfrak{u}_{\Theta}^{\mathrm{opp}}$ and $\mathfrak{p}_\Theta^{\mathrm{opp}} = \mathfrak{l}_\Theta \oplus \mathfrak{u}_\Theta^{\mathrm{opp}}$ its Lie algebra. Those are the standard parabolic subgroups associated with $\Theta$, and the corresponding Levi subgroup is $L_{\Theta} = P_{\Theta}\cap P_{\Theta}^{\mathrm{opp}}$. The subgroup $U_{\Theta} = \mathrm{exp}(\mathfrak{u}_{\Theta})$ (resp. $U^{\mathrm{opp}}_{\Theta} = \mathrm{exp}(\mathfrak{u}^{\mathrm{opp}}_{\Theta})$) is the unipotent radical of $P_\Theta$ (resp. $P_\Theta^{\mathrm{opp}}$). The Lie algebra of $L_{\Theta}$ is $\mathfrak{l}_\Theta$, however $L_\Theta$ may not be connected so denote by $L_\Theta^0$ the connected component of the identity in $L_\Theta$. 

In order to understand the action of $L_\Theta$ on $\mathfrak{u}_\Theta$, since $L_\Theta$ is a reductive group, we first need to understand its center and its Cartan subalgebra. First, $\mathfrak{l}_\Theta$ splits into its center $Z(\mathfrak{l}_\Theta)$ and its semisimple part $\mathfrak{l}'_\Theta = \left[ \mathfrak{l}_\Theta,\mathfrak{l}_\Theta\right] $:
\[ \mathfrak{l}_\Theta = Z(\mathfrak{l}_\Theta) \oplus \mathfrak{l}'_\Theta\; .\]
The Lie algebra ${{\mathfrak{z}}_{\Theta }}$ of the subgroup $Z(\mathfrak{l}_\Theta)$ once again splits into two parts: $\mathfrak{z}_\Theta = Z(\mathfrak{l}_\Theta)\cap {\mathfrak{k}}^\perp = Z(\mathfrak{l}_\Theta)\cap \mathfrak{a}$ and $Z(\mathfrak{l}_\Theta)\cap \mathfrak{k}$. 

The Cartan subalgebra of $\mathfrak{l}'_\Theta$ is $\mathfrak{a}_\Theta = \mathfrak{a} \cap \mathfrak{l}'_\Theta$, its restricted root system is the restriction to $\mathfrak{a}_\Theta$ of $\Sigma\cap\mathrm{Span}(\Delta\setminus\Theta)$, a possible choice of simple roots is the restriction to $\mathfrak{a}_\Theta$ of $\Delta\setminus\Theta$, and the corresponding Dynkin diagram is the largest subgraph of the Dynkin diagram of $G$ with vertices $\Delta\setminus\Theta$.

Note that since $\mathfrak{a}\subset\mathfrak{l}_\Theta\cap \mathfrak{k}^\perp$, we have 
\[ \mathfrak{a} = \mathfrak{a}_\Theta\oplus \mathfrak{z}_\Theta\; .\] 
In particular, $\mathfrak{z}_\Theta$ acts on $\mathfrak{u}_\Theta$ by restriction of the action of $\mathfrak{a}$ on $\mathfrak{g}$, so $\mathfrak{u}_\Theta$ decomposes into weight spaces for this action as
\begin{equation}\labelx{eq:uTheta} 
\mathfrak{u}_\Theta = \bigoplus_{\beta\in \mathfrak{z}^{*}_{\Theta}} \mathfrak{u}_\beta\; , 
\end{equation}
where 
\begin{equation}\labelx{eq:ubeta}
\mathfrak{u}_\beta = \left\{ x\in \mathfrak{u}_\Theta \vert \text{ for all } z\in \mathfrak{z}_\Theta, \mathrm{ad}(z).x = \beta(z)x \right\}
\end{equation}
is $L_\Theta^0$-invariant.
    
The inclusion $\mathfrak{z}_\Theta\subset \mathfrak{a}$ induces a projection $i^*:\mathfrak{a}^*\to \mathfrak{z}_{\Theta}^{*}$ given by the restriction of a linear form on $\mathfrak{a}$ to $\mathfrak{z}_\Theta$. Thus we can write $\mathfrak{u}_\beta$ as a sum of weight spaces for the action of $\mathfrak{a}$:
\[ \mathfrak{u}_\beta = \bigoplus_{\alpha\in \Sigma_{\Theta}^{+}, \alpha \vert_{\mathfrak{z}_\Theta} = \beta} \mathfrak{g}_\alpha\; .\]

In fact, one has that $\mathfrak{u}_\beta$ is irreducible:
 	
\begin{theorem}[{\cite[Thm. 0.1]{Kostant}}]\labelx{thm:ubeta irred}
For all $\beta \in \mathfrak{z}_{\Theta}^{*}$, $\mathfrak{u}_\beta$ is an irreducible representation of $L_{\Theta}^{0}$, and 
\[ \left[ \mathfrak{u}_\beta, \mathfrak{u}_{\beta'} \right] = \mathfrak{u}_{\beta+\beta'}\; .\]
 \end{theorem}

Suppose that $\beta$ is the restriction of a root $\alpha\in\Theta$. 
Then $\alpha$ is the only element of $\mathrm{Span}(\Theta)$ which restricts to $\beta$, allowing us the following slight abuse of notation:
\begin{equation}\labelx{eq:ualpha}
\mathfrak{u}_\alpha:=\mathfrak{u}_\beta=\bigoplus_{\gamma\in \mathrm{Span}_{\mathbb{N}}(\Delta\setminus\Theta)}\mathfrak{g}_{\alpha+\gamma}\; .
\end{equation}

\begin{definition}\labelx{def:sharp1}
Let $V$ be a finite dimensional real vector space. A \emph{cone}\index{cone} in $V$ is a subset $C$ which is stable by positive scalar multiplication, that is, for all $x\in C$ and for all $\lambda >0$, $\lambda x\in C$. The cone $C$ is said to be \emph{sharp}\index{cone!sharp} if it contains no affine line.
\end{definition}
 
\begin{remark}The sharpness condition ensures the cone is not in a sense too ``wide'', for instance the half plane $\mathbb{H}^2 \subset \mathbb{R}^2$ is a cone, but not sharp since it contains many horizontal affine lines. We are interested in sharp \emph{convex}\index{cone!convex} cones as they will give an appropriate analog of $\mathbb{R}_{+}^{*}$ in $\mathbb{R}$, which is our first example of sharp convex cone. A second example of a sharp convex cone is the set of positive definite symmetric real matrices in the space of symmetric real matrices.
\end{remark}

We now finally have all the tools and notation in our possession to give the definition of $\Theta$-positivity, unifying and generalizing the notions of positivity we described in detail in Sections \ref{sec:Hitchin_reps} and \ref{sec:Maximal_reps} for the case of split real Lie groups and Hermitian Lie groups of tube type respectively. 
 	
\begin{definition}[Guichard--Wienhard, \cite{GW, GW22}]\labelx{defn_theta_positivity}
 Let $G$ be a real semisimple Lie group with finite center and let $\Theta \subset \Delta$ be a subset of simple roots. Then $G$ is said to admit a \emph{$\Theta$-positive structure}\index{$\Theta$-positive!structure} if for all $\beta\in\Theta$, the action of $L_\Theta^0$ on $\mathfrak{u}_\beta$ preserves a sharp convex cone.
\end{definition}
  
A central result in \cite{GW,GW22} provides that the semisimple Lie groups $G$ admitting a $\Theta$-positive structure are classified as follows: 
\begin{theorem}[Guichard--Wienhard, Thm. 3.4 in \cite{GW22}]\labelx{classification}
A semisimple Lie group $G$ admits a $\Theta$-positive structure if and only if the pair $\left( G, \Theta \right)$ belongs to one of the following four cases:
\begin{enumerate}
\item $G$ is a split real form and $\Theta = \Delta$.
\item $G$ is a Hermitian symmetric Lie group of tube type and $\Theta =\left\{ {{\alpha }_{r}} \right\}$.
\item $G$ is a Lie group locally isomorphic to a group $\mathrm{SO}\left( p,q \right)$, for $p\ne q$, and $\Theta =\left\{ {{\alpha }_{1}},\ldots ,{{\alpha }_{p-1}} \right\}$.
\item $G$ is a real form of the groups ${{F}_{4}}$, ${{E}_{6}}$, ${{E}_{7}}$, ${{E}_{8}}$ with restricted root system of type ${{F}_{4}}$, and $\Theta =\left\{ {{\alpha }_{1}},{{\alpha }_{2}} \right\}$.
\end{enumerate}
\end{theorem}

One may consider positive triples in the generalized flag variety  ${G}/{{{P}_{\Theta }}}\;$ as follows: 

\begin{definition}
Fix ${{E}_{\Theta }}$ and  ${{F}_{\Theta }}$ to be the standard flags in ${G}/{{{P}_{\Theta }}}\;$ such that $\mathrm{Stab}_{G}\left( {{F}_{\Theta }} \right)={{P}_{\Theta }}$ and $\mathrm{Stab}_{G}\left( {{E}_{\Theta }} \right)=P_{\Theta }^{\mathrm{opp}}$. For any  ${{S}_{\Theta }}\in {G}/{{{P}_{\Theta }}}\;$ transverse to  ${{F}_{\Theta }}$, there exists ${{u}_{{{S}_{\Theta }}}}\subset {{U}_{\Theta }}$ such that  ${{S}_{\Theta }}={{u}_{{{S}_{\Theta }}}}{{E}_{\Theta }}$. The triple  $\left( {{E}_{\Theta }},{{S}_{\Theta }},{{F}_{\Theta }} \right)$ in the generalized flag variety  ${G}/{{{P}_{\Theta }}}\;$ will be called \emph{$\Theta $-positive}\index{flag!$\Theta$-positive triple of }, if  ${{u}_{{{S}_{\Theta }}}}\in U_{\Theta }^{>0}$, for  $U_{\Theta }^{>0}$ the $\Theta $-positive semigroup of  ${{U}_{\Theta }}$ defined in a similar way as the positive semigroup $U_{\geq 0}$ from Lusztig's definition of the positive elements in the case of a split real group $G$ (see Section 4 in \cite{GW22} for the precise notion). 
\end{definition}

The definition of a $\Theta$-positive fundamental group representation is now the following: 

\begin{definition}
Let $S_{g}$ be a closed connected and oriented topological surface of genus $g\ge 2$ and let $G$ be a semisimple Lie group admitting a $\Theta $-positive structure. A representation of the fundamental group of  $S_g$ into $G$ will be called \emph{$\Theta$-positive}\index{$\Theta$-positive!representation}\index{representation!$\Theta$-positive}, if there exists a  $\rho $-equivariant positive map  $\xi\colon \partial {{\pi }_{1}}\left( S_g \right)=\mathbb{R}{{\mathbb{P}}^{1}}\to {G}/{{{P}_{\Theta }}}\;$ sending positive triples in  $\mathbb{R}{{\mathbb{P}}^{1}}$ to  $\Theta $-positive triples in  ${G}/{{{P}_{\Theta }}}\;$. 
\end{definition}

In \cite{GLW}, Guichard, Labourie and Wienhard show that $\Theta$-positive representations are $\Theta$-Anosov, thus discrete and faithful, and that, in fact, for the four families of semisimple Lie groups $G$ listed in \Cref{classification} above, there are higher-rank Teichm\"{u}ller spaces in the character variety:

\begin{theorem}[Guichard--Labourie--Wienhard, Thm. A in \cite{GLW}]\labelx{GLW_conjecture}
Let $G$ be a semisimple Lie group that admits a $\Theta$-positive structure. Then there exists a connected component of the representation variety $\mathcal{R}(S_g,G)$ that consists solely of discrete and faithful representations.
\end{theorem}

\section{\texorpdfstring{$\Theta$}{Theta}-positivity in \texorpdfstring{$G=\mathrm{SL}_n(\mathbb{R})$}{G = SL(n,R)}}
	
We now exhibit the $\Theta$-positive structure in detail in the case of the group $\mathrm{SL}_n(\mathbb{R})$. Its Lie algebra is the set of traceless matrices.
	
Following the general notation introduced earlier, take now  $G=\mathrm{SL}_n(\mathbb{R})$, $K=\mathrm{SO}_n(\mathbb{R})$ a maximal compact subgroup of $G$ with Lie algebra $\mathfrak{k}$, the set of skew-symmetric matrices. The orthogonal complement  ${\mathfrak{k}}^{\perp}$ of $\mathfrak{k}$ with respect to the Killing form is the set of traceless symmetric matrices, and a maximal abelian Cartan subspace $\mathfrak{a}$ of ${\mathfrak{k}}^{\perp}$ is the set of traceless diagonal matrices. The restricted root system $\Sigma = \Sigma(\mathfrak{g},\mathfrak{a})$ is in this case 
$$\Sigma = \left\{ \alpha_{ij} := \epsilon_i-\epsilon_j \mid i,j \in \left\{ 1,\dots,n\right\}, i\neq j \right\}\; ,$$
where $\epsilon_i :(a_{pq})_{p,q \in \left\{ 1,\dots,n\right\}} \mapsto a_{ii}$. The standard choice of simple roots $\Delta\subset \Sigma$  is $\Delta = \{\alpha_{i, i+1}\}_{i\in\left\{ 1,\dots,n-1\right\} }$, and the set of positive roots with respect to $\Delta$ is $\Sigma^+ = \{\alpha_{ij}\}_{i<j}$. The weight space $\mathfrak{g}_{\alpha_{ij}}$ is the $1$-dimensional subspace of $\mathfrak{sl}_n(\R)$ spanned by $E_{ij}$.
 
\newcommand{\UP}[2]{\makebox[0pt]{\smash{\raisebox{1.5em}{$\phantom{#2}#1$}}}#2}
\newcommand{\LF}[1]{\makebox[0pt]{$#1$\hspace{4.5em}}}
\renewcommand{\arraystretch}{1.5}
\begin{example}
Let us first describe a choice of $\Theta\subset\Delta$ that does not yield a $\Theta$-positive structure, but will allow us to better understand the different Lie algebras associated to $\Theta$. We choose $\Theta = \left\{\alpha_{i, i+1}, \alpha_{j, j+1}\right\} $ with $1\leq i<j<n$. Then the Lie algebra $\mathfrak{sl}_n(\mathbb{R})$ admits the following block decomposition
\vspace{1em}
\[
 			\left(
 			\begin{array}{c@{}ccc|ccc|ccc}
 				           &  \hspace{1ex}*\hspace{1ex}  &  \dots  &  \UP{i}{\hspace{1em}*\hspace{1em}}  &  \UP{i+1}{\hspace{1em}*\hspace{1em}}  &  \dots  &  \UP{j}{\hspace{1em}*\hspace{1em}}  &  \UP{j+1}{\hspace{1em}*\hspace{1em}}  &  \dots  &  \hspace{1ex}*\hspace{1ex}  \\
 					       & \vdots & \ddots & \vdots & \vdots & \ddots & \vdots & \vdots & \ddots & \vdots \\ 
 				\LF{i} & *    &  \dots &   *               & *                     &\dots     &*                  &*                      &\dots     &* \\ \hline
 				 \LF{i+1}          &  *  &  \dots  &  *  & *  &  \dots  &  *  & *  &  \dots  &  *  \\
 				           & \vdots & \ddots & \vdots & \vdots & \ddots & \vdots & \vdots & \ddots & \vdots \\
 				\LF{j} & *    &  \dots &   *               & *                     &\dots     &*                  &*                      &\dots     &* \\ \hline
 				 \LF{j+1}  &  *  &  \dots  &  *  & * &  \dots  &  *  & *  &  \dots  &  *  \\
 				           & \vdots & \ddots & \vdots & \vdots & \ddots & \vdots & \vdots & \ddots & \vdots \\
 			               & *    &  \dots &   *               & *                     &\dots     &*                  &*                      &\dots     &* \\
 			\end{array}\right)\;  ~~\text{ with }\mathrm{Tr} = 0\; .\]
 		\renewcommand{\arraystretch}{1}

In the sequel we shall refer to this block decomposition when speaking about matrices defined block-wise. The Levi subalgebra $\mathfrak{l}_\Theta$ is the set of traceless block-diagonal matrices, the algebra $\mathfrak{u}_\Theta$ is the set of block strictly upper triangular matrices and $\mathfrak{p}_\Theta=\mathfrak{l}_\Theta \oplus \mathfrak{u}_\Theta$ is the set of traceless block upper triangular matrices:
\vspace{1pt}
 		\[
 		\renewcommand{\arraystretch}{1}
 		\mathfrak{l}_\Theta = \left\{ \left(
 		\begin{array}{c|c|c}
 			*& & \\ \hline
 			 & *& \\ \hline
 			  & & *
 		\end{array}\right), \mathrm{Tr} = 0\right\} , ~~
 	\mathfrak{u}_\Theta = \left\{\left(
 	\begin{array}{c|c|c}
 	\;\,& *&* \\ \hline
 	& & *\\ \hline
 	& & 
	 \end{array}\right), \mathrm{Tr} = 0\right\}\; ,\]
  \[
	\mathfrak{p}_\Theta = \left\lbrace \left(
	\begin{array}{c|c|c}
	*& *&* \\ \hline
	& *& *\\ \hline
	& & *
	\end{array}\right), \mathrm{Tr} = 0\right\} 
 		\renewcommand{\arraystretch}{1}\; .
 		\]
 		The parabolic subgroup $P_\Theta$ is the stabilizer of the standard (partial) flag of $\mathbb{R}^n$ with subspaces of dimensions $(0,i,j,n)$.

For our choice of simple roots $\Theta = \left\{\alpha_{i, i+1}, \alpha_{j, j+1}\right\} $ with $1\leq i<j<n$, the center $Z(\mathfrak{l}_\Theta)$ of $\mathfrak{l}_\Theta$ is the set of matrices of the form 
\vspace{1pt}\[  \left(
 		\begin{array}{c|c|c}
 			\lambda I_i& & \\ \hline
 			& \mu I_{j-i}& \\ \hline
 			& & \nu I_{n-j}
 		\end{array}\right)\]
   \vspace{1pt}with $i\lambda+(j-i)\mu + (n-j)\nu = 0$. The semisimple part of $\mathfrak{l}_\Theta$ is then the set of block-diagonal matrices for which every block is traceless. In this situation, $Z(\mathfrak{l}_\Theta)$ is entirely contained in $\mathfrak{a}$, so $\mathfrak{z}_\Theta = Z(\mathfrak{l}_\Theta)$. The weight spaces $\mathfrak{u}_\beta$ are the block-upper triangular matrices in $\mathfrak{u}_\Theta$ for which only one of the three blocks is non-zero, and the $\mathfrak{u}_\beta$ with $\beta\in\Theta$ are the blocks just above the diagonal. The third $\mathfrak{u}_\beta$ (whose only non-zero block lies at the top right of the matrix) is associated to the restriction of the root $\alpha_{i,j+1} = \alpha_{i, i+1} + \dots +\alpha_{j, j+1}$ to $\mathfrak{z}_\Theta$, and all the roots $\alpha_{i+1, i+2},\dots,\alpha_{j-1, j}$ vanish on $\mathfrak{z}_\Theta$. The group $L^0_\Theta$ is the set of matrices of the form
   \vspace{1pt}\[  \left(
 		\begin{array}{c|c|c}
 			\lambda A& & \\ \hline
 			& \mu B& \\ \hline
 			& & \nu C
 		\end{array}\right)\]
    with $A,B,C$ invertible matrices of positive determinant.
   In this case, the action of $L^0_\Theta$ on $\mathfrak{u}_\beta$ does not preserve any sharp convex cone.
\end{example}

We now demonstrate how total positivity is a case of the more general notion of $\Theta$-positivity for a certain choice of simple roots. 
For the group $G=\mathrm{SL}_n(\mathbb{R})$, choose $\Theta = \Delta$ (with the usual choice of $\Delta$ for $\mathrm{SL}_n(\mathbb{R})$). The Levi subgroup $L_\Theta$ is the subgroup of diagonal matrices of determinant 1, and all weight spaces associated to elements of $\Theta$ are one-dimensional: 
\[ \text{ for all } \beta = \alpha_{i, i+1}\in \Theta, \mathfrak{u}_\beta = \mathfrak{g}_\beta = \left\langle E_{i,i+1}\right\rangle\; .\]
In Lusztig's total positivity, the semigroup of positive elements of a split real group is generated by elements of the form $u_i(t) = I_n + tE_{i,i+1} = \mathrm{exp}(tE_{i,i+1})$, with $t>0$. Here the sharp convex cone in $\mathfrak{u}_\beta = \mathfrak{g}_{\alpha_{i,i+1}}$ is the cone of positive scalar multiples of $E_{i,i+1}$.

\section{\texorpdfstring{$\Theta$}{Theta}-positivity in \texorpdfstring{$G=\mathrm{Sp}_{2n}(\mathbb{R})$}{G = Sp(2n,R)}}\labelx{sec:HLG_positivity}

For Hermitian Lie groups of tube type there is another $\Theta$-positive structure. We introduce this structure for the real symplectic group $\mathrm{Sp}_{2n}(\mathbb{R})$ keeping in mind that the construction for other Hermitian Lie groups of tube type is very similar.

In this case, the notion of $\Theta$-positivity reduces to the notion introduced and studied in \cite{BILW,BIW}. Choose simple roots $\Theta = \left\{ \alpha_n\right\}$, where $\alpha_n$ is the root at the end of the Dynkin diagram with two arrows pointing out. There is only one space $\mathfrak{u}_\beta$, namely 
$$\mathfrak{u}_{\alpha_n}=\left\{ \begin{pmatrix}0 & B \\ 0 & 0\end{pmatrix} \mid B\in\mathrm{Sym}_n(\mathbb R) \right\}\; .$$
Further, $\mathfrak l_{\alpha_n}$ and $\mathfrak p_{\alpha_n}$ agree with the Lie algebra of the Levi subgroup $L$ and of the parabolic subgroup $P$ respectively from \Cref{sec:lagr}, and are explicitly given by 
$$\mathfrak l_{\alpha_n}=\left\{ \begin{pmatrix}A & 0 \\ 0 & -A^t\end{pmatrix} \mid A\in\mathrm{Mat}_n(\mathbb R) \right\}, \text{ and }$$
$$\mathfrak p_{\alpha_n}=\left\{ \begin{pmatrix}A & B \\ 0 & -A^t\end{pmatrix} \mid A\in\mathrm{Mat}_n(\mathbb R),\;B\in\mathrm{Sym}_n(\mathbb R) \right\}\; .$$
There is an open sharp convex cone
$$\mathfrak{u}^+_{\alpha_n}=\left\{ \begin{pmatrix}0 & B \\ 0 & 0\end{pmatrix} \mid B\in\mathrm{Sym}^+_n(\mathbb R) \right\}\subset \mathfrak{u}_{\alpha_n}\; ,$$
which is invariant under the action of $L$ by conjugation. So it defines a positive structure on $\mathrm{Sp}_{2n}(\mathbb{R})$. Notice that, if $n\geq 2$, this cone is not 1-dimensional, so this positive structure is different from the totally positive structure arising from the fact that $\mathrm{Sp}_{2n}(\mathbb{R})$ is a split real group. If $n=1$, then $\mathrm{Sp}_{2}(\mathbb R)=\mathrm{SL}_{2}(\mathbb R)$ and this notion of $\Theta$-positivity coincides with the total positivity discussed in the previous section.

\newpage
\part{Spectral networks in Geometry}\labelx{partIV}

\vspace{2cm}

We first motivate the introduction of spectral networks as a geometric tool for studying spaces of fundamental group representations, defining them in a purely topological and combinatorial manner. Alongside this, we present the necessary concepts for a comprehensive study of spectral networks. Next, we employ holomorphic differentials to approach spectral networks from a holomorphic and analytic perspective, discussing the (non-)abelianization of local systems. We also include several purely mathematical applications of spectral networks that have emerged in the literature, particularly the connection between spectral networks and the WKB method for solving differential equations on Riemann surfaces.

\vspace{2cm}

\parttoc


\chapter{Non-degenerate spectral networks}


\abstract{The introduction of spectral networks is motivated from the point of view of the construction of a non-abelianization map. 
Then spectral networks on ciliated surfaces are defined. We start with the simplest case of \textit{small spectral networks} and then give a more general definition of non-degenerate spectral networks. Two different approaches to the construction of spectral networks are presented: the first is purely topological and combinatorial and works for ciliated surfaces $S$ with punctures without any additional structure. The second, called WKB-spectral network, is rather analytic, with a choice of complex structure on $S$. Not all general spectral networks can be realized using this analytic construction. Further, we describe a path lifting rule using spectral networks which generalizes the usual path lifting property for non-ramified coverings, and which is homotopy invariant for ramified coverings.}

\section{Towards a non-abelianization map}

Let $S=S_{g,\vec{p}}$ be a ciliated surface. We consider the space of fundamental group representations 
$$\mathcal{E}(S,\mathrm{GL}_n(\mathbb C)):=\mathrm{Hom}(\pi_1(S),\mathrm{GL}_n(\mathbb{C}))/\mathrm{GL}_n(\mathbb{C})\; ,$$
where $\mathrm{GL}_n(\mathbb{C})$ acts by conjugation on the space of homomorphisms.

\begin{example}\labelx{ex:line_bdle}
In the simplest, non-trivial case $n=1$, the group $\mathrm{GL}_1(\mathbb{C})=\mathbb{C}^\times$ is abelian, and the action of $\mathrm{GL}_1(\mathbb{C})$ by conjugation on the space of homomorphisms is trivial. We obtain:
$$\mathcal{E}(S,\mathrm{GL}_1(\mathbb{C})):=\mathrm{Hom}(\pi_1(S),\mathbb{C}^\times)\; .$$
Since $\mathbb{C}^\times$ is abelian, any homomorphism $\rho\in \mathrm{Hom}(\pi_1(S),\mathbb{C}^\times)$ sends all commutators in $\pi_1(S)$ to the identity, i.e. it factors through the first homology group $\mathrm{H}_1(S,\mathbb{Z})=\pi_1(S)/[\pi_1(S),\pi_1(S)]$:
$$\mathrm{Hom}(\pi_1(S),\mathbb{C}^\times)=\mathrm{Hom}(\mathrm{H}_1(S,\mathbb{Z}),\mathbb{C}^\times)\; .$$
Since $\mathrm{H}_1(S,\mathbb{Z})$ is a free module over $\mathbb{Z}$, there is a basis $\mathbf{e}:=(e_1,\dots,e_s)$ of $\mathrm{H}_1(S,\mathbb{Z})$. Then the map
$$\begin{matrix}
\mathrm{Hom}(\mathrm{H}_1(S,\mathbb{Z}),\mathbb{C}^\times) & \to & (\mathbb{C}^\times)^s \\
\rho & \mapsto & (\rho(e_1),\dots,\rho(e_s))
\end{matrix}$$
provides a homeomorphism between the two spaces.
\end{example}

The example above illustrates that if $G=\mathbb{C}^\times$ (or more generally any abelian group), then the topology of the space $\mathcal{E}(S,G)$ is easy to understand. In order to study the general case when $G$ is not abelian, we want to reduce it to the abelian case. More precisely, we are going to find another  surface $\Sigma$ and a map
$$\mathcal{E}(\Sigma,\mathbb{C}^\times)\to\mathcal{E}(S,\mathrm{GL}_n(\mathbb{C}))$$
that is ``good enough'' to understand the space $\mathcal{E}(S,\mathrm{GL}_n(\mathbb{C}))$, e.g. a map that is generically finite-to-one with an open and dense image. We call any such map a \emph{non-abelianization map}\index{non-abelianization!map}. We will also often consider maps that are locally inverse to non-abelianization maps. Such maps are called \emph{abelianization maps}\index{abelianization!map} $$\mathcal{E}(S,\mathrm{GL}_n(\mathbb{C}))\to\mathcal{E}(\Sigma,\mathbb{C}^\times)\; .$$

Let us consider an example of a map that tends to be a non-abelianization map, but is not ``good enough'' in the sense described above.
\begin{example}\labelx{unbranched_example}
As we have seen in \Cref{Riemann-Hilbert correspondence}, the space $\mathcal{E}(S,\mathrm{GL}_n(\mathbb{C}))$ can be identified with the set of equivalence classes of flat $\mathbb{C}$-vector bundles of rank $n$.

\noindent Let $\pi\colon\Sigma\to S$ be an (unramified) $n:1$-covering map. This map induces a map:
$$\begin{matrix}
\pi_*\colon & \mathcal{E}(\Sigma,\mathbb{C}^\times) & \to & \mathcal{E}(S,\mathrm{GL}_n(\mathbb{C}))\\
& (L,\nabla) & \mapsto & (\pi_*L,\pi_*\nabla)
\end{matrix}$$
where $L$ is a $\mathbb{C}$-bundle of rank $1$ over $\Sigma$ with a flat connection $\nabla$ and $\pi_*L$ is a rank $n$ bundle over $S$ equipped with the flat connection $\pi_*\nabla$ that we are going to describe below.

\noindent  For a point $z\in S$, let $\pi^{-1}(z)=\{z^{(1)},\dots,z^{(n)}\}$ be its pre-image. We define a fiber of the bundle $\pi_*L$ over the point $z\in S$ as $$(\pi_*L)_z:=\oplus_{i=1}^n L_{z^{(i)}}\; ,$$ where  $L_{z^{(i)}}$ is the fiber of $L$ over $z^{(i)}\in\Sigma$.

\noindent  Let now $\gamma\colon [0,1]\to S$ be a path in $S$ starting at $z$, i.e. $\gamma(0)=z$. To define a connection on $S$, we need to describe the parallel transport along $\gamma$. For this, we lift $\gamma$ to $\Sigma$. There are precisely $n$ lifts $\gamma^{(1)},\dots,\gamma^{(n)}$ such that for every $i\in\{1,\dots n\}$, $\gamma^{(i)}(0)=z^{(i)}$.

\noindent  The parallel transport along every $\gamma^{(i)}$ is defined by $\nabla$, so we obtain $n$ linear maps $P_{\gamma^{(i)}}\colon L_{z^{(i)}}\to L_{\gamma^{(i)}(1)}$. We define the parallel transport along $\gamma$ as follows:
$$\begin{matrix}
P_\gamma:=\oplus_{i=1}^n P_{\gamma^{(i)}}\colon & \oplus_{i=1}^n L_{z^{(i)}} & \to & \oplus_{i=1}^n L_{\gamma^{(i)}(1)}\\
& (v_1,\dots, v_n) & \mapsto & (P_{\gamma^{(1)}}v_1,\dots, P_{\gamma^{(n)}}v_n)\; .
\end{matrix}$$
This map defines a connection $\pi_*\nabla$ on $\pi_*L$ that is automatically flat since $\nabla$ is flat (exercise).
\end{example}

The map $\pi_*$, considered more generally for fundamental group representations which are not necessarily reductive, could be a good candidate for a non-abelianization map, but a more detailed calculation shows that the complex dimension of the space $\mathcal{E}(\Sigma,\mathbb{C}^\times)$ is $n(2g-2)$ where $g$ is the genus of $S$, while the dimension of $\mathcal{E}(S,\mathrm{GL}_n(\mathbb{C}))$ is $n^2(2g-2)$. So $\pi_*$ has no chance to have an open image. However, this example gives an idea on how non-abelianization maps can be constructed. In the next section, we will see that if we use ramified coverings instead of unramified ones, the non-abelianization map can be indeed constructed. This will lead to the notion of a spectral network.

\section{Path algebra}\labelx{Sec:pathalgebra}

As announced in the previous section, we will use ramified coverings to construct non-abelianization maps. The problem with ramified coverings is that the standard path lifting rule for them is not homotopy invariant. We will illustrate this problem in the next example:

\begin{example}
Recall the example given in \Cref{rem:ram_path_lifting} which illustrated that the standard path lifting rule is not homotopy invariant for ramified coverings. This example also illustrates that if we try to push forward a flat connection from the ramified covering $\Sigma$ to $S$ outside of the branch locus $B\subset S$ as in \Cref{unbranched_example}, the connection obtained on $S\setminus B$ will not extend to a flat connection on $S$ since it will have non-trivial monodromies around the branch points.

\end{example}
We will modify the construction from \Cref{unbranched_example} so that it applies for ramified coverings. For this, we need the notion of a \emph{path algebra} on surfaces.

\begin{definition}
The \emph{space of paths}\index{path!space of} on a surface $S$ is defined as
$$\mathcal P(S):=\{\gamma\colon [0,1]\to S\;\text{continuous}\}/\sim \; ,$$ 
where two paths $\gamma_0$, $\gamma_1$ are equivalent (denoted by $\gamma_0\sim\gamma_1$), if $\gamma_0$ and $\gamma_1$ are homotopic relative to their endpoints, that is, $\gamma_0(0)=\gamma_1(0)$, $\gamma_0(1)=\gamma_1(1)$ and there exists a continuous map $H\colon [0,1]\times [0,1]\to S$ such that 
\begin{itemize}
    \item $H(0,t)=\gamma_0(t)$, $H(1,t)=\gamma_1(t)$ for all $t\in[0,1]$,
    \item $H(s,0)=\gamma_0(0)=\gamma_1(0)$, $H(s,1)=\gamma_0(1)=\gamma_1(1)$ for all $s\in[0,1]$.
\end{itemize}
\end{definition}

\begin{definition}
The \emph{path algebra}\index{path!algebra} on $S$ is the free abelian group $(\mathbb{Z}[\mathcal P(S)],+)$ over $\mathcal P(S)$ endowed with the multiplication $\cdot$ defined as follows (\Cref{fig:concatination}): $\gamma_2\cdot\gamma_1=0$ if $\gamma_1(1)\neq\gamma_2(0)$ and $\gamma_2\cdot\gamma_1$ is the concatenation of $\gamma_1$ and $\gamma_2$ otherwise, i.e. $\gamma_2\cdot\gamma_1=\gamma$ where
$$\gamma(t) =
\begin{cases}
\gamma_1(2t), & \text{if } t\in[0,\frac{1}{2}]\; ;\\
\gamma_2(2t-1), & \text{if } t\in[\frac{1}{2},1]\; .
\end{cases}$$
\end{definition}
\begin{figure}[h]
\centering
\includegraphics[scale=1]{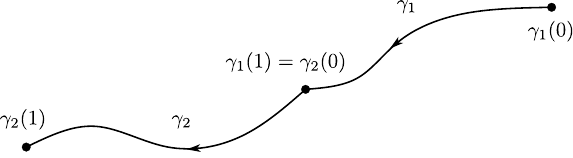}
\caption{Multiplication in the path algebra.}
\labelx{fig:concatination}
\end{figure}
The set $\mathcal P(S)$ naturally embeds into $\mathbb{Z}[\mathcal P(S)]$.

Let $\pi\colon\Sigma\to S$ be a ramified covering. We denote by $B'$ the set of ramification points in $\Sigma$ and $B:=\pi(B')$. Let $b\in B'$, $D$ be an open contractible neighbourhood of $b$ that does not contain other ramification points, and let $z\in D\setminus\{b\}$. Let $\alpha_z^b\colon [0,1]\to D\setminus\{b\}$ be a loop at $z$ which goes once around $b$ and which is contractible in $D$. The loop $\alpha_z^b$ is not unique as a path but all such $\alpha_z^b$ are homotopic, i.e. all such $\alpha_z^b$ define the same element of $\mathcal P(\Sigma\setminus B')$ that we, slightly abusing the notation, denote again by $\alpha_z^b$. Further, let $e_z\colon [0,1]\to D\setminus\{b\}$ be a loop at $z$ which is contractible in $D\setminus\{b\}$. Similarly, the loop $e_z$ is not unique as a path but all such $e_z$ are homotopic, i.e. all such $e_z$ define the same element of $\mathcal P(\Sigma\setminus B')$ that we denote again by $e_z$.

The element $e_z$ can be interpreted as ``the identity element at the point $z$'', i.e. for every element $\gamma\in\mathcal P(\Sigma\setminus B')$ such that $\gamma(0)=z$, then $\gamma\cdot e_z=\gamma$.

We define the ideal $\mathcal I$ that is generated by all $\alpha_z^b+e_z$, when $z\in \Sigma$ and $b\in B'$.
\begin{definition}
The \emph{twisted path algebra}\index{path!algebra!twisted} is the quotient algebra: $$\mathbb{Z}^{\text{tw}}_{B'}[\mathcal P(\Sigma)]=\mathbb{Z}[\mathcal P(\Sigma\setminus B')]/\mathcal I\; .$$
\end{definition}

We will write elements of the twisted path algebra $\mathbb{Z}^{\text{tw}}_{B'}[\mathcal P(\Sigma)]$ as elements of $\mathbb{Z}[\mathcal P(\Sigma\setminus B')]$ keeping in mind that we have to identify some elements. For example, in $\mathbb{Z}^{\text{tw}}_{B'}[\mathcal P(\Sigma)]$ we have the equality $\alpha_z^b=-e_z$.

\section{Small spectral networks}\labelx{sec:small_SN}

Let $\bar{S}$ be a compact surface with a finite set of marked points  $P=\{s_1,...,s_n\}$, for $n\in \mathbb{Z}_{>0}$. Let $S=\bar S\setminus P$ be the corresponding ciliated surface. 

Let $\pi\colon \bar\Sigma\to \bar S$ be a ramified $k$-fold covering of $\bar S$. We also consider the corresponding restriction $\pi\colon \Sigma\to S$ where $\Sigma=\pi^{-1}(S)=\bar\Sigma\setminus P'$ and $P'=\pi^{-1}(P)$ is the set of punctures of $\Sigma$. We assume that the covering is not ramified over the punctures. As before, we denote by $B'$ the set of ramification points of $\Sigma$, and $B=\pi(B')\subset S$.

We assume that every branch point is \emph{simple}. That means that every $b\in B$ has an open neighborhood $U$ such that
$$\pi^{-1}(U)=\coprod_{i=1}^{k-1}V_i\; ,$$
where $V_i$ are open neighborhoods of one of the pre-images of $b$ in $\Sigma$, $\pi\vert_{V_i}\colon V_i\to U$ is a homeomorphism for $i=1,...,k-2$, and the map $\pi\vert_{V_{k-1}}\colon V_{k-1}\to U$ is up to isotopy the map $z\mapsto z^2$.

We also assume that for each $s\in P$ an order on $\pi^{-1}(s)=\{s_i^{(1)},...,s_i^{(k)}\}$ is given by $s_i^{(1)}<s_i^{(2)}<...<s_i^{(k)}$. This order on the fiber over $s$ is an additional structure on the covering.

Now we are ready to define small spectral networks, which are a simplified but still useful version of general spectral networks. We follow the definition from \cite{Alessandrini_vid}, which is equivalent to the original definition by Gaiotto--Moore--Neitzke in \cite{Gaiotto:2012rg}.

\begin{definition}\labelx{ssn}
A \emph{small spectral network}\index{spectral network!small}\index{spectral!network} $\mathcal{W}$ of rank $k$ associated to the covering $\pi$ is a graph on $\bar\Sigma$, i.e. a finite collection of smooth simple oriented paths $\{p_h\colon [-1,1]\to\bar\Sigma\}_{h\in H}$, where $H$ is a finite index set (\Cref{sn1}), that may intersect only transversely, and satisfy the following conditions:
\begin{enumerate}
\item $\pi(p_h(t))=\pi(p_h(-t))$ for all $t\in[-1,1]$;
\item $p_h(0)$ is a ramification point on $\Sigma$;
\item $\pi(p_h(1))=\pi(p_h(-1))=s_j\in P$ (for some $j$) is a puncture and $p_h(-1)<p_h(1)$ with respect to the order on $\pi^{-1}(s_j)$;
\begin{figure}[!ht]
\centering
\includegraphics[scale=1]{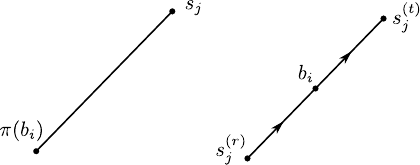}
\caption{Line of spectral network: in $\bar S$ (left), in $\bar\Sigma$ (right).}
\labelx{sn1}
\end{figure}

\item Every branch point in $S$ and every ramification point in $\Sigma$ has a neighborhood as in \Cref{sn2}.
\begin{figure}[!ht]
\centering
\includegraphics[scale=1]{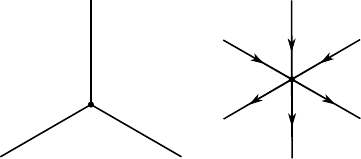}
\caption{A neighborhood of a branch point in $S$ (left), and of a ramification point in $\Sigma$ (right).}
\labelx{sn2}
\end{figure}

\item Different paths $p_h$ and $p_{h'}$ can only intersect at ramification points, at punctures or intersect as shown in \Cref{Fig:def-small-spectral-network-crossing}. In the last case, when the intersection is at $p_h(t)=p_{h'}(t')$ with $t,t'\in (-1,1)\setminus\{0\}$, we require that no line of $\mathcal{W}$ goes through the opposite points $p_h(-t)$ or $p_{h'}(-t')$.
\begin{figure}[!ht]
\centering
\includegraphics[scale=0.83]{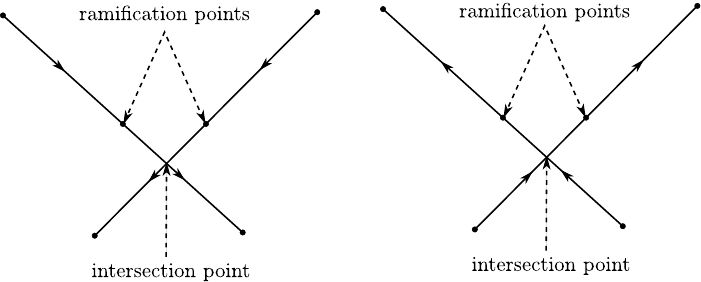}
\caption{Intersection of lines of spectral network. The picture in $\bar\Sigma$.}
\labelx{Fig:def-small-spectral-network-crossing}
\end{figure}
\end{enumerate}

We denote by $$\mathcal{W}_\Sigma:=\bigcup_{h\in H}p_h((-1,1))\subseteq \Sigma$$
the spectral network on $\Sigma$ and by $$\mathcal{W}_S:=\pi(\mathcal{W}_\Sigma)\subseteq S$$ the spectral network on ${S}$.
\end{definition}

\begin{remark}
    When there is no ambiguity, we will use $\mathcal{W}$ to denote the spectral network on either $S$ or $\Sigma$.
\end{remark}

\begin{remark}
\labelx{small}
For small spectral networks, the intersection of paths as in \Cref{fig:joint} (called \emph{joint}\index{spectral network!joint}) is forbidden.
\begin{figure}[!ht]
\centering
\includegraphics[scale=1]{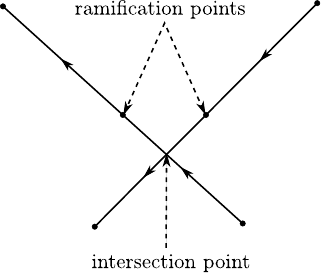}
\caption{Forbidden path intersection. The picture in $\bar\Sigma$.}
\labelx{fig:joint}
\end{figure}
We will define in \Cref{sec:non-degen_SN} a more general class of spectral networks in which this type of intersection is allowed.
\end{remark}

\begin{proposition}\labelx{n_joint}
The intersection described in \Cref{small} and in \Cref{ssn}~(5) cannot occur in spectral networks of rank 2.
\end{proposition}

\begin{proof}
This fact follows directly from the axiom (5) in \Cref{ssn}.
\begin{figure}[!ht]
\centering
\includegraphics[scale=1]{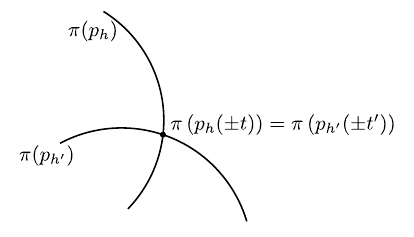}
\caption{The picture in ${S}$.}
\labelx{sn10}
\end{figure}
If we assume that there are two paths $p_h$ and $p_{h'}$ such that there exist $t,t'\in (0,1)\setminus\{0\}$ and
$$\pi(p_h(t))=\pi(p_h(-t))=\pi(p_{h'}(t'))=\pi(p_{h'}(t'))\; ,$$
then because $\pi\colon\Sigma\to S$ has exactly two sheets, there are only two possibilities: $$p_h(t)=p_{h'}(t')\text{ and }p_h(-t)=p_{h'}(-t')$$ or $$p_h(t)=p_{h'}(-t')\text{ and } p_h(-t)=p_{h'}(t')\; ,$$ see \Cref{sn10} and \Cref{sn11}.
\begin{figure}[!ht]
\centering
\includegraphics[scale=1]{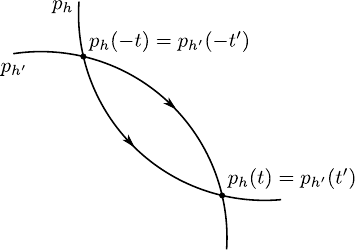}\;\;
\includegraphics[scale=1]{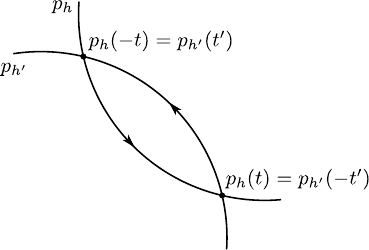}
\caption{The picture in ${\Sigma}$.}
\labelx{sn11}
\end{figure}
Both of them are prohibited by axiom (5) of \Cref{ssn}.
\end{proof}

\begin{remark}
From \Cref{n_joint}, it follows that all spectral networks of rank 2 are small.
\end{remark}

\section{Non-degenerate spectral networks}\labelx{sec:non-degen_SN}

As we have seen in \Cref{small}, there are intersections called \emph{joints} of lines of a spectral network that are forbidden for small spectral networks. To account for this kind of intersection, we need to add additional paths to the spectral network every time such an intersection occurs.

Let as before $\pi\colon\bar\Sigma\to \bar S$ be a ramified $k$-fold covering. We assume $k\geq 3$ because, as we have seen, joints cannot occur for spectral networks of rank $2$. 
\begin{definition}\labelx{ndsn}
A \emph{non-degenerate spectral network}\index{spectral network!non-degenerate} $\mathcal{W}$ of rank $k$ associated to $\pi$ is consisted of two finite collections of smooth simple oriented paths on $\Sigma$: $\{p_h\colon [-1,1]\to\bar\Sigma\}_{h\in H_1}$ and $\{p_h\colon [-1,1]\setminus\{0\}\to\bar\Sigma\}_{h\in H_2}$, where $H_1$ and $H_2$ are two disjoint index sets. The paths may intersect only transversely and satisfy the following conditions:
\begin{enumerate}
\item for all $h\in H_1\cup H_2$ and for all $t\in(0,1]$, $\pi(p_h(t))=\pi(p_h(-t))$;

\item for all $h\in H_1$, $\pi(p_h(0))\in B$ is a branch point;

\item for all $h\in H_2$, there exist limits: $p_h(+0):=\lim_{t\to +0}p_h(t)$ and $p_h(-0):=\lim_{t\to -0}p_h(t)$. Further, $p_h(+0)\neq p_h(-0)$ but $\pi(p_h(+0))=\pi(p_h(-0))$;

\item for all $h\in H_1\cup H_2$, $\pi(p_h(1))=\pi(p_h(-1))=s_j\in P$ (for some $j$) is a puncture and $p_h(-1)<p_h(1)$ with respect to the order on $\pi^{-1}(s_j)$;

\item A neighborhood of any ramification point is as in \Cref{sn2+a}.
\begin{figure}[!ht]
\centering
\includegraphics[scale=1]{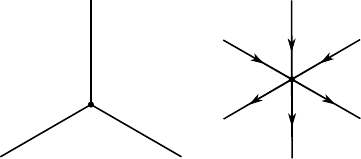}
\caption{A neighborhood of a branch point in $ S$ (left), and a ramification point in $\Sigma$ (right).}
\labelx{sn2+a}
\end{figure}

\item Different paths $p_h$ and $p_{h'}$ in $\bar\Sigma$ can meet only at a point $p_h(t)=p_{h'}(t')$, and one of the following holds:
\begin{enumerate}
\item $h,h'\in H_1$, $t=t'=0$, and $p_h(t)=p_{h'}(t')$ is a ramification point;

\item $h,h'\in H_1\cup H_2$, $t,t'\in\{-1,1\}$, and $p_h(t)=p_{h'}(t')$ is a puncture;

\item $h,h'\in H_1\cup H_2$, $t\cdot t'> 0$, then we also require that no line of $\mathcal{W}$ goes through $p_h(-t)$ or $p_{h'}(-t')$;
\item $h,h'\in H_1\cup H_2$, $t\cdot t'< 0$, then without loss of generality, we assume that $t<0$, $t'>0$. In this case, there exists a path $p_{h''}$ with $h''\in H_2$ such that $p_{h''}(+0)=p_h(-t)$ and $p_{h''}(-0)=p_{h'}(-t')$. Moreover, no other line of $\mathcal{W}$ goes through $p_h(-t)$ or $p_{h'}(-t')$. This type of intersection is called a \emph{joint}\index{spectral network!joint}
\end{enumerate}
\end{enumerate}

We denote by $$\mathcal{W}_\Sigma:=\bigcup_{h\in H_1\cup H_2}p_h((-1,1))\subseteq \Sigma$$
the spectral network on ${\Sigma}$ and by $$\mathcal{W}_S:=\pi(\mathcal{W}_\Sigma)\subseteq S$$ the spectral network on ${S}$.
\end{definition}

\begin{example}
    The example given in \Cref{fig:sn_example} is a non-degenerate spectral network that is not small, since it contains two joints.

    \begin{figure}[!ht]
    \centering
    \includegraphics[scale=0.8]{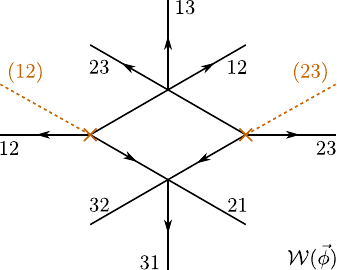}
    \caption{A non-degenerate spectral network that is not small.}
    \labelx{fig:sn_example}
    \end{figure}
\end{example}

\begin{remark}
    A small spectral network is in particular non-degenerate.
\end{remark}

\begin{remark}
In contrast to small spectral networks, a non-degenerate spectral network need not be finite. Given the scope of this book, we will always assume that a compact subset of $S$ intersects the spectral network only finitely many times. If $H_1$ and $H_2$ are finite sets, we call the corresponding spectral network \emph{finite}.
\end{remark}

\section{Combinatorial construction}
In this subsection, we present in detail a combinatorial construction of the spectral surface and some small spectral network associated to a triangulation of the underlying surface; the construction is first demonstrated in rank 2 and then generalized for higher rank.

\subsection{Rank two spectral networks}\labelx{sec:Combinatorial_rank_2}

Let $S$ be a ciliated surface and let $\Delta$ be an ideal triangulation of $S$. The complement of $\Delta$ in $S$ is a collection of contractible open sets that are incident to precisely three punctures. We call the closures of such sets in $S$, \emph{triangles} of $\Delta$. We denote by $\Delta_3$ the set of all triangles of $\Delta$. Note that triangles are closed contractible subsets of $S$ and the boundary of each triangle consists of three edges of $\Delta$.

For every triangle $T\in\Delta_3$, we fix a homeomorphism $\phi_T\colon T\to \mathbb T$, where $\mathbb T$ is the compact regular Euclidean triangle in $\mathbb{C}$ with the vertices $1$, $\E^{2\I\pi/3}$ and $\E^{4\I\pi/3}$ removed. We call $\mathbb T$ the \emph{standard triangle}.

Let now $\mathbb H$ be the hexagon in $\mathbb{C}$ removing the vertices $1$, $\E^{\I\pi/3}$, $\E^{2\I\pi/3}$, $-1$, $\E^{4\I\pi/3}$ and $\E^{5\I\pi/3}$. The map $z\mapsto z^2$ on $\mathbb{C}$ maps $\mathbb H$ to $\mathbb T$ (see \Cref{fig:HT-covering}). This map is a ramified covering with only one ramification point $z=0$. Now, for every $T\in\Delta_3$, we consider a pair $H_T:=(\mathbb H,T)$ and define maps $\psi_T\colon H_T\to\mathbb H$ by $(z,T)\mapsto z$. So we obtain trivially the following commutative diagram and the map $\theta \colon H_T\to T:=\psi_T\circ (z\mapsto z^2) \circ \phi_T^{-1}$ that is a ramified covering: 

$$\begin{CD}
H_T @>\psi_T>> \mathbb H\\
@V \theta_T VV @VV z\mapsto z^2 V \\
T @>\phi_T>> \mathbb T
\end{CD}$$

In $\mathbb H$, we take three oriented paths: 
$$\begin{array}{rccl}
p_k\colon & (-1,1) & \to & \mathbb H\\
& t & \mapsto & t \E^{2\I k\pi/3}\; ,
\end{array}$$
for $k\in\{0,1,2\}$, and we project each of these paths using the map $z\mapsto z^2$. Obviously, these paths satisfy the conditions of the spectral network for the covering $\mathbb H\to\mathbb T$.
\begin{figure}[!ht]
\centering
\includegraphics[scale=1]{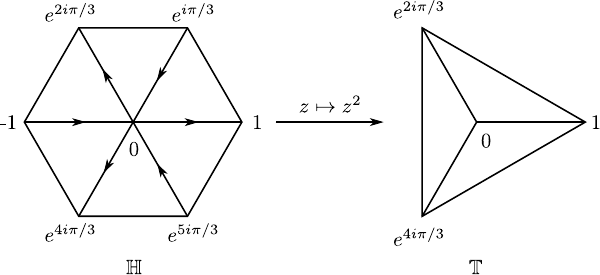}
\caption{The covering $\mathbb H\to\mathbb T$.}
\labelx{fig:HT-covering}

\end{figure}

We want to obtain $\Sigma$ by gluing hexagons $H_T$ along edges in the following way: Let $e$ be an edge of $\Delta$ that separates two triangles $T_1$ and $T_2$. Let $\alpha_e\colon (0,1)\to e$ be a parameterization of $e$. Since $\theta$ is not ramified over $e$, the lift of $\alpha_e$ under $\theta_{T_1}$ consists of two disjoint paths $\alpha^1_{e_1}, \alpha^1_{e_2}\colon (0,1)\to H_{T_1}$. We assume that the lift $\alpha^1_{e_1}$ is the lift that goes from the sink to the source with respect to the spectral network on $H_{T_1}$, and the lift $\alpha^1_{e_2}$ is the lift that goes from the source to the sink with respect to the spectral network on $H_{T_1}$.  We have the same for $T_2$: $\alpha^2_{e_1}, \alpha^2_{e_2}\colon (0,1)\to H_{T_2}$. Finally, we glue hexagons $H_{T_1}$ and $H_{T_2}$ identifying $\alpha^1_{e_1}(t)$ and $\alpha^2_{e_1}(t)$, for all $t\in(0,1)$, as well as $\alpha^1_{e_2}(t)$ and $\alpha^2_{e_2}(t)$, for all $t\in(0,1)$, see \Cref{fig:gluing}. The space constructed via this gluing construction is a surface, which we denote by $\Sigma$. The maps $\theta_T$ give rise to the ramified covering $\theta\colon\Sigma\to S$ with the spectral network on it.

\begin{figure}[!ht]
\centering
\includegraphics[scale=1]{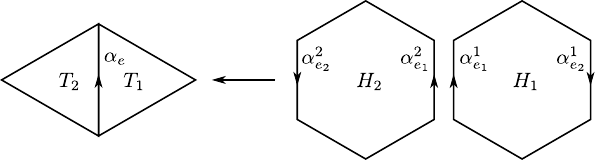}
\caption{Gluing along the edge $e$ of the ideal triangulation $\Delta$.}
\labelx{fig:gluing}

\end{figure}

\subsection{Higher-rank spectral networks}\labelx{sec:Combinatorial_higher_rank}

In this section, we will use a combinatorial construction of the spectral covering for bundles of rank $k\geq 2$ introduced in \cite{GK} by Goncharov and Kontsevich to give a combinatorial construction of a small spectral network associated to this $k$-fold ramified covering. This construction yields a very specific type of spectral network called \emph{minimal spectral network} in \cite{Gaiotto:2012db}, but this restricted class of spectral networks will nonetheless be very relevant in the following, in particular for its ties with Fock--Goncharov coordinates (see \Cref{Sec:SNandsnakes}). We first give the construction of the spectral cover given in \cite{GK}. Let $S$ be an oriented ciliated surface, endowed with an ideal triangulation $\Delta$. For any integer $k\geq 2$, we define a bipartite graph $\Gamma_k$ on $S$ by gluing, for all triangles $t$ of $S$, elementary pieces $\Gamma_k^{(t)}$ as described in \Cref{fig:spectral_graph}.
\begin{figure}[!ht]
		\centering
			\includegraphics{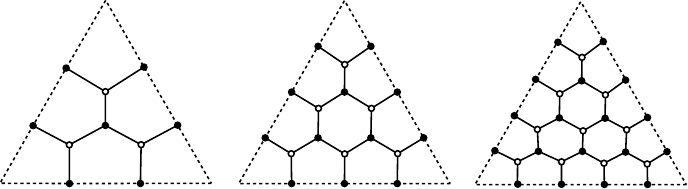}
			\caption{The graphs $\Gamma_2^{(t)}$, $\Gamma_3^{(t)}$ and $\Gamma_4^{(t)}$ restricted to a triangle $t$ of $\Delta$.}
			\labelx{fig:spectral_graph}
		
	\end{figure}

Then for every pair of triangles $t$ and $t'$ that share an edge $e$ of the triangulation, we identify the corresponding black vertices of $\Gamma_k^{(t)}$ and $\Gamma_k^{(t')}$ on $e$; see \Cref{fig:spectral_graph_gluing}.
\begin{figure}[!ht]
		\centering
			\includegraphics{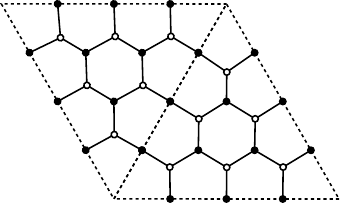}
			\caption{The gluing of $\Gamma_3^{(t)}$ and $\Gamma_3^{(t')}$ for two adjacent triangles $t$ and $t'$.}
			\labelx{fig:spectral_graph_gluing}
		
	\end{figure}
The white vertices are trivalent and the black vertices are either monovalent, bivalent or trivalent: the monovalent and bivalent black vertices will be called \emph{external} and the trivalent black vertices will be called \emph{internal}. The monovalent black vertices are exactly the vertices of $\Gamma_k$ lying on the boundary of $S$.
	
The graph $\Gamma_k$ is embedded into $S$ so every vertex inherits a cyclic order on the edges incident to it from the orientation of $S$. Let $\gamma=(v_1,\dots,v_r)$ be a path on $\Gamma_k$ and $1<i<k$. We say that $\gamma$ \emph{turns left} (resp. \emph{turns right}) at $v_{i}$ if the edges $(v_{i-1},v_i)$ and $(v_i,v_{i+1})$ (resp. $(v_i,v_{i+1})$ and $(v_{i-1},v_i)$) are consecutive in that order with respect to the cyclic ordering at $v_i$. Since the graph $\Gamma_k$ is embedded into $S$, a path in $\Gamma_k$ will be identified with its image in $S$.
	
A \emph{zig-zag path} on $\Gamma_k$ is a path $(v_1,\dots,v_r)$ that turns right at every white vertex and turns left at every black vertex. In particular, every oriented edge of $\Gamma_k$ gives rise to exactly one zig-zag path starting with this edge. Every zig-zag path can thus be extended maximally in both directions. Since the graph $\Gamma_k$ is finite, a zig-zag path can be extended either to a path connecting two boundary components, or to a cycle on $\Gamma_k$, every extension of this cycle being periodic. In the following, we will assume all zig-zag paths to be simple cycles or paths joining boundary components.

	\begin{proposition}
		Let $\gamma$ be a zig-zag path on $\Gamma_k$. Then the image of $\gamma$ in $S$ is homeomorphic either to a circle $\mathbb{S}^1$ with an orientation matching the one of $S$ or to a closed interval, and $S\setminus \gamma$ has two connected components, one being homeomorphic to either an open disk with one puncture or an open half-disk with one puncture on the boundary. We denote this particular connected component by $S_\gamma$ and the puncture inside it by $s_\gamma$.
	\end{proposition}

	Let $\gamma$ be a zig-zag path on $\Gamma_k$ and $\bar{S_\gamma}$ the closure of $S_\gamma$ in $S$. The boundary of $S_\gamma$ is the image of $\gamma$ in $S$, in particular it does not contain any puncture, so $\bar{S_\gamma}$ is homeomorphic to a closed (half-)disk with one puncture. 
	
	\begin{lemma}\labelx{lem:punct_disks}
		Let $k\geq 2$. For every puncture $s$ of $S$, there are exactly $k$ many zig-zag paths $\gamma_{s}^{(1)},\dots,\gamma_s^{(k)}$ in $\Gamma_k$ such that $s = s_{\gamma_s^{(i)}}$, for all $1\leq i\leq k$. Moreover, for all $1\leq i<j\leq k$, we have $\gamma_s^{(i)}\subset S_{\gamma_s^{(j)}}$ (\Cref{fig:zig-zag_paths}). 
	\end{lemma}
	
	\begin{figure}[!ht]
		\centering
			\includegraphics{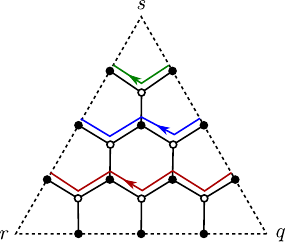}
			\caption{The three zig-zag paths $\gamma_{s}^{(1)}, \gamma_{s}^{(2)}$ and $\gamma_{s}^{(3)}$ in $\Gamma_3^{(t)}$ that circle around the puncture $s$. The subset of $t$ above a path $\gamma_{s}^{(i)}$ is contained in $S_{\gamma_{s}^{(i)}}$.}
			\labelx{fig:zig-zag_paths}
		
	\end{figure}
	
	We define $\Sigma_k$ as the topological surface obtained by gluing punctured (half-)disks to $\Gamma_k$: along every zig-zag path $\gamma$ in $\Gamma_k$ we glue a copy of $\bar{S_\gamma}$.
    We define $\pi : \Sigma_k\to S$ as follows: the image of a point in $\Gamma_k$ is the immersion in $S$, and every cell $S_\gamma$ is endowed with a map $S_\gamma\to S$ given by the inclusion. We denote by $s_i$ the puncture inside $S_{\gamma_s^{(i)}}$ in $\Sigma_k$.
	
	\begin{proposition}
		The map $\pi : \Sigma_k\to S$ is a ramified $k$-fold covering, with simple ramification points. The ramification points are the internal black vertices of $\Gamma_k$.
	\end{proposition}
	
	\begin{proof}
		Every edge of $\Gamma_k$ is part of exactly two zig-zag paths so $\Sigma_k$ is a topological surface in the neighborhood of a point in the interior of every edge of $\Gamma_k$ and in the neighborhood of an external black vertex. Every white vertex is on the boundary of exactly three zig-zag paths, each cell $S_\gamma$ being glued to a pair of adjacent edges (see \Cref{fig:vertex_neighborhood}). The neighborhood of a white vertex is then a topological surface. In all these cases, we also see that the covering $\pi$ is regular. In the neighborhood of every internal black vertex, $\Sigma_k$ is also a topological surface, but the covering $\pi$ is simply ramified at every internal black vertex (see \Cref{fig:vertex_neighborhood}).
		\begin{figure}[!ht]
			\centering
				\includegraphics{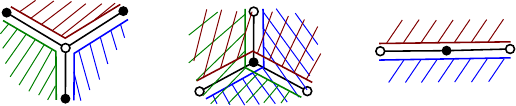}
				\caption{The neighborhood of a white vertex, an internal black vertex and an external black vertex, together with the different cells glued in the neighborhood of those vertices.}
				\labelx{fig:vertex_neighborhood}
			
		\end{figure}
	\end{proof}

	\begin{remark}
		The surface $\Sigma_2$ and the covering $\pi : \Sigma_2 \to S$ are homeomorphic to the ones constructed in \Cref{sec:Combinatorial_rank_2}.
	\end{remark}

	\begin{figure}[!ht]
		\centering
			\includegraphics[width=\textwidth]{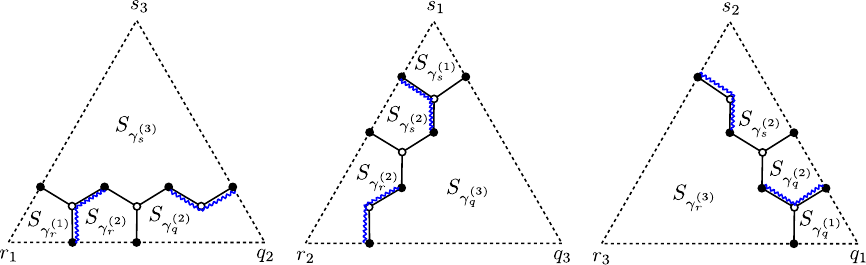}
			\caption{The $3$-fold ramified covering of a triangle of $S$. The wavy blue lines are branch cuts.}
			\labelx{fig:Sigma_3}
		
	\end{figure}
	
	We can now construct a small spectral network on $\Sigma_k$ associated with the $k$-fold ramified covering $\pi : \Sigma_k\to S$. For this, we will describe the spectral network on the covering of a triangle of $S$, the complete spectral network on $\Sigma_k$ being the union of all spectral networks on the coverings of the triangles. First, notice that every puncture $s$ of $S$ is unramified, i.e. $\vert\pi^{-1}(s)\vert = k$. Those $k$ lifts $s_1,\dots,s_k$ of $s$ each lies in one of the punctured disks $S_{\gamma_s^{(i)}}$ defined in \Cref{lem:punct_disks}, and inherit the order given by the inclusion of these punctured disks $s_1<\dots<s_k$. Let $t$ be a triangle of $S$ and $s$ a vertex of this triangle, and let $2\leq i\leq k$. There are $k-i$ internal black vertices on the boundary of $S_{\gamma_s^{(i)}}\cap t$ and there are $k-i$ white vertices on the boundary of $S_{\gamma_s^{(i-1)}}\cap t$. Each of those black vertices is connected to exactly one of those white vertices via an edge of $\Gamma_k^{(t)}$, and those black vertices are ramification points. On $\Sigma_k$, we define for each of those black vertices a line of the spectral network as follows: let $v_b$ be an internal black vertex on the boundary of $S_{\gamma_s^{(i)}}\cap t$ and $v_w$ the white vertex on the boundary of $S_{\gamma_s^{(i-1)}}\cap t$ associated to $v_b$. Let $p_1$ a path from $s_i$ to $v_b$ in $S_{\gamma_s^{(i)}}\cap t$, let $p_2$ be the path from $v_b$ to $v_w$ obtained by following the edge of the graph $\Gamma_k^{(t)}$, and let $p_3$ be the path from $v_w$ to $s_{i-1}$ in $S_{\gamma_s^{(i-1)}}\cap t$. The line of the spectral network is the concatenation $p_3. p_2. p_1$ (see \Cref{fig:Sigma_3} and \Cref{fig:Spectral_network_GL3} for $k=3$).
	
	\begin{figure}[!ht]
		\centering
			\includegraphics[width=\textwidth]{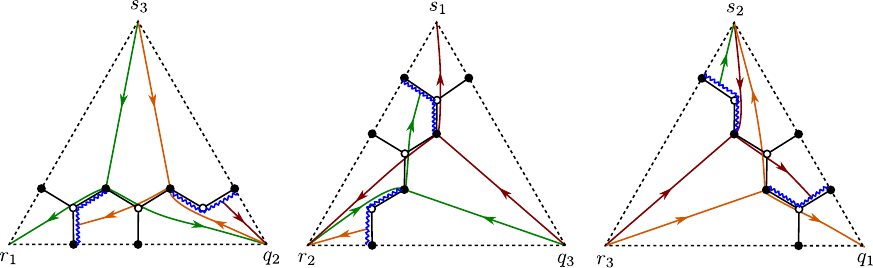}
			\caption{The small spectral network on the lift of a triangle to $\Sigma_3$. The lines of the spectral network are colored differently depending on the branch point they are going through.}
			\labelx{fig:Spectral_network_GL3}
		
	\end{figure}
	
	\begin{proposition}
		The set $\mathcal{W}$ of all paths on $\Sigma_k$ constructed as above is a small spectral network associated to the $k$-fold ramified covering $\pi \colon \Sigma_k\to S$.
	\end{proposition}

	\begin{proof}
		Every line $p \colon [-1,1]\to\bar\Sigma_k$ satisfies $\pi(p(t)) = \pi(p(-t))$, for $t\in[-1,1]$, and $\pi(p(-1)) = \pi(p(1))$ is a puncture in $\bar S$ and $p(0)$ is an internal black vertex, i.e. a ramification point. Every internal black vertex is part of three zig-zag paths, so it is on the boundary of three cells of the form $S_{\gamma_s^{(i)}}$. There are three lines of $\mathcal{W}$ going through each black vertex, with alternating directions (see \Cref{fig:Black_vertex_intersection}).
		
		\begin{figure}[!ht]
			\centering
				\includegraphics{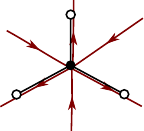}
				\caption{The intersection of three lines of the spectral network at an internal black vertex.}
				\labelx{fig:Black_vertex_intersection}
			
		\end{figure}
	
	Each edge of $\Gamma_k$ incident to an internal black vertex is followed by exactly one line of $\mathcal{W}$, and the edges incident to an external black vertex do not meet any lines of $\mathcal{W}$. There are no intersections between lines of $\mathcal{W}$ inside a punctured disk of the form $S_{\gamma_s^{(i)}}$, so the only intersections between lines of $\mathcal{W}$ (except those at ramification points) lie on white vertices of $\Gamma_k$. That means all those intersections occur between lines having already crossed their ramification points, thus the set $\mathcal{W}$ is a small spectral network.
	\end{proof}

\section{Analytic construction: WKB spectral networks}\labelx{sec:analytic_con}

Let now a surface with punctures $S$ equipped with a complex structure in the sense of \Cref{complex_str_ciliated} and consider its compactification $\bar{S}:=S \cup P$ (cf. \Cref{Sec:ciliatedsurfaces}). Let $\pi\colon T^*\bar S\to\bar S$ be the holomorphic cotangent bundle over $\bar S$, the holomorphic structure of which is canonically induced by the complex structure of $S$. Consider a tuple $\vec\phi:=(\phi_1,\dots,\phi_k)$, where $\phi_i$ is a meromorphic section of $(T^*\bar S)^{\otimes i}$, for all $i\in\{1,\dots, k\}$, with $P\subset \bar S$ being the set of all poles of $\vec\phi$.

The \emph{spectral cover}\index{spectral!cover} associated to $\vec\phi$ is defined as 
$$\Sigma_{\vec\phi}:=\left\{\sum_{i=1}^k\phi_i y^{k-i}=0\mid y\in T^*S\right\}\subset T^*S\; ,$$
where $T^*S$ is the cotangent bundle of $S$. If we choose the sections $\phi_i$ above to be generic enough, then the covering $\pi\vert_{\Sigma_{\vec\phi}}\colon\Sigma_{\vec\phi}\to S$ is a smooth ramified $k$-fold covering with only simple ramifications. As before, let $B$ be the set of branch points in $S$.

\begin{remark}
For simplicity, we will abuse our notation by writing $\pi\colon\Sigma_{\vec\phi}\to S$ instead of $\pi\vert_{\Sigma_{\vec\phi}}\colon\Sigma_{\vec\phi}\to S$.
\end{remark}

\begin{definition}
The covering $\pi\colon\Sigma_{\vec\phi}\to S$ is called \emph{simple} if $\Sigma_{\vec\phi}$ is smooth and has only simple ramification points.
\end{definition}

\begin{remark}
In case $n=2$ and $\phi_1=0$, then $\Sigma_{\vec\phi}:=\left\{y^2+\phi_2=0\mid y\in T^*S\right\}$. In particular, $\Sigma_{\vec\phi}$ is simple if and only if $\phi_2$ has only simple zeroes.
\end{remark}

Assume $\pi\colon\Sigma\to S$ is a smooth spectral covering with only simple ramifications as before. We define the space
$$\hat \Sigma:=\overline{\{(y,y')\in\Sigma\times \Sigma\mid \pi(y)=\pi(y'),\;y\neq y'\}}\subset T^*S\times T^*S\; ,$$
called the \emph{root covering}\index{covering!root}\index{root!covering} of the covering $\pi\colon\Sigma \to S$. It is a smooth $(k^2-k)$-sheeted ramified covering of $S$. There is a natural involution  $\mu\colon \hat\Sigma\to\hat\Sigma$ defined by $\mu(y,y'):=(y',y)$, and two projection maps $p_1,p_2\colon\hat\Sigma\to\Sigma$ with $p_1(x,y):=x$ and $p_2(x,y):=y$ respectively. Clearly, $p_1=p_2\circ\mu$ and $\pi\circ p_1=\pi\circ p_2:=\hat\pi$.

On the root covering there is a natural abelian differential $\rho$ defined as follows: Let $\lambda$ be the canonical Liouville 1-form on $T^*S$. We define $\rho:=p_1^*\lambda-p_2^*\lambda$ with zeroes over ramification points. If we take the oriented foliation $F_\rho$ corresponding to $\hat\Sigma$, then $\mu$ preserves the leaves of $F_\rho$, but reverses orientation, because $\mu^*\rho=-\rho$.

We construct a spectral network $\mathcal W(\vec\phi)$ on $\hat\Sigma$, which is associated with a tuple of differentials $\vec\phi$, as a set of leaves of $\mathcal F_\rho$ using the following two steps:
\begin{enumerate}
\item The critical graph of $\rho$ is in $\mathcal W(\vec\phi)$;
\item Let $\alpha_1,\alpha_2\colon (-1,1)\to\hat\Sigma$ be two leaves of $\mathcal W(\vec\phi)$ such that there exist $t_1,t_2\in(-1,1)$ such that $\alpha_1(t_1)$ is in the sheet $(i,j)$, $\alpha_2(t_2)$ is in the sheet $(j,k)$ and $p_j(\alpha_1(t_1))=p_j(\alpha_2(t_2))$, that is, the projections of these two leaves to $\Sigma$ intersect in the sheet $j$.  Let $p_0$ be the point in the sheet $(i,k)$ and $p_0'$ be the points in the sheet $(k,i)$ such that $\hat\pi(p_0)=\hat\pi(p_0')=\hat\pi(\alpha_1(t_1))\in S$. Then we add two half-leaves of $\mathcal F_\rho$ to $\mathcal W(\vec\phi)$, one in the sheet $(i,k)$ starting at the point $p_0$ and such that this point is the source of this leaf, and another leaf in the sheet $(k,i)$ starting at the point $p_0'$ and such that this point is the sink of this leaf; see \Cref{fig:0}.
\end{enumerate}

\begin{figure}[!ht]
\centering
\includegraphics{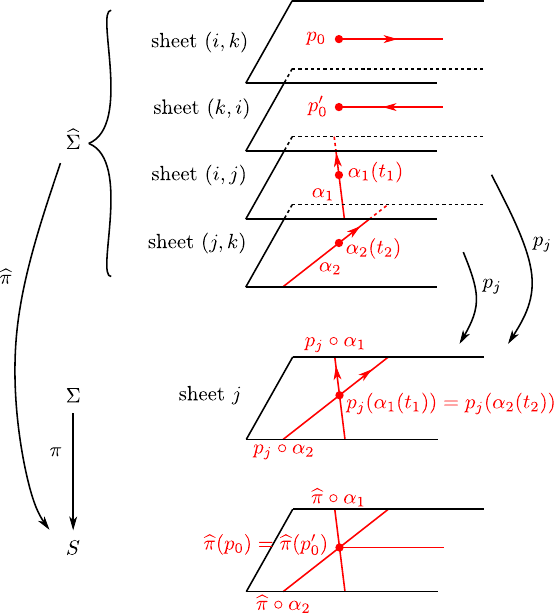}
\caption{The construction described in Step 2. above.}
\labelx{fig:0}

\end{figure}

The procedure of constructing $\mathcal W(\vec\phi)$ described above is generically infinite because adding a leaf in the second step usually produces new intersections of leaves projected to $\Sigma$. But sometimes this procedure ends in finitely many steps. In this latter case, $\mathcal W(\vec\phi)$ projected to $\Sigma$ and to $S$ satisfies the conditions of spectral network. We call a spectral network corresponding to a meromorphic differential, a \emph{WKB spectral network}\index{spectral network!WKB}\index{WKB!spectral network}.

\begin{example} If $k=2$, then $p_1$ and $p_2$ are diffeomorphisms of surfaces. 

If $k=3$, $p_1\circ\pi\colon\hat\Sigma\to X$ is a $6:1$-ramified covering, $\pi\colon \Sigma\to X$ is a $3:1$-ramified covering and $p_i\colon\hat\Sigma\to \Sigma$, for $i\in\{1,2\}$, are $2:1$-ramified coverings.
\end{example}

\begin{example}\labelx{ex:CP1SN}
Let $\bar S=\mathbb{CP}^1$ and $k=3$. We take $\vec\phi=\E^{\frac{\I\pi}{2}}(0,-\D z^2,z\D z^3)$. The only pole of $\vec\phi$ is the point at infinity, so $S=\mathbb{CP}^1\setminus\{\infty\}=\mathbb{C}$. Then 
$$\Sigma=\{(z,y)\in T^*S\mid y^3-y+z=0\}\; .$$
The branch locus agrees with the set of points with zero discriminant: $\Delta(z)=4z^2-27=0$ and consists of two points. The spectral network $W(\vec\phi)$ is the one illustrated on \Cref{fig:sn_example}.
\end{example}

\begin{remark}
    Further examples of spectral networks are given in \Cref{Chap:BPSstatesclassS}, where the physical essence of WKB spectral networks is also discussed.
\end{remark}

\begin{remark}
    All spectral networks of rank 2 are isotopic to a WKB spectral network. Conversely, since a quadratic differential on a ciliated surface defines an ideal triangulation (see \Cref{sec:Crit_graph_quadratic}), any WKB spectral network of rank 2 is of the form described in \Cref{sec:Combinatorial_rank_2}. In contrast, for spectral networks of rank $k\geq 3$, neither of these two statements hold: not all non-degenerate spectral networks are isotopic to a WKB spectral network, and not all WKB spectral networks are of the form described in \Cref{sec:Combinatorial_higher_rank}.
\end{remark}

\section{Path lifting using non-degenerate spectral networks}\labelx{sec:path_lifting}

Now that we have the definition of small spectral networks in hand, we can describe the path lifting rule which allows later on the construction of non-abelianization maps.

Let $\pi\colon \Sigma\to S$ be a ramified $k$-fold covering satisfying the conditions above, and let $\mathcal{W}$ be a small spectral network of rank $k$ on $\Sigma$. We want to construct a map $\pi_1(S,p)\to \mathbb{Z}^{\text{tw}}_B[\mathcal P(\Sigma)]$, where $p\in S$ and $p$ is not contained in any line of $\mathcal{W}$.

We consider a path $\gamma\colon [0,1]\to S$ with endpoints not lying on any line of the spectral network, and not going through any branch point. To lift this path to $\bar{\Sigma}$, we split it in pieces $\gamma=\gamma_n\cdot...\cdot\gamma_1$ such that each $\gamma_i$ intersects the spectral network $\mathcal{W}$ at most once. Here we denote the concatenation of paths by $\cdot$ as in the definition of a path algebra.

If $\gamma_i$ does not intersect the spectral network, then we lift it in the usual way and get $k$-many lifts $\gamma_i^{(1)}, ...,\gamma_i^{(k)}$ in ${\Sigma}$ that also do not intersect the spectral network.

If $\gamma_i$ intersects a line $p_h$ of the spectral network, then there are only two standard lifts of $\gamma_i$ that intersect $p_h$ in $ \Sigma$. One of these two lifts $\gamma_i^{(j)}$  intersects $p_h$ before it goes through its ramification point, and the other lift $\gamma_i^{(\ell)}$ occurs after $p_h$ passes through its ramification point. In this case, we have to add to the standard lifts a new path $\gamma_i'$, as in \Cref{sn5}. The path $\gamma_i'$ is obtained by following $\gamma_i^{(j)}$ from its starting point until its intersection with the line $p_h$, then following $p_h$ without crossing it until it meets $\gamma_i^{(\ell)}$, and then following  $\gamma_i^{(\ell)}$ until its ending point.

\begin{figure}[!ht]
\centering
\includegraphics[scale=1]{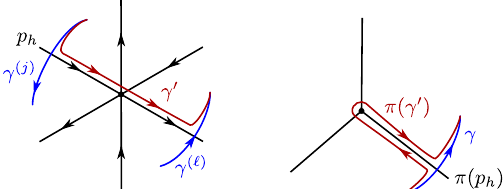}
\caption{Path lifting using a spectral network. The picture in $\Sigma$ (left), and in $S$ (right).}
\labelx{sn5}

\end{figure}

\begin{definition}
The \emph{lift} $\tilde\gamma$ in ${\Sigma}$ with respect to the spectral network $\mathcal{W}$ of the curve $\gamma$ is the product of all lifts of $\gamma_i$ in $\mathbb{Z}^{\text{tw}}_B[\mathcal P({\Sigma})]$:
$$\tilde\gamma=\left(\sum_{i=1}^{k}\gamma_1^{(i)}+\gamma_1'\right)\cdot...\cdot\left(\sum_{i=1}^{k}\gamma_n^{(i)}+\gamma_n'\right)\; ,$$
where $\gamma_i'=0$ for pieces $\gamma_i$ which do not intersect the spectral network.
\end{definition}

\begin{theorem}
This path lifting rule is homotopy invariant, i.e. if two paths $\gamma$ and $\delta$ are homotopic relative to their endpoints, then $\tilde{\gamma}=\tilde{\delta}$ in $\mathbb{Z}^{\text{tw}}_B[\mathcal P(\Sigma)]$.
\end{theorem}

\noindent The theorem is a consequence of the two following lemmas:
	
	\begin{lemma}
		Let $\gamma$ be a path on ${S}$ that intersects exactly twice the same line $p_h$ of the spectral network and no other line of $\mathcal{W}$, as in \Cref{fig:SN5}. Then $\tilde\gamma=\gamma^{(1)} +\dots+ \gamma^{(k)}$ where $\gamma^{(1)},\dots,\gamma^{(k)}$ are the $k$ standard lifts of $\gamma$ to ${\Sigma}$.
	\end{lemma}
	
	\begin{figure}[!ht]
		\centering
			\includegraphics[scale = 1]{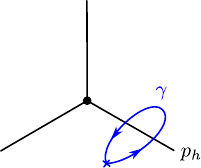}
			\caption{A loop intersecting twice the same line of $\mathcal{W}$.}
			\labelx{fig:SN5}
		
	\end{figure}
	
	\begin{proof}
		There are two additional lifts $\gamma'$ and $\gamma''$ created by the spectral network. These two special lifts are homotopic up to crossing a ramification point, hence $\gamma'+\gamma'' =0$ in $\mathbb{Z}^{\text{tw}}_B[\mathcal P(\Sigma)]$; see \Cref{fig:sn3}.
		\begin{figure}[!ht]
			\centering
				\includegraphics[scale = 0.6]{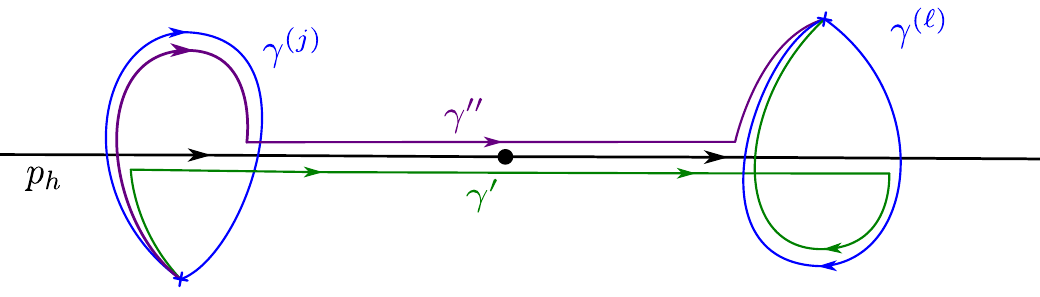}
				\caption{Spectral network lifts of $\gamma$.}
                \labelx{fig:sn3}
			
		\end{figure}
		
	\end{proof}
	
	\begin{lemma}
		Let $\gamma$ be a contractible loop based at $m\in S$ that loops around the branch point $b$ in $S$, intersecting exactly once each of the three lines of $\mathcal{W}$ going out of $b$, as in \Cref{fig:SN4}. Then $\tilde\gamma=e_{m_1} +\dots+ e_{m_k}$, where $m_1,...,m_k$ are the $k$ lifts of $m$ to $\Sigma$  and $e_{m_i}$ denotes the trivial loop based at $m_i$.
	\end{lemma}
	
	\begin{figure}[!ht]
		\centering
			\includegraphics{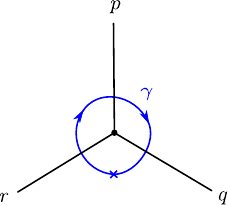}
			\caption{A small loop around a branch point.}
			\labelx{fig:SN4}
		
	\end{figure}
	
	\begin{proof}
		By applying the spectral network lifting rule to $\gamma$, we get $k+6$ paths: the $k$ standard lifts $\gamma^{(1)},\dots,\gamma^{(k)}$ (with $\gamma^{(j)}$ and $\gamma^{(\ell)}$ intersecting $\mathcal{W}$ as before), and six additional paths $\gamma'_1,\dots,\gamma'_6$ shown in \Cref{fig:SN6}. Up to relabeling, we suppose that the endpoints of $\gamma^{(j)}$ and $\gamma^{(\ell)}$ are $m_1$ and $m_2$.
		\begin{figure}[!ht]
			\centering
				\includegraphics[scale=0.55]{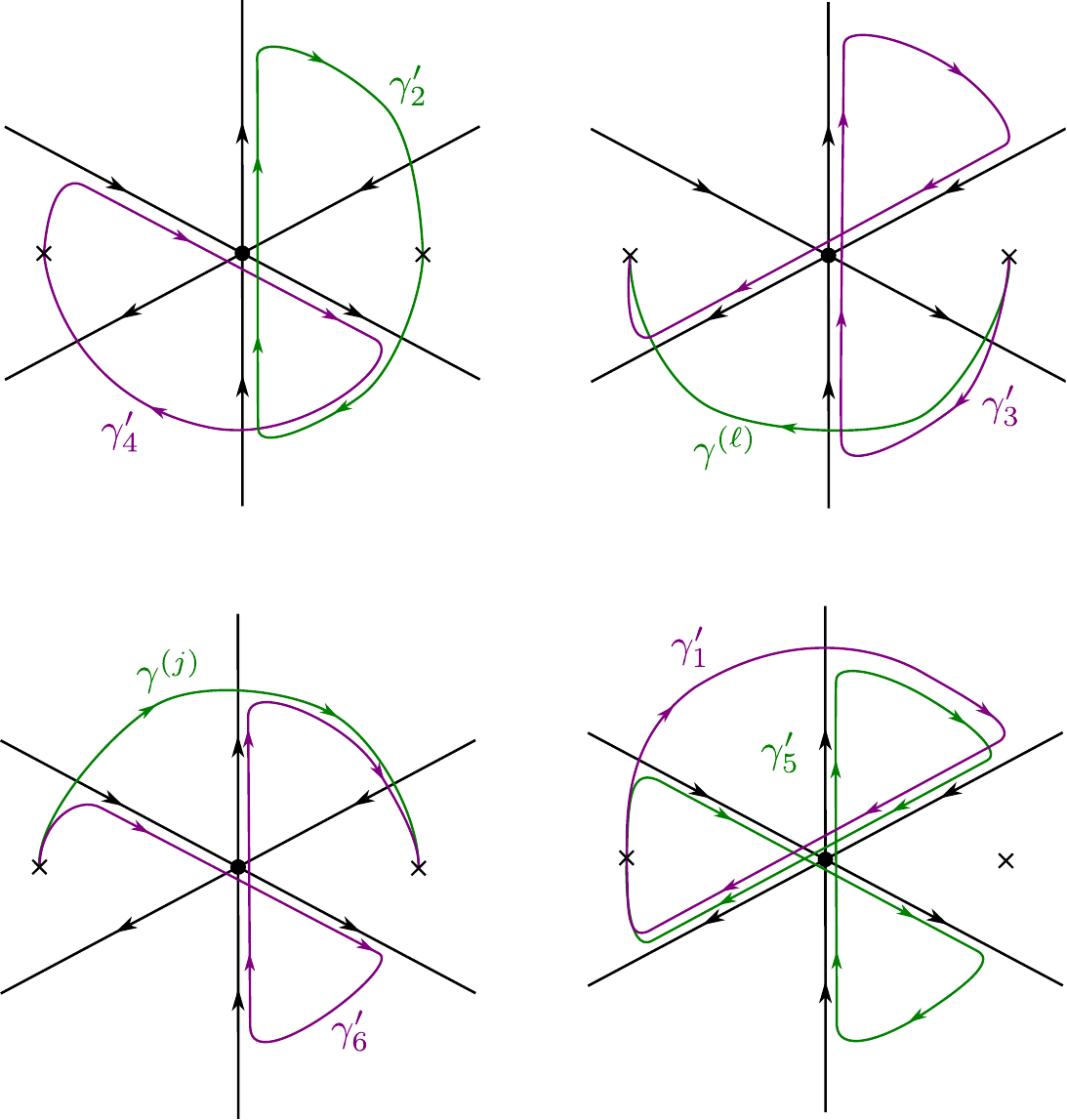}
				\caption{All six paths added by intersections with the spectral network, together with the standard lifts $\gamma^{(j)}$ and $\gamma^{(\ell)}$. On the upper left picture are the paths homotopic to trivial paths, and on the other pictures are the remaining lifts, grouped as pairs of paths canceling out each other in $\mathbb{Z}^{\text{tw}}_B[\mathcal P(\Sigma)]$.}\labelx{fig:SN6}
			
		\end{figure}
		
		\noindent In $\mathbb{Z}^{\text{tw}}_B[\mathcal P(\Sigma)]$, we have:
            \begin{align*}
			\gamma^{(\ell)} +\gamma'_3 = 0\; ,\\
			\gamma^{(j)} + \gamma'_6 = 0\; ,\\
			\gamma'_1 +\gamma'_5 =0\; ,\\
			\gamma'_4 = e_{m_1}\; ,\\
			\gamma'_2 = e_{m_2}\; .
		\end{align*}
		Therefore, we may write:
		$$\tilde\gamma = \sum_{i=1}^{k} \gamma^{(i)}+\gamma'_1+\gamma'_2+\gamma'_3+\gamma'_4+\gamma'_5+\gamma'_6= \sum_{i=1}^{k}e_{m_i}\; .$$
	\end{proof}

\textbf{Exercise}:
The path lifting rule can be extended to non-degenerate spectral networks, and it is homotopy invariant.

We give a hint on how to lift a path that intersects a line $p_h$, for $h\in H_2$. On \Cref{fig:joint_lift} we draw in red a curve in $S$ that we are lifting and their standard lifts in $\Sigma$, and in blue we draw the additional lift.

\begin{figure}[!ht]
\centering
\includegraphics{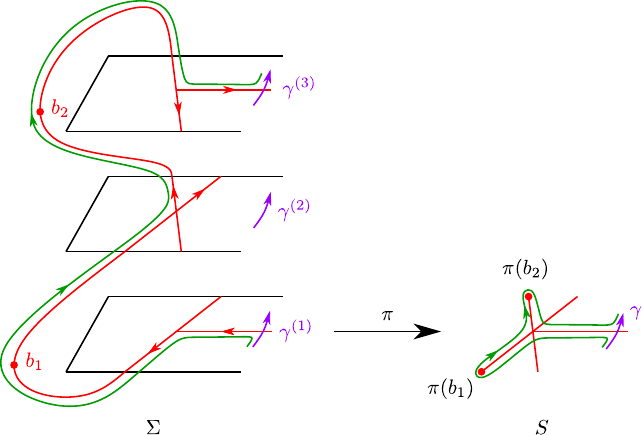}
\caption{Path lifting along a joint.}
\labelx{fig:joint_lift}

\end{figure}
\chapter{General and Fenchel--Nielsen spectral networks}

\abstract{In this chapter we study more general (degenerate) spectral networks which are very relevant for a number of physical applications. A special class of such general networks arises when all leaves of the critical graph for a Strebel differential have both ends on branch points. Such spectral networks are called Fenchel--Nielsen spectral networks due to their relationship with certain length-twist parameters, a construction that will be demonstrated in \Cref{sec:FN_SN_and_FN_coord}.}

\section{General spectral networks}

We will extend even further the definition of spectral networks to account for lines \emph{overlapping} and for lines ending at branch points instead of punctures. This level of generality will be very relevant to \Cref{partV}, but we will not describe here their path-lifting map as the general construction is more technical. We refer to \cite{horn2025} for a detailed description of the path-lifting map.

Let $\pi\colon\bar\Sigma\to \bar S$ be a ramified $k$-fold covering. 
\begin{definition}\labelx{dsn}
A \emph{general (or degenerate) spectral network}\index{spectral network!degenerate}\index{spectral network!general} $\mathcal{W}$ of rank $k$ associated to $\pi$ is consisted of two finite collections of smooth simple oriented paths on $\Sigma$: $\{p_h\colon [-1,1]\to\bar\Sigma\}_{h\in H_1}$ and $\{p_h\colon [-1,1]\setminus\{0\}\to\bar\Sigma\}_{h\in H_2}$, where $H_1$ and $H_2$ are two disjoint index sets. The paths may intersect only transversely \emph{or overlap} and satisfy the following conditions:
\begin{enumerate}
\item for all $h\in H_1\cup H_2$ and for all $t\in(0,1]$, $\pi(p_h(t))=\pi(p_h(-t))$;

\item for all $h\in H_1$, $\pi(p_h(0))\in B$ is a branch point;

\item for all $h\in H_2$, there exist limits: $p_h(+0):=\lim_{t\to +0}p_h(t)$ and $p_h(-0):=\lim_{t\to -0}p_h(t)$. Further, $p_h(+0)\neq p_h(-0)$ but $\pi(p_h(+0))=\pi(p_h(-0))$;

\item for all $h\in H_1\cup H_2$, $\pi(p_h(1))=\pi(p_h(-1))=s_j\in P$ (for some $j$) is either:
\begin{enumerate}
    \item a puncture, in which case $p_h(-1)<p_h(1)$ with respect to the order on $\pi^{-1}(s_j)$
    \item a ramification point $b'$, in which case there is also a line $p_{h'}$ going through $b'$ and overlapping with $p_h$ in the opposite direction (see \Cref{fig:resolutions}). This pair of overlapping lines is called a \emph{double wall}\index{double wall}\index{spectral network!double wall} or a \emph{two-way street}\index{spectral network!two-way street};
\end{enumerate}

\item A neighborhood of any ramification point is as in \Cref{sn2+}.
\begin{figure}[!ht]
\centering
\includegraphics[scale=1]{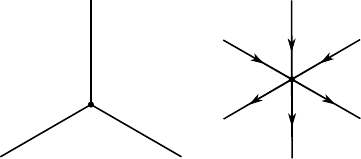}
\caption{A neighborhood of a branch point in $S$ (left), and a ramification point in $\Sigma$ (right). Note that any line can be a double wall.}
\labelx{sn2+}

\end{figure}

\item Different paths $p_h$ and $p_{h'}$ in $\bar\Sigma$ can meet only at a point $p_h(t)=p_{h'}(t')$, and one of the following holds:
\begin{enumerate}
\item $h,h'\in H_1$, $t=t'=0$, and $p_h(t)=p_{h'}(t')$ is a ramification point;

\item $h,h'\in H_1\cup H_2$, $t,t'\in\{-1,1\}$, and $p_h(t)=p_{h'}(t')$ is a puncture;

\item $h,h'\in H_1\cup H_2$, $t\cdot t'> 0$, then we also require that no line of $\mathcal{W}$ goes through $p_h(-t)$ or $p_{h'}(-t')$ apart from a line overlapping $p_h$ or $p_{h'}$;
\item $h,h'\in H_1\cup H_2$, $t\cdot t'< 0$, then without loss of generality, we assume that $t<0$, $t'>0$. In this case, there exists a path $p_{h''}$ with $h''\in H_2$ such that $p_{h''}(+0)=p_h(-t)$ and $p_{h''}(-0)=p_{h'}(-t')$. Moreover, no other line of $\mathcal{W}$ goes through $p_h(-t)$ or $p_{h'}(-t')$ apart from a line overlapping $p_h$, $p_{h'}$ or $p_{h''}$.
\end{enumerate}
\end{enumerate}
As before, we denote by $$\mathcal{W}_\Sigma:=\bigcup_{h\in H_1\cup H_2}p_h((-1,1))\subseteq \Sigma$$
the spectral network on ${\Sigma}$ and by $$\mathcal{W}_S:=\pi(\mathcal{W}_\Sigma)\subseteq S$$ the spectral network on ${S}$.
\end{definition}

Given a general spectral network containing at least one double wall, we call a \emph{resolution}\index{double wall!resolution} the choice of either "British" or "American" for each double wall. We think of this choice as a way to infinitesimally split apart the double wall, the terminology coming from the road usage conventions in England and the USA. See \Cref{fig:resolutions} for the two types of resolution of a double wall. Whenever we will be dealing with general spectral networks, we will always assume that a choice of resolution has been made.

\begin{figure}[!ht]
\centering
\includegraphics[scale=1]{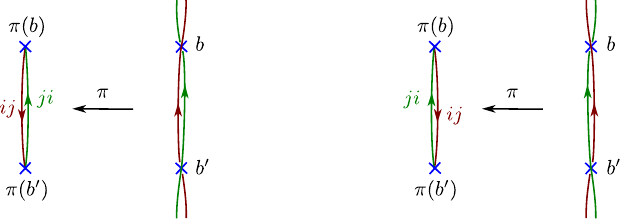}
\caption{An American (left) and a British (right) resolution of a double wall, drawn on $S$. A line labeled $ij$ goes from sheet $i$ to sheet $j$ of $\Sigma$. Notice that each line of the double wall passes through exactly one of the two branch points by definition, the other one being the limit when $t\to \pm 1$.}
\labelx{fig:resolutions}

\end{figure}

\section{Fenchel--Nielsen spectral networks}\labelx{sec:FN_spectral_networks}

The definition of a general spectral network allows for lines to end on branch points instead of punctures as long as they are double walls. One could then imagine a spectral network with only double walls, having no lines ending at punctures. Such a spectral network could even be defined on a closed surface since it does not need the existence of any puncture. Such spectral networks do exist, and we will now define a specific class of them, called \emph{Fenchel--Nielsen spectral networks}. This particular type of spectral networks was introduced by Hollands and Neitzke in \cite{HoNe}, and is called this way because such spectral networks are strongly related to the Fenchel--Nielsen coordinates defined in \Cref{sec: Fenchel--Nielsen}. This relation will be further investigated in \Cref{sec:FN_SN_and_FN_coord}.
These special WKB-spectral networks emerge when considering the critical graph of a Strebel differential as per \Cref{sec:Strebel_difs}.

For a closed Riemann surface $\bar{S}$ and for $P=\{s_1,...,s_n\}$ a finite collection of marked points on $\bar{S}$ with $n\in \mathbb{Z}_{>0}$, let $S=\bar{S}\backslash P$. Let $\omega$ be a meromorphic quadratic differential on $\bar{S}$ with double poles at the points of $P$. Choose a phase $\vartheta \in \mathbb{R}/ 2\pi \mathbb{Z}$, such that $\phi:=\E^{-2\I\vartheta}\omega$ is a Strebel differential in the sense of \Cref{defn:Strebel_dif}. This would then mean that there is a pants decomposition of $S$ such that the period of $\sqrt{\omega}$ around each leg has phase $\vartheta$. The square roots of the meromorphic quadratic  differential $\omega$ define a double covering $\pi\colon \Sigma_{\omega} \to S$ with
\[\Sigma_{\omega}:=\{ (z, \lambda)\in S \times T_z^*S : \lambda^2=\omega(z)\}\; ,\]
where $T_z^*S$ is the fiber at $z$ of the cotangent bundle of $S$. Here, $\pi(z, \lambda)=z$, and the branch points of the covering $\pi \colon \Sigma \to S$ are the zeroes of $\omega$. Similarly to the construction of a WKB-spectral network as in \Cref{sec:analytic_con}, we construct a spectral network $\mathcal{W}(\phi)$ as a set of leaves of the critical graph $\hat{\mathcal{F}}_{\phi}$, where all leaves have both ends on branch points; thus, each leaf gives a double wall of the network $\mathcal{W}(\phi)$. Such spectral networks are called \emph{Fenchel--Nielsen spectral networks}\index{spectral network!Fenchel--Nielsen}\index{Fenchel--Nielsen!spectral network}.

\begin{remark}
    By perturbing the phase $\vartheta$ to $\vartheta_+ = \vartheta+\epsilon$ (resp. $\vartheta_-=\vartheta-\epsilon$) for some small $\epsilon>0$, the critical leaves of $\phi$ split into two regular leaves connecting a branch point and a puncture. The limit when $\epsilon\to 0$ then forms a double wall, corresponding to the British (resp. American) resolution of $\mathcal{W}(\phi)$.
\end{remark}

Each component of $S\backslash \mathcal{W}(\phi)$ is a punctured disk swept out by closed $\vartheta$-trajectories and the boundary of each such component is a polygon consisting of one or more saddle connections. For each such punctured disc, the coefficient $m_j\E^{-\I\vartheta}$ of the Strebel differential at the puncture $s_j$, for $j=1,...,n$, equivalently, the length of any closed horizontal $\vartheta$-trajectory in the punctured disc, is a positive real number that labels each component of $S\backslash \mathcal{W}(\phi)$. 

A Fenchel--Nielsen spectral network induces a pants decomposition of the surface $S$, and on each pair of pants the spectral network is of one of two types, called \emph{molecule I} and \emph{molecule II} in \cite{HoNe}, see \Cref{fig:molecules}.
\begin{figure}[!ht]
\centering
\includegraphics[scale=1]{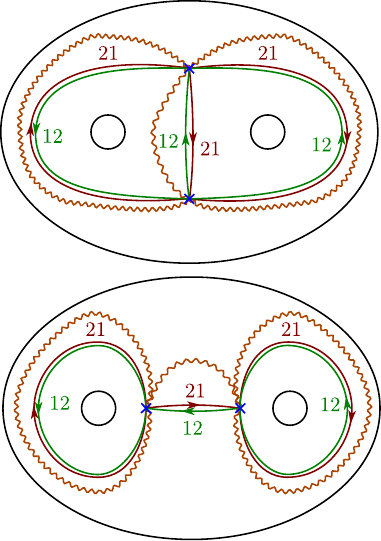}
\caption{The two possible Fenchel--Nielsen spectral networks on a pair of pants: molecule I (top) and molecule II (bottom). Wavy lines are branch cuts, and both spectral networks use the British resolution.}
\labelx{fig:molecules}

\end{figure}

\begin{remark}
There is also another non-generic case for the meromorphic quadratic differential $\omega$ when interesting spectral networks can be constructed. Namely, a WKB-spectral network can be constructed in the same way containing both single and double walls. Such spectral networks are called \emph{mixed}\index{spectral network!mixed}; examples of those can be found in \cite{HoNe}, where such networks were first introduced. Another interesting case of spectral networks, also introduced in \cite{HoNe} involves the so-called \emph{contracted spectral networks}\index{spectral network!contracted}, which are obtained by collapsing some double walls together in a Fenchel--Nielsen network, in other words, by considering a degeneration of an ordinary Fenchel--Nielsen spectral network where some annulus shrinks to zero size.  
\end{remark}

\chapter{Non-abelianization and abelianization}\labelx{sec:non-abel}

\abstract{As an important application of the spectral networks we have so far introduced, we present in this chapter the construction of non-abelianization and abelianization maps. These maps describe the equivalence between flat connections on a rank $n$ bundle over a Riemann surface $S$ and flat connections on a line bundle over the spectral curve $\Sigma$, an $n$-sheeted ramified covering of $S$. The monodromies of the abelianized local system are cluster $\mathcal{X}$-coordinates. The process of \mbox{(non-)}abelianization thus provides very strong motivation for using spectral networks in the study of higher-rank Teichm\"{u}ller spaces. In particular, we show how one can use spectral networks and abelianization to induce Fock--Goncharov coordinates or Fenchel--Nielsen coordinates on spaces of flat connections. Suitable references demonstrating the link between spectral networks and higher-rank Teichm\"{u}ller theory include \cite{Gaiotto:2012rg}, \cite{HoKi}, \cite{HoNe},  \cite{hollands2020exact}, \cite{Nho},  \cite{Nikolaev_abelianization}.}

\section{Non-abelianization map for non-degenerate spectral networks}\labelx{sec:non-ab}

Let $S$ be a ciliated surface, and $\pi\colon \bar\Sigma \to \bar S$ be a ramified $k$-fold covering with a non-degenerate spectral network $\mathcal W$. As before, the ramification locus of $\pi$ on $\Sigma$ is denoted by $B'$ and $B:=\pi(B')\subset S$. Let $\mathcal{E}^{\text{tw}}(\Sigma,\mathbb{C}^\times)$ be the space of flat line bundles over $\Sigma\setminus B'$ having monodromy $-1$ around the ramification points.

We take a point in the space $\mathcal{E}^{\text{tw}}(\Sigma,\mathbb{C}^\times)$ which yields a line bundle $L\to \Sigma\setminus B'$ over $\Sigma\setminus B'$ with a flat connection $\nabla$. For $q\in\Sigma\setminus B'$, we denote $L_q$ the fiber of $L\to \Sigma\setminus B'$ over $q$ which is a 1-dimensional $\mathbb{C}$-vector space.

Further, we fix $p\in S\setminus \pi(\mathcal{W})$. We construct a representation $\rho\colon \pi_1(S,p)\to \mathrm{GL}_k(\mathbb{C})$ as follows. We consider the following vector space of dimension $k$:
$$V_p:=\bigoplus_{i=1}^{k}L_{p_i}\; ,$$
where $\{p_1, \dots p_k\} = \pi^{-1}(p)$. We have the natural basis of $V_p$, namely $(e_1,...,e_k)$, where $(e_i)$ is a basis of $L_{p_i}$, for $i\in \{1,...,k\}$.

Every loop $\gamma$ on $S$ such that $[\gamma]\in \pi_1(S,p)$ can be lifted  to $\Sigma$ with respect to the spectral network $\mathcal{W}$. So we get an element $x\in \mathbb{Z}^{\text{tw}}_{B'}[\mathcal P(\Sigma)]$, which is by construction a finite sum of curves $\gamma_r$ on $\Sigma$, $r\in I$, for a finite index set $I$ such that
\begin{equation}\labelx{gmma}
\gamma_r(0),\gamma_r(1)\in\pi^{-1}(p)\; ,
\end{equation}
and
$x=\sum_{r\in I}\gamma_r$.
We consider a vector $w=\sum_{i=1}^k a_ie_i\in V_p$, $i\in\{1,..,k\}$. Because of (\ref{gmma}), the parallel transport along $\gamma$ can be defined by
$$T_\gamma:=\sum_{r\in I} T_{\gamma_r}\in\bigoplus_{\tilde p,\tilde q\in\pi^{-1}(p)} L(E_{\tilde p},E_{\tilde q})\; ,$$
where $T_{\gamma_r}$ are the parallel transports along $\gamma_r$ given by the connection $\nabla$ on $\Sigma$.

For each $\gamma_r$ such that $\gamma_r(0)=p_i$, $\gamma_r(1)=p_j$, $i,j\in\{1,...,k\}$ we can consider $T_{\gamma_r}e_i=t_re_j\in E_{p_j}$. We can extend $T_{\gamma_r}$ to a linear map on $V_p$ by the following rule:
$$T_{\gamma_r}e_l=\left\{\begin{matrix}
t_re_j, & l=i\\
0, & l\neq i\; .
\end{matrix}\right.$$
We define the linear map $T_\gamma\colon V_p\to V_p$ by the rule $T_\gamma:=\sum_{r\in I} T_{\gamma_k}$. By construction this map is linear and, since the path lifting using the spectral network $\mathcal W$ is homotopically invariant, and the connection on $\Sigma$ is flat, the map $T_\gamma$ depends only on the homotopy class of $\gamma$ in $\pi_1(S,p)$.

We thus get a map $\rho\colon \pi_1(S,p)\to \mathrm{Aut}(V_p)$ which is a group homomorphism. After an identification of $\mathrm{Aut}(V_p)$ with $\mathrm{GL}_k(\mathbb{C})$  we obtain an element of the space $\mathcal{E}(S,\mathrm{GL}_k(\mathbb{C}))$.
Every element of $\mathcal{E}^{\text{tw}}(\Sigma,\mathbb{C}^\times)$ together with the covering $\pi\colon\Sigma\to S$ and the spectral network $\mathcal{W}$ associated to $\pi$ yield an element of the space $\mathcal{E}(S,\mathrm{GL}_k(\mathbb{C}))$:
$$\pi_1(S,p)\xrightarrow[{\text{network}}]{\text{spectral}} \mathbb{Z}^{\text{tw}}_{B'}[\mathcal P(\Sigma)]\xrightarrow[\text{on }\Sigma]{\text{connection}}\mathrm{GL}_k(\mathbb{C})\; .$$

We have thus constructed a map $\mathcal{E}^{\text{tw}}(\Sigma,\mathbb{C}^\times)\to\mathcal{E}(S,\mathrm{GL}_k(\mathbb{C}))$. Furthermore, the $\mathrm{GL}_k(\mathbb{C})$-local system obtained admits a natural framing given by $$ F(p):=(L_{p_1},L_{p_1}\oplus L_{p_2},\dots,L_{p_1}\oplus\dots \oplus L_{p_k})\; ,$$
hence we obtain a map $\mathcal{E}^{\text{tw}}(\Sigma,\mathbb{C}^\times)\to\mathcal{E}^{\mathrm{fr}}(S,\mathrm{GL}_k(\mathbb{C}))$. To show that this is indeed a non-abelianization map\index{non-abelianization!map}, we need to check that it covers an open and dense subset of $\mathcal{E}^{\mathrm{fr}}(S,\mathrm{GL}_k(\mathbb{C}))$. In the next section, we prove this for spectral networks adapted to ideal triangulations of $S$.

\section{Abelianization map for non-degenerate spectral networks}\index{abelianization!map}
\subsection{Rank two}\labelx{sec:ab_rk_2}

As before, let $S$ be a ciliated surface. We fix an ideal triangulation $\Delta$ on $S$ and let $\pi\colon \bar\Sigma \to \bar S$ be the ramified $2:1$-covering constructed out of $\Delta$ with a spectral network $\mathcal W$.

We take an element of $\mathcal{E}^{\mathrm{fr}}_\Delta(S,\mathrm{GL}_2(\mathbb{C}))$ seen again as a flat vector bundle $(V,\nabla)$ of rank 2 with a framing, i.e. for every puncture $p$ in a small (contractible to $p$) neighborhood $U_p$ of $p$ there is a $\nabla$-invariant line subbundle $l_p$ of $V$.
By the $\Delta$-transversality property defined in \Cref{sec:FG_coord}, for two punctures connected by an edge of $\Delta$, the $\nabla$-invariant line bundles (transported along the edge to some common point) are in the direct sum.

\begin{remark}
A representation into $\mathrm{GL}_2(\mathbb{C})$ may not admit any $\Delta$-transverse framing. For example, this property fails for connections that come from representations that factor through the group of upper-triangular matrices. However, connections admitting a $\Delta$-transverse framing are generic, i.e. they form an open dense subset of $\mathcal{E}(S,\mathrm{GL}_2(\mathbb C))$.   
\end{remark}

Now we take a connected component $C$ of $S\setminus\pi(\mathcal W)$ called a \emph{cell} (see \Cref{fig:cells} (right)). $C$ is a quadrilateral, two vertices of $C$ are branch points, and two other vertices are punctures, say, $p_1$ and $p_2$.  Since $C$ is simply connected, we can take any point $z\in C$ and transport $\nabla$-invariant subbundles $l_1:=l_{p_1}$ and $l_2:=l_{p_2}$ from neighborhoods $U_{p_1}$ and $U_{p_2}$ to $z$. Since $C$ is simply connected, this parallel transport does not depend on a path.

\begin{figure}[!ht]
\centering
\includegraphics[scale=0.8]{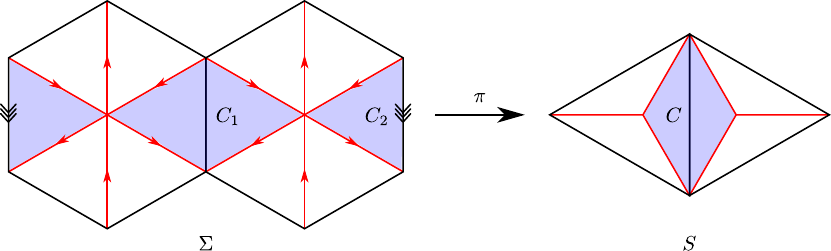}
\caption{Cells in $\Sigma$ and in $S$.}
\labelx{fig:cells}

\end{figure}

The lift of $C$ with respect to $\pi$ consists of two disjoint copies $C_1$ and $C_2$ of $C$ (see \Cref{fig:cells} (left)). We assume that $C_1$ contains the lift of $p_1$ which is a sink and $C_2$ contains the lift of $p_2$ which is a sink. We define the line bundle $L$ over $C_1$ as $l_1$ and over $C_2$ as $l_2$.  

To construct $L$ globally, we need to glue $L$ along lines of the spectral network in a way that preserves the parallel transport. For this, we take two cells $C$ and $C'$ that share an edge $w$ of $\mathcal W$. Then $C$ and $C'$ share a puncture; without loss of generality, we assume that it is $p_2$. The second puncture of $C'$ is denoted by $p'_2$. The corresponding $\nabla$-invariant line bundles over $C'$ are denoted by $l'_1$ (corresponding to $p_1$) and $l'_2$ (corresponding to $p'_2$). We choose sections $s_1$ and $s_2$ of $l_1$ and $l_2$ over $C$ and $s'_1$ and $s'_2$ of $l'_1$ and $l'_2$ over $C'$. We can assume that $s_1$ agrees with $s_1'$ in a neighborhood $U_w$ of $w$ and that $s'_2(z)=s_2(z)+\alpha(z)s_1(z)$, for some function $\alpha\colon U_w\to\mathbb{C}$.

Let now $\nabla s_i=d_i s_i$ and $\nabla s'_i=d'_i s_i$ in $U_w$ for some 1-forms $d_1,d_2,d'_1,d'_2$. We glue $L\vert_C$ and $L\vert_{C'}$ along $U_w$ identifying $s_i$ with $s'_i$, for $i\in\{1,2\}$. This gluing preserves the parallel transport  if and only if $d_1=d'_1$ and $d_2=d'_2$. Since $s_1=s'_1$ in $U_w$, then $d_1=d'_1$. Further, on one hand $\nabla s'_2=\nabla(s_2+\alpha s_1)=\nabla s_2+ \mathrm{d}\alpha\cdot s_1+\alpha\nabla s_1=d_2s_2+ \mathrm{d}\alpha\cdot s_1+\alpha d_1 s_1$. On the other hand, $\nabla s'_2=d'_2s'_2=d'_2s_2+d'_2\alpha s_1$. Since $s_1$ and $s_2$ are linearly independent, then $d_2=d'_2$. Therefore, in this way we can construct a flat line bundle over $\Sigma\setminus B'$ which has monodromy $-1$ around points of $B'$, hence an element of $\mathcal{E}^{\text{tw}}(\Sigma,\mathbb{C}^\times)$. A direct computation shows, that this procedure is inverse to the non-abelianization procedure from \Cref{sec:non-ab}.

\subsection{Higher rank}

Let $S$ be a ciliated surface and let $\Delta$ be an ideal triangulation of $S$. Let $k\geq 2$. Let $\pi : \Sigma_k\to S$ be the ramified $k$-fold covering constructed in \Cref{sec:Combinatorial_higher_rank}. Let $(V,\nabla,F)\in \mathcal{E}^{\mathrm{fr}}_\Delta(S,\mathrm{GL}_k(\mathbb C))$. Once again, we have a framing $F$ of $(V,\nabla)$, i.e. a nested sequence $F^{p}_0\subset F^{p}_1\subset\dots\subset F^{p}_{n}$ of flat subbundles of $V$ in a small neighborhood around every puncture $p$, where $\mathrm{dim} F^{p}_i = i$ for all $1\leq i\leq n-1$.

Let $t$ be a triangle of $\Delta$ and let $(p,q,r)$ be the punctures at the vertices of $t$. Since $t$ is simply connected, we can transport using $\nabla$ the three flag subbundles $F^{p}, F^{q}$ and $F^{r}$ to any common point in $t$. Recall that since $\nabla$ is $\Delta$-transverse, for any triple of non-negative integers $a,b,c$ such that $a+b+c = k$, we have
$$F^{p}_a\oplus F^{q}_b\oplus F^{r}_c = V\; .$$
In particular, this assumption ensures that for any triple of non-negative integers $a,b,c$ such that $a+b+c = 2k+1$, the canonical projection from $F^{p}_a\cap F^{q}_b\cap F^{r}_c$ to $F^{p}_a/F^{p}_{a-1}$ (resp. $F^{q}_b/F^{q}_{b-1}$ and $F^{r}_c/F^{r}_{c-1}$) is an isomorphism.

\begin{remark}
The subset of connections admitting a $\Delta$-transverse framing is open and dense in $\mathcal{E}(S,\mathrm{GL}_k(\mathbb C))$.
\end{remark}

We construct the line bundle $L$ on $\Sigma_k$ as follows. Let $p$ be a puncture of $S$ and let $1\leq a\leq n$. Let $S_{\gamma_p^{(a)}}$ be the punctured disk defined in \Cref{lem:punct_disks}. On this punctured disk the line bundle $L$ is the trivial bundle $F^{p}_a/F^{p}_{a-1}$. Let $e$ be an edge of $\Gamma_k$ and let $t = (p,q,r)$ be the triangle containing $e$. Along $e$ are glued exactly two punctured disks in $\Sigma_k$. Let $S_{\gamma_p^{(a)}}$ and $S_{\gamma_q^{(b)}}$ be those two disks. Inside those disks the line bundle $L$ is respectively $F^{p}_a/F^{p}_{a-1}$ and $F^{q}_b/F^{q}_{b-1}$. They are glued along $e$ via the isomorphism
$$F^{p}_a/F^{p}_{a-1} \cong F^{p}_a\cap F^{q}_b\cap F^{r}_{2k+1-a-b} \cong F^{q}_b/F^{q}_{b-1}\; .$$
This construction of $L$ has locally constant transition function so it defines a flat line bundle $L$ on $\Sigma_k$ minus the vertices of $\Gamma_k$. A simple linear algebra computation shows that the monodromy of $L$ around white vertices of $\Gamma_k$ is trivial, while the monodromy around internal black vertices (the ramification locus of $\pi:\Sigma_k\to S$) is $-1$. We have thus constructed an element of $\mathcal{E}^{\text{tw}}(\Sigma,\mathbb{C}^\times)$. Again, a direct computation shows that the non-abelianization of $L$ is $(V,\nabla)$.

\section{Spectral networks and Fock--Goncharov coordinates}\labelx{Sec:SNandsnakes}

Let $S$ be a ciliated surface, and let $\Delta$ be an ideal triangulation of $S$. The non-abelianization and abelianization maps define a homeomorphism between the moduli space $\mathcal E^{\mathrm{fr}}_\Delta(S,\GL_k(\mathbb{C}))$ of $\Delta$-transverse framed flat vector bundles of rank $k$ on $S$ and the moduli space $\mathcal E^{\text{tw}}(\Sigma_k,\mathbb{C}^\times)$ of flat line bundles on $\Sigma_k$ with monodromy $-1$ around ramification points. The latter moduli space is much easier to study: there is no framing (hence no transversality condition) and the space of line bundles admits a very convenient parameterization as the set of characters of the first homology group $\mathrm{H}_1(\Sigma_k)$ (see \Cref{rk:line_bundles} and \Cref{ex:line_bdle}). In particular, the space $\mathcal E(\Sigma_k,\mathbb{C}^\times)$ is homeomorphic to $(\mathbb{C}^\times)^n$, where $n$ is the dimension of $\mathrm{H}_1(\Sigma_k)$. This homeomorphism is given by the choice of a basis of $\mathrm{H}_1(\Sigma_k)$, which we can choose to be a collection of $n$ curves on $\Sigma_k$. The choice of the holonomy of those $n$ curves then determines uniquely a flat line bundle on $\Sigma_k$, hence a $\Delta$-transverse framed flat vector bundle of rank $k$ on $S$ through the non-abelianization map. This is not the first time we are parameterizing $\mathcal E^{\mathrm{fr}}_\Delta(S,\GL_k(\mathbb{C}))$; we already described one parameterization of this space by Fock--Goncharov coordinates in \Cref{sec:FG_coord_PGLn}. The aim of this section is to show that for a suitable choice of spectral surface $\Sigma_k$ and basis of $\mathrm{H}_1(\Sigma_k)$, those two parameterizations are the same: each Fock--Goncharov coordinate corresponds to the holonomy of a specific curve on $\Sigma_k$, and the collection of all these curves form a basis of $\mathrm{H}_1(\Sigma_k)$.

The relations between spectral networks and Fock--Goncharov coordinates\index{Fock--Goncharov coordinates} is studied by Gaiotto, Moore and Neitzke in \cite{Gaiotto:2012db}. In this article, they use a correspondence between connected components of the complement of a well-chosen spectral network on the base surface $S$ and snakes associated to a framed representation $(V,\nabla,F)$ (as discussed in \Cref{sec:snakes}) to show that the monodromies of the abelianized local system around specific curves on the spectral cover are the $\mathcal{X}$-coordinates of $V$. We follow here a slightly different approach, due to the way we constructed the abelianization map. 

We will first focus on the case of a triple of transverse flags, i.e. when $S$ is an ideal triangle with punctures $p,q,r$. Let $n\geq 2$ and let $(A,B,C)$ be a triple of transverse flags in $\mathbb{C}^n$. As discussed in \Cref{sec:FG_coord_PGLn}, the triple of flags is determined by the data of $\frac{(n-1)(n-2)}{2}$ triple ratios which are called the $\mathcal{X}$-coordinates of the triple. We constructed in \Cref{sec:Combinatorial_higher_rank} a $k$-fold spectral cover $\Sigma_k$ of $S$ together with a spectral network $\mathcal{W}$ on $\Sigma_k$. The trivial local system on $S$ framed by the triple of flags $(A,B,C)$ can then be abelianized to a flat line bundle $L$ on $\Sigma_k$. The surface $\Sigma_k$ has $\frac{n}{2}$ boundary components when $n$ is even and $\frac{n+1}{2}$ boundary components when $n$ is odd, and there are $\frac{n(n-1)}{2}$ simple ramification points on $\Sigma_k$. By the Riemann--Hurwitz formula, the genus of $\Sigma_k$ is $\frac{(n-2)^2}{4}$ when $n$ is even and $\frac{(n-1)(n-3)}{4}$ when $n$ is odd. In both cases, the dimension of the first homology group $\mathrm{H}_1(\Sigma_k)$ (which is also the dimension of the moduli space of flat line bundles on $\Sigma_k$) is $\frac{(n-1)(n-2)}{2}$. Using the notations of \Cref{sec:Combinatorial_higher_rank}, let $a,b,c \geq 0$ such that $a+b+c = 2k+3$, and let $v_{a-1,b,c}$ (resp. $v_{a,b-1,c}$, $v_{a,b,c-1}$) be the black vertex of $\Gamma_k$ sitting on the intersection $\gamma_p^{(a-1)}\cap \gamma_q^{(b)}\cap\gamma_r^{(c)}$ (resp. $\gamma_p^{(a)}\cap \gamma_q^{(b-1)}\cap\gamma_r^{(c)}$,$\gamma_p^{(a)}\cap \gamma_q^{(b)}\cap\gamma_r^{(c-1)}$). Let $\tilde{x}_{a,b,c}$ be the loop on $S$ going around those three ramification points (see \Cref{fig:curve_triple_ratio}).

\begin{figure}[!ht]
\centering
\includegraphics{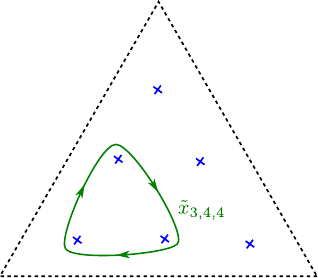}
\caption{The path $\tilde{x}_{3,4,4}$ associated to $\Gamma_4$. The ramification points (i.e. the internal black vertices) are represented by blue crosses.}
\labelx{fig:curve_triple_ratio}

\end{figure}

Lifting the path $\tilde{x}_{a,b,c}$ to $\Sigma_k$ (using the standard path-lifting rule, not the spectral network one), we obtain $k$ different lifts, $k-2$ of which are closed loops in $\Sigma_k$ and two of which are not loops, but form a loop once concatenated. We denote by $x_{a,b,c}$ the loop obtained by concatenating those two lifts. The following proposition follows from the definition of the abelianized local system $L$ and the one of the triple ratio:

\begin{proposition}
    The monodromy of $L$ along $x_{a,b,c}$ is equal to the triple ratio $r_{a,b,c}(F(p),F(q),F(r))$.
\end{proposition}

Let us now consider two triangles $t_1$ and $t_2$ sharing an edge with punctures $p$ and $q$. Let $r$ and $s$ be the two remaining punctures such that $t_1=(p,q,r)$ and $t_2 = (p,s,q)$, and let $(F(p),F(s),F(q),F(r))$ be a transverse framing of the trivial local system on $S = t_1\cup t_2$. Let $\Gamma_k$ be the graph on $S$ constructed in \Cref{sec:Combinatorial_higher_rank}. For $a,b\geq 0$ such that $a+b = k+2$, let $\tilde{x}_{a,b}$ be a loop going around the two internal black vertices $v_{a,b,n}^{(t_1)}$ and $v_{a,n,b}^{(t_2)}$, as in \Cref{fig:curve_cross_ratio}.

\begin{figure}[!ht]
\centering
\includegraphics{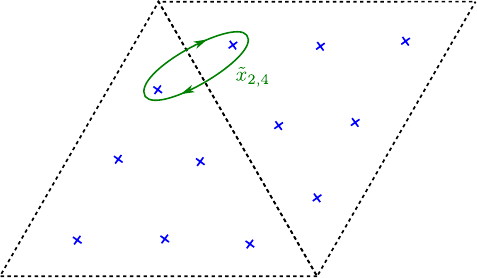}
\caption{The path $\tilde{x}_{2,4}$ associated to $\Gamma_4$. The ramification points (i.e. the internal black vertices) are represented by blue crosses.}
\labelx{fig:curve_cross_ratio}

\end{figure}

Again, there are $k$ standard lifts of $\tilde{x}_{a,b}$ to $\Sigma_k$, exactly two of which are not contractible loops. We choose one of these lifts $x_{a,b}$ on $\Sigma_k$. From our construction of the abelianized local system $L$ on $\Sigma_k$, we have the following proposition: 

\begin{proposition}
    The monodromy of $L$ along $x_{a,b}$ is equal to the cross ratio $r_{a,b}(F(p),F(s),F(q),F(r))$.
\end{proposition}

From a thorough study of the topology of the spectral cover $\Sigma_k$, one can show that the family obtained by taking all the possible loops $x_{a,b,c}$ and $x_{a,b}$ forms a basis of $\mathrm{H}_1(\Sigma_k)$. Thus, the Fock--Goncharov coordinates are exactly the monodromies around a basis of the first homology group of the spectral surface $\Sigma_k$.

\begin{remark}
    This correspondence is specific to the particular choice of the spectral surface $\Sigma_k$. It is in general not easy to relate the Fock--Goncharov coordinates to holonomies of the abelianization over a different spectral surface. It is also not known whether the other cluster variables admit a similar geometric description.
\end{remark}

\begin{remark}
    In \cite{Gaiotto:2012db}, Gaiotto, Moore and Neitzke show that for $k\leq 5$, the spectral surface $\Sigma_k$ and its spectral network are WKB-spectral networks, in the sense of \Cref{sec:analytic_con}. This conjecturally stays true for $k>5$. They also show that these spectral networks can be obtained by perturbation of the ``level $k$ lift'' of the Argyres--Douglas theory $AD_1$.
\end{remark}

\section{Abelianization map for Fenchel--Nielsen spectral networks}\labelx{sec:abelian_FNsn}

We will now construct an abelianization map associated to a Fenchel--Nielsen spectral network $\mathcal{W}$ (with a chosen American or British resolution) on a \emph{closed} surface $S$. We follow a very similar approach to the one presented in \Cref{sec:ab_rk_2} for rank 2 non-degenerate networks. As for the non-degenerate case, we will need some extra data to the one of a flat $\GL_2$-bundle over $S$, which we call a \emph{framing} as for non-closed surfaces. For this, we note that the complement of $\mathcal{W}$ in $S$ is a disjoint union of finitely many connected components each with the topology of an annulus, and the lift to $\Sigma$ of any of those components $C$ is a disjoint union of two annuli $C_1,C_2\subset \Sigma$. We use the resolution of the spectral network to make a choice of orientation of $C$:  the boundary of either $C_1$ or $C_2$ is made out of first halves of spectral lines, i.e. images of $p_h((-1,0])$ for some lines $p_h\in\mathcal{W}$. Up to relabeling, we call this lift $C_1$, and orient $C$ according to the orientation of the boundary of $C_1$ induced by the spectral lines. We call this orientation of $C$ \emph{positive} (or $+$) and the opposite orientation \emph{negative} (or $-$).  
\begin{definition}
    Let $V\in \mathcal{E}(S,\GL_2(\mathbb C))$ be a flat $\GL_2(\mathbb C)$-bundle over $S$.  A \emph{framing}\index{framing} of $V$ is the choice for each such oriented annulus $C$ of an eigenline $\ell_+(C)$ of the monodromy $M_+$ of $V$ around $C$ in the $+$ direction, and an eigenline $\ell_-(C)$ of the monodromy $M_{-}=M_{+}^{-1}$ in the $-$ direction. We ask that $\ell_+\neq \ell_-$, and that the framing associated to both sides of a double wall are transverse.
\end{definition}

\begin{remark}
    For the flat $\GL_2(\mathbb C)$-bundle to admit a framing, the monodromy around every annulus must be diagonalisable. This is the case in particular if the associated representation is fuchsian, as the monodromy around every closed curve will be a hyperbolic element of $\SL_2(\mathbb C)$. 
\end{remark}

Let $V\in \mathcal{E}(S,\GL_2(\mathbb C))$ and let $\ell$ be a framing of $V$. We want to construct a flat line bundle $L$ over $\Sigma$. We first construct this bundle over the complement of $\mathcal{W}$. Each connected component $C$ of $S\setminus \mathcal{W}$ has two lifts $C_1$ and $C_2$ labeled as described above. We define $L$ to be the trivial (flat) bundle equal to $\ell_+(C)$ on $C_1$ and equal to $\ell_-(C)$ on $C_2$. We now want to glue the flat line bundle we just constructed along the lines of the spectral network, which are all double walls. Let $C,C'$ be two connected components of $S\setminus \mathcal{W}$ separated by a double wall. The two lifts of double wall on $\Sigma$ separate $C_1$ and $C'_2$ (resp. $C_2$ and $C'_1$), and we glue $\ell_+(C)$ to $\ell_-(C')$ using the isomorphism:

$$\begin{array}{lcccr}
     \ell_+(C)& \to & \ell_+(C')\oplus\ell_-(C')&\to&\ell_-(C') \\
      x & \mapsto & x = x'_+ + x'_- & \mapsto & x'_- \; .
\end{array}$$

We glue similarly $\ell_-(C)$ to $\ell_+(C')$. The gluing maps are locally constant, hence the resulting line bundle $L\to \Sigma$ carries a flat connection extending the trivial connection on the lift of each annuli. The monodromy around a ramification point is $-1$, thus we obtain an element of $\mathcal{E}^{\text{tw}}(\Sigma,\mathbb{C}^\times)$ which we call the \emph{abelianization} of $V$.

\section{Non-abelianization map for Fenchel--Nielsen spectral networks}

We define the non-abelianization map for Fenchel--Nielsen spectral networks in a slightly different way than the one we used for non-degenerate networks. This is due to the fact that a path-lifting rule adapted to a Fenchel--Nielsen spectral network will in general create infinitely many lifts, hence would require us to make sense of an infinite formal sum of paths on the spectral cover. The reader who wants to follow this path-lifting approach can read the details in \cite{horn2025}. We choose here a different approach not only for the sake of simplicity but also to demonstrate another way to carry this construction: solving for soliton content.

Let $S$ be a closed surface, let $\mathcal{W}$ be a Fenchel--Nielsen spectral network on $S$ (together with a resolution) and $\pi \colon \Sigma\to S$ be the spectral cover. Let $L \in \mathcal{E}^{\text{tw}}(\Sigma,\mathbb{C}^\times)$ be a flat line bundle over $\Sigma$. We choose a branch cut and a numbering of the two sheets of $\Sigma$, each line of the spectral network is then labeled by an ordered pair $(1,2)$ or $(2,1)$ given by the sheet they start at then the sheet they end to, and we orient the lines on $S$ such that they point away from the branch point they pass through. We first push forward the line bundle $L$ to $S\setminus \mathcal{W}$ and obtain a rank 2 flat vector bundle $\pi_*L$ on $S\setminus \mathcal{W}$. We now want to associate to each line $p_h$ labeled $(i,j)$ of $\mathcal{W}$ a map $e_h : L(z_i)\to L(z'_j)$ where $z$ (resp. $z'$) is a point close to $p_h$ on its right (resp. its left). Crossing a double wall on $S$ will then induce a map of the form:
$$\begin{pmatrix}
    1 & e_h\\
    0 & 1
\end{pmatrix}\begin{pmatrix}
    1 & 0\\
    -e_{h'} & 1
\end{pmatrix}\; .$$

Now we impose the additional constraint that the monodromy of the rank 2 bundle on $S$ around any branch point must be trivial. For each of the two elementary pieces of a Fenchel--Nielsen spectral network (molecule I and II, see \Cref{fig:molecules}), this gives two 2-by-2 matrix-valued equations with 6 undetermined variables, the 6 maps $e_h$ associated to the 6 lines of the spectral network in an elementary piece. These equations admit a unique solution. The reader can find detailed computations of these solutions in \cite{HoNe}. Gluing the bundle $\pi_*L$ along the maps associated to each double wall results in a flat rank 2 bundle $V$ over $S$. Since this bundle restricted to a connected component $C$ of $S\setminus \mathcal{W}$ is a trivial bundle given by $\pi_*L|_{C_1\cup C_2}$, it is endowed with a splitting into two flat line subbundles, both preserved by the monodromy around $C$. This shows that the bundle $V$ obtained is naturally endowed with a framing. We have thus constructed an element of $\mathcal{E}(S,\GL_2(\mathbb C))$ which we call the \emph{non-abelianization} of $L$.

\section{Spectral networks and Fenchel--Nielsen coordinates }\labelx{sec:FN_SN_and_FN_coord}

For a Fenchel--Nielsen spectral network, we will see next how we can obtain complexified Fenchel--Nielsen coordinates on moduli spaces of flat $\mathrm{SL}_2(\mathbb{C})$-connections, following \cite[Section 8.4]{HoNe}. 

Remember that in \Cref{Sec:SNandsnakes} we used a $k$-fold spectral covering of a ciliated surface $S$ with an ideal triangulation $\Delta$ and a spectral network on $\Sigma_k$ to obtain Fock--Goncharov coordinates on the space $\mathcal{E}_{\Delta}^{\mathrm{fr}}(S, \mathrm{GL}_k(\mathbb{C}))$ as monodromies around a basis of $\mathrm{H}_1(\Sigma_k)$. 

Here, we take a closed surface $S$ and a Fenchel--Nielsen spectral network $\mathcal{W}$ on $S$ with a chosen American or British resolution. Consider a point $\nabla$ in the space of flat $\mathrm{GL}_2(\mathbb{C})$-bundles over $S$ equipped with a framing. The abelianization procedure of \Cref{sec:abelian_FNsn} is providing an element $\nabla^{\mathrm{ab}}$ of $\mathcal{E}^{\text{tw}}(\Sigma,\mathbb{C}^\times)$. The holonomy of $\nabla^{\mathrm{ab}}$ along a lift of a path going around the annulus in the positive direction is equal to the monodromy eigenvalue of $M_+$. Thus, this monodromy eigenvalue is a complexified version of a square root of the exponentiated \emph{length} coordinate of $\nabla$. Note that in the light of the abelianization map, such Fenchel--Nielsen length coordinates are canonically obtained given a Fenchel--Nielsen spectral network. Yet, these only give half of the basis elements for $\mathrm{H}_1(\Sigma_2)$ (unlike the case of Fock--Goncharov coordinates where the collection of all related curves was providing a basis for $\mathrm{H}_1(\Sigma_k)$). In other words, the choice of a Fenchel--Nielsen spectral network and of a pants decomposition alone do not suffice for giving a complete coordinate system on the moduli space of flat $\mathrm{PSL}_2(\mathbb{C})$-connections.

The other half of the basis elements are not canonically obtained and emerge from considering the holonomy of $\nabla^{\mathrm{ab}}$ along a cycle $\gamma$, under a certain modification of the connection $\nabla$, called the \emph{twist flow}. This modification appears when cutting the surface $S$ along an annulus, and gluing back the connections over the boundary components using automorphisms that preserve the monodromy eigenspaces. Note that in the way we obtained the abelianization in \Cref{sec:abelian_FNsn}, the trivial such automorphism was chosen.

Therefore, provided this way of cutting and gluing, the original connection $\nabla$ can be always abelianized to a connection $\nabla^{\mathrm{ab}}$ but the parallel transport is affected by the ambiguity of an automorphism as described above. In fact, one obtains a 1-parameter family of modified connections $\nabla (\lambda)$, and in terms of sections $s_1, s_2$ of the bundle over the annulus, the action of such an automorphism can be written as
\[s_1 \to \lambda s_1, \quad s_2\to \lambda^{-1}s_2\; .\]
For a cycle $\gamma \in \mathrm{H}_1(\Sigma_2)$ as in \Cref{fig:cycle_twist}, one checks that the twist flow acts on the holonomy $\mathcal{X}_{\gamma}$ of $\nabla^{\mathrm{ab}}$ by 
\[\mathcal{X}_{\gamma}\mapsto \lambda^2 \mathcal{X}_{\gamma}\; .\]
This transformation law under the twist flow gives a complexified exponentiated Fenchel--Nielsen \emph{twist} coordinate. Finally, changing the choice of the cycle $\gamma$ while keeping the network $\mathcal{W}$ fixed, multiplies the twist coordinate by a power of the exponentiated length coordinate. Detailed examples of Fenchel--Nielsen spectral networks leading to such length-twist coordinates can be found in \cite[Sect. 9.3]{HoNe}.

\begin{figure}[!ht]
\centering
\includegraphics{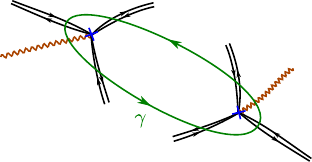}
\caption{The cycle $\gamma$ whose lift's monodromy represents the twist parameter.}
\labelx{fig:cycle_twist}

\end{figure}

\begin{remark}
In fact, holonomies of the abelianizations $\nabla^{\mathrm{ab}}$ above contain slightly more information than the Fenchel--Nielsen coordinates from \Cref{sec: Fenchel--Nielsen}, since the length-twist coordinates described here via the abelianization procedure give a full $\mathrm{SL}_2(\mathbb{C})$-connection up to equivalence, while the complexified Fenchel--Nielsen coordinates determine only its projection to $\mathrm{PSL}_2(\mathbb{C})$.    
\end{remark}

\begin{remark}
A strong physical motivation for studying Fenchel--Nielsen spectral networks $\mathcal{W}(\phi)$, for a Strebel differential $\phi$ as above, is that the corresponding complexified Fenchel--Nielsen coordinates are linked to a significant ``real'' subspace inside the Coulomb branch of the related $\mathcal{N}=2$ supersymmetric field theory. This provides a new approach in understanding the AGT correspondence from a rather geometric viewpoint; we refer to \cite[Sect. 1.2]{HoNe} and the references therein for more information on this perspective.
\end{remark}

\chapter{WKB method and Stokes graphs}\labelx{sec:WKB_Stokes}


\abstract{Spectral networks appear in the theory of linear differential equations on Riemann surfaces. A special approach to solve these differential equations is the WKB method giving formal solutions. Using Borel resummation, these formal solutions can be converted into exact solutions on the complement of a graph -- the spectral network. We explore this approach and link it to the abelianization process. We will mainly stick to the case of $\mathrm{SL}_2(\C)$.}

\section{Overview}

In this chapter, we will see another viewpoint on the abelianization process of flat connections and the appearance of spectral networks from the analysis of differential equations. 

The very rough line of reasoning is the following:

\vspace{0.5cm}

\begin{center}
\begin{tikzpicture}
\node[draw, align=center] (C) at (-4.5,0) {Flat connection \\ in rank $n$ bundle}; 
\node[draw, align=center] (G) at (-0.8,0) {Differential equation \\ for flat sections};
\node[draw] (T) at (4,0) {Abelianization};
\draw[->, >=latex] (C)--(G); 
\draw[->, >=latex] (G)--(T); 
\draw[above] (1.8,0) node {WKB};
\draw[below] (1.8,0) node {method};
\end{tikzpicture}
\end{center}

\vspace{0.5cm}

The character variety describes gauge-classes of flat connections by the Riemann--Hilbert correspondence (see \Cref{sec:RH}).
An $\mathrm{SL}_n(\C)$-connection on a rank $n$ bundle over a Riemann surface $S$ is flat if and only if it admits local bases of flat sections. The equation for a flat section leads to a differential equation which allows a global meaning on $S$. Applying the WKB method to this differential equation leads naturally to the spectral curve with a line bundle and an abelian connection. Hence we recover the abelianization process from the previous \Cref{sec:non-abel}.

From a physical perspective, \emph{the WKB method makes a link between classical physics to quantum physics}. More precisely, it links the wavefunction in quantum mechanics, which is the solution to some differential equation (typically the Schr\"odinger equation), to the generating function of some Lagrangian submanifold of the classical phase space. This process is invertible: starting from the classical data, we can recover the wavefunction.

The WKB method can be performed formally or exactly. The formal WKB method reconstructs the wavefunction as a formal power series, while the exact WKB method gives a well-defined analytic function. The main concepts involved are summarized in the following diagram:

\vspace{0.5cm}

\begin{center}
\begin{tikzpicture}
\node[draw] (W) at (0,0) {WKB method}; 
\node[draw, align=center] (K) at (-3,-1.6) {Hamilton--Jacobi \\ equation};
\node[draw, align=center] (B) at (3,-1.6) {Borel resummation \\ method};
\node[draw] (G) at (-3,-3.4) {recursion};
\node[draw, align=center] (M) at (3,-3.4) {Stokes graph = \\ spectral network};
\draw[->, >=latex] (W)--(K); 
\draw[->, >=latex] (W)--(B); 
\draw[->, >=latex] (K)--(G); 
\draw[->, >=latex] (B)--(M); 
\draw[left] (-1.4,-0.55) node {formal};
\draw[right] (1.4,-0.55) node {exact};
\end{tikzpicture}
\end{center}

\vspace{0.5cm}

The WKB method gives formal solutions which are obtained by solving a Hamilton--Jacobi equation (which in the case of a Riemann surface is simply a polynomial equation) and a recursion. These solutions have typically a radius of convergence zero. To find exact solutions, Borel resummation is used. For $\mathrm{SL}_2(\C)$, it turns out that the Borel resummation method converges in $S\setminus \Gamma$, where $\Gamma$ is a special graph, called the \emph{Stokes graph}; it is a spectral network of type $A_1$. Conjecturally this stays true for higher rank.

In order to stay within the scope of the book, we can only sketch the main ideas without giving the proofs which are very technical. The research area of Borel summability and exact WKB is very active, and many conjectures remain open, namely, for differential equations of order at least 3.

\medskip
We invite the interested reader to consult the following references for more details. We only give a very incomplete list of achievements; a more complete literature overview can be found in \cite{Nikolaev}.

The original \emph{WKB method}, intended to give approximate solutions to equations appearing in quantum mechanics, was developed in 1926 independently by Wentzel \cite{Wentzel}, Kramers \cite{Kramers} and Brillouin \cite{Brillouin}. 
In the early 1980s, Voros \cite{voros1983return} introduced the \emph{exact WKB method}. Around the same time, Ecalle \cite{Ecalle:1992} introduced the space of resurgent functions and developed the so-called alien calculus. We also recommend Sauzin \cite{Sauzin:2016} for a nice introduction to Ecalle's work on resurgence. For more physical-minded readers, we recommend the lecture notes of Dorigoni \cite{Dorigoni} and Serone \cite{Serone:resurgence}, as well as the book by Mari{\~n}o \cite[Chap. 2]{Marino-book}. More recent developments include the link to spectral networks, abelianization and cluster Fock--Goncharov coordinates. In particular, the relation between the exact WKB method and spectral networks was first given in detail by Hollands--Neitzke \cite{hollands2020exact}; we also recommend the foundational paper by Gaiotto--Moore--Neitzke \cite{Gaiotto:2012rg}, Hollands--R\"uter--Szabo \cite[Sect. 4.4]{HRS21}, the work of Iwaki--Nakanishi \cite{iwaki2014exact}, the book by Kawai--Takei \cite{KT} and the recent geometric viewpoint of Nikolaev \cite{Nikolaev_resurge}.

\section{Formal local WKB}

Named after G. Wentzel, H. Kramers and L. Brillouin, the \emph{WKB method}\index{WKB!method} was developed in 1926 as a way to find approximate solutions to the Schrödinger equation. More generally, it provides a tool to solve any linear differential equation with parameter $h$ (thought as being small), coming typically from quantum mechanics where $h$ is Planck's constant. 

Consider a linear differential equation. The historical example is the time-independent (or stationary) 1-dimensional \emph{Schrödinger equation}\index{Schrödinger equation}:
\begin{equation}\labelx{Eq:schroedinger1}
(h^2\partial^2-\hat{t}(z,h))\psi(z,h)=0\; .
\end{equation}
Here, $z$ is a local coordinate on an open set of $\mathbb{C}$, $\partial=\tfrac{\D}{\D z}$ is the derivative, $h$ is a (small) parameter, $\hat{t}(z,h)$ is a given function (a Taylor series in $h$) and $\psi$ is the function we want to solve for.

More generally, we can consider any linear differential equation in $n$ variables $x_1, x_2, ..., x_n$ with derivatives $\partial_i:=\tfrac{\partial}{\partial x_i}$. We can write such an equation in the following compact form:

\begin{equation}\labelx{Eq:equa-diff}
P(h\partial_1,...,h\partial_n)\psi(x_1,...,x_n,h) = 0\; ,
\end{equation}
where $P$ is a polynomial evaluated at the operators $h\partial_i$. Again $h\in \C^*$ is a (small) parameter and the coefficients of $P$ are functions of the $x_i$ and of $h$. In the formal setting we will assume that the coefficients are formal power series in $h$. Evaluating $P$ at partial derivatives means to replace the variables by the derivatives where the coefficients of $P$ are put to the left. 

For the Schrödinger equation \eqref{Eq:schroedinger1}, we have $n=1$, $x_1=z$ and $P(X)=X^2-\hat{t}(z,h)$. Indeed, replacing $X$ by $h\partial$ we get \Cref{Eq:schroedinger1}. In the sequel we will write $\psi(x_i,h)$ as a shorthand notation for $\psi(x_1,...,x_n,h)$.

The WKB method studies the following ansatz to get formal solutions: 
\begin{equation}\labelx{WKB-ansatz}
\psi(x_i,h) = \exp\left(\tfrac{1}{h}(s_0(x_i)+h s_1(x_i)+h^2 s_2(x_i)+...)\right)\; .
\end{equation}
We illustrate the method on the Schrödinger equation.

\begin{example}
Consider again the Schrödinger equation in one variable $x_1=z$:
\begin{equation}\labelx{Schroedinger}
(h^2\partial^2-\hat{t}(z,h))\psi(z,h)=0\; ,
\end{equation}
where $z\in U$ an open set in $\C$, $\partial=\frac{\partial}{\partial z}$ and $\hat{t}(z,h)$ is a formal power series with coefficients being holomorphic functions on $U$. Plugging in the ansatz $\psi=\exp\left(\frac{1}{h}s(z)\right)$ with $s(z)=s_0(z)+h s_1(z)+...$, we get 
\begin{equation}\labelx{Schroedinger-WKB}
(\del s)^2+h \del^2s - \hat{t}(z,h) = 0\; .
\end{equation}
Write $\hat{t}=t +\mathcal{O}(h)$, then the $h^0$-term of \Cref{Schroedinger-WKB} gives 
$$(\del s_0)^2-t = 0\; ,$$
to which we get two solutions $\del s_0=\pm \sqrt{t}$. The important feature of the WKB method is that knowing $\hat{t}(z,h)=\sum_i h^i t^{(i)}(z)$ to all orders, we get all $s_i$ (for $i>0$) recursively from $s_0$. For example, the $h$-term of \Cref{Schroedinger-WKB} reads $2\del s_0\del s_1+\del^2 s_0-t^{(1)}=0$. In general, the equation for $s_i$ is $$2\partial s_0\partial s_i +\textstyle\sum_{k=1}^{i-1}\partial s_k\partial s_{i-k} +\partial^2s_{i-1}-t^{(i)}=0\; .$$
Eventually, we get two fundamental formal solutions, one for each choice of $\del s_0$; this is in compliance with the fact that the Schr\"{o}dinger equation is of second order. The integration constants for the $s_i$ can be absorbed in an overall factor (a formal power series in $h$) which scales the fundamental solutions.
\end{example}

In general, plugging in the WKB ansatz \eqref{WKB-ansatz} into the differential equation \eqref{Eq:equa-diff}, the lowest order term gives a non-linear differential equation for $s_0$. All other $s_i$ for $i>0$ are then recursively determined from $s_0$.
The equation for $s_0$ reads
\begin{equation}\labelx{Eq:HamJac}
P(x_i,\del_i s_0)=0\; .
\end{equation}
This is a so-called \emph{Hamilton--Jacobi equation}\index{Hamilton--Jacobi equation}. This equation has a nice interpretation in symplectic geometry, giving a general solution scheme \cite[Chap. 9]{arnold-class-mechanics}.

\section{Interlude: Hamilton--Jacobi equation}

We describe how to reduce the Hamilton--Jacobi equation to a pair of equations, one polynomial and the other an ordinary differential equation. We assume some knowledge from symplectic geometry. Since we will work in one variable afterwards, this section is not necessary in the sequel. We recommend Arnold's book \cite{arnold-class-mechanics}, especially its Chapter 9, for more details.

Consider a symplectic manifold $(M,\omega)$ of dimension $2n$. Locally we can find Darboux coordinates $(x_i,y_i)_{1\leq i\leq n}$. Let $K\subset M$ be a hypersurface locally given by the equation $P(x_i,y_i)=0$.
The aim is to find a Lagrangian submanifold $L$ inside $K$.

This amounts to solving the Hamilton--Jacobi equation. Indeed, locally any Lagrangian submanifold can be described via a generating function $s(x_i)$ which only depends on half of the variables. The Lagrangian $L$ is then defined by $y_i=\partial_i s$. The condition $L\subset K$ then becomes
\begin{equation}\labelx{Eq:HJ-2}
P\left(x_1,...,x_n,\partial_1 s,...,\partial_n s\right)=0\; ,
\end{equation}
which is precisely the Hamilton--Jacobi equation \eqref{Eq:HamJac}.

We can construct $L$ in another way which leads to a solution of the Hamilton--Jacobi equation. First, we must ensure that the initial condition of the differential equation lies in $K$, which turns out to be a polynomial equation. Second, we integrate the symplectic gradient flow of $P$, which is equivalent to solving an ordinary differential equation. The result is the Lagrangian $L$.

Using suitable coordinates, the initial condition of \Cref{Eq:HJ-2} can be described by $s\!\mid_{\{x_1=0\}}=f(x_2,...,x_n)$, where $f$ is some given function. We have to ensure that
$$P\left(0,x_2,...,x_n,\partial_1 s,\partial_2 f,...,\partial_n f\right)=0\; ,$$
which is a polynomial equation for $\partial_1 s\!\mid_{\{x_1=0\}}$. The solution of the latter determines an $(n-1)$-dimensional isotropic submanifold $L'$ in $K$.

Then, consider the flow generated by the symplectic gradient of $P$. Physically, this corresponds to the time evolution of a classical system with Hamiltonian $P$. Since $K$ is a level set of $P$, the flow will stay inside $K$. Integrating this flow amounts to solving an ordinary differential equation. The image of $L'$ under this flow is a Lagrangian submanifold $L$. Indeed, the image of an isotropic subspace under a Hamiltonian flow stays isotropic and the flow direction of $P$ is the kernel of $\omega\mid_K$, since $K$ is a level set of $P$. Hence, $\omega\mid_L=0$ and $\dim(L)=n$, so $L$ is Lagrangian.

\section{Formal global WKB}

Consider a closed Riemann surface $S$. We will make sense of the differential equation \eqref{Eq:equa-diff} on $S$ and then apply the formal WKB method.

The setting simplifies considerably since we work now with a single (complex) variable $z$. Locally \Cref{Eq:equa-diff} now reads
\begin{equation}\labelx{Eq:equa-diff-2}
(h^n\del^n + a_{1}(z,h)h^{n-1}\del^{n-1}+...+a_n(z,h))\psi(z,h) = 0\; ,
\end{equation}
where $\del = \frac{\D}{\D z}$, the $a_i$ are holomorphic functions in $z$ and formal power series in $h$, and $\psi$ is the unknown function we want to solve for.

To give a global meaning to this differential equation, we first rewrite it as a first-order differential equation:
\begin{equation}\labelx{Eq:equa-diff-global}
(h\partial-A)\Psi=0\; ,
\end{equation}
where
\begin{equation}\labelx{Eq:matrix-diff-eq}
A = \begin{pmatrix} 0&1&&\\ &\ddots & \ddots &\\ &&0&1\\ -a_n & -a_{n-1} &\cdots & -a_{1}\end{pmatrix} \;\text{ and }\; \Psi=\begin{pmatrix}\psi\\ h\partial \psi \\ \vdots \\ h^{n-1}\partial^{n-1}\psi\end{pmatrix}\; .
\end{equation}

We interpret $h\del-A$ as an $h$-connection in some holomorphic bundle over $S$. Recall from \Cref{Sec:connections} that a connection is a way to define directional derivatives of sections, i.e. it associates to a section $s$ and a vector field $X\in \Gamma(TS)$ a new section $\nabla_X(s)$. An $h$-connection is very similar, the only difference being in a modified Leibniz rule:
\begin{definition}
    An \emph{$h$-connection}\index{connection!$h$-} on a vector bundle $E$ over $S$ is a map $\nabla\colon \Gamma(E)\times \Gamma(TS)\to\Gamma(E), (s,v)\mapsto \nabla_v(s)$ satisfying
\begin{enumerate}
\item linearity in fibers: $\nabla_v(s+t)=\nabla_v(s)+\nabla_v(t)$,
\item $\mathcal{C}^\infty(S)$-linearity in $v$: $\nabla_{fv+gw}=f\nabla_v+g\nabla_w$,
\item $h$-Leibniz rule: $\nabla_v(fs)=h \, \mathrm{d}f_v(s)+f\nabla_v(s)$,
\end{enumerate}
where $s,t\in \Gamma(E), f,g\in\mathcal{C}^\infty(S)$ and $v,w\in\Gamma(TS)$.
\end{definition}
We recover the notion of a usual connection for $h=1$.

\begin{example}\labelx{Ex:psi-nature}
    Consider again the Schrödinger equation
$$(h^2\del^2+t(h,z))\psi(h,z)=0\; .$$
Rewrite it as a first-order differential equation
$$\left(h\del-\begin{pmatrix} 0 & 1 \\ -t & 0\end{pmatrix}\right)\binom{\psi}{h\del \psi}=0\; .$$
The operator is an $h$-connection in a holomorphic bundle $E$ over $S$. Since we wish to have two linear independent solutions to the differential equation, the connection must be flat. This implies $\delbar t=0$, explaining why we consider holomorphic connections. In addition, the bundle $E$ is the trivial bundle $\mathcal{O}$ since we can use a basis of flat sections to trivialize it. 

Let us determine the global meaning of $\psi$. It is a section of some line bundle $L$. Then $h\del\psi$ is a section of $L\otimes K$ where $K$ denotes the canonical bundle of holomorphic 1-forms. Hence $\Psi$ is a section of $L\oplus L\otimes K=E\cong\mathcal{O}$. The associated determinant bundle is also trivial: $L^2\otimes K\cong \mathcal{O}$. Hence $L=K^{-1/2}$, where we have to choose a square root of $K$ (a line bundle $\ell$ such that $\ell \otimes \ell \cong K$).
\end{example}

The arguments from the example above can be generalized. The global meaning of \Cref{Eq:equa-diff-global} is as follows: $h\del-A$ is a holomorphic $h$-connection in the trivial holomorphic bundle $\mathcal{O}$ of rank $n$ and we look for a flat holomorphic section $\Psi$. 

The initial indeterminate $\psi$ is a holomorphic section of the line bundle $K^{(1-n)/2}$. Indeed, it is a section of some line bundle $L$. Then $h^k\del^k\psi$ is a section of $L\otimes K^k$, so $\Psi$ is a section of $\bigoplus_{k=0}^{n-1} L\otimes K^k$. The associated determinant bundle is given by $L^n\otimes K^{(n-1)n/2}$ which is the trivial bundle $\mathcal{O}$. Hence $L=K^{(1-n)/2}$.

\bigskip
We have seen in \Cref{WKB-ansatz} that the WKB method looks for formal local solutions of the form
\begin{equation}\labelx{Eq:WKB-ansatz-3}
\psi(h,z)=\exp\left(\tfrac{1}{h} s(z,h)\right)\; ,
\end{equation}
where $s(z,h)=s_0(z)+hs_1(z)+h^2s_2(z)+...$ is a formal power series. We will give the global meaning of this ansatz.

The lowest order term gives the Hamilton--Jacobi equation $P(s_0,z)=0$, which is a polynomial equation in this case. This equation should be understood as defining a ramified covering $\pi:\Sigma \to S$, called the \emph{spectral curve}\index{spectral!curve}, on which $s_0$ is well-defined:
\begin{equation}\labelx{Eq:formal.WKB.sol}
\Sigma = \{(p,z)\in T^*S\mid P(p,z)=0\}\; ,
\end{equation}
where $p$ denotes a holomorphic coordinate on the fiber of $T^*S$. 

The spectral curve $\Sigma$ is a ramified covering of degree $n$, the degree of the polynomial $P$ (and of the initial linear differential equation). Locally and away from the ramification points, we can label the leaves from 1 to $n$. We can define global labels using branch cuts.

The global meaning of the WKB ansatz is the following. We look for $n$ different formal solutions of the form
\begin{equation}\labelx{Eq:WKB-global-12}
\psi^{(i)}_{\text{formal}}(z,h)=\exp\left(\tfrac{1}{h}\int_{y_0}^{y_i} \tilde{\lambda}(h)\right)\; ,
\end{equation}
where $y_i\in\pi^{-1}(z)$ is the pre-image of $z$ in the $i$-th sheet of $\Sigma$, $y_0\in \Sigma$ is a basepoint and $\tilde{\lambda}(h)=\lambda_0+h\lambda_1+...$ is a formal power series of 1-forms. 

The important observation is that $\lambda_0$ is the Liouville 1-form of $T^*S$ restricted to the spectral curve $\Sigma$. In other words: using only the Liouville 1-form in \Cref{Eq:WKB-global-12} solves the initial differential equation to highest order. The other 1-forms $\lambda_i$ for $i\geq 1$ are determined in a recursive way.

\begin{example}
    In case of the Schrödinger equation 
    $$(h^2\del^2-t(z,h))\psi(h,z)=0$$
    with $t(z,h)=t_0(z)+ht_1(z)+...$, the WKB method \eqref{Eq:WKB-ansatz-3} gives $$s_0^2(z)-t_0(z)=0\; .$$
The spectral curve $\Sigma$ is defined by $p^2=t_0$. Choosing locally a branch of the square root $\sqrt{t_0(z)}$ is the same as choosing a branch of $\Sigma\to S$. 

The Liouville 1-form restricted to $\Sigma$ is then given by $\lambda_0=\sqrt{t_0}$. Following \Cref{Eq:WKB-global-12}, one can directly check that
$$\exp(h^{-1}\int^z \sqrt{t_0(z)})$$
solves the Schrödinger equation to leading order. The subleading terms of $\tilde{\lambda}$ are given by
$$\tilde{\lambda}=\sqrt{t_0}-h\frac{\partial t_0}{4t_0}-h^2\sqrt{t_0}\frac{5(\partial t_0)^2-4t_0\partial^2 t_0}{32t_0^3}+... \; .$$
\end{example}

In general, we can easily verify that the formal solution $\psi^{(i)}_{\text{formal}}$ is a solution to
\begin{equation}\labelx{Eq:abelian}
(h\del-\lambda^{(i)})\psi^{(i)}=0\; ,
\end{equation}
where $\lambda^{(i)}$ is the restriction of $\tilde{\lambda}$ to the $i$-th sheet of $\Sigma$.
We can interpret this equation as follows: $h\del-\lambda^{(i)}$ is an $h$-connection on a line bundle $L$ over an open set $U$ of the $i$-th leaf of the spectral curve $\Sigma$ and $\psi^{(i)}$ is a section of $L$. Hence, we see that \Cref{Eq:abelian} is an abelianization of the flat connection $h\del-A$ from \Cref{Eq:equa-diff-global}.

The WKB ansatz \eqref{Eq:formal.WKB.sol} gives $n$ formal solutions on $S\setminus P$, where $P$ denotes the set of poles and branch points.
There seems to be a contradiction, since around a branch point, sheets get exchanged, but from the differential equation perspective, these are smooth points, so there should be no monodromy. The resolution of this apparent contradiction is that the WKB solutions are only formal solutions with zero radius of convergence.

The naive hope would be that there is an analytic solution $\psi^{(i)}_{\text{an}}$ to the differential equation for which the WKB formal solution $\psi^{(i)}_{\text{formal}}$ is an asymptotic expansion when $h\to 0$. Unfortunately, such an exact solution does not exist in general. We have to restrict the way $h$ can tend to 0. Due to the exponential function in the WKB ansatz, we have to restrict $h$ to stay in a half-plane as the following example shows.

\begin{example}
Consider the function $f(z)=\E^{-1/z}$. It is a holomorphic function on $\mathbb{C}^*$ with an essential singularity at the origin. For $\mathrm{Re}(z)>0$, the limit of $f(z)$ for $z\to 0$ (while staying in the half plane defined by $\mathrm{Re}(z)>0$) is 0. For $\mathrm{Re}(z)<0$, the same limit diverges.
\end{example}

In the sequel, we will modify the WKB formal solutions to exact solutions which are generically asymptotic to the formal solution in a half-plane. The points on the surface where this asymptotic behavior fails form a combinatorial object - the spectral network.

\section{Exact WKB via Borel resummation}

An important question concerning the WKB method is the convergence. In general, the formal sum in \Cref{WKB-ansatz} never converges; it is merely an asymptotic series. In the 1980's, Voros \cite{voros1983return} introduced the exact WKB method which combines the formal WKB method with the Borel resummation to get effective solutions. 
After some basics on the Laplace transform, we expose the Borel resummation method following \cite{iwaki2014exact}. Physics regorges of examples, where these methods are known under the name of \emph{resurgence} or \emph{alien calculus}, see for example \cite{Marino} or \cite{Dorigoni} for a more mathematical treatment.

\bigskip 
The \emph{Laplace transform}\index{Laplace transform} of a function $f\colon \R^+ \to \C$ is given by
\begin{equation}\labelx{Eq:Laplace-transfo}
\mathcal{L}[f](s)=\int_0^\infty f(t)\E^{-st}\D t\; ,
\end{equation}
where $s\in\C$. This is well-defined for $\mathrm{Re}(s)>0$ and functions $f$ such that $\lvert f(t)\rvert \leq C_1 \E^{C_2t}$, for some positive constants $C_1,C_2$.

One can think of transformation \eqref{Eq:Laplace-transfo} as an analytic continuation of the Fourier transform which we almost\footnote{The Fourier transform is the \emph{bilateral} Laplace transform at $s=\I\xi$, i.e. a version of the Laplace transform where one integrates over $\mathbb{R}$.} get for $s=\I\xi$.

\begin{example}
    A direct computation shows that the Laplace transform of $f(t)=t^q$ with $q\in \mathbb{C}$, $\mathrm{Re}(q)>-1$ is given by $$\mathcal{L}[t^q](s)=\frac{\Gamma(q+1)}{s^{q+1}}\; ,$$ where $\Gamma$ is the Gamma function. In particular, for $q=n\in \mathbb{N}$:
\begin{equation}\labelx{Eq:laplace-t-n}
\mathcal{L}[t^n](s)=\frac{n!}{s^{n+1}}\; .
\end{equation}
\end{example}

As the Fourier transform, the Laplace transform is very useful to solving differential equations. Using integration by parts one sees that
$$\mathcal{L}[f'](s)=s\mathcal{L}f(s)-f(0)\; .$$
Hence applying $\mathcal{L}$ to a linear differential equation transforms it into a polynomial equation. 

Further the Laplace transform is invertible via a contour integral:
$$(\mathcal{L}^{-1}f)(t)=\int_\gamma \E^{st}f(s)\D s\; .$$
The path of integration $\gamma$ is typically of the form $\{z\in \mathbb{C}\mid \mathrm{arg}(z)=\theta_0\}$ for some fixed $\theta_0\in [0,2\pi)$, and must lie inside the region of convergence of the integrand.

The inverse Laplace transform can be used to solve inhomogeneous linear differential equations with constant coefficients. To solve 
\begin{equation}\labelx{Eq:ODE-scheme}
Df(t)=g(t) \; ,
\end{equation}
where $D$ is a linear differential operator with constant coefficients, it is sufficient to find the Green functions $G$, i.e. the solutions to $DG(t)=\delta_0(t)$, where $\delta_0$ is the Dirac distribution supported at zero. The general solution to \Cref{Eq:ODE-scheme} is then the convolution product $G\star g$. 
Applying the Laplace transform to $DG(t)=\delta(t)$ gives $P(t)\mathcal{L}(G)=1$, where $P$ is the polynomial such that $D=P(\tfrac{\D}{\D t})$. Therefore, 
$$G(t)=\mathcal{L}^{-1}(1/P)\; .$$ 
When $P$ has simple zeros, there are $n$ poles of $1/P$, so $n+1$ different (homotopy types of) integration paths for the inverse Laplace transform. This gives $n$ linear independent Green's functions, since the sum of all residues is zero. Note that the Fourier transform gives only one Green function. 
The Green functions coming from the inverse Laplace transform differ by the residues of $\E^{st}/P(s)$. These residues are exactly the fundamental solutions to the homogeneous equation $Df(t)=0$. Such a residue is a constant multiple of $\E^{x_0 t}$ with $x_0$ a root of $P$. Hence we recover the solution scheme for homogeneous differential equations with constant coefficients.

\begin{example}
    Consider $f''(t)+f(t)=\delta_0(t)$. Applying the Laplace transform and putting $F=\mathcal{L}f$ gives $(s^2+1)F=1$. The residues of $\tfrac{\E^{st}}{1+s^2}$ at $\pm \I$ are given by $\pm \tfrac{\E^{\pm \I t}}{2\I}$, which form a basis of solutions to the homogeneous equation $f''(t)+f(t)=0$.
\end{example}

\bigskip
Now we are ready to explain Borel resummation, which is a two-step procedure. The first step is the Borel transform.

\begin{definition}
    A formal power series $f(s)=\sum_{n\geq 0} f_n s^{-n}$ is \emph{Borel summable}\index{Borel!summability} if
    \begin{equation}\labelx{Eq:def-borel-transfo}
    f_B(y)=\sum_{n\geq 1} f_n\frac{y^{n-1}}{(n-1)!}
    \end{equation}
converges near $y=0$ and can be analytically continued to a domain $\Omega$ containing the positive real axis, where $\lvert f_B(y)\rvert \leq C_1 \E^{C_2\lvert y\rvert}$, for some positive constants $C_1, C_2$.
The map $f\mapsto f_B$ is called the \emph{Borel transform}\index{Borel!transform}.
\end{definition}

Comparing the definition \eqref{Eq:def-borel-transfo} with the Laplace transformation of the function $t\mapsto t^n$ from \Cref{Eq:laplace-t-n}, we see why the \emph{Borel transformation is a formal term-wise inverse Laplace transform}.

A typical example of Borel summable functions are formal power series $\sum_{n\geq 0} f_n s^{-n}$ where the coefficients satisfy $\lvert f_n\rvert \leq Cn!\rho^n$ for some positive constants $C$ and $\rho$. Such formal power series are called \emph{Gevrey-1}. For Gevrey-1 power series, the Borel transform has strictly positive radius of convergence. 

Importantly, Borel summable functions appear ubiquitously in perturbative methods in Physics. A typical example are expansions via Feynman diagrams, since it is known that there are roughly $n!$ Feynman diagrams with $n$ nodes.

\medskip
The second step of the Borel resummation procedure amounts to applying the Laplace transform to the (analytically continued) Borel transform of the first step:
\begin{definition}
    The \emph{Borel sum}, or \emph{Borel resummation}\index{Borel!resummation}, of a Borel summable $f(s)=\sum_{n\geq 0}f_n s^{-n}$ is the function in $s$ given by
\begin{equation}\labelx{Eq:def-borel-resum}
    S[f](s) = f_0+\int_0^\infty \E^{-sy}f_B(y) \D y\; .
\end{equation}
\end{definition}
Note that this definition is well-defined for $\mathrm{Re}(s)>0$ since $\R_+$ is contained in the domain of $f_B$, together with an exponential bound there.

\begin{example}
    Consider $f(s)=\sum_{n\geq 0} s^{-n}$ which converges for $\lvert s^{-1}\rvert < 1$. The Borel transform gives $f_B(y)=\sum_{n\geq 1}\frac{y^{n-1}}{(n-1)!}=\E^y$. The Borel sum is then given by $S[f](s)=1+\int_0^\infty \E^{y(1-s)}\D y = 1+\frac{1}{1-s}=\frac{1}{1-s^{-1}}$, for $\mathrm{Re}(s^{-1})<1$. Note that $S[f]$ is an analytic continuation of $f$.
\end{example}

\begin{example}
    Consider the differential equation $$-\tfrac{\D f}{\D s}+f(s)=1/s\; ,$$ known as Euler's differential equation of first order. A formal solution is given by the power series $f(s)=\sum_{n\geq 0} (-1)^n n!s^{-n-1}$. It is easy to check that $f$ is Borel summable with $$S[f](s)=\int_0^\infty \frac{\E^{-su}}{1+u}\D u\; ,$$ giving an exact solution to the Euler equation.
\end{example}

An important feature of Borel resummation is that it is the identity map for series which are already converging:

\begin{proposition}
If $f(s)$ converges to a holomorphic function near $s=\infty$, Borel resummation is the identity: $S[f]=f$.
\end{proposition}
The proof is simply to see that if $f$ converges around $s=\infty$, then the first step is truly the inverse Laplace transform, which gets compensated by the second step.

In the case where the function $f$ is not converging, it gives an asymptotic series to the Borel resummation:
\begin{proposition}[{\cite[Prop. 2.11(b)]{iwaki2014exact}}]
    For $s\to \infty$, we have $S[f](s)\sim f(s)$.
\end{proposition}

Finally, some properties following directly from the definition are the following:
\begin{proposition}[{\cite[Prop. 2.11(a,c)]{iwaki2014exact}}]\labelx{Prop:Borel-sum}
\begin{enumerate}
    \item $S[f+g]=S[f]+S[g]$ and $S[fg]=S[f]S[g]$, for Borel summable $f$ and $g$;
    \item If $A(t)=\sum_{k\geq 0}A_k t^k$ converges near $t=0$ and $f(s)=\sum_{n\geq 1}f_ns^{-n}$ has no constant term, then $S[A\circ f]=A\circ S[f]$.
\end{enumerate}
\end{proposition}

\begin{remark}
    There is a class of functions, called \emph{resurgent}, which is closed under the operations we have used above: Laplace transforms, Borel transforms, composition, but also under analytic continuation and convolution. Roughly speaking, a resurgent function is Gevrey-1 and its Borel transform has only finitely many singularities along rays starting from the origin. See \cite{Dorigoni,Ecalle:1992,Sauzin:2016} for more details.
\end{remark}

There are many variations and extensions of Borel resummation. 
For example, we can define the resummation for functions more general than formal power series. For a formal function $g$ of the form $g(s)=\E^{s\mu}s^{-\rho}f(s)$ for $\mu,\rho\in\C$ and $f$ as above, we define 
\begin{equation}\labelx{Eq:Borel-sum-factor}
    S[g](s):= \E^{s\mu}s^{-\rho}S[f](s)\; .
\end{equation}

Another generalization is to include some angle in step two. Instead of integrating over $\R_+$, we can define the \emph{Borel sum in direction $\theta$}\index{Borel!sum}, where $\theta\in[0,2\pi)$ by:
$$S_\theta[f](s):=f_0+\int_0^{\infty \E^{-\I\theta}} \E^{-sy}f_B(y) \D y\; ,$$
where we integrate over a half-line starting at the origin with angle $-\theta$ to the real axis. There is an adequate notion for $f$ being Borel summable in direction $\theta$. The Borel sum $S_\theta[f]$ is analytic in $\{s\in \C\mid \lvert \mathrm{arg}(s)-\theta \rvert < \pi/2, \lvert s\rvert \gg 1\}$.

Borel resummation with angle $\theta$ is particularly important whenever $f_B$ has a pole along the integration path. There are two choices then: going around the pole above or below (see \Cref{fig:contours}). The two results differ by the residue around the pole. This is the so-called \emph{Stokes phenomenon}\index{Stokes!phenomenon}.

\begin{figure}[!ht]
\centering
\includegraphics[width=\textwidth]{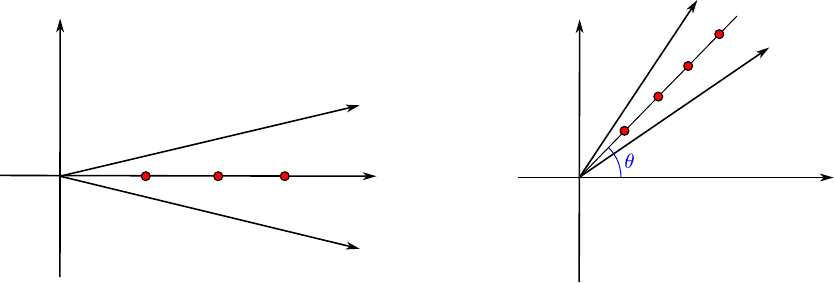}
\caption{Contours for Borel resummation (red dots represent poles).}
\labelx{fig:contours}

\end{figure}

The Stokes phenomenon comes from the non-uniqueness of an analytic function with given asymptotic series. 
This is expressed by the fact that Borel resummation depends on the integration path (in the Laplace transform in the second step). Two non-homotopic paths give in general different results. 

\begin{example}
    Consider the differential equation $$f'(s)+f(s)=\tfrac{1}{s}\; .$$
    A formal solution is given by $f(s)=\sum_{n\geq 0}n! s^{-n-1}$, which is Borel summable. The Borel transform is $f_B(y)=\tfrac{1}{1-y}$, which has a pole on the real axis for $y=1$. One computes that the difference between $S_{0^+}[f]$, the Borel sum going above the pole, and $S_{0^-}[f]$, the Borel sum going below the pole, is given by $$S_{0^+}[f](s)-S_{0^-}[f](s)=2\pi \I \E^{-s}\; .$$
    Note that the difference is exponentially small.
\end{example}

We can slightly generalize the above example. Assume that the Borel transform $f_B(y)$ has a pole at $y=y_0\in \R$. Then the residue of the function $\E^{-sy}f_B(y)$ at $y=y_0$ is given by $\E^{-sy_0}\mathrm{Res}_{y_0}(f_B)$ which is exponentially small when $s\to \infty$. Using an integration path going above or below $y_0$ in the Borel sum gives two different solutions that have the same asymptotics for $s\to\infty$ since their difference is exponentially small.

\begin{remark}
    The Stokes phenomenon can be captured by the so-called \emph{alien derivative}, see for example  \cite[Sects. 4 and 6]{Dorigoni}. It is a derivation linked to the automorphism describing the jump across a Stokes line. The alien derivative can be linked to the usual derivative using \emph{Ecalle's bridge equation}. 

\noindent The theory of resurgence can be seen as a systematic study of Stokes phenomena. It can give meaning to non-Borel summable asymptotic functions. The key idea is to look for a solution to a differential equation of the form
$$\widetilde{F}(h)=\sum_{n=0}^\infty \sigma_n \E^{-a_n/h}f_n(h)\; ,$$
where $\sigma_n$ are constants (called Stokes constants), $a_n$ are the singularities of the Borel transform, and $f_n$ are formal power series. The function $\widetilde{F}$ is a typical example of a \emph{trans-series}. A single $f_n$ might not be Borel summable, but the whole trans-series $\widetilde{F}$ is Borel summable in many situations. The exact resummation of trans-series often encode non-perturbative information although the initial ansatz is perturbative. A nice and concrete example, the Airy function, is discussed in detail in \cite[Sect. 3.1]{Serone:resurgence}.
\end{remark}

\section{Stokes curves for the Schr\"{o}dinger equation}\labelx{Sec:StockescurvesSchroding}

We now restrict to the Schr\"{o}dinger equation and analyze the exact WKB method leading to the so-called \emph{Stokes curves}, which turn out to be a spectral network of type $A_1$. We follow \cite{iwaki2014exact}.

Consider a compact Riemann surface $S$ and choose a square root $K^{1/2}$ of the canonical bundle. We slightly change notation and consider the equation
\begin{equation}
    (\del^2-\eta^2t(z,\eta))\psi(z,\eta)=0\; ,
\end{equation}
where $\psi\in \mathrm{H}^0(S,K^{-1/2})$ and $t(z,\eta)=t_0(z)+\eta^{-1}t_1(z)+...$. We recover the Schr\"{o}dinger equation from above via $\eta=h^{-1}$.

One can check that $\varphi(z)=t_0(z)\D z^2$ is a meromorphic quadratic differential on $S$. This quadratic differential governs the behavior of the WKB method. Note that the poles of $\varphi$ are singular points for the differential equation. Also the zeros of $\varphi$, called \emph{turning points}\index{turning point}, play an important role. Denote by $P_0$ and $P_\infty$ the set of zeros and poles of $\varphi$ and put $P=P_0\cup P_\infty$. 

Finally, we impose some technical conditions which simplify the analysis of the possible trajectories and the connection formula \eqref{Eq:wall-crossing} below. We assume that $\varphi$ has at least one zero and at least one pole, that all zeros are simple and all poles have order at least 2.

As before we consider the WKB ansatz $\psi=\exp\left(\int^z s(z,\eta)\D z\right)$ which leads to two formal local solutions $\psi_{\pm}$. 

We can get formal \emph{global} solutions as follows. Define the odd and even parts of $s$ by $s_{\text{odd}}=\frac{1}{2}(s_+-s_-)$ and $s_{\text{even}}=\frac{1}{2}(s_+ +s_-)$. One can check that $s_{\text{even}}=-\frac{1}{2}\del(\ln s_{\text{odd}})$. In addition $s_{\text{odd}}\D z$ is well-defined globally on the spectral curve $\Sigma$ away from the ramification points \cite[Cor. 2.17]{KT}. Hence,
\begin{equation}\labelx{Eq:global-formal-sol}
\psi_{\pm}(z,\eta) = \frac{1}{\sqrt{s_{\text{odd}}}}\exp\left(\pm \int^z s_{\text{odd}} \D z\right)
\end{equation}
is a formal global solution. Indeed, $\del\psi_+ = (s_{\text{odd}}-\frac{1}{2}\del(\ln s_{\text{odd}}))\psi_+ = s\psi_+$, thus $\del^2 \psi_+ = (s^2+\del s)\psi_+=\eta^2 t\psi_+$. A similar computation holds for $\psi_-$. 
We can further expand this expression into
\begin{equation}\labelx{Eq:WKB-Borel-sum}
    \psi_{\pm}(z,\eta)=\exp\left(\pm \int^z \sqrt{t_0(z)}\D z\right)\eta^{-1/2}\sum_{k=0}^\infty \eta^{-k}\psi_{\pm,k}(z)\; .
\end{equation}
The principal term of this expression is given by
\begin{equation}\labelx{Eq:WKB-approx}
    \psi_{\pm}(z,\eta) = t_0^{-1/4}\eta^{-1/2}\exp\left(\pm \int^z \eta\sqrt{t_0}+\frac{t_1}{2\sqrt{t_0}}\right)\mathcal{O}(1+\eta^{-1})\; ,
\end{equation}
which is often called the \emph{WKB approximation} of the Schr\"{o}dinger equation. It comes from $s_{\text{odd}}= \sqrt{t_0}+\eta^{-1}\frac{t_1}{2\sqrt{t_0}}+\mathcal{O}(\eta^{-2})$.

Note that $s_{\text{odd}}\D z$ is not integrable at a pole of $\varphi$. To remedy this, define the regular part of $s_{\text{odd}}$ to be
$$s^{\text{reg}}_{\text{odd}}(z,\eta)\D z := (s_{\text{odd}}(z)-\eta \sqrt{t_0(z)})\D z\; .$$

\bigskip
Now we use Borel resummation. Fix an angle $\theta\in[0,2\pi)$. Use \Cref{Eq:Borel-sum-factor} to define Borel summability with angle $\theta$ of \Cref{Eq:WKB-Borel-sum}. This will depend on the point $z\in S$. 

As we have seen in \Cref{Sec:Differentials}, to the meromorphic differential $\varphi$ and the angle $\theta$ we can associate a foliation on $S$ by $$F_\theta := \mathrm{ker} \,\mathrm{Re}\left(\E^{\I\theta}\sqrt{\varphi}\right)\; .$$
Recall that this means the following: at every point $z\in S\setminus P_0$ which is not a zero of $\varphi$, there is a unique direction $L(z)$ for which $\E^{\I\theta}\varphi(v)$ is real for all $v\in L(z)$. Integrating the leaves $L(z)$ over $z\in S$ gives the foliation $F_\theta$. In other words, a leaf $\gamma$ of $F_\theta$ is characterized by the fact that the argument of $\int_\gamma \sqrt{\varphi}$ is constant.

From \Cref{Sec:Differentials} we know the different trajectories of the foliation. Recall that a leaf is called \emph{critical} if one endpoint is in $P_0$ (i.e. a zero of $\varphi$). A critical leaf is also called a \emph{Stokes curve}\index{Stokes!curve} or \emph{WKB curve}\index{WKB!curve}. A \emph{saddle}\index{saddle} links two points in $P_0$ and a \emph{regular critical leaf} (also called \emph{separating leaf}) links a point in $P_0$ to a point in $P_\infty$. 
We assume that there is at most one saddle (later on we even assume no saddles at all). 

The \emph{critical graph}\index{critical!graph} or \emph{Stokes graph}\index{Stokes!graph} is the graph with vertex set $P$ and Stokes curves as edges.

The complement of the Stokes graph has three types of regions \cite[Sect. 3.4]{BS13}: strips, half-planes or a disk; see \Cref{fig:regions}. 
These are exactly the critical trajectories of meromorphic quadratic differentials on a Riemann surface (cf. \Cref{Sec:Differentials}).

 \begin{figure}[!ht]
\centering
\includegraphics[scale=0.6]{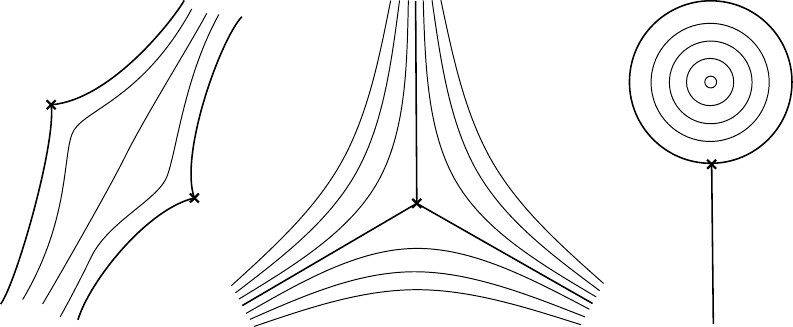}
\caption{The complement of a Stokes graph.}
\labelx{fig:regions}

\end{figure}

The next theorem gives an important link between the exact WKB method and the Stokes graph, which is a spectral network of type $A_1$:

\begin{theorem}[Koike--Sch\"{a}fke]
    The regular part $s^{\text{reg}}_{\text{odd}}(z,\eta)$ is Borel summable in direction $\theta$ for each $z\in S$ in the complement of the Stokes graph. The Borel sum is holomorphic on $\{\eta\in \C\mid \lvert \mathrm{arg}(\eta)-\theta\rvert < \pi/2, \lvert\eta\rvert \gg 1\}$ and around $z$. Finally, if $p\in P_\infty$ is a pole, then $\int_p^z s^{\text{reg}}_{\text{odd}}(z,\eta)\D z$ is Borel summable with the same properties.
\end{theorem}
This theorem has been communicated by T. Koike at the RIMS workshop ``Exact WKB analysis – Borel summability of WKB solutions'' in September 2010. The authors never published the result. In case of a quadratic potential $t(z,\eta)=t_0(z)+\eta^{-1}t_1(z)+\eta^{-2}t_2(z)$, this has been proven in \cite{Nemes}. The general case has recently been proven by Nikolaev \cite{Nikolaev, Nikolaev_resurge}, using geometric methods.

Eventually we want to prove Borel summability of $\psi_\pm$. Since Borel summability passes to products and exponentials (by \Cref{Prop:Borel-sum}), we only have to analyze $\int^z s^{\text{reg}}_{\text{odd}}(z,\eta)\D z$, where we integrate over a path in the spectral curve $\Sigma$.

Define an \emph{admissible path} to be a path in $\Sigma$ whose projection onto $S$ does intersect the Stokes graph only at poles.

\begin{proposition}[{\cite[Lem. 2.20]{iwaki2014exact}}]
    If the endpoints of a path $\beta$ do not project to zeros and if the latter never intersects a saddle, then $\beta$ is a sum of admissible paths.
\end{proposition}

\begin{corollary}
    If $\beta$ is a sum of admissible paths, then $\int_\beta s^{\text{reg}}_{\text{odd}}(z,\eta)\D z$ is Borel summable.
\end{corollary}
In particular, if there is no saddle trajectory, then the normalized WKB solutions are Borel summable.

\bigskip
To sum up, fixing an angle $\theta$ gives a Stokes graph which is a spectral network associated to the quadratic differential $\varphi(z)=t_0(z)\D z^2$ and angle $\theta$. In the complementary regions, the formal solutions from the WKB method are Borel summable, hence giving exact analytic solutions to the Schr\"odinger equation. 

This picture stays conjecturally true for higher-order differential equations: 
the formal solutions from the WKB method are Borel summable in the complementary regions of a spectral network. Unfortunately. the analysis for higher-order differential equations is much harder to carry out. A more precise statement can be made in the special case of opers, see \Cref{Sec:opers} below.

There are two interesting phenomena to study: what happens when one crosses a Stokes curve and what happens when we vary the angle $\theta$?

\section{Connection formula and Voros symbols}\labelx{Sec:connformulaVorossymb}

We continue in the same setting as in the previous subsection. 

Suppose the Stokes graph has no saddles. Then we can connect the Borel sum between two adjacent regions $D_1$ and $D_2$. Denote by $\Psi_\pm^{D_i}$ the two exact solutions in $D_i$, for $i=1,2$. Denote by $C$ the Stokes curve between $D_1$ and $D_2$, see \Cref{fig:connection-formula}.

\begin{figure}[!ht]
\centering
\includegraphics{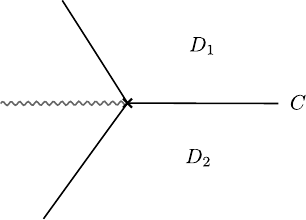}
\caption{Connection formula.}
\labelx{fig:connection-formula}

\end{figure}

The solutions in different regions can be linked to each other by the following proposition.
This kind of formulas are called \emph{connection formulas}\index{connection!formula}.

\begin{proposition}[{\cite[Thm. 2.25]{iwaki2014exact}}]
    If $\mathrm{Re}\left(\E^{\I\theta}\int \sqrt{\varphi}\right)\geq 0$ on $C$ then
\begin{equation}\labelx{Eq:wall-crossing}
    \Psi_+^{D_1}=\Psi_+^{D_2}+\I\Psi_-^{D_2} \;\;\text{ and }\;\; \Psi_-^{D_1}=\Psi_-^{D_2}\; .
\end{equation}
Otherwise, we get:
\begin{equation}
    \Psi_+^{D_1}=\Psi_+^{D_2}\;\;\text{ and }\;\; \Psi_-^{D_1}=\Psi_-^{D_2}+\I\Psi_+^{D_2}\; .
\end{equation}
\end{proposition}

\bigskip
Let us analyze the change of the Stokes graph and the exact solutions when we vary the angle $\theta$.

When there is at most one saddle, we have already seen that the complementary regions are strips, half-planes or disks. Varying $\theta$ can introduce extra saddle connections. There are two possible changes in the topology of the Stokes graph: a \emph{flip}\index{flip} and a \emph{pop}\index{pop}, see \Cref{fig:WKB-flip} and \Cref{fig:WKB-pop}. 

\begin{figure}[!ht]
\centering
\includegraphics[scale=0.6]{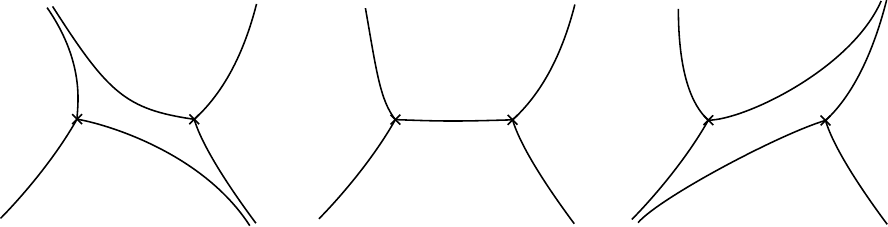}
\caption{Changes in Stokes graph: flip.}
\labelx{fig:WKB-flip}

\end{figure}

\begin{figure}[!ht]
\centering
\includegraphics[scale=0.6]{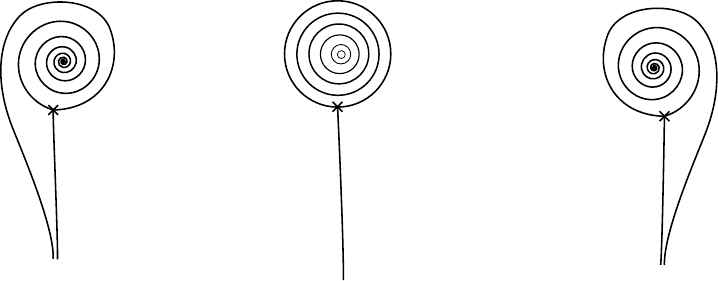}
\caption{Changes in Stokes graph: pop.}
\labelx{fig:WKB-pop}

\end{figure}

To describe how the exact WKB solutions change, we consider special functions called \emph{Voros symbols}\index{Voros symbols}. Denote by $\tilde{P}, \tilde{P}_0$ and $\tilde{P}_\infty$ the pre-images of $P=P_0\cup P_\infty, P_0$ and $P_\infty$ under the projection map $\pi:\Sigma\to S$. Let $\gamma\in \mathrm{H}_1(\Sigma\setminus \tilde{P};\Z)$ and $\beta\in \mathrm{H}_1(\Sigma\setminus \tilde{P}_0,\tilde{P}_\infty;\Z)$, i.e. a path in $\Sigma\setminus \tilde{P}_0$ with endpoints in $\tilde{P}_\infty$.

The \emph{Voros symbols} or \emph{period coordinates}\index{period coordinates} associated to $\beta$ and $\gamma$ are defined to be
$$W_\beta := \exp\left(\int_\beta s_{\text{odd}}^{\text{reg}}\D z\right) \;\;\text{ and }\;\; V_\gamma:= \exp\left(\int_\gamma s_{\text{odd}}^{\text{reg}}\D z\right)\; .$$
Here, we use the exact WKB method to make sense of $W_\beta$ and $V_\gamma$. Note that the Voros symbols are well-defined since the integrant $s_{\text{odd}}^{\text{reg}}\D z$ is invariant under the involution on $\Sigma$ (which is a ramified double cover of $S$). Especially important are the Voros symbols $V_\gamma$ where $\gamma$ is a cycle around a saddle trajectory; see \Cref{fig:Voros-symbols}.

\begin{figure}[!ht]
\centering
\includegraphics[scale=0.83]{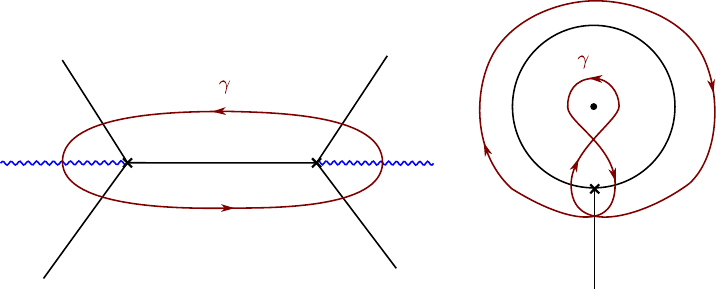}
\caption{Voros symbols.}
\labelx{fig:Voros-symbols}

\end{figure}

The Voros symbols are coordinates on the space of flat connections. Under a flip, they change via the $\mathcal{X}$-mutation rule (\Cref{Eq:cluster-X-mutation} in \Cref{sec:Clusters}). Hence they can be identified with the \emph{Fock--Goncharov coordinates}\index{Fock--Goncharov coordinates} for the $\mathrm{SL}_2(\mathbb{C})$-character variety; see \Cref{Sec:FockGoncharov}. This is the main result of Iwaki--Nakanishi \cite{iwaki2014exact}.

This stays conjecturally true in higher rank: it is known that Fock--Goncharov coordinates can be described by periods of a spectral curve. This spectral curve should be identical to the one appearing in the WKB method and the exact WKB solutions around special closed curves should be identified with the Fock--Goncharov coordinates.

\section{Opers}\labelx{Sec:opers}

An important particular example of differential equations on Riemann surfaces are given by opers. They generalize the Schrödinger equation to higher-order differential equations and allow to formulate precise conjectures.

Start again from a differential equation of order $n$ in one variable $z$ defined in an open set $U\subset \mathbb{C}$:
$$(a_0(z,h)\partial^n + a_{1}(z,h)\partial^{n-1}+...+a_{n-1}(z,h)\partial+a_n(z,h))\psi(z,h)=0\; ,$$
where the coefficients are holomorphic functions (we can also consider more general functions) and $a_0$ is nowhere vanishing.
The space of holomorphic functions acts by pre- and postcomposition on the space of these operators. Using these two actions allows to put $a_0=1$ (divide by $1/a_0$) and $a_{1}=0$ (by an appropriate shift).

The remaining parameters $(a_2,a_3,...,a_n)$ do not transform as tensors when changing the coordinate $z$ to $w(z)$. Their transformation behavior is complicated, but there is a way to combine these coefficients and their derivatives to get tensors. For that, consider again the matrix version of the differential operator (as in \Cref{Eq:matrix-diff-eq}):
\begin{equation}\labelx{Eq:matrix-diff-eq-2}
\partial+\begin{pmatrix} 0&1&0&0\\ \vdots&\ddots & \ddots &0\\ 0&\cdots&0&1\\ -a_n & -a_{n-1} &\cdots & -a_{1}\end{pmatrix}\; .
\end{equation}

\begin{proposition}{\cite[Thm. 1]{DiFrancesco:1991W}}\labelx{Prop:opers-para}
There is a unipotent gauge transformation transforming the connection in \Cref{Eq:matrix-diff-eq-2} to a connection of the form
\begin{equation}\labelx{Eq:matrix-diff-eq-3}
\partial+\begin{pmatrix} 0&1&0&\cdots&0\\ -u_2 & 0 & 1 &\ddots&\vdots\\
-u_3&-u_2&0&\ddots &0\\ \vdots &&\ddots& \ddots &1\\ -u_n & \cdots & -u_3 & -u_2 & 0 \end{pmatrix}\; .
\end{equation}
The $u_i$ above, for $i=2,...,n$, are linear combinations of the $(a_j)_{j\geq 3}$ and their derivatives and polynomials in $a_2$ and its derivatives. For $i\geq 3$, the coordinate $u_i$ is a tensor of type $(i,0)$, i.e. it is an $i$-differential. The coordinate $u_2$ transforms like a complex projective structure:
\begin{equation}\labelx{Eq:CP1-structure}
u_2(w)=\left(\tfrac{\D z}{\D w}\right)^2 u_2(z) + S(z,w)\; ,
\end{equation}
where $S(z,w)$ denotes the Schwarzian derivative.
\end{proposition}

We invite the reader to do the computation for $n=2$: compute how the Schrödinger operator $\partial_z^2+a_2(z)$ changes under a coordinate change $z\mapsto w(z)$. The double action of holomorphic functions has to be used to put the resulting operator again in the form of a Schrödinger operator. Then, the transformation rule \eqref{Eq:CP1-structure} appears for $a_2$. The \emph{Schwarzian derivative}
$$S(f,z)=\frac{f'''}{f'}-\frac{3}{2}\left(\frac{f''}{f'}\right)^2$$
is a differential operator invariant under M\"obius transformations. We recommend \cite{Ovsienko-Tabachnikov} for details on the Schwarzian derivative.

 \Cref{Prop:opers-para} parameterizes the space of differential equations by a tuple of differentials of degree from 3 to n, and a special parameter encoding a complex projective structure.

A \emph{complex projective structure}, or $\mathbb{CP}^1$-structure in short, on a surface $S$ is an atlas where the charts map into open subsets of $\mathbb{CP}^1$ and the transition functions are M\"obius transformations. A complex projective structure induces a complex structure (since M\"obius transformations are holomorphic). It is known that the space of all $\mathbb{CP}^1$-structures inducing the same complex structure on $S$ forms an affine space modeled on the vector space of holomorphic quadratic differentials $\mathrm{H}^0(C,K_C^{\otimes 2})$. 

Hence, if we start from a Riemann surface $C$ and choose a complex projective structure compatible with the complex structure on $C$, we can describe any other compatible $\mathbb{CP}^1$-structure by a holomorphic quadratic differential. Thus, on a Riemann surface with $\mathbb{CP}^1$-structure, a differential equation of order $n$ can be parameterized by a tuple of holomorphic differentials $(u_2,u_3,...,u_n)$, where $u_i\in \mathrm{H}^0(C,K^{\otimes i})$.

We can invert the unipotent gauge transformation given by \Cref{Prop:opers-para}, to write the differential equation in terms of the holomorphic differentials. For small values of $n$, we get (see also \cite[Tab. 1]{DiFrancesco:1991W}):
\begin{align*}
n=2: \;\; & \partial^2+u_2 \\
n=3: \;\; & \partial^3+u_2\partial+u_3+\tfrac{1}{2}\partial u_2 \\
n=4: \;\; & \partial^4+u_2\partial^2+(u_3+\partial u_2)\partial+(u_4+\tfrac{1}{2}\partial u_3+\tfrac{1}{20}\partial^2 u_2+\tfrac{9}{100}u_2^2)
\end{align*}

Differential operators of this form on a Riemann surface $C$ are called \emph{opers}. They were introduced by Beilinson and Drinfeld in \cite{Beilinson:opers}. As we have seen after \Cref{Ex:psi-nature}, these differential operators act on the space $K^{(1-n)/2}$. For $n$ even, this means that we also have to fix a square root of the canonical bundle, i.e. a spin structure on $C$.

To sum up, we consider a Riemann surface $C$ together with a compatible $\mathbb{CP}^1$-structure and a spin structure. For $u=(u_2,...,u_n)$ a tuple of holomorphic differentials of degree 2 to $n$ on $C$, we denote by $D_u: K^{(1-n)/2}\to K^{(n-1)/2}$ the oper parameterized by $u$.

For the WKB analysis, we introduce a parameter $h\in\mathbb{C}^*$, we scale the differentials by
$$h.u = (h^{-2}u_2,h^{-3}u_3,...,h^{-n}u_n)\; ,$$
and we consider the oper $D_{h.u}$. By multiplying by some power of $h$, this amounts to replace in $D_u$ all derivatives $\partial$ by $h\partial$. So we get an $h$-connection.

We can now formulate a precise conjecture about the convergence of the exact WKB method in the case of opers.

\begin{conjecture}\labelx{conj:conv_exact_WKB}
Let $C$ be a Riemann surface with a compatible $\mathbb{CP}^1$-structure and a spin structure. To a tuple of differentials $u=(u_2,...,u_n)\in\bigoplus_{i=2}^n \mathrm{H}^0(C,K^{\otimes i})$, associate the spectral network $\mathcal{W}_u^\theta$ with angle $\theta$ and the oper $D_{h.u}$ as above. 

Then, the exact WKB method in direction $\theta$ applied to the differential equation associated to $D_{h.u}$ converges on the complement of $\mathcal{W}_u^\theta$. Further, for $\gamma\in \mathrm{H}_1(\Sigma)$, the holonomy $\mathcal{X}^\theta_\gamma(h)\in\mathbb{C}^*$ of the abelianized local system on the spectral curve $\Sigma$ behaves asymptotically for $h\to 0$ as
\begin{equation}\labelx{Eq:Conj-asymp-periods}
\mathcal{X}^\theta_\gamma(h) \sim \exp(Z_\gamma / h)\; ,
\end{equation}
where $Z_\gamma=\int_\gamma \lambda\in\mathbb{C}$ is the period of the Liouville form $\lambda$ around $\gamma$.
\end{conjecture}

\begin{remark}
The Hitchin section is another way to associate a flat connection to a tuple of differentials $u$ as above. Looking for flat sections gives again a differential equation parameterized by $u$. This differential equation is not explicit since the Hitchin section uses the highly transcendental non-abelian Hodge correspon\-dence. Nevertheless, it is expected that the statement of the conjecture above should still hold for the Hitchin section and that the asymptotic behavior of \Cref{Eq:Conj-asymp-periods} is exponentially good. Partial results can be found for example in \cite{Neitzke24}.
\end{remark}

\newpage
\part{Counting BPS states}\labelx{partV}

\vspace{2cm}

This part is devoted to the study of spectral networks and their role in analyzing BPS states in four-dimensional $\mathcal{N}=2$ supersymmetric gauge theories, assuming a basic familiarity with quantum field theory and supersymmetry. We begin with a review of supersymmetric gauge theories in four dimensions, including the $\mathcal{N}=1$ case as preparatory background, before specializing to $\mathcal{N}=2$ theories. We then examine various electric–magnetic dualities, emphasizing that the natural realization of solitonic BPS states in $\mathcal{N}=2$ gauge theories coincides with the notion of BPS states arising from the representation theory of the super-Poincaré algebra. The exposition of Seiberg–Witten theory constitutes the central theme of this part, and is followed by an introduction to class~$\mathcal{S}$ theories. The final chapter streamlines the logic that led to the development of spectral networks as a tool for analyzing BPS spectra in class~$\mathcal{S}$ theories, with extensive references provided for further study.

\vspace{2cm}

\parttoc


\chapter{Four-dimensional supersymmetric quantum field theories}\labelx{Chapter:N=1theories}




\abstract{This chapter focuses on the general framework of $4d$ gauge theories with $\mathcal N=1$ supersymmetry, of which $\mathcal N=2$ theories form a special subset.
We first discuss the symmetries that govern $4d$ quantum field theories, offering essential background for understanding supersymmetric models. Following this, we explore the classical properties of $4d$ $\mathcal N=1$ gauge theories, with particular emphasis on the construction of $\mathcal N=1$ Lagrangian densities. Lastly, we address the quantum aspects of these theories, delving into key topics such as renormalization, anomalies, holomorphy, and quantum corrections to moduli spaces. This groundwork will prepare us for a deeper study of $\mathcal N=2$ dynamics and, ultimately, spectral networks.}

\vspace{0.5cm}

For readers seeking a deeper dive into four-dimensional supersymmetry, there is a rich collection of textbooks \cite{Wess:1992cp,Freed:1999mn,Deligne:1999Supersolutions,Weinberg:2000cr,Varadarajan:2004yz,Terning:2006bq} and lecture notes \cite{Lykken:1996xt,Martin:1997ns,Bilal:2001nv,Argyres:2001eva,Strassler:2003qg,Argurio:2003ym,Dine:2007zp,Bertolini} that provide accessible introductions, detailed discussions, and numerous applications.

\section{Symmetries}

\subsection{Super-Poincaré algebras}\labelx{Sec:superPoincarealgebras}

The symmetries of any four-dimensional $(4d)$ quantum field theory\footnote{We will consider local and unitary quantum field theories only.} (QFT) in Minkowski spacetime $\mathcal M_4=(\mathbb{R}^{1,3},\eta_{\mu\nu})$ include the \emph{Poincaré group}\index{Poincaré group}, which is the semi-direct product of spacetime translations with the \emph{Lorentz group}\index{Lorentz group} $\mathrm{O}(1,3)$. More precisely, the \emph{Coleman--Mandula theorem}\index{Coleman--Mandula theorem} states that if $G$ is the Lie group of symmetries of a non-trivial unitary relativistic $S$-matrix satisfying some assumptions (e.g. particle finiteness and weak elastic analycity), then $G$ is the direct product of the Poincaré group with an internal symmetry group \cite{Coleman:1967ad,Weinberg:2000cr}. 

The Lorentz group has four connected components. The component containing the identity is known as the restricted Lorentz group, denoted as $\mathrm{SO}(1,3)^+$. The quotient group $\mathrm{O}(1,3)/\mathrm{SO}(1,3)^+$ is isomorphic to $\mathbb{Z}_2 \times \mathbb{Z}_2$, which corresponds to the discrete symmetries ${1, \mathcal{P}, \mathcal{T}, \mathcal{PT}}$. Here, $\mathcal{P}$ represents parity (spatial inversion), $\mathcal{T}$ represents time reversal, and $\mathcal{PT}$ is the combination of both parity and time reversal. At the level of infinitesimal symmetries, one has the isomorphism of Lie algebras:
\begin{equation*}
    \mathfrak{so}(1,3)\simeq\mathfrak{su}(2)\times\mathfrak{su}(2)^*\; ,
\end{equation*}
where the superscript $^*$ indicates that complex conjugation exchanges the two copies of $\mathfrak{su}(2)$. The universal cover of the restricted Lorentz group is the spin group $\mathrm{Spin}(1,3)^+$, which is isomorphic---as a real Lie group---to $\mathrm{SL}_2(\mathbb{C})\simeq\mathrm{Sp}_2(\mathbb{C})$. It is a double cover of $\mathrm{SO}(1,3)^+$, i.e. $\mathrm{SO}(1,3)^+$ is isomorphic to the \emph{Möbius group}\index{Möbius group} $\mathrm{PSL}_2(\mathbb{C})$. The group $\mathrm{SL}_2(\mathbb{C})$ has two inequivalent (real) irreducible representations of complex dimension 2, corresponding to the two non-equivalent spinor\footnote{For a pedagogic introduction to spinors and their relation to the Lorentz group, see for instance \cite{Steane:2013wra}.} representations of $\mathfrak{so}(1,3)$. The components of the elements in the first representation are denoted $\psi_\alpha\in\mathbb{C}$ with $\alpha=1,2$, and in the second one, $\psi_{\dot\alpha}\in\mathbb{C}$ with $\dot\alpha=\dot 1,\dot 2$ (this is the \emph{van der Waerden notation}\index{van der Waerden notation}). A matrix $A\in\mathrm{SL}_2(\mathbb{C})$ acts as:
\begin{equation*}
\begin{pmatrix}
    \psi_1 \\
    \psi_2
\end{pmatrix}\rightarrow A \begin{pmatrix}
    \psi_1 \\
    \psi_2
\end{pmatrix}\; ,\quad \begin{pmatrix}
    \psi_{\dot 1} \\
    \psi_{\dot 2}
\end{pmatrix}\rightarrow A^* \begin{pmatrix}
    \psi_{\dot 1} \\
    \psi_{\dot 2}
\end{pmatrix}\; .
\end{equation*}
The $\psi_\alpha$'s (resp. $\psi_{\dot\alpha}$'s) are called left-handed (resp. right-handed) \emph{Weyl spinors}\index{Weyl!spinor}\index{spinor!Weyl}, and satisfy $\psi_\alpha^\dagger = \overline{\psi}_{\dot\alpha}$ under hermitian conjugation.

Correspondingly,  $\mathrm{SL}_2(\mathbb{C})$ acts on $\mathcal M_4$ in two different ways called \emph{Weyl representations}\index{Weyl!representation}. Let $(\sigma_\mu) = (\sigma_0,\sigma_1,\sigma_2,\sigma_3)$ with $\sigma_0=\mathrm{Id}_{2\times 2}$ and the $\sigma_i$ the usual Pauli matrices:
\begin{equation*}
    \sigma_1 = \begin{pmatrix}
        0 & 1 \\
        1 & 0
    \end{pmatrix}\; ,\quad \sigma_2 = \begin{pmatrix}
        0 & -i \\
        i & 0
    \end{pmatrix}\; ,\quad \sigma_3 = \begin{pmatrix}
        1 & 0 \\
        0 & -1
    \end{pmatrix}\; ,
\end{equation*}
and let $(\sigma^\mu) = (\sigma_0,-\sigma_1,-\sigma_2,-\sigma_3)$. A point $x=(x^\mu)\in\mathbb{R}^{1,3}$ is recast as a Hermitian $2\times 2$ matrix in two ways:
\begin{equation*}
    X = x^\mu\sigma_\mu\; ,\quad\overline{X} = x_\mu\sigma^\mu\; ,
\end{equation*}
on which a matrix $A\in\mathrm{SL}_2(\mathbb{C})$ acts by conjugation:
\begin{equation*}
    X\rightarrow AXA^\dagger\; ,\quad\overline{X}\rightarrow (A^{-1})^\dagger \overline{X}A^{-1}\; .
\end{equation*}
These maps describe the two independent irreducible chiral\index{spinor!chiral} spinor representations of the Lorentz group: $X$ (resp. $\overline{X}$) corresponds to left-handed (resp. right-handed) spinors. The matrices $\sigma_\mu$ (and $\sigma^\mu$) naturally have one undotted and one dotted index $\sigma^\mu_{\alpha\dot\alpha}$, and one can build invariants such as:
\begin{equation*}
    \psi\sigma^\mu\overline{\chi} := \psi^\alpha\sigma^\mu_{\alpha\dot\alpha}\overline{\chi}^{\dot\alpha}\; .
\end{equation*}

Since $\mathrm{SL}_2(\mathbb{C})\simeq\mathrm{Sp}_2(\mathbb{C})$, there exists an antisymmetric invariant tensor:
\begin{equation*}
    \epsilon_{\alpha\beta} = \epsilon_{\dot\alpha\dot\beta} = \begin{pmatrix}
        0 & -1 \\
        1 & 0
    \end{pmatrix}\quad \text{and}\quad \epsilon^{\alpha\beta} = \epsilon^{\dot\alpha\dot\beta} = \begin{pmatrix}
        0 & 1 \\
        -1 & 0
    \end{pmatrix}\; ,
\end{equation*}
in terms of which, e.g. $\psi^\alpha = \epsilon^{\alpha\beta}\psi_\beta$, and which induces a scalar product: $\psi\chi = \psi^\alpha\chi_\alpha = \chi\psi$. 

In the Weyl representation, Dirac matrices write:
\begin{equation*}
    \gamma^\mu = \begin{pmatrix}
        0 & \sigma^\mu \\
        \overline{\sigma}^\mu & 0 
    \end{pmatrix}\; , \quad \gamma_5 = \begin{pmatrix}
        \mathbf{1} & 0 \\
        0 & -\mathbf{1} 
    \end{pmatrix}\; ,
\end{equation*}
and a Dirac spinor decomposes as $(\psi_\alpha,\overline{\chi}^{\dot\alpha})$. A \emph{Dirac spinor}\index{spinor!Dirac} is \emph{Majorana}\index{spinor!Majorana} (i.e. real) if and only if $\psi=\chi$.

\vspace{0.3cm}

In some QFTs, additional symmetries exist beyond those allowed by the Coleman–Mandula theorem. An important case where this occurs is when the symmetries of the theory form a super Lie group, rather than a conventional Lie group, i.e. there are fermionic (odd) symmetry generators on top of the bosonic ones. The corresponding algebra of infinitesimal symmetries becomes $\mathbb{Z}_2$-graded:
\begin{equation*}
	\mathfrak g = \mathfrak g_0\oplus\mathfrak g_1\; ,
\end{equation*}
where $\mathfrak g_0$ is the even subalgebra, containing bosonic generators, and $\mathfrak g_1$ is the odd part, containing fermionic generators. The super Lie algebras compatible with physical constraints, such as the inclusion of supersymmetry, are governed by the \emph{Haag–Lopuszanski–Sohnius theorem}\index{Haag--Lopuszanski--Sohnius theorem} \cite{Haag:1974qh}, which extends the Coleman–Mandula result to include these fermionic symmetries.

\begin{definition}
Let $\mathcal N\in\mathbb{N}_{>0}$. The \emph{$\mathcal N$-extended super-Poincaré algebra in $4d$} is generated by the usual Poincaré algebra bosonic generators as well as by odd generators $Q_\alpha^I$ and $\overline{Q}_{\dot\alpha}^I$ for $I=1,\dots,\mathcal N$, satisfying:
\begin{equation}\labelx{Eq:SUSY=squareroot}
    \begin{split}
	[P_\mu,Q_\alpha^I] &= 0\; , \\
        [P_\mu,\overline{Q}_{\dot\alpha}^I] &= 0\; , \\
	[M_{\mu\nu},Q_\alpha^I] &= i\tensor{(\sigma_{\mu\nu})}{_\alpha^\beta} Q_\beta^I\; , \\
        [M_{\mu\nu},\overline{Q}^{I\dot\alpha}] &= i\tensor{(\overline{\sigma}_{\mu\nu})}{^{\dot\alpha}_{\dot\beta}} \overline{Q}^{I\dot\beta}\; , \\
	\{Q_\alpha^I, \overline{Q}^J_{\dot\beta}\} &= 2\sigma^\mu_{\alpha\dot{\beta}}P_\mu\delta^{IJ}\; , \\
        \{Q_\alpha^I,Q_\beta^J\} &= 2\sqrt{2}\epsilon_{\alpha\beta}\mathcal{Z}^{IJ}\; ,\\
	\{\overline{Q}_{\dot\alpha}^I, \overline{Q}_{\dot\beta}^J\} &= 2\sqrt{2}\epsilon_{\dot\alpha \dot\beta}\left(\mathcal{Z}^{IJ}\right)^*\; .
    \end{split}
\end{equation}
Here $P_\mu$ and $M_{\mu\nu}$ are the usual bosonic generators of spacetime translations and infinitesimal Lorentz transformations, and:
\begin{equation*}
    (\sigma^{\mu\nu})_\alpha^\beta = (\sigma^\mu_{\alpha\dot\gamma}(\overline{\sigma}^\nu)^{\dot\gamma \beta} - \sigma^\nu_{\alpha\dot\gamma}(\overline{\sigma}^\mu)^{\dot\gamma \beta})/4\; .
\end{equation*} 
The integer $\mathcal N$ is the number of supersymmetries of the algebra; the number of supercharges $Q_\alpha^I$ and $\overline{Q}_{\dot\alpha}^I$ in the superalgebra is $4\mathcal N$. The generators $\mathcal{Z}^{IJ}$ are central and called \emph{central charges}\index{central charge}; $\mathcal{Z}^{IJ}$ is antisymmetric in its indices. Additionally, we impose the existence of an $\mathrm{SU}(\mathcal N)_R$-symmetry rotating the supercharges according to their index $I$ (the notation will be explained in \Cref{Subsec:internalsym}), i.e. that all $\mathcal N=1$ subalgebras are equivalent.
\end{definition}

The super-Poincaré algebra with $\mathcal N$ supersymmetries exponentiates to the \emph{super-Poincaré group}\index{super-Poincaré!group} $\overline{\mathrm{OSp}(4\vert\mathcal N)}$. The $\mathcal N=1$ superspace is the homogeneous space: 
\begin{equation*}
	\mathcal M_{4\vert 1} = \overline{\mathrm{OSp}(4\vert 1)}/\mathrm{SO}(1,3)\; ,
\end{equation*}
with usual affine coordinates $(x^\mu,\theta_\alpha,\overline{\theta}_{\dot\alpha})$, $\mu=0,1,2,3$ and $\alpha,\dot{\alpha}=1,2$, where the $\theta$'s and $\overline{\theta}$'s are anti-commuting Grassmann numbers.

This discussion generalizes in arbitrary dimension, leading to $d$-dimensional Poincaré superalgebras with $\mathcal N$ supersymmetries. Spinor representations are crucially dimension-dependent; in particular their properties depend on the dimension modulo 8 (\emph{Bott periodicity}). 

We will be mostly interested in QFTs containing only elementary particles of spin/helicity less than one (i.e. non-gravitational quantum field theories). This requirement sets an upper bound on the number $\mathcal N$ of supersymmetries, depending on $d$. For instance, when $d=4$, the bound is $\mathcal N \leq 4$.

\subsubsection*{Representations of the supersymmetry algebras}\labelx{Par:repSUSY}

One of the postulates of quantum mechanics is that a quantum system is associated to a \emph{Hilbert space of states}\index{Hilbert space} $\mathcal H$, and that the (pure) states of the system correspond bijectively to the complex lines in $\mathcal H$, i.e. to the elements of the projective space $\mathbb{P}(\mathcal H)$. Complex lines in $\mathcal H$ are in one-to-one correspondence with rank-one projection operators (density matrices of pure states). The transition amplitude between two states is given by the inner product in $\mathcal H$, and symmetries of the system are required to preserve transition amplitudes. \emph{Wigner's theorem}\index{Wigner's theorem} states that every symmetry of a quantum system, defined as an isometry of the ray space $\mathbb{P}(\mathcal H)$, can be represented as a linear unitary or an antilinear antiunitary operator on $\mathcal H$.

\begin{definition}
    In a QFT, a \emph{one-particle state} or \emph{multiplet} is a unitary finite-dimensional irreducible representation of the group of spacetime symmetries.
\end{definition} 

In the case of a generic $4d$ QFT, the group of spacetime symmetries is the four-dimensional Poincaré group. The corresponding algebra has two \emph{Casimirs}\index{Casimir element}: $P^2$ and $W^2$, where the $P_\mu$'s generate spacetime translations, and $W^\mu=\frac{1}{2}\epsilon^{\mu\nu\rho\sigma}P_\mu M_{\rho\sigma}$ is the \emph{Pauli--Lubanski vector}. Massive particles are determined by their mass ($P^2=m^2$) and spin $j$. One has: $W^2=-m^2j(j+1)$. Massless particles satisfy $P^2=W^2=0$ and $W^\mu=\pm jP^\mu$ in the frame in which $P^\mu=(E,0,0,E)$, where $j$ is the helicity. Note that particles are defined as representations of the Poincaré group and not only of the Poincaré algebra; in other words, the corresponding irreducible representations must be \emph{$\mathcal{CPT}$-invariant}\index{$\mathcal{CPT}$-invariance}, where $\mathcal C$ denotes charge conjugation. Massive one-particle states are always $\mathcal{CPT}$ invariant, but massless ones are not. For instance, $\mathcal{CPT}$ conjugation always flips the sign of the helicity of massless one-particle states. For this reason, by massless particle one generally means the combination of \emph{two} massless representations of the Poincaré algebra with opposite helicity. By the \emph{spin-statistics theorem}\index{spin-statistics theorem}, particles with integer spin/helicity are bosons, whereas those with (strictly) half-integer spin/helicity are fermions.

By analogy, in the case of supersymmetric QFTs, one sets the following:

\begin{definition}
    A superparticle or \emph{supermultiplet}\index{supermultiplet} is an irreducible finite-dimensional unitary representation of the super-Poincaré algebra at hand, invariant under $\mathcal{CPT}$ conjugation. 
\end{definition}

The operator $P^2$ is still a Casimir in $4d$ super-Poincaré algebras, but $W^2$ is not. This follows from the fact that the supercharges $Q_\alpha^I$ and $\overline{Q}_{\dot\alpha}^I$ (which generate supersymmetry transformations) do not commute with the generators of the Lorentz group. As a result, supermultiplets generally contain particles of identical mass and internal charges, but different spins/helicities since acting with supercharges changes the spin/helicity. 

Supersymmetric QFT satisfy the following important property:
\begin{proposition}
   Every state in a supersymmetric QFT has non-negative energy.
\end{proposition}

\begin{proof}
Fix some $I=1,\dots,\mathcal N$. One has:
\begin{equation}\labelx{Eq:positivenessSUSY}
    \begin{split}
        0\leq \sum_{\alpha} \left\vert\left\vert (Q_\alpha^I)^\dagger \ket{\phi}\right\vert\right\vert^2+\left\vert\left\vert Q_\alpha^I \ket{\phi}\right\vert\right\vert^2 &= \sum_{\alpha = \beta} \bra{\phi} \{Q_\alpha^I,\overline{Q}_{\dot\beta}^I\}\ket{\phi} \\
        &= \sum_{\alpha = \beta} 2\sigma^\mu_{\alpha\dot\beta}\bra{\phi}P_\mu\ket{\phi} \\
        &= 4\bra{\phi}P_0\ket{\phi}= 4\bra{\phi}H\ket{\phi}\; ,
    \end{split}
\end{equation}
where $H=P_0$ is the Hamiltonian. 
\end{proof}

In practice, one obtains irreducible representations of super-Poincaré algebras by recasting the supersymmetry generators $Q_\alpha^I$'s and $Q_{\dot\alpha}^J$'s
as fermionic creation ($(a_{1,2}^{\dagger})^I\propto \overline{Q}_{\dot1,\dot2}^I$) and annihilation ($a_{1,2}^I\propto Q_{1,2}^I$) operators. 

\begin{definition}
    A Poincaré multiplet is a \emph{Clifford vacuum}\index{Clifford vacuum} if it is annihilated by all $a_I$'s.
\end{definition} 

One constructs the representation corresponding to a given Clifford vacuum by acting on it with creation operators $(a^\dagger)^I$'s in all possible ways. The Clifford vacuum generates the lowest weight space of this representation.

\subsubsection*{Massless supermultiplets} 

There are two important properties which characterize massless supermultiplets among all possible supermultiplets. First, in the frame in which the four-momentum of a massless particle reads $P_\mu = (E,0,0,E)$, one has:
\begin{equation*}
    \left\{Q_\alpha^I,\overline{Q}_{\dot\beta}^J\right\} = \begin{pmatrix}
        0 & 0 \\
        0 & 4E
    \end{pmatrix}_{\alpha\dot\beta}\delta^{IJ}\; .
\end{equation*}
This implies that only half of the creation operators (the $(a^\dagger_2)^I$'s) enter in the construction of the supermultiplet. In other words, massless supermultiplets are \emph{shorter} than generic ones: they contain less Poincaré particles than what one would generically expect. Each creation operator which is not trivially realized raises the helicity by one-half, meaning that if the Clifford vacuum has helicity $j_0$ then the corresponding supermultiplet consists of $\binom{\mathcal N}{k}$ states of helicity $j_0+k/2$ for $k=0,\dots,N$.

The second distinctive feature of massless supermultiplets relates to $\mathcal{CPT}$ invariance. We have already emphasized that massless representations of a super-Poincaré \emph{algebra} usually need to be paired with their helicity-flipped counterpart in order to preserve $\mathcal{CPT}$ symmetry. This means that, typically, massless particles of helicity $h$ must be paired with particles of helicity $-h$ within the same supermultiplet. Note that when $j_0=-\mathcal N/4$, there exists the possibility of achieving $\mathcal{CPT}$ invariance without pairing; however this does not always work as other constrains depending on the specific structure of the supermultiplet may not be satisfied just so.

Massless supermultiplets that will be relevant later in our discussion are among the following types, where the bold numbers represent the helicity of the Poincaré constituents, while the non-bold numbers before parentheses indicate a multiplicity:
\begin{itemize}
    \item $\mathcal N=1$ chiral multiplet: $(\mathbf{0},\mathbf{1/2})\oplus_{\mathcal{CPT}}(\mathbf{0},\mathbf{-1/2})$,
    \item $\mathcal N=1$ vector multiplet: $(\mathbf{1/2},\mathbf{1})\oplus_{\mathcal{CPT}}(\mathbf{-1/2},\mathbf{-1})$,
    \item $\mathcal N=2$ hypermultiplet: $(\mathbf{-1/2},2(\mathbf{0}),\mathbf{1/2})\oplus_{\mathcal{CPT}}(\mathbf{1/2},2(\mathbf{0}),\mathbf{-1/2})$. Here $j_0=-\mathcal N/4$, however pairing is nonetheless needed for $\mathcal{CPT}$ invariance,
    \item $\mathcal N=2$ vector multiplet: $(\mathbf{0},2(\mathbf{1/2}),\mathbf{1})\oplus_{\mathcal{CPT}}(\mathbf{0},2(\mathbf{-1/2}),\mathbf{-1})$,
    \item $\mathcal N=4$ vector multiplet: $(\mathbf{-1},4(\mathbf{-1/2}),6(\mathbf{0}),4(\mathbf{1/2}),\mathbf{1})$. This supermultiplet is self $\mathcal{CPT}$-conjugate.
\end{itemize}

\subsubsection*{Massive supermultiplets} In contrast, generically none of the creation operators is realized trivially on massive supermultiplets, and these representations of super-Poincaré algebras are dubbed \emph{long}. The $(a^\dagger_{\dot1})^I$'s lower the spin whereas the $(a^\dagger_{\dot2})^I$'s raise it, again by increments of $1/2$. This leads to $2^{2\mathcal N}$ states, half of them bosonic and the others fermionic. Two massive supermultiplets of prime interest in the non-extended case $\mathcal N=1$ are the following (where now the bold numbers in parentheses denote the spin and the non-bold ones in front, the multiplicities):
\begin{itemize}
    \item $\mathcal N=1$ (massive) matter multiplet: $(-\mathbf{1/2},2(\mathbf{0}),\mathbf{1/2})$,
    \item $\mathcal N=1$ (massive) vector multiplet: $(\mathbf{-1},2(\mathbf{-1/2}),2(\mathbf{0}),2(\mathbf{1/2}),\mathbf{1})$. This matter content matches the combination of a massless $\mathcal N=1$ chiral supermultiplet and a massless $\mathcal N=1$ vector supermultiplet, as expected from the \emph{super-Higgs mechanism}\index{super-Higgs mechanism}. 
\end{itemize}

\subsubsection*{BPS states}

Let us now consider the case of extended supersymmetry ($\mathcal N>1$) and assume, for simplicity and because these are the cases in which we will be mostly interested, that $\mathcal N$ is even. Recall that $\{Q_\alpha^I,Q_\beta^J\} = 2\sqrt{2}\epsilon_{\alpha\beta} \mathcal{Z}^{IJ}$, with $\mathcal{Z}^{IJ}$ the antisymmetric matrix of complex central charges of the algebra. Up to conjugation by an element of $\mathrm{U}(\mathcal N)$, one can assume that $\mathcal{Z}^{IJ}$ is block-diagonal with $2\times 2$ blocks $\mathrm{Mat}(0,\mathcal{Z}_i,-\mathcal{Z}_i,0)$ with $Z_i$ real positive, for $i=1,\dots,\mathcal N/2$. For all $i=1,\dots,\mathcal N/2$, let:
\begin{equation*}
    a^i_\alpha = Q^{2i-1}_\alpha +\epsilon_{\alpha\beta}(Q^{2i}_\beta)^\dagger\; , \quad \text{and} \quad b^i_\alpha = Q^{2i-1}_\alpha -\epsilon_{\alpha\beta}(Q^{2i}_\beta)^\dagger\; .
\end{equation*}
In the rest frame of a supermultiplet of mass $m$, one has:
\begin{equation*}
    \begin{split}
        \{a_\alpha^i,(a_\beta^j)^\dagger\} &= 4(m+\sqrt{2}\mathcal{Z}_i) \delta_{ij}\delta_{\alpha\beta}\; , \\
        \{b_\alpha^i,(b_\beta^j)^\dagger\} &= 4(m-\sqrt{4}\mathcal{Z}_i) \delta_{ij}\delta_{\alpha\beta}\; ,
    \end{split}
\end{equation*}
with all other anticommutators vanishing. 

\begin{proposition}
    The mass of any representation of the $\mathcal N$-extended super-Poincaré algebra satisfies:
\begin{equation}\labelx{Eq:BPSSUSY}
    m\geq \sqrt{2} \mathcal{Z}_i\; , \quad i=1,\dots,\mathcal N/2\; .
\end{equation}
\end{proposition}

\begin{proof}
    This follows from replacing the $Q_\alpha^I$'s in \Cref{Eq:positivenessSUSY} by the $b_\alpha^i$. More generally, if $\mathcal Z^{IJ}$ is taken to be block diagonal as above, but with complex coefficients $\mathcal Z_i$, one obtains the inequality $m\geq\sqrt{2}\vert \mathcal Z_i\vert$ by considering linear combinations of $a_\alpha^i$ and $b_\beta^j$ with complex coefficients. From now on, we consider the $\mathcal Z_i$ to be arbitrary complex numbers.
\end{proof}

When $m$ equals $\sqrt{2}\vert \mathcal{Z}_i\vert$ for some values of $i$, some creation operators are trivially realized. Such non-generic\footnote{Here, by \emph{generic}, we mean that $m> \sqrt{2}\mathcal{Z}_i$ for all $i = 1,\dots,\mathcal{N}/2$.} massive supermultiplets will be \emph{shorter} than generic ones.

\begin{definition}
    Massive supermultiplets for which $m=\sqrt{2}\vert \mathcal{Z}_i\vert$ for some values of $i$ are called \emph{BPS states}\index{BPS!state}, or \emph{short representations}. More precisely, if there are $k$ values of $i$ for which $m=\sqrt{2}\vert \mathcal{Z}_i\vert$, the supermultiplet is called a $\frac{k}{\mathcal N}$-BPS state.\footnote{The origin of this terminology will be clarified in \Cref{Sec:EMdualityinSUSY}.} Conversely, when $m > \sqrt{2}\mathcal Z_i$ for all $i$, the supermultiplet is called \emph{long}. 
\end{definition}

In a $\frac{k}{\mathcal N}$-BPS state, out of the $2\mathcal N$ fermionic creation operators constructed with the supercharges (that is, the $(a_\alpha^i)^\dagger$ and $(b_\alpha^i)^\dagger$ when the $\mathcal Z_i$ are real), $2k$ of them (among the $(b_\alpha^i)^\dagger$) are trivially realized. Thus, the corresponding representation consists of $2^{2(N-k)}$ (Poincaré) states, half of them bosonic and the others fermionic. Note that $\frac{1}{2}$-BPS supermultiplets contain as many states as massless supermultiplets.

Examples of massive representations of $4d$ super-Poincaré algebras with extended supersymmetry include:
\begin{itemize}
    \item $\mathcal N=2$ long vector multiplet: $(\mathbf{-1},4(\mathbf{-1/2}),6(\textbf{0}),4(\mathbf{1/2}),\textbf{1})$. A long $\mathcal N=2$ multiplet whose Clifford vacuum is not of spin zero necessarily contain particles of spin strictly greater than 1.
    \item $\mathcal N=2$ half-BPS massive hypermultiplet: $(2(\mathbf{-1/2}),4(\mathbf{0}),2(\mathbf{1/2}))$. The doubling of states is imposed by $\mathrm{SU}(2)_R$ invariance: in order to preserve $\mathrm{SU}(2)_R$, the scalars in the hypermultiplet must transform as a doublet (quaternionic) representation of $\mathrm{SU}(2)_R$. However, the two scalars produced by the creation-annihilation algebra acting on the Clifford vacuum are real and therefore cannot fulfill this requirement (more on this in \Cref{Chap:N=2}). This also explains the doubling of states in massless $\mathcal N=2$ hypermultiplets alluded to above, since $\mathcal{CPT}$ self-conjugation would impose the reality of the two scalars. 
    \item $\mathcal N=2$ half-BPS vector multiplet: $(\mathbf{-1},2(\mathbf{-1/2}),2(\mathbf{0}),2(\mathbf{1/2}),\mathbf{1})$.
    \item $\mathcal N=4$ quarter-BPS vector multiplet: $(\mathbf{-1},4(\mathbf{-1/2}),6(\mathbf{0}),4(\mathbf{1/2}),\mathbf{1})$.
\end{itemize}

\subsection{Conformal and superconformal algebras}\labelx{Sec:conformal-algebras}

\subsubsection*{Conformal invariance}

Another way to extend Poincaré symmetry in flat space is to consider scale-invariant theories. Scale invariance is incompatible with the definition of an $S$-matrix and thus such theories are not constrained by the Coleman--Mandula theorem. A broad class of scale-invariant theories is that of \emph{Conformal Field Theories}\index{conformal!field theory} (CFTs), for which the bosonic symmetry group contains the conformal group of Minkowsky spacetime $\mathcal M_d$. Reciprocally, most of the known scale-invariant theories we know of are actually CFTs. We refer to \cite{Ginsparg:1988ui,DiFrancesco:1997nk,Minwalla:1997ka,Aharony:1999ti} for more details.

\begin{definition}
    The \emph{conformal group} of $\mathcal M_d$ is the group of transformations which preserve the metric up to a scale factor: $\eta_{\mu\nu}\rightarrow \Omega^2(x)\eta_{\mu\nu}$. It contains Poincaré transformations, dilations $x^\mu\rightarrow \lambda x^\mu$ with $\lambda\in\mathbb{R}_{>0}$, inversion $x^\mu\rightarrow x^\mu/x^2$ and special conformal transformations obtained by conjugating translations with the inversion. At the level of the infinitesimal symmetries, the conformal algebra $\mathfrak{so}(d,2)$ is generated by infinitesimal translations $P_\mu$, Lorentz transformations $M_{\mu\nu}$, dilations $D$ and special conformal transformations $K_\mu$, which satisfy in particular:
\begin{equation*}
    [D,P_\mu] = -\I P_\mu\; , \quad [D,K_\mu] = \I K_\mu\; .
\end{equation*}
\end{definition}

In the conformal algebra, $P^2$ is not a Casimir. Fields that are eigenfunctions of the dilation operator, i.e. $\phi(x)\rightarrow \lambda^\Delta \phi(\lambda x)$n are called \emph{quasi-primary fields}\index{quasi-primary fields} of dimension $\Delta$, and the quasi-primary fields that are also annihilated by the $K_\mu$'s are called \emph{primary fields}\index{primary fields}. Representations of the conformal algebra are constructed via the action of Poincaré generators on primary fields (``highest weight'' vectors).

In unitary CFTs there are lower bounds on the dimension of fields depending on which representation of Poincaré algebra they sit in. For example in general dimension $d$: 
\begin{equation*}
    \begin{split}
        \Delta(\text{scalar}~\phi) &\geq (d-2)/2\; ,\\
        \Delta(\text{spinor}~\psi) &\geq (d-1)/2\; ,\\
        \Delta(\text{vector}~J_\mu) &\geq d\; ,
    \end{split}
\end{equation*}
with equality if $\partial^2\phi=0$ (free scalar), $\slashed{\partial}\psi=0$ (free fermion field) and $\partial^\mu J_\mu=0$ (conserved current), respectively. These inequalities are called \emph{unitary bounds}\index{unitary bound}.

\subsubsection*{Superconformal algebras}

One obtains \emph{superconformal algebras}\index{superconformal!algebra} by combining supersymmetry with conformal invariance \cite{Ginsparg:1988ui,Minwalla:1997ka,Aharony:1999ti}. More precisely, a superconformal algebra in $d$ dimensions is required to contain both the conformal algebra $\mathfrak{so}(d,2)$ and the $d$-dimensional Poincaré superalgebra with $\mathcal N$ supersymmetries, and to have its fermionic generators in spinor representations of $\mathfrak{so}(d,2)$.

When $d=4$ for example, on top of the conformal generators $P_\mu$, $M_{\mu\nu}$, $D$, $K_{\mu}$ and the supersymmetry generators $Q_\alpha$ and $\overline{Q}_{\dot\alpha}$, the superconformal algebra admits superconformal generators $S_\alpha$ and $\overline{S}_{\dot\alpha}$ and $R$-symmetry generators in the commutator of $Q$ and $S$, where $R$-symmetries are defined in \Cref{Subsec:internalsym}.

In general, superconformal algebras exist only in dimensions $d\leq 6$, and are very constrained when $d\geq 3$. For $d=4$, the only superconformal algebras with $\mathcal N$ supersymmetries are:
\begin{itemize}
    \item $\mathcal N\leq 3$: $\mathfrak{su}(2,2\vert\mathcal N) \supset \mathfrak{su}(2,2)\times\mathfrak{u}(\mathcal N)_R$,
    \item $\mathcal N=4$: $\mathfrak{psu}(2,2\vert 4) \supset \mathfrak{su}(2,2)\times\mathfrak{su}(4)_R$.
\end{itemize}
The number of Poincaré supersymmetries is again bounded from above in SCFTs, and in $d=4$ the maximal number is $\mathcal N=4$ corresponding to $32$ supercharges (8 $Q_\alpha^I$'s, 8 $\overline{Q}_{\dot\alpha}^I$'s, 8 $S_\alpha^I$'s and 8 $\overline{S}_{\dot\alpha}^I$'s). \emph{Superconformal primary fields}\index{superconformal!primary field} are quasi-primary fields annihilated by all $K_\mu$'s and superconformal generators, and chiral primary fields are primary fields annihilated by all $Q_{\dot\alpha}$.

\subsection{Internal symmetries}\labelx{Subsec:internalsym}

The group of internal symmetries in a quantum field theory encompasses all non-spacetime symmetries. Within this group, we can identify \emph{$R$-symmetries}\index{$R$-symmetry}, which are defined as those symmetries that do not commute with the supercharges. In contrast, all non-$R$ symmetries commute with the super-Poincaré algebra.

The group of internal symmetries characterizes a quantum field theory and remains invariant regardless of the specific formulation of the theory. This is in contrast to gauge symmetries, which are features of the description of the theory rather than of the theory itself. For instance, a single quantum field theory can have dual descriptions as different gauge theories with varying gauge groups. We will encounter our first example of this when discussing Seiberg duality at the end of this chapter, with further instances to be explored later. However, dual descriptions of the same theory must always have the same internal symmetries.

\subsubsection*{Global versus local symmetries}

Internal symmetries are often referred to as \emph{global}, as they apply simultaneously to all spacetime points. For example, the $4d$ theory of a free complex scalar field $\phi(x)$ of mass $m$, defined by the action:
\begin{equation}\labelx{Eq:freescalar}
    S = \int\mathrm{d}^4x~\partial_\mu\overline{\phi}\partial^\mu\phi+m^2\vert\phi\vert^2\; ,
\end{equation}
has a $\mathrm{U}(1)$ global symmetry given by $\phi(x)\rightarrow \E^{\I\alpha}\phi(x)$, where $\alpha$ is a constant. 

One can \emph{gauge} this symmetry by allowing it to depend non-trivially on spacetime, i.e. $\phi(x)\rightarrow \E^{\I \alpha(x)}\phi(x)$ with $\alpha(x)$ a function of $x$. Such a spacetime-dependent symmetry is said to be \emph{local}. Gauging requires the introduction of a $\mathrm{U}(1)$ vector field (or gauge field) $A_\mu(x)$ to compensate the additional terms this induces. If in the action the derivatives $\partial_\mu$ are replaced by covariant ones $D_\mu = \partial_\mu+\I A_\mu$ and if $A_\mu$ varies as $A_\mu\rightarrow A_\mu-\partial_\mu\alpha$ as $\phi(x)\rightarrow \E^{\I\alpha(x)}\phi(x)$, then $D_\mu\phi\rightarrow \E^{\I\alpha(x)}D_\mu\phi$, and the modified action
\begin{equation*}
    S = \int\mathrm{d}^4x~D_\mu\overline{\phi}D^\mu\phi+m^2\vert\phi\vert^2
\end{equation*}
is invariant under the \emph{local} $\mathrm{U}(1)$ symmetry. Adding the standard kinetic term $\propto F_{\mu\nu}F^{\mu\nu}$ for the vector field $A_\mu$ where $F_{\mu\nu}=\partial_\mu A_\nu-\partial_\nu A_\mu$, yields the action of scalar electrodynamics. Geometrically, $A_\mu$ is a connection $1$-form. One distinguishes \emph{small} and \emph{large} gauge transformations: the former are continuously connected to the identity, whereas the latter are not. 

Contrarily to global symmetries, gauge symmetries are not genuine symmetries: all states and observables of the system must be gauge-invariant. Gauging a global symmetry is, in essence, similar to quotienting states and observables by this symmetry, which leads to a new theory without the original symmetry. Nevertheless, gauge symmetries do have physical consequences. For instance, gauge symmetries induce \emph{Gauss-like laws}, and are essential to construct Lagrangian asymptotically-free theories, a result known as the \emph{Coleman--Gross theorem} \cite{Coleman:1973sx}.

\begin{remark}
    Gauge symmetries frequently occur in $d$-dimensional quantum field theories involving differential forms $A^{(k)}$ of degree $k\geq 1$ entering the action through their exterior derivative $\mathrm{d}A^{(k)}$. Due to the property $\mathrm{d}\circ\mathrm{d}=0$, the action remains invariant under the transformation $A^{(k)}\rightarrow A^{(k)}+\mathrm{d}B^{(k-1)}$, where $B^{(k-1)}$ is any differential form of degree $(k-1)$. This symmetry is \emph{local}, rather than global, because $B^{(k-1)}$ is not necessarily closed (the higher-form analogue of being constant). When $k>1$, this is referred to as a \emph{higher-form gauge symmetry}\index{higher-form gauge symmetry}.
\end{remark}

\subsubsection*{Global symmetries as topological operators}

The historical understanding of global symmetries is deeply rooted in Noether's first theorem. According to this theorem, each continuous global symmetry group $G$ gives rise to a conserved current: a 1-form $j^{(1)}_\mu(x)$ that satisfies the conservation equation $\partial^\mu j_\mu^{(1)}(x)=0$ (see e.g. \cite[Chap. 2]{DiFrancesco:1997nk}). For instance, in the theory defined by \Cref{Eq:freescalar}, the conserved current associated with the global $\mathrm{U}(1)$ symmetry previously discussed takes the following form:
\begin{equation*}
j_\mu^{(1)}(x)= \phi(x)\partial_\mu\overline{\phi}(x)-\overline{\phi}(x)\partial_\mu\phi(x)\; .
\end{equation*}

The conservation equation for the current $j_\mu^{(1)}(x)$ can be rewritten more naturally in the language of differential forms: $\mathrm{d}\left(\star_dj^{(1)}\right)=0$, where $\star_d$ is the $d$-dimensional Hodge star. As a result, $\star_dj^{(1)}$ is a $(d-1)$-form. It can be integrated on codimension-1 submanifolds $M_{(d-1)}$, leading to the associated \emph{conserved charge}\index{conserved charge}:
\begin{equation*}
    Q(M_{(d-1)})=\int_{M_{(d-1)}} \star_dj^{(1)}\; .
\end{equation*}
The conservation equation implies that $Q(M_{(d-1)})=Q(N_{(d-1)})$ whenever $M_{(d-1)}$ and $N_{(d-1)}$ can be continuously deformed into each other; it is in this sense that the charge $Q$ is conserved.

If the symmetry at hand corresponds to a Lie group $G$, the current $j_\mu^{(1)}$ is in fact valued in the Lie algebra $\mathfrak{g}=\mathrm{Lie}(G)$ and there is one scalar conserved charge for each generator of $\mathfrak{g}$. By exponentiation one can construct \emph{topological operators}, or \emph{defects}\index{topological!defect}, $\mathcal T_g(\mathcal M_{(d-1)})$, associated to the submanifold $\mathcal M_{(d-1)}$ and any $g\in G$. The charged operators for the symmetry $G$ are fields $\mathcal O^i(x)$ (which are \emph{local}, i.e. functions on spacetime) characterized by a representation $R$ of $G$. The symmetry translates into \emph{Ward-Takahashi identities} between correlation functions:
\begin{equation}\labelx{Eq:WardTakahashi}
\left\langle\mathcal T_g(S_{(d-1)})\mathcal O^i(x)\right\rangle = (\mathrm{Mat}_R(g))^i_j\left\langle\mathcal O^j(x)\right\rangle\; ,
\end{equation}
where $\mathrm{Mat}_R(g)$ is the image of $g$ in the representation $R$, and  $S_{(d-1)}$ is a topological $(d-1)$-dimensional sphere enclosing the point $x$. Topological defects supported on homotopy-equivalent submanifolds can be fused in a way which corresponds to the product in $G$. In fact, one can define ordinary symmetries as follows:
\begin{definition}\labelx{Def:globalsymmetry}
    A \emph{symmetry} corresponding to a group $G$ in a quantum field theory is a set of topological defects $\mathcal T_g([\mathcal M_{(d-1)}])$ indexed by the elements $g$ of $G$, with $[\mathcal M_{(d-1)}]$ any homotopy class of codimension-1 submanifolds  of spacetime. 
\end{definition}

While the above construction relies on the presence of conserved currents which exist only when $G$ is continuous, topological defects corresponding to global symmetries exist even when $G$ is discrete, and implement elements of $G$ on homotopy classes of codimension-1 submanifolds of spacetime. The group law in $G$ reflects in the fusion rules of these topological operators.

\subsubsection*{Generalized global symmetries}

Topological defects corresponding to usual global symmetries, to which we will refer as \emph{0-form symmetries}\index{symmetry!0-form} for the rest of this section, are supported on codimension-1 manifolds and act on local operators. One can relax the condition on the codimension of topological defects, allowing them to be supported of codimension $(p+1)$-manifolds, where $0\leq p\leq d-1$. This perspective, first advocated in \cite{Gaiotto:2014kfa}, has inspired fruitful generalizations of the standard notion of symmetry in QFT. We briefly review them here, referring the reader to e.g. \cite{Bhardwaj:2023kri,Shao:2023gho} and references therein for a more detailed account.

\begin{definition}
    Such operators encode so-called \emph{$p$-form symmetries}\index{symmetry!$p$-form}; they act on extended $p$-dimensional objects, generalizing \Cref{Eq:WardTakahashi}. 
\end{definition}

\begin{remark}
    A version of the Eckmann--Hilton argument shows that $p$-form symmetries with $p\geq 1$ are always abelian: the fusion rules satisfied by the set of topological defects implementing the symmetry corresponds to the product in some abelian group $G$. 
\end{remark}

Continuous $p$-form symmetries correspond to a conserved Noether $(p+1)$-form current $j^{(p+1)}$ satisfying $\mathrm{d}\star_d j^{(p+1)}=0$; integrating $\star_d j^{(p+1)}$ on codimension-$(p+1)$ manifolds defines topological charges, and after exponentiation, topological defects. As above, the existence of a conserved current is not tantamount to the symmetry, which might well be discrete. 

Let us consider $d=4$ pure $\mathrm{U}(1)$ gauge theory (i.e. Maxwell) with action:
\begin{equation*}
    S[A^{(1)}] \propto \int\mathrm{d}^4x~F^{(2)}\wedge \star_4 F^{(2)} = \frac{1}{2}\int\mathrm{d}^4x~F_{\mu\nu}F^{\mu\nu}\; .
\end{equation*}
The equation of motion $\mathrm{d} \star_4 F^{(2)}=0$ can be regarded as the conservation equation of a $2$-form current; the corresponding $\mathrm{U}(1)^{(1)}_e$ 1-form symmetry is dubbed \emph{electric}\index{electric 1-form symmetry} and encodes the freedom in shifting the gauge field $A^{(1)}$ by a closed 1-form $\lambda^{(1)}$ (i.e. $\mathrm{d}\lambda^{(1)}=0$). As alluded to above, being closed is the higher-degree differential form analogue of being constant for a function; in this sense, $\mathrm{U}(1)^{(1)}_e$ is a global symmetry. The $2$-form $\star_4 F^{(2)}$ can be integrated on codimension-2 manifolds, leading to conserved charges (Gauss law) and codimension-2 topological defects.

This generalizes to pure $\mathrm{U}(1)$ higher-gauge theories in $d$ dimensions, defined by the action:
\begin{equation*}
    S[A^{(k)}] \propto \int\mathrm{d}^dx ~F^{(k+1)}\wedge\star_d F^{(k+1)}\; ,
\end{equation*}
where $F^{(k+1)}=\mathrm{d}A^{(k)}$: such theories admit an electric $k$-form symmetry $\mathrm{U}(1)^{(k)}_e$.

Such theories display another conservation equation, of topological nature: 
\begin{equation*}
    \mathrm{d}F^{(k+1)}\propto \mathrm{d}\star_d\left(\star_d F^{(k+1)}\right)=0\; .
\end{equation*} 
The conserved current is a $(d-k-1)$-form and the symmetry is $\mathrm{U}(1)^{(d-k-2)}_m$, dubbed \emph{magnetic}\index{magnetic $(d-k-2)$-form symmetry}. It corresponds to the freedom in shifting the magnetic dual gauge field $A^{(d-k-2)}_D$ by a closed $(d-k-2)$-form. We will discuss electric-magnetic duality in more details in \Cref{Sec:Sduality}. 

In pure $\mathrm{U}(1)$ gauge theory in $4d$, the topological defects implementing the $\mathrm{U}(1)^{(1)}_e$ electric 1-form symmetry are supported on codimension-2 manifolds. The charged objects are Wilson operators $W_L(q) = \exp(\I q\oint_L~A^{(1)})$, with $L$ a 1-dimensional submanifold of spacetime and $q$ the electric charge. Upon $A^{(1)}\rightarrow A^{(1)}+\lambda^{(1)}$ with $\lambda^{(1)}$ closed one has $W_L(q)\rightarrow e^{iqM(\lambda^{(1)},L)} W_L(q)$, where $M(\lambda^{(1)},L)$ is the monodromy of $\lambda^{(1)}$ along $L$. Likewise, the charged objects for the magnetic 1-form symmetry are 't Hooft operators. 

In the presence of electrically and/or magnetically charged fields, the 1-form symmetry $\mathrm{U}(1)^{(1)}_e\times\mathrm{U}(1)^{(1)}_m$ is broken to a subgroup by screening. For example, in the presence of electrically charged fields $\phi$ of charge $n$, the electric 1-form symmetry is broken to its discrete subgroup $(\mathbb{Z}_{n})_e^{(1)}$ since Wilson lines of charge $n$ can end on $\phi(x)$. A cartoon of this is shown in \Cref{Fig:highersymbreak}.

\begin{figure}[!ht]
    \centering
    \includegraphics[width=\textwidth]{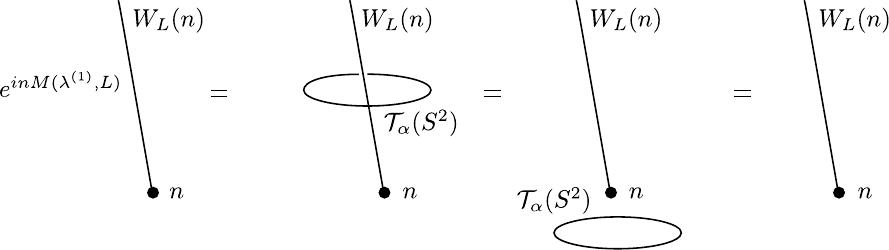}
    \caption{Breaking of the $1$-form symmetry in the presence of charged operators.}
    \labelx{Fig:highersymbreak}
\end{figure}

\begin{remark}
Non-abelian Yang--Mills theories in $4$ dimensions also admit electric and magnetic one-form symmetries with charged objects Wilson--'t Hooft line operators. They encode subtle information about the global structure of the gauge group, to which the action of pure Yang--Mills: 
\begin{equation*}
    S = \frac{-1}{2g^2}\int\mathrm{d}^4x~\tr_\mathfrak{g} (F\wedge F)\; ,    
\end{equation*}
is insensitive, since it only depends on the Lie algebra $\mathfrak g$. For example, $\mathrm{SU}(N)$ and $\mathrm{PSU}(N)$ Yang--Mills theories share the same action: their local dynamics is the same. The choice of global structure for the gauge group $G$ corresponding to the algebra $\mathfrak g$ is equivalent to the choice of a maximal collection of local Wilson--'t Hooft operators \cite{Aharony:2013hda}, which in turn determines the 1-form global symmetry of the theory \cite{Gaiotto:2014kfa}.
\end{remark}

In the last years, generalized symmetries of quantum field theories and gauge theories has been scrutinized and interesting mathematical structures have emerged from their study. Global symmetries of different ranks can combine, leading to higher-groups (in a categorical sense); besides, many gauge theories admit non-invertible topological defects which generalize the \emph{Kramers--Wannier duality}\index{Kramers--Wannier duality} defect of the $2d$ Ising model.

\subsubsection*{Backgrounds and gauging}

Let us consider a $d$-dimensional quantum field theory described by an action $S$, with a continuous $p$-form global symmetry $G^{(p)}$ and corresponding Noether current $j^{(p+1)}$ defined by the transformation law
\begin{equation}
    S \rightarrow S +\I\int \mathrm{d}\lambda^{(p)}\wedge \star_d j^{(p+1)}\; ,
\end{equation}
for a general (spacetime dependent) $p$-form symmetry parameter $\lambda^{(p)}$.

One can consider the theory in the presence of a background field $A^{(p+1)}$ for the global symmetry $G^{(p)}$, meaning that one considers the modified action:
\begin{equation}\labelx{Eq:backgroundglobalsym}
    S[A^{(p+1)}] = S -\I\int A^{(p+1)}\wedge \star_d j^{(p+1)}\; .
\end{equation}
With background fields, the theory is invariant under shifts by any (not necessarily closed) $p$-form $\lambda^{(p)}$ if one imposes that $A^{(p+1)}$ transforms as $A^{(p+1)}\rightarrow A^{(p+1)}+\mathrm{d}\lambda^{(p)}$. Note however that $A^{(p+1)}$ is a \emph{parameter} of the modified theory and not a dynamical field: one does not sum over $A^{(p+1)}$, thus the local symmetry is \emph{not} a symmetry of the modified theory with fixed background $A^{(p+1)}$.

\begin{definition}
    Summing over all possible backgrounds $A^{(p+1)}$ is the way to go to obtain yet another theory in which the local symmetry is really a gauge symmetry; this is one way to describe the \emph{gauging}\index{gauging} procedure. 
\end{definition}

Topological defects corresponding to $G^{(p)}$ write:
\begin{equation*}
	\begin{split}
		\mathcal T_\alpha(\mathcal{M}_{(d-p-1)}) &= \exp{\I\alpha\int_{\mathcal{M}_{(d-p-1)}} \star_d j^{(p+1)}} \\
		&= \exp{\I\alpha\int \Pi_{\mathcal{M}}^{(d-p-1)}  \wedge \star_d j^{(p+1)}}\; ,
	\end{split}
\end{equation*}
where $\Pi_{\mathcal{M}}^{(d-p-1)}$ is the differential form Poincaré dual to $\mathcal{M}_{(d-p-1)}$. This shows that inserting a topological defect implementing $G^{(p)}$ corresponds to turning on a background gauge field. Therefore, the gauging procedure can be equivalently described as summing over all possible networks of symmetry defects in spacetime.

\begin{definition}
    There exist obstructions to gauging: \emph{'t~Hooft anomalies}\index{'t~Hooft!anomaly}. A generalized symmetry with conserved current $j^{(p+1)}$ has an 't~Hooft anomaly if $d\star j^{(p+1)}$ depends on background fields in an essential way. Since the latter are non-dynamical, 't~Hooft anomalies are features of global symmetries. 't~Hooft anomalies also exist for discrete symmetries, and can involve two or more distinct global symmetries, in which case one speaks of \emph{mixed 't~Hooft anomalies}\index{'t~Hooft!mixed anomaly}.
\end{definition}

't~Hooft anomalies lead to stringent constraints on the dynamics of quantum field theories because they are conserved under the renormalization group flow: this result is known as the \emph{'t~Hooft anomaly matching condition}\index{'t~Hooft!anomaly matching condition}. A famous consequence is that a quantum field theory in which an ordinary global symmetry has a non-zero 't Hooft anomaly cannot have a mass gap (meaning that the difference between the energy of any state in the QFT and the energy of the vacuum is bounded from below by a strictly positive number), as it would be then impossible to match the anomaly in the low-energy effective theory. 't Hooft anomalies are intrinsic properties of quantum field theories, contrarily to other types of anomalies (such as \emph{chiral global or gauge anomalies}\index{anomaly!chiral}) which are rather features of a specific description of a theory. We will return to a more detailed discussion of anomalies in \Cref{Sec:anomalies}.

\section{Classical field theories }\labelx{Subsubsec:ClassdynN=1}

In this section, we explore the construction of Lagrangian densities for $4d$ $\mathcal N=1$ gauge theories, focusing on their classical structure. These theories exhibit key features, such as the presence of a \emph{classical moduli space of supersymmetric vacua}\index{moduli space!classical}. Classical $4d$ $\mathcal N=1$ gauge theories often lead to well-defined quantum versions of the same theories. While not all quantum field theories (QFTs) can be described by Lagrangians, when a Lagrangian exists, it serves as a powerful tool for studying the theory. In particular, we emphasize renormalizable QFTs, as they will be of primary importance later on. Accordingly, we will differentiate between two types of Lagrangian densities: those that are renormalizable and those that are not. Although renormalizability is a quantum concept, it can be inferred from the Lagrangian structure when there is one. Renormalizable Lagrangians describe UV (microscopic) theories, while non-renormalizable ones typically encode effective low-energy theories that are only valid below a certain energy scale.

Throughout this section and the next, we focus on $4d$ $\mathcal N=1$ theories, following primarily \cite{Bertolini}.

\subsection{Superfields}

The constraints imposed by combining special relativity with quantum mechanics lead to the need for considering fields that carry representations of the symmetry algebra, rather than just focusing on the one-particle representations discussed earlier in \Cref{Par:repSUSY}. Fields provide a more flexible framework, as they describe systems of multiple particles and can account for the creation and annihilation processes, all while respecting the underlying symmetries of spacetime and supersymmetry (see e.g. \cite{Alvarez-Gaume:2005ojg}). The generators of the Poincaré algebra are represented in field space as differential operators, e.g. $P_\mu\leftrightarrow -\I\partial_\mu$: at first order, under an infinitesimal translation $x\rightarrow x+a$, a field $\phi(x)$ varies as:
\begin{equation*}
    \phi(x)\rightarrow \phi(x+a) = \phi(x)+a^\mu\partial_\mu\phi(x) = \phi(x)+\I a^\mu P_\mu\phi(x)\; .
\end{equation*}

Similarly, in supersymmetric (SUSY) relativistic quantum theory one is interested in representations of the super-Poincaré algebra on fields. The supermultiplets of \Cref{Par:repSUSY} are associated to \emph{superfields}\index{superfield} which (after quantization) describe creation and annihilation of the corresponding supermultiplets.

The construction of the superfields and supersymmetric Lagrangian densities simplifies a lot in $4d$ $\mathcal N=1$ superspace $\mathcal{M}_{4\vert1}$. Let $x^\mu,\theta_\alpha,\overline{\theta}_{\dot\alpha}$ be the standard coordinates on $\mathcal{M}_{4\vert1}$. The supercharges $Q_\alpha$ and $\overline{Q}_{\dot\alpha}$ embody as differential operators in superspace:
\begin{align*}
    Q_\alpha = -\I \partial_\alpha -\sigma^\mu_{\alpha\dot\beta}\overline{\theta}^{\dot\beta}\partial_\mu\; ,   \quad\overline{Q}_{\dot\alpha} = \I \overline{\partial}_{\dot\alpha} +\theta^\beta\sigma^\mu_{\beta\dot\alpha}\partial_\mu\; , \\
    \Rightarrow \{Q_\alpha,\overline{Q}_{\dot\beta}\} = -2\I \sigma^\mu_{\alpha\dot{\beta}}\partial_\mu = 2\sigma^\mu_{\alpha\dot{\beta}}P_\mu\; .
\end{align*}

\begin{definition}
    A \emph{superfield} is a function $Y(x,\theta,\overline{\theta})$ on $\mathcal{M}_{4\vert 1}$, which, under infinitesimal super-translations of spinor parameters $\epsilon_\alpha,\overline{\epsilon}_{\dot\alpha}$, transforms as:
\begin{equation*}
    Y(x,\theta,\overline{\theta}) \rightarrow Y(x+\delta x,\theta+\delta\theta,\overline{\theta}+\delta\overline{\theta}) = (\I \epsilon Q+\I \overline{\epsilon}\overline{Q})Y(x,\theta,\overline{\theta})\; ,
\end{equation*}
where $\delta\theta=\epsilon$, $\delta\overline{\theta}=\overline{\epsilon}$ and $\delta x^\mu=i\theta\sigma^\mu\overline{\epsilon}-i\epsilon\sigma^\mu\overline{\theta}$, as imposed by the relation $\{Q_\alpha,\overline{Q}_{\dot\beta}\}\propto P_\mu$ in \Cref{Eq:SUSY=squareroot}.
\end{definition} 

The $\theta_\alpha$'s and $\overline{\theta}_{\dot\alpha}$'s being Grassmann variables, one can expand any superfield in finitely many field components:
\begin{align*}
    Y(x,\theta,\overline{\theta}) = &f(x)+\theta\psi(x)+\overline{\theta}\overline{\chi}(x)+\theta\theta m(x)+\overline{\theta}\overline{\theta} n(x) + \theta \sigma^\mu\overline{\theta}v_\mu(x) \\ +&\theta\theta\overline{\theta}\overline{\lambda}(x)+\overline{\theta}\overline{\theta}\theta\rho(x)+\theta\theta\overline{\theta}\overline{\theta} d(x)\; .
\end{align*}
Products and sums of superfields are again superfields.

Superfields corresponding to specific irreducible representations of the $4d$ $\mathcal N=1$ SUSY algebra, particularly chiral and vector supermultiplets, are obtained by imposing constraints which commute with $Q_\alpha$ and $\overline{Q}_{\dot\alpha}$.

\subsubsection*{Chiral superfields}

Consider the differential operators:
\begin{equation}\labelx{Eq:covderSUSY}
	D_\alpha = \partial_\alpha +\I \sigma^\mu_{\alpha\dot\beta}\overline{\theta}^{\dot\beta}\partial_\mu~,~~~\overline{D}_{\dot\alpha} = \overline{\partial}_{\dot\alpha} +\I \theta^\beta\sigma^\mu_{\beta\dot\alpha}\partial_\mu.
\end{equation}
They satisfy the following anticommutation relations:
\begin{align*}
	\{D_\alpha,\overline{D}_{\dot\beta}\} = -2\sigma^\mu_{\alpha\dot\beta} P_\mu\; ,   \quad &\{D_\alpha,D_\beta\} = \{D_\alpha,Q_\beta\} = \{D_\alpha,\overline{Q}_{\dot\beta}\}=0\; ,\\
    &\{\overline{D}_{\dot\alpha},\overline{D}_{\dot\beta}\} = \{\overline{D}_{\dot\alpha},Q_\beta\} = \{\overline{D}_{\dot\alpha},\overline{Q}_{\dot\beta}\}=0\; .
\end{align*}

\begin{definition}
    A \emph{chiral}\index{superfield!chiral} superfield is a superfield $\Phi$ which satisfies for every $\dot\alpha$, $\overline{D}_{\dot\alpha}\Phi = 0$. An \emph{antichiral}\index{superfield!antichiral} superfield is a superfield $\Psi$ which satisfies for every $\alpha$, $D_\alpha\Psi = 0$.
\end{definition}
In coordinates, a chiral superfield writes
\begin{equation}\labelx{Eq:expchiral}
    \begin{split}
        \Phi(x,\theta,\overline{\theta}) = &\phi(x)+\sqrt{2}\theta\psi(x) + \I \theta\sigma^\mu\overline{\theta}\partial_\mu\phi(x)-\theta\theta F(x) \\
        &-\frac{\I}{\sqrt{2}}\theta\theta\partial_\mu\psi(x)\sigma^\mu\overline{\theta}-\frac{1}{4}\theta\theta\overline{\theta}\overline{\theta}\square\phi(x)\; ,
    \end{split}
\end{equation}
and corresponds to the chiral supermultiplet of the $4d$ $\mathcal N=1$ super-Poincaré algebra. Acting with $(\I \epsilon Q+\I \overline{\epsilon}\overline{Q})$, one deduces the following SUSY transformations:
\begin{align*}
    \delta\phi &= \sqrt{2}\epsilon\psi\; ,\\
    \delta\psi_\alpha &= \sqrt{2}\I (\sigma^\mu\overline{\epsilon})_\alpha\partial_\mu\phi - \sqrt{2}\epsilon_\alpha F\; ,\nonumber \\
    \delta F &= \I \sqrt{2}\partial_\mu\psi \sigma^\mu\overline{\epsilon}\; .
\end{align*}

Products and sums of chiral superfields are again chiral superfields. Moreover, if $\Phi$ is a chiral superfield then its conjugate $\overline{\Phi}$ is an anti-chiral superfield. Anti-chiral superfields admit an expansion in components similar to the one of \Cref{Eq:expchiral}, on which one can deduce the supersymmetry transformations just as for chiral superfields.

\subsubsection*{Vector superfields}

\begin{definition}
    An $\mathcal N=1$ \emph{vector superfield}\index{superfield!vector} is a superfield $V$ which satisfies the reality condition $\overline V=V$.
\end{definition} 

In components, one can check explicitly that the term $\theta\sigma^\mu\overline{\theta} v_\mu$ appears, where $v_\mu$ is a real vector field. Moreover, if $\Phi$ is a chiral superfield, then $\Phi+\overline{\Phi}$ is a vector superfield whose corresponding vector field is $-2\partial_\mu\Im\phi$. Therefore, the real vector field appearing in the component decomposition of $V+\Phi+\overline{\Phi}$ is $v_\mu-2\partial_\mu\Im\phi$: the transformation $V\rightarrow V+\Phi+\overline{\Phi}$ is a supersymmetric generalization of abelian gauge transformations.

\begin{remark}
    There always exists a chiral superfield $\Phi$ such that $V\rightarrow V_\mathrm{WZ} = V+\Phi+\overline{\Phi}$ is in normal form called \emph{Wess--Zumino gauge}\index{Wess--Zumino!gauge}, where it decomposes as:
\begin{equation*}
    V_\mathrm{WZ}(x,\theta,\overline{\theta})= \theta\sigma^\mu\overline{\theta}v_\mu(x)+\I \theta\theta\overline{\theta}\overline{\lambda}(x)-\I \overline{\theta}\overline{\theta}\theta\lambda(x)+\frac{1}{2}\theta\theta\overline{\theta}\overline{\theta}D(x)\; .
\end{equation*}
The vector field $v_\mu(x)$ of $V$ is a usual \emph{gauge field}, the Majorana fermion $\lambda$ is called the \emph{gaugino}\index{gaugino} associated to $v_\mu$, and $D(x)$ is the \emph{auxiliary field}\index{auxiliary field} of the $\mathcal N=1$ vector superfield. The superfield $V$ embodies the vector supermuliplet of the $4d$ $\mathcal N=1$ super-Poincaré algebra.
\end{remark}

\subsubsection*{(Super)current superfields}

The Noether current $j_\mu$ corresponding to a global continuous (non-$R$) symmetry is a component of a vector superfield $\mathcal J$ satisfying $D^2\mathcal J=\overline{D}^2\mathcal J=0$; this additional constraint supersymmetrizes the conservation law $\partial^\mu j_\mu=0$. 

Supersymmetries also define conserved currents $S_{\alpha\mu}$ which can be supersymmetrized. The \emph{Ferrara--Zumino completion}\index{Ferrara--Zumino completion} consists of a pair of a vector superfield $\mathcal J_\mu$ and a chiral superfield $X$ satisfying $2\overline{D}^{\dot\alpha}\sigma^\mu_{\alpha\dot\alpha}\mathcal J_\mu = D_\alpha X$. 

Lastly, the conserved current of an $R$-symmetry is supersymmetrized into a pair of a vector superfield $\mathcal R_\mu$ and a chiral superfield $\chi_\alpha$ satisfying $2\overline{D}^{\dot\alpha}\sigma^\mu_{\alpha\dot\alpha}\mathcal R_\mu = \chi_\alpha$ as well as $\overline{D}_{\dot\alpha}\overline{\chi}^{\dot\alpha}-D^\alpha\chi_\alpha=0$.

\subsection{Lagrangians}\labelx{Sec:N=1Lagrangians}

One can easily obtain $4d$ actions invariant under $\mathcal N=1$ SUSY by integration of a superfield $\mathcal A(x,\theta,\overline{\theta})$ over superspace:
\begin{equation*}
    S = \int\mathrm{d}^4x\mathrm{d}^2\theta\mathrm{d}^2\overline{\theta}~\mathcal{A}(x,\theta,\overline{\theta}) = \int\mathrm{d}^4x~\mathcal L(\phi(x),\psi(x),\lambda(x),A_\mu(x),\dots)\; .
\end{equation*}
Invariance of the action follows from the one of the measure, since:
\begin{equation*}
    \delta_{\epsilon,\overline{\epsilon}} S = \int\mathrm{d}^4x\mathrm{d}^2\theta\mathrm{d}^2\overline{\theta}~\delta_{\epsilon,\overline{\epsilon}}\mathcal{A}(x,\theta,\overline{\theta}) = 0\; ,
\end{equation*}
where the second equality follows from
\begin{equation*}
    \int\mathrm{d}^2\theta\mathrm{d}^2\overline{\theta}~ \delta_{\epsilon,\overline{\epsilon}}\mathcal{A}(x,\theta,\overline{\theta})
\end{equation*} 
being a total derivative. Physically, $\mathcal A$ is required to satisfy additional constraints in order for the Lagrangian density to be meaningful. Note that one can rewrite
\begin{equation*}
    \int\mathrm{d}^4x\mathrm{d}^2\theta\mathrm{d}^2\overline{\theta}~\mathcal{A}(x,\theta,\overline{\theta}) = \int\mathrm{d}^4x\mathrm{d}^2\theta~\overline{D}^2 \mathcal{A}(x,\theta,\overline{\theta})\; ,
\end{equation*}
and $\overline{D}^2 \mathcal{A}$ is a chiral superfield since $\overline{D}^3=0$: in general, integrating chiral superfields against $\mathrm{d}^2\theta$ also yields Lagrangian densities invariant under $\mathcal N=1$ supersymmetry. However, not all terms of the form
\begin{equation*}
    \int\mathrm{d}^4x\mathrm{d}^2\theta~\Phi\; ,
\end{equation*}
with $\Phi$ a chiral superfield, can be rewritten as integrals over the full superspace. 

\begin{definition}
    Integral terms that cannot be expressed as integrals over the entire superspace are referred to as \emph{F-terms}\index{F-term}, while those that can be written in this way are known as \emph{D-terms}\index{D-term}.
\end{definition}

\subsubsection*{Chiral superfields}

\begin{definition}
  The \emph{kinetic term} for a set of chiral superfields $\Phi_i$ is a D-term of the form:
\begin{equation}\labelx{Eq:kintermchiral}
	\int \mathrm{d}^2\theta\mathrm{d}^2\overline{\theta}~K(\Phi_i,\overline{\Phi_i})\; ,
\end{equation}
where $K$ is a real analytic (scalar) function of $\Phi_i$ and $\overline{\Phi}_i$ of mass dimension 2, called \emph{Kähler potential}\index{Kähler potential}.  
\end{definition} 

In the case of a single chiral superfield $\Phi$:
\begin{equation*}
	K(\Phi,\overline{\Phi}) = \sum_{m,n=1}^\infty c_{mn}\Phi^m\overline{\Phi}^n \quad \text{with} \quad c_{mn}=\overline{c}_{nm}\; ,
\end{equation*}
and the theory it defines is renormalizable only if $c_{mn}=0$ for $(m,n)\neq(1,1)$, in which case integration against $\mathrm{d}^2\theta\mathrm{d}^2\overline{\theta}$ leads to the canonical kinetic terms for the scalar and Weyl fermion components of $\Phi$:
\begin{equation*}
    \int\mathrm{d}^2\theta\mathrm{d}^2\overline{\theta} ~\overline{\Phi}\Phi = \partial_\mu\overline{\phi}\partial^\mu\phi+\frac{\I}{2}(\partial_\mu\psi\sigma^\mu\overline{\psi}-\psi\sigma^\mu\partial_\mu\overline{\psi})+\overline{F}F+\text{total derivative}\; .
\end{equation*}

In general, \Cref{Eq:kintermchiral} describes an $\mathcal N=1$ \emph{supersymmetric sigma model}\index{sigma model} into a Riemannian manifold coordinatized by the scalar components of the chiral superfields. SUSY implies that this manifold is in fact Kähler, with Kähler potential $K$ (hence the name). Such supersymmetric sigma models involving solely chiral superfields arise as \emph{low-energy effective theories}\index{low-energy effective theory} (cf. \Cref{sec:N=1quantumdyn}). The target space is the moduli space of the gauge theory, and non-renormalizability give away the fact that the sigma model description is only valid at low energies.

\begin{definition}
    For $W$ an analytic function of the $\Phi_i$'s one can also consider the $F$-terms:
\begin{equation}\labelx{Eq:SUPOterm}
    \int\mathrm{d}^2\theta~W(\Phi_i)+\text{c.c.} = -\frac{\partial W}{\partial\phi_i}F_i -\frac{1}{2}\frac{\partial^2W}{\partial\phi_i\partial\phi_j}\psi_i\psi_j+\text{c.c.}+\text{total derivative}\; ,
\end{equation}
which lead to mass and interaction terms. $W$ is called \emph{superpotential}\index{superpotential}. The superpotential term is renormalizable only if $W$ is at most cubic.
\end{definition}

All in all, $\mathcal N=1$ Lagrangian densities for chiral superfields are of the form:
\begin{equation}\labelx{Eq:chiralsupLagSUSY}
    \int \mathrm{d}^2\theta\mathrm{d}^2\overline{\theta}~K(\Phi_i,\overline{\Phi_i}) + \int\mathrm{d}^2\theta~W(\Phi_i) + \int\mathrm{d}^2\overline{\theta}~\overline{W}(\overline{\Phi_i})\; .
\end{equation}

\subsubsection*{Vector superfields}

\begin{definition}
    Let $V$ be an $\mathcal N=1$ vector superfield. The \emph{supersymmetric field strength}\index{supersymmetric!field strength} of $V$ is:
\begin{equation}\labelx{Eq:SUSYfieldstrength}
    W_\alpha = -\frac{1}{4}\overline{D}^2D_\alpha V\; ,
\end{equation}
where the $D$'s on the right hand side are defined in \Cref{Eq:covderSUSY}.
\end{definition}

The superfield field strength $W_\alpha$ is invariant under gauge transformations $V\rightarrow V+\Phi+\overline{\Phi}$, and:
\begin{equation}\labelx{Eq:kintermabegauge}
    \int\mathrm{d}^2\theta W^\alpha W_\alpha +\text{c.c.} = -F_{\mu\nu}F^{\mu\nu}- 4\I \lambda\sigma^\mu\partial_\mu\overline{\lambda}+2D^2\; ,
\end{equation}
with $F_{\mu\nu} = \partial_\mu v_\nu-\partial_\nu v_\mu$, and where here and henceforth we omit the `$+$total derivative' terms. Moreover:
\begin{equation*}
    \int\mathrm{d}^2\theta W^\alpha W_\alpha -\text{c.c.} = \frac{\I}{2}\epsilon^{\mu\nu\rho\sigma}F_{\mu\nu}F_{\rho\sigma}=\I F_{\mu\nu}(\star F)^{\mu\nu}\; ,
\end{equation*}
where $\star$ is Hodge star in $4d$ Minkowski space. Thus:
\begin{definition}
    The $\mathcal N=1$ generalization of $4d$ Maxwell theory, or equivalently the kinetic (and $\Theta$) term for an $\mathcal N=1$ vector superfield, reads: 
    \begin{equation}\labelx{Eq:abeliankineticSUSY}
    \frac{e^2}{4\pi}\Im\left(\tau_M\int\mathrm{d}^2\theta~W^\alpha W_\alpha\right) = -\frac{1}{4}F_{\mu\nu}F^{\mu\nu}-\I \lambda\sigma^\mu\partial_\mu\overline{\lambda}+\frac{1}{2}D^2+\frac{\Theta e^2}{16\pi^2}F_{\mu\nu}(\star F)^{\mu\nu}\; ,
    \end{equation}
    where 
    \begin{equation}\labelx{eq:defcomplexgaugecoupling}
        \tau_M=\frac{2\pi \I}{e^2}+\frac{\Theta}{2\pi}
    \end{equation}
    is the \emph{complexified abelian gauge coupling}\index{complexified gauge coupling}.
\end{definition}

This generalizes to non-abelian Yang--Mills theories as follows. Let $(T^a)_{a=1}^{\dim\mathfrak g}$ be the hermitian generators of a compact simple Lie algebra $\mathfrak g$ normalized in such a way that $2\tr(T^aT^b)=\delta^{ab}$, where $\tr$ denotes a Killing form on $\mathfrak g$ such that the long roots of $\mathfrak{g}$ have squared length 2. Let $V=2gV_aT^a$, with the $V_a$'s usual vector superfields and $g$ the non-abelian coupling constant. 

\begin{definition}
    The supersymmetric field strength for a non-abelian $\mathcal N=1$ vector superfield reads:
    \begin{equation}
        W_\alpha = -\frac{1}{4}\overline{D}^2(\E^{-V}D_\alpha \E^V)\; .
    \end{equation}
    The kinetic (and $\Theta$) term for a non-abelian $\mathcal N=1$ vector superfield writes:
    \begin{equation}\labelx{Eq:nonabeliankineticSUSY}
        \begin{split}
            \frac{1}{16\pi}\Im\left(\tau\int\mathrm{d}^2\theta \tr W^\alpha W_\alpha\right) = &\tr\left[-\frac{1}{2}F_{\mu\nu}F^{\mu\nu}-2\I \lambda\sigma^\mu D_\mu \overline{\lambda}
        +D^2\right]\\
        +&\frac{\Theta g^2}{16\pi^2}\tr F_{\mu\nu}(\star F)^{\mu\nu}\; ,
        \end{split}
    \end{equation}
    with:
    \begin{equation}\labelx{eq:complexgaugecouplingYM}
        \tau = \frac{4\pi \I}{g^2}+\frac{\Theta}{2\pi}\; , \quad F_{\mu\nu} = \partial_\mu v_\nu-\partial_\nu v_\mu-\I g[v_\mu,v_\nu]\; , \quad \text{and} \quad D_\mu = \partial_\mu-\I g[v_\mu,\cdot]\; ,
    \end{equation}
    respectively the \emph{complexified non-abelian gauge coupling}\index{complexified gauge coupling}, non-abelian field strength\footnote{Geometrically, the antisymmetric tensor $F_{\mu\nu}$ is better viewed as a $\mathfrak{g}$-valued 2-form $F=F_{\mu\nu}\mathrm{d}x^\mu\otimes\mathrm{d}x^{\nu}=\sum_{\mu<\nu}F_{\mu\nu}\mathrm{d}x^\mu\wedge\mathrm{d}x^\nu$, where $\mathrm{d}x^\mu\wedge\mathrm{d}x^\nu=(\mathrm{d}x^\mu\otimes\mathrm{d}x^\nu-\mathrm{d}x^\nu\otimes\mathrm{d}x^\mu)$. The volume form reads $\mathrm{d}^4x=\mathrm{d}x^0\wedge\mathrm{d}x^1\wedge\mathrm{d}x^2\wedge\mathrm{d}x^3$. With this notation $\mathrm{d}^4x~F_{\mu\nu}(\star F)^{\mu\nu} = 2 F\wedge F$ and the theta term can be recast as
    \begin{equation*}
        k\theta g^2=\frac{\theta g^2}{16\pi^2}\int\mathrm{d}^4x~\tr F_{\mu\nu}(\star F)^{\mu\nu} = \frac{\theta g^2}{8\pi^2}\int \tr F\wedge F\; .
    \end{equation*}
    The number $k$ is the instanton number of the principal $G$-bundle of which $F$ is the curvature. Let us consider $\mathfrak g = \mathfrak{su}(N)$ here for concreteness; the normalization is such that if the gauge group with Lie algebra $\mathfrak g$ is connected and simply-connected, i.e. $G=\mathrm{SU}(N)$, then $k$ is an integer \cite{Vafa:1994tf,Witten:2000nv,Gukov:2008Langlands}. This implies that $\theta$ has periodicity $\mathrm{2\pi}/g^2$; for other gauge groups $G=\mathrm{SU}(N)/\mathbb{Z}_k$ where $\mathbb{Z}_k$ is a subgroup of the center $\mathbb{Z}_N$ of $\mathrm{SU}(N)$, the angle $\theta$ naturally has an enlarged periodicity, for instance $\theta\sim \theta+4\pi N/g^2$ when $k=N$. Instantons will be discussed in \Cref{Subsec:instantons}.}, and covariant derivative (acting on the vector bundle associated to the adjoint representation). The Lagrangian density of \Cref{Eq:nonabeliankineticSUSY} defines $4d$ $\mathcal N=1$ super Yang--Mills (SYM) theories with gauge algebra $\mathfrak g$.
\end{definition}

The D-term of an abelian vector superfield $V$ transforms as a total derivative under (abelian) gauge transformations, allowing additional contributions to the Lagrangian density, of the form:
\begin{equation}\labelx{Eq:FayetIlioSUSY}
    \xi \int\mathrm{d}^2\theta\mathrm{d}^2\overline{\theta}~V(x,\theta,\overline{\theta}) = g\xi D(x)\; , \quad \xi\in\mathbb{R}\; .
\end{equation}
These are called \emph{Fayet--Iliopoulos terms}\index{Fayet--Iliopoulos terms}.

\subsubsection*{$\mathcal N=1$ gauge-matter Lagrangians}

To couple gauge symmetries to matter, one needs to consider chiral superfields $\Phi_i$ transforming in non-trivial representations $R_i$ of the gauge group. The combined transformations:
\begin{equation}\labelx{Eq:NonabSUSYgaugetrans}
    \begin{split}
        \E^V &\rightarrow \E^{\I\overline{\Lambda}_aT^a}\E^V\E^{\I\Lambda_aT^a}\; , \\
        \Phi_i &\rightarrow \E^{\I \Lambda_aT^a_{R_i}}\Phi_i\; ,
    \end{split}
\end{equation}
form the supersymmetric generalization of gauge symmetries, where the $\Phi_i$'s are chiral superfields transforming in representations $R_i$ of the gauge group, the $T^a_{R_i}$'s are the gauge generators in the representation $R_i$ and the $\Lambda_a$'s ($a=1,\dots,\dim\mathfrak g$) are chiral superfields embodying the parameters of the symmetry transformation. The \emph{Dynkin index}\index{Dynkin index} $I(R_i)$ of the representation $R_i$ is defined as $\tr_{R_i}(T^a_{R_i}T^b_{R_i})=I(R_i)\delta^{ab}$ (see e.g. \cite{DiFrancesco:1997nk}).

Gauge invariance requires the kinetic term for charged chiral superfields to be modified to:
\begin{equation}\labelx{Eq:kinchargedmatterSUSY}
    \int \mathrm{d}^2\theta\mathrm{d}^2\overline{\theta}~K(\Phi_i,(\overline{\Phi}\E^{2gV_R})_i)\; .
\end{equation}
In particular, the kinetic term (with canonical Kähler potential) of a chiral superfield $\Phi$ transforming in a representation $R$ of $G$ becomes:
\begin{equation}\labelx{Eq:chargechiralN=1kinetic}
    \begin{split}
        \int \mathrm{d}^2\theta\mathrm{d}^2\overline{\theta}~\overline{\Phi}\E^{2gV_R}\Phi = &(\overline{D}_\mu \overline{\phi}) D^\mu\phi -\I \psi\sigma^\mu D_\mu\overline{\psi}+\overline{F}F \\
    +&\I \sqrt{2}g\overline{\phi}\lambda\psi-\I \sqrt{2}g\overline{\psi}\overline{\lambda}\phi+g\overline{\phi}D\phi\; ,
    \end{split}
\end{equation}
with $D_\mu=\partial_\mu-\I gv^a_\mu T^a_R$ the covariant derivative on the vector bundle associated to the representation $R$ (whereas $D$ is the auxiliary field in the vector multiplet $V$). Note that every term in the second row of the previous equations involves one field in $R$, one in $\overline R$ and another in $\mathfrak{g}_R\simeq\mathrm{End}(R)$ or $\mathfrak{g}_{\overline{R}}$. The superpotential $W$ must be gauge invariant. 

\begin{definition}
    General $4d$ $\mathcal N=1$ gauge-matter Lagrangian densities are obtained by summing terms obtained above: if the gauge group $G$ decomposes as: 
    \begin{equation*}
	G=\prod_{a=1}^k\mathrm{U}(1)_a\times\prod_{j=1}^l G_j\; ,
    \end{equation*} 
    where the $G_i$'s are compact simple Lie groups, for each abelian or simple factor there is a term of the form of \Cref{Eq:abeliankineticSUSY} or \Cref{Eq:nonabeliankineticSUSY}. Abelian factors come with Fayet--Iliopoulos terms as in \Cref{Eq:FayetIlioSUSY}. The contribution of chiral superfields $\Phi_i$ is as in \Cref{Eq:chiralsupLagSUSY}, with kinetic terms modified as in \Cref{Eq:kinchargedmatterSUSY} if they are charged under the gauge group. The superpotential $W$ must be a $G$-invariant analytic function; this implies in particular that only chiral superfields in real representations of $G$ can be massive.
\end{definition}

\subsection{Moduli spaces}

Let us consider a $4d$ $\mathcal N=1$ $G$-gauge theory with renormalizable Lagrangian density and canonical kinetic term for the chiral superfields. A chiral superfield $\Phi$ in a representation $R$ of $G$ is considered as a vector of $(\dim R)$ chiral superfields $\Phi^i$ (and likewise $\overline{\Phi}=(\overline{\Phi}_i)_{i=1}^{\dim R}$). Let $(T^a)_{a=1}^{\dim\mathfrak g}$ be the generators of the gauge group $G$ and let $A$ be the subset of $[\vert1,\dim G\vert]$ such that for each $a\in A$ the generator $T^a$ corresponds to an abelian factor of $G$. The expansion of the Lagrangian density in components makes it clear that the fields $F_i$ and $D_a$ are auxiliary (i.e. they do not correspond to any kinetic term): the only terms in the Lagrangian involving $D$ and $F$ are:
\begin{equation*}
	\tr_\mathfrak g(D^2)+g\sum_{a\in A} \xi^a D_a+\overline{F}_iF^i+g\overline{\phi_i}D_a(T^a_{R})^i_j\phi^j-\frac{\partial W}{\partial \phi^i}F^i-\frac{\partial \overline W}{\partial \overline \phi_i}\overline F_i\in\mathcal L\; .
\end{equation*} 

The equations of motion for $\overline{F}_i$, $F^i$ and $D_a$ are:
\begin{equation*}
    \overline{F}_i = \frac{\partial W}{\partial \phi^i}~,~~~F^i = \frac{\partial \overline{W}}{\partial \overline\phi_i}~,~~~D^a = -g\xi^a-g\overline{\phi}_i( T^a_{R_i})^i_j\phi^j\; .
\end{equation*}
Integrating the $D_a$'s and $F_i$'s out leads to an \emph{on-shell Lagrangian}\index{Lagrangian!on-shell}, in which the terms that only depend on the scalar fields $\phi^i$ (and $\overline{\phi}_i$) are:
\begin{equation*}
    \mathcal L \ni \overline{D}_\mu\overline{\phi}D^\mu\phi-V(\phi^i,\overline\phi_i)\; ,
\end{equation*}
with $V(\phi^i,\overline{\phi}_i)$ the (semi-positive definite) \emph{scalar potential}:
\begin{equation}\labelx{Eq:scalarpotential}
	V(\phi^i,\overline{\phi}_i) = \frac{\partial W}{\partial\phi_i}\frac{\partial\overline{W}}{\partial\overline{\phi}_i}+\frac{g^2}{2}\sum_a \left\vert \overline{\phi}_i(T^a)^i_j\phi^j+\xi^a\right\vert^2\; ,
\end{equation}
where we have set $\xi^a=0$ whenever $a\notin A$. 

\begin{definition}
   A \emph{vacuum}\index{vacuum} of a quantum field theory (QFT) is a Lorentz-invariant state in which the kinetic energy of all fields vanishes. Being Lorentz invariant means that only Lorentz-invariants operators can have a non-zero \emph{vacuum expectation value}\index{vacuum!expectation value} (VEV), denoted as $\langle \phi^i \rangle$. A vacuum is meta-stable if the associated set of VEVs minimizes the scalar potential $V$ locally, and stable if it does so globally.
\end{definition}

A metastable vacuum preserves supersymmetry if and only if all supercharges act trivially on it. Recalling \Cref{Eq:positivenessSUSY}, this means that the scalar potential $V$ evaluated at the vacuum expectation values (VEVs) vanishes. Consequently, any metastable supersymmetric vacuum is genuinely stable, as the potential $V$ is semi-positive definite.

\Cref{Eq:scalarpotential} shows that the supersymmetric vacua of a $4d$ $\mathcal N=1$ gauge theory correspond bijectively to collections of scalar VEVs solving the \emph{F-term}\index{F-term!equation} and \emph{D-term equations}\index{D-term!equation}:
\begin{equation*}
	\overline F_i(\phi) = 0\; , \quad D^a(\phi,\overline\phi)=0\; .
\end{equation*} 

Gauge-equivalent vacua must be physically identified, which leads to the following:

\begin{definition}
    The \emph{classical moduli space}\index{moduli space!classical} (of supersymmetric vacua), denoted as $\mathcal M_\mathrm{cl}$, is the submanifold of solutions to the $F$-term and $D$-term equations within the complex affine space of scalar vacuum expectation values (VEVs), modulo gauge symmetries. In more formal terms, $\mathcal M_\mathrm{cl}$ can be described as the Kähler quotient of the solutions to the $F$-term equations by the group $G$. Alternatively, one can define $\mathcal M_\mathrm{cl}$ using Geometric Invariant Theory, as the space parametrized by the VEVs of gauge-invariant scalar operators, factoring in classical relations and $F$-terms. The equivalence of these two approaches is established by the Kempf–Ness theorem; see also \cite{Luty:1995sd}.
\end{definition}

\subsection{Four-dimensional \texorpdfstring{$\mathcal N=1$}{N=1} SQCD (I)}\labelx{Sec:classicalSQCD}

We conclude the section with the example of $\mathcal N=1$ SQCD (\emph{Super Quantum ChromoDynamics}\index{$\mathcal N=1$ SQCD}), which is the $\mathcal N=1$ generalization of quantum chromodynamics as well as a rich illustration of the general discussion. Here we consider SQCD as a classical $4d$ $\mathcal N=1$ field theory and defer the study of quantum effects to the next section. 

SQCD is a (renormalizable) Lagrangian gauge theory with gauge group $\mathrm{SU}(N)$ ($N$ is the number of \emph{colors}\index{symmetry!color}), $F>1$ pairs of chiral superfields $(Q_i,\widetilde Q^i)$ for $i=1,\dots,F$ called \emph{flavors}\index{symmetry!flavor}, such that each $Q_i$ (resp. $\widetilde Q^i$) transforms in the fundamental (resp. antifundamental) representation of $\mathrm{SU}(N)$, and no superpotential. Explicitly in $\mathcal N=1$ superspace, the Lagrangian writes:
\begin{equation*}
    \mathcal L = \frac{1}{16\pi}\mathrm{Im}\left(\tau \int\mathrm{d}^2\theta \tr W^\alpha W_\alpha\right)+\int\mathrm{d}^2\theta\mathrm{d}^2\overline{\theta}~\left(\overline Q_i \E^{2gV}Q_i+\overline{\widetilde{Q}^i} \E^{-2gV}\widetilde{Q^i}\right)\; .
\end{equation*}

The simple or abelian (0-form) global symmetry factors of $\mathcal L$ are:
\begin{equation}\labelx{Eq:globsymclassSQCD}
	\begin{array}{c||c|c|c|c|c|}
		& \mathrm{SU}(F)_L & \mathrm{SU}(F)_R & \mathrm{U}(1)_B & \mathrm{U}(1)_A & \mathrm{U}(1)_R \\
		\hline
		Q_i & \overline{F} & - & 1  & 1 & 1 \\
		\widetilde{Q}^i & - & F & -1 & 1 & 1 \\
		\lambda & - & - & - & - & 1 
	\end{array}
\end{equation}
As before, $\lambda$ is the \emph{gaugino}\index{gaugino}, i.e. the real spinor in the vector supermultiplet. Under an $R$-symmetry $\lambda$ must have charge $1$ (this follows from the fact that $W_\alpha \sim -\I \lambda_\alpha + \dots$, together with $R[\mathrm{d}\theta]=-1$), and the $R$-charges of $Q_i$ and $\widetilde{Q}^i$ must be equal since the $Q_i$'s and $\widetilde{Q}^i$'s are mutually charge-conjugate (charge conjugation commutes with the supercharges). Note that combining $\mathrm{U}(1)_A$ and $\mathrm{U}(1)_R$ yields other (though not independent) $R$-symmetries under which $\lambda$ always has charge $1$ but the $Q_i$'s and $\widetilde Q^i$'s can have any charge $a\in\mathbb{R}$. The group $\mathrm{SU}(F)_L\times\mathrm{SU}(F)_R$ is the \emph{group of flavor symmetries}\index{flavor symmetry group} of SQCD. Flavor indices will be denoted $i,j,\dots=1,\dots,F$ and color indices $a,b,\dots=1,\dots,N$.

\vspace{0.3cm}

Let $q_i^a(x)$ (resp. $\widetilde{q}^i_a(x)$) be the scalar components of the chiral superfields $Q_i^a(x,\theta,\overline{\theta})$ (resp. $\widetilde{Q}^i_a(x,\theta,\overline{\theta})$). The $q_i^a(x)$'s form an $N\times F$ matrix $q(x)$ of scalar fields, and the $\widetilde{q}^i_a(x)$'s an $F\times N$ matrix $\overline{q}(x)$. Under a gauge transformation $G(x)\in\mathrm{SU}(N)$ and flavor symmetry $(L,R)\in\mathrm{SU}(F)_L\times\mathrm{SU}(F)_R$, these matrices transform as:
\begin{equation*}
    \widetilde{q}(x) \rightarrow L\widetilde{q}(x) G^\dagger(x)\; ,\quad q(x)\rightarrow G(x) q(x) R\; .
\end{equation*} 

The moduli space of SQCD is parameterized by the VEVs of $q$ and $\widetilde q$, which are $N\times F$ and $F\times N$ matrices with constant coefficients. These VEVs will still be denoted $q_i^a$ and $q_a^i$. The group of gauge and global symmetries acting on them is $\mathrm{SU}(N)\times\mathrm{SU}(F)_L\times\mathrm{SU}(F)_R$ where $\mathrm{SU}(N)$ consists of all gauge transformations constant in spacetime. Since the superpotential vanishes in SQCD, the classical moduli space of vacua is the subvariety of $\mathbb{C}^{2NF}$ defined by D-terms only:
\begin{equation*}
	(D^A)^i_j = q^{i\dagger}_a(T^A)^a_bq_j^b - \widetilde{q}^i_a(T^A)^a_b\widetilde{q}^{b\dagger}_j = 0\; , \quad A=1,\dots,N^2-1\; ,
\end{equation*}
and modulo gauge transformations. As before, the $T^A$'s are the generators of $\mathfrak{su}(N)$ in the fundamental representation (the ones in the antifundamental representation are $-T^A$). Equivalently, the classical moduli space is the subvariety of $\mathbb{C}^{2NF}$ whose ring of functions consists of the VEVs of scalar gauge invariant operators.

\vspace{0.3cm}

One distinguishes the cases $F<N$ and $F\geq N$ \cite{Affleck:1983mk,Intriligator:1995au,Argyres:2001eva,Bertolini}. 

\vspace{0.3cm}

\begin{itemize}
    \item $\mathbf{F<N}.$ Under $\mathrm{SU}(N)\times\mathrm{SU}(F)_R\times\mathrm{U}(1)_B$ one can assume that $q$ is decomposed in singular values, i.e. $q=\mathrm{Diag}(v_1,\dots,v_F)$ with zeros away from the diagonal and $v_i\in\mathbb{R}_{\geq0}$ for all $i$. With the help of $\mathrm{SU}(F)_L$ one can moreover QR-decompose $\widetilde q$, i.e. assume that $\widetilde q$ is upper-triangular. Vanishing of the D-terms then in fact imposes that $\widetilde q=q^T$. At generic values of $q$, all $v_i$'s are strictly positive and distinct. Such a choice of VEVs breaks the gauge group to $\mathrm{SU}(N-F)$, and by the (super-)Higgs mechanism each broken gauge generator gives a non-zero mass to a previously massless complex scalar. Therefore the number of remaining massless complex scalars, i.e. the dimension of the classical moduli space, is: 
    \begin{equation*}
	2NF - [(N^2-1) - ((N-F)^2-1)] = F^2\; .
    \end{equation*} 
	
    The gauge-invariant scalar operators parameterizing the moduli space in this case are the entries of the \emph{meson matrix}\index{meson matrix} $M=(m^i_j)$:
    \begin{equation*}   
        m^i_j =  \widetilde{q}_a^iq_j^a\; , \quad a=1,\dots,N\; , \quad i,j=1,\dots,F\; .
    \end{equation*}
    Generically $m$ has rank $F$, consistently with the the moduli space having complex dimension $F^2$. 
	
    \item $\mathbf{F\geq N}$. As before, up to $\mathrm{SU}(N)\times\mathrm{SU}(F)_R\times\mathrm{U}(1)_B$ one can assume that $q$ is diagonal with real positive diagonal elements $v_1,\dots,v_n$, and up to $\mathrm{SU}(F)_L$, that $\widetilde{q}$ is upper-triangular. Vanishing of D-terms implies that $\widetilde{q}$ is diagonal with diagonal elements $\widetilde{v}_1,\dots,\widetilde{v}_N$ such that for all $i$, $\vert \widetilde{v}_i\vert^2-v_i^2=\rho$ for some $i$-independent $\rho$. Generically the gauge group is now spontaneously fully broken, i.e. the moduli space has complex dimension:
    \begin{equation*}
        2NF- (N^2-1)\; .
    \end{equation*}
    The gauge invariant operators parameterizing the moduli space are the elements of the meson matrix $\widetilde{q}q$, as well as \emph{baryons}\index{baryons} $b$ and \emph{anti-baryons} $\widetilde{b}$:
    \begin{align*}
        b_{i_1 i_2\dots i_N} &= \epsilon_{a_1 a_2\dots a_N}q_{i_1}^{a_1} q_{i_2}^{a_2} \dots q_{i_N}^{a_N}\; ,\\
        \widetilde b^{i_1 i_2\dots i_N} &= \epsilon^{a_1 a_2\dots a_N}\widetilde q_{a_1}^{i_1} \widetilde q_{a_2}^{i_2} \dots \widetilde q_{a_N}^{i_N}\; .
    \end{align*}
    The matrices $b$ and $\widetilde b$ are fully antisymmetric, therefore in total there are
    \begin{equation*}
	2 {\binom{F}{N}} + F^2
    \end{equation*}
    mesons and baryons, among which only $2NF-(N^2-1)$ are independent: the relations follow from the definitions and $m, b$ and $\widetilde{b}$: the product of a baryon and an anti-baryon is the minor of the meson matrix $m$ defined by the color indices of the (anti-)baryon.
\end{itemize}

\section{Quantum supersymmetric gauge dynamics}\labelx{sec:N=1quantumdyn}

In QFTs, one of the most significant quantum effects is the renormalization of fields and couplings. This process means that the physics encoded by a QFT is dependent on the energy scale (or equivalently, the length scale) at which it is observed. Our focus will typically be on asymptotically free gauge theories, which transition from a classical behavior at very high energies (in the UV) to very quantum, strongly-coupled dynamics at low-energies (in the IR). We are particularly interested in the low-energy effective physics and how it is influenced by the choice of vacuum.

Low-energy effective physics refers to phenomena occurring at energies below any non-zero scale present in the theory. In this context, we distinguish between \emph{gapped vacua}\index{vacuum!gapped}, where there is a significant energy gap between the vacuum state and the lowest non-vacuum state, and \emph{gapless vacua}\index{vacuum!gapless}, which allow massless excitations with arbitrarily small energy.

Renormalization in supersymmetric (SUSY) QFTs is generally more manageable than in non-supersymmetric theories, as the former satisfy various \emph{supersymmetric non-renormalization theorems}\index{non-renormalization theorem}. In these theories, the low-energy effective theory at any given supersymmetric vacuum remains supersymmetric. Therefore, at a gapless SUSY vacuum, if the low-energy effective physics can be described by a Lagrangian, it will consist of massless chiral and vector superfields. The scalar components of their gauge-invariant combinations parameterize the moduli space of supersymmetric vacua $\mathcal M$.

\emph{Instantons}\index{instanton}, i.e. critical Euclidean gauge field configurations, are important non-perturbative quantum effects that significantly influence the semi-classical analysis of gauge theories. \emph{Anomalies}\index{anomaly} are also crucial in the study of quantum dynamics. Lastly, we discuss \emph{holomorphy}\index{holomorphy}, a key property of $4d$ $\mathcal N=1$ theories signifying that some effective couplings depend holomorphically on the UV parameters. This will serve as our primary tool for investigating the effective low-energy physics of SUSY gauge theories.

Quantum effects can drastically modify the structure of the moduli space $\mathcal M$ of a Lagrangian QFT. For this reason, one distinguishes between the classical moduli space $\mathcal M_\mathrm{cl}$ and the genuine \emph{quantum moduli space}\index{moduli space} $\mathcal M$. The metric, topology, or even the dimension of $\mathcal M$ may differ from those of $\mathcal M_\mathrm{cl}$. Singularities in the moduli space are particularly significant, as they indicate breakdowns in the low-energy effective description of the theory, usually associated with generically massive fields becoming massless. We will delve into the specifics of quantum $4d$ $\mathcal N=1$ SQCD and conclude with a discussion of \emph{Seiberg duality}\index{Seiberg duality}.

\subsection{Renormalization}\labelx{Subsec:N=1renorm}

The one-loop renormalization of the gauge coupling in any $4d$ Yang--Mills theory with simple compact gauge group $G$ together with scalars and fermions charged under $G$ reads \cite{Gross:1973ju,Politzer:1973fx}:
\begin{equation}\labelx{Eq:rengen}
    \frac{\mathrm{d}}{\mathrm{d}\log\mu} g(\mu)  = -\frac{g^3(\mu)}{(4\pi)^2}\left[\frac{11}{3}C(\mathrm{Adj})-\frac{2}{3}n_fI(R_f)-\frac{1}{3}n_sI(R_s)\right]\; ,
\end{equation}
where $g(\mu)$ is the gauge coupling at the energy scale $\mu$, and $n_f$ (resp. $n_s$) is the number of Weyl/Majorana fermions (resp. scalars) in the representation $R_f$ (resp. $R_s$) of $G$. For any representation $R$ of $G$ the \emph{Casimir number}\index{Casimir element} $C(R)$ is defined by:
\begin{equation*}
    T_{R,a}\otimes T_R^a = C(R)\mathbf{1}_R\; .
\end{equation*}
It satisfies: 
\begin{equation*}
    C(R)\dim(R)=I(R)\dim(G)\; ,
\end{equation*}
where $I(R)$ is the \emph{Dynkin index}\index{Dynkin index} of the representation $R$ defined below \Cref{Eq:NonabSUSYgaugetrans} (in this normalization, $C(\mathrm{Adj})=I(\mathrm{Adj})$ equals the dual Coxeter number of $G$, e.g. $C(\mathrm{Adj})=N$ when $G=\mathrm{SU}(N)$).

For $n_f$ and $n_s$ small enough, non-abelian gauge dynamics implies asymptotic freedom; moreover, under some conditions, non-abelian gauge interactions are necessary for asymptotic freedom (Coleman--Gross theorem \cite{Coleman:1973sx}). Rewriting \Cref{Eq:rengen} as:
\begin{equation*}
    \frac{\mathrm{d}}{\mathrm{d}\log\mu}\left(\frac{8\pi^2}{g^2(\mu)}\right) = b_1\; ,
\end{equation*}
the solutions to this differential equation read:
\begin{equation*}
    \frac{8\pi^2}{g^2(\mu)} = -b_1\log\frac{\widetilde\Lambda}{\mu}\; ,
\end{equation*}
with $\widetilde\Lambda$ a $\mu$-independent integration constant referred to as the (one-loop) \emph{dynamical scale}\index{dynamical scale} of the theory. It is the scale at which the one-loop renormalization of the gauge coupling diverges.

Lagrangian $\mathcal N=1$ gauge theories with gauge group $G$ consist of a vector superfield in the adjoint representation of $G$ and chiral superfields $\chi$ in representations $R_\chi$. Therefore, \Cref{Eq:rengen} simplifies to:
\begin{equation*}
    \frac{\mathrm{d}}{\mathrm{d}\log\mu} \left(\frac{8\pi^2}{g^2(\mu)}\right)  = 3C(\mathrm{Adj})-n_\chi I(R_\chi)\; .
\end{equation*}
For instance, $b_1=3N-F$ for SQCD with $N$ colors -- i.e. $G=\mathrm{SU}(N)$ -- and $F$ flavors.

The theta term in Yang--Mills theory introduced in \Cref{Eq:abeliankineticSUSY} being a total derivative, $\Theta$ is perturbatively exact. Thus at one loop in perturbation theory one can write:
\begin{equation}\labelx{Eq:0renN=1}
    \frac{\mathrm{d}}{\mathrm{d}\log\mu}\tau = \frac{\mathrm{d}}{\mathrm{d}\log\mu} \left(\frac{4\pi \I}{g^2(\mu)}+\frac{\Theta}{2\pi}\right) = -\frac{b_1}{2\pi \I}\; ,
\end{equation}
where $\tau$ is the complexified gauge coupling defined in \Cref{eq:defcomplexgaugecoupling}. This equation is solved by:
\begin{equation}\labelx{Eq:renN=1}
    \tau(\mu) = \frac{b_1}{2\pi \I}\log\frac{\Lambda}{\mu}\; ,
\end{equation}
with the $\mu$-independent \emph{holomorphic dynamical scale}\index{holomorphic!dynamical scale} $\Lambda$ satisfying $\widetilde\Lambda=\vert\Lambda\vert$. In terms of the coupling $\tau(\mu)$ at any scale $\mu$ one has:
\begin{equation*}
    \Lambda = \mu\exp(\frac{2\pi \I\tau(\mu)}{b_1})\; .
\end{equation*}

If an $\mathcal N=1$ asymptotically free theory has coupling $\tau_\mathrm{UV}$ at some UV scale $\mu_\mathrm{UV}$, this equation yields:
\begin{equation*}
    \tau(\mu) = \tau_\mathrm{UV} - \frac{b_1}{2\pi \I}\log\frac{\mu}{\mu_\mathrm{UV}}\; ,
\end{equation*}
where $\mu$ is any other energy scale for which the one-loop perturbative approximation of the theory is justified all the way from $\mu_\mathrm{UV}$ to $\mu$.

\subsubsection*{SUSY non-renormalization theorems}

One of the most important features of SUSY field theories is that they are better behaved under renormalization than generic quantum field theories. One can prove that perturbative loop corrections in $d=4$ $\mathcal N=1$ gauge theories can only generate D-terms and not F-terms. 
In particular, the superpotential is tree-level exact in perturbation theory (however, kinetic terms are D-terms and thus generically receive perturbative quantum corrections). Such SUSY non-renormalization theorems can uniformly be derived from another fundamental property of supersymmetric field theories: \emph{holomorphy}\index{holomorphy}, to be discussed in \Cref{Sec:HolomorphySUSY}.

\subsection{Instantons}\labelx{Subsec:instantons}

In the canonical normalization, the Lagrangian of pure Yang--Mills theory with simple compact gauge group $G$ reads:
\begin{equation}\labelx{Eq:YMTheta}
    \mathcal L[A_\mu] = -\frac{1}{2g^2}\tr F_{\mu\nu}F^{\mu\nu} + \frac{\Theta}{16\pi^2}\tr F_{\mu\nu}(\star F)^{\mu\nu}\; .
\end{equation}
Every connected simple compact group $G$ satisfies $\pi_3(G)\simeq \mathbb{Z}$. This follows from every such $G$ being homotopy equivalent to a bouquet of odd-dimensional spheres of dimension $d_i\geq 3$ (or a quotient thereof by a finite group), where among these spheres, exactly one is of dimension 3. This allows countably many topologically-inequivalent classical vacua in pure Yang--Mills theory. In the quantum theory, tunneling occurs; the quantum vacuum of the theory (called \emph{$\Theta$-vacuum}\index{vacuum!$\Theta$-}) has an intricate structure. Tunneling amplitudes are handily computed with the help of \emph{instantons}\index{instanton} \cite{Vafa:1994tf,donaldson1997geometry,Witten:2000nv,Dorey:2002ik,Tong:2005un}. 

\begin{definition}
    An \emph{instanton} is a classical solution of the Euclidean theory with finite action:
\begin{equation}
    S_E[A_\mu] = \int_\mathcal{M}\mathrm{d}^4x~\frac{1}{2g^2}\tr F_{\mu\nu}F^{\mu\nu} -\frac{\I\Theta}{16\pi^2}\tr F_{\mu\nu}(\star F)^{\mu\nu}\; ,
\end{equation}
where, in this Section, $\mathcal M$ denotes the Wick-rotated Euclidean spacetime.
\end{definition}

One has:
\begin{equation}\labelx{Eq:ineqinst}
    \begin{split}
        \int_\mathcal{M}\mathrm{d}^4x~\frac{1}{2g^2}\tr F^2 &= \int_\mathcal{M}\mathrm{d}^4x~\left(\frac{1}{4g^2}\tr (F\mp\star F)^2 \pm \frac{1}{2g^2}\tr F_{\mu\nu}(\star F)^{\mu\nu}\right) \\
        &\geq \pm\int_\mathcal{M}\frac{1}{2g^2}\tr F_{\mu\nu}(\star F)^{\mu\nu} = \pm\frac{8\pi^2}{g^2}n\; ,  
    \end{split}
\end{equation}
where 
\begin{equation*}
    n= \frac{1}{16\pi^2}\int_\mathcal{M} \mathrm{d}^4x~\tr F_{\mu\nu}(\star F)^{\mu\nu} = \frac{1}{8\pi^2}\int_\mathcal{M}\tr F\wedge F
\end{equation*}
is the instanton number of the gauge configuration, or in more mathematical terms, the first \emph{Pontryagin number}\index{Pontryagin!number} of the (real) vector bundle associated to the $G$-principal bundle on which $A_\mu$ is a connection, and the fundamental representation of $G$. The inequality in \Cref{Eq:ineqinst} saturates when $F=\pm\star F$, i.e. when $A_\mu$ is a self-dual (resp. anti-self-dual) $G$-connection on $\mathcal M$. Self-dual (resp. anti-self-dual) connections on $\mathcal M$ satisfy $n\geq 0$ (resp. $n\leq 0$). Thus:

\begin{definition}
    An \emph{instanton} (resp. anti-instanton) with instanton number $n$ when $n\geq 0$ (resp. $n\leq 0$) is a Euclidean gauge configuration which saturates the above bound.
\end{definition}

The contribution to the path integral of a configuration with instanton number $n\geq0$ is:
\begin{equation}\labelx{Eq:actioninstanton}
	\exp(-S_E) = \exp(-n\left(\frac{8\pi^2}{g^2} -\I \Theta\right)) = \exp(2\pi \I n\tau) = \left(\frac{\Lambda}{\mu}\right)^{b_1n}\; .
\end{equation}
This shows that instanton contributions are exponentially weak when $g^2$ is small (or equivalently when $\Im\tau$ is large); they are non-perturbative effects.

\begin{remark}
    When $G$ is simply connected, e.g. $G=\mathrm{SU}(N)$, the instanton number $n$ is always an integer, provided the trace in \Cref{Eq:YMTheta} is a Killing form on $\mathfrak g =\mathrm{Lie}(G)$ such that the long roots of $\mathfrak{g}$ have squared length 2 (see e.g. \cite{Gukov:2008Langlands}). When $G$ is not simply connected, $n$ can be rational. Let us restrict to groups $G$ such that $\mathrm{Lie}(G)=\mathfrak{su}(N)$ for concreteness, and let $X$ be an arbitrary smooth Euclidean four-manifold. Up to isomorphism, an $\mathrm{SU}(N)$-bundle over $X$ is completely determined by $n\in\mathbb{Z}$. In contrast, an $\mathrm{SU}(N)/\mathbb{Z}_p$-bundle over $X$ (where $p$ divides $N$) is determined by the instanton number $n$ as well as a \emph{discrete magnetic flux}\index{discrete magnetic flux} $w_2\in \mathrm{H}^2(X,\mathbb{Z}_N)$. These two invariants are related:
\begin{equation*}
    n = \frac{N(p-1)}{2p^2}\int_X \mathcal{P}(w_2) \quad \text{mod. } 1\; ,
\end{equation*}
where $\mathcal P$ is the \emph{Pontryagin square}\index{Pontryagin!square} operation. A gauge configuration for which $n$ is not integral is called \emph{fractional instanton}\index{instanton!fractional}. 
\end{remark}

Even if the topology of $\mathcal M$ is trivial, gauge theories describe extended objects. 't Hooft lines, for example, correspond to singular gauge field configurations; the effective topology one should consider when studying instantons in presence of such a line is not $\mathcal M$ but rather $\mathcal M\setminus\{\text{line}\}$, which has non-trivial topology. Therefore, even if it naively seems that one cannot distinguish an $\mathrm{SU}(N)$ gauge theory from an $\mathrm{SU}(N)/\mathbb{Z}_p$ theory in flat spacetime, the difference between the two is captured by the spectrum of line operators in the theory \cite{Aharony:2013hda}. In fact, there even exist distinct $\mathrm{SU}(N)/\mathbb{Z}_p$ gauge theories called \emph{global variants}\index{global variant}: since $n$ is rational rather than integral, the $\Theta$-angle generally has periodicity larger than $2\pi$. Shifting $\Theta\rightarrow\Theta+2\pi$ in an $\mathrm{SU}(N)/\mathbb{Z}_p$ gauge theory is often not a symmetry of the theory, but rather maps one global variant of the $\mathrm{SU}(N)/\mathbb{Z}_p$ theory to another.

Instantons in supersymmetric gauge theories have an intimate relationship with supersymmetry and BPS states, as (anti)-self dual gauge fields are invariant under half of the supersymmetry generators. Supersymmetry helps computing instanton contributions which play a role in \emph{gaugino condensation}\index{gaugino!condensation} at low energies in $\mathcal N=1$ super Yang--Mills theories. We will come back to gaugino condensation in $\mathcal N=1$ SYM in \Cref{Sec:quantSQCD}.

\subsection{Anomalies}\labelx{Sec:anomalies}

\begin{definition}\labelx{def:anomaly}
    A classical symmetry of a Lagrangian field theory is said to be \emph{anomalous}\index{anomaly} if it fails to be a symmetry of the corresponding quantum theory.
\end{definition}
If the symmetry is continuous with classically conserved 1-form current $j$, the anomaly means that the classical conservation equation $\mathrm{d}(\star j) = \partial_\mu j^\mu=0$ does not hold in the quantum theory, or, more precisely, that the Ward--Takahashi identity corresponding to the classical conservation equation does not hold.

\begin{remark}
    When quantizing a classical gauge theory, the gauge symmetry is crucial to obtain a quantum theory without negative-norm states: gauge symmetries cannot be anomalous in a consistent quantum field theory. This constrains the matter content of the theory. On the other hand, an anomalous classical global symmetry does not conflict with the overall consistency of the theory; it merely means that the actual symmetry of the quantum theory is smaller than what one naively expects from the classical description (even though sometimes part of an anomalous symmetry revives as a non-invertible symmetry \cite{Choi:2022jqy}).
\end{remark}

\subsubsection*{Anomalies of continuous global chiral symmetries}

Anomalies famously occur in gauge theories with chiral fermions (such as $\mathcal N=1$ SQCD of \Cref{Sec:classicalSQCD}); we refer to the reviews \cite{Harvey:2005it,Bilal:2008qx} for in-depth discussions. 

\begin{definition}
    A symmetry is \emph{chiral}\index{symmetry!chiral} if it acts differently on left-handed and right-handed Weyl fermions, i.e. if the corresponding current (if they are continuous) is chiral.
\end{definition} 

Chiral symmetries are often anomalous because under the symmetry, the transformations of the path integral measures of right-handed and left-handed fermions do not cancel out unless the matter content of the theory is tuned properly \cite{Fujikawa:1979}. Anomalies of chiral currents are usually referred to as Adler--Bell--Jackiw (ABJ) anomalies after \cite{Adler:1969gk} and \cite{Bell:1969ts}. They are exact at one-loop in perturbation theory, and computed (in a $d$-dimensional quantum field theory with $d$ even) by evaluating one-loop $(d/2+1)$-gon Feynman diagrams only (this is the Adler--Bardeen theorem; see e.g. \cite{Adler:2004qt} for an account). Hence in $4d$, the chiral anomaly is computed from triangle diagrams of fermions with one global and two gauge current insertions.

In a Yang--Mills theory with gauge group $G$, if the chiral symmetry group $\widetilde G$ is continuous with current~$\widetilde j_a$, then:
\begin{equation}\labelx{Eq:anomglobalchiralcurrent}
    \partial_\mu (\widetilde j^a)^\mu = \frac{1}{8\pi^2} \tr_{\widetilde R}(T^a)\tr_{R}\{T^b,T^c\}F_b\wedge F_c\; ,
\end{equation} 
for every Weyl fermion in the representation $(\widetilde R,R)$ of $\widetilde G\times G$. The right-hand side of \Cref{Eq:anomglobalchiralcurrent} vanishes unless $\widetilde G$ is abelian, i.e. only abelian global symmetry factors can be anomalous. Summing over fermions yields the total anomaly $\mathcal A$, which for $\widetilde G=\mathrm{U}(1)$ reads:
\begin{equation}\labelx{Eq:ABJanomaly}
    \mathcal A = \sum_i \widetilde q_i I(R_i) \quad \text{with} \quad \partial_\mu \widetilde j^\mu = \frac{\mathcal A}{8\pi^2} F_b\wedge F^b\; .
\end{equation}
The sum runs over all Weyl fermions in the theory, $\widetilde q_i$ is the $\widetilde G$-charge of the $i$-th fermion, and $I(R_i)$ is the Dynkin index of the representation $R_i$ of $G$ under which the $i$-th fermion transforms. The amount of charge violation due to the anomaly is:
\begin{equation}\labelx{Eq:ABJanomalycharge}
    \Delta Q = \int \mathrm{d}^4x ~\partial_\mu\widetilde j^\mu = \frac{2 \mathcal A}{8\pi^2} \int \tr F\wedge F\; \; ,
\end{equation}
where the trace is again taken in the fundamental representation. This means that in an instanton background of instanton number $n$ (defined in \Cref{Subsec:instantons}), the symmetry is broken by the amount $\Delta Q = 2\mathcal A n$. This shows that the ABJ anomaly associated to a $\mathrm{U}(1)$-transformation of parameter $\alpha$ amounts to a shift of the Yang--Mills theta angle $\Theta\rightarrow \Theta+2\alpha \mathcal A$. If the theory one considers is invariant under $\Theta\rightarrow \Theta+2\pi k$ with $k$ the smallest such integer, the ABJ anomaly implies that only the subgroup $\mathbb{Z}_{2\mathcal A/k}$ of the classical $\mathrm{U}(1)$ global symmetry is a genuine global symmetry of the quantum theory.

\begin{example}
    Let us consider $4d$ $\mathcal N=1$ SQCD (cf. \Cref{Sec:classicalSQCD}). Chiral global symmetries of the theory were listed in \Cref{Eq:globsymclassSQCD}; among all possible $R$-symmetries we consider a general one $\mathrm{U}(1)_R^a$ under which the $Q_i$'s and $\widetilde Q^i$ have charge $a$. Non-abelian global symmetry factors are never anomalous. The $\mathrm{U}(1)_B$ factor is not anomalous either since $I(R_\mathrm{fund})=I(R_\mathrm{antifund})=1/2$. As far as $\mathrm{U}(1)_A$ and $\mathrm{U}(1)_R^a$ are concerned, one finds:
\begin{equation*}
	\mathcal A_A = F\; , \quad \mathcal A_R = F(a-1)+N\; .
\end{equation*}
This follows from $I(\text{fund})=I(\text{antifund})=1/2$, $I(\text{adj})=N$, and from the fact that a fermion in a chiral supermultiplet of $R$-charge $a$ has $R$-charge $(a-1)$. The contribution $+N$ to $\mathcal A_R$ comes from the gaugino which always has $R$-charge $1$ (because of the form of $\mathcal N=1$ gauge-matter Lagrangian densities) and transforms in the adjoint representation of the gauge group. Therefore $\mathcal U(1)_A$ is always anomalous (only its $\mathbb{Z}_{2F}$ subgroup is a(n ordinary) symmetry of the quantum theory because $\Theta$ is $2\pi$ periodic in SQCD), and the $R$-symmetry is non-anomalous only if $F\neq 0$ and $a=(F-N)/F$. Thus, for $F\neq 0$, the global 0-form symmetries of quantum SQCD are: 

\begin{equation*}
    \begin{array}{c||c|c|c|c|c|}
	& \mathrm{SU}(F)_L & \mathrm{SU}(F)_R & \mathrm{U}(1)_B & (\mathbb{Z}_{2F})_A & \mathrm{U}(1)_R \\
        \hline
	Q_i & \overline{F} & - & 1  & 1 & (F-N)/F \\
	\widetilde{Q}^i & - & F & -1 & 1 & (F-N)/F \\
	\lambda & - & - & - & - & 1 
    \end{array}
\end{equation*}
\end{example}

\vspace{0.25cm}

The appearance of the instanton number in \Cref{Eq:ABJanomalycharge} hints for a link between chiral ABJ anomalies and topology. One can show that the axial anomaly in Yang--Mills theory with gauge group $G$ and fermions is twice the index of the Dirac operator in a given gauge background. More generally, the chiral anomaly embodies \emph{Atiyah--Singer index theorem}\index{Atiyah--Singer index theorem} and detects the instanton number $n$ of a given principal $G$-bundle. In quantum gauge theories every possible gauge field configuration contributes to the path integral, thus an anomalous chiral symmetry is never a symmetry of the quantum theory, even if it is symmetry of the equations of motion.

Chiral continuous global symmetries might be anomalous when put on curved manifolds, in which case one speaks of \emph{gravitational chiral anomaly}\index{anomaly!chiral!gravitational}. The classical conservation equation for the chiral current is modified by an ABJ-like term in which the Riemann curvature $R$ appears in place of the gauge curvature $F$ (and with a different numerical factor). 

\subsubsection*{Gauge anomalies}

As already emphasized, the gauge symmetries of a classical field theory must be non-anomalous in order for the corresponding QFT to be consistent. Requiring gauge symmetries to be non-anomalous constrains the matter content of QFTs. One can distinguish between small and large gauge anomalies depending on whether the anomaly comes from small or large gauge transformations (in the literature such anomalies are sometimes referred to as local and global, even though they both correspond to gauge (i.e. local) symmetries).

\begin{proposition}
    Small gauge anomalies occur in chiral theories in even spacetime dimension, similarly to ABJ anomalies of chiral global symmetries. They are computed in $d=4$ from triangle diagrams of fermions with three gauge current insertions. One finds that a single Weyl fermion in a representation $R$ of the Yang--Mills gauge group $G$ contributes to the anomaly by a factor:
    \begin{equation*}
	   \partial_\mu (j^a)^\mu \propto \tr_R(T^a\{T^b,T^c\})F_b\wedge F_c\; .
    \end{equation*}  
\end{proposition}

When $G=\mathrm{SU}(N)$ with $N>2$, the relative contributions of Weyl fermions in the fundamental,  antifundamental, symmetric or antisymmetric representations of $G$ are respectively $1$, $-1$, $N+4$, $N-4$. Quantum consistency imposes the fermion content to be such that the total anomaly vanishes. For example, $\mathrm{SU}(N)$ Yang--Mills theory with an antisymmetric Weyl fermion only is inconsistent, but the gauge anomaly cancels in $\mathrm{SU}(N)$ Yang--Mills with $N-4$ Weyl fermions in the anti-fundamental representation of $\mathrm{SU}(N)$ and a single antisymmetric one. 

In general, $\tr_R(T^a\{T^b,T^c\})$ vanishes when $R$ is real or pseudo-real; this implies in particular that adjoint Weyl fermions or massive fermions never contribute to the anomaly. Sometimes mild modifications are enough to cure a theory from its small gauge anomalies; the \emph{Green--Schwarz mechanism}\index{Green--Schwarz mechanism} for example cancels the anomaly by adding terms to the Lagrangian.

\begin{example}
    $\mathcal N=1$ $\mathrm{SU}(N)$ super Yang--Mills with $N-4$ antifundamental chiral superfields and one antisymmetric chiral superfield is not anomalous (the gauginos are adjoint and thus do not contribute to the anomaly). Likewise, in SQCD with $N$ colors there are as many chiral superfields in the fundamental representation of $\mathrm{SU}(N)$ as in the antifundamental, hence the gauge symmetry is not anomalous.
\end{example} 

The \emph{Wess--Zumino descent}\index{Wess--Zumino!descent} procedure relates the variation of a $d$-dimensional action under a gauge transformation to a $(d+2)$-dimensional polynomial called the \emph{anomaly polynomial}\index{anomaly!polynomial}, which is the Atiyah--Singer index density of the $d+2$-dimensional Dirac operator. This correspondence between gauge anomalies in $d$-dimensions and the chiral anomaly in $d+2$-dimensions gives rise to the \emph{anomaly inflow}\index{anomaly!inflow} mechanism: if the $d$-dimensional theory is seen as living on a $d$-dimensional defect in the $(d+2)$-dimensional one (called the bulk), the chiral anomaly in the bulk cancels the gauge anomaly on the defect in such a way that the total anomaly vanishes. 

\vspace{0.5cm}

Even if a $G$-gauge theory is free from small gauge anomalies, it might be afflicted by large ones. These arise from a lack of invariance of the correlation functions under gauge transformations not continuously connected to the identity. Such large gauge transformations exist (in flat spacetime) only when $\pi_4(G)\neq 0$. It turns out that $\pi_4(G)=0$ for all compact simple Lie groups $G$ but for $G=\mathrm{USp}(N)$ ($N\geq 2$), for which $\pi_4(\mathrm{USp}(N))=\mathbb{Z}_2$. The path integral measure of a Weyl fermion in a representation $R$ of $G$ is mapped to minus itself under a large gauge transformation when $I(R)$ is half-integral. Thus, a $\mathrm{USp}(N)$-gauge theory with a fundamental Weyl fermion is for instance inconsistent quantum mechanically. This is called \emph{Witten global anomaly}\index{Witten!anomaly}\index{anomaly!Witten} \cite{Witten:1982fp}.

This type of anomaly generalizes to so-called \emph{Dai--Freed anomaly}\index{anomaly!Dai--Freed} after \cite{Dai:1994kq} (cf. \cite{Garcia-Etxebarria:2018ajm} for a recent discussion). When the $d$-dimensional spacetime manifold $X$ is the boundary of a $(d+1)$-dimensional manifold $Y$, the anomalous transformation of the partition function under large gauge transformations can be encoded in a topological quantum field theory on $Y$. This point of view pervades the modern perspective on anomalies (see e.g. \cite{Monnier:2019ytc}).

\subsubsection*{'t Hooft anomalies}

As discussed at the end of \Cref{Subsec:internalsym}, for every ordinary or higher-form global continuous or discrete symmetry $G$ in a QFT, one can consider the theory in the background of a non-trivial $G$-bundle. The classical conservation equation of currents associated to continuous global symmetries can be modified by functions of the background fields. Concretely, if $G^{(p)}$ is a continuous global symmetry with corresponding conserved current $j^{(p+1)}$, a non-trivial background for $G^{(p)}$ is a $(p+1)$-form $A^{(p+1)}$ on which $G^{(p)}$ acts by gauge transformations. If the QFT is defined by an action $S$, the theory in the background $A^{(p+1)}$ corresponds to the modified action:
\begin{equation*}
    S[A^{(p+1)}] = S -\I \int A^{(p+1)}\wedge\star j^{(p+1)}\; , 
\end{equation*}
as in \Cref{Eq:backgroundglobalsym}. 

\begin{definition}
    The symmetry $G^{(p)}$ has an \emph{'t~Hooft anomaly}\index{'t~Hooft!anomaly}\index{anomaly!'t~Hooft} if, in the presence of the background $A^{(p+1)}$, the conservation equation for the current, $\mathrm{d}\star j^{(p+1)}$, depends on $A^{(p+1)}$ in a non-trivial and irreducible way. Two continuous global symmetries have a mixed 't Hooft anomaly when the conservation equation of one of the current is modified by a function of the background field for the other symmetry, in an irreducible way. The notion of 't~Hooft anomaly generalizes to discrete symmetries, as briefly discussed in \Cref{Subsec:internalsym}.
\end{definition} 

In $4d$ QFTs, 't Hooft anomalies of continuous global 0-form symmetries can be computed via triangle diagrams of fermions with three global current insertions.

\begin{remark}
    't Hooft anomalies of global symmetries are not anomalies as defined in \Cref{def:anomaly}: a global symmetry with an 't Hooft anomaly is still a genuine symmetry of the quantum theory. Rather, 't Hooft anomalies encode obstructions to gauging: if a global symmetry $G$ has an 't Hooft anomaly then gauging $G$ makes the quantum theory inconsistent. If two global symmetry factors $G_1$ and $G_2$ have a mixed 't Hooft anomaly but no self 't Hooft anomalies, it is possible to gauge either $G_1$ or $G_2$, but doing so necessarily breaks the remaining symmetry.
\end{remark}

Since 't Hooft anomalies are not genuine anomalies, they do not constraint the matter content of the theory nor the quantum global symmetries of the theory. Their interest is rather dynamical\index{'t~Hooft!anomaly!matching condition}: 

\begin{proposition}['t~Hooft anomaly matching condition]
    't Hooft anomalies are scale invariant quantities.
\end{proposition}
A famous application is that a theory with continuous global 0-form symmetry $G$ with non-zero 't~Hooft anomaly can never be gapped, as otherwise it would be impossible to match the anomaly in the low-energy effective theory. 't Hooft anomalies are intrinsic features of a given QFT and provide important tests of \emph{dualities}\index{duality}.

\subsection{Holomorphy}\labelx{Sec:HolomorphySUSY}

Consider an asymptotically free Lagrangian gauge theory. Such a theory is defined by the data of fields and couplings appearing in the Lagrangian (as well as the Lagrangian itself) at a given cutoff scale $\mu$ at which the QFT is weakly coupled, i.e. with gauge coupling $g\ll 1$. Under a small change of the energy scale (i.e. $\mu\rightarrow\mu'$) the fields and the couplings renormalize in order for the physics at any scale to be unchanged. The variation of the couplings is encoded in Wilsonian renormalization group equations:
\begin{equation*}
   \mu\frac{\mathrm{d}}{\mathrm{d}\mu}c_i = \beta(c)\; .
\end{equation*}

\begin{definition}
    The \emph{Wilsonian effective action}\index{Wilsonian action} at the scale $\mu'<\mu$ is the action obtained by integrating all degrees of freedom of momentum greater than $\mu'$; it is such that its bare couplings correspond to the renormalized couplings at the scale $\mu'$ (see for example \cite[Sec. 3.1]{DHoker:1999yni}).
\end{definition} 
There is a working assumption: it must be possible to describe the theory at scales $\mu$ and $\mu'$ in terms of the same degrees of freedom, which is why $\mu'$ is restricted to be close to $\mu$ (we take this as the definition of what ``close'' means).     

\begin{proposition}
    For $4d$ $\mathcal N=1$ gauge theories, there exists a \emph{holomorphic renormalization scheme} in which the effective gauge and superpotential couplings at the scale $\mu'$ depend only holomorphically on the couplings and fields at $\mu$ \cite{Seiberg:1993vc}.
\end{proposition} 

\begin{proof}
    Any superpotential is holomorphic in the fields, and in particular the effective one at scale $\mu'$ is holomorphic in the fields at scale $\mu$ if $\mu'$ is close to $\mu$. The couplings under scrutiny can be thought of as VEVs of \emph{spurious chiral superfields}\index{superfield!spurious} which also appear holomorphically in the Lagrangian, which proves the proposition. The theory $\mathcal T'$ defined by promoting the couplings of the original theory $\mathcal T$ to superfields still has $\mathcal N=1$ supersymmetry, and generally a larger global symmetry, broken to the global symmetry of the theory $\mathcal T$ by the choice of spurious VEVs.
\end{proof}

This fundamental property of $4d$ $\mathcal N=1$ theories is called \emph{holomorphy} \cite{Seiberg:1993vc,Seiberg:1994bp,Intriligator:1995au,Argyres:2001eva,Bertolini}. Together with symmetry considerations and the imposed form of renormalizations in some limits of the parameter space, holomorphy starkly constrains the Wilsonian effective action. 

\begin{example}
    Consider the holomorphic gauge coupling $\tau$ in $4d$ $\mathcal N=1$ SYM. Holomorphy implies that:
\begin{equation*}
    \frac{\mathrm{d}}{\mathrm{d}\log\mu}\tau = -\frac{b_1}{2\pi\I}+\sum_{n=2}^\infty f_n(\tau^{1-n})+\text{non-perturbative},
\end{equation*}
where for all $n\geq 2$, $f_n(\cdot)$ is a \emph{holomorphic} function which encodes the contribution to the perturbative renormalization of $\tau$ at $n$ loops. Since the Yang--Mills theta term is topological, $\Theta = 2\pi\Re(\tau)$ does not renormalize in perturbation theory, i.e.:
\begin{equation*}
    \frac{\mathrm{d}}{\mathrm{d}\log\mu}\tau = -\frac{b_1}{2\pi\I}+\sum_{n=2}^\infty f_n((\Im\tau)^{1-n})+\text{non-perturbative}.
\end{equation*}
Now, $(\Im\tau)^{1-n}$ is a holomorphic function of $\tau$ only when $n=1$, thus holomorphy implies that all the $f_n$'s are identically zero, and that the renormalization of $\tau$ is exact at one-loop in perturbation:
\begin{equation*}
    \frac{\mathrm{d}}{\mathrm{d}\log\mu}\tau = -\frac{b_1}{2\pi\I}+\text{non-perturbative}.
\end{equation*}
\end{example}

\begin{example}
    The \emph{Wess--Zumino model}\index{Wess--Zumino!model} of a single chiral superfield $\Phi$ is defined by the weakly-coupled Lagrangian:
\begin{equation*}
    \mathcal L = \int\mathrm{d}^2\theta\mathrm{d}^2\overline{\theta}~\Phi^\dagger\Phi + \int\mathrm{d}^2\theta~\left(m\Phi^2+g\Phi^3\right) + \mathrm{c.c.}\; ,
\end{equation*}
at some UV scale $\mu_\mathrm{UV}$. Promoting the mass $m$ and coupling $g$ to chiral superfields $M$ and $G$, the resulting theory has global symmetry $\mathrm{U}(1)\times\mathrm{U}(1)_R$, under which the (spurious) fields have charges:
\begin{equation*}
    \begin{array}{c||c|c|}
	& \mathrm{U}(1) & \mathrm{U}(1)_R \\
        \hline
	\Phi & 1 & 1 \\
	M & -2 & 0 \\
	G & -3 & -1 
    \end{array}
\end{equation*}
The effective superpotential $W_{\mu'}$ at a scale $\mu'$ close to $\mu$ must have $\mathrm{U}(1)\times\mathrm{U}(1)_R$ charge $(0,2)$, hence it must be of the form:
\begin{equation*}
    W_{\mu'} = m\Phi^2f\left(\frac{G\Phi}{M}\right)\; ,
\end{equation*}
where $f$ is a holomorphic function. In the limit where $g,m\rightarrow 0$, the above theory becomes free. Therefore, $W_{\mu'}\sim m\Phi^2+g\Phi^3$. This implies $f(t)=1+t$, and hence $W_{\mu'}=m\Phi^2+g\Phi^3$ at any $\mu'$.
\end{example}

Holomorphy also constrains non-perturbative quantum effects, as shown in the following:
\begin{example}
    The holomorphic gauge coupling $\tau$ of SQCD at a scale $\mu'$ close to the defining scale $\mu_\mathrm{UV}$ satisfies: 
\begin{equation*}
    \tau(\mu') = \frac{3N-F}{2\pi \I}\log\frac{\Lambda}{\mu'}+ f(\Lambda,\mu')\; , \quad \Lambda = \mu_\mathrm{UV}\exp(\frac{2\pi \I\tau(\mu_\mathrm{UV})}{3N-F})\; ,
\end{equation*}
where the first term in the expression of $\tau(\mu')$ is the perturbative contribution at one loop and $f(\cdot,\mu')$ is a holomorphic function. The theta angle $\Theta$ of SQCD is $2\pi$-periodic, i.e. $\tau\rightarrow\tau+1$, or equivalently:
\begin{equation}\labelx{Eq:shiftThetaSQCD}
    \Lambda\rightarrow \exp(\frac{2\pi \I}{3N-F})\Lambda\; ,
\end{equation}
is a symmetry of the theory. This forces $f$ to be invariant under the shift of \Cref{Eq:shiftThetaSQCD}. Moreover $f$ must have zero constant term so that the classical limit ($\Lambda\rightarrow 0$) gives back the perturbative one-loop result. The only solutions to these requirements are of the form:
\begin{equation}\labelx{Eq:renormtauSQCD}
    \tau(\Lambda,\mu') = \frac{3N-F}{2\pi \I}\log\frac{\Lambda}{\mu'} + \sum_{n=1}^{\infty} a_n\left(\frac{\Lambda}{\mu'}\right)^{(3N-F)n}\; ,
\end{equation} 
where the $a_n$'s are complex coefficients. The contributions in the sum are non-perturbative and correspond to gauge instantons, cf. \Cref{Eq:actioninstanton}. Thus, the renormalization of $\tau$ is perturbatively exact at one-loop, and moreover, \Cref{Eq:renormtauSQCD} is the exact renormalization of the holomorphic coupling, at least when SQCD is weakly coupled.
\end{example}

The existence of anomalous global symmetries in the weakly-coupled descriptions of SYM and SQCD (the axial symmetry $\mathrm{U}(1)_A$ in SQCD and the $R$-symmetry $\mathrm{U}(1)_R$ in SYM) further constrains the renormalization of $\tau$. The form of the ABJ anomaly shows that a transformation of parameter $\alpha$ under the anomalous symmetry can be compensated by an appropriate shift of the theta angle: $\Theta\rightarrow \Theta+2N\alpha$ in the case of SYM, and $\Theta\rightarrow \Theta+2F\alpha$ in the case of SQCD with $F$ flavors. Promoting $\Lambda$ to a spurious chiral superfield, the anomalous symmetries of SYM and SQCD become genuine symmetries of the enlarged theories, under which $\Lambda$ as well as $\tau(\Lambda,\mu')$ in \Cref{Eq:renormtauSQCD} are charged. The only way for $\tau(\Lambda,\mu')$ to transform correctly is if $a_n=0$ for all $n\geq 1$. This means that higher-order corrections beyond the one-loop contribution, such as instanton corrections, do not contribute. As a result, the one-loop running of the holomorphic coupling constant is exact at the quantum level, meaning no further corrections arise from instanton effects or higher-loop terms.

\subsubsection*{Holomorphic and physical couplings}

The holomorphic gauge coupling $\tau$ is associated with the normalization of the Yang--Mills Lagrangian of \Cref{Eq:nonabeliankineticSUSY}. This is not the standard (physical) normalization used in Yang–Mills theory, where the term $\tr F_{\mu\nu}F^{\mu\nu}$ is instead multiplied by $(2g_p^2)^{-1}$, with the subscript $_p$ indicating the ``physical'' coupling. The two normalizations are connected through a rescaling of the non-Abelian vector superfield, specifically $V\rightarrow g_pV$. 

The measure in the path integral is not invariant under this rescaling and the relationship between $g$ and $g_p$ is not analytic. This is to be expected, as the renormalization of $g_p$ in SYM and SQCD is known to receive perturbative contributions at all loops. For example, in SQCD with $N$ colors and $F$ flavors, the renormalization of the physical gauge coupling $g_p$ is given by the \emph{Novikov--Shifman--Vainshtein--Morozov (NSVZ) formula}\index{NSVZ formula}:
\begin{equation}\labelx{Eq:NSVZ}
    \frac{\mathrm{d}}{\mathrm{d}\log\mu}g_p = -\frac{g^3_p}{(4\pi)^2}\frac{3N-\displaystyle\sum(1-\gamma_\chi)}{1-Ng_p^2/8\pi^2}\; ,
\end{equation}
where the sum runs over the set of chiral superfields $\chi$ and where $\gamma_\chi$ is the \emph{anomalous dimension}\index{anomalous dimension} of $\chi$. Taking $F=0$ yields the renormalization of the physical gauge coupling $g_p$ in $\mathrm{SU}(N)$ SYM.

The non-analytic transformation law between $g$ and $g_p$ relates the exact running of $g$ to the one of $g_p$ given by the NSVZ formula. The appearance of the anomalous dimensions $\gamma_\chi$ of the chiral superfields in the NSVZ formula follow from the fact that the running of $g$ in SQCD is one-loop exact only with non-canonical kinetic terms for the chiral superfields in the effective Wilsonian action at scale $\mu'<\mu$ \cite{Arkani-Hamed:1997qui}.

\subsubsection*{An important caveat}\labelx{par:importantcaveat}

The statement that the one-loop running of the gauge coupling in SQCD is exact depends on the assumption that the energy scales $\mu'$ and $\mu$ are close enough for the renormalization to be expressed solely in terms of the fields at $\mu$. SQCD is asymptotically free when $3F < N$, meaning the gauge coupling increases as the energy scale decreases. Consequently, SQCD eventually becomes strongly coupled at low energies, near the dynamical scale $\Lambda$. At strong coupling, it no longer makes sense to describe the theory using the weakly coupled UV degrees of freedom. Certain operators, irrelevant at high energies, may become relevant along the renormalization group (RG) flow, undermining the validity of the weakly coupled analysis.

If one integrates the differential equation for the renormalized gauge coupling $g_p$ in supersymmetric Yang-Mills (SYM) in terms of the energy scale, a threshold appears, below which the equation has no solution—the RG flow of the gauge coupling effectively halts. This does not, of course, imply that SYM ceases to exist at lower energy scales; rather, it signals that the equation is only reliable when the theory is weakly coupled. In other words, holomorphy alone cannot resolve the strong-coupling dynamics of supersymmetric gauge theories. For a more detailed discussion, see \cite[Sect. 2.4.4]{Argyres:2001eva}.

\subsection{Four-dimensional \texorpdfstring{$\mathcal N=1$}{N=1} SQCD (II)}\labelx{Sec:quantSQCD}

Whereas it is difficult to say anything about non-perturbative quantum effects in general gauge theories, supersymmetry --- and in particular, holomorphy --- is of great help in that regard. We will illustrate this with a very brief overview of the quantum dynamics of SQCD with $N$ colors and $F$ flavors\index{$\mathcal N=1$ SQCD} (see e.g. \cite{Seiberg:1993vc,Seiberg:1994bz,Seiberg:1994bp,Intriligator:1995au,Peskin:1997qi,Shifman:1995ua,Argyres:2001eva,Bertolini} for more details).

\subsubsection*{Pure $\mathrm{SU}(N)$ SYM ($F=0$)} 

At low energies, pure $\mathrm{SU}(N)$ SYM is believed to confine with a mass gap, just as pure $\mathrm{SU}(N)$ Yang--Mills theory (the gaugini are adjoint fermions and thus cannot break flux lines as the quarks in QCD do). There is a lot of evidence in favor of this, of theoretical, experimental and computational (on the lattice) natures. 

Assuming that pure $\mathrm{SU}(N)$ and $\mathrm{PSU}(N)$ SYM share the same Lagrangian density. Nevertheless, one can show that the local physics in the $N$  indeed confines with a mass gap, the low-energy physics should only depend on $\tau$ (or rather, on a spurious chiral superfield whose VEV is $\tau$) through an effective superpotential $W_\mathrm{eff}$. The extended theory obtained from SYM by promoting $\tau$ to a spurious chiral superfield $T$ has a non-anomalous $R$-symmetry under which $\exp(2\pi iT/N)$ has charge 2. This (and dimensional analysis) constrains $W_\mathrm{eff}$ to be of the form:
\begin{equation*}
    W_\mathrm{eff} = \widetilde{c}_N\mu^3\E^{2\I \pi\tau/N} = \widetilde{c}_N\Lambda^3\; ,
\end{equation*}
with $\widetilde{c}_N$ a constant. The auxiliary field in $T$ sources the gaugino bilinear $\tr \lambda\lambda$, thus:
\begin{equation*}
    \langle \tr\lambda\lambda\rangle = -32\pi^2 \E^{\frac{2\I k\pi}{N}}\Lambda^3\; ,
\end{equation*}
i.e. the gaugini condensate (if $\widetilde{c}_N\neq 0$) in the vacuum. Assigning the VEV $\tau$ to $T$ breaks the $R$-symmetry to $\mathbb{Z}_{2N}$, and the gaugino condensate further breaks $\mathbb{Z}_{2N}$ spontaneously to $\mathbb{Z}_2$. Hence:
\begin{proposition}
    $\mathrm{SU}(N)$ SYM has $N$ isolated gapped vacua labeled $k=0,\dots,N-1$; in the $k$-th vacuum the phase of the condensate is $2k\pi/N$.
\end{proposition} 

\begin{remark}
    In fact, the Lagrangian of pure SYM only depends on the gauge algebra and does not determine the global form of the gauge group. For example, pure $\mathrm{SU}(N)$ and $\mathrm{PSU}(N)$ SYM share the same Lagrangian density, and both theories, if they indeed confine, display $N$ gapped vacua. However, the effective low-energy physics in the vacua depends on the global form of the gauge group \cite{Aharony:2013hda}. 
\end{remark}

\subsubsection*{SQCD with $0<F<N$ flavors} 

We have shown in \Cref{Sec:classicalSQCD} that SQCD with $0<F<N$ flavors has a classical moduli space of complex dimension $F^2$ parameterized by the entries of the $F\times F$ meson matrix $M$. Holomorphy determines the possible non-perturbative contributions to the superpotential (at weak coupling). Promoting $\tau$ to a spurious chiral superfield, one obtains an extended theory in which $\mathrm{U}(1)_A$ is a genuine global symmetry. Effective superpotentials $W_\mathrm{eff}$ must be uncharged under the gauge and global symmetry group $\mathrm{SU}(N)\times\mathrm{SU}(F)_L\times\mathrm{SU}(F)_R\times\mathrm{U}(1)_A\times\mathrm{U}(1)_B$, and of charge $2$ under $\mathrm{U}(1)_R$. This, together with dimensional analysis and the study of the weak coupling limit, constraints $W_\mathrm{eff}$ to be of the form:
\begin{equation*}
    W_\mathrm{eff} = c_{N,F}\left(\frac{\Lambda^{3N-F}}{\det M}\right)^{\frac{1}{N-F}}\; ,
\end{equation*}
with $c_{N,F}=N-F$ some constant. This non-perturbative contribution is called \emph{Affleck--Dine--Seiberg (ADS) superpotential}\index{superpotential!Affleck--Dine--Seiberg} \cite{Affleck:1983rr,Affleck:1983mk}.

Consider a point in the (classical) moduli space at which every meson VEVs has a value $v\sim\det M^{1/2F}$, with the gauge group maximally broken to $\mathrm{SU}(N-F)$, and far away from the origin of the moduli space (i.e. $v$ large) so that the breaking occurs in the weakly-coupled regime of the UV SQCD theory. At energies above $v$ the holomorphic gauge coupling $\tau$ of the $\mathrm{SU}(N)$ gauge group satisfies:
\begin{equation*}
    \tau(\mu) = \frac{3N-F}{2\pi \I}\log\frac{\Lambda}{\mu}\; ,
\end{equation*}
where $\Lambda$ is the dynamical scale of the UV theory. In contrast, below $v$ the holomorphic coupling $\tau'$ of the $\mathrm{SU}(N-F)$ gauge group runs as:
\begin{equation*}
    \tau'(\mu) = \frac{3(N-F)}{2\pi \I}\log\frac{\Lambda'}{\mu}\; \; ,
\end{equation*}
where $\Lambda'$ is the dynamical scale of the effective $\mathrm{SU}(N-F)$ SYM theory. At $\mu=v$ one must have $\tau(v)=\tau'(v)$ (this is \emph{scale matching}), from which one derives:
\begin{equation*}
    (\Lambda')^3 = \left(\frac{\Lambda^{3N-F}}{\det M}\right)^{\frac{1}{N-F}}\; .
\end{equation*}
Assuming that SYM condense with a mass gap, this implies $c_{N,F}=\widetilde{c}_{N-F}$.

If one of the chiral superfields of the UV theory is given a large mass $m$, at scales $\mu>m$ the theory is the UV SQCD theory with $N$ colors and $F$ flavors, whereas when $\mu<m$ the massive superfield decouples and one is effectively left with SQCD with $N$ colors and $F-1$ flavors. Another scale matching argument implies that:
\begin{equation*}
    (c_{N,F-1})^{N-F+1} = (N-F+1)\left(\frac{c_{N,F}}{N-F}\right)^{N-F}\; ,
\end{equation*}
leading to $c_{N,F}=(N-F)c^{1/(N-F)}$ with $c_{N,0}=\widetilde{c}_N$. When $F=N-1$, the ADS superpotential has the form of an instanton contribution; at weak coupling one can compute such instanton contributions explicitly and one finds $c=1$, i.e. $c_{N,F}=N-F$. Thus:
\begin{proposition}
    An ADS superpotential is indeed generated:
 \begin{equation*}
    W_\mathrm{eff} = (N-F)\left(\frac{\Lambda^{3N-F}}{\det M}\right)^{\frac{1}{N-F}}\; .
 \end{equation*}
 Moreover, one has $\widetilde{c}_N=N$, hence the gaugini in pure $\mathrm{SU}(N)$ SYM indeed condense, as anticipated above. However, it is important to note that this is by no means a proof that pure SYM confines.
\end{proposition}

\begin{remark}
    The existence of the ADS non-perturbative contribution to the superpotential has dramatic consequences for the moduli space: the scalar potential (including the contribution of $W_\mathrm{ADS}$) is a strictly positive function of the gauge-invariant scalar VEVs. Hence (while there is a classical moduli space of complex dimension $F^2$) the quantum moduli space is empty: there exists no supersymmetric vacuum at finite distance in field space. The potential slopes to zero at infinity; this behaviour is dubbed \emph{runaway}\index{runaway}. 
\end{remark}

\subsubsection*{SQCD with $F=N$ flavors} 

In that case as well as when $F>N$, there exists no effective superpotential compatible with all spurious symmetries and other physical constraints. When $F=N$, the classical moduli space is parameterized by the VEVs of the meson matrix $M$ and two baryons $B$, $\widetilde B$ satisfying:
\begin{equation*}
    \det M = B\widetilde B\; .
\end{equation*}

The quantum moduli space is a deformation of the classical one; it is still parameterized by (the VEV of the scalar components of) $B$, $\widetilde B$ and the matrix elements of $M$, however now satisfying:
\begin{equation}\labelx{Eq:defSQCDf=N}
    \det M - B\widetilde{B} = \Lambda^{2N}\; .
\end{equation}
This is the only way one can retrieve the ADS superpotential if one of the chiral superfields is taken to be massive. \Cref{Eq:defSQCDf=N} is a highly non-trivial statement about the strongly-coupled regime of SQCD with $F=N$ flavors.

The quantum moduli space of SQCD with $F=N$ flavors is not empty; however $\mathcal M$ (contrarily to the classical moduli space $\mathcal M_\mathrm{cl}$) has no singularity at the origin. At generic points the global symmetry $\mathrm{SU}(F)_L\times\mathrm{SU}(F)_R\times\mathrm{U}(1)_B\times\mathrm{U}(1)_R$ is fully broken, however there exist submanifolds of enhanced global symmetry. On the \emph{mesonic branch}\index{branch of moduli space!mesonic}\index{moduli space!mesonic branch} defined as $M^i_j=\Lambda^2\delta^i_j$ and $B=\widetilde B=0$, the global symmetry is $\mathrm{SU}(F)_D\times\mathrm{U}(1)_B\times\mathrm{U}(1)_R$, whereas on the \emph{baryonic branch}\index{branch of moduli space!baryonic}\index{moduli space!baryonic branch} defined as $M=0$ and $B=-\widetilde{B}=\Lambda^N$, the global symmetry is $\mathrm{SU}(F)_L\times\mathrm{SU}(F)_R\times\mathrm{U}(1)_R$. Using mesons and baryons as low-energy degrees of freedom everywhere along $\mathcal M$ (even close to the origin, where the gauge theory strongly coupled) is a reasonable assumption, in particular, because of the matching of 't Hooft anomalies of global symmetries in the UV, and on the mesonic and baryonic branches with mesons and baryons as degrees of freedom.

\subsubsection*{SQCD with $F\geq N+1$ flavors} 

For $F\geq N+1$, it turns out that $\mathcal M=\mathcal M_\mathrm{cl}$ as Kähler manifolds: there are no non-perturbative contributions to the superpotential and no deformations. This is imposed by compatibility with the results of the previous paragraphs.  However, the interpretation of the singularities differs in the classical and quantum moduli spaces. 

Consider for example the singularity at the origin of $\mathcal M$ when $F=N+1$. Classically it corresponds to gluons becoming massless, however in this region of the moduli space the gauge theory is strongly coupled and gluons are not appropriate as low-energy degrees of freedom. In the quantum theory one can argue that $\mathcal M$ is parameterized by mesons and baryons; the singularity at the origin gives away the appearance of new massless degrees of freedom: there, all relations between mesons and baryons are trivially satisfied.

When $F\geq N+2$, the 't Hooft anomalies of continuous global symmetries cannot match between the ultraviolet (UV) and infrared (IR) regimes if mesons and baryons are assumed to be the low-energy degrees of freedom. For $F>3N$, supersymmetric QCD (SQCD) becomes IR-free, implying that the low-energy theory at the origin of the moduli space $\mathcal M$ is a free non-abelian gauge theory. This observation suggests that the low-energy effective theory near the origin of $\mathcal M$, for all values of $F\geq N+2$, could be in such a non-abelian Coulomb phase \cite{Seiberg:1994bz}. 

\subsection{Seiberg duality} 

At the origin of the classically-exact moduli space of SQCD with $F\geq N+1$ flavors, the continuous global symmetry $\mathrm{SU}(F)_L\times\mathrm{SU}(F)_R\times\mathrm{U}(1)_B\times\mathrm{U}(1)_R$ is unbroken. This suggests that there are $F$ massless flavors in low-energy effective theory. In general, low-energy limits of QFTs are fixed points of the renormalization group flow (scale-invariant theories) which can be either trivial (i.e. free) or non-trivial. Most of the time supersymmetric scale-invariant $4d$ $\mathcal N=1$ QFTs are $4d$ $\mathcal N=1$ superconformal field theories (SCFTs).

When $N$ is large and $\epsilon = 3-F/N\ll 1$, the NSVZ beta function for the physical gauge coupling $g_p$ of SQCD with $N$ colors and $F$ flavors reads: 
\begin{equation*}
    \beta(g_p) = -\frac{g_p^3}{(4\pi)^2}\frac{3N-F(1-\gamma(g_p^2))}{1-Ng_p^2/8\pi^2}\; ,
\end{equation*}
where
\begin{equation*}
    \gamma(g_p^2)=-\frac{g_p^2}{8\pi^2}\frac{N^2-1}{N}+\mathcal{O}(g_p^4)
\end{equation*}
is the anomalous dimension of the chiral superfields. The value
\begin{equation*}
    g_p^* = \frac{8\pi^2}{3}\frac{N}{N^2-1}\epsilon
\end{equation*}
corresponds to a (non-trivial) fixed point of the renormalization group flow i.e. $\beta(g_p^*)=0$. Thus one expects that when $N$ is large and $\epsilon$ small, SQCD flows to a non-trivial SCFT at low-energies \cite{Seiberg:1994pq}.

The superconformal $R$-symmetry can only be the $R$-symmetry of SQCD. As in any $4d$ SCFT, the scaling dimension of chiral operators equals $3/2$ their $R$-charge; from this one deduces the following scaling dimensions:
\begin{equation*}
    \Delta(M)=\frac{3}{2}R(M)=3\frac{F-N}{F}\; , \quad \Delta(B)=\Delta(\widetilde{B})=\frac{3N(F-N)}{2F}\; .
\end{equation*}
The unitary bound for scalar fields when $d=4$ is $\Delta\geq 1$; it is satisfied for the mesons when $3N/2\leq F< 3N$ and saturates when $F=3N/2$. Hence SQCD might flow to such an interacting SCFT not only for small values of $\epsilon$, but in the whole range $3N/2< F< 3N$. Assuming this, one obtains a consistent and compelling picture. The range $3N/2< F< 3N$ is called \emph{conformal window}\index{conformal!window}.

In SQCD with $N+2\leq F<3N/2$ flavors, one expects the low-energy effective field theory at the origin of the moduli space to be described by $F$ flavors of chiral superfields charged under some gauge group. A consistent choice is an $\mathrm{SU}(F-N)$ theory with $F$ fundamental (resp. antifundamental) chiral superfields $q$ (resp. $\widetilde q$), a neutral chiral superfield $M$ and tree-level superpotential $W=\widetilde{q}Mq$ \cite{Seiberg:1994pq}. The global charges of $q$, $\widetilde q$ and $M$ are: 
\begin{equation*}
    \begin{array}{c||c|c|c|c|c|}
	& \mathrm{SU}(F)_L & \mathrm{SU}(F)_R & \mathrm{U}(1)_B & \mathrm{U}(1)_R \\
        \hline
	q^i & F & - & N/(F-N) & N/F \\
	\widetilde{q}_i & - & \overline{F} & -N/(F-N) & N/F \\
	M_i^j & \overline{F} & F & - & 2(F-N)/F 
    \end{array}
\end{equation*}
This auxiliary theory is denoted mSQCD. Global symmetries as well as 't Hooft anomalies for continuous global symmetries in SQCD and mSQCD match. The relation between SQCD and mSQCD is called \emph{Seiberg duality}\index{Seiberg duality}. 

\begin{proposition}[Seiberg duality]
    For $F\geq N+2$ both SQCD (with $N$ colors and $F$ flavors) and mSQCD (with $F-N$ colors and $F$ flavors) flow to the same renormalization fixed point at low energies. Note that $(N,F)\rightarrow (N',F')=(F-N,F)$ exchanges the range $N+2\leq F\leq 3N/2$ with $F'\geq 3N'$ and vice-versa, and maps the conformal window of SQCD $3N/2< F< 3N$ to the conformal window $3N'/2<F'<3N'$ of mSQCD.
\end{proposition}

Thus:
\begin{itemize}
    \item When $N+2\leq F\leq 3N/2$, the low-energy effective physics of SQCD close to the origin of the moduli space is better described as the low-energy physics of the corresponding mSQCD theory, which is IR free: SQCD is in a \emph{free magnetic phase}\index{free phase of SQCD!magnetic},
    \item When $3N/2< F< 3N$, close to the origin of the moduli space, SQCD flows to an interacting conformal theory which can be equivalently described as the low-energy fixed points of the dual theory mSQCD, 
    \item When $F\geq 3N$, SQCD is IR free. This is the \emph{free electric phase}\index{free phase of SQCD!electric}.
\end{itemize}

\begin{remark}
    Seiberg duality is an electric-magnetic duality in the sense that it maps the weakly coupled regime of one theory to the strongly coupled regime of the other: the dynamical scales $\Lambda$ and $\Lambda_m$ of SQCD and mSQCD satisfy $\Lambda^{3N-F}\Lambda_m^{2F-3N} = c$, where $c$ is a constant. 
\end{remark}

Although we will not delve into this in detail, it is worth noting that Seiberg duality extends to many other types of $4d$ $\mathcal N=1$ gauge theories. This includes supersymmetric SQCD with orthogonal gauge groups \cite{Seiberg:1994pq}, as well as a broader range of quiver gauge theories \cite{Berenstein:2002fi}. In the latter case, Seiberg duality acts as a mutation of the quiver, and transformation rules for the ranks of the gauge group at the quiver nodes follow the tropical cluster $\mathcal A$-mutation rules (cf. \Cref{sec:Clusters}). 
\chapter{Four-dimensional \texorpdfstring{$\mathcal N=2$}{N=2} gauge theories}\labelx{Chap:N=2}

\abstract{We investigate four-dimensional quantum field theories with $\mathcal N=2$ supersymmetry. Our analysis begins with the kinematic structure of these theories, including supermultiplets, Lagrangians, and classical moduli spaces, and subsequently addresses quantum $\mathcal N=2$ gauge dynamics.}

\vspace{0.5cm}

Four-dimensional $\mathcal N=2$ quantum field theories (QFTs) possess the $\mathcal N=2$ super-Poincaré algebra of \Cref{Eq:SUSY=squareroot}, along with internal symmetries, as their infinitesimal symmetry algebra. The $\mathcal N=2$ super-Poincaré algebra naturally includes $\mathcal N=1$ super-Poincaré algebras as subalgebras, making $4d$ $\mathcal N=2$ QFTs a more specific subset of $4d$ $\mathcal N=1$ QFTs. The main difference from the general $\mathcal N=1$ case is the presence of eight supercharges, $Q_\alpha^I$ and $\overline Q_{\dot\alpha}^I$, with $I=1,2$. Furthermore, $\mathcal N=2$ supersymmetry in four dimensions allows for a complex central charge $\mathcal{Z}$, appearing in the following anticommutation relations: 
\begin{equation*}
    \{Q_\alpha^I,Q_\beta^J\} = 2\sqrt{2}\epsilon_{\alpha\beta}\epsilon^{IJ}\mathcal{Z} \quad \text{and} \quad \{\overline{Q}_{\dot\alpha}^I, \overline{Q}_{\dot\beta}^J\} = 2\sqrt{2}\epsilon_{\alpha\beta}\epsilon^{IJ}\mathcal{Z}^*\; .
\end{equation*} 
We also include in the definition of the $\mathcal N=2$ theories that they possess a global $\mathrm{SU}(2)_R$ $R$-symmetry rotating the supercharges as:
\begin{equation*}
	\text{for all } \alpha=1,2, \quad\begin{pmatrix} a & b \\ c & d \end{pmatrix} \begin{pmatrix} Q_\alpha^1 \\ Q_\alpha^2 \end{pmatrix} = \begin{pmatrix} a Q_\alpha^1+b Q_\alpha^2  \\ c Q_\alpha^1+d Q_\alpha^2 \end{pmatrix}\; ,
\end{equation*}
with a similar action on $\overline Q_{\dot\alpha}^I$ in the conjugate representation. In addition to $\mathrm{SU}(2)_R$, the $R$-symmetry of a $4d$ $\mathcal N=2$ theory may also include a $\mathrm{U}(1)_R$ factor under which the $Q_\alpha^I$'s have charge $-1$ and the $\overline{Q}_{\dot\alpha}^I$'s charge $1$. However, this $\mathrm{U}(1)_R$ symmetry is not strictly part of the $\mathcal N=2$ super-Poincaré algebra and can be explicitly broken or anomalous: only a (possibly trivial) subgroup of $\mathrm{U}(1)_R$ may be realized.

\section{Classical field theories}

\subsection{Supermultiplets}

In $4d$ $\mathcal N=2$ supersymmetric theories, the gauge and matter content consists of $\mathcal N=2$ supermultiplets that only contain Lorentz particles of spin (or helicity) less than one. These supermultiplets are constructed by acting with creation operators derived from the supercharges, on a Clifford vacuum (cf. \Cref{Sec:superPoincarealgebras}). Moreover, they can be decomposed into sums of $\mathcal N=1$ supermultiplets when viewed relative to the $\mathcal N=1$ subalgebra with supercharges $Q_\alpha^1$ and $\overline Q_{\dot\alpha}^1$. Note that only the abelian subgroup 
\begin{equation*}
    \mathrm{U}(1)_J=\left\{\begin{pmatrix} \E^{\I\vartheta} & 0 \\ 0 & \E^{-\I\vartheta}\end{pmatrix},~\vartheta\in[0,2\pi)\right\}<\mathrm{SU}(2)_R
\end{equation*} 
preserves this specific $\mathcal N=1$ subalgebra.

\subsubsection*{$\mathcal N=2$ vector multiplets}

\emph{Massless $\mathcal N=2$ vector multiplets}\index{$\mathcal N=2$ vector multiplet} split as a massless $\mathcal N=1$ vector multiplet $V=(A_\mu,\lambda_\alpha)$ and a massless $\mathcal N=1$ chiral multiplet $\Phi=(\phi,\widetilde{\lambda}_\alpha)$, with respect to any $\mathcal N=1$ subalgebra:
\begin{equation*}
    (\mathbf{0},2(\mathbf{1/2}),\mathbf{1}) \oplus_{\mathcal{CPT}} (\mathbf{0},2(\mathbf{-1/2}),\mathbf{-1})\; .
\end{equation*}
If the $\mathcal N=1$ vector supermultiplet is abelian, the corresponding $\mathcal N=2$ supermultiplet is also called abelian.

The $\mathrm{SU}(2)_R$ symmetry acts on $A_\mu$ and $\phi$ trivially, and on $(\lambda_\alpha,\widetilde{\lambda}_\alpha)$ as a doublet. It is convenient to represent the Poincaré content of $\mathcal N=2$ supermultiplets as diamonds; in the case of vector supermultiplets the diamond is:
\begin{equation*}
    \begin{array}{ccc}
    & A_\mu & \\
    \lambda_\alpha & & \widetilde{\lambda}_\alpha \\
    & \phi & 
    \end{array}\; ,
\end{equation*}
where the SW-NE (resp. SE-NW) direction corresponds to the action of $Q_{\alpha}^1$ and $\overline{Q}_{\dot \alpha}^1$ (resp. $Q_{\alpha}^2$ and $\overline{Q}_{\dot \alpha}^2$), and where $\mathrm{SU}(2)_R$ acts on the rows.

Just as in the $\mathcal N=1$ case, given a simple compact Lie algebra $\mathfrak g$ with generators $T^a$ one can construct non-abelian $\mathcal N=2$ $\mathfrak g$-vector supermultiplets:
\begin{equation*}
    V = 2g(V_a\oplus \Phi_a)T^a\; ,
\end{equation*}
where $g$ is the gauge coupling. In other words, a massless $\mathcal N=2$ $\mathfrak g$-vector supermultiplet consists of a massless $\mathcal N=1$ $\mathfrak g$-vector supermultiplet together with a massless $\mathcal N=1$ chiral multiplet in the adjoint representation of $\mathfrak g$.

Massive $\mathcal N=2$ vector supermultiplets split as a massive $\mathcal N=1$ vector and a massive $\mathcal N=1$ chiral multiplet, with Poincaré content:
\begin{equation*}
    (\mathbf{-1},4 (\mathbf{-1/2}), 6(\mathbf{0}), 4(\mathbf{1/2}), \mathbf{1})\; .
\end{equation*}
Consistently with the $\mathcal N=2$ super-Higgs mechanism, this corresponds to the sum of a massless $\mathcal N=2$ vector supermultiplet and a massless $\mathcal N=2$ hypermultiplet, to be described shortly. Massive $\mathcal N=2$ vector supermultiplets can be (half-)BPS, in which case their Poincaré content is $(\mathbf{-1},2 (\mathbf{-1/2}), 2(\mathbf{0}), 2(\mathbf{1/2}), \mathbf{1})$.

\subsubsection*{Hypermultiplets}

In this paragraph and below, we will often denote $\Phi^\dagger$ the hermitian adjoint of a (super)field $\Phi$, instead of $\overline{\Phi}$, to lighten notation. A \emph{massless $\mathcal N=2$ hypermultiplet}\index{$\mathcal N=2$ hypermultiplet} splits as the sum of a massless $\mathcal N=1$ chiral multiplet\footnote{We follow the convention that chiral multiplets in hypermultiplets are denoted $Q$ (for Quark). These denote $\mathcal N=1$ chiral superfields, not supercharges --- a distinction that will always be clear from context.} $Q=(q,\psi_\alpha)$ and a massless $\mathcal N=1$ antichiral multiplet $\widetilde{Q}^\dagger =(\widetilde{q}^\dagger,\widetilde{\psi}_\alpha^\dagger)$. In order to belong to the same $\mathcal N=2$ hypermultiplet, the chiral multiplets $Q$ and $\widetilde{Q}$ must have opposite charges. Thus, a massless hypermultiplet has Poincaré content:
\begin{equation}\labelx{eq:mattercontenthyper}
    (\mathbf{-1/2},2(\mathbf{0}), \mathbf{1/2})\oplus_{\mathcal{CPT}}(\mathbf{1/2},2(\mathbf{0}),\mathbf{-1/2})\; ,
\end{equation}
which can be recast as the diamond: 
\begin{equation*}
    \begin{array}{ccc}
    & \psi_\alpha & \\
    q~ & & \widetilde{q}^\dagger \\
    & \widetilde{\psi}_{\dot\alpha}^\dagger & 
    \end{array}\; .
\end{equation*}
The need to add the $\mathcal{CPT}$ conjugate follows from the constraint that $\mathrm{SU}(2)_R$ must act on the scalar components of the hypermultiplets as a doublet, which can never be the case is for the two complex scalars in one of the two copies appearing in \Cref{eq:mattercontenthyper}. There is, however, a way to halve the number of degrees of freedom for hypermultiplets in quaternionic representations\footnote{A complex representation $R$ of a group $G$ is \emph{quaternionic}\index{representation!quaternionic} if it admits an antilinear $G$-equivariant map $j:R\rightarrow R$ such that $j^2=-\mathrm{Id}_R$, or equivalently, an anti-symmetric invariant tensor, and \emph{real}\index{representation!real} if it admits an antilinear $G$-equivariant map $j:R\rightarrow R$ such that $j^2=\mathrm{Id}_R$, or equivalently, a symmetric invariant tensor. Irreducible representations cannot simultaneously be real and quaternionic.} of the gauge group $G$.

In general, the flavor symmetry for $n$ massless hypermultiplets $Q_i\oplus \widetilde{Q}^{i\dagger}$ transforming in an irreducible complex representation $R$ of the gauge group $G$, is $\mathrm{SU}(n)$. However, if $R$ is isomorphic to its conjugate representation $\overline R$, the flavor symmetry is enhanced. Precisely, the flavor symmetry is $\mathrm{SO}(2n)$ when $R$ is quaternionic, and $\mathrm{USp}(2n)$ when $R$ is real. When $Q\oplus\widetilde{Q}^\dagger$ is in a quaternionic representation $R$ with antisymmetric invariant tensor $\epsilon_{ab}$, one can impose:
\begin{equation}\labelx{eq:defhalfhyper}
    \widetilde{Q}_a=\epsilon_{ab}Q^b\; .
\end{equation}

\begin{definition}
    A \emph{half-hypermultiplet}\index{half-hypermultiplet} in the quaternionic representation $R$ of $G$ is a hypermultiplet $Q\oplus\widetilde{Q}^\dagger$ satisfying the constraint of \Cref{eq:defhalfhyper}.
\end{definition}

Massive $\mathcal N=2$ hypermultiplets must be (half-)BPS in order to contain particles of spin less than one only, i.e. they must satisfy $m\geq\sqrt{2}\vert Z\vert$. The Poincaré content of half-BPS hypermuliplets is:
\begin{equation*}
    (2(\mathbf{-1/2}),4(\mathbf{0}), 2\mathbf{(1/2)})\; ,
\end{equation*}
where the doubling is again imposed by $\mathrm{SU}(2)_R$-invariance. 

\subsection{Lagrangians}\labelx{Subsec:N=2Lagrangians}

We are primarily interested in studying the low-energy behavior of $\mathcal N=2$ gauge theories. Theories that admit a Lagrangian description will be easier to analyze, though not all $\mathcal N=2$ theories allow for this. Among the Lagrangian $\mathcal N=2$ gauge theories, renormalizable theories are suitable UV theories, while non-renormalizable ones typically arise as low-energy effective descriptions.

We will write all $\mathcal N=2$ Lagrangians in $\mathcal N=1$ superspace formalism, and hence construct them from $\mathcal N=1$ chiral and vector superfields as in \Cref{Sec:N=1Lagrangians}, with respect to the $\mathcal N=1$ subalgebra generated by the supercharges $Q_\alpha^1$ and $\overline{Q}_{\dot\alpha}^1$. To ensure that the theory retains $\mathcal N=2$ supersymmetry, the additional condition of $\mathrm{SU}(2)_R$-invariance must be imposed. This requirement of $\mathrm{SU}(2)_R$-symmetry turns out to be both necessary and sufficient for an $\mathcal N=1$ Lagrangian to also possess $\mathcal N=2$ supersymmetry.

\subsubsection*{Renormalizable $\mathcal N=2$ Lagrangians}

\begin{proposition}
    The Lagrangian for pure $\mathcal N=2$ SYM with simple compact gauge algebra $\mathfrak g$ involves of a $\mathfrak{g}$-non-abelian $\mathcal N=1$ vector superfield $V$ with supersymmetric field strength $W_\alpha$, and an $\mathcal N=1$ massless chiral superfield $\Phi$ in the adjoint representation of $\mathfrak{g}$. It reads:
\begin{equation}\labelx{Eq:N=2gauge}
    \begin{split}
        &\mathcal L_\mathrm{SYM}^{\mathcal N=2} = \frac{1}{16\pi}\Im\left(\tau \int\mathrm{d}^2\theta~\tr W^\alpha W_\alpha\right)+2\tr\int\mathrm{d}^2\theta\mathrm{d}^2\overline{\theta}~\overline{\Phi}\E^{2g[V,\cdot]}\Phi \\
        &= \tr\left[ -\frac{1}{2}F_{\mu\nu}F^{\mu\nu} - 2\I\lambda\sigma^\mu D_\mu\overline{\lambda} - 2\I\widetilde{\lambda}\sigma^\mu D_\mu\overline{\widetilde{\lambda}} + 2\overline{D_\mu\phi}D^\mu\phi + \frac{\Theta g^2}{16\pi^2}F_{\mu\nu}\star F^{\mu\nu} \right. \\
        &\left.+D^2+ 2\overline{F}F + 2\I\sqrt{2}g\overline{\phi} \{\lambda,\widetilde{\lambda}\} - 2\I\sqrt{2}g\{\overline{\widetilde \lambda},\overline{\lambda}\}\phi + 2gD[\phi,\overline\phi]\right]\; ,
    \end{split}
\end{equation}
with the notation of \Cref{Eq:nonabeliankineticSUSY}, \Cref{eq:complexgaugecouplingYM} and \Cref{Eq:chargechiralN=1kinetic}.
\end{proposition}

$\mathcal N=2$ Maxwell theory is defined by a similar Lagrangian, however with $V$ an abelian $\mathcal N=1$ vector multiplet and $\Phi$ a massless neutral chiral superfield. Therefore, the structure of the Lagrangian changes accordingly: the term $\E^{2g[V,\cdot]}$ in the chiral kinetic term disappears, the coupling $\tau$ becomes $\tau_M$ defined in \Cref{eq:defcomplexgaugecoupling}, and the normalization changes by $e^2/8\pi$ as in \Cref{Eq:abeliankineticSUSY}. 

In \Cref{Eq:N=2gauge}, $\tr$ is, as in \Cref{Eq:nonabeliankineticSUSY}, to be understood as a Killing form on $\mathfrak g$ with respect to which long roots have squared length two, and the generators $T^a$ of $\mathfrak g$ in this representation are normalized so that: 
\begin{equation}
    \tr(T^a T^b) = \delta^{ab}/2\; .
\end{equation}
The non-degenerate metric $\tr$ on $\mathfrak g$ induces an isomorphism between $\mathfrak g$ and its dual, with dual basis $T_a=2\tr(T^a~\cdot)\in\mathfrak{g}^*$. That explains the term $2\tr$ in front on the second superspace integral in \Cref{Eq:N=2gauge}. The commutators and anti-commutators in the component expansion follows from the fact that $\Phi$ transforms in the adjoint representation:
\begin{equation*}
    \begin{split}
        \overline{\phi}\lambda\psi = \overline{\phi}^c\lambda_a(T^a)^b_c\psi_b &=\overline{\phi}_c\lambda_a(T^a)^{bc}\psi_b \nonumber\\
        &= 2\overline{\phi}_c\tr(T^c[T^a,T^b])\psi^b = 2\tr \overline\phi\{\lambda,\psi\}\; .
    \end{split}
\end{equation*}

Moreover, if this Lagrangian were treated as an $\mathcal N=1$ Lagrangian, the relative normalization of the terms would be free. However, $\mathcal N=2$ supersymmetry requires $\mathrm{SU}(2)_R$-invariance, which dictates that the kinetic terms for $\lambda$ and $\widetilde \lambda$ must have matching normalizations. This symmetry constraint uniquely fixes the relative normalization of the kinetic terms in the $\mathcal N=2$ Lagrangian. Additionally, due to $\mathrm{SU}(2)_R$ invariance, no superpotential term can be introduced for $\Phi$.

\vspace{0.3cm}

In $\mathcal N=2$ theories, matter fields consist of hypermultiplets $Q_i\oplus \widetilde {Q}^{i\dagger}$ (where $i$ is a flavor index) in some representation $R$ of $\mathfrak g$. As before, $V\oplus \Phi$ is the $\mathcal N=2$ vector superfield corresponding to $\mathfrak 
g$. 
\begin{proposition}
    The most general renormalizable $\mathcal N=2$ Lagrangian density for such hypermultiplets follows from \Cref{Eq:chargechiralN=1kinetic} and reads:
\begin{equation}\labelx{Eq:N=2hyper}
    \begin{split}
        \mathcal L = \int \mathrm{d}^2\theta\mathrm{d}^2\overline{\theta}~&\left(Q^{i\dagger} \E^{2gV_R} Q_{i}+\widetilde{Q}_{i}^\dagger \E^{-2gV_R}\widetilde {Q}^{i}\right)+\sqrt{2}g\left(\int\mathrm{d}^2\theta~ \widetilde{Q}^i\Phi_R Q_i +c.c. \right) \\
	+&\left(\int\mathrm{d}^2\theta~\tensor{\mu}{^i_j} \widetilde{Q}^jQ_i +c.c. \right)\; .
    \end{split}
\end{equation}
\end{proposition}

Invariance under $\mathrm{SU}(2)_R$ symmetry imposes several important constraints on the structure of the Lagrangian, including the relative normalization of the terms, the form of the mass term, and specific conditions on the mass matrix $\mu$. One crucial result is that $\mathrm{SU}(2)_R$-invariance requires the mass matrix $\mu$ to satisfy the condition: 
\begin{equation}\labelx{eq:massmatrixN=2normal}
    [\mu,\mu^\dagger]=0\; .
\end{equation}
This implies that the mass matrix $\mu$ must be normal meaning it can be diagonalized by a flavor rotation. Consequently, this leads to diagonal mass terms of the form:
\begin{equation*}
    \mu\widetilde{Q}_a Q^a\; ,    
\end{equation*}
where $a$ is the gauge index (which was omitted in \Cref{Eq:N=2hyper}). For a half-hypermultiplet $Q$, which satisfies the condition $\widetilde{Q}_a=\epsilon_{ab}Q^b$, the mass term must vanish: $\mu=0$. This implies that half-hypermultiplets cannot possess a mass on their own. Furthermore, as discussed in \Cref{Sec:superPoincarealgebras}, massive hypermultiplets in $\mathcal N=2$ theories, which include only particles with spin less than 1, are necessarily BPS states. Lastly, any cubic superpotential terms in the Lagrangian must vanish. This is because $\mathrm{SU}(2)_R$ does not have a cubic invariant, and thus such terms are incompatible with $\mathrm{SU}(2)_R$-invariance, even though they are allowed by renormalizability considerations.

\subsubsection*{Non-renormalizable $\mathcal N=2$ Lagrangians}

If one drops the renormalizability requirement for $\mathcal N=1$ Lagrangians, one obtains in general an $\mathcal N=1$ gauged non-linear sigma model describing abelian and non-abelian $\mathcal N=1$ vector superfields $V=2gV_a T^a$ as well as chiral superfields $\Phi^i$ charged under the gauge group, with Lagrangian of the form:
\begin{equation*}
    \begin{split}
        \mathcal L = \frac{1}{16\pi} \Im&\left[\int \mathrm{d}^2\theta~ \tau_{ab}(\Phi) W^{\alpha,a}W_{\alpha}^b\right]+\int\mathrm{d}^2\theta\mathrm{d}^2\overline{\theta}~K(\Phi^i,(\overline{\Phi}\E^{2gV})_i) \\
    &+ \int \mathrm{d}^2\theta~ W(\Phi^i)+\int \mathrm{d}^2\overline\theta~ \overline W(\overline\Phi^i)\; .
    \end{split}
\end{equation*}
The gauge coupling $\tau$ has been promoted to an analytic function of the chiral superfields $\Phi^i$, thus it is a chiral superfield itself. If the Lagrangian is invariant under $\mathcal N=2$ supersymmetry, some of the chiral superfields $\Phi^i$ belong to $\mathcal N=2$ vector superfields whereas the others are paired into hypermultiplets. $\mathrm{SU}(2)_R$-invariance constrains the relative normalization of the gauge kinetic terms and the kinetic terms for chiral superfields in $\mathcal N=2$ vector multiplets, since the kinetic terms of the gauginos $\lambda$ and of the Weyl fermions $\widetilde \lambda$ in the adjoint chiral fields must match. 

\vspace{0.3cm}

In what follows, we will focus primarily on non-renormalizable $\mathcal N=2$ Lagrangians that emerge as low-energy effective theories on the Coulomb branch of the moduli space of renormalizable $\mathcal N=2$ theories (to be defined shortly). These low-energy theories are gapless by definition and exist specifically in abelian Coulomb phases. 

\begin{proposition}
    Such theories can generically be described by (non-renormalizable) Lagrangian sigma models involving only abelian gauge fields and massless gauge-neutral matter fields.
\end{proposition} 

\begin{proof}
    Massive fields, by definition, do not contribute to the low-energy effective theory. As for massless charged matter fields, the abelian gauge coupling for the gauge group they are charged under goes to zero in the infrared (IR), causing the gauge group and the associated charged matter fields to decouple. Furthermore, massless charged fields cannot parameterize the moduli space, as any non-zero vacuum expectation value (VEV) would spontaneously break the gauge group they are charged under. Consequently, these fields, as well as the abelian gauge groups, do not appear in the low-energy effective Lagrangian. Thus, we are left with $\mathcal N=2$ Lagrangian sigma models of abelian gauge fields and massless matter fields uncharged under the gauge group.
\end{proof}

We begin by considering non-linear sigma models composed solely of $\mathcal N=2$ abelian vector superfields $A_i=V_i\oplus \Phi_i$, where $i=1,\dots,r$ with $r$ the rank of the theory (in particular, $i$ is not a flavor index here), following the discussion of \cite[Sect. 2.4]{Tachikawa:2013kta}. In preparation for what follows, we will henceforth denote the scalar component of the chiral superfields as $a_i$ instead of $\phi_i$. Ensuring that the kinetic terms for $\lambda_i$ and $\widetilde \lambda_i$ match leads to the following relation:
\begin{equation*}
    -\I(\tau^{ij}-\overline{\tau}^{ij}) = 2 \frac{\partial^2 K}{\partial a_i\partial\overline{a}_j}\; .
\end{equation*}
Since $\tau^{ij}$ is analytic in the $a_i$'s, it follows that for any $i,j,k$, the following relation holds:
\begin{equation*}
    -\I\frac{\partial \tau^{ij}}{\partial a_k} = 2\frac{\partial^3 K}{\partial a_k\partial a_i\partial\overline{a}_j}\; ,
\end{equation*}
which is symmetric in $i,j$ and $k$.  Therefore, locally one can write:
\begin{equation}\labelx{eq:lowenergycouplingsN=2}
    \tau^{ij} = \int \mathrm{d}a_k \frac{\partial \tau^{ij}}{\partial a_k} = \frac{\partial}{\partial a_i}\int\mathrm{d}a_k~\tau^{kj} = \frac{\partial^2}{\partial a_i\partial a_j} \int\mathrm{d}a_k\mathrm{d}a_l~\tau^{kl} =: \frac{\partial^2 \mathcal F}{\partial a_i\partial a_j}\; , 
\end{equation}
where the \emph{prepotential}\index{prepotential} $\mathcal F$ is a holomorphic function in the $a_i$. In fact, $\mathcal F$ can be considered as an analytic function of the $A_i$'s, so that the superfield
\begin{equation*}
    T^{ij}=\frac{\partial^2 \mathcal F}{\partial A_i\partial A_j}
\end{equation*}
has $\tau_{ij}$ as scalar component.
The prepotential is generally only locally defined and, as a result, is better understood as a holomorphic section of a holomorphic line bundle over the manifold parameterized by the $a_i$ rather than as a genuine holomorphic function. In terms of the prepotential, the Kähler potential $K$ can be expressed as:
\begin{equation}\labelx{eq:Kählerpotentialspecialcoor}
    K = \Im\left(\frac{\partial \mathcal F}{\partial a^i}\overline{a}^i\right)\; .
\end{equation}

\begin{definition}\labelx{def:specialKählermanifold}
    A Kähler manifold for which a Kähler potential can be written like this is called \emph{special Kähler manifold}\index{special!Kähler manifold} \cite{Freed:1999special}.
\end{definition}

\begin{proposition}\labelx{prop:specialKählermetric}
    $\mathcal N=2$ supersymmetry implies that the target manifold of an $\mathcal N=2$ non-linear sigma model of vector superfields only is a special Kähler manifold, with Kähler metric:
\begin{equation*}
    \mathrm{d}s^2 = \frac{\partial^2 K}{\partial a^i\partial \overline{a}^j}\mathrm{d}a^i\mathrm{d}\overline{a}^j = \Im(\tau_{ij})\mathrm{d}a^i\mathrm{d}\overline{a}^j\; .
\end{equation*}
\end{proposition}

Now, let us turn our attention to non-renormalizable $\mathcal N=2$ Lagrangians involving only massless, uncharged hypermultiplets, as discussed in \cite[Sect. 7.1]{Tachikawa:2013kta}. Imposing $\mathrm{SU}(2)_R$ invariance places constraints on the target manifold of the sigma model, specifically requiring it to admit not just one, but a whole family of complex structures---forming a $\mathbb{CP}^1$ worth of such structures. 

\begin{proposition}
    The target manifold of any $4d$ $\mathcal N=2$ sigma model of massless, uncharged hypermultiplets is \emph{hyperkähler}\index{hyperkähler!manifold}. 
\end{proposition}
We refer to \cite{Hitchin:1986ea} for a general discussion on hyperkähler manifolds.

\begin{proof}
    This follows from the fact that for each supersymmetry transformation $\delta_\alpha$ and for each chiral superfield $\Phi=(\lambda,\phi)$ the following holds: 
\begin{equation*}        
\delta_{\dot\alpha}^\dagger\delta_\alpha \left(\begin{array}{c}
     \Re \phi  \\
      \Im \phi
\end{array}\right)= \sigma^\mu_{\alpha\dot\alpha}\partial_\mu I\left(\begin{array}{c}
     \Re \phi  \\
      \Im \phi
\end{array}\right)\; ,
\end{equation*}
where 
\begin{equation*}
    I=\begin{pmatrix}
    0 & 1 \\
      -1 & 0
    \end{pmatrix}\; ;
\end{equation*}
in particular, $I^2=-1$. Let $\delta_\alpha^1$ and $\delta_\alpha^2$ be the two independent supersymmetries appearing in the definition of the $\mathcal N=2$ super-Poincaré algebra. Then, for each $c=(c_1,c_2)\in\mathbb{C}^2$ such that $\vert c_1\vert^2+\vert c_2\vert^2=1$, $\delta_\alpha^c = c_1\delta_\alpha^1+c_2\delta_\alpha^2$ also generates an $\mathcal N=1$ subalgebra. Moreover, the matrices:
\begin{equation*}
    I^{(c)}= I_an^a\; , \quad \text{where} \quad n^a=\begin{pmatrix} \overline{c}_1 & \overline{c}_2 \end{pmatrix}\sigma^a\begin{pmatrix}
        c_1  \\
        c_2
    \end{pmatrix}\; ,
\end{equation*}
all square to $-1$, and $I_1$, $I_2$, $I_3$ satisfy the relations of the quaternions.    
\end{proof}

General $\mathcal N=2$ Lagrangians for abelian vector multiplets and massless neutral hypermultiplets read:
\begin{equation}\labelx{Eq:generalN=2sigmamodel}
    \begin{split}
        \int \mathrm{d}^2\theta\frac{-i}{8\pi}\frac{\partial^2\mathcal F}{\partial \Phi_i\partial \Phi_j} W_{i,\alpha}W^\alpha_j+\text{c.c.}+i&\int\mathrm{d}^2\theta\mathrm{d}^2\overline{\theta} \left(\frac{\partial\overline{\mathcal F}}{\partial \Phi^\dagger_i}\Phi_i-\Phi^\dagger_i\frac{\partial\mathcal F}{\partial \Phi_i}\right)\\
    +&\int\mathrm{d}^2\theta\mathrm{d}^2\overline{\theta} K_h(Q^\dagger,Q)\; ,
    \end{split}
\end{equation}
where $Q$ represents all chiral multiplets within the hypermultiplets, and $K_h$ is the \emph{hyperkähler potential}\index{hyperkähler!potential}, a real analytic function of the $Q^\dagger$'s and $Q$'s. The metric on the hypermultiplet scalar manifold is derived from this potential. 

\begin{proposition}\labelx{prop:productmetric}
    The metric on the target space of the sigma model, which is parameterized by the vacuum expectation values (VEVs) of the scalars in both the $\mathcal N=2$ vector superfields and the hypermultiplets, takes on a local product form, $\mathcal M_V\times\mathcal M_H$. Here $\mathcal M_V$ is a special Kähler manifold with metric derived from the prepotential $\mathcal F$, and $\mathcal M_H$ is a hyperkähler manifold, whose metric is determined by $K_h$.
\end{proposition}

\begin{proof}
    This follows from the fact that the Lagrangian terms for the vector multiplets and hypermultiplets are completely decoupled, as the hypermultiplets are uncharged.
\end{proof}

\section{Classical moduli spaces.}\labelx{Sec:classN=2modulispaces}

As for general $\mathcal N=1$ QFTs (cf. \Cref{Subsubsec:ClassdynN=1}), the scalar potential in $\mathcal N=2$ gauge theories is expressed as a sum of non-negative functions of the scalar fields. The \emph{classical moduli space of supersymmetric vacua}\index{moduli space!classical} in an $\mathcal N=2$ gauge theory is defined as the subvariety of the complex affine space parameterized by scalar vacuum expectation values (VEVs) where the scalar potential vanishes, modulo gauge symmetries. While in general $\mathcal N=1$ QFTs all scalar fields correspond to components of chiral superfields, in $\mathcal N=2$ gauge theories it is possible to consistently distinguish between two types of scalar fields: those that belong to vector multiplets and those that belong to hypermultiplets.

For a renormalizable $\mathcal N=2$ theory characterized by the Lagrangian terms in \Cref{Eq:N=2gauge} and \Cref{Eq:N=2hyper}, the equations that define the classical moduli space arise from the variations of three different auxiliary fields. In this section, we restore the previous notation: $a$ is a gauge index, and $i,j,\dots$, flavor indices. 

Firstly, the auxiliary field $D$ from the $\mathcal N=1$ vector superfield appears in the component expansion of the kinetic terms for the vector superfield, the adjoint chiral superfield in the $\mathcal N=2$ vector, and the chiral superfields in hypermultiplets. This leads to the following $D$-term equations:
\begin{equation}\labelx{eq:DtermN=2}
    -D^a = [\phi,\overline\phi]^a + (q^i)^\dagger T^a_R q^i - \widetilde{q}_i^\dagger T^a_R \widetilde{q}_i = 0\; , \quad \text{for every} \; a=1,\dots,\dim\mathfrak g.
\end{equation}
Secondly, the variation of the auxiliary field in the adjoint chiral superfield from the $\mathcal N=2$ vector gives rise to the $F$-term equations:
\begin{equation}\labelx{eq:Fterms1N=2}
    F_\mathrm{adj}^a = \widetilde{q}^i T^a_R q_i= 0\; , \quad \text{for every} \; a=1,\dots,\dim\mathfrak g.
\end{equation}
Lastly, the variation of the auxiliary field in the $\mathcal N=1$ chiral superfields contained in the hypermultiplets leads to the following $F$-term equations:
\begin{equation}\labelx{Fterms2N=2}
    F_{\mathrm{hyper},i} = \phi q_i+\mu_i^jq_j=0\; ,    \quad \widetilde{F}_{\mathrm{hyper}}^i = \widetilde{q}^i\phi +\mu_j^i\widetilde{q}^j=0\; , \quad \text{for every} \; i.
\end{equation}

\begin{definition}
    The \emph{classical moduli space} of the theory is defined as the subvariety of the affine space spanned by the VEVs of the fields $\phi$, $q_i$, $\widetilde{q}^i$, which are constant vectors in $\mathfrak g$, $R$ and $\overline{R}$ respectively, satisfying the condition:
    \begin{equation*}
        \sum_{a=1}^{\dim\mathfrak{g}} \frac{g^2}{2}\vert D^a\vert^2 + \vert F_\mathrm{adj}^a\vert^2+ \sum_{i} \vert F_{\mathrm{hyper},i} \vert ^2 +  \vert \widetilde{F}_{\mathrm{hyper}}^i \vert ^2 = 0\; .
    \end{equation*}
\end{definition} 

These equations are invariant under $\mathrm{SU}(2)_R$, thus:
\begin{proposition}
    The classical $\mathcal N=1$ moduli space of supersymmetric vacua in any $\mathcal N=2$ theory is always $\mathcal N=2$ supersymmetric.
\end{proposition}

The equations that define the classical moduli space of vacua in any $\mathcal N=2$ gauge theory can be reorganized into three distinct sets (see, for instance, \cite{Argyres:1996eh} for the case of $\mathcal N=2$ SQCD). In the first set, the equations depend solely on the adjoint scalars contained in $\mathcal N=1$ vector superfields:
\begin{equation}\labelx{Eq:Coulombbranch}
	[\phi^\dagger,\phi] = 0\; .
\end{equation}
In the second one, they only depend on the scalars in hypermultiplets:
\begin{equation}\labelx{Eq:hypmodulispace}
	\left(q^{i\dagger} T^a_R q_i-\widetilde{q}^i T^a_R \widetilde{q}^\dagger_i\right) = 0\; , \quad \left(\widetilde{q}^i T^a_R q_i\right) =0\; , \quad \text{ for all } a =1,\dots,\dim\mathfrak g\; ,
\end{equation}
while in the third one, they depend on both kinds of scalar fields:
\begin{equation}\labelx{Eq:mixedmodulispace}
    \begin{split}
        \phi q_i+\mu_i^jq_j=0\; ,& \quad  \widetilde{q}^i\phi +\mu_j^i\widetilde{q}^j=0\; , \\
	\phi^\dagger q_i+\mu_i^{\dagger j}q_j=0\; ,& \quad \widetilde{q}^i\phi^\dagger +\mu_j^{\dagger i}\widetilde{q}^j=0\; .
    \end{split}
\end{equation}

This splitting of the equations defining the classical moduli space leads to the definition of interesting subvarieties of the moduli space called \emph{branches}\index{branch of moduli space}. 

\subsubsection*{Coulomb branch}

\begin{definition}\labelx{def:Coulomb}
    The \emph{Coulomb branch}\index{branch of moduli space!Coulomb}\index{moduli space!Coulomb branch} is the subvariety of the moduli space satisfying $q_i=\widetilde q^i=0$.
\end{definition}

It follows from the definition that the Coulomb branch is entirely parameterized by the VEVs of the scalar component $\phi$ of the adjoint chiral superfield in the $\mathcal N=2$ vector, and that \Cref{Eq:Coulombbranch} is the only constraint. Moreover, gauge-equivalent configurations must be identified. It turns out that:

\begin{proposition}\labelx{prop:CartansubalgebraCoulomb}
    Up to gauge transformations one can always assume that $\phi$ belongs to a fixed Cartan subalgebra $\mathfrak h$ of $\mathfrak g$. 
\end{proposition}

Since the normalizer of $\mathfrak h$ in $\mathfrak g$ modulo $\mathfrak h$ is isomorphic to the Weyl group $W(\mathfrak g)$, the classical Coulomb branch of an $\mathcal N=2$ $\mathfrak g$-gauge theory is $\mathfrak h/W$. At a generic point in $\mathcal C$, the gauge algebra is broken to $\mathfrak{u}(1)^r\rtimes W$, where $r$ is the rank of $\mathfrak g$. This explains the name \emph{Coulomb branch}: classically, the effective low-energy theory at any generic point of $\mathcal C$ is a (non-renormalizable) abelian $4d$ $\mathcal N=2$ gauge theory. \Cref{Subsec:N=2Lagrangians} shows that the low-energy physics on the Coulomb branch is completely determined by a prepotential\index{prepotential} $\mathcal F$, and that $\mathcal C$ is a special Kähler manifold\index{special!Kähler manifold}.

One can parameterize $\mathcal C$ by the set of $\mathfrak g$-invariant polynomials in the $\phi_a$'s (see e.g. \cite{fulton2013representation}), also called \emph{Coulomb branch operators}\index{Coulomb branch operator}. For any finite-dimensional simple Lie algebra $\mathfrak g$, the ring of $W$-invariant polynomials is finitely generated and there are no relations between the generators. Hence $\mathcal C$ is parameterized by the Coulomb branch operators without relations.

\begin{example}
    Let us consider pure $\mathcal N=2$ $\mathfrak{su}(2)$ SYM, for which the moduli space consists only of $\mathcal C$. Modulo gauge transformations one can always assume that the VEV of the scalar $\phi$ in the $\mathcal N=2$ vector multiplet is of the form
\begin{equation}\labelx{Eq:exampleSU2Coulomb}
    \langle\phi\rangle = \alpha\sigma_3 = \alpha\begin{pmatrix} 1 & 0 \\ 0 & -1 \end{pmatrix}\; ,
\end{equation}
for $\alpha$ a complex number, and where the vacua corresponding to $\alpha$ and $-\alpha$ are identified by the gauge action of the Weyl group $W\simeq \mathbb{Z}_2$. One has $\mathcal C \simeq \mathbb{C}/\mathbb{Z}_2$, and the ring of Coulomb branch operators is $\mathbb{C}[u]$ with $u=\alpha^2=\tr\langle\phi\rangle^2$. The gauge algebra breaks spontaneously to $\mathfrak{u}(1)\rtimes W$ when $\alpha\neq 0$. 

When $\alpha\neq0$, the kinetic term $2\tr\vert D_\mu\phi\vert^2= 2\tr\vert \partial_\mu\phi+[A_\mu,\phi]\vert^2 $ for the scalar $\phi$ with VEV $\langle\phi\rangle$ yields the mass term
\begin{equation*}
    2\tr\left\vert[A_\mu,\langle\phi\rangle] \right\vert^2
\end{equation*}
for the vector field. Since
\begin{equation*}
    \left[\begin{pmatrix}
        A^0/2 & W^+/2 \\
        W^-/2 & -A^0/2
    \end{pmatrix}_\mu,\begin{pmatrix}
        \alpha & 0 \\
        0 & -\alpha
    \end{pmatrix}\right] = \begin{pmatrix}
        0 & -\alpha W^+ \\
        \alpha W^- & 0
    \end{pmatrix}_\mu\; ,
\end{equation*}
the $W^\pm$ bosons become massive and the gauge group indeed breaks spontaneously to $\mathfrak{u}(1)\rtimes W$. Note that this computation is classical and holds only if quantum effects can be neglected. 

The classical prepotential reads:
\begin{equation}\labelx{eq:classprepot}
    \mathcal F_\mathrm{cl} = \frac{1}{2}\tau_\mathrm{cl} A^2\; ,
\end{equation}
since:
\begin{equation*}
    \tau_\mathrm{cl} = \frac{\partial^2\mathcal F}{\partial A\partial A}\; .
\end{equation*}
The prepotential can (and in general does) receive quantum corrections.
\end{example}

When fundamental hypermultiplets $(Q, \widetilde Q)$ are added, the Coulomb branch remains unchanged, as it is independent of these hypermultiplets. However, the choice of vacuum expectation value (VEV) given in \Cref{Eq:exampleSU2Coulomb} affects the masses of $Q$ and $\widetilde Q$ due to the presence of the term $\widetilde{Q}\Phi Q$ in the Lagrangian:
\begin{equation}\labelx{eq:masseshypersvacuumCoulomb}
    \sqrt{2}\widetilde Q\langle \Phi\rangle Q +\mu\widetilde Q Q = \begin{pmatrix}
        \widetilde{Q}_1 & \widetilde{Q}_2 \end{pmatrix} \begin{pmatrix}
        \sqrt{2}\alpha+\mu & 0 \\ 0 & -\sqrt{2}\alpha+\mu \end{pmatrix}\begin{pmatrix}
        Q_{1} \\ Q_{2}
    \end{pmatrix}\; .
\end{equation}

\subsubsection*{Higgs branch}

Let us assume for simplicity that the mass matrix $\mu$ vanishes (in fact, Higgs branches exist whenever there are two or more hypermultiplets of the same mass, see e.g. \cite{Seiberg:1994aj}). 

\begin{definition}
    The \emph{Higgs branch}\index{branch of moduli space!Higgs}\index{moduli space!Higgs branch} $\mathcal H$ is defined by $\phi=0$ and hence is parameterized by the VEVs of $h$ and $\tilde h$ only, satisfying \Cref{Eq:hypmodulispace} and modulo gauge transformations.
\end{definition} 

For generic choices of $h$ and $\tilde h$, the gauge group $G$ is completely broken, hence the name of that branch. Classically, one expects that low energy effective theory on $\mathcal H$ is generically a non-renormalizable $\mathcal N=2$ sigma model of massless neutral hypermultiplets only, and hence that the Higgs branch is hyperkähler \index{hyperkähler!manifold}. We refer to \cite{Argyres:1996eh} and \cite[Chap. 7]{Tachikawa:2013kta} for more details.

In an $\mathcal N=2$ gauge theory with gauge algebra $\mathfrak g$ and $k$ hypermultiplets in a representation $R$ of $\mathfrak g$, each hypermultiplet consists of two complex scalar fields, resulting in a total of $2k\dim R$ complex scalars. One can show that the classical Higgs branch is the \emph{hyperkähler quotient} of $\mathbb{C}^{2k\dim R}$ by the complexified gauge algebra $\mathfrak g_\mathbb{C}$, of complex dimension $2(k\dim R-\dim\mathfrak g_\mathbb{C})$. 

\begin{example}
    In $\mathcal N=2$ $\mathfrak{su}(2)$ SYM with $F$ fundamental hypermultiplets, the classical Higgs branch is of complex dimension $4F-2\dim\mathfrak{su}(2)=4F-6$. In particular, $\mathcal{H}$ is empty when $F=0,1$, and of complex dimension $2,6,10$ when $F=2,3,4$, respectively.
\end{example}

\begin{example}
    Consider $\mathcal N=2$ Maxwell theory with two hypermultiplets $(Q_i\oplus Q^{i\dagger})$ for $i=1,2$, of charge $1$, as in \cite[Sect. 7.3.1]{Tachikawa:2013kta}. The Higgs branch is determined by the equations:
    \begin{equation*}
        q_1\widetilde{q}^1+q_2\widetilde{q}^2=0 \quad \text{and} \quad \vert q_1\vert^2+\vert q_2\vert^2-\vert \widetilde q^1\vert^2-\vert \widetilde q^2\vert^2=0\; ,
    \end{equation*}
    up to $\mathrm{U}(1)$ gauge rotations. The first equation can be solved by setting $(q_1,\widetilde q^1)=(z,\widetilde{z}t)$ and $(q_2,\widetilde q^2)=(\widetilde z,-zt)$. The second equation imposes the condition $\vert z\vert^2+\vert\widetilde z\vert^2=\vert t\vert^2(\vert z\vert^2+\vert\widetilde z\vert^2)$, leading to $\vert t\vert =1$. The $\mathrm{U}(1)$ symmetry acts as $(z,\widetilde z,t)\rightarrow (e^{i\theta}z,e^{i\theta}\widetilde z,e^{-2i\theta}t)$, allowing us to fix $t=1$. Thus:
    \begin{equation*}
        \mathcal H = \{(z,\widetilde z)\in\mathbb{C}^2\}/\sim\; ,
    \end{equation*}
    where $(z,\widetilde z)\sim(-z,-\widetilde z)$, implying that $\mathcal H \simeq \mathbb{C}^2/\mathbb{Z}_2$ with its natural hyperkähler metric. 
\end{example}

\subsubsection*{Mixed branches}

In the classical moduli space, generically neither $\phi$, $h$ nor $\tilde h$ vanishes, and all three sets of equations defining the moduli space (including \Cref{Eq:mixedmodulispace}) play a role. If the low-energy degrees of freedom correspond to a non-renormalizable $\mathcal N=2$ abelian gauge non-linear sigma model with neutral massless matter, it can be described by a Lagrangian density of the form given in \Cref{Eq:generalN=2sigmamodel}. As underlined in \Cref{prop:productmetric}, the metric on the target manifold is necessarily locally a product of the special Kähler metric on the Coulomb branch and the hyperkähler metric on the Higgs branch. The Higgs and Coulomb branches intersect only at the origin of the moduli space, where all VEVs vanish.

\subsubsection*{Symmetry definitions of the branches}

There exist many $4d$ $\mathcal N=2$ known gauge theories without a known Lagrangian description\index{non-Lagrangian $4d$ $\mathcal N=2$ theory}, to which the above analysis of the moduli space does not apply. However, even in the absence of a Lagrangian or precise knowledge of the fundamental degrees of freedom, one can argue for the existence of a moduli space of supersymmetric vacua and extract definitions of branches\index{branch of moduli space} based on symmetries.

The distinction between Higgs\index{branch of moduli space!Higgs} and Coulomb branches\index{branch of moduli space!Coulomb} hinges particularly on the fate of the global $\mathrm{SU}(2)_R$ symmetry factor, which depends on which type of scalar fields acquire VEVs. The scalar $\phi$ in any $\mathcal N=2$ vector multiplet is uncharged under this symmetry, while the $q$'s and $\widetilde q^\dagger$'s in hypermultiplets transform as a doublet. If the hypermultiplet fields acquire non-zero VEVs in a vacuum, $\mathrm{SU}(2)_R$ is spontaneously broken, allowing it to act non-trivially by mapping this vacuum to another. In contrast, on the Coulomb branch, $\mathrm{SU}(2)_R$ remains unbroken and acts trivially on the vacua. Therefore, \Cref{def:Coulomb} extends to:

\begin{definition}
    The Coulomb branch is the subvariety of the moduli space where $\mathrm{SU}(2)_R$ is unbroken. By analogy with Lagrangian cases, one defines the \emph{rank} of any $\mathcal N=2$ gauge theory as the dimension of its Coulomb branch.
\end{definition}

The Higgs branch can be defined for non-Lagrangian theories in a similar way. By studying global symmetries other than $\mathrm{SU}(2)_R$ (such as baryonic symmetries) one can even define interesting sub-branches of the Higgs branch \cite{Argyres:1996eh}. Defining the branches of moduli spaces according to the fate of global symmetries solidifies the intuition that these branches represent vacua where the low-energy physics exhibits uniform properties.

Generalizing the remark made below \Cref{prop:CartansubalgebraCoulomb}, it is believed that the Coulomb branch of any $4d$ $\mathcal N=2$ theory, not necessarily Lagrangian, is the quotient of a finite-dimensional complex affine space by a finite group $G$, with the ring of \emph{Coulomb branch operators}\index{Coulomb branch operator} being freely generated \cite{Tachikawa:2013kta}. The finite groups $G$ that satisfy this condition are \emph{complex reflection groups}\index{complex!reflection groups}, as established by the \emph{Chevalley–Shephard–Todd theorem}\index{Chevalley–Shephard–Todd theorem}, of which Weyl groups are special instances. Notably, there are somewhat exotic $4d$ $\mathcal N=2$ theories for which the Coulomb branch is the quotient of a complex finite-dimensional vector space by a complex reflection group that is not a Weyl group (for example, \emph{$\mathcal N=3$ S-folds}\index{$\mathcal N=3$ S-fold} \cite{Garcia-Etxebarria:2015wns, Aharony:2016kai, Garcia-Etxebarria:2016erx}).

\section{Quantum gauge dynamics}

\subsection{Renormalization}\labelx{Sec:RenormN=2}

We emphasized in \Cref{Subsec:N=1renorm}, perturbative quantum corrections in supersymmetric $4d$ QFTs can only generate D-terms, not F-terms. Consequently, the renormalization of any coupling in the superpotential can be expressed through the renormalization of chiral superfields. Additionally, gauge invariance ensures that the term $gV$ (where $V$ is any vector superfield and $g$ the corresponding gauge coupling) does not renormalize. In an $\mathcal N=2$ Lagrangian gauge theory, the renormalizations of the $\mathcal N=1$ vector and chiral superfields belonging to the same $\mathcal N=2$ vector superfield must be identical, and similarly, for the chiral and antichiral superfields in the same hypermultiplet. The superpotential term $\sqrt{2}g\widetilde{Q}\Phi Q$ imposed by $\mathrm{SU}(2)_R$-invariance further implies that the chiral superfields in hypermultiplets do not renormalize. As a result:
\begin{proposition}
    The hyperkähler metric on the Higgs branch is tree-level exact.
\end{proposition} 

In contrast, the special Kähler metric on the Coulomb branch can, and in general does, receive quantum corrections.

Since $\mathcal N=2$ SYM theories are special cases of $\mathcal N=1$ theories, the holomorphic gauge coupling of any $\mathcal N=2$ SYM theory with compact simple gauge group is perturbatively exact at one-loop:
\begin{equation}\labelx{eq:renormN=2}
    \frac{\mathrm{d}}{\mathrm{d}\log\mu}\tau = -\frac{b_1}{2\pi i} + \text{non-perturbative}\; ,
\end{equation}
with $b_1=3C(\mathrm{Adj})-n_\chi I(R_\chi)$ the one-loop renormalization coefficient as in \Cref{Eq:0renN=1}. The (scale-independent) holomorphic dynamical scale\index{dynamical scale} reads:
\begin{equation}\labelx{eq:N=2dynamicalscale}
    \Lambda = \mu\exp(\frac{2\pi \I\tau}{b_1})\; .
\end{equation}
In $\mathcal N=2$ theories, each $\mathcal N=1$ vector superfield is paired with a chiral superfield in the adjoint representation. Therefore, the expression of $b_1$ simplifies to:
\begin{equation*}
    b_1=2C(\mathrm{Adj})-n_\chi I(R_\chi)\; ,
\end{equation*} 
where now the sum over $\chi$ runs over the $\mathcal N=1$ chiral superfields in hypermultiplets only. 

\begin{example}\labelx{ex:b1N=2SYM}
    In $\mathcal N=2$ $\mathfrak{su}(N)$ SYM with $F$ fundamental hypermultiplets, i.e. $\mathcal N=2$ SQCD with $N$ colors and $F$ flavors, one has $b_1=2N-F$. In particular, this theory is asymptotically free when $2F<N$ and IR-free when $2F>N$.
\end{example}

\begin{example}
    In $\mathcal N=2$ $\mathfrak{su}(N)$ SYM with an adjoint hypermultiplet, one has $b_1=0$. When the hypermultiplet is massive these theories, often referred to as $\mathfrak{su}(N)$ $\mathcal N=2^*$ theories, are scale-dependent, as the hypermultiplet decouples at energies smaller than its mass. In contrast, when the mass of the adjoint hypermultiplet vanishes, the theory enjoys not only $\mathcal N=2$ but in fact $\mathcal N=4$ supersymmetry. These theories are called $\mathcal N=4$ $\mathfrak{su}(N)$ SYM. They are scale-invariant \cite{Seiberg:1988ur}, and even superconformal.
\end{example}

In \Cref{Sec:HolomorphySUSY}, we highlighted the distinction between holomorphic and physical gauge couplings in $\mathcal N=1$ gauge theories. Due to the non-analytic nature of the transformation between the two, the physical gauge coupling in a general $\mathcal N=1$ gauge theory receives corrections at all orders in perturbation theory. This phenomenon, however, does not occur in $\mathcal N=2$ theories, thanks to the presence of the adjoint chiral superfield in the $\mathcal N=2$ vector multiplet \cite{Arkani-Hamed:1997qui}. 

\begin{proposition}\labelx{prop:N=2holomandphysicalgaugecouplings}
    In $\mathcal N=2$ gauge theories, the physical gauge coupling coincides with the holomorphic one, making it exact at one loop in perturbation theory, and tree-level exact when the one-loop coefficient vanishes.
\end{proposition} 

This implies for example that the gauge coupling in $\mathcal N=2$ SQCD with $N$ colors and $2F$ flavors is tree-level exact in perturbation theory; moreover, it is widely believed that non-perturbative quantum corrections also vanish. These theories form another standard class of $\mathcal N=2$ SCFTs \cite{Seiberg:1994aj,Gaiotto:2009we}.

\begin{remark}
    On the Coulomb branch of an asymptotically free Lagrangian $4d$ $\mathcal N=2$ theory, the renormalization of the gauge coupling in \Cref{eq:renormN=2} can be recast as a correction to the prepotential that governs the low-energy dynamics, see for instance \cite{Seiberg:1988ur}. Moreover, this renormalization propagates to the low-energy abelian gauge coupling. As the special Kähler metric on the Coulomb branch depends on the latter, one can see that the Coulomb branch metric indeed receives quantum corrections.
\end{remark}

\subsection{Anomalies}\labelx{sec:AnomaliesinN=2theories}

\subsubsection*{Gauge anomalies}

\begin{proposition}
    There are no small chiral gauge anomalies in Lagrangian $4d$ $\mathcal N=2$ gauge theories.
\end{proposition}

\begin{proof}
    In Lagrangian $4d$ $\mathcal N=2$ gauge theories, gauge theories, all matter fields belong either to adjoint chiral superfields within $\mathcal N=2$ vector superfields or to chiral superfields in hypermultiplets. These fields do not contribute to perturbative chiral gauge anomalies\index{anomaly!chiral}. This is because, firstly, the adjoint representation is real, and secondly, each hypermultiplet consists of two chiral superfields in conjugate representations of the gauge group, leading to their combined contribution to the anomaly canceling out.
\end{proof}

However, there can be global gauge anomalies, such as Witten's anomaly when $G=\mathrm{USp}(2N)$. For instance, $\mathcal N=2$ $\mathrm{SU}(2)$ SYM with a half-hypermultiplet in the fundamental representation is inconsistent as a quantum theory \cite{Witten:1982fp}.

\subsubsection*{Anomalies of abelian global symmetries}

Chiral $\mathrm{U}(1)$ global symmetries group can be anomalous, cf. \Cref{Sec:anomalies}. The anomaly reads:
\begin{equation*}
	\mathcal A = \sum_i q_i I(R_i)\; ,
\end{equation*} 
where the sum runs over Weyl fermions of charge $q_i$ under the chiral symmetry and in representations $R_i$ of the gauge group.

Consider the $\mathrm{U}(1)_R$ $R$-symmetry (distinct from the $\mathrm{SU}(2)_R$ symmetry) in pure $\mathrm{SU}(N)$ $\mathcal N=2$ SYM, under which the fields $A_\mu$, $\lambda$, $\widetilde \lambda$ and $\phi$ carry charge 0, 1, 1 and 2, respectively. This symmetry is anomalous:
\begin{equation*}
    \partial_\mu j^\mu_R = \frac{\mathcal A}{16\pi^2}F^{\mu\nu}_a\widetilde{F}^a_{\mu\nu}\; ,
\end{equation*}
with $\mathcal A = 2I(\text{adj})=2N$. A $\mathrm{U}(1)_R$ transformation with parameter $\alpha$ has the same effect as a shift of the $\Theta$-angle: $\Theta\rightarrow\Theta+2\alpha\mathcal A = \Theta+4N\alpha$. Hence, only the $(\mathbb{Z}_{4N})_R$-subgroup of $\mathrm{U}(1)_R$ is a genuine symmetry of the QFT, as it corresponds to shifts of $\Theta$ by multiples of $2\pi$. Furthermore, since $\phi$ has charge $2$ under $(\mathbb{Z}_{4N})_R$, this symmetry is spontaneously broken to $\mathbb{Z}_2$ on the Coulomb branch.

In contrast to $\mathcal N=1$ SQCD, adding matter does not necessarily restore the full $R$-symmetry. Since $R(\phi)=2$, the superpotential term $H_1\Phi H_2$ forces the hypermultiplets to have $R$-charge 0. As a result, their Weyl components carry charge $-1$ and contribute to the ABJ anomaly. For example, the quantum $R$-symmetry in $\mathcal N=2$ SQCD with $N$ colors and $F$ fundamental flavors is $\mathbb{Z}_{4N-2F}$. However, when $F=2N$ the $R$-symmetry is non-anomalous, consistent with the fact that this theory is an SCFT, where a $\mathrm{U}(1)_R$ factor is part of the $\mathcal N=2$ $4d$ superconformal algebra.

\subsection{Moduli spaces}\labelx{Sec:quantN=2modulispaces}

Consider an asymptotically free $\mathcal N=2$ gauge theory with massless hypermultiplets. In a supersymmetric vacuum far from the origin of the moduli space, the gauge group breaks spontaneously in the weak-coupling regime of the UV theory, either completely or to an abelian subgroup. In this case, the classical analysis provides a good approximation of the genuine dynamics of the theory at scales above the breaking one. On the Coulomb branch\index{branch of moduli space!Coulomb}, the resulting low-energy theory is typically a free theory of abelian $\mathcal N=2$ vector superfields and neutral massless hypermultiplets, that does not run under renormalization group flow. Therefore, the classical description of the moduli space is a good approximation far away from its origin. However, as one approaches the origin, strongly coupled non-abelian gauge interactions come into play, invalidating the classical analysis. This reasoning also applies to theories with massive hypermultiplets (which decouple below their mass scale) and to SCFTs (which are weakly coupled when the gauge coupling is small).

We have seen in \Cref{prop:productmetric} that the classical moduli space of any $4d$ $\mathcal N=2$ QFT is locally a product of the Coulomb and Higgs branches. This is also true for the quantum moduli space: if the low-energy effective theory consists of $\mathcal N=2$ vector superfields and neutral massless hypermultiplets, the quantum moduli space remains a local product of a special Kähler manifold (the Coulomb branch) and a hyperkähler one (the Higgs branch\index{branch of moduli space!Higgs}) \cite{Argyres:1996eh}. However, the quantum metrics on these branches can differ from their classical counterparts due to quantum corrections. However, one has the important property:

\begin{proposition}
    Higgs branches are classically exact.
\end{proposition}
In other words, the quantum hyperkähler metric on a Higgs branch coincides with the classical one. This can be understood through $\mathcal N=2$ holomorphy\index{holomorphy}, where the UV couplings are promoted to background spurious superfields \cite{Argyres:1996eh}. The essential point is that the Higgs branch metric does not depend on $\mathcal N=2$ vector multiplets. Since the gauge coupling $\tau$ (or equivalently, the dynamical scale $\Lambda$) as well as the masses $\mu^i_j$ can be interpreted as VEVs of spurious $\mathcal N=2$ vector superfields, the Higgs branch metric is independent of $\Lambda$ and $\mu^i_j$. In particular, if a Higgs branch exists classically, then it also exists in the full quantum theory.

In contrast, the metric on the Coulomb branch generally receives quantum corrections. Nevertheless:
\begin{proposition}\labelx{prop:quantumCoulombnotlifted}
    The Coulomb branch is not lifted by quantum effects.
\end{proposition}

In the semi-classical region far from the origin of the Coulomb branch, the classical analysis provides a good approximation of the quantum dynamics. The flat directions are parameterized by the VEVs of the scalars in the low-energy abelian $\mathcal N=2$ vector superfields, and the only mechanism that could lift the Coulomb branch would involve these scalars acquiring mass. However, this can only occur via the Higgs mechanism, and since the low-energy theory consists only of abelian vector fields and massless neutral matter, the abelian gauge group cannot be further Higgsed. Thus, the Coulomb branch survives in the quantum theory, at least far from the origin. By analytic continuation, this implies that the Coulomb branch persists even when the VEVs of the adjoint scalar in the UV $\mathcal N=2$ vector superfield are small. Therefore, Lagrangian $\mathcal N=2$ gauge theories admit a quantum Coulomb branch, which is a special Kähler manifold. Its metric approaches the classical one asymptotically at large distances. How quantum effects modify the Coulomb branch metric will be explored in detail in \Cref{Chap:SWtheory}.
\chapter{Electric-magnetic duality}\labelx{Sec:Sduality}

\abstract{We investigate electric–magnetic duality in Maxwell theory, beginning with vanishing theta angle $\Theta$ and extending to arbitrary values. Monopoles and dyons can be constructed as solitonic configurations of the Georgi–Glashow model, following the constructions of ’t~Hooft–Polyakov and Julia–Zee. Their masses are bounded from below by the Bogomol’nyi bound, which can saturate in the Bogomol’nyi–Prasad–Sommerfield (BPS) limit, giving rise to the notion of BPS states. These ideas extend naturally to non-abelian Yang–Mills theories, leading to Montonen–Olive duality and its connection to Langlands dual Lie groups. Yet, the strongly coupled regime of pure Yang–Mills theory remains elusive, and renormalization further complicates the quantum formulation of duality.
$\mathcal N=2$ supersymmetry substantially refines the study of electric–magnetic duality by granting BPS states remarkable stability properties, thereby providing a window into strongly coupled gauge dynamics. In four dimensions, $\mathcal N=4$ super-Yang–Mills (SYM) theory is both superconformal and scale-invariant, features that render the conjectural Montonen–Olive duality — referred to as S-duality in that context — particularly natural. Extending S-duality to $\mathcal N=2$ theories, which encompass a broader and more intricate class of quantum field theories, has yielded significant insights, albeit with considerably greater technical challenges than in the $\mathcal N=4$ case.}

\section{Maxwell theory without \texorpdfstring{$\Theta$}{Theta}-term}\labelx{Subsec:EMdualityMaxwell}

\subsection{Electric-magnetic duality}

The action for \emph{Maxwell theory}\index{Maxwell theory} without $\Theta$-term reads:
\begin{equation}\labelx{Eq:Maxzerotheta}
    S_M[A;e]=-\frac{1}{4e^2}\int\mathrm{d}^4x~F_{\mu\nu}F^{\mu\nu} = -\frac{1}{2e^2}\int F\wedge \star F\; .
\end{equation}
The electromagnetic tensor $F=\mathrm{d}A$ is the curvature of a $\mathrm{U}(1)$-connection $A$ on a principal $\mathrm{U}(1)$ bundle over spacetime. It satisfies the \emph{Bianchi identity}\index{Bianchi identity} $\mathrm{d}F=0$, which accounts for the first half of Maxwell's equations. The second half consists of the equation of motion, which reads $\mathrm{d}\star F=0$, where $\star$ is the Hodge star. When electrically charged matter is present, producing an electric one-form current $j_e$,  the equation of motion is modified to $d\star F=\star j_e$. In components, $F$ and its dual $\star F$ write:
\begin{equation*}
    F^{\mu\nu} = \begin{pmatrix}
        0 & -E_x & -E_y & -E_z \\
        E_x & 0 & -B_z & B_y \\
        E_y & B_z & 0 & -B_x \\
        E_z & -B_y & B_x & 0 
    \end{pmatrix}\; , \quad \star F^{\mu\nu} = \begin{pmatrix}
        0 & -B_x & -B_y & -B_z \\
        E_x & 0 & E_z & -E_y \\
        E_y & -E_z & 0 & E_x \\
        E_z & E_y & -B_x & 0 
    \end{pmatrix}\; ,
\end{equation*}
where $\Vec{E}$ and $\Vec{B}$ are the electric and magnetic fields, respectively. Maxwell's action is invariant under gauge transformations of the form $A\rightarrow A+\I g^{-1}\mathrm{d}g$, where $g$ is a map from spacetime to the gauge group $\mathrm{U}(1)=\{\E^{\I\alpha}~\vert~\alpha\sim\alpha+2\pi\}$. In terms of the angle $\alpha$, the gauge transformation becomes $A\rightarrow A-\mathrm{d}\alpha$.

When $j_e=0$, Maxwell's equations are invariant under the exchange of $F$ and $\star F$, or equivalently $(\Vec{E},\Vec{B})\mapsto (\Vec{B},-\Vec{E})$. In the presence of matter, this invariance still holds if one introduces a magnetic one-form current $j_m$ on top of the usual electric one $j_e$, and if Maxwell equations are modified to:
\begin{align*}
    \mathrm{d}\star F &= \star j_e\; , \\ \mathrm{d} F &= \star j_m\; .
\end{align*}

\begin{definition}
    \emph{Electric-magnetic duality of Maxwell theory} consists of the combined transformations:
    \begin{align*}
        F &\longmapsto \star F\; , \\ 
        (j_e, j_m) &\longmapsto (j_m,-j_e)\; .
    \end{align*}
\end{definition}

The charge $q_\mathrm{el}=n_ee$ $(n_e\in\mathbb{Z})$ of a point-like source can be determined using \emph{Gauss law}\index{Gauss law}. For any 2-sphere $S^2$ that encloses the source at a fixed time, the charge is given by:
\begin{equation}\labelx{Eq:Gauss}
    \int_{S^2} \frac{1}{e}\Vec{E}\cdot\mathrm{d}\Vec{s} = \frac{1}{e}\int_{S^2} \star F = q_\mathrm{el}\; .
\end{equation}

\begin{remark}
    The integer $n_e$ labels the representation of the gauge group $\mathrm{U}(1)$ in which the corresponding field sits. Ultimately, the quantization of electric charge arises because the gauge group $\mathrm{U}(1)$ is compact. In contrast, in a similar theory with non-compact gauge group $\mathbb{R}$, the electric charge would not necessarily be quantized.
\end{remark}

\begin{definition}
    \emph{Magnetic monopoles}\index{magnetic monopole} are point-like sources for the magnetic field $\Vec B$. In analogy to the Gauss law for electric charge of \Cref{Eq:Gauss}, for any 2-sphere $S^2$ surrounding a magnetic monopole with magnetic charge $q_\mathrm{mag}$ at a fixed time, the magnetic charge can be determined by the following equation:
    \begin{equation}\labelx{Eq:Gaussmag}
        \int_{S^2} \frac{1}{e}\Vec{B}\cdot\mathrm{d}\Vec{s} = \frac{1}{e}\int_{S^2} F = q_\mathrm{mag}\; .
    \end{equation}
\end{definition}

The magnetic charge is also quantized, given by $q_\mathrm{mag}=2\pi n_m/e$, where $n_m\in\mathbb{Z}$. This quantization arises from topology: the integer $n_m$ represents the first Chern class of the principal $\mathrm{U}(1)$-bundle defined by the monopole, restricted to $S^2$. In contrast, for gauge group $\mathbb{R}$, every principal $\mathbb{R}$-bundle over $S^2$ is trivial. Therefore, in a theory with gauge group $\mathbb{R}$, there are no magnetic monopoles.

\begin{proposition}[Dirac quantization condition]\index{Dirac!quantization condition}\labelx{prop:Diracquant}
    In Maxwell theory, if there exists a magnetic monopole of charge $q_\mathrm{mag}$ and an electrically-charged particle of charge $q_\mathrm{el}$, then necessarily:
    \begin{equation*}
        q_\mathrm{el}q_\mathrm{mag}\in 2\pi\mathbb{Z}\; .
    \end{equation*}
\end{proposition}

\begin{proof}
    This requirement arises, among other reasons, in order for the phase of an electric source with charge $q_\mathrm{el}$ to be well-defined in the presence of a magnetic monopole with charge $q_\mathrm{mag}$ (see e.g. \cite{tong2018gauge}).
\end{proof}

Similarly to magnetic monopoles, one can also introduce \emph{dyons}\index{dyon} in Maxwell theory:
\begin{definition}
    A \emph{dyon} is a particle that carries both electric and magnetic charges.
\end{definition}

Generalizing \Cref{prop:Diracquant}, one has the:
\begin{proposition}[Dirac--Schwinger--Zwanziger (DSZ) quantization condition]\index{DSZ!quantization condition}\labelx{prop:DSZquant}
    For any two dyons of charges $(q_\mathrm{el}^1,q_\mathrm{mag,1})$ and $(q_\mathrm{el}^2,q_\mathrm{mag,2})$, it must be the case that:
    \begin{equation*}
     q_\mathrm{el}^1q_\mathrm{mag,2}-q_\mathrm{el}^2q_\mathrm{mag,1} \in 2\pi\mathbb{Z}\; .
    \end{equation*}
    Writing $q_\mathrm{el}^1=n_e^i e$ and $q_\mathrm{mag,i} = 2\pi n_{m,i}/e$ for $i=1,2$, the DSZ quantization condition rewrites: 
    \begin{equation*}
        n_e^1n_{m,2}-n_e^2n_{m,1}\in\mathbb{Z}\; .
    \end{equation*}
\end{proposition}

To preserve the normalization of \Cref{Eq:Gauss,Eq:Gaussmag} under the exchange of $F$ and $\star F$, one sets:
\begin{equation*}
    F_D = \frac{2\pi}{e^2}\star F \quad \text{and} \quad e_D^2=\frac{4\pi^2}{e^2}\; .
\end{equation*}

\begin{definition}
    \emph{Electric-magnetic duality} is the statement that the actions $S_M[A;e]$ and $S_M[A_D;e_D]$ describe the same physics, where $S_M[A;e]$ is given in \Cref{Eq:Maxzerotheta} and:
    \begin{equation*}
        S_M[A_D;e_D] = -\frac{1}{2e_D^2}\int F_D\wedge \star F_D\; ,
    \end{equation*}
    with $A_D$ a $\mathrm{U}(1)$ connection with curvature $F_D$. 
\end{definition}

This operation is in general a duality rather than a symmetry, as it relates Maxwell theory with coupling $e$ to Maxwell theory with coupling $e_D$. Note that under electric-magnetic duality, the charges of dyons transform as $(n_e,n_m)\longrightarrow (-n_m,n_e)$; in particular, electrons become monopoles and vice-versa. Importantly, the \emph{DSZ pairing}\index{DSZ!pairing} $n_e^1n_{m,2}-n_e^2n_{m,1}$ is preserved by the duality.

\subsection{Monopoles and dyons from non-abelian gauge theories}

In the previous paragraph, we discussed how magnetic monopoles in Maxwell theory can be interpreted as topologically non-trivial abelian gauge field configurations. These monopoles emerge naturally as solitons in theories with non-abelian compact gauge groups, where electromagnetism arises from the spontaneous breaking of gauge symmetry. Here we outline the logic behind this construction and refer to \cite{Coleman:1975qj, Coleman:1982cx, Alvarez-Gaume:1996ohl, Harvey:1996ur, Weinberg:2006rq, tong2018gauge} for further details.

\vspace{0.3cm}

Consider the $\mathrm{SU}(2)$ Yang--Mills theory with a scalar field $\phi$ in the adjoint representation defined by the Lagrangian:
\begin{equation}\labelx{Eq:GGmodel}
    \mathcal L = \frac{-1}{2g^2}\tr F_{\mu\nu}F^{\mu\nu}+\frac{1}{g^2}\tr D_\mu \phi D^\mu\phi -\frac{\lambda}{4}\left(\tr \phi^2-\frac{v^2}{2}\right)^2\; ,
\end{equation}
where $v$ is a real parameter. This theory is referred to as the \emph{Georgi--Glashow model}\index{Georgi--Glashow model}. This theory has a classical moduli space of vacua parameterized by the scalar VEVs $\langle \phi\rangle$ such that $\tr\langle\phi\rangle^2=v^2/2$. Writing $\phi=\phi_i\sigma^i/2$, the equation reads $\sum \phi_i^2 = v^2$ and hence defines a 2-sphere $S^2_\mathrm{vac}$ of radius $v$ in $\mathbb{R}^3$. One can show that any such vacuum is gauge equivalent to:
\begin{equation*}
    \langle \phi\rangle = \frac{1}{2}\begin{pmatrix}
    v & 0 \\ 0 & -v \end{pmatrix}\; ,
\end{equation*}
therefore the gauge group always spontaneously breaks to $\mathrm{U}(1)$, with abelian gauge coupling $e=g$.

Let us consider static field configurations of finite energy. In order for the configuration to have finite energy, the scalar potential must go to zero fast enough at infinity; in particular $\phi$ needs to satisfy $\tr\phi^2 \sim v^2/2$ at infinity. This yields a map
\begin{equation}\labelx{Eq:GGmodelinfinity}
    S^2_\infty \longrightarrow \mathrm{S}^2_\mathrm{vac}\; ,
\end{equation}
where $S^2_\infty$ is the sphere of asymptotic spatial directions. Topologically, such maps are classified by their index $\nu\in\pi_2(S^2)\simeq \mathbb{Z}$. When the index $\nu$ is non-zero the unbroken subgroup $\mathrm{U}(1)\subset\mathrm{SU}(2)$ depends on the asymptotic direction; the low-energy abelian gauge field $A_\mu^{\mathrm{U}(1)}$ can be expressed as:
\begin{equation}\labelx{Eq:GGunbrokenU1}
    A_\mu^{\mathrm{U}(1)} = \frac{1}{v}\tr(\phi A_\mu)\; ,
\end{equation}
where $A_\mu$ is the $\mathrm{SU}(2)$ gauge field. Moreover the scalar $\phi$ must be asymptotically covariantly constant:
\begin{equation*}
    D_\mu\phi\sim_\infty 0 \quad \Rightarrow \quad A_\mu \sim_{\infty} \frac{\I}{v^2}[\phi,\partial_\mu\phi]+\frac{\phi A_\mu^{\mathrm{U}(1)}}{v}\; .
\end{equation*}
The field strength $F_{\mu\nu}^{\mathrm{U}(1)}$ of the unbroken abelian $A_\mu^{\mathrm{U}(1)}$ reads:
\begin{equation*}
    F_{\mu\nu}^{\mathrm{U}(1)} = \frac{1}{v}\tr(\phi F_{\mu\nu}) \sim_\infty \partial_\mu A_\mu^{\mathrm{U}(1)} - \partial_\nu A_\mu^{\mathrm{U}(1)} +\frac{\I}{v^3}\tr(\phi[\partial_\mu\phi,\partial_\nu\phi])\; ,
\end{equation*} 
and satisfies $\mathrm{d}F^{\mathrm{U}(1)}=\mathrm{d}\star F^{\mathrm{U}(1)}=0$. The last term in this expression follows from the dependence of the embedding $\mathrm{U}(1)\subset \mathrm{SU}(2)$ on the asymptotic direction. One can compute that
\begin{equation*}
    \int_{S_\infty^2} F^{\mathrm{U}(1)} = \int_{S_\infty^2} \Vec{B}^{\mathrm{U}(1)}\cdot\mathrm{d}\Vec{s} = 4\pi \nu\; ,
\end{equation*}
where $\nu$ is the index of the map of \Cref{Eq:GGmodelinfinity}. In other words:
\begin{proposition}
    Finite energy static configurations of the Georgi--Glashow model with $\nu\neq 0$ yield magnetic monopoles for the unbroken abelian $\mathrm{U}(1)$ gauge group of magnetic charge $q_\mathrm{mag} = 2\nu/e$.
\end{proposition} 

One can write an ansatz for such solitons, and the resulting monopoles are called \emph{'t~Hooft--Polyakov monopoles}\index{'t~Hooft--Polyakov monopole} \cite{tHooft:1974kcl,Polyakov:1974ek}.

\begin{remark}
    The fact that the magnetic number $n_m$ of a finite energy static configuration is twice the index $\nu$ is related to the fact that the non-abelian gauge group we started with is $\mathrm{SU}(2)$ instead of $\mathrm{SO}(3)$. Matter in the fundamental representation of $\mathrm{SU}(2)$ would yield fields of electric charge $1/2$ in the low-energy theory, with respect to which $m=2$ is the smallest allowed by Dirac quantization condition. In contrast, fields in representations of $\mathrm{SO}(3)$ always lead to integer electric charges compatible with the existence of magnetic monopoles of magnetic charge $1$. We will come back to this in \Cref{Sec:MontonenOlive}.
\end{remark}

In the Georgi--Glashow model defined in \Cref{Eq:GGmodel}, one can also construct finite-energy static configurations carrying both electric and magnetic charge for the low-energy unbroken $\mathrm{U}(1)$. These are called \emph{Julia--Zee dyons}\index{Julia--Zee dyon} \cite{Julia:1975ff}. As in the special case of 't Hooft--Polyakov monopoles, the field strength of the abelian unbroken gauge field writes:
\begin{equation*}
    F^{\mathrm{U}(1)}_{\mu\nu} = \frac{1}{v}\tr(\phi F_{\mu\nu})\; .
\end{equation*}

\subsection{Bogomol’nyi bound and the BPS limit}

Writing $E_i=F_{0i}$ and $2B_i=-\epsilon_{ijk}F^{jk}$, the mass of a Julia--Zee dyon in the Georgi--Glashow model satisfies \cite{Julia:1975ff,Bogomolny:1975de}:
\begin{equation}\labelx{eq:massdyons}
    \begin{split}
        M = &\frac{1}{g^2}\int \mathrm{d}^3x  \left[\tr(E_i^2+B_i^2+(D_i\phi)^2+(D_0\phi)^2)+V(\phi)\right] \\
    = &\frac{1}{g^2}\int \mathrm{d}^3x \left[\tr((E_i- \sin(\theta)D_i\phi)^2+(B_i- \cos(\theta)D_i\phi)^2+(D_0\phi)^2)+V(\phi)\right] \\
    + &\frac{2}{g^2}\int \mathrm{d}^3x~\left[\sin(\theta)\tr(E_iD^i\phi) + \cos(\theta)\tr(B_iD^i\phi)\right]\; , 
    \end{split}
\end{equation}
with $\theta$ an arbitrary angle. Moreover:
\begin{equation}\labelx{eq:defIRchargesGGmodel}
    \frac{1}{g}\int \mathrm{d}^3x~\tr(E_iD^i\phi) = vq_\mathrm{el}\; , \quad \frac{1}{g}\int \mathrm{d}^3x~\tr(B_iD^i\phi) = vq_\mathrm{mag}\; ,
\end{equation}
where $(q_\mathrm{el},q_\mathrm{mag})$ are the electric-magnetic charges of the dyon. Since $V(\phi)$ is positive definite, for $\theta$ such that $\tan\theta = q_\mathrm{el}/q_\mathrm{mag}$ one deduces the:

\begin{proposition}[Bogomol'nyi bound]\index{Bogomol'nyi bound}
    The mass $M_{q_\mathrm{el},q_\mathrm{mag}}$ of a dyon of charges $(q_\mathrm{el},q_\mathrm{mag})$ satisfies the bound \cite{Bogomolny:1975de}:
    \begin{equation}\labelx{Eq:Bogolbound}
    M_{q_\mathrm{el},q_\mathrm{mag}}\geq \frac{2v}{g}\sqrt{q_\mathrm{el}^2+q_\mathrm{mag}^2}\; .
\end{equation}
\end{proposition}

A direct consequence of the bound is that magnetic monopoles are very heavy when $g=e$ is small, and dyons even heavier \cite{Julia:1975ff,Alvarez-Gaume:1996ohl}. For comparison, the mass of low-energy electrons, i.e. $W^\pm$-bosons, is typically of order $M_W\sim ve$.

The terms $(D_0\phi)^2$ and $V(\phi)$ in \Cref{eq:massdyons} imply that the Bogomol'nyi bound can be saturated only when:
\begin{equation*}
    D_0\phi=0\; , \quad V(\phi)=0\; .
\end{equation*}
This, in turn, is possible only if the coupling $\lambda$ appearing in \Cref{Eq:GGmodel} vanishes identically. 

\begin{definition}
    This limit of the Georgi--Glashow model is called \emph{Bogomol'nyi--Prasad--Sommerfield (BPS) limit} \cite{Prasad:1975kr}. In this case, the Lagrangian of the theory is simply:
    \begin{equation*}
        \mathcal L = \frac{-1}{2g^2}\tr F_{\mu\nu}F^{\mu\nu}+\frac{1}{g^2}\tr D_\mu \phi D^\mu\phi\; ,
    \end{equation*} 
    i.e. the theory is simply $\mathrm{SU}(2)$ YM minimally coupled to a massless scalar field in the adjoint representation.
\end{definition}

Even if the potential $V(\phi)$ vanishes in the BPS limit, configurations with zero gauge and scalar kinetic energies, and where $\tr\langle \phi\rangle^2=v^2/2$, are valid vacuum states of the theory. The Bogomol'nyi bound saturates when:
\begin{equation*}
    E_i = \sin(\theta)D_i\phi \quad \text{and} \quad B_i = \cos(\theta)D_i\phi\; , \quad \text{for}\; \tan(\theta) = q_\mathrm{el}/q_\mathrm{mag}\; .
\end{equation*}

\begin{definition}
    These equations are called \emph{BPS equations}\index{BPS!equations}, and configurations which saturate the Bogomol'nyi bound are called \emph{BPS states}\index{BPS!state}.
\end{definition}

For a direct analysis of the Georgi--Glashow model in the BPS limit, see \cite[Sect. 1.3]{Tachikawa:2013kta}.

\section{Witten effect}

The $\Theta$-term:
\begin{equation*}
    \frac{\Theta}{16\pi^2}\int \mathrm{d}^4x~F_{\mu\nu}(\star F)^{\mu\nu} = \frac{\Theta}{8\pi^2}\int  F\wedge F\; ,
\end{equation*}
is quadratic in the field strength, hence relevant, therefore it can be present at low-energies. Let us consider adding it to \Cref{Eq:Maxzerotheta}, so that the full action now reads: 
\begin{equation*}
    S_M = \int\mathrm{d}^4x~\frac{-1}{4e^2} F_{\mu\nu}F^{\mu\nu}+\frac{\Theta}{8\pi^2} \int F\wedge F\; . 
\end{equation*}
Recalling that $F$ is the curvature of a unitary connection on a complex line bundle $E$, one has:
\begin{equation*}
    \frac{1}{8\pi^2}\int F\wedge F = \frac{1}{2}\int c_1(E)^2 \in \frac{1}{2}\mathbb{Z}\; ,
\end{equation*}
where $c_1(E)$ is the first Chern class of $E$. 

\begin{remark}
    The last equation implies that $\Theta$ is $4\pi$-periodic. However, if spacetime is a \emph{spin manifold}\index{spin manifold}, the integral of $c_1(E)^2$ is actually an even integer. We will focus on spin spacetimes, where $\Theta$ is in fact $2\pi$-periodic.
\end{remark}

\begin{remark}
    The discrete Poincaré transformation $\mathcal{CP}$ maps $\Theta$ to $-\Theta$, so $\mathcal{CP}$ (or equivalently, time-reversal $\mathcal{T}$) is broken except when $\Theta=0$ or $\Theta=\pi$.
\end{remark}

When $\Theta\neq0$, Maxwell equations are modified as follows:
\begin{align}
    \mathrm{d} F &= 0\; , \labelx{Eq:MaxthetaEOM1}\\
    \mathrm{d}\left(\frac{2\pi}{e^2}\star F+\frac{\Theta}{2\pi}F\right) &= 0\; .\labelx{Eq:MaxthetaEOM2}
\end{align}
As a result, the field that enters the electric Gauss law is not the electric field $\Vec{E}$, but instead:
\begin{equation*}
    \frac{2\pi}{e^2}\Vec{E}+\frac{\Theta}{2\pi}\Vec{B}\; .
\end{equation*}

Electric charges remain quantized, and for any two-sphere $S^2$ at a fixed time surrounding a local operator with electric charge $ne$ (where $n\in\mathbb{Z}$), the following holds:
\begin{equation*}
    \int_{S^2} \left(\frac{2\pi}{e^2}\star F+\frac{\Theta}{2\pi} F \right) = \int_{S^2}\mathrm{d}\Vec{s}\cdot \left(\frac{2\pi}{e^2}\Vec{E}+\frac{\Theta}{2\pi}\Vec{B} \right)= 2\pi n\; .
\end{equation*}
If $\Theta$ is varied adiabatically, one observes that $\Vec{E}$ must acquire a contribution proportional to $\Theta\Vec{B}$ in order for Gauss's law to remain valid. Specifically, when $\Theta\neq0$, a magnetic monopole generates a radial electric flux corresponding to an electric charge of $e\Theta/2\pi$. This phenomenon is known as the \emph{Witten effect}\index{Witten!effect}. 

\begin{proposition}
    Under the transformation $\Theta\rightarrow\Theta+2\pi$ dyons transform in the following way:
    \begin{equation}\labelx{eq:defT}
        T\colon (q_\mathrm{el},q_\mathrm{mag}) \longmapsto  (q_\mathrm{el}+q_\mathrm{mag},q_\mathrm{mag})\; ,
    \end{equation}
    where the $T$ representing this transformation should not be confused with time-reversal symmetry.
\end{proposition}

\begin{remark}
    Note that the fact that the electric charge of dyons receives a contribution proportional to $q_\mathrm{mag}\Theta$ when $\Theta\neq 0$ does not contradict the DSZ quantization condition. For an electric charge $e$ and a dyon with charge $(q_\mathrm{el},q_\mathrm{mag})$, the DSZ condition requires $q_\mathrm{mag}=2k\pi/e$ where $k\in\mathbb{Z}$. In contrast, the quantization condition for two dyons with charges $(q_\mathrm{el},q_\mathrm{mag})=(q_\mathrm{el,i},2\pi/e)$, for $i=1,2$, is:
    \begin{equation*}
        q_\mathrm{el,1}-q_\mathrm{el,2}\in e\mathbb{Z}\; .
    \end{equation*}
    This shows that in a general $\mathrm{U}(1)$ gauge theory with a particle of charge $e$, the DSZ quantization condition only constrains the difference between the electric charges of any two dyons to be an integer multiple of $e$. 
    
    Since the $\mathcal{CP}$ transformation acts as $\mathcal{CP}:(q_\mathrm{el},q_\mathrm{mag})\rightarrow(-q_\mathrm{el},q_\mathrm{mag})$, $\mathcal{CP}$-invariance restricts the electric charges of dyons to be either all integers (for $\Theta=0$) or all half-integers (for $\Theta=\pi$). Dyons in general $\mathrm{U}(1)$ gauge theories can be analyzed using the methods discussed in the previous paragraphs if a $\mathcal{CP}$-violating $\Theta$-term is added to the Georgi–Glashow-like models. As expected, one finds that the magnetic charge of dyons contributes to their electric charge by a factor of $e\Theta g_\mathrm{mag}/2\pi$ \cite{Witten:1979ey}.
\end{remark}

\subsection{Electric-magnetic duality in the presence of a \texorpdfstring{$\Theta$}{Theta}-term}

When $\Theta\neq 0$, electric-magnetic duality exchanges \Cref{Eq:MaxthetaEOM1} and \Cref{Eq:MaxthetaEOM2}. Denoting
\begin{equation}\labelx{eq:definitionFdual}
    F_D = \frac{2\pi}{e^2}\star F+\frac{\Theta}{2\pi}F\; ,
\end{equation}
one can rewrite Maxwell equations as:
\begin{equation*}
    \mathrm{d}\left(\frac{2\pi}{e_D^2}\star F_D+\frac{\Theta_D}{2\pi}F\right) = 0\; , \quad 
    \mathrm{d} F_D = 0\; ,
\end{equation*}
provided: 
\begin{equation}\labelx{Eq:EMdualconsist}
    -\frac{4\pi^2}{e^2e_D^2}+\frac{\Theta\Theta_D}{4\pi^2} = - 1\; , \quad \frac{\Theta}{e_D^2}+\frac{\Theta_D}{e^2}=0\; .
\end{equation}
Let us define the complexified gauge coupling and dual gauge coupling as:
\begin{equation*}
    \tau = \frac{2\pi \I}{e^2}+\frac{\Theta}{2\pi}\; , \quad \tau_D = \frac{2\pi \I}{e_D^2}+\frac{\Theta_D}{2\pi}\; .
\end{equation*}

\begin{proposition}
    Electric-magnetic duality acts on the complexified gauge coupling as:
    \begin{equation*}
        \tau_D = S(\tau) := \frac{-1}{\tau}\; ,
    \end{equation*}
    and on the charges, as:
    \begin{equation*}
        S:(q_\mathrm{el},q_\mathrm{mag})\longmapsto (-q_\mathrm{mag},q_\mathrm{el})\; .
    \end{equation*}
\end{proposition}

In terms of the complexified gauge coupling, \Cref{eq:defT} rewrites:
\begin{equation*}
    T \colon \tau \longmapsto \tau+1\; .
\end{equation*}

The transformations $S$ and $T$ generate the group $\mathrm{SL}_2(\mathbb{Z})$. The action of the center $-\mathrm{Id}\in \mathrm{SL}_2(\mathbb{Z})$ on the complexified gauge coupling $\tau$ is trivial, although its action on the charges is not. Therefore, the relevant group as far as the coupling only is concerned, is $\mathrm{PSL}_2(\mathbb{Z})=\mathrm{SL}_2(\mathbb{Z})/\{\pm\mathrm{Id}\}$.

\section{Non-abelian electric-magnetic duality}\labelx{Sec:MontonenOlive}\index{Montonen--Olive duality}

\subsection{Non-abelian quantization condition}

't~Hooft--Polyakov and Julia--Zee's constructions extend to gauge theories where the non-abelian UV gauge group $G_0$ breaks spontaneously to a non-abelian subgroup $G$. In this setup, there exist classical solitons in the $G_0$ Yang--Mills theory, corresponding to (non-abelian) magnetic monopoles and dyons associated with the unbroken subgroup $G$. These finite-energy field configurations were studied and classified by Goddard, Nyuts, and Olive in \cite{Goddard:1976qe}. 

Specifically, let $G_0$ be simple, compact, connected and simply-connected, and consider $G_0$ Yang--Mills with a scalar field $\phi$ in a representation $R$. We focus on the subspace $\mathcal M$ of the classical moduli space of vacua, in which $G_0$ is spontaneously broken to a simple, compact and connected subgroup $G$. In general, $G_0$ acts transitively on $\mathcal M$, and $\mathcal M \simeq G_0/G$. As in \Cref{Eq:GGmodelinfinity}, in any finite-energy $G_0$-configuration, the scalar $\phi$ defines a map from the sphere at infinity $S_\infty$ to $\mathcal M$, which is topologically characterized by its class in $\pi_2(G_0/G)$. 

\begin{remark}
    Any simple, compact and connected $G_0$ satisfies $\pi_2(G_0)=0$. Assuming moreover that $G_0$ is simply-connected, the exact sequence of homotopy groups:
    \begin{equation*}
        \dots \rightarrow \pi_2(G_0)\rightarrow \pi_2(G_0/G) \rightarrow \pi_1(G) \rightarrow \pi_1(G_0) \rightarrow \pi_1(G_0/G) \rightarrow \dots
    \end{equation*}
    implies that $\pi_2(G_0/G)\simeq \pi_1(G)$.
\end{remark} 

\begin{definition}
    Let $F_{\mu\nu}$ be the field strength of the unbroken group $G$. A \emph{$G$-magnetic monopole}\index{magnetic monopole!non-abelian} in its rest frame is defined by the condition that, at spatial infinity, the spatial components of $F$ satisfy:
    \begin{equation*}
        F_{ij}(\Vec{r}) = \frac{\epsilon_{ijk}\Vec{r}^k}{r^3} F(\Vec{r}) \in\mathfrak{g}\; ,
    \end{equation*}
    where $\Vec{r}^k$ is the $k$-th component of the unit spatial radial vector, and $r=\vert\Vec{r}\vert$ is the radial distance.
\end{definition}

\begin{proposition}
    In order for the field theory to be consistent, the following quantization condition must be satisfied:
\begin{equation}\labelx{Eq:quantcondnonab}
    \exp(4\pi \I F(\Vec{r})) = 1 \in G\; .
\end{equation}
\end{proposition}

As usual one must identify gauge-equivalent configurations. Up to conjugation by the unbroken subgroup $G$ of $G_0$, one can always assume that $F$ lies in a fixed \emph{Cartan subgroup}\index{subgroup!Cartan} $H$ of $G$. Thus, one can reformulate the quantization condition of \Cref{Eq:quantcondnonab} as a geometric condition: 

\begin{proposition}
    The field strength $F(\Vec{r})$ must lie in the $G$-\emph{cocharacter lattice}\index{cocharacter lattice} $\Gamma_{w}^G\subset \mathfrak h$, where $\mathfrak h$ is the Lie algebra of $H$. Therefore, $G$-monopoles are characterized by their magnetic charge, which takes values in $\Gamma_{w}^G/W$, where $W$ is the Weyl group of $\mathfrak g=\mathrm{Lie}(G)$. 
\end{proposition}

Fields that are ``electrically'' charged under the low-energy gauge group $G$ must transform in finite-dimensional irreducible representations of $G$. These representations correspond one-to-one with the dominant weights of $G$, i.e. with the elements of $\Lambda_w^G/W$, where $\Lambda_w^G\subset \mathfrak h^*$ is the $G$-\emph{character lattice}\index{character!lattice}. In other words, electric charges are classified by $\Lambda_w^G/W$. Finally, dyons in the low-energy theory carry both electric and magnetic charges under $G$, and their allowed charges are in one-to-one correspondence with $(\Lambda_w^G\times\Gamma_w^G)/W$. 

\begin{remark}
    The roots of $\mathfrak g$ are special weights, i.e. the root lattice $\Lambda_r$ of $\mathfrak g$ is a sublattice of the weight lattice $\Lambda_w$ of $\mathfrak g$. For a compact, connected Lie group $G$ with $\mathrm{Lie}(G)=\mathfrak g$, not all representations of $\mathfrak g$ necessarily lift to representations of $G$. For instance, the fundamental $N$-dimensional representation of $\mathfrak{su}(N)$ does not lift to a representation of $\mathrm{PSU}(N)$. However, the adjoint representation of $\mathfrak g$ always lifts to a representation of $G$. In fact, finite-dimensional irreducible representations of $G$ correspond bijectively to elements of $\Lambda_w^G/W$, where $\Lambda_w^G$ is the lattice of $G$-characters, satisfying $\Lambda_r\subset\Lambda_w^G\subset\Lambda_w\subset \mathfrak h^*$.
    
    Dually, there are two fundamental lattices $\Gamma_r\subset\Gamma_w\subset \mathfrak h$ in the Cartan subalgebra $\mathfrak h$ of $\mathfrak g$: the coroot lattice $\Gamma_r$ and the coweight lattices $\Gamma_w$. The choice of group $G$ defines the so-called $G$-cocharacter lattice $\Gamma_w^G$, which satisfies $\Gamma_r\subset\Gamma_w^G\subset \Gamma_w$. For example, when $\mathfrak g=\mathfrak{su}(N)$, one has $\Lambda_w^{\mathrm{SU}(N)}=\Lambda_w$ and $\Gamma_w^{\mathrm{SU}(N)}=\Gamma_r$, whereas $\Lambda_w^{\mathrm{PSU}(N)}=\Lambda_r$ and $\Gamma_w^{\mathrm{PSU}(N)}=\Gamma_w$. Fundamentals of Lie theory are gathered in \Cref{app:Lietheory}.
\end{remark}

\subsection{Montonen--Olive duality conjecture}

The fact that electric-magnetic duality must exchange electrically charged particles and magnetic monopoles led Montonen and Olive to conjecture the following.
\begin{conjecture}[Montonen--Olive duality\index{Montonen--Olive duality} \cite{Montonen:1977sn}]
    Electric-magnetic duality $S$\index{$S$-duality} maps pure Yang--Mills theory with gauge group $G$ and gauge coupling $\tau$ to pure Yang--Mills theory with gauge group $G^\vee$ and gauge coupling $-1/\tau$, where $G^\vee$ is the group with Lie algebra $\mathrm{Lie}(G^\vee)=\mathfrak{g}=\mathrm{Lie}(G)$ such that $\Lambda_w^{G^\vee}=\Gamma_w^G$ (or dually, $\Gamma_w^{G^\vee}=\Lambda_w^G$).
\end{conjecture}

The group $G^\vee$ is called \emph{GNO-dual}  \cite{Goddard:1976qe}, or \emph{magnetic gauge group}\index{gauge!group!magnetic} (in contrast to $G$, referred to as \emph{electric gauge group}\index{gauge!group!electric}) or \emph{Langlands dual group}\index{Langlands!dual group}, as it plays a central role in the geometric Langlands correspondence. The relationship between electric-magnetic duality in gauge theories and the geometric Langlands correspondence has been thoroughly explored in \cite{Kapustin:2006pk}.

\begin{remark}
    At the classical level, the Montonen--Olive conjecture is natural, particularly because in any Yang--Mills theory obtained from a Georgi--Glashow-like model in the BPS limit, all classical states, including W-bosons, monopoles, and dyons, saturate the BPS bound. However, unlike Maxwell theory, the gauge coupling in non-abelian Yang--Mills theories runs, and quantum effects could very well spoil the BPS bound. Moreover, the Montonen--Olive conjecture is very difficult to test due to the lack of tools to analyze strongly-coupled Yang--Mills theories: since $\tau_D=-1/\tau$, one side of the duality is necessarily strongly coupled. We will see shortly that the situation improves significantly with the help of supersymmetry: one can provide non-trivial evidence in favor of the Montonen--Olive conjecture, though not in pure Yang--Mills theories, but rather in $\mathcal N=4$ super Yang--Mills theories. 
\end{remark}

\subsection{Global variants of gauge theories}

The gauge group $G$ of a gauge theory is not determined by its matter content; rather, it is the choice of the gauge group that dictates in which representations the charged matter fields can transform. For instance, since the adjoint representation is a genuine representation of both $\mathrm{SU}(N)$ and $\mathrm{PSU}(N)$, fields in the adjoint representation of $G$ can appear in both $\mathrm{SU}(N)$ and $\mathrm{PSU}(N)$ gauge theories. However, fundamental fields can only exist in $\mathrm{SU}(N)$ gauge theories, because the fundamental representation of $\mathrm{SU}(N)$ is not a genuine representation of $\mathrm{PSU}(N)$. Moreover, even if the matter content of a Lagrangian is compatible with two distinct gauge groups $G$ and $G'$ sharing the same Lie algebra---for example, $G=\mathrm{SU}(N)$ and $G'=\mathrm{PSU}(N)$---one can still distinguish between the gauge theory corresponding to $G$ and that corresponding to $G'$ \cite{Aharony:2013hda}.

This follows from the fact that gauge theories possess \emph{line operators}\index{line operator}, particularly \emph{Wilson lines}\index{Wilson line} and \emph{'t~Hooft lines}\index{'t~Hooft!line}. We refer to \cite{tong2018gauge} for an introduction to Wilson and 't~Hooft lines in gauge theories. 

\begin{proposition}
    Wilson lines are labeled by the representations of $G$, or equivalently, by $\Lambda_w^G/W$. Conversely, 't~Hooft lines are labeled by $\Gamma_w^G/W$.
\end{proposition}

Note that this implies, in particular, that pure $\mathrm{SU}(N)$ YM and pure $\mathrm{PSU}(N)$ YM theories differ, as they do not have the same line operator content. 

More generally, gauge theories admit \emph{dyonic line operators}\index{dyonic line operators}, also called \emph{Wilson--'t~Hooft lines}\index{Wilson--'t~Hooft line}, with charges in $(\Gamma_w^G\times\Lambda_w^G)/W$ \cite{Kapustin:2005py}. There is a version of Dirac--Schwinger--Zwanziger (DSZ) quantization condition between line and local operators (a line satisfying it is \emph{local} with respect to dynamical fields) and between two line operators (any two lines satisfying it are said to be \emph{mutually local}\index{mutually local}), ensuring the consistency of the theory. Wilson lines and 't Hooft lines for a $G$-gauge theory are always mutually local.

\begin{definition}[\cite{Aharony:2013hda}]\labelx{def:globalvariant}
    The definition of a $G$-gauge theory includes the set of its line operators, which is required to be a maximal collection of local and mutually local dyonic lines (including Wilson and 't~Hooft lines). Theories differing only in their line operator content are dubbed \emph{global variants}\index{global variant}.
\end{definition}

While the Wilson and 't Hooft lines in the theory are fully determined by the choice of gauge group $G$ as explained above, in general, they can be completed with dyonic lines into a maximal set of mutually local lines in more than one way. For example, there exist $N$ distinct global variants of pure $\mathrm{PSU}(N)$ YM theory, denoted as $\mathrm{PSU}(N)_0$, $\dots$, $\mathrm{PSU}(N)_{N-1}$.

The charges of line operators are acted upon by $S$ and $T$, where $T$ corresponds to a shift of the $\Theta$-angle by $2\pi$. In general, $S$ and $T$ shuffle the global variants. For instance, in a gauge theory with non-simply connected gauge group, $\Theta$ is not necessarily $2\pi$-periodic, as fractional instantons may exist. Consequently, $T$ maps the original global variant to another.

\section{Electric-magnetic duality in supersymmetric gauge theories}\labelx{Sec:EMdualityinSUSY}

\subsection{Witten--Olive--BPS states}
 
Consider a $4d$ $\mathcal N=2$ asymptotically free or conformal gauge theory defined by a Lagrangian of the form given in \Cref{Eq:N=2gauge}. At a generic point of its Coulomb branch, the low-energy effective theory is an abelian $4d$ $\mathcal N=2$ gauge theory. Finite-energy gauge configurations of the microscopic theory give rise to monopoles and dyons in the low-energy effective one. In fact, Lagrangian $4d$ $\mathcal N=2$ gauge theories are special instances of Georgi--Glashow theories---with fermions---in the BPS limit. Moreover, the BPS bound saturates in the classical limit.

There are two independent conserved supercurrents corresponding to the fact that the theory has $\mathcal N=2$ supersymmetry. When integrated on spatial slices, these supercurrents give rise to the conserved supercharges $Q_\alpha^I$, $Q_{\dot\alpha}^I$, for $I=1,2$, which appear as odd generators in the $4d$ $\mathcal N=2$ super-Poincaré algebra and define the central charge $\mathcal Z$ of the algebra as:
\begin{equation*}
    \{Q_\alpha^I,Q_\beta^J\} = 2\sqrt{2}\epsilon_{\alpha\beta}\epsilon^{IJ}\mathcal Z\; ,
\end{equation*}
as in \Cref{Eq:SUSY=squareroot}. The expression of the supercurrents $Q_\alpha^I$ and $Q_{\dot\alpha}^I$ typically involves the UV Yang--Mills field strength $F$. 

Recalling \Cref{eq:defIRchargesGGmodel}, the electric-magnetic charges of solitonic gauge configurations of the UV theory are integrals of $F$ at spatial infinity. In the seminal article \cite{Witten:1978mh}, Witten and Olive showed that the central charge $\mathcal{Z}$ of the $4d$ $\mathcal N=2$ super-Poincaré algebra writes as a linear combination of the low-energy charges. Moreover, the Bogomol'nyi bound in such theories reads $m\geq \sqrt{2}\vert Z\vert$, meaning that it matches exactly the SUSY bound of \Cref{Eq:BPSSUSY}, derived from the superalgebra. This explains why short representations of extended super-Poincaré algebra are called \emph{BPS states}\index{BPS!state}:

\begin{proposition}[\cite{Witten:1978mh}]
    In quantum field theories with extended supersymmetry, BPS states transform in \emph{short} representations of the supersymmetry algebra. Thus, the two definitions of BPS states, as states saturating the Bogomol'nyi bound on the first hand, and as short representation of the $4d$ $\mathcal N=2$ super-Poincaré algebra on the other, coincide.
\end{proposition}

Fermionic zero modes in the background of classical monopoles and dyons generate $\mathcal N=2$ BPS hypermultiplets in which the scalar component is the complex scalar corresponding to these solitons, see e.g. \cite{Osborn:1979tq} and \cite[Chap. 1]{Tachikawa:2013kta}. The identification of BPS states and short supermultiplets implies the following important property:

\begin{proposition}
    In $4d$ $\mathcal N=2$ gauge theories, the BPS bound cannot be violated by quantum effects. Therefore, any state saturating the Bogomol'nyi bound classically is also BPS in the quantum theory.
\end{proposition} 

\begin{proof}
    This follows from the fact that long supermultiplets display 16 degrees of freedom whereas short ones, only 8.
\end{proof}

\begin{example}
    Pure $\mathcal N=2$ $\mathrm{SU}(2)$ SYM is asymptotically free. At a generic point of its Coulomb branch $\mathcal C$ far away from the origin, the UV theory never becomes strongly coupled before the spontaneous gauge symmetry breakdown to $\mathrm{U}(1)$. Classically, the 't Hooft--Polyakov--Julia--Zee construction yields a monopole with charges $(q_\mathrm{el},q_\mathrm{mag})=\pm(0,1)$ and dyons with charges $(q_\mathrm{el},q_\mathrm{mag})=\pm(2k,1)$ for any $k\in\mathbb{Z}$. Therefore, the low-energy theory at such a point of the Coulomb branch displays massive BPS $\mathcal N=2$ hypermultiplets whose scalar components are a monopole with charges $(q_\mathrm{el},q_\mathrm{mag})=\pm(0,1)$ and dyons with charges $(q_\mathrm{el},q_\mathrm{mag})=\pm(2k,1)$ for any $k\in\mathbb{Z}$.
\end{example}

Therefore, considering $4d$ $\mathcal N=2$ gauge theories, rather than more general gauge theories, addresses one of the two issues related to the quantum interpretation of Montonen–Olive duality, specifically the concern that quantum effects could violate the BPS bound.

\subsection{\texorpdfstring{$S$}{S}-duality of \texorpdfstring{$4d$ $\mathcal N=4$}{4d N=4} SYM theories}

The second issue related to the quantum interpretation of Montonen–Olive duality in pure Yang–Mills theories is the running of the gauge coupling. This can be addressed by considering scale-invariant theories instead. \emph{Four-dimensional $\mathcal N=4$ super Yang--Mills (SYM) theories}\index{$4d$ $\mathcal N=4$ SYM} are special cases of $4d$ $\mathcal N=2$ theories, specifically $4d$ $\mathcal N=2$ theories with a massless hypermultiplet in the adjoint representation. They are superconformal and, hence, scale-invariant, meaning in particular that the gauge coupling does not run. Therefore, Montonen–Olive conjecture as the duality between $\mathcal N=4$ SYM with gauge group $G$ and coupling $\tau$, and $\mathcal N~=~4$ SYM with dual gauge group $G^\vee$ and coupling $-1/\tau$, makes sense. In fact, the existence of distinct global variants of these theories necessitates a refinement of this conjecture, which will be discussed shortly. Assuming that Montonen–Olive duality holds, the two dual theories should be viewed as distinct but equivalent descriptions of the same underlying quantum field theory, rather than as two separate QFTs.

On the Coulomb branch of $\mathcal N=4$ $\mathrm{SU}(2)$ SYM, which is $\mathcal N=2$ $\mathrm{SU}(2)$ SYM with a massless adjoint hypermultiplet, the low-energy states---electrons, monopoles, and dyons---are all quarter-BPS hypermultiplets of the $4d$ $\mathcal N=4$ super-Poincaré algebra. Thus, $S$-duality can map electrons to monopoles, and vice-versa. Consequently, this theory is a compelling candidate for the realization of Montonen–Olive duality, a notion initially proposed in \cite{Osborn:1979tq}. Subsequent studies, particularly \cite{Sen:1994fa} and \cite{Vafa:1994tf}, have further substantiated the embodiment of Montonen–Olive duality within $4d$ $\mathcal N=4$ SYM theories, leading to a robust consensus in the theoretical community regarding this phenomenon.

Electric-magnetic duality $S$ and the non-abelian transformation $T$ generate the \emph{duality group} $\mathrm{SL}_2(\mathbb{Z})$\footnote{Or rather, its metaplectic extension $\mathrm{Mp}_2(\mathbb{Z})$ \cite{Pantev:2016nze}.}. Its action on the gauge coupling reads:
\begin{equation*}
    \tau \longmapsto \frac{a\tau+b}{c\tau+d}\; , \quad \text{for} \; \begin{pmatrix}
    a & b \\ c & d \end{pmatrix}\in \mathrm{SL}_2(\mathbb{Z})\; ,
\end{equation*}
thus the action on $\tau$ reduces to that of $\mathrm{PSL}_2(\mathbb{Z})$, since $\{-\mathrm{Id}\}<\mathrm{SL}_2(\mathbb{Z})$ acts trivially. However, $\{-\mathrm{Id}\}$ acts non-trivially on the charges. 

\begin{remark}
    The $S$-duality of $4d$ $\mathcal N=4$ super Yang-Mills (SYM) theory implies that the low-energy effective theory encompasses BPS states for all electromagnetic charges $(p,q)$ with $p$ and $q$ coprimes. This assertion has been directly supported in \cite{Sen:1994yi}.
\end{remark}

The existence of distinct global variants of gauge theories with gauge group $G$ implies that the statement of $S$-duality must be refined. For example, there are two global variants of $\mathcal N=4$ SYM with gauge group $\mathrm{SO}(3)$, denoted $\mathrm{SO}(3)_+$ and $\mathrm{SO}(3)_-$. The $\Theta$-angle of $\mathcal N=4$ $\mathrm{SU}(2)$ SYM is $2\pi$-periodic, thus this theory is $T$-invariant. In contrast, $T$ exchanges the global variants $\mathrm{SO}(3)_+$ and $\mathrm{SO}(3)_-$. On the other hand, studying how $S$ acts on line operators shows that it maps the $\mathrm{SU}(2)$ theory to $\mathrm{SO}(3)_+$, while $\mathrm{SO}(3)_-$ is $S$-self-dual. Therefore, at $\tau=\I$, $S$-duality enhances to a symmetry of $\mathrm{SO}(3)_-$. The \emph{duality graph}\index{duality!graph} of $\mathfrak{su}(2)$ $\mathcal N=4$ SYM theories is depicted in \Cref{fig:dualitygraphsu2SYM}.

\begin{figure}[!ht]
    \centering
    \includegraphics[scale=0.83]{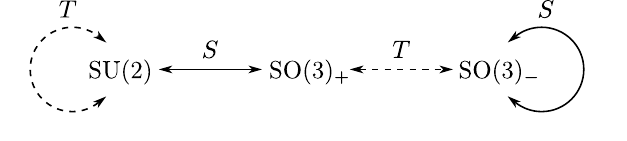}
    \caption{Duality graph of $\mathfrak{su}(2)$ $\mathcal N=4$ SYM theories.}
    \labelx{fig:dualitygraphsu2SYM}
\end{figure}

\begin{remark}
    Combining $S$ with the gauging of the one-form symmetry (which also shuffles the global variants) at self-dual values of the coupling $\tau$ yields non-invertible Krammers--Wannier-like duality defects \cite{Kaidi:2021xfk,Choi:2022zal,Kaidi:2022uux}.
\end{remark}

\subsection{\texorpdfstring{$S$}{S}-duality of \texorpdfstring{$4d$ $\mathcal N=2$}{4d N=2} SYM theories}

In order to study $S$-duality in a context closer to that of pure Yang--Mills theories, one can break half of the supersymmetries and consider $4d$ $\mathcal N=2$ theories instead of $\mathcal N=4$ SYM. In this context, the relationship between BPS states and small representations of the super-Poincaré algebra still holds, however not all $4d$ $\mathcal N=2$ theories are scale-invariant. 

Another issue arises, for example in pure $\mathcal N=2$ $\mathrm{SU}(2)$ SYM. In this theory, the only electrically charged particles present at low energies along the Coulomb branch are the $W$-bosons, which are part of $1/2$-BPS vector supermultiplets. In contrast, magnetic monopoles and dyons fall into $1/2$-BPS hypermultiplets. Since $S$-duality is expected to relate electrically charged particles with their magnetically charged counterparts, this discrepancy poses a challenge. This indicates that $S$-duality in $\mathcal N=2$ SYM theories, if it exists, is more intricate than what the Montonen–Olive conjecture suggests. The Seiberg–Witten analysis of pure $\mathcal N=2$ $\mathrm{SU}(2)$ SYM, to be studied in \Cref{Sec:pure4dN=2SYM}, reveals that a refined notion of $S$-duality does indeed exist. This refined duality involves the concept of \emph{wall-crossing}\index{wall-crossing}, a phenomenon absent as far as the half-BPS states of $4d$ $\mathcal N=4$ SYM are concerned.

Before delving into Seiberg--Witten theory, let us discuss the interplay between the special Kähler geometry of the Coulomb branch $\mathcal C$ of $4d$ $\mathcal N=2$ gauge theories, and electric-magnetic duality of the abelian low-energy effective theory on $\mathcal C$. Recall from \Cref{prop:quantumCoulombnotlifted} that the Coulomb branch of any rank-$r$ $\mathcal N=2$ theory is a special Kähler manifold of complex dimension $r$, meaning that the Kähler potential can be written as in \Cref{def:specialKählermanifold}:
\begin{equation*}
    K(a) = \Im\left(\frac{\partial\mathcal F}{\partial a_i}\overline{a}_i\right)\; , \quad i=1,\dots,r\; ,
\end{equation*}
where $a_1$, $\dots$, $a_n$ are local holomorphic coordinates on $\mathcal C$, and where $\mathcal F$ is a local holomorphic function of $a_i$ called \emph{prepotential}\index{prepotential}. The low-energy abelian gauge couplings $\tau^{ij}$ write as in \Cref{eq:lowenergycouplingsN=2}:
\begin{equation}\labelx{eq:LEcouplings}
    \tau^{ij}(a) = \tau^{ji}(a) = \frac{\partial^2\mathcal F}{\partial a_i\partial a_j},
\end{equation}
and the special Kähler metric, as in \Cref{prop:specialKählermetric}: 
\begin{equation}\labelx{eq:specKahlmetric}
    \mathrm{d}s^2 = \Im \tau^{ij}(a)~\mathrm{d}a_i\mathrm{d}\overline{a}_j\; .
\end{equation}
For any $i=1,\dots,r$, let:
\begin{equation}\labelx{eq:LEdualvar}
   a_D^i = \frac{\partial\mathcal F}{\partial a_i}\; . 
\end{equation}
In terms of both $a_i$ and $a_D^i$, the potential $K$ rewrites:
\begin{equation*}
    K = 2\I\left(a_i\overline{a}_D^i-\overline{a}_ia_D^i\right)\; ,
\end{equation*}
and the special Kähler metric:
\begin{equation*}
    \mathrm{d}s^2 = \Im 
    \mathrm{d}a_D^i \mathrm{d}\overline{a}_i = -\frac{\I}{2}\left(\mathrm{d}a_D^i \mathrm{d}\overline{a}_i-\mathrm{d}\overline{a}_D^i \mathrm{d}a_i\right)\; .
\end{equation*}
These expressions are symmetric in $a_i$ and $a_D^i$. Consequently, replacing each $a_i$ with $a_D^i$ would result in a special Kähler metric analogous to that in \Cref{eq:specKahlmetric}, albeit with a different harmonic function substituted for $\Im \tau(a)$. 

\begin{definition}
    The variables $a_i$ and $a_D^i$ are referred to as \emph{special coordinates}\index{special!coordinates} on the special Kähler manifold $\mathcal C$.
\end{definition} 

\noindent The bosonic part of the low-energy effective action on a Coulomb branch reads:
\begin{equation*}
    \mathcal L_B = \frac{\Im \tau^{ij}}{4\pi} \partial_\mu\overline{a}_i\partial^\mu a_j + \frac{\Im \tau^{ij}}{8\pi} F_{\mu\nu, i} F^{\mu\nu}_j + \frac{\Re \tau^{ij}}{8\pi} F_{\mu\nu, i} (\star F)^{\mu\nu}_j\; .
\end{equation*}
Generalizing \Cref{eq:definitionFdual}, let:
\begin{equation*}
    F_{D\mu\nu}^i = \Im \tau^{ij} F_{\mu\nu,j} + \Re \tau^{ij} (\star F)_{\mu\nu,j}, \quad \text{and} \quad \tau_{Dij} = (-\tau^{-1})_{ij}.
\end{equation*}
One can then rewrite $\mathcal L_B$ as:
\begin{equation}\labelx{eq:defEMduals}
    \mathcal L_B = \frac{\Im \tau^{ij}}{4\pi} \partial_\mu\overline{a}_i\partial^\mu a_j + \frac{\Im \tau_{D ij}}{8\pi} F_{D\mu\nu}^i F_D^{\mu\nu j} + \frac{\Re \tau_{D ij}}{8\pi} F_{D\mu\nu}^i (\star F)_D^{\mu\nu j}\; .
\end{equation}
Now, since:
\begin{equation*}
    \begin{split}
        \frac{\Im \tau^{kl}}{4\pi}\partial_\mu \overline{a}_k\partial^\mu a_l &= \frac{-1}{8\pi \I}(\overline{\tau}^{kl}-\tau^{kl})\partial_\mu \overline{a}_k \partial^\mu a_l \\
        &= \frac{-1}{8\pi \I}(\overline{\tau}^{ki}\delta_i^l-\delta_j^k\tau^{jl})\partial_\mu \overline{a}_k \partial^\mu a_l \\
        &= \frac{1}{8\pi \I}(\tau_{Dij}-\overline{\tau}_{Dij})\overline{\tau}^{ki}\tau^{jl}\partial_\mu\overline{a}_k\partial^\mu a_l \\
        &= \frac{1}{4\pi} \Im \tau_{Dij} \overline{\tau}^{ki}\partial_\mu\overline{a}_k \tau^{jl}\partial^\mu a_l \\
        &= \frac{1}{4\pi} \Im \tau_{Dij} \partial_\mu \overline{a}_D^i \partial^\mu a_D^j\; ,
    \end{split}
\end{equation*}
where we used \Cref{eq:LEcouplings,eq:LEdualvar,eq:defEMduals}, one obtains:
\begin{equation*}
    \mathcal L_B = \frac{\Im \tau_{Dij}}{4\pi} \partial_\mu\overline{a}_D^i\partial^\mu a_D^j + \frac{\Im \tau_{D ij}}{8\pi} F_{D\mu\nu}^i F_D^{\mu\nu j} + \frac{\Re \tau_{D ij}}{8\pi} F_{D\mu\nu}^i (\star F)_D^{\mu\nu j}\; .
\end{equation*}
Thus:
\begin{proposition}
    The scalar $a_D^i$ is in the same $\mathcal N=2$ vector multiplet as the magnetic dual gauge field $A_{D\mu}^i$ whose field is strength $F_{D\mu\nu}^i$. In other words, $a_D^i$ is dual to $a_i$ in the sense of $\mathcal N=2$ electric-magnetic duality \cite{Seiberg:1994rs,Tachikawa:2013kta}. 
\end{proposition}

The group of transformations on $a_i$ and $a_D^i$ leaving the general form of the metric of \Cref{eq:specKahlmetric} invariant is derived in \cite[Sect. 3]{Seiberg:1994rs}. Let $X\simeq \mathbb{C}^{2r}$ with coordinates $a_i,a_D^i$ for $i=1,\dots,r$, endowed with a symplectic form $\omega$ and a holomorphic 2-form $\omega_h$:
\begin{equation*}
    \begin{split}
        \omega &= \frac{\I}{2}\left(\mathrm{d}a_i\wedge\mathrm{d}\overline{a}_D^i-\mathrm{d}a_D^i\wedge\mathrm{d}a_i\right)\; , \\
    \omega_h &= \mathrm{d}a_i\wedge\mathrm{d}a_D^i\; .
    \end{split}
\end{equation*}
Let also $u_1,\dots,u_r$ be a set of $r$ arbitrary local holomorphic coordinates on $\mathcal C$. Therefore, locally a holomorphic map $f\colon \mathcal C\rightarrow X$ is described by $2r$ functions $a_i(u)$ and $a_D^i(u)$. Fix such an $f$, and endow $\mathcal C$ with the metric whose Kähler form is $f^*(\omega)$:
\begin{equation*}
    \mathrm{d}s^2 = \Im \frac{\partial a_D^i}{\partial u_k}\frac{\partial \overline{a}_i}{\partial \overline{u}_l} \mathrm{d}u_k\mathrm{d}\overline{u}_l\; .
\end{equation*}
One moreover requires that $f^*(\omega_h)=0$, so that when for all $i$, $~u_i = a_i$, the resulting metric is of the form of \Cref{eq:specKahlmetric}. Writing $v~=~(a_1,\dots,a_r,a_D^1,\dots,a_D^r)^T$, the metric on $\mathcal C$ keeps its special form under all transformations 
\begin{equation}\labelx{eq:generaldualityN=2}
    v \longmapsto Mv+c\; ,
\end{equation}
where $M\in\mathrm{Sp}_{2r}(\mathbb{R})$ and where $c$ is a constant $2r$-vector.

\subsection{\texorpdfstring{$\mathcal N=2$}{N=2} central charge, BPS states, stability and wall-crossing}

In pure $4d$ $\mathcal N=2$ SYM with gauge group $G$ of rank $r$, at a generic point of the Coulomb branch $\mathcal C$, low-energy states are characterized by their electromagnetic charges $(n_e^i,n_{m,i})$, where $i=1,\dots,r$. The central charge $\mathcal{Z}$, which is invariant under renormalization, is given by \cite{Witten:1978mh,Seiberg:1994rs,Seiberg:1994aj}:
\begin{equation*}
    Z = n_e^ia_i+n_{m,i} a_D^i\; ,
\end{equation*}
where $(a_i,a_D^i)$ are the special coordinates on the Coulomb branch $\mathcal C$. If the UV theory also contains matter hypermultiplets, low-energy states can also carry $\mathrm{U}(1)$ flavor charges. These flavor charges contribute additional terms to the central charge $\mathcal{Z}$, cf. \Cref{eq:centralchargeSU2F=1,eq:centralchargegeneralcase}.

\begin{proposition}\labelx{prop:BPSprotectedagainstdecay}
    BPS states are generically protected against decay.
\end{proposition}

\begin{proof}
    Consider a BPS state with central charge $Z_1+Z_2$; its mass reads $m = \sqrt{2}\vert Z_1+Z_2\vert$. Generically, the vectors $Z_1$ and $Z_2$ are not aligned in the complex plane, hence: 
    \begin{equation}\labelx{eq:stabilityBPS}
        m_1+m_2 \geq \sqrt{2}\left(\vert Z_1\vert+\vert Z_2\vert\right) > \sqrt{2} \vert Z_1+Z_2\vert = m\; ,
    \end{equation}
    and the decay of this state into its constituents of respective central charges $Z_1$ and $Z_2$ and masses $m_1$ and $m_2$ is energetically unfavorable.
\end{proof}

\begin{remark}
    When $Z_1$ and $Z_2$ are aligned the strict inequality of \Cref{eq:stabilityBPS} becomes an equality. Then, a BPS state of central charge $Z_1+Z_2$ is only \emph{marginally stable} against decay into its BPS constituents of central charge $Z_1$ and $Z_2$. 
\end{remark}

\begin{definition}
    The codimension-1 submanifolds in $\mathcal C$ defined by imposing that $Z_1$ and $Z_2$ are aligned are called \emph{walls of marginal stability}\index{wall of marginal stability}.
\end{definition}

Away from the wall of marginal stability, a BPS state of central charge $Z_1+Z_2$ is protected against decay. However, the closer one is to the wall where $Z_1$ and $Z_2$ are aligned, the smaller is the energy gap $\vert Z_1\vert + \vert Z_2\vert -\vert Z_1-Z_2\vert$. On the wall, the gap closes completely and the BPS state can decay. On the other side of the wall, the BPS state would be stable again, however it might not exist anymore in this region of the moduli space. 

\begin{definition}
    This phenomenon is called \emph{wall-crossing}\index{wall-crossing!phenomenon}.
\end{definition}
\chapter{Seiberg--Witten theory}\labelx{Chap:SWtheory}

\abstract{Seiberg–Witten theory yields exact solutions for the quantum dynamics on the Coulomb branch of $\mathcal N=2$ supersymmetric gauge theories and uncovers the phenomenon of \emph{wall-crossing}\index{wall-crossing!phenomenon}. These developments enable a consistent realization of $\mathcal N=2$ S-duality, even in situations where the BPS spectrum appears to lack the necessary symmetries. We begin with the Seiberg–Witten analysis of four-dimensional $\mathcal N=2$ $\mathfrak{su}(2)$ SYM theory, both in the pure case and with up to four fundamental hypermultiplets, before outlining its generalizations to other $\mathcal N=2$ gauge theories. Our discussion follows the original works \cite{Seiberg:1994rs,Seiberg:1994aj,Donagi:1995cf} and key reviews \cite{Alvarez-Gaume:1996ohl,Lerche:1997sm,DHoker:1999yni,Tachikawa:2013kta}, while more mathematically oriented perspectives are provided in \cite{Witten:1999Dynamics,Freed:1999special}.}

\section{Pure \texorpdfstring{$\mathfrak{su}(2)$}{su(2)} super-Yang--Mills theories}\labelx{Sec:pure4dN=2SYM}

The Lagrangian density of pure $4d$ $\mathcal N=2$ $\mathfrak{su}(2)$ SYM is of the form of \Cref{Eq:N=2gauge}. By \Cref{prop:N=2holomandphysicalgaugecouplings} one can safely rescale the $\mathcal N=2$ vector superfield by $1/2g$ in \Cref{Eq:N=2gauge}, yielding:
\begin{equation}\labelx{eq:pureN=2SU2Lag}
    \begin{split}
        \mathcal L &= \frac{1}{4\pi}\Im\left(\widetilde{\tau} \int\mathrm{d}^2\theta~\tr W^\alpha W_\alpha\right)+\frac{1}{g^2}\tr\int\mathrm{d}^2\theta\mathrm{d}^2\overline{\theta}~\overline{\Phi}\E^{[V,\cdot]}\Phi \\
        &= \frac{1}{g^2}\tr\left[ -\frac{1}{2}F_{\mu\nu}F^{\mu\nu} - 2\I\lambda\sigma^\mu D_\mu\overline{\lambda} - 2\I\widetilde{\lambda}\sigma^\mu D_\mu\overline{\widetilde{\lambda}} + \frac{\Theta g^2}{16\pi^2}F_{\mu\nu}(\star F)^{\mu\nu} \right. \\
        &\left.+ 2\overline{D_\mu\phi}D^\mu\phi  +D^2+ 2\overline{F}F + 2\I\sqrt{2}g\overline{\phi} \{\lambda,\widetilde{\lambda}\} - 2\I\sqrt{2}g\{\overline{\widetilde \lambda},\overline{\lambda}\}\phi + 2gD[\phi,\overline\phi]\right]\; ,
    \end{split}
\end{equation}
where the complexified gauge coupling is now denoted by $\widetilde{\tau}$. It runs logarithmically with the scale:
\begin{equation}\labelx{eq:runningUVcoupling}
    \widetilde{\tau}(\mu) = - \frac{b_1}{2\pi \I}\log(\frac{\mu}{\Lambda})-\frac{\I}{\pi}\sum_{k} c_k \left(\frac{\Lambda}{\mu}\right)^{4k}\; ,
\end{equation}
with $b_1 = 2N-F = 4$ as in \Cref{ex:b1N=2SYM}, $\Lambda$ the \emph{dynamical scale}\index{dynamical scale} of \Cref{eq:N=2dynamicalscale}, and where the general form of non-perturbative corrections is constrained by \emph{holomorphy}\index{holomorphy}. In the sum of \Cref{eq:runningUVcoupling}, each $c_k$ is a complex number that encodes the contribution of gauge configurations with instanton number $k$ to the running of $\widetilde \tau$. Since $b_1>0$, pure $4d$ $\mathcal N=2$ $\mathfrak{su}(2)$ SYM is asymptotically free.

\begin{proposition}
    The classical moduli space of pure $4d$ $\mathcal N=2$ $\mathfrak{su}(2)$ SYM coincides with its classical Coulomb branch, defined by the condition $[\langle\phi^\dagger\rangle,\langle\phi\rangle]=0$. 
\end{proposition}
Every solution is gauge equivalent to $\langle\phi\rangle = \alpha\sigma_3$, where $\alpha$ is a complex number. This vacuum expectation value (VEV) spontaneously breaks the gauge group down to $\mathrm{U}(1)\times\mathbb{Z}_2$, where $\mathbb{Z}_2$ is the Weyl group of $\mathfrak{su}(2)$. 

By \Cref{prop:quantumCoulombnotlifted}, the theory possesses a quantum Coulomb branch $\mathcal C$, which forms the entire quantum moduli space. The Coulomb branch $\mathcal C$ can still be parameterized by the complex parameter $\alpha$, up to gauge equivalence. Specifically, the remaining discrete gauge symmetry $\mathbb{Z}_2$ acts as $\alpha\mapsto -\alpha$. Therefore:
\begin{proposition}
    A suitable gauge-invariant coordinate on $\mathcal C$ is $u=\langle\tr\phi^2\rangle/2 = \alpha^2$. 
\end{proposition}
At energy scales below the mass of the massive excitations charged under $\mathrm{U}(1)$, these excitations decouple from the theory. For instance, the $W^\pm$-bosons acquire a mass proportional to $\alpha$ due to this spontaneous symmetry breaking, causing them to decouple at energies lower than $\vert\alpha\vert$. As a result, at very low energies, the effective theory reduces to pure $\mathcal N=2$ Maxwell theory, which is free. However, at non-generic points on the Coulomb branch, there can be massless excitations that are charged under the low-energy gauge group $\mathrm{U}(1)$. These excitations then appear in the low-energy effective theory.

The Coulomb branch $\mathcal C$ is a \emph{special Kähler manifold}\index{special!Kähler manifold}. Generically, the low-energy effective theory on $\mathcal C$ is an $\mathcal N=2$ sigma model of a single $\mathcal N=2$ vector superfield $A$ into $\mathcal C$. The VEVs $a$ of its scalar component is a local holomorphic coordinate on $\mathcal C$, i.e. a local holomorphic function of $u$. Schematically, the bosonic part of the low-energy effective Lagrangian reads: 
\begin{equation*}
    \mathcal L_\mathrm{eff} = \Im(\tau(u))\left(\partial_\mu \overline{a} \partial^\mu a + F_\mathrm{\mu\nu}^{\mathrm{U}(1)}F^{\mu\nu}_{\mathrm{U}(1)}\right)+\Re(\tau(u)) F_{\mu\nu}^{\mathrm{U}(1)}(\star F^{\mu\nu}_{\mathrm{U}(1)})\; ,
\end{equation*}
where $\tau(u)$ is the low-energy abelian complexified gauge coupling at $u\in\mathcal C$. This Lagrangian is fully determined by the \emph{prepotential}\index{prepotential} $\mathcal F$, which is a local holomorphic function of $A$. In particular: 
\begin{equation*}
    \tau = \frac{\partial^2 \mathcal F}{\partial a^2} = \frac{\partial a_D}{\partial a}\; ,
\end{equation*}
where $a_D$ is the magnetic dual coordinate to $a$, defined in \Cref{eq:LEdualvar}.

\begin{figure}[!ht]
    \centering
    \includegraphics[scale=0.83]{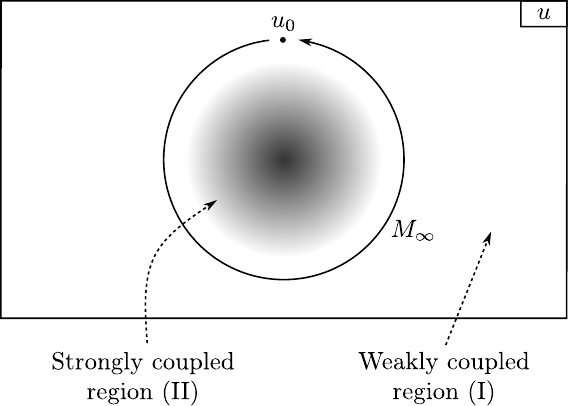}
    \caption{Structure of the Coulomb branch of pure $\mathcal N=2$ $\mathrm{SU}(2)$ SYM.}
    \labelx{fig:CoulombbranchquantumSW}
\end{figure}

Recalling the important caveat discussed at the end of \Cref{par:importantcaveat}, the running of the UV gauge coupling $\widetilde \tau$, given in \Cref{eq:runningUVcoupling}, is reliable only when $\widetilde \tau$ has a large imaginary part, or equivalently when the gauge coupling $g^2$ is small. This condition is satisfied at energy scales much larger than $\vert\Lambda\vert$. Consequently, two distinct regions in the Coulomb branch $\mathcal C$ can be identified, as illustrated in \Cref{fig:CoulombbranchquantumSW}. Precisely:

\begin{itemize}
    \item[(I)]\; The \emph{weakly-coupled region} is the region where $\vert u \vert \gg\vert\Lambda^2\vert$. In this region, the gauge symmetry breaking at the scale $\vert\alpha\vert$ occurs before the UV non-abelian gauge dynamics enters the strongly-coupled regime. Therefore, the behavior of the UV theory in this region can be accurately described by perturbation theory from the UV down to the scale $\vert\alpha\vert$. Below this scale, the effective theory becomes essentially free, and the running of the gauge coupling can be depicted as shown in \Cref{Fig:runningSWSU2}.

    \begin{figure}[!ht]
        \centering
        \includegraphics[scale=0.83]{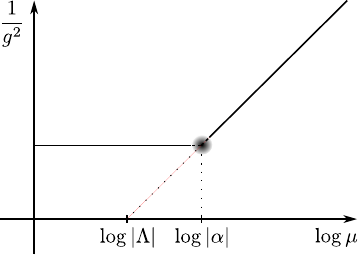}
        \caption{Running of the gauge coupling in the weakly-coupled region (I).}\labelx{Fig:runningSWSU2}
    \end{figure}
    
    \item[(II)]\; The \emph{strongly-coupled region} is the region where $\vert u\vert \lesssim \vert\Lambda^2\vert$. In this regime, the UV non-abelian gauge dynamics becomes strongly coupled before the gauge symmetry is broken. As a result, the running of the UV coupling cannot be described by the perturbative expression in \Cref{eq:runningUVcoupling} from the UV down to the scale $\vert\alpha\vert$.
\end{itemize}
We will analyze regions (I) and (II) in turn.

\subsection{Weakly-coupled region}

Asymptotically at infinity in the quantum Coulomb branch $\mathcal C$, the quantum corrections become small, and the branch approaches the classical moduli space, where $a=\alpha$. Therefore, in the weakly-coupled region of the Coulomb branch, we have the relation:
\begin{equation*}
    a = \sqrt{u} + \dots\; ,
\end{equation*}
where the dots represent quantum corrections. Classically, the field strength for the unbroken $\mathrm{U}(1)$ gauge symmetry can be expressed as:
\begin{equation*}
    F_{\mu\nu} = \begin{pmatrix}
        F_{\mu\nu}^{\mathrm{U}(1)}  & 0 \\
        0 & -F_{\mu\nu}^{\mathrm{U}(1)}
    \end{pmatrix} = F_{\mu\nu}^{\mathrm{U}(1)} \sigma^3\; ,
\end{equation*}
which serves as our definition of the low-energy $F_{\mu\nu}^{\mathrm{U}(1)}$\footnote{Here, we assume the UV gauge group is $G=\mathrm{SU}(2)$. The discussion of alternative global variants such as $\mathrm{SO}(3)_+$ and $\mathrm{SO}(3)_-$ will be addressed later in this section.}. Given this, we have the following relation:
\begin{equation*}
    \frac{1}{2g^2} \tr(F_{\mu\nu}F^{\mu\nu}) = \frac{1}{g^2} F_{\mu\nu}^{\mathrm{U}(1)}F^{\mu\nu}_{\mathrm{U}(1)}\; ,
\end{equation*}
which shows that $g^2 = 4e^2$, where $e$ is the low-energy abelian gauge coupling. Similarly, the topological $\Theta$-term becomes:
\begin{equation*}
    \frac{\Theta}{16\pi^2}\tr F_{\mu\nu}(\star F)^{\mu\nu} = \frac{\Theta}{8\pi^2} F_{\mu\nu}^{\mathrm{U}(1)}\left(\star F^{\mathrm{U}(1)}\right)^{\mu\nu}\; ,
\end{equation*}
from which we deduce that the $\mathrm{U}(1)$ $\Theta$-angle is $\Theta^{\mathrm{U}(1)}=2\Theta$. 

\begin{proposition}
    In terms of complexified couplings, this gives:
    \begin{equation*}
        \tau := \frac{2\pi \I}{e^2}+\frac{\Theta^{\mathrm{U}(1)}}{2\pi} = 2\left(\frac{4\pi \I}{g^2}+\frac{\Theta}{2\pi}\right) = 2\widetilde\tau\; ,
    \end{equation*}
    where $\tau$ is the complexified low-energy abelian gauge coupling, and $\widetilde{\tau}$ is the UV coupling. Since the UV gauge coupling stops running below the scale $\vert \alpha\vert$, the low-energy abelian coupling $\tau(\alpha)$  in the extreme infrared satisfies:
    \begin{equation}\labelx{Eq:renormpureSU(2)andU(1)}
        \tau(\alpha) \approx 2\widetilde{\tau}(\alpha) = - \frac{b_1}{\pi \I}\log(\frac{\alpha}{\Lambda})-\frac{2\I}{\pi}\sum_{k} c_k \left(\frac{\Lambda}{\alpha}\right)^{4k}\; ,
    \end{equation}
    where the non-perturbative instanton corrections become negligible as $\vert\alpha\vert$ (or equivalently, $\vert u\vert$) increases.
\end{proposition}

In the weakly-coupled region (I), the perturbative one-loop renormalization of the UV gauge coupling is a reliable approximation. By substituting $\alpha$ with $a$ in \Cref{Eq:renormpureSU(2)andU(1)} and integrating once with respect to $a$, we obtain the following expression for $a_D$:
\begin{equation*}
    a_D = -\frac{a b_1}{\pi \I}\log(\frac{a}{\Lambda})+\frac{b_1a}{\pi \I}+\dots\; .
\end{equation*}

\begin{proposition}
    Since $u\simeq a^2$, starting at a point $u_0$ in the weakly-coupled region (I) and performing a full loop around region (II) without leaving region (I), as sketched in \Cref{fig:CoulombbranchquantumSW}, induces the following transformation on the special coordinates $(a,a_D)$:
\begin{equation*}
    \begin{pmatrix}
        a \\ a_D
    \end{pmatrix} \longmapsto \begin{pmatrix}
        -a  \\ -a_D + 4a
    \end{pmatrix}\; .
\end{equation*}
This transformation can be written in matrix form as:
\begin{equation*}
    \begin{pmatrix}
        a & a_D
    \end{pmatrix} \longmapsto   \begin{pmatrix}
        a & a_D
    \end{pmatrix}\begin{pmatrix} -1 & 4 \\ 0 & -1 \end{pmatrix} = \begin{pmatrix}
        -a & 4a - a_D
    \end{pmatrix}\; .
\end{equation*}
\end{proposition}

Repeating this process multiple times yields an infinite number of distinct, yet physically equivalent, descriptions at $u_0\in\mathcal C$. Importantly, all intrinsic features of the theory must remain invariant under this transformation. In particular, the mass spectrum of particles charged under the low-energy $\mathrm{U}(1)$ gauge group must be preserved. 

The BPS mass formula on the Coulomb branch of $4d$ $\mathcal N=2$ SYM theories of rank 1 is given by:
\begin{equation}\labelx{Eq:BPSmass}
    M = \sqrt{2}\vert n_ea+n_ma_D \vert = \sqrt{2}\left\vert \begin{pmatrix}
        a & a_D
    \end{pmatrix}\begin{pmatrix}
        n_e \\ n_m
    \end{pmatrix} \right\vert\; .
\end{equation}
For the spectrum to remain unchanged under this transformation, the low-energy charges $(n_e,n_m)$ must also transform accordingly:
\begin{equation*}
    \begin{pmatrix}
        n_e \\ n_m
    \end{pmatrix} \longmapsto \begin{pmatrix}
         -1 & -4 \\
           0 & -1
    \end{pmatrix}   
       \begin{pmatrix}
        n_e \\ n_m
    \end{pmatrix}=\begin{pmatrix}
        -n_e-4n_m \\ -n_m
    \end{pmatrix}\; .
\end{equation*}
The matrix:
\begin{equation*}
    M_\infty = \begin{pmatrix}
           -1 & 4 \\
           0 & -1
       \end{pmatrix} \in\mathrm{SL}_2(\mathbb{Z}) \simeq \mathrm{Sp}_2(\mathbb{Z})\; ,
\end{equation*} 
is called \emph{monodromy at infinity}. The fact that $M_\infty\in\mathrm{Sp}_2(\mathbb{Z})$  ensures that the matrix preserves the antisymmetric \emph{Dirac--Schwinger--Zwanziger (DSZ) pairing}\index{DSZ!pairing}.

Thus, even with just the knowledge of the perturbative one-loop running of the UV coupling, we can determine that the electric-magnetic charges in the low-energy theory are not globally defined. Instead, they undergo monodromy when encircling infinity in the compactified Coulomb branch $\overline{\mathcal C}\simeq \mathrm{CP}^1$. Mathematically, this means that the lattice of low-energy charges forms a \emph{local system}\index{local system}\footnote{A local system on a topological space $X$ is a locally constant sheaf. More geometrically, it can be described as a bundle with a flat connection or a representation of the fundamental group $\pi_1(X)$---this is the Riemann--Hilbert correspondence discussed in \Cref{Sec:bundlesandconnections}. For lattices, a local system is a fibration of lattices.} $\Gamma(u)$ endowed with the DSZ pairing. 

At each point $u\in\overline{\mathcal{C}}$, except for a finite number of singularities, $\Gamma(u)$ is a rank-2 lattice of electric-magnetic charges $(n_e,n_m)$, equipped with the DSZ pairing. This lattice undergoes non-trivial monodromy when encircling the singular points, one of which is located at infinity with monodromy $M_\infty$. Additionally, $\Gamma(u)$ varies holomorphically with $u$, as implied by the holomorphicity of $a$ and $a_D$, and by the BPS mass formula. The compactified Coulomb branch is illustrated schematically in \Cref{fig:Coulombbranchcompactified}.

\begin{figure}[!ht]
    \centering
    \includegraphics[scale=0.83]{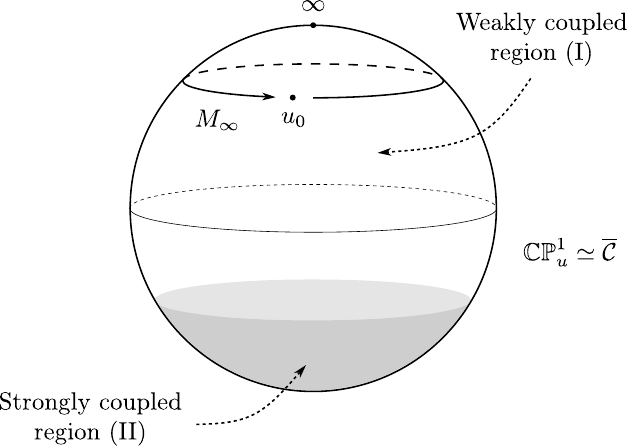}
    \caption{Structure of the Coulomb branch of pure $\mathcal N=2$ $\mathrm{SU}(2)$ SYM.}
    \labelx{fig:Coulombbranchcompactified}
\end{figure}    

\subsection{Strongly-coupled region}

In this region, one can neither rely on \Cref{Eq:renormpureSU(2)andU(1)} nor easily relate the natural coordinate $u$ on the Coulomb branch—defined in terms of the UV adjoint scalar VEVs—to the natural low-energy variable $a$. However, whatever occurs in the strongly-coupled region must be consistent with our findings in the weakly-coupled region, which imposes significant constraints.

\begin{proposition}\labelx{prop:RsymmetrySU2}
    The Coulomb branch is $\mathbb{Z}_2$-symmetric.
\end{proposition}

\begin{proof}
    Pure $\mathcal N=2$ $\mathrm{SU}(2)$ SYM has a $(\mathbb{Z}_8)_R$ discrete $R$-symmetry, cf. \Cref{sec:AnomaliesinN=2theories}, under which the adjoint scalar $\phi$ has charge $2$. Taking into account the discrete $\mathbb{Z}_2$ gauge (Weyl) symmetry, $(\mathbb{Z}_8)_R$ induces a $\mathbb{Z}_2$ symmetry on the Coulomb branch:
    \begin{equation*}
        u\mapsto -u\; .
    \end{equation*}
\end{proof}

Note that $\Im\tau(a)$ must be positive and single-valued on $\mathcal C$ in order for 
\begin{equation}\labelx{eq:metricCoulomb}
        \mathrm{d}s^2 = \Im \tau(a)~ \mathrm{d}a\mathrm{d}\overline{a}\; ,
\end{equation}
to be a genuine metric. Since $\tau$ is a holomorphic function of $a$, its imaginary part $\Im \tau$ is harmonic. Besides, it must be single-valued otherwise \Cref{eq:metricCoulomb} would not make sense. Harmonicity thus implies that $\Im \tau$ cannot be globally-defined, or it would be unbounded. Moreover, one has the following:

\begin{proposition}\labelx{prop:atleasttwosing}
    The structure of the classical Coulomb branch $\mathcal{C}_\mathrm{cl}$, featuring a single singularity at the origin of $\mathcal C$, cannot be the correct structure for the genuine Coulomb branch $\mathcal C$.
\end{proposition}

\begin{proof}
    We know from the perturbative renormalization of the UV gauge coupling that there is a non-trivial monodromy $M_\infty$ at infinity for the local system of charge lattices $\Gamma(u)$. This implies that the local system of charges cannot be trivial in $\mathcal C$, as $\pi_1(\mathcal C)=\pi_1(\overline{\mathcal C}-\{\infty\})=\{\mathrm{id}\}$. Among the various topological complications that can arise, perhaps the simplest solution is to assume the presence of additional singularities in $\mathcal C$, around which the local system $\Gamma(u)$ undergoes monodromy. This solution presents a very compelling picture, strongly suggesting its correctness. Let us assume that there are additional singularities in $\mathcal C$ at some points $u_1,\dots,u_n$, and let $M_1,\dots,M_n\in\mathrm{SL}_2(\mathbb{Z})$ be the monodromies around them, given a choice of base point $u_0\in\mathcal C$ and loops based at $u_0$ circling each singularity $u_i$ counterclockwise. The triviality of $\pi_1(\overline{\mathcal C})$ implies that 
    \begin{equation*}
        M_\infty^{-1} = \prod_{i=1}^n M_i\; .
    \end{equation*}
    
    If there were only one such additional singularity (as is the case in $\mathcal{C}_\mathrm{cl}$, where the $W$-bosons become massless and where the full $\mathrm{SU}(2)$ gauge symmetry is restored) it would be at the origin because of \Cref{prop:RsymmetrySU2}. The low-energy coupling $\tau$, as a function of the VEV $a$, would therefore be of the form
    \begin{equation*}
        \tau(a) = -\frac{1}{\pi \I}\log\frac{a}{\Lambda}+f(a)\; ,
    \end{equation*}
    with $f$ a meromorphic function of $a$ with a single pole at the origin. This in turn implies that $\Im\tau$ is unbounded and therefore not positive definite.
\end{proof}

The simplest singularity structure for $\mathcal C$ therefore consists of two singularities positioned at opposite values, $\pm u_1$. Since $\Lambda$ is the only scale in the theory, it is reasonable to expect $u_1\propto \Lambda^2$. It turns out that assuming the existence of two singularities at $\pm2\Lambda^2$ results in a very compelling framework. The proposed structure of $\mathcal C$ is illustrated in \Cref{fig:CoulombbranchquantumSW2}, where $M_1$ and $M_2$ have been renamed as $M_\pm$.

\begin{figure}[!ht]
    \centering
    \includegraphics[scale=0.83]{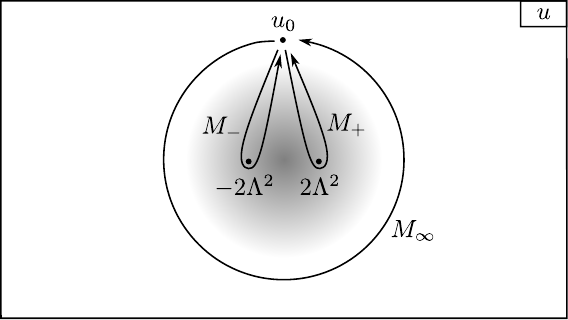}
    \caption{Coulomb branch of pure $\mathcal N=2$ $\mathrm{SU}(2)$ SYM.}
    \labelx{fig:CoulombbranchquantumSW2}
\end{figure}

Let $M_+$ and $M_-$ be the matrices encoding the monodromy of $\begin{pmatrix}
    a & a_D
\end{pmatrix}$, or equivalently of the local system of charges, at $2\Lambda^2$ and $-\pm2\Lambda^2$, respectively. These matrices must satisfy
\begin{equation*}
    M_\infty = M_+M_-\; .
\end{equation*}
Moreover, as $2\Lambda^2$ and $-\pm2\Lambda^2$ are exchanged by the $\mathbb{Z}_2$ symmetry, $M_+$ and $M_-$ must be conjugate in $\mathrm{SL}_2(\mathbb{Z})$. One can explicitly check the following:
\begin{proposition}
    The matrices:
    \begin{equation}\labelx{eq:monodstronglycoupled}
        M_+ = \begin{pmatrix}
            1 & 0 \\
            -1 & 1
        \end{pmatrix} \quad \text{and} \quad M_- = \begin{pmatrix}
            -1 & 4 \\
            -1 & 3
        \end{pmatrix}\; ,
    \end{equation}
    solve both of these constraints. 
\end{proposition}

A natural interpretation of singularities in a moduli space of supersymmetric vacua is that they signal a breakdown of the low-energy description due to the appearance of additional massless particles, as discussed in the context of $\mathcal N=1$ theories in \Cref{Chapter:N=1theories}. For example, the singularity at the origin of the classical Coulomb branch $\mathcal C_\mathrm{cl}$ arises because the $W^\pm$-bosons become massless at this point. However, the proposed scenario, in which the singularities occur at non-zero VEVs $\langle \tr \phi^2\rangle\propto \Lambda^2$ conflicts with the idea that these singularities are associated with massless gauge bosons \cite{Seiberg:1994rs}. Instead, it suggests that the two singularities on the quantum Coulomb branch $\mathcal C$ are due to BPS hypermultiplets becoming massless. 

Let $M_{(n_e,n_m)}\in\mathrm{SL}_2(\mathbb{Z})$ be the monodromy induced by low-energy BPS hypermultiplet with electric-magnetic charges $(n_e,n_m)$ becoming massless. This monodromy must leave $(n_e,n_m)$ invariant. 

The BPS mass formula $M = \sqrt{2}\vert n_ea+n_ma_D \vert$ indicates that a magnetic monopole becomes massless at points $u_s\in\mathcal C$ where $a_D=0$. Since magnetic monopoles couple in a non-local way to the photon field~$A$, there cannot be any effective low-energy Lagrangian containing both $A$ and a monopole field. However, monopoles couple locally to $A_D$. Thus, one can perform a duality transformation of the form of \Cref{eq:generaldualityN=2} and express the low-energy Lagrangian in terms of the magnetic dual field $A_D$ and the monopole field. 

Note that the low-energy coupling $e_D$ goes to zero as one approaches $u_s$, while $e$ diverges there. Equivalently, $\Im\tau_D \rightarrow \infty$ as $a_D\rightarrow 0$. Therefore, the low-energy theory close to $u_s$ is $\mathcal N=2$ Maxwell theory, however with gauge field $A_D$, and with light matter field the monopole. This indicates that $a_D$ is better suited than $a$ as a local coordinate on $\mathcal C$ close to $u_s$. Moreover, close to $u_s$ one can write:
\begin{equation*}
    \tau_D \simeq -\frac{\I}{\pi}\ln a_D\; ,
\end{equation*}
where:
\begin{equation*}
    a_D = c_0(u-u_s)\; .
\end{equation*}
Since $\tau_D = -\mathrm{d}a/\mathrm{d}a_D$, this leads to the following expression of $a(u)$:
\begin{equation*}
    a(u) \simeq a_0 + \frac{\I}{2\pi}c_0(u-u_s)\ln(u-u_s)\; ,
\end{equation*}
for some constant $a_0$ which cannot be zero, otherwise electric particles would also be massless at $u_s$, which would invalidate the low-energy Lagrangian description in terms of $A_D$ and monopoles. As $u$ circles $u_s$ counterclockwise, the special coordinates $(a,a_D)$ undergo the monodromy:
\begin{equation*}
    \begin{pmatrix}
        a & a_D
    \end{pmatrix} \longmapsto \begin{pmatrix}
        a - a_D & a_D
    \end{pmatrix} = \begin{pmatrix}
        a & a_D
    \end{pmatrix}M_+\; ,
\end{equation*}
where $M_+$ is defined in \Cref{eq:monodstronglycoupled}. Thus: 
\begin{proposition}
    The monodromy $M_+$ is induced by monopoles becoming massless at $2\Lambda^2$.
\end{proposition}

Consider now a low-energy BPS dyon with charges $(2,1)$. Its mass vanishes at points $u_s\in\mathcal C$ where $2a+a_D=0$. Close to such a point, $2a+a_D$ is a good coordinate on $\mathcal C$:
\begin{equation*}
    2a+a_D = c_0'(u-u_s)\; ,
\end{equation*}
and $\Im\tau_- = \Im (2\tau+1)/(\tau+1)\rightarrow \infty$ as $2a+a_D\rightarrow 0$. Since $\tau_- = \mathrm{d}(a+a_D)/\mathrm{d}(2a+a_D)$, one deduces:
\begin{equation*}
    (a+a_D)(u) \simeq b_0 - \frac{\I}{2\pi}c_0'(u-u_s)\ln(u-u_s)\; .
\end{equation*}
Thus, as one circles $u_s$ counterclockwise, the special coordinates $(a,a_D)$ undergo the monodromy 
\begin{equation*}
    \begin{pmatrix}
    a & a_D
\end{pmatrix} \longmapsto \begin{pmatrix}
    a & a_D
\end{pmatrix}M_- = \begin{pmatrix}
    -a-a_D & 4a+3a_D
\end{pmatrix}\; ,
\end{equation*}
where $M_-$ is defined in \Cref{eq:monodstronglycoupled}. Thus:
\begin{proposition}
    The monodromy $M_-$ is induced by dyons with charges $(2,1)$ becoming massless.
\end{proposition} 

Generalizing the above reasoning, one finds:

\begin{proposition}
    The monodromy of $\begin{pmatrix}
    a & a_D
\end{pmatrix}$ induced by a dyon of charges $(n_e,n_m)$ becoming massless at some point in $\mathcal C$ is:
\begin{equation*}
    M_{(n_e,n_m)} = \begin{pmatrix}
        1 - n_en_m & n_e^2 \\
        -n_m^2 & 1+n_en_m
    \end{pmatrix}\; .
\end{equation*}
\end{proposition}

\begin{remark}
    Note that the monodromies depend on the choice of loops based at $u_0$ that circle each singularity. For instance, if one circles the strongly coupled region (II) counterclockwise before and clockwise after encircling the singularity at $2\Lambda^2$, the monodromy appears as $M_\infty^{-1}M_+M_\infty$ instead of $M_+$. Since:
    \begin{equation*}
        M_\infty^{-1}M_+M_\infty = \begin{pmatrix}
        -3 & 16 \\ -1 & 5
        \end{pmatrix}\; ,
    \end{equation*}
    the singularity at $2\Lambda^2$ now seems to correspond to a dyon with charges $(4,1)$ becoming massless, rather than a magnetic monopole. Ultimately, more important than the specific monodromies $M_\infty,M_\pm$ is the global structure of the flat $\mathrm{SL}_2(\mathbb{Z})$ bundle, of which $(a,a_D)$ is a section.
\end{remark}

\subsection{Breaking \texorpdfstring{$\mathcal N=2$}{N=2} to \texorpdfstring{$\mathcal N=1$}{N=1}}

Adding a mass term $m\tr\Phi^2$ for the chiral superfield in the $\mathcal N=2$ vector superfield breaks one of the two supersymmetries of pure $\mathcal N=2$ $\mathrm{SU}(2)$ SYM, reducing the theory at low energies to pure $\mathcal N=1$ $\mathfrak{su}(2)$ SYM. As discussed in \Cref{Sec:quantSQCD}, pure $\mathcal N=1$ $\mathrm{SU}(2)$ SYM has a chiral $\mathbb{Z}_4$ $R$-symmetry and is believed to confine, developing a mass gap and exhibiting spontaneous symmetry breaking of the chiral $\mathbb{Z}_4$ $R$-symmetry down to $\mathbb{Z}_2$, leading to two distinct vacua. 

If we adopt the conjecture that the Coulomb branch $\mathcal C$ of pure $\mathcal N=2$ $\mathfrak{su}(2)$ SYM features two singularities at $\pm2\Lambda^2$, we can understand how the introduction of the $\mathcal N=1$-preserving mass term lifts the moduli space $\mathcal C$ but retains the two critical points corresponding to additional massless degrees of freedom \cite{Seiberg:1994rs}. At these points, the low-energy gauge field $A$ acquires a mass due to the Higgs mechanism, and the theory develops a mass gap. Moreover, the $\mathbb{Z}_4$ $R$-symmetry of pure $\mathcal N=2$ $\mathrm{SU}(2)$ indeed breaks spontaneously down to $\mathbb{Z}_2$, as the $\mathbb{Z}_2$ subgroup acting non-trivially on $\mathcal C$ exchanges $2\Lambda^2$ and $-2\Lambda^2$.

This reasoning offers a compelling verification of the conjectured singularity structure in the quantum Coulomb branch $\mathcal C$.

\subsection{Seiberg--Witten curve\index{Seiberg--Witten!curve}}

We would like to construct holomorphic functions $a$ and $a_D$ on $\mathcal C$ such that $(a,a_D)$ undergo monodromies $M_\pm$ at $\pm u_0$ and $M_\infty$ at infinity. This is a Riemann--Hilbert problem on $\mathcal C$ for a flat $\mathrm{SL}_2(\mathbb{Z})$ bundle. With the asymptotic conditions fixed by the semi-classical analysis of the weakly-coupled region of $\mathcal C$, the solution is unique. It turns out that $a$ and $a_D$ can be neatly interpreted in terms of period integrals \cite{Seiberg:1994rs}. 

\begin{proposition}
    The monodromies $M_{\pm}$ and $M_\infty$ generate the subgroup $\Gamma_0(4)$ of $\mathrm{PSL}_2(\mathbb{Z})$:
    \begin{equation*}
        \Gamma_0(4) = \left\{\left.\begin{pmatrix}
        a & b \\
        c & d
        \end{pmatrix} \in\mathrm{SL}_2(\mathbb{Z})~\right\vert~b\equiv 0~[4]\right\}\; ,
    \end{equation*}
    which classifies complex tori along with a cyclic subgroup of order 4. The quotient of the upper half-plane $\H$ by $\Gamma_0(4)$\footnote{This quotient is called classical modular curve, and denoted $X_0(4)$.} has the same singularity structure as the one suggested for $\mathcal C$.
\end{proposition}

This suggests constructing $a$ and $a_D$ out of the family of complex tori over $\mathcal C$ given by the tautological fibration over the moduli space $\mathbb{H}/\Gamma_0(4)$. Consider the following equation:
\begin{equation}\labelx{eq:SWpureSU2}
    (\Sigma_u)\colon \quad \Lambda^2 z + \frac{\Lambda^2}{z} = x^2 - u\; ,
\end{equation}
where $z\in\mathbb{C}^\times$ and $x\in\mathbb{C}$. This defines a complex surface in $\mathbb{C}^\times_z\times\mathbb{C}_x\times\mathbb{C}_u$ which can be compactified by homogenizing the equation, allowing in particular $z\in\mathbb{CP}^1$. The family has an obvious $\mathbb{Z}_4$ symmetry under which $(u,x,z)\mapsto (-u,ix,-z)$, which projects to the $\mathbb{Z}_2$ symmetry $u\mapsto -u$ on $\mathcal C$.

For all generic $u$, \Cref{eq:SWpureSU2} defines an elliptic curve $\Sigma_u\subset \mathbb{CP}^1_z\times\mathbb{C}_x$, which is a degree-2 ramified covering of $\mathbb{CP}^1_z$, with four ramification points of order 2 at $0,\infty$, and $z_\pm$, where:
\begin{equation}\labelx{eq:ramificationpoints}
     z_\pm = \frac{-u\pm\sqrt{u^2-4\Lambda^4}}{2\Lambda^2}\; .
\end{equation}

Recall that the metric on the Coulomb branch is given by:
\begin{equation*}
    \mathrm{d}s^2 = \Im(\tau)\mathrm{d}u\mathrm{d}\overline{u}\; ,
\end{equation*}
where
\begin{equation}\labelx{Eq:couplinginu}
    \tau(u) = \frac{\mathrm{d} a_D}{\mathrm{d} a} =\frac{\mathrm{d}a_D/\mathrm{d}u}{\mathrm{d}a/\mathrm{d}u} 
\end{equation}
must have a positive-definite imaginary part. 

There is a canonical way to construct such $\tau(u)$ on $\mathcal C$, from the family of elliptic curves defined in \Cref{eq:SWpureSU2}: for each $u$, one lets $\tau(u)\in\mathbb{H}$ be the complex modulus of $\Sigma_u$. Concretely, each $\Sigma_u$ admits a unique (up to multiplicative constant) holomorphic differential $\omega_u = \mathrm{d}z/xz$. Let $A$ and $B$ be a symplectic basis for $\mathrm{H}_1(\Sigma_u,\mathbb{Z})$ (i.e. one requires $A\cdot B=1$, where $\cdot$ is the intersection pairing) and let:
\begin{equation*}
    \omega_A = \oint_A \omega\; , \quad \omega_B = \oint_B \omega\; .
\end{equation*}
The ratio $\omega_B/\omega_A$ is the complex modulus of $\Sigma_u$, which lives in $\mathbb{H}$ and therefore has a positive imaginary part. This is a specific case of the \emph{Riemann bilinear relations}\index{Riemann!bilinear relations}. 

\begin{definition}
    Let $\lambda$ be a meromorphic family of differentials, with $\lambda_u$ a meromorphic differential on $\Sigma_u$ with vanishing residues, such that $\mathrm{d}\lambda/\mathrm{d}u = \omega_u$. Then, letting:
\begin{equation}\labelx{eq:periodspureSU2}
    a = \oint_A \lambda \nonumber\; , \quad a_D = \oint_B \lambda\; ,
\end{equation}
one can express $\tau(u)$ as in \Cref{Eq:couplinginu}.
\end{definition} 
Therefore, the periods of \Cref{eq:periodspureSU2} are valid candidates for the special coordinates $a$ and $a_D$ on $\mathcal C$. In fact, $\lambda$ is fully determined, modulo exact forms, by the asymptotic behavior of $a$ and $a_D$:
\begin{equation*}
    \lambda = x\frac{\mathrm{d}z}{z}\; .
\end{equation*}
This differential satisfies $\mathrm{d}\lambda/\mathrm{d}u = \omega_u$. We refer to \cite{Seiberg:1994rs,Seiberg:1994aj,Tachikawa:2013kta} for more explicit checks showing that the periods $a$ and $a_D$ defined in \Cref{eq:periodspureSU2} can indeed be identified with the special coordinates $a$ and $a_D$ on $\mathcal C$, which we henceforth assume. 

\begin{remark}
    This identification notably implies that electrically-charged particles never become massless on $\mathcal C$, as it can be computed that $a$ never vanishes, especially not at $u=0$. Consequently, the classical singularity at the origin of $\mathcal C_\mathrm{cl}$ actually no longer exists in the quantum Coulomb branch $C$, consistently with the analysis of the strongly-coupled region above and its conjectural singularity structure.
\end{remark}

The fibration in \Cref{eq:SWpureSU2} clearly shows that the points $\pm 2\Lambda^2$ are singular, as the ramification points $z_+$ and $z_-$ of \Cref{eq:ramificationpoints} collide there. At these singularities, the monodromy of the 1-cycles $A$ and $B$, and consequently of $a$ and $a_D$, is given by the \emph{Picard--Lefschetz monodromy formula}\index{Picard--Lefschetz monodromy formula}: at a point where a 1-cycle $\nu\in \mathrm{H}_1(\Sigma_u,\mathbb{Z})$ vanishes, the monodromy acts on $\mathrm{H}_1(\Sigma_u,\mathbb{Z})$ as:
\begin{equation*}
	\text{for all } \gamma\in \mathrm{H}_1(\Sigma_u,\mathbb{Z})\; , \quad \gamma\longmapsto \gamma - (\gamma\cdot\nu)\nu\; .
\end{equation*}
Given a choice of 1-cycles $A$ and $B$ consistent with the asymptotic behavior of $a$ and $a_D$ at infinity, one can easily check that the vanishing cycle at $2\Lambda^2$ (resp. $-2\Lambda^2$) is $B$ (resp. $2A+B$), yielding the matrices $M_+$ and $M_-$ of \Cref{eq:monodstronglycoupled}.

\subsection{BPS spectrum and wall-crossing}\labelx{Sec:BPSwallcrossSU2}

At any point $u\in\mathcal C$, the central charge of a BPS state with low-energy electric-magnetic charges $(n_e,n_m)$ is given by the renormalization-group invariant expression:
\begin{equation*}
    \mathcal Z = n_ea + n_m a_D\; ,
\end{equation*}
leading to the BPS mass formula:
\begin{equation*}
    \vert M\vert = \sqrt{2}\vert n_ea + n_m a_D\vert\; .
\end{equation*}

The Seiberg--Witten solution allows for the calculation of the central charge, and thus the mass, of any hypothetical low-energy BPS particle with electric-magnetic charges $(n_e,n_m)$. However, while the central charge can be computed for potential BPS states, this does not guarantee their existence in the low-energy theory. Contrarily to $\mathcal N=4$ SYM theories where one expects BPS states for all possible low-energy charges, in pure $4d$ $\mathcal{N}=2$ $\mathrm{SU}(2)$ SYM, the spectrum of BPS states is more restricted. 

In the weakly-coupled region (I), the BPS spectrum can be computed semi-classically using the methods of \emph{'t~Hooft--Polyakov}\index{'t~Hooft--Polyakov monopole} and \emph{Julia--Zee}\index{Julia--Zee dyon}. It consists of $W^\pm$-bosons with charges $(\pm 2,0)$, monopoles with charges $(0,\pm1)$, and dyons with charges $\pm (2k,\pm1)$, where $k\in\mathbb{N}$. The $W^\pm$-bosons are part of BPS vector multiplets, whereas all other monopoles and dyons are in BPS hypermultiplets. 

\begin{proposition}
    The semi-classical BPS spectrum is globally invariant under the monodromy $M_\infty$\footnote{However, individual particles are not. For instance, under the process depicted in \Cref{fig:CoulombbranchquantumSW}, a monopole transforms into a dyon with charges $(-4,-1)$, and a dyon with charges $(2,-1)$ transforms into a dyon with charges $(2,1)$.}. However, it is \textbf{not} invariant under $M_\pm$ and therefore also not invariant under $\Gamma_0(4)$.
\end{proposition} 

For example, under $M_+$, a dyon with charges $(2,1)$ transforms into a dyon with charges $(2,3)$, which does not exist in the weakly-coupled region. This obviously poses a challenge for the realization of $S$-duality in $\mathcal N=2$ SYM theories. Nevertheless, this conclusion is premature, as \emph{wall-crossing}\index{wall-crossing!phenomenon} occurs within the Coulomb branch $\mathcal C$, allowing the BPS spectrum to change as one moves across different regions.

Recall from \Cref{prop:BPSprotectedagainstdecay} that the triangle inequality:
\begin{equation*}
    \vert \mathcal Z_1+\mathcal Z_2 \vert \leq \vert\mathcal Z_1\vert + \vert \mathcal Z_2\vert\; ,
\end{equation*}
implies that BPS states with charges $(n_e,n_m)$, where $n_e$ and $n_m$ are coprime, are generally stable against decay into two or more particles. This is because $a/a_D\notin \mathbb{R}$ generically, implying that the inequality is strict. However, when $a/a_D\in \mathbb{R}$, the inequality saturates, and the BPS states are only marginally stable. In this situation, decay is possible, as the BPS states can split into multiple particles while still preserving the BPS condition. This \emph{wall-crossing phenomenon}\index{wall-crossing!phenomenon} was first identified in the context of two-dimensional quantum field theories \cite{Cecotti:1992qh, Cecotti:1992rm}. The work of Seiberg and Witten later made it clear that wall-crossing also occurs in $4d$ $\mathcal N=2$ theories.
\begin{definition}
   The condition $a/a_D\in\mathbb{R}$ defines real codimension 1 submanifolds of $\mathcal C$ known as \emph{walls of marginal stability}\index{wall of marginal stability}. 
\end{definition}
On either side of these walls, the spectrum of BPS states remains constant, though the central charge and hence the mass of a specific BPS state can vary with the point in $\mathcal C$. As one crosses a wall of marginal stability, decay can occur, leading to a jump in the BPS spectrum. In pure $\mathcal N=2$ $\mathrm{SU}(2)$ SYM, there is such a wall: it is a curve passing through $\pm 2\Lambda^2$, as illustrated in \Cref{fig:wallsmarginalpureSU2}.

\begin{figure}[!ht]
    \centering
    \includegraphics[width=\textwidth]{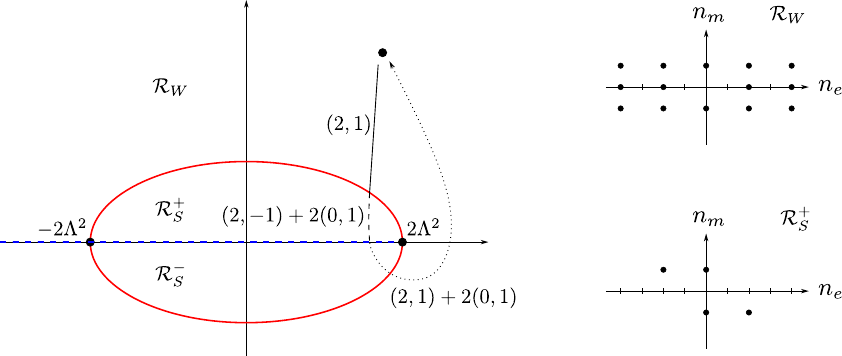}
    \caption{The wall of marginal stability on the Coulomb branch of pure $\mathcal N=2$ $\mathrm{SU}(2)$ SYM.}
    \labelx{fig:wallsmarginalpureSU2}
\end{figure}

Massless BPS states can only exist on $\mathcal C$, since $n_ea+n_ma_D=0$ implies that $a/a_D\in\mathbb{Q}\subset\mathbb{R}$. Moreover, it can be shown that $a_D/a$ varies monotonically from $-2$ to $2$ as one follows the wall from $-2\Lambda^2$ to itself through $2\Lambda^2$, first in the upper half-plane and then in the lower one. This implies that a BPS state with charges $(n_e,n_m)$ must become massless somewhere on the curve if $n_e/n_m\in[-2,2]$. If a BPS state becomes massless on the wall, it exists on both sides of the wall since it is the only massless particle that appears when crossing the wall, hence it is stable. Consequently, since we know that the monopole and dyons with charges $\pm(2,\pm1)$ become massless on $\mathcal C$, they are stable BPS states both in the weakly-coupled region and within the wall of marginal stability \cite{Ferrari:1996sv}.

\begin{proposition}
    Outside the curve of marginal stability, the BPS spectrum consists of $W^\pm$-bosons, monopoles, and all dyons with charges $\pm(2k,\pm1)$ where $k\in\mathbb{Z}$. Inside the curve, the BPS spectrum consists only of the monopole and specific dyons \cite{Ferrari:1996sv}. One can choose as branch cut for $M_+$ the segment from $-\infty$ to $1$ along the real axis. With the notation of \Cref{fig:wallsmarginalpureSU2}, the spectra are as follows:
\begin{itemize}
    \item $\mathcal R_W$: $W^\pm$-bosons $\pm(2,0)$, monopoles $\pm(0,1)$, dyons $\pm(2k,\pm1)$ for $k\in\mathbb{Z}$,
    \item $\mathcal R_S^+$: monopoles $\pm(0,1)$, dyons $\pm(2,-1)$,
    \item $\mathcal R_S^-$: monopoles $\pm(0,1)$, dyons $\pm(2,1)$.
\end{itemize}
\end{proposition}

The difference between the BPS spectra in $\mathcal R_W$ and $\mathcal R_S$ demonstrate that wall-crossing does occur. For instance, consider a dyon with charges $(2,1)$ at a point outside the wall of marginal stability as in \Cref{fig:wallsmarginalpureSU2}. As one circles around $2\Lambda^2$ counterclockwise and first crosses the wall, the dyon $(2,1)$ no longer exists within $\mathcal R_S^+$. Consequently, it must decay as:
\begin{equation*}
    (2,1)\mapsto (2,-1)+2\cdot(0,1)\; .
\end{equation*} 
Since the segment $]-\infty,-1]$ represents the branch cut for $M_+$, crossing it results in:
\begin{equation*}
    (2,-1)+2\cdot(0,1) \mapsto (2,1)+2\cdot(0,1)\; .
\end{equation*}
Thus, even if no BPS dyon with charges $(2,3)$ exists in $\mathcal R_S$, wall-crossing implies that a single BPS particle can become a pair of BPS particles under $M_\pm$. Thus, although more subtle than in $4d$ $\mathcal N=4$ SYM, the specific form of the weakly-coupled BPS spectrum does not contradict the realization of $S$-duality in $4d$ $\mathcal N=2$ pure $\mathfrak{su}(2)$ SYM.

\subsection{Global variants}

In pure $4d$ $\mathcal N=2$ $\mathfrak{su}(2)$ SYM theories, the matter content resides solely in the adjoint representation of the gauge group, and the Lagrangian of \Cref{eq:pureN=2SU2Lag} remains identical for both $G=\mathrm{SU}(2)$ and $G=\mathrm{SO}(3)=\mathrm{SU}(2)/\mathbb{Z}_2$. Furthermore, there is not just one $\mathrm{SO}(3)$ theory, but rather two distinct ones, denoted $\mathrm{SO}(3)_+$ and $\mathrm{SO}(3)_-$, which differ based on their line operator content. These are called global variants, as defined in \Cref{def:globalvariant}.

\begin{proposition}
    The distinction between these global variants of pure $4d$ $\mathcal N=2$ $\mathfrak{su}(2)$ SYM reflects in the definition of $a$ and $a_D$, and consequently, in the choice of the $A$ and $B$ cycles on the Seiberg–Witten curve. This ensures that the low-energy electric-magnetic charges correspond to the appropriate embedding of the low-energy $\mathrm{U}(1)$ into the UV gauge group \cite[Sect. 4.5]{Tachikawa:2013kta}.
\end{proposition}

When $G=\mathrm{SU}(2)$, electric particles (or Wilson lines) can be in any representation of $\mathrm{SU}(2)$, resulting in any integer-valued electric charge with respect to the low-energy $\mathrm{U}(1)$. The fact that the $W^\pm$-bosons have low-energy charges of $\pm(2,0)$ indicates that their charges are twice the minimal allowed value, consistent with their transformation under the adjoint representation of $\mathrm{SU}(2)$, which has N-ality 2. Magnetic monopoles (or 't Hooft lines) in this case correspond to the coweight lattice of $\mathrm{SU}(2)$, with the elementary monopole having a low-energy magnetic charge of 1, which is the minimal value allowed.

When $G=\mathrm{SO}(3)$, electric particles (or Wilson lines) must transform under a representation with N-ality 2 of $\mathrm{SU}(2)$. This redefinition of low-energy charges ensures that an electric charge of 1 is the smallest corresponding to a physically realizable particle (or Wilson line) in the theory. Accordingly, the $W^\pm$-bosons now carry charges of $\pm(1,0)$. Magnetic monopoles (or 't Hooft lines) for $G=\mathrm{SO}(3)$ correspond to the coweight lattice of $\mathrm{SO}(3)$, which is dual to the weight lattice of $\mathrm{SU}(2)$. In this case, the monopole that condenses at $2\Lambda^2$ has twice the minimum allowed magnetic charge, resulting in low-energy charges of $\pm(0,2)$. Specifically (cf. \cite{Closset:2023pmc} for a detailed discussion):
\begin{itemize}
    \item The global variant $\mathrm{SO}(3)_+$ is realized by redefining the variables $(a',a_D') = (2a, a_D/2)$, or equivalently, the cycles as $(A',B') = (2A,B/2)$. This leads to a redefinition of the low-energy charges as $(n_e',n_m') = (n_e/2, 2n_m)$. With this new framework, the monopole becoming massless at $2\Lambda^2$ has charges $(0,2)$, and the corresponding monodromy matrix $M_+$ is given by
\begin{equation*}
    M_+^{\mathrm{SO}(3)_+} = \begin{pmatrix}
        1 & 0 \\ -4 & 1
    \end{pmatrix}\; .
\end{equation*}
Similarly, with this redefinition one has
\begin{equation*}
    M_-^{\mathrm{SO}(3)_+} = \begin{pmatrix}
        -1 & 1 \\ -4 & 3
    \end{pmatrix}\; , \quad M_\infty^{\mathrm{SO}(3)_+} = \begin{pmatrix}
        -1 & 1 \\ 0 & -1
    \end{pmatrix}\; .
\end{equation*}

    \item The global variant $\mathrm{SO}(3)_-$ is realized by redefining the variables $(a'',a_D'') = (2a, a_D/2+a)$, or equivalently, the cycles as $(A'',B'') = (2A,B/2+A)$. This corresponds to a redefinition of the low-energy charges as $(n_e'',n_m'') = (n_e/2-n_m, 2n_m)$. The monodromy matrices become:
\begin{equation*}
    M_+^{\mathrm{SO}(3)_-} = \begin{pmatrix}
        3 & 1 \\ -4 & -1
    \end{pmatrix}\; , \quad M_-^{\mathrm{SO}(3)_-} = \begin{pmatrix}
        1 & 0 \\ -4 & 1
    \end{pmatrix}\; , \quad M_\infty^{\mathrm{SO}(3)_-} = \begin{pmatrix}
        -1 & 1 \\ 0 & -1
    \end{pmatrix}\; .
\end{equation*}
\end{itemize}

\section{\texorpdfstring{$\mathrm{SU}(2)$}{SU(2)} super-Yang--Mills theories with fundamental matter}\labelx{Sec:SdualitySU2flavors}

The renormalization of the complexified gauge coupling in Lagrangian $4d$ $\mathcal N=2$ gauge theories is exact at one-loop, and the one-loop coefficient for $4d$ $\mathcal N=2$ $\mathrm{SU}(2)$ SYM with $F>0$ fundamental hypermultiplets\footnote{Here, the global form of the theory is unambiguous since the fundamental representation of $\mathrm{SU}(2)$ is not a representation of $\mathrm{SO}(3)$.} is $b_1 = 4 - F$ (cf. \Cref{ex:b1N=2SYM}). Consequently, for $1 \leq F \leq 3$, the theory is asymptotically free. When $F = 4$, the perturbative renormalization of the gauge coupling becomes zero. This suggests that $4d$ $\mathcal{N}=2$ $\mathrm{SU}(2)$ SYM theory with four (massless) fundamental hypermultiplets could be conformally invariant---a hypothesis supported by substantial evidence. For $F>4$, the theory is infrared-free and is not well-defined as a UV theory. Therefore, we focus on $4d$ $\mathcal{N}=2$ $\mathrm{SU}(2)$ SYM theories with $1 \leq F \leq 4$ fundamental hypermultiplets, and for $F\leq 3$ we denote the dynamical scale of the theory by $\Lambda_F$.

In $\mathcal{N}=1$ language, each hypermultiplet consists of two chiral superfields $Q^a_i$ and $\widetilde{Q}_{a}^i$, where $a=1,2$ are color indices and $i=1,\dots,F$ are flavor indices. These hypermultiplets can be massive, with bare mass terms $m^i_j$ transforming in the adjoint representation of the flavor group. Invariance under $\mathrm{SU}(2)_R$ imposes the condition $[m,m^\dagger]=0$ as in \Cref{eq:massmatrixN=2normal}, which in turn implies that the mass matrix is diagonalizable. Thus, without loss of generality, the superpotential can be written as:
\begin{equation}\labelx{eq:SuPomassiveSU2withmatter}
    W = \sqrt{2}\widetilde{Q}_i^a\Phi_a^b Q^i_b + \sum_i m_i\widetilde{Q}_i^a Q^i_a\; , \quad m_i\in\mathbb{C}\; ,
\end{equation}
where $\Phi$ is the $\mathcal N=1$ adjoint chiral superfield in the $\mathcal N=2$ vector superfield. Moreover, $[m,m^\dagger]=0$ implies that a generic mass matrix $m$ breaks the flavor symmetry group to its Cartan subgroup.

\subsection{Global symmetries}

Let us assume that all hypermultiplets are massless. Then:
\begin{proposition}\labelx{prop:flavorsymmSU2}
    The classical flavor symmetry group of the theory is $\mathrm{O}(2F)$.
\end{proposition}

\begin{proof}
    This follows from the fact that the fundamental representation of the gauge group $\mathrm{SU}(2)$ in which the fields $Q_i^a$ transform, is isomorphic to the antifundamental representation in which the fields $\widetilde{Q}^i_a$ transform. Specifically, the fundamental representation of $\mathrm{SU}(2)$ is \emph{quaternionic}\index{representation!quaternionic}. More generally, $F$ massless hypermultiplets transforming in a quaternionic representation of any gauge group~$G$ lead to an $\mathrm{O}(2F)$ flavor symmetry. Similarly, hypermultiplets transforming in\emph{ real representations}\index{representation!real} give rise to symplectic flavor symmetry groups. Furthermore, the flavor symmetry is $\mathrm{O}(2F)$ rather than~$\mathrm{SO}(2F)$ due to the presence of the $\mathbb{Z}_2$ factor $\rho$ that exchanges $Q_1$ with $\widetilde{Q}^1$.
\end{proof}

\begin{remark}
    When the mass matrix $m$ is non-zero, the classical flavor symmetry group is reduced to a subgroup of $\mathrm{O}(2F)$. In extreme cases where the mass matrix is generic, the symmetry group is further broken down to the Cartan subgroup of $\mathrm{O}(2F)$.
\end{remark}

The group of global symmetries of the classical theory are given by a quotient of the product of $\mathrm{O}(2F) \times \mathrm{SU}(2)_R \times \mathrm{U}(1)_R$ with the Lorentz group \cite[Sect. 3]{Seiberg:1994aj}. This quotient arises because the element $-1 = (-1)^F \in \mathrm{U}(1)_R$ is actually part of the Lorentz group, and because the combination of this element $-1 \in \mathrm{U}(1)_R$ with the generator of the center of $\mathrm{SU}(2)_R$ must be identified with the generator of the $\mathbb{Z}_2$ center of the flavor group $\mathrm{O}(2F)$.

All $4d$ $\mathcal N=2$ $\mathrm{SU}(2)$ theories with $1\leq F\leq 4$ fundamental hypermultiplets admit a complex 1-dimensional classical Coulomb branch on which the gauge group breaks spontaneously to $\mathrm{U}(1)$. By \Cref{prop:quantumCoulombnotlifted}, these theories also admit a quantum \emph{Coulomb branch}\index{branch of moduli space!Coulomb} $\mathcal C^{(F)}$, which is a \emph{special Kähler manifold}\index{special!Kähler manifold}. As in \Cref{Sec:pure4dN=2SYM}, the parameter $u=\langle\tr\phi^2 \rangle/2$ serves as a natural coordinate on $\mathcal C^{(F)}$. Since $4d$ $\mathcal N=2$ $\mathrm{SU}(2)$ SYM with $1\leq F\leq 3$ fundamental hypermultiplets is asymptotically free, one again distinguishes a weakly coupled region and a strongly coupled region in $\mathcal C^{(F)}$.

When $F\geq 2$ and at least two hypermultiplets share the same mass, the theory also displays a \emph{Higgs branch}\index{branch of moduli space!Higgs}---a \emph{hyperkähler manifold}\index{hyperkähler!manifold}. Higgs branches are always classically exact. For $F=2$ massless hypermultiplets, the Higgs branch consists of two components $\mathbb{C}^2/\mathbb{Z}_2$, intersecting each other and the Coulomb branch at their apex and the origin of $\mathcal C^{(2)}$. For $F\geq 3$ massless hypermultiplets, the Higgs branch consists of a single component, which intersects the Coulomb branch $\mathcal C^{(F)}$ at the origin. Since our primary interest lies on BPS states and wall-crossing phenomena, which happen on Coulomb branches, we will not explore Higgs branches further, and refer instead to \cite[Sect. 3]{Seiberg:1994aj} and \cite[Chap. 7]{Tachikawa:2013kta} for more details.

\begin{proposition}[{\cite[Sect. 4]{Seiberg:1994aj}}]
    Quantum mechanically, the $(\mathbb{Z}_2)_\rho\times\mathrm{U}(1)_R$ factor of the classical global symmetry group is anomalous when $1\leq F\leq 3$: only a $\mathbb{Z}_{4(4-F)}$ subgroup survives as a genuine symmetry of the quantum theory. The generator acts as:
\begin{equation*}
    \Phi \mapsto \exp(\frac{\I\pi}{4-F})\Phi\; , \quad Q^i \mapsto  \exp(\frac{-\I\pi}{2(4-F)}) Q^i\; , \quad \widetilde{Q}^i \mapsto  \exp(\frac{-\I\pi}{2(4-F)}) \widetilde{Q}^i\; .
\end{equation*}
Note that since $\Phi$ has charge 2 under $\mathbb{Z}_{4(4-F)}$, $u$ has charge 4; therefore, $\mathbb{Z}_{4(4-F)}$ breaks spontaneously to $\mathbb{Z}_4$ at generic points of $\mathcal C^{(F)}$, and $\mathcal C^{(F)}$ is $(\mathbb{Z}_{4-F})$-symmetric.
\end{proposition}

\subsection{Central charge and monodromies}

Recall that the central charge $\mathcal Z$ of the $\mathcal N=2$ super Poincaré algebra is a linear combination of abelian conserved charges. On the Coulomb branch $\mathcal C^{(F)}$, the low-energy electric and magnetic charges $(n_e,n_m)$ contribute to $\mathcal Z$, as previously discussed. Moreover, additional abelian conserved charges---such as those arising from massive hypermultiplets---also contribute to $\mathcal Z$. 

For instance, in $4d$ $\mathcal N=2$ $\mathrm{SU}(2)$ SYM with $F=1$ massive hypermultiplet of bare mass $m_1\gg\Lambda_1$ in the weakly-coupled region of the Coulomb branch $\mathcal C^{(F)}$, i.e. $u\gg\Lambda_1^2$, the hypermultiplet becomes massless when $a=\pm m_1/\sqrt{2}$, cf. \Cref{eq:masseshypersvacuumCoulomb}. This BPS state has low-energy electric-magnetic charges $(n_e,n_m)=\pm(1,0)$ and flavor charge $\pm1$, thus one learns that:
\begin{equation}\labelx{eq:centralchargeSU2F=1}
    \mathcal Z = n_e a+ n_m a_D +S\frac{m_1}{\sqrt{2}},
\end{equation}
where $S$ is the flavor charge. Close to the point in $\mathcal C^{(F)}$ where $a=\pm m_1/\sqrt{2}$, one has:
\begin{align*}
    a &\simeq m_1/\sqrt{2}\; ,\\
    a_D &\simeq c-\frac{\I}{2\pi}(a-m_1/\sqrt{2})\ln(a-m_1/\sqrt{2})\; \; ,
\end{align*}
because the light hypermultiplet contributes logarithmically to $a_D$. Therefore, the monodromy reads:
\begin{equation}\labelx{eq:monodwithflavor}
    \begin{split}
        a &\mapsto a\; , \\
        a_D &\mapsto a_D + a-m_1/\sqrt{2}\; .
    \end{split}
\end{equation}
The special coordinates $(a,a_D)$ do not simply transform by $\mathrm{SL}_2(\mathbb{Z})$ under monodromy: they also pick up additive constants. This is consistent with the general structure of duality transformations that preserve the general form of the metric form, as discussed in \Cref{eq:generaldualityN=2}. In matrix form:
\begin{equation*}
    \begin{pmatrix}
        a & a_D & m_1/\sqrt{2}
    \end{pmatrix}\mapsto \begin{pmatrix}
        a & a_D & m_1/\sqrt{2}
    \end{pmatrix}\begin{pmatrix}
        1 & 1 & 0 \\
        0 & 1 & 0 \\
        0 & -1 & 1
    \end{pmatrix}\; ,
\end{equation*}
where the structure of the last column ensures that $m_1/\sqrt{2}$ remains invariant under monodromy ($m_1$ is a non-dynamical parameter of the theory). The action on the charges is given by the inverse matrix:
\begin{equation*}
    \begin{pmatrix}
        n_e \\ n_m \\ S
    \end{pmatrix} \mapsto \begin{pmatrix}
        1 & -1 & 0 \\
        0 & 1 & 0 \\
        0 & 1 & 1
    \end{pmatrix}\begin{pmatrix}
        n_e \\ n_m \\ S
    \end{pmatrix} = \begin{pmatrix}
        n_e-n_m \\ n_m \\ S+n_m
    \end{pmatrix}\; .
\end{equation*}
The invariance of the bare masses $m_i$ under monodromies implies that while gauge charges can contribute to flavor charges, flavor charges never contribute to gauge charges. 

\begin{proposition}
    For $1\leq F\leq 4$ fundamental hypermultiplets with generic bare masses $m_1,\dots,m_F$, the flavor symmetry is the Cartan subgroup $\mathrm{U}(1)^F$ of $\mathrm{O}(2F)$. The central charge takes the form:
    \begin{equation*}
        \mathcal Z = n_ea + n_m a_D + \sum_{i=1}^F S_i \frac{m_i}{\sqrt{2}}\; ,
    \end{equation*}
    where $S_1,\dots,S_F$ represent the charges under the abelian flavor symmetry group $\mathrm{U}(1)^F$. 
\end{proposition}

This shows that under monodromy around a point where a BPS particle becomes massless, the coordinates $(a,a_D)$ generally undergo not just an $\mathrm{SL}_2(\mathbb{Z})$ transformation, but also acquire additive constants that are integral linear combinations of $m_i/\sqrt{2}$. 

\subsection{Structure of the Coulomb branches}

The structure of $\mathcal C^{(F)}$ can be studied as in \Cref{Sec:pure4dN=2SYM}. Consider first $F=1$, with $\widetilde\tau$ the UV gauge coupling, and $\tau(u)$ the low-energy abelian one at $u\in\mathcal C^{(1)}$.

In the weakly-coupled region of $\mathcal C^{(1)}$, i.e. when $\vert u\vert\gg\vert\Lambda_1^2\vert$, one has $u \simeq a^2$, and:
\begin{equation*}
    \tau(a) \simeq 2\widetilde{\tau}(a) = -\frac{3}{\pi \I}\log\left(\frac{a}{\Lambda_1}\right)+\dots\; ,
\end{equation*}
where the dots represent instanton corrections. This implies:
\begin{equation*}
    a_D = -\frac{3a}{\pi \I}\log\left(\frac{a}{\Lambda_1}\right)+\frac{3a}{\pi \I}+\dots\; ,
\end{equation*}
leading to the monodromy at infinity:
\begin{equation*}
    \begin{pmatrix}
        a & a_D 
    \end{pmatrix}\mapsto \begin{pmatrix}
        a & a_D
    \end{pmatrix}\begin{pmatrix}
        -1 & 3 \\ 0 & -1
    \end{pmatrix}\; .
\end{equation*}

Let us now turn to the strongly-coupled region, first for $m_1\gg\Lambda_1$. At scales below $\vert m_1\vert$, the massive hypermultiplet decouples, effectively yielding pure $4d$ $\mathcal N=2$ $\mathrm{SU}(2)$ SYM. The dynamical scale $\Lambda_0$ of the latter theory can be expressed in the one-loop approximation in terms of $\Lambda_1$:
\begin{equation*}
    \Lambda_0^4 \simeq m_1\Lambda_1^3\; .
\end{equation*}

We know from \Cref{Sec:pure4dN=2SYM} that two singularities arise at $\pm 2\Lambda_0^2$ on $\mathcal C^{(F)}$: at $2\Lambda^2$, a monopole with charges $(n_e,n_m,S) = (0,1,0)$ becomes massless, while at $-2\Lambda^2$, a dyon with charges $(2,1,0)$ becomes massless. There is a third singularity at $u\simeq m_1^2$, where the BPS hypermultiplet $(1,0,1)$ becomes massless. 

Let now $m_1=0$. The $\mathbb{Z}_{12}$ symmetry of the quantum theory reduces to a $\mathbb{Z}_3$ symmetry on $\mathcal C^{(1)}$. As in the pure theory, $\mathcal C^{(1)}$ cannot exhibit a unique singularity at the origin since $\Im\tau$ must be positive definite. The simplest generalization, consistent with the analysis for $m_1\gg\Lambda_1^2$, suggests the presence of three singularities at $u\propto \zeta_3^k\Lambda_1^2$, where $\zeta_3=\exp(2\I\pi/3)$ and $k=0,1,2$. Since these singularities are permuted by $\mathbb{Z}_3$, the associated monodromies are conjugate. The structure of $\mathcal C^{(1)}$ is shown in \Cref{fig:CoulombbranchquantumSWF=1}.

\begin{figure}[!ht]
    \centering
    \includegraphics[width=\textwidth]{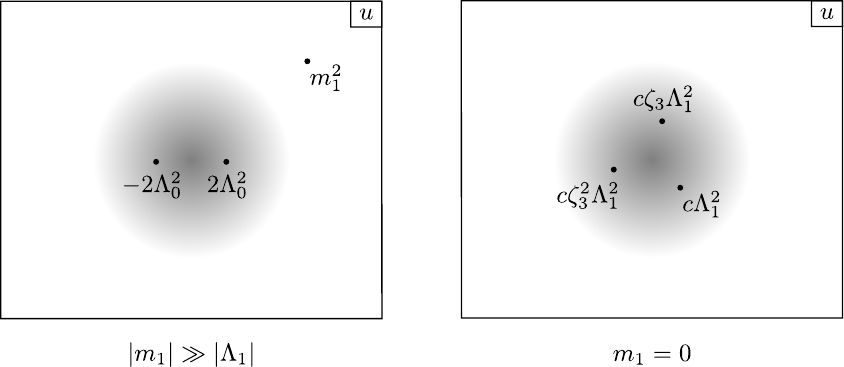}
    \caption{Coulomb branch of $\mathcal N=2$ $\mathrm{SU}(2)$ SYM with $F=1$ fundamental hypermultiplet.}
    \labelx{fig:CoulombbranchquantumSWF=1}
\end{figure}

We simply state the structure of the Coulomb branches $\mathcal{C}^{(F)}$ for $2\leq F\leq 4$ and refer to \cite{Seiberg:1994aj} and \cite[Chaps. 5,8,9]{Tachikawa:2013kta} for further details:
\begin{itemize}
    \item When $F=2$, the same reasoning results in a Coulomb branch structure with four singularities on $\mathcal C^{(2)}$ when the bare masses are generic. When all the masses vanish, these four singularities merge in pairs into two singularities. These two singularities are exchanged by the $\mathbb{Z}_2$ symmetry acting on $\mathcal C^{(2)}$, which descends from the $\mathbb{Z}_{8}$ symmetry of the quantum theory. 
    \item When $F=3$, the Coulomb branch $\mathcal C^{(3)}$ contains five singularities for generic values of the masses. When all masses vanish, four of these singularities coalesce, leaving two singularities. In this case, the $\mathbb{Z}_4$ symmetry of the quantum theory acts trivially on the Coulomb branch.
    \item Finally, for $F=4$, the Coulomb branch $\mathcal C^{(4)}$ exhibits six singularities when the hypermultiplet bare masses are generic. When all masses vanish, these six singularities merge at the origin of the Coulomb branch. This scenario is not problematic because the monodromy at infinity acts simply as charge conjugation, transforming $(n_e,n_m)\rightarrow -(n_e,n_m)$, or equivalently $(a,a_D)\rightarrow -(a,a_D)$. In this case, the $\mathrm{U}(1)_R$ symmetry is not anomalous.
\end{itemize}

\begin{remark}
    For any $1\leq F\leq 4$, taking the limit where the bare mass $m_F$ becomes infinitely large yields the Coulomb branch of the theory with $F-1$ flavors. This approach offers an alternative method for determining the Coulomb branch structure of these theories, and serves as a non-trivial consistency check for the listed results. 
    
    As in \Cref{Sec:pure4dN=2SYM}, one can add an $\mathcal N=1$ mass term for the adjoint chiral superfield in the $\mathcal N=2$ vector superfield, leading to $\mathcal N=1$ SQCD theories. The Coulomb branch is completely lifted, except its singularities, whereas the Higgs branches yield the moduli spaces described in \Cref{Sec:quantSQCD}.
\end{remark}

\subsection{Seiberg--Witten curves\index{Seiberg--Witten!curve}}

As in \Cref{Sec:pure4dN=2SYM}, the singularity structure determined from the analysis of the Coulomb branch fully determines the holomorphic (local) functions $a(u)$ and $a_D(u)$. These can be defined as the periods of a meromorphic one-form $\lambda$ on a complex torus $\Sigma_u$, where $(\Sigma_u)_{u\in\mathcal C^{(F)}}$ forms a holomorphic family of complex tori over $\mathcal C^{(F)}$. Additionally, each $\Sigma_u$ is a degree-2 ramified covering of a $u$-independent Riemann sphere $\mathbb{CP}^1_z$. 

For instance, when $F=1$, the pair $(\Sigma_u,\lambda_u)$ can be described by the equation:
\begin{equation}\labelx{eq:SWcurveF=1}
    (\Sigma_u)\colon \quad \Lambda_1^2 z + \frac{2\Lambda_1(x-m_1)}{z} = x^2 - u\; , \quad \lambda_u = x\frac{\mathrm{d}z}{z}\; .
\end{equation}
The meromorphic one-form $\lambda_u$ is required to satisfy:
\begin{equation*}
    \frac{\mathrm{d} \lambda_u}{\mathrm{d}u} = \omega_u\; ,
\end{equation*}
where $\omega_u$ is the unique holomorphic one-form (up to normalization) on $\Sigma_u$. This condition along with \emph{Riemann bilinear relations}\index{Riemann!bilinear relations} implies that the imaginary part of the gauge coupling:
\begin{equation*}
    \tau(u) = \frac{\partial a_D}{\partial a} = \frac{\partial a_D/\partial u}{\partial a/\partial u}\; ,
\end{equation*}
is positive-definite, and in turn, that $\mathrm{d}s^2 = \Im\tau(u)\mathrm{d}a\mathrm{d}\overline{a}$ is a well-defined metric on $\mathcal C^{(F)}$.

\begin{remark}
    In pure $\mathrm{SU}(2)$ SYM, the meromorphic differential $\lambda$ is required to have vanishing residues. This condition ensures that the definition of the special coordinates $a$ and $a_D$ depended solely on the homology classes of the cycles $A$ and $B$ on $\Sigma_u$. However, for $F\geq 1$, the meromorphic differential $\lambda$ must have non-zero residues to account for the contributions of the hypermultiplet bare masses to the central charge. These residues in turn allow the more general monodromies for $a$ and $a_D$ which appear when $F\geq 1$, as in \Cref{eq:monodwithflavor}. For example, in \Cref{eq:SWcurveF=1} the point $z=0$ is not a branch point of the covering $\Sigma_u\rightarrow \mathbb{CP}^1_z$, contrarily to what happens in pure SYM. The differential $\lambda_u$ has a non-zero residue $\pm m_1$ at the preimages of $z=0$ in $\Sigma_u$, which is necessary to account for the monodromy of \Cref{eq:monodwithflavor}.
\end{remark}

The appropriate decoupling limits applied to the equation defining the Seiberg--Witten curve for a theory with $1\leq F\leq 4$ flavors will reproduce the Seiberg--Witten curve for any theory with $F'\leq F$ flavors. Again, we refer to \cite{Seiberg:1994aj} and \cite[Chaps. 5,8,9]{Tachikawa:2013kta} for the computation of the Seiberg--Witten curves of $4d$ $\mathcal N=2$ $\mathrm{SU}(2)$ SYM with $1\leq F\leq 4$ fundamental hypermultiplets and the analysis of their properties.

\subsection{\texorpdfstring{$S$}{S}-duality of massless \texorpdfstring{$\mathrm{SU(2)}$}{SU(2)} SYM with four fundamental flavors}

For $F=4$, the vanishing of the one-loop renormalization coefficient for the gauge coupling implies that the coupling is perturbatively constant. While instanton corrections could in principle contribute to the renormalization of $\widetilde{\tau}$, such contributions would result in an exponentially small breaking of both scale invariance and $\mathrm{U}(1)_R$ symmetry. This, in turn, would cause $\Im\tau\mathrm{d}a\mathrm{d}\overline{a}$ to lose its positive-definiteness \cite{Seiberg:1994aj}. Although these problems might be resolved through additional non-perturbative effects, simplicity and consistency favor the assumption that the theory is truly scale-invariant. This is a widely accepted viewpoint, which we henceforth assume. 

\begin{remark}\labelx{rem:scaleinvariance}
    Scale invariance implies that $\tau = 2\widetilde{\tau}_\mathrm{cl}$ is independent of $u$, and that:
    \begin{align*}
        a(u) &= \sqrt{u}\; ,\\
        a_D(u) &= 2\widetilde{\tau}_\mathrm{cl} a(u)\; ,
    \end{align*}
    are exact equations, as in $\mathcal N=4$ SYM theories. This drastically simplifies the realization of electric-magnetic duality compared to asymptotically free theories. In particular the BPS spectrum is $\mathrm{SL}_2(\mathbb{Z})$-invariant, very much like in $\mathcal N=4$ SYM theories. However, unlike $\mathcal N=4$ SYM, $\mathrm{SU}(2)$ SYM with $F=4$ fundamental hypermultiplets possesses a non-trivial flavor symmetry, $\mathrm{SO}(8)$. This makes electric-magnetic duality richer; in particular, it mixes with the flavor symmetry.
\end{remark}

For any $F\geq 1$, the fundamental hypermultiplets induce $2F$ fermion zero modes in the background of dyons, which transform in the vector representation of the flavor symmetry group $\mathrm{SO}(2F)$. Quantizing these zero modes converts monopoles and dyons into spinors of $\mathrm{SO}(2F)$, meaning that the quantum flavor symmetry is actually the universal cover $\mathrm{Spin}(2F)$ of $\mathrm{SO}(2F)$. When $F=4$, the center of $\mathrm{Spin}(8)$ is $\mathbb{Z}_2\times\mathbb{Z}_2$, and representations of $\mathrm{Spin}(8)$ are classified into classes labeled by the elements of $\mathbb{Z}_2\times\mathbb{Z}_2$. In particular, the adjoint and trivial representations correspond to $o=(0,0)$, the vector representation to $v=(1,0)$, the spinor representation of positive chirality to $s=(0,1)$ and the conjugate spinor representation to $c=(1,1)$. Importantly, $\mathrm{Spin}(8)$ exhibits triality: its group of outer automorphisms is $\mathfrak{S}_3$, the permutation group on three objects, which here are the vector representation (defining the class $v$), the spinor representation ($s$) and the conjugate spinor one ($c$). All of them are 8-dimensional.

\begin{proposition}[{\cite[Sect. 6]{Seiberg:1994aj}}]
    The charges of low-energy particles under the center of $\mathrm{Spin}(8)$ are determined by the reduction modulo 2 of their electric and magnetic charges $(n_e,n_m)$.
\end{proposition}

For example, the fundamental hypermultiplets lead to low-energy BPS states with charges $(n_e,n_m)=(1,0)$, which transform in the vector representation of $\mathrm{Spin}(8)$. The monopoles with charges $(0,1)$ rather transform in the spinor representation of $\mathrm{Spin}(8)$. Therefore, since S-duality exchanges fundamental hypermultiplets and monopoles, it must also permute the representations of $\mathrm{Spin}(8)$. Furthermore, S-duality implies that for any given pair of coprime charges $(n_e,n_m)$, there are eight BPS states transforming in the eight-dimensional representation of $\mathrm{Spin}(8)$ determined by the reduction modulo 2 of $(n_e,n_m)$. 

The $W^\pm$-bosons have low-energy charges $(2,0)$. They are marginally stable against decay into fundamental hypermultiplets. S-duality requires the existence of BPS states with charges $(2n_e,2n_m)$ for all coprime integers $n_e,n_m$. 

The Seiberg--Witten curve for massless $4d$ $\mathcal N=2$ $\mathrm{SU}(2)$ $F=4$ SYM is triality and $\mathrm{SL}_2(\mathbb{Z})$-invariant, providing significant evidence supporting the realization of S-duality in this theory.

\vspace{0.3cm}

With massive fundamental hypermultiplets, both scale invariance and the $\mathrm{U}(1)_R$ symmetry are explicitly broken. The masses of the hypermultiplets transform in the adjoint representation of the flavor symmetry group, hence they are acted upon by the triality group $\mathfrak{S}_3$ as follows \cite[Sect. 10]{Seiberg:1994aj}. The transformation that exchanges the spinor representations $s$ and $c$ while leaving $v$ invariant, acts by flipping the sign of $m_4$, i.e. $m_4\rightarrow -m_4$ while leaving the other masses $m_1,m_2,m_3$ unchanged. The transformation that exchanges the vector $v$ with the spinor $s$ while leaving $c$ invariant, acts as:
\begin{equation}\labelx{eq:transfomasses}
    \begin{split}
        m_1 &\mapsto \frac{1}{2}(m_1+m_2+m_3+m_4)\; , \\
    m_2 &\mapsto \frac{1}{2}(m_1+m_2-m_3-m_4)\; , \\
    m_3 &\mapsto \frac{1}{2}(m_1-m_2+m_3-m_4)\; , \\
    m_4 &\mapsto \frac{1}{2}(m_1-m_2-m_3+m_4)\; . 
    \end{split}
\end{equation}
These two transformations together generate the full triality group $\mathfrak{S}_3$.

\subsection{Wall-crossing}

A careful analysis of Seiberg--Witten solutions for $4d$ $\mathcal N=2$ $\mathrm{SU}(2)$ SYM with $1\leq F\leq 3$ massless fundamental flavors, reveals that wall-crossing occurs in these theories \cite{Bilal:1996sk}. The wall of marginal stability is again diffeomorphic to a circle and passes through all singularities in the strongly-coupled region of the moduli space $\mathcal C^{(F)}$:
\begin{itemize}
    \item For $F=1$, outside the wall in $\mathcal C^{(1)}$, the BPS spectrum consists of dyons with charges $\pm(n_e,1)$ for all $n_e\in\mathbb{Z}$, the elementary hypermultiplets $\pm(1,0)$ and the $W^\pm$-bosons $\pm(2,0)$. The curve of marginal stability passes through the three singularities in the strongly-coupled region. Inside the wall, the BPS spectrum reduces to the states responsible for the strongly-coupled singularities: the monopole $\pm(0,1)$ and the dyons $\pm(-1,1)$ and $\pm(-2,1)$. 
    \item When $F=2$, the weakly-coupled BPS spectrum consists of all dyons $\pm(n_e,1)$ for all $n_e\in\mathbb{Z}$, the elementary hypermultiplets $\pm(1,0)$ and the $W^\pm$-bosons $\pm(2,0)$. Inside the wall, the BPS spectrum consists of only the monopole $\pm(0,1)$ and the dyon $\pm(1,1)$.
    \item When $F=3$, the weakly-coupled BPS spectrum includes all dyons $\pm(n_e,1)$ and $\pm(2n+1,2)$ for all $n_e,n\in\mathbb{Z}$, along with the elementary hypermultiplets $\pm(1,0)$ and the $W^\pm$-bosons $\pm(2,0)$. Inside the wall, the BPS spectrum consists of only the monopole $\pm(0,1)$ and the dyon $\pm(-1,2)$.
\end{itemize}

\begin{remark}
    In all three cases, wall-crossing reconciles $S$-duality with the fact that the BPS spectrum is not invariant under the group generated by the monodromies. As one crosses a wall, typically when circling the strongly-coupled singularities, certain BPS states decay.
\end{remark}

When the fundamental hypermultiplets are massive, both the singularities and the wall of marginal stability are displaced within the moduli space $\mathcal C^{(F)}$. For example, if the first hypermultiplet has bare mass $m_1\gg \Lambda_F$, there is a singularity at $u\simeq m_1^2/2$ in the weakly-coupled region, where this fundamental hypermultiplet becomes massless. The corresponding monodromy is given by \Cref{eq:monodwithflavor}, or equivalently:
\begin{equation}\labelx{eq:monodwithflavorcharges}
    \begin{pmatrix}
        n_e \\ n_m \\ S
    \end{pmatrix} \mapsto \begin{pmatrix}
        1 & -1 & 0 \\
        0 & 1 & 0 \\
        0 & 1 & 1
    \end{pmatrix}\begin{pmatrix}
        n_e \\ n_m \\ S
    \end{pmatrix} = \begin{pmatrix}
        n_e -n_m \\ n_m \\ S+n_m
    \end{pmatrix}\; .
\end{equation}
In the weakly-coupled region, the flavor charge $S$ of BPS states is bounded. Applying \Cref{eq:monodwithflavorcharges} to the monopole with charges $(0,1,1)$ reveals that there is a wall of marginal stability passing through this singularity. Thus, in $4d$ $\mathcal N=2$ $\mathrm{SU}(2)$ SYM theories with very massive fundamental hypermultiplets, wall-crossing occurs even in the weakly-coupled region of the Coulomb branch.

\section{Seiberg--Witten analysis: general case}\labelx{Sec:SWgeneral}

The analysis of Seiberg and Witten has rapidly been extended to $4d$ $\mathcal N=2$ pure SYM theories with various gauge algebras, including $\mathfrak{su}(N)$ \cite{Argyres:1994xh,Klemm:1994qs}, $\mathfrak{so}(2N+1)$ \cite{Danielsson:1995is}, $\mathfrak{so}(2N)$ \cite{Brandhuber:1995zp} as well as to $\mathcal N=2$ SQCD theories with fundamental matter \cite{Hanany:1995na, Argyres:1995wt, Argyres:1995fw, Hanany:1995fu}.

As for the $\mathfrak{su}(2)$ theories discussed in detail in \Cref{Sec:pure4dN=2SYM,Sec:SdualitySU2flavors}, the exact non-perturbative dynamics of these theories can be entirely captured by a family of Riemann surfaces called \emph{Seiberg--Witten curves}\index{Seiberg--Witten!curve}. These curves are parameterized by the Coulomb branch of the theory and come equipped with a meromorphic differential called \emph{Seiberg--Witten differential}\index{Seiberg--Witten!differential}, defined up to exact 1-forms. The underlying mathematical structure is that of an \emph{algebraic completely integrable system}\index{algebraic integrable system}  \cite{Donagi:1995cf}, to be discussed shortly.

In a general asymptotically free or conformal $4d$ $\mathcal N=2$ quantum field theory, the Coulomb branch $\mathcal C$ is a complex $r$-dimensional affine space, where $r$ is the rank of the theory. When the theory is a Lagrangian SYM theory with gauge algebra $\mathfrak{g}$, natural coordinates on $\mathcal C$ arise from the $\mathfrak{g}$-invariant polynomials of the adjoint scalar $\phi$, as discussed in \Cref{Sec:classN=2modulispaces}. For instance, when $\mathfrak{g}=\mathrm{su}(N)$ the rank is $r=N-1$ and the Coulomb branch $\mathcal C$ can be parameterized by $u_2=\langle\tr\phi^2\rangle, u_3=\langle\tr\phi^3\rangle, \dots, u_N=\langle\tr\phi^N\rangle$. 

At generic points of $\mathcal C$, the low-energy theory is a $4d$ $\mathcal N=2$ generalized Maxwell theory, with gauge group $\mathrm{U}(1)^r$. The theory may also have an abelian flavor symmetry $\mathrm{U}(1)^f$. The low-energy charges read:
\begin{equation*}
    (n_e^1,n_{m,1}, \dots, n_e^r, n_{m,r}, S_1, \dots, S_f)\; ,
\end{equation*}
and form a lattice $\Gamma\simeq\mathbb{Z}^{2r+f}$ endowed with the Dirac--Schwinger--Zwanziger (DSZ) pairing:
\begin{equation*}
    \left\langle \left(n_e^i,n_{m,i},S_k\right) , \left((n_e^{i})',n_{m,i}',S_k'\right) \right\rangle = \left(\sum_{i=1}^r\right) n_e^in_{m,i}'-n_{m,i}(n_e^{i})'\; .
\end{equation*}

For each $i=1,\dots,r$, the VEV $a_i$ of the $i$-th $\mathcal N=2$ abelian vector superfield is a local holomorphic function on $\mathcal C$. Generically, the set $(a_1,\dots,a_r)$ serves as a set of local holomorphic coordinates on $\mathcal C$. The special Kähler metric on $\mathcal C$ can be expressed in terms of the low-energy abelian gauge couplings $\tau^{ij}$, as:
\begin{equation}\labelx{eq:generalmetricCoulomb}
    \mathrm{d}s^2 = \Im \tau^{ij}\mathrm{d}a_i\mathrm{d}\overline{a}_j\; .
\end{equation}
The matrix $(\tau^{ij})$ is symmetric due to $\mathcal N=2$ supersymmetry. This in turn implies the existence of a locally holomorphic \emph{prepotential}\index{prepotential} $\mathcal F$ such that:
\begin{equation*}
    \tau^{ij} = \frac{\partial^2\mathcal F}{\partial a_i\partial a_j}\; .
\end{equation*}
Additionally, one defines:
\begin{equation*}
    a_{D}^i = \frac{\partial\mathcal F}{\partial a_i}\; ,
\end{equation*}
where $a_i$ and $a_D^i$ are called \emph{special coordinates}\index{special!coordinates} on the \emph{special Kähler manifold}\index{special!Kähler manifold} $\mathcal C$. One has:
\begin{equation*}
    \mathrm{d}s^2 = \Im \mathrm{d}a_{D}^i\mathrm{d}\overline{a}_i\; .
\end{equation*}

The $4d$ $\mathcal N=2$ central charge reads:
\begin{equation}\labelx{eq:centralchargegeneralcase}
    \mathcal Z = \left(\sum_{i=1}^r\right) n_e^ia_i+n_{m,i}a_{D}^i + \left(\sum_{k=1}^f\right) S_k \frac{m_k}{\sqrt{2}}\; ,
\end{equation}
where the $m_k$ are constants associated with the flavor charges. The mass $M$ of any one-particle state is constrained by the \emph{BPS inequality}\index{BPS!inequality}:
\begin{equation*}
    M \geq \vert \mathcal Z\vert\; ,
\end{equation*}
with equality holding if and only if the state is BPS. The charges of a given one-particle state correspond to a point $\gamma\in\Gamma$ in the charge lattice, but not all $\gamma$ necessarily correspond to BPS states. 

\begin{definition}
    The \emph{BPS spectrum} of the theory is the set of charges $\gamma$ for which BPS states exist, along with their multiplicities.
\end{definition} 
Spectral networks, which we will discuss in the next chapter, form a very powerful method for determining the BPS spectra of many $4d$ $\mathcal N=2$ theories.

Around metric singularity on $\mathcal C$, the special coordinates and the constants $m_k$ undergo monodromies: 
\begin{equation}\labelx{eq:monodgeneralform}
    \begin{pmatrix}
        a_1 & a_{D}^1 & \dots & a_r & a_D^r & m_1/\sqrt{2} & \dots & m_f/\sqrt{2}
    \end{pmatrix} \begin{pmatrix}
        M & 0_{2r\times f} \\ 
        C & \mathrm{Id}_{f}
    \end{pmatrix}\; ,
\end{equation}
where $M\in\mathrm{Sp}(2r,\mathbb{Z})$, ensuring that the DSZ pairing is preserved, and $C\in\mathcal{M}_{f\times 2r}(\mathbb{Z})$. The electric, magnetic, and flavor charges transform under the inverse of such monodromy matrices:
\begin{equation*}
    \begin{pmatrix}
        n_e^1 \\ n_{m,1} \\ \dots \\ n_r^1 \\ n_{m,r} \\ S_1 \\ \dots \\ S_f
    \end{pmatrix} \mapsto \begin{pmatrix}
        M^{-1} & 0_{2r\times f} \\ \widetilde{C} & \mathrm{Id}_{k\times f}
    \end{pmatrix} \begin{pmatrix}
        n_e^1 \\ n_{m,1} \\ \dots \\ n_r^1 \\ n_{m,r} \\ S_1 \\ \dots \\ S_f
    \end{pmatrix}\; ,
\end{equation*}
so as to preserve the central charge $\mathcal Z$. This form indicates that while flavor charges $S_k$ can receive contributions from the abelian gauge charges, gauge charges cannot be modified by flavor charges under these transformations.

\subsection{Seiberg--Witten solution}

In $4d$ $\mathcal{N}=2$ SYM theories with various gauge algebras of general rank---either pure \cite{Argyres:1994xh,Klemm:1994qs,Danielsson:1995is,Brandhuber:1995zp} or with fundamental matter \cite{Hanany:1995na, Argyres:1995wt, Argyres:1995fw, Hanany:1995fu}---it has been shown that the special coordinates $a_i$ and $a_D^i$ on the Coulomb branch $\mathcal C$ can be derived as period integrals of a meromorphic differential $\lambda$ on certain hyperelliptic curves. This generalizes the original Seiberg--Witten analysis of $\mathfrak{su}(2)$ gauge theories to higher-rank gauge algebras, providing a powerful and exact, non-perturbative description of the quantum dynamics in these theories.

Denoting $\Vec{u} = (u_1,\dots,u_r)$ the Coulomb branch operators, i.e. the gauge-invariant coordinates on $\mathcal C$, one has: 
\begin{proposition}
    Each such $4d$ $\mathcal{N}=2$ theory corresponds to a holomorphic family $(\Sigma_{\Vec{u}})$ of Riemann surfaces parameterized by $\mathcal C$, called \emph{Seiberg--Witten curves}\index{Seiberg--Witten!curve}. These surfaces are defined by a polynomial equation involving the complex variables $z\in\mathbb{CP}^1$, $x\in\mathbb{C}$, the Coulomb branch coordinates $\Vec{u}$, and the bare masses $\Vec{m}$ of the hypermultiplets, which are considered fixed parameters of the theory. 
\end{proposition}

At a fixed generic point in the Coulomb branch, i.e., for a fixed generic value of $\Vec{u}$, the Riemann surface $\Sigma_{\Vec{u}}$ in $\mathbb{CP}^1_z\times\mathbb{C}_x$ is of genus $r$ with $f$ punctures. Additionally, $\Sigma_{\Vec{u}}$ is a ramified covering of $\mathbb{CP}^1_z$ of degree $r$. As $\Vec{u}$ varies, the surface may degenerate at non-generic points $\Vec{u_0}$ in the Coulomb branch, accounting for the metric singularities on $\mathcal C$. This simplest type of degeneration occurs when two branch points of the covering $\Sigma_{\Vec{u}}\rightarrow \mathbb{CP}^1_z$ collide. Such degenerations are called \emph{nodal singularities}\index{singularity!nodal}.

\begin{proposition}
    The special coordinates $a_i$, $a_D^i$ and the masses $m_k$ are obtained as period integrals of a meromorphic differential $\lambda_{\Vec{u}}$ called \emph{Seiberg--Witten differential}\index{Seiberg--Witten!differential}, on $\Sigma_{\Vec{u}}$ with compactified punctures. The poles of $\lambda_{\Vec{u}}$ are at the punctures of $\Sigma_{\Vec{u}}$. At a fixed value of $\Vec{u}$, the homology group $\mathrm{H}_1(\Sigma_{\Vec{u}},\mathbb{Z})$ forms a lattice of rank $2r+f$. Over $\mathcal C$, this homology group forms a local system of lattices, equipped with the antisymmetric intersection pairing $(\cdot,\cdot)$ which has the cycles around punctures in its kernel. 
\end{proposition}

\begin{definition}
    A \emph{symplectic basis} of $\mathrm{H}_1(\Sigma_{\Vec{u}},\mathbb{Z})$ consists of cycles $A_1,B_1,\dots,A_r,B_r,C_1,\dots,C_f$ on $\Sigma_{\Vec{u}}$, where the cycles $C_k$ enclose punctures, and such that the intersection pairing satisfies:
\begin{equation*}
    (A_i,B_j) = -(B_j,A_i) = \delta_{ij}\; ,
\end{equation*}
with all other intersections vanishing.
\end{definition}

Locally on $\mathcal C$, one can choose a symplectic basis of $\mathrm{H}_1(\Sigma_{\Vec{u}},\mathbb{Z})$, allowing the identification of the local system of homology lattices $\mathrm{H}_1(\Sigma_{\Vec{u}},\mathbb{Z})$, endowed with the intersection pairing $(\cdot,\cdot)$, with the local system of charge lattices $\Gamma_{\Vec{u}}$, endowed with the DSZ pairing $\langle\cdot,\cdot\rangle$. Then:

\begin{proposition}
    For each $i=1,\dots,r$ the special coordinates $a_i$ and $a_D^i$ are given by the period integrals of the Seiberg–Witten differential $\lambda$ along the corresponding homology cycles:
    \begin{equation*}
        a_i = \oint_{A_i} \lambda\; , \quad \text{and} \quad a_D^i = \oint_{B_i} \lambda\; .
    \end{equation*}
    The flavor parameters $m_k/\sqrt{2}$ are obtained as the residues of the Seiberg–Witten differential $\lambda$ at the punctures of $\Sigma_{\Vec{u}}$. 
\end{proposition}

The singularities in the fibration $\Sigma\rightarrow \mathcal C$ induce monodromies in the homology group  $\mathrm{H}_1(\Sigma_{\Vec{u}}, \mathbb{Z})$, which take the form described earlier and can be derived from the \emph{Picard--Lefschetz monodromy formula}\index{Picard--Lefschetz monodromy formula}. These monodromies give rise to the monodromies of the charge lattice described in \Cref{eq:monodgeneralform}.

An important constraint comes from the fact that the gauge couplings $\tau^{ij} = \partial a_D^i/\partial a_j$ must be non-degenerate and have a positive-definite imaginary part, ensuring that the metric on the Coulomb branch in \Cref{eq:generalmetricCoulomb} is well-defined and positive-definite. The compactifications of the Riemann surfaces $\Sigma_{\Vec{u}}$ are compact and of genus $r$, hence they admit $r$ independent holomorphic (abelian) differentials $\omega_1,\dots,\omega_r$. For $A_1,B_1,\dots,A_r,B_r$ a symplectic basis of $\mathrm{H}_1(\Sigma_{\Vec{u}})$ and the period matrices:
\begin{equation*}
    u_{ij} = \oint_{A_i} \omega_j\; , \quad \text{and} \quad v_{ij} = \oint_{B_i} \omega_j\; ,
\end{equation*}
\emph{Riemann bilinear relations}\index{Riemann!bilinear relations} imply that the matrix $(v_{ij}/u_{ij})$ is non-degenerate with positive-definite imaginary part. Thus, one imposes that the derivatives
\begin{equation*}
    \frac{\partial \lambda}{\partial u_i}
\end{equation*}
form a basis of abelian differentials on $\Sigma_{\Vec{u}}$. 

\begin{remark}
    The Seiberg--Witten differential $\lambda$ is not uniquely determined: it can still be shifted by the addition of exact differentials, i.e., $\lambda\mapsto \lambda+\mathrm{d}\alpha$, as this does not affect the periods of $\lambda$.
\end{remark}

\subsection{Integrable systems}

\begin{definition}
    \emph{$\mathcal N=2^*$ theory}\index{$\mathcal N=2^*$ theory} with gauge group $G$ is the Lagrangian $4d$ $\mathcal N=2$ theory consisting of a $G$-vector superfield and a massive hypermultiplet in the adjoint representation of $G$. Sending the mass of the hypermultiplet to zero yields $4d$ $\mathcal N=4$ SYM theory with gauge group $G$.
\end{definition}

\begin{remark}
     As in $\mathcal N=4$ SYM theories, depending on the gauge group $G$ there might be distinct global variants of $\mathcal N=2^*$ theories.
\end{remark}

$\mathcal N=2^*$ theories can also be described within the Seiberg--Witten framework, although the Seiberg--Witten curve in this case is not a hyperelliptic curve, as it is for pure SYM and SYM with fundamental hypermultiplets, but a ramified covering of a punctured torus \cite{Donagi:1995cf}. The study of $\mathcal N=2^*$ theories by Donagi and Witten led to deep insights into the underlying mathematical structures describing the low-energy physics of $4d$ $\mathcal N=2$ quantum field theories.

\begin{proposition}[\cite{Donagi:1995cf}]
    In general, the low-energy effective Lagrangian on the Coulomb branch of any rank-$r$ $4d$ $\mathcal N=2$ theory defines a family $X\rightarrow \mathcal C$ of compact, complex, $r$-dimensional tori $X_{\Vec{u}}$ over the Coulomb branch $\mathcal C$, i.e. \emph{abelian varieties}\index{abelian variety}. As with previous cases, the special coordinates on $\mathcal C$, along with the mass parameters, can be obtained by taking the period integrals of a meromorphic differential $\lambda$ on $X$, where $\lambda$ is closed when restricted to the fibers of the fibration $X\rightarrow \mathcal C$, and is defined up to exact one-forms. The exterior derivative $\mathrm{d}\lambda$ of $\lambda$ is a non-ambiguous non-degenerate, closed, and holomorphic two-form $\omega$ on $X$, which vanishes when restricted to the fibers. 
\end{proposition}

The DSZ pairing naturally induces a polarization on the fibers $X_{\Vec{u}}$, resulting in a family of polarized abelian varieties over $\mathcal C$, equipped with the holomorphic two-form $\omega$. This form defines a complex symplectic structure, and consequently, a Poisson structure on $X$. The Coulomb branch coordinates $u_1,\dots,u_r$ constitute a maximal set of mutually commuting and algebraically independent functions with respect to this Poisson bracket. In other words:

\begin{proposition}
    The low-energy effective Lagrangian on the Coulomb branch of any rank-$r$ $4d$ $\mathcal N=2$ theory defines a complex completely integrable system.
\end{proposition}

The commuting Hamiltonians of the integrable system encoding the low-energy physics of a $4d$ $\mathcal N=2$ theory are the VEVs of the scalar fields in the low-energy $\mathcal N=2$ abelian vector fields. The orbits they generate encode the kinetic energy of the low-energy abelian vector superfields.

\begin{example}
    In the previously discussed cases, the Seiberg--Witten curves are ramified coverings of the Riemann sphere $\mathbb{CP}^1$, with associated abelian varieties the \emph{Jacobian varieties} $J(\Sigma_{\Vec{u}})$ of the Seiberg--Witten curves $\Sigma_{\Vec{u}}$.
\end{example}

There exists a broad class of complex completely integrable systems, known as \emph{Hitchin integrable systems}\index{Hitchin!system} \cite{Hitchin87,Hitchin87b}, encompassing all previously discussed cases as well as more general ones, such as $4d$ $\mathcal N=2^*$ theories. Hitchin systems are associated with an arbitrary Riemann surface $C$ of genus $g$ and a reductive compact Lie group $G$, and arise from the dimensional reduction of the $4d$ self-dual (instanton) Yang–Mills equations with gauge group $G$ on $\mathcal C$. In more high-energy physics terms, Hitchin systems are constructed from $2d$ adjoint QCD theories. The Hitchin systems for non-compact Riemann surfaces were constructed in \cite{Markman:1994sp,Donagi:1995non}. The reader is referred to Hitchin's original articles \cite{Hitchin87,Hitchin87b} and the book \cite{Hitchin:2013in} for general aspects of Hitchin integrable systems.

A point in a Hitchin system consists of a holomorphic $G_\mathbb{C}$-bundle $P_{G_\mathbb{C}}$ on $C$, where $G_\mathbb{C}$ is the complexification of the real Lie group $G$, and an adjoint-valued holomorphic one-form $\phi$ on $C$. The integrable structure of this system becomes more explicit with the notion of \emph{spectral curve}\index{spectral!curve} or \emph{spectral cover}\index{spectral!cover}. The integrable system structure involves the Jacobian of the spectral curve $\Sigma$.

\begin{definition}
    A \emph{spectral cover} of $C$ is a ramified covering $\Sigma\subset T^*C$ of $C$, where $T^*C$ is the cotangent bundle of $C$, determined by the pair $(P_{G_\mathbb{C}},\phi)$.
\end{definition} 

For $G=\mathrm{SU}(N)$, where $G_\mathbb{C} = \mathrm{SL}_N(\mathbb{C})$,  the spectral cover $\Sigma$ is defined by the determinant equation:
\begin{equation}\labelx{eq:spectralcover}
    \det(\eta-\phi) = \eta^{\otimes N} + \eta^{\otimes (N-2)}W_2(\phi)+ \eta^{\otimes (N-3)}W_3(\phi) + W_N(\phi)= 0\; ,
\end{equation}
where $\eta$ is the tautological 1-form on the cotangent bundle $T^*C$. This equation generically defines a Riemann surface $\Sigma\subset T^*C$ of genus $\widetilde{g}=(N^2-1)(g-1)+g$, which is a ramified covering of $C$ of degree $N$. The holomorphic $i$-differentials $W_i(\phi)$ on $C$ are determined by $\phi$. 

The Hitchin system structure, $X\rightarrow \mathcal B$, where fibers are abelian varieties, is as follows: the base $\mathcal B$ is the space of differentials $W_i(\phi)$, forming a complex affine space of dimension $r=(N-1)$. Each fiber is the Jacobian variety of the spectral curve $\Sigma$ associated with the point on $\mathcal B$. There is also a distinguished section: a generic point on $C$ corresponds to $N$ distinct points on $\Sigma$, the eigenvalues of \Cref{eq:spectralcover}. These eigenvalues define a holomorphic line bundle on $\Sigma$, and hence a point in the Jacobian $J(\Sigma)$. 

\begin{example}
    For SYM theories, either pure or with fundamental matter, the curve $C$ is the Riemann sphere $\mathbb{CP}^1$. For $\mathcal N=2^*$ theories, $C$ is a punctured torus. The group $G$ is the gauge group of the $4d$ $\mathcal N=2$ theory under consideration. The base $\mathcal B$ of the Hitchin system is identified with the Coulomb branch $\mathcal C$, and the holomorphic $k$-differentials $W_k(\phi)$ correspond to the Coulomb branch operators. The Seiberg–Witten curve $\Sigma$ is identified with the spectral cover determined by a point on $\mathcal C$. 
    
    The Hitchin systems encoding the low-energy physics of pure SYM theories or $\mathcal N=2^*$ theories correspond to well-known families of complex, completely integrable systems. Pure $\mathrm{SU}(N)$ SYM theories are associated with the so-called \emph{periodic Toda chains}\index{periodic Toda chain}, while $\mathcal N=2^*$ theories correspond to \emph{elliptic Calogero--Moser systems}\index{elliptic Calogero--Moser system}. For more details on the connections between these integrable systems and Seiberg–Witten theory, we refer to the pedagogical introduction in \cite{DHoker:1999yni} and references therein.
\end{example}

\begin{remark}
    This discussion prompts the question of whether all Hitchin systems encapsulate the low-energy physics on the Coulomb branch of specific $4d$ $\mathcal N=2$ theories. We will delve into this and provide an answer in the next chapter.
\end{remark}

\subsection{Remarkable \texorpdfstring{$4d$ $\mathcal N=2$}{4d N=2} SCFTs}

Let us highlight some intriguing and somewhat exotic theories that have been uncovered through the formalism developed by Seiberg and Witten.

One notable class of examples are the so-called \emph{Argyres–Douglas theories}\index{Argyres--Douglas theory} \cite{Argyres:1995jj,Argyres:1995xn}. These $4d$ $\mathcal N=2$ superconformal field theories (SCFTs) emerge as low-energy descriptions at specific singularities on the Coulomb branch $\mathcal C$ of $4d$ $\mathcal N=2$ theories, particularly where multiple branch points of the covering $\Sigma\rightarrow C$ collide. A prototypical case occurs in $\mathrm{SU}(2)$ SYM with one massive fundamental hypermultiplet. The Seiberg–Witten curve for this setup is given in \Cref{eq:SWcurveF=1}, which we reproduce here:
\begin{equation*}
    (\Sigma_{u,m})\colon \quad \Lambda^2 z + \frac{2\Lambda(x-m)}{z} = x^2 - u\; ,
\end{equation*}
where $u\in\mathcal C\simeq \mathbb{C}$, $\Lambda$ is the dynamical scale, $m$ is the bare mass of the fundamental hypermultiplet, $z\in\mathbb{CP}^1$ and $x\in\mathbb{C}$. This Seiberg--Witten curve $\Sigma_{u,m}$ is a branched covering of degree 2 of $\mathbb{CP}^1_z$, with ramification points determined by the discriminant equation:
\begin{equation*}
    \Delta = z^3+\frac{uz^2}{\Lambda^2}-\frac{2 mz}{\Lambda}+1 = 0\; ,
\end{equation*}
along with the point $z=\infty$. Singularities in $\mathcal C$ occur when two branch points collide, specifically at values of $u$ where the discriminant of the discriminant:
\begin{equation*}
    \Delta' = \Delta(\Delta) = u^3-m^2u^2+9\Lambda^3mu+\frac{27}{4}\Lambda^6-8\Lambda^3m^3\; ,
\end{equation*}
considered as a function of $u$, vanishes. There are, as expected, three solutions in $\mathcal C$. The discriminant of $\Delta'$ reads:
\begin{equation*}
    \Delta(\Delta') = \Delta(\Delta(\Delta)) = 8m^3+27\Lambda^3\; .
\end{equation*}
Thus, when $m^3=-27\Lambda^3/8$, two singularities collide, leading to a more intricate singularity where two independent cycles of the Seiberg–Witten curve shrink, leading to a \emph{cusp singularity}\index{singularity!cusp} (in contrast to simple singularities, which are \emph{nodal points}\index{singularity!nodal}).

Remarkably, at such specific singular points in $\mathcal C$ the low-energy theory contains mutually non-local massless particles, such as quarks and monopoles, indicating that this SCFT cannot be described by a Lagrangian. For further exploration of Argyres–Douglas theories and their properties, we refer the reader to \cite[Chap. 10]{Tachikawa:2013kta} and references therein.

\vspace{0.3cm}

Another set of significant $4d$ $\mathcal N=2$ SCFTs, whose existence was predicted through Seiberg–-Witten theory, are the so-called \emph{Minahan--Nemeschansky exceptional theories}\index{Minahan--Nemeschansky theory} \cite{Minahan:1996fg, Minahan:1996cj}. They play a remarkable role in the landscape of such theories.

In the context of Seiberg–Witten theory, rank-1 $4d$ $\mathcal N=2$ SCFTs emerge as low-energy descriptions of the physics near singularities on the Coulomb branch $\mathcal C$ of rank-1 $4d$ $\mathcal N=2$ systems \cite{Argyres:1995xn}. Away from singularities, the low-energy behavior is governed by $4d$ $\mathcal N=2$ Maxwell theory. The simplest types of singularities on $\mathcal C$ are \emph{nodal}\index{singularity!nodal}, meaning that the Seiberg–Witten curve develops a double point there. At such singularities, the low-energy theory is $4d$ $\mathcal N=2$ Maxwell theory coupled to massless charged hypermultiplets.

However, more intricate singularities can occur, for example when multiple simpler singularities coalesce. We have seen that near these more complicated singularities, the low-energy dynamics is described by Argyres–Douglas theories. Another pathway to rank-1 $4d$ $\mathcal N=1$ SCFTs arises when considering the low-energy theory at the origin of the Coulomb branch in massless $4d$ $\mathcal N=2$ $\mathrm{SU}(2)$ supersymmetric Yang--Mills theory with $F\geq 4$ fundamental flavors. Specifically, when $F=4$ the theory is conformal, while for $F>4$, the theory becomes infrared-free.

The SCFT describing the physics near a singularity depends only on the local geometric features near the singular point, leading to a classification problem for the possible singularities that can appear. In Seiberg–Witten theory, each rank-1 $4d$ $\mathcal N=2$ theory is associated with a fibration of complex tori $\Sigma\rightarrow\mathcal C$ over the Coulomb branch. The total space of this fibration forms a complex two-dimensional variety. All possible singularities of such varieties have been classified by Kodaira \cite{Kodaira63}, and they are locally modeled by quotients $\mathbb{C}^2/\Gamma$, where $\Gamma$ is a finite subgroup of $\mathrm{SU}(2)$.

Finite subgroups $\Gamma$ of $\mathrm{SU}(2)$ admit an \emph{$ADE$ classification}. Notably, the Dynkin type of the flavor symmetry Lie group of the
$4d$ $\mathcal N=2$ SCFT corresponding to
$\mathbb{C}^2/\Gamma$ matches that of $\Gamma$. While the $A$ and $D$ types correspond to previously discussed theories---such as $4d$ $\mathcal N=2$ Maxwell theory with matter, Argyres--Douglas theories, or $\mathrm{SU}(2)$ SYM with $F\geq 4$ fundamental flavors---the exceptional $E_6$, $E_7$ and $E_8$ cases led to the discovery of new $4d$ $\mathcal N=2$ SCFTs. These SCFTs have flavor symmetry groups of type $E_6$, $E_7$ and $E_8$ respectively, and are known as \emph{Minahan--Nemeschansky theories}\index{Minahan--Nemeschansky theory} \cite{Minahan:1996fg, Minahan:1996cj}.

\subsection{Instanton counting}

We conclude this chapter by highlighting a major development contemporaneous with Seiberg–Witten theory, namely the program of \emph{instanton counting}\index{instanton!counting}. A central achievement of Seiberg–Witten theory is the indirect determination of the low-energy prepotential, which in turn encodes the non-perturbative instanton corrections to the renormalization of the gauge coupling.\footnote{By virtue of $\mathcal N=2$ supersymmetry, the holomorphic and physical gauge couplings coincide, and holomorphy constrains the renormalization of the complexified gauge coupling to consist solely of one-loop perturbative contributions and non-perturbative instanton effects (\Cref{Sec:RenormN=2}).} For an explicit discussion in the case of pure $\mathcal N=2$ $\mathrm{SU}(2)$ SYM, see \cite{Lerche:1997sm}.

It is natural to ask whether the prepotential determined by Seiberg–Witten theory can be derived more directly by evaluating path integrals over instanton configurations. This question spurred significant advances in the study of multi-instanton moduli spaces (i.e., configurations with instanton number $n>1$), culminating in the comprehensive review \cite{Dorey:2002ik}. The modern program of \emph{instanton counting} was initiated in Nekrasov’s seminal work \cite{Nekrasov:2002qd}, which demonstrated that the multi-instanton integrals of \cite{Dorey:2002ik}, yielding the low-energy prepotential of Seiberg–Witten theory, can be interpreted as equivariant integrals and thus computed via \emph{equivariant localization}\index{equivariant localization} techniques. 

Physically, the equivariant integrals that yield a direct derivation of the low-energy prepotential for $\mathcal N=2$ gauge theories admit a natural interpretation in terms of a specific deformation of the theory, parametrized by two real constants $\epsilon_1$ and $\epsilon_2$. Concretely, one considers the $5d$ $\mathcal N=1$ theory associated to the given $4d$ $\mathcal N=2$ gauge theory, placed on a $\mathbb{C}^2_{z_1,z_2}$ fibration over a circle $S^1_{\xi}$ of radius $\beta$, with the identification:
\begin{equation*}
(z_1,z_2,\xi) \sim \left(\E^{\I\beta\epsilon_1}z_1, \E^{\I\beta\epsilon_2}z_2, \xi+\beta\right)\; .
\end{equation*}
The resulting geometry is referred to as the \emph{$\Omega$-background}\index{$\Omega$-background}. In this setting, the supersymmetric partition function $Z^{5d}(\beta,\epsilon_1,\epsilon_2,\dots)$ admits a well-defined four-dimensional limit,
\begin{equation*}
Z^{4d}(\epsilon_1,\epsilon_2,\dots) = \lim_{\beta\to 0} Z^{5d}(\beta,\epsilon_1,\epsilon_2,\dots)\; ,
\end{equation*}
where the dots denotes possible fugacities associated with global symmetries of the gauge theory, when present. Crucially, this partition function encodes the non-perturbative low-energy dynamics: the Seiberg–Witten prepotential is recovered in the limit
\begin{equation*}
\mathcal F(\dots) = \lim_{\epsilon_1,\epsilon_2\to 0} \epsilon_1\epsilon_2 \log Z^{4d}(\epsilon_1,\epsilon_2,\dots)\; .
\end{equation*}

The supersymmetric partition function $Z^{4d}(\epsilon_1,\epsilon_2,\dots)$ can be computed for a wide class of gauge theories, including $\mathcal N=2$ SYM theories with arbitrary simple gauge groups and matter content, as well as unitary quiver gauge theories. 

We will not pursue the subject of instanton counting further in this monograph, and instead refer the reader to the comprehensive reviews \cite{Shadchin:2005hp,Bianchi:2007ft,Tachikawa:2014dja}, which examine the subject in detail and highlight its connections to a broad range of developments in gauge theory and string theory. A particularly notable development is the \emph{AGT correspondence}\index{AGT correspondence} (see \cite{LeFloch:2020uop} for a review), which establishes a deep relation between four-dimensional $\mathcal N=2$ theories of class $\mathcal S$ (to be discussed in \Cref{Chap:theoriesofclassS}) and two-dimensional conformal field theories. The appearance of \emph{W-algebras}\index{W-algebra} (which themselves are connected to higher-rank Teichmüller theory \cite{Bilal:1990wn}) in this context is reviewed in \cite{Tachikawa:2014dja}. More directly related to the themes of the present monograph is the interplay between instanton counting and spectral networks via the exact WKB method, which has been explored in the recent work \cite{HRS21}.

\chapter{Class \texorpdfstring{$\mathcal S$}{S} theories}\labelx{Chap:theoriesofclassS}



\abstract{We review the construction of Seiberg–Witten curves for a wide class of four-dimensional $\mathcal N=2$ theories using techniques from superstring theory and M-theory, following the seminal work of \cite{Witten:1997sc} and the detailed treatment in \cite[Sect. 3.2]{Gaiotto:2009hg}. After a brief overview of superstring and M-theory with emphasis on their brane content, we describe the type IIA brane configurations that realize broad families of $\mathcal N=2$ gauge theories. Their M-theory uplift yields the corresponding Seiberg–Witten curves, providing a powerful framework for constructing these curves systematically.
We then introduce the framework of class $\mathcal S$ theories, drawing on \cite{Gaiotto:2009we,Tachikawa:2013kta}. We first show how the $S$-duality properties of four-dimensional $\mathcal N=2$ $\mathrm{SU}(2)$ SYM with four flavors give rise to generalized $\mathrm{SU}(2)$ quivers, which capture weak-coupling limits of a broad class of $\mathcal N=2$ superconformal field theories, namely class $\mathcal S$ theories of type $A_1$. This construction naturally generalizes from $\mathrm{SU}(2)$ to $\mathrm{SU}(3)$ and, more generally, to $\mathrm{SU}(N)$.}

\vspace{0.5cm}

The central objects of this monograph, \emph{spectral networks} (introduced in the following \Cref{Chap:BPSstatesclassS} from the perspective of BPS state counting in class~$\mathcal S$ theories) have found significant applications in higher-rank Teichmüller theory, as discussed in \Cref{partIV}. This connection stems from a defining feature of class~$\mathcal S$ theories: their intrinsic relation to Riemann surfaces ($\mathcal S$ stands for “surface”). In particular, the Seiberg–Witten curve of a class~$\mathcal S$ theory naturally arises as a \emph{ramified covering} (cf. \Cref{Sec:coveringspaces}) of another Riemann surface, referred to as the UV curve. Moreover, the \emph{Teichmüller space} of the UV curve as well as its \emph{mapping class group}, introduced in \Cref{subsec_Teich_sp} and \Cref{sec:hyptriang} respectively, will acquire concrete physical interpretations.

\section{Superstring theories and M-theory}

In \emph{string theory}\index{string!theory}, the fundamental objects are strings rather than point-like particles. Technically, this means that string theories are defined perturbatively as 2d quantum field theories whose fields are maps from a Riemann surface into a (pseudo-)Riemannian target space $X$, interpreted as spacetime. At scales larger than the string scale, a string appears as an ordinary particle, with its mass, charge, and other properties determined by its vibrational state. In this way, string theory defines an effective quantum field theory on $X$ at low energies.

To describe both fermions and bosons in $X$, one needs to consider supersymmetric string theories. There are five known supersymmetric string theories: two heterotic string theories (denoted HO and HE), two versions of type II string theories (type IIA and type IIB), and type I superstring theory. In what follows, we focus mainly on \emph{type II string theories}\index{string!theory!type II}, particularly type IIA and its strong coupling limit, known as \emph{M-theory}\index{M-theory}.

The quantization of superstring theories imposes constraints on the fields. On natural way to solve these is to fix the dimension of the target space $X$ to the so-called \emph{critical dimension}\index{critical!dimension}. The critical dimension for type II superstring theories is 10. At low energies, type IIA and IIB superstring theories are well approximated by type IIA and type IIB supergravity, respectively. Type A (resp. B) admits $\mathcal N =(1,1)$ (resp. $\mathcal N=(2,0)$) supersymmetry in 10-dimensional Minkowski spacetime $\mathbb{R}^{1,9}$. M-theory is not a string theory, but rather a theory of membranes; is naturally defined in 11 dimensions, and well-approximated at low energies by 11-dimensional supergravity.

The five string theories are related by a web of \emph{dualities}\index{string!duality}, meaning that they are better thought of as different descriptions of a single underlying theory rather than of distinct theories. For instance, T-duality relates type IIA and type IIB string theories. There are numerous textbooks on string theory, superstring theory, and M-theory, such as \cite{Green:2012oqa, Green:2012pqa, Polchinski:1998rq, Polchinski:1998rr, Johnson:2003glb, Becker:2006dvp}.

The fields of type IIA supergravity include a scalar field $\phi$, a metric $g$, a 2-form $B^{(2)}$, a 1-form $\lambda^{(1)}$ and a 3-form $\lambda^{(3)}$, along with their supersymmetric partners. These fields have different origins: $\phi$, $g$ and $B^{(2)}$ arise from the \emph{Neveu--Schwarz (NS) sector}\index{Neveu--Schwarz sector} of the closed string, while $\lambda^{(1)}$ and $\lambda^{(3)}$ come from the \emph{Ramond--Ramond (RR) sector}\index{Ramond--Ramond sector}. The field $\lambda^{(1)}$ is a $\mathrm{U}(1)$ gauge field, with D0-branes electrically charged and \emph{D6-branes}\index{brane!D6-} magnetically charged under it. The higher-rank gauge field $B^{(2)}$ couple electrically to fundamental strings---sometimes referred to as NS1-branes---and magnetically to \emph{NS5-branes}\index{brane!NS5-}, and $\lambda^{(3)}$ couple electrically to D2-branes, and magnetically to \emph{D4-branes}\index{brane!D4-}. Fundamental strings, NS5-branes, D2-branes, and D4-branes are respectively 2-, 6-, 3-, and 5-dimensional objects in $X$. All of these objects are thought to exist in the full type IIA superstring theory, not just in its low-energy limit. Crucially, branes are BPS, meaning they preserve a portion of the supersymmetry in $X$.

In superstring theory, \emph{D-branes}\index{brane!D-} (where D stands for Dirichlet) can also be described as boundary conditions for open strings, with the quantization of open strings in their presence defining a quantum field theory on the brane at low-enough energies, referred to as the \emph{worldvolume theory}\index{worldvolume} on the brane. For instance, D4-branes in $\mathbb{R}^{1,9}$ support a $5d$ $\mathcal N=2$ super Yang–Mills theory with gauge group $\mathrm{U}(1)$. When~$N$~D4-branes coincide, the gauge algebra enhances to $\mathfrak{u}(N)$. NS5-branes, on the other hand, support a gauge theory of self-dual antisymmetric tensors, rather than a Yang–Mills theory. For $N$ coincident NS5-branes, one obtains a superconformal theory of non-abelian self-dual antisymmetric tensors of rank~$N-1$.

The defining property of D-branes that fundamental strings can end on them implies more generally, through various sequences of dualities, that genuine branes can often end on other branes \cite{Strominger:1995ac, Townsend:1995af}. For example, D4-branes can end on NS5-branes, allowing the construction of webs of D4- and NS5-branes in $X$ that preserve some fraction of the supersymmetry of type IIA superstring theory.

In 11-dimensional supergravity, the fields include a metric $g$ and a 3-form $C^{(3)}$, along with their superpartners. M-theory admits BPS \emph{M2-branes}\index{brane!M2-} (electrically charged under $C^{(3)}$) and \emph{M5-branes}\index{brane!M5-} (magnetically charged under $C^{(3)}$). M2- and M5-branes can end on M5-branes. The relationship between type IIA superstrings and M-theory is as follows:
\begin{proposition}
    Type IIA on $X$ with string coupling constant $\lambda$ is equivalent to M-theory on $X\times S^1$, where the radius of $S^1$ is proportional to $\lambda$. 
\end{proposition}
The circle $S^1$ is referred to as the \emph{M-theory circle}\index{M-theory!circle}. In the limit $\lambda\rightarrow \infty$, the M-theory circle decompactifies. In other words, M-theory in flat 11-dimensional Minkowski space $\mathbb{R}^{1,10}$ can be understood as the infinite-coupling limit of type IIA string theory. Branes of M-theory on $X\times\mathbb{S}^1$ descend to type IIA branes on $X$: fundamental strings correspond to M2-branes wrapping $S^1$ and extending along a surface in $X$, while D2-branes in $X$ correspond to M2-branes not wrapping $S^1$. Similarly, D4-branes correspond to M5-branes wrapping $S^1$, while NS5-branes correspond to M5-branes that do not wrap the circle. 

In contrast, $D_0$ and $D_6$ branes in type IIA superstring theory do not seem to descend from any M-theory brane; in fact, $D6$-branes uplift to singularities of the $S^1$-fibration over $X$, on which M-theory is defined. In the presence of $D6$-branes, a configuration of branes in type IIA superstring theory uplifts to a configuration of M-theory branes not in $X\times S^1$, but rather in an $S^1$-fibration over $X$ with singularities. These fibrations are called generalized \emph{Taub--NUT geometries}\index{Taub--NUT geometry}.

\section{Brane configurations and \texorpdfstring{$\mathcal N=2$}{N=2} quivers}\labelx{Subsec:braneconfig}

\subsection{Simple setups} 

Let us first consider a configuration of $N$ coincident D4-branes stretching between two NS5-branes in type IIA superstring theory on $\mathbb{R}^{1,9}$, with coordinates $x^0,\dots,x^9$. Specifically, the NS5-branes extend along the directions $0,1,2,3,4,5$ and are (classically) located at $x^6=0$ and $x^6=L$, respectively, while the D4-branes extend along $0,1,2,3$ and the segment~$[0,L]$ along $x^6$, and are located at $x^{4,5}=0$. Moreover, all branes are located at $x^{7,8,9}=0$. This setup is depicted on the left of \Cref{fig:simpleIIAbraneconfiguration}. 

\begin{figure}[!ht]
    \centering
    \includegraphics[scale=0.83]{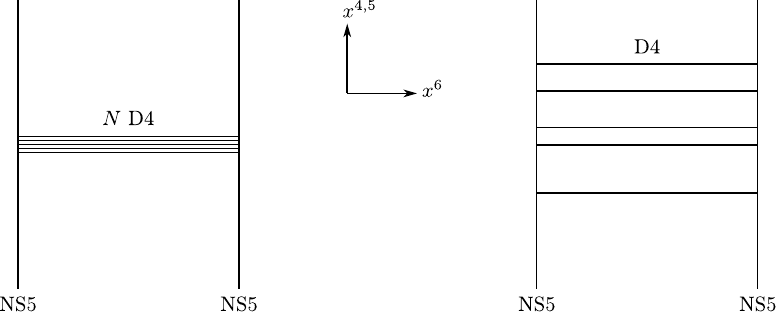}
    \caption{Two type IIA brane configurations}
    \labelx{fig:simpleIIAbraneconfiguration}
\end{figure}

\begin{fact}[see \cite{Hanany:1996ie,Witten:1997sc}]
    Such brane configurations preserve one fourth of the type IIA supersymmetries in $\mathbb{R}^{1,9}$, i.e., 8 supercharges out of 32.
\end{fact}

Due to the D4-branes having finite extension along $x^6$, their worldvolume theory effectively reduces to four dimensions at low-enough energies: specifically, $4d$ $\mathcal N=2$ $\mathfrak{u}(N)$ (or rather, $\mathfrak{su}(N)$, as we will explain shortly) super-Yang-Mills (SYM) theory along $0,1,2,3$. The energy threshold under which the worldvolume theory on the D4-branes effectively becomes four-dimensional is inversely proportional to $L$.

Let us now consider a related setup in which the D4-branes are separated along the $x^{4,5}$ directions in a generic way, as shown on the right in \Cref{fig:simpleIIAbraneconfiguration}. The worldvolume theory on the D4-branes becomes a $4d$ $\mathcal N=2$ theory with a $\mathrm{U}(1)^N$ (or rather, $\mathrm{U}(1)^{N-1}$) gauge group, instead of $\mathrm{U}(N)$. This brane configuration is interpreted as corresponding to a specific point on the Coulomb branch of $4d$ $\mathcal N=2$ $\mathrm{U}(N)$ SYM, as separating the stack of D4-branes amounts to breaking $\mathrm{SU}(N)$ down to its Cartan subgroup $\mathrm{U}(1)^{N-1}$.

\subsection{More general setups} 

We will be interested in generalizations of the simple setups we have introduced, depicted in \Cref{fig:involvedIIAbraneconfiguration}. These brane configurations consist of $k$ NS5-branes positioned at fixed values of $x^6$, together with D4-branes stretching between them. There can be semi-infinite D4-branes, attached to an NS5-brane at one end and extending to $\pm\infty$ along $x^6$. Moreover, one can add D6-branes extending along $0,1,2,3,7,8,9$ without breaking additional supersymmetries. Though we will come back to this possibility shortly, we first consider configurations of D4- and NS5-branes only.

\begin{figure}[!ht]
    \centering
    \includegraphics[width=\textwidth]{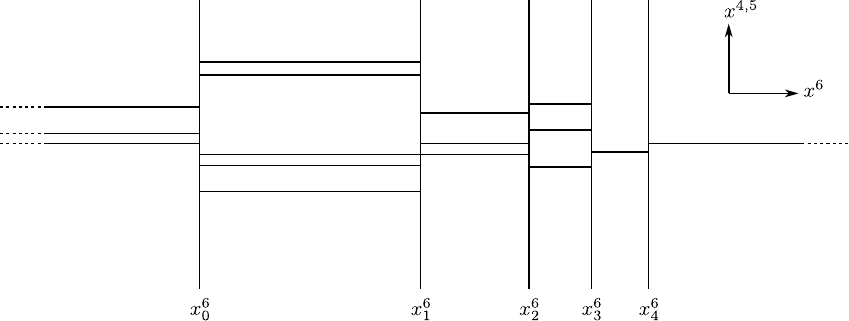}
    \caption{A more involved type IIA brane configuration}
    \labelx{fig:involvedIIAbraneconfiguration}
\end{figure}

\subsection{Backreactions} 

One can improve our depiction of the brane configurations we are considering by taking into account the backreaction of the D4-endpoints within the NS5-branes. These endpoints are charged codimension-2 objects in the NS5-worldvolumes. More precisely, the D4-endpoints are magnetic monopoles, or vortices, for the worldvolume theory on the NS5-branes.

These sources cause NS5-branes to bend logarithmically. A direct consequence of this bending is that, in general, the NS5-branes are not located at a fixed value of $x^6$. Denoting $v=x^4+\I x^5\in\mathbb{C}$, let us consider a single D4-brane at $v=0$ ending on an NS5-brane that is classically positioned at $x^6=0$. The NS5-brane tends to minimize its volume. This, together with the presence of a D4-endpoint behaving as a codimension-2 source, implies that far enough from the D4-endpoint, the position of the NS5-brane along $x^6$ obeys the two-dimensional Laplace equation:
\begin{equation*}
    \nabla^2x^6(v) = 0\; .
\end{equation*}
This implies that the NS5’s position along $x^6$ takes the form $x^6(v) = k\ln(\vert v\vert)+c$, where $k$ and $c$ are constants. 

More generally, for $q_L$ D4-branes located at $v=a_1,\dots,a_{q_L}$ ending on the left side of the NS5-brane, and $q_R$ D4-branes $v=b_1,\dots,b_{q_R}$ ending on the right side of the NS5-brane, the position of NS5-brane position along $x^6$ reads:
\begin{equation}\labelx{eq:backreactionD4endpoints}
    x^6(v) = k \sum_{i=1}^{q_L} \ln\vert v-a_i\vert - k \sum_{j=1}^{q_R} \ln\vert v-b_j\vert + c\; .
\end{equation}
Thus, the NS5-brane approaches a specific value of $x^6$ as $\vert v\vert \rightarrow \infty$ only if $q_L = q_R$. Only in these cases can NS5-branes be assigned a distinct $x^6$-value, and this value indicates only that the position of the NS5-brane along $x^6$ asymptotically tends toward this fixed value as $\vert v\vert \rightarrow \infty$.

Let us now consider the motion of the D4-branes along $v$, i.e., variations in the complex parameters $a_i$ and $b_j$. As the D4-branes move, they induce deformations in the NS5-branes, which, in turn, affect the kinetic energy of the D4-branes. One can distinguish between deformations of the D4-brane stack that incur a finite energy cost, referred to as \emph{normalizable deformations}\index{normalizable deformation}, and others. As shown in \cite[Sect. 2]{Witten:1997sc}, a deformation is normalizable if the following condition holds: 
\begin{equation}\labelx{eq:freezing}
\sum_i a_i - \sum_j b_j = q \in\mathbb{C}\; ,
\end{equation} 
where $q$ is constant. This constraint implies that there is no relative displacement between the center of mass of the two D4-brane sets across the NS5. When considering only normalizable deformations (as we will), $q\in\mathbb{C}$ acts as a constant associated with the NS5-brane in the configuration.

\subsection{Gauge-theoretic interpretation}

Consider a general configuration as in \Cref{fig:involvedIIAbraneconfiguration} with $k+1$ NS5-branes labeled as $\mathrm{NS5}_\alpha$ for $\alpha=0,\dots,k$, and, for each $1\leq\alpha \leq k$, $N_\alpha$ D4-branes stretching between the consecutive $\mathrm{NS5}_{\alpha-1}$ and $\mathrm{NS5}_\alpha$. We extend the definition of $N_\alpha$ to include $N_0$, the number of semi-infinite D4-branes to the left of the configuration, and $N_{k+1}$ the number of semi-infinite D4-branes to the right of the configuration. 

Naively, for each $1\leq \alpha\leq k$ each stack of $N_\alpha$ D4-branes contributes a $\mathrm{U}(N_\alpha)$ factor to the gauge group of the worldvolume theory on the D4-branes. The D4-branes that extend to infinity do not contribute to the gauge group, however, they give rise to $N_0$ hypermultiplets in the antifundamental representation of $\mathrm{U}(N_1)$, and $N_{k+1}$ hypermultiplets in the fundamental representation of $\mathrm{U}(N_k)$. More precisely:

\begin{proposition}
    \Cref{eq:freezing} implies that for each $2\leq \alpha\leq n-1$, the diagonal $\mathrm{U}(1)$ factor in the product $\mathrm{U}(N_{\alpha-1})\times\mathrm{U}(N_\alpha)$ is frozen. Therefore, the gauge group $G$ of the worldvolume theory on the D4-branes is:
    \begin{equation*}
        G = \prod_{\alpha=1}^k \mathrm{SU}(N_\alpha)\; .
    \end{equation*}
\end{proposition}

The semi-infinite D4-branes still give rise to $N_0$ hypermultiplets in the antifundamental of $\mathrm{SU}(N_1)$, and $N_{k+1}$ hypermultiplets in the fundamental of $\mathrm{SU}(N_k)$. These hypermultiplets contribute to $\mathrm{U}(N_0)$ and $\mathrm{U}(N_{k+1})$ flavor factors, respectively. For instance, $M$ hypermultiplets in the fundamental of $\mathrm{SU}(N)$ can be denoted as:
\begin{equation*}
    (Q^a_i,\widetilde{Q}^i_a), \quad a=1,\dots,N\; , \; \;\text{and} \; i=1,\dots,M\; .
\end{equation*}
The flavor symmetry\footnote{We are implicitly assuming that $N$ is greater than 2. For $N=2$, the corresponding flavor factor would enhance to $\mathrm{SO}(2N)$ (cf. \Cref{prop:flavorsymmSU2}).} here is $\mathrm{U}(M) \simeq \mathrm{SU}(M) \times \mathrm{U}(1)$, where the simple factor $\mathrm{SU}(M)$ acts on the flavor index $i$. Specifically, the $Q$'s transform in the antifundamental representation of $\mathrm{SU}(M)$, while the $\widetilde{Q}$'s transform in the fundamental. Under the abelian factor $\mathrm{U}(1)$, the $Q$'s have charge $+1$, and the $\widetilde{Q}$'s have charge $-1$. Similarly, for every NS5-brane $\mathrm{NS}5_\alpha$, $2\leq \alpha \leq k-1$, there is a hypermultiplet in the \emph{bifundamental representation}\index{representation!bifundamental} of $\mathrm{SU}(N_{\alpha-1})\times \mathrm{SU}(N_\alpha)$, i.e.:
\begin{equation*}
    (Q^a_b,\widetilde{Q}^b_a), \quad  a=1,\dots,N_{\alpha}\; , \;\; \text{and} \; b=1,\dots,N_{\alpha-1}\; .
\end{equation*}
Such bifundamental hypermultiplets can be massive (BPS) hypermultiplets, with bare mass identified with the constant $q_\alpha$ appearing in \Cref{eq:freezing}. It is convenient to represent such gauge theories as \emph{$\mathcal N=2$ quivers}\index{quiver!$\mathcal N=2$}, as in \Cref{fig:N=2quiver}.

\begin{figure}[!ht]
    \centering
    \includegraphics[scale=0.83]{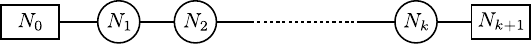}
    \caption{Representation of the theory at hand as a linear $\mathcal N=2$ quiver.}
    \labelx{fig:N=2quiver}
\end{figure}

\begin{definition}
    An \emph{$\mathcal{N}=2$ quiver} is a finite, connected graph with two types of nodes: round nodes, corresponding to $\mathrm{SU}$ gauge factors, and square nodes, representing $\mathrm{U}$ flavor factors. Each node, whether round or square, is labeled with a positive integer: a round node labeled by $N$ depicts an $\mathrm{SU}(N)$ gauge factor, while a square node labeled by $N$ depicts a $\mathrm{U}(N)$ flavor factor. The edges are unoriented and connect either two round nodes or a round node with a square node, they represent bifundamental hypermultiplets.
\end{definition}

More generally, $\mathcal N=1$ gauge theories with gauge group a product $G=\prod G_i$ and chiral superfields in bifundamental representations of the gauge factors can be represented as a quiver defined as in \Cref{def:quiver}---namely, as finite oriented graphs. In this framework, the nodes of the quiver represent the factors $G_i$, while an arrow from $G_i$ to $G_j$ indicates a bifundamental chiral superfield. A chiral superfield in the adjoint representation of $G_i$ is represented by a loop at the node $G_i$. One can adapt the definition to account for flavor groups.
    
A hypermultiplet in representation $R$ corresponds to a pair of chiral superfields, one in $R$ and the other in the conjugate representation $\overline{R}$. Thus, the $\mathcal{N}=1$ quiver for an $\mathcal{N}=2$ gauge theory with gauge group $G = \prod G_i$ and bifundamental hypermultiplets contains arrows that always occur in pairs with opposite orientations. By convention, in $\mathcal{N}=2$ quivers, these pairs are depicted as a single undirected edge. Each $\mathcal{N}=2$ vector superfield also includes an $\mathcal{N}=1$ chiral superfield in the adjoint representation, which is generally omitted in $\mathcal{N}=2$ quiver diagrams.

In some definitions of quivers, one excludes loops and pairs of oppositely oriented arrows. However, loops and such pairs are prevalent in $\mathcal{N}=2$ quivers. This is compatible with the structure of a \emph{quiver with potential}\index{quiver!with potential}, which permits these elements (see, for instance, \cite{Derksen:2008qui}).

\vspace{0.3cm}

On general grounds, one expects that the gauge coupling $\mathrm{SU}(N_\alpha)$ gauge theory, corresponding to a stack of $N_\alpha$ D4-branes stretching between NS5-branes classically positioned at $x^6_{\alpha-1}$ and $x^6_{\alpha}$, is given by:
\begin{equation}\labelx{eq:gaugecouplings}
    \frac{1}{g_\alpha^2} = \frac{x^6_\alpha-x^6_{\alpha-1}}{\lambda}\; , 
\end{equation}
where $\lambda$ is the type IIA string coupling. However, we have seen that in general $x^6_{\alpha-1}$ and $x^6_\alpha$ depend on the complex coordinate $v$, which parameterizes the position of the D4-branes in the $v=x^4+\I x^5$-plane. As before, let $N_{\alpha-1}$ be the number of D4-branes on the left of the NS5-brane at $x^6_{\alpha-1}$, and $N_{\alpha+1}$ the number of D4-branes on the right of the NS5-brane at $x^6_\alpha$. If both $N_{\alpha-1}$ and $N_{\alpha+1}$ are less than $N_\alpha$, at large $v$, $x^6_\alpha-x^6_{\alpha-1}$ grows logarithmically as $(2N_\alpha-N_{\alpha-1}-N_{\alpha+1}) k \ln\vert v\vert$. 

From the perspective of the $\mathrm{SU}(N_\alpha)$ gauge theory, the neighboring stacks of $N_{\alpha-1}$ and $N_{\alpha+1}$ D4-branes effectively contribute to the renormalization of the gauge coupling $g_\alpha$ in the same way as $N_{\alpha-1}+N_{\alpha+1}$ fundamental hypermultiplets. Thus:

\begin{proposition}
    The corrected equation
\begin{equation}\labelx{eq:gaugecouplingscorrected}
    \frac{1}{g_\alpha^2}(v) = \frac{x^6_\alpha(v)-x^6_{\alpha-1}(v)}{\lambda} 
\end{equation}
reflects the one-loop beta function of the gauge group $\mathrm{SU}(N_\alpha)$, for which $b_1=2N_\alpha-N_{\alpha-1}-N_{\alpha+1}$. Correspondingly, the distance $\vert v\vert$ is interpreted as an energy scale: large values of~$\vert v\vert$ correspond to the UV, where $g_\alpha^2$ tends to $0$, while small values of $\vert v\vert$ correspond to the IR.
\end{proposition}

\subsection{More general \texorpdfstring{$\mathcal N=2$}{N=2} quiver gauge theories}

One can add infinite D6-branes extending along the directions $x^{0,1,2,3,7,8,9}$, while being located at fixed values of $x^6$ and $v=x^4+\I x^5$, without breaking any additional supersymmetry compared to the configurations involving only D4-branes and NS5-branes. Therefore, the worldvolume theory on the D4-branes remains a $4d$ $\mathcal N=2$ theory. Adding D6-branes introduces fundamental hypermultiplets for some of the $\mathrm{SU}(N_\alpha)$ gauge factors, leading to more general linear quiver gauge theories depicted in \Cref{fig:N=2quiverD6}. The quiver in \Cref{fig:N=2quiverD6} corresponds to a brane configuration where $F_\alpha$ D6-branes are placed between the NS5-branes $\mathrm{NS5}_{\alpha-1}$ and $\mathrm{NS5}_{\alpha}$.

\begin{figure}[!ht]
    \centering
    \includegraphics[scale=0.83]{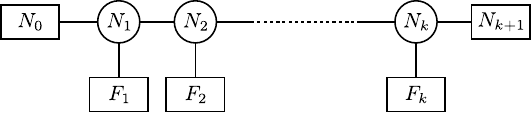}
    \caption{More general linear $\mathcal N=2$ quivers.}
    \labelx{fig:N=2quiverD6}
\end{figure}

One can also consider setups for which the $x^6$ direction is taken to be compact, i.e. forming a circle. There can be no semi-infinite D4-branes in such setups. The quiver that encodes the structure of the low-energy theory on the D4-branes forms a closed loop, and is referred to as a \emph{necklace quiver}\index{quiver!$\mathcal N=2$!necklace}. Necklace quivers are examples of $\mathcal N=2$ quivers which are not linear: the gauge group is the product of $k$ $\mathrm{SU}(N_\alpha)$ factors\footnote{The gauge group of the worldvolume theory on the D4-branes also contains a $\mathrm{U}(1)$ factor---the diagonal $\mathrm{U}(1)$ subgroup of $\prod\mathrm{U}(N_\alpha)$. This $\mathrm{U}(1)$ factor is free and decouples in the IR.}, each corresponding to a stack of D4-branes stretching between consecutive NS5-branes. Additionally, if D6-branes are introduced in this setup, they contribute extra flavor groups as before. The resulting theories are called \emph{elliptic models}\index{elliptic model} for reasons that will become clear shortly.

\subsubsection*{Conformal quivers}

In Lagrangian $4d$ $\mathcal{N}=2$ theories, the renormalization of each gauge coupling is perturbatively exact at one loop. The one-loop beta function coefficient for the gauge factor $\mathrm{SU}(N_\alpha)$ in the linear quiver theory of \Cref{fig:N=2quiverD6} is given by: 
\begin{equation*} 
b_{1,\alpha} = 2N_\alpha - N_{\alpha-1} - N_{\alpha+1} - F_\alpha\; . 
\end{equation*} 
If, for each $\alpha$, one has $N_\alpha \geq N_{\alpha-1} + N_{\alpha+1}$, it is possible to adjust $F_\alpha$ such that $b_{1,\alpha} = 0$. Quivers satisfying this condition are believed to be conformal, and are hence called \emph{conformal quivers}\index{quiver!$\mathcal N=2$!conformal}.

\begin{example}
    Taking $N_\alpha = N$ for $0\leq \alpha\leq k+1$ and $F_\alpha=0$ for $1\leq \alpha\leq k$ defines conformal quivers with gauge group $\mathrm{SU}(N)^k$, where the first and last gauge groups have $N$ fundamental hypermultiplets. When $k=1$, the theory is $\mathrm{SU}(N)$ SYM theory with $2N$ fundamental flavors.
\end{example}

\begin{figure}[!ht]
    \centering
    \includegraphics[width=\textwidth]{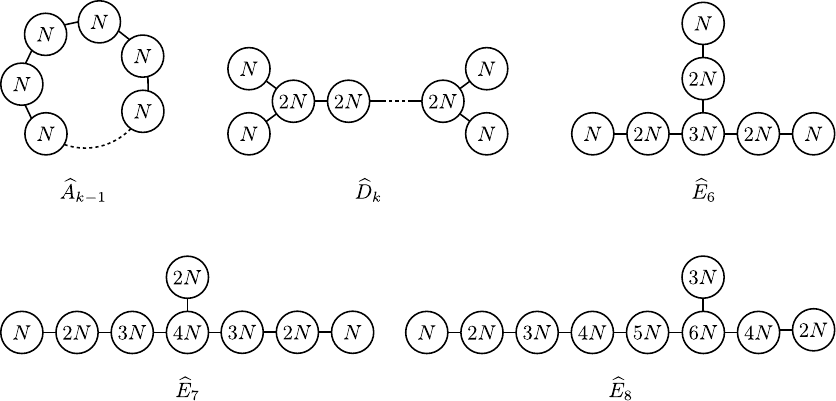}
    \caption{Balanced quivers.}
    \labelx{fig:conformalquivers}
\end{figure}

\begin{example}
    The quivers shown in \Cref{fig:conformalquivers}---which correspond to affine $A$, $D$, $E$ Dynkin diagrams\footnote{In fact, these ADE quivers are the only conformal quivers without flavor groups \cite{Gaiotto:2008ak}. This follows from the observation that the conditions for conformality define affine Cartan matrices.}---are conformal. They are sometimes called \emph{balanced quivers}\index{quiver!$\mathcal N=2$!balanced}. The $\widehat{A}_{k-1}$ cases are necklace quivers. Affine quivers of type $\widehat{D}_n$ can also be realized in type IIA superstrings by introducing NS-\emph{orientifold planes}\index{orientifold plane} \cite{Kapustin:1998fa}. Some $4d$ $\mathcal N=2$ quivers do not admit any known brane realization; for instance, no brane construction is known for the $\widehat{E}_6$, $\widehat{E}_7$ or $\widehat{E}_8$ quiver theories at present.
\end{example}

\section{M-theory uplifts and Seiberg--Witten curves}\labelx{Sec:Mtheoryuplifts}

We have identified a broad class of $4d$ $\mathcal{N}=2$ gauge theories that can be realized as worldvolume theories on D4-branes in type IIA superstring theory, with configurations involving D4-, NS5-, and possibly D6-branes. When each stack of ``gauge'' D4-branes consists of coincident branes, the theory is an $\mathcal{N}=2$ quiver with a gauge group that is a product of simple factors. The difference in position of neighboring stacks along $v$ encodes the bare mass of bifundamental hypermultiplets, while the separation of semi-infinite D4-branes or D6-branes encodes the bare mass of (anti-)fundamental hypermultiplets. Normalizable deformations of such configurations, where D4-branes are separated in the $v = x^4 + \I x^5$ plane, correspond to the Coulomb branch of these $\mathcal{N}=2$ quivers. We are now going to explain how the low-energy dynamics on $\mathcal{C}$, encoded in the Seiberg--Witten fibration over $\mathcal{C}$, arises naturally from string theory, or more specifically, from the uplift of type IIA brane configurations to \emph{M-theory}\index{M-theory}.

\subsection{M-theory uplift}

The M-theory limit of type IIA superstring theory corresponds to $\lambda \rightarrow \infty$. \Cref{eq:gaugecouplings,eq:gaugecouplingscorrected} show that one can take this limit without altering the $4d$ gauge couplings, if one simultaneously rescales the $x^6$ direction. Therefore, the $\mathcal N=2$ quivers that can be realized as brane configurations in type IIA superstring theory can equivalently be realized as brane configurations in M-theory, called \emph{M-theory uplifts}\index{M-theory!uplift} of the type IIA configurations.

Let us denote by $x^{10}\sim x^{10}+2\pi R$ the periodic coordinate on the M-theory circle $S^1$ (where $R$ is proportional to $\lambda$). Both D4-branes and NS5-branes descend from M5-branes in M-theory: a D4-brane extending along $x^{0,1,2,3,6}$ descends from an M5-brane extending along $x^{0,1,2,3,6,10}$, while an NS5-brane along $x^{0,1,2,3,4,5}$, placed at a specific value of $x^6$, descends from an M5-brane extending along the same coordinates, and placed at fixed values of $x^6$ and $x^{10}$. For example, $N$ parallel D4-branes intersecting two NS5-branes uplift to $N$ M5-branes extending along $x^{0,1,2,3,6,10}$, hence wrapping the cylinder parameterized by $x^{6,10}$, along with two M5-branes extending along $x^{0,1,2,3,4,5}$ and placed at specific points on the cylinder, as illustrated in \Cref{fig:Mtheoryuplift}. This M-theory configuration corresponds to $\mathcal N=2$ $\mathrm{SU}(N)$ SYM with $F=2N$ fundamental flavors at the origin of the Coulomb branch.

\begin{figure}[!ht]
    \centering
    \includegraphics[width=\textwidth]{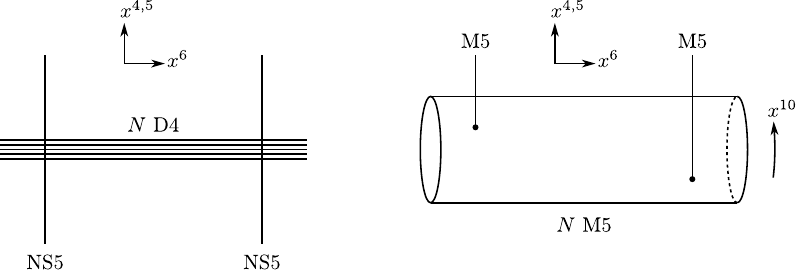}
    \caption{Uplifting type IIA brane configurations to M-theory.}
    \labelx{fig:Mtheoryuplift}
\end{figure}

In general type IIA configurations of D4-branes stretching between NS5-branes, the position of NS5-branes along the $x^6$ direction depends on $v$. The analogous statement in M-theory is that the position along $x^6 + ix^{10}$ of the M5-brane uplift of any NS5-brane, also depends on $v$. In order for the M-theory configuration to be supersymmetric, $x^6 + \I x^{10}$ must depend holomorphically on $v$. This leads to the following generalization of \Cref{eq:backreactionD4endpoints}, where $s:=(x^6+\I x^{10})/R$:
\begin{equation}\labelx{eq:bendingcomplexified}
    s(v) = \sum_{i=1}^{q_L} \ln (v-a_i) - \sum_{j=1}^{q_R} \ln (v-b_j) + c\; .
\end{equation}

Letting $\tau_\alpha = 4\pi \I/g^2_\alpha + \theta_\alpha/(2\pi)$ be the usual complexified gauge coupling for the $\mathrm{SU}(N_\alpha)$ gauge theory on the worldvolume of $N_\alpha$ D4-branes stretched between two NS5-branes, $\mathrm{NS5}_{\alpha-1}$ and $\mathrm{NS5}_\alpha$, the expression in \Cref{eq:gaugecouplingscorrected} generalizes to:
\begin{equation*}
    \I \tau_\alpha(v) = s_\alpha(v) - s_{\alpha-1}(v)\; .
\end{equation*}
This generalization is interpreted as the perturbative renormalization of the coupling $\tau_\alpha$. One can see that the uplift to M-theory naturally incorporates theta angles into the discussion from \Cref{Subsec:braneconfig}.

The M-theory uplift of a generic configuration of D4-branes and NS5-branes as the one depicted in \Cref{fig:involvedIIAbraneconfiguration}, must correspond to a Riemann surface $\Sigma$ in the complex two-dimensional space $Q = \mathbb{C}_v\times \mathbb{C}^\times_s$, in order to preserve the same amount of supersymmetry as the original type IIA brane setup. Normalizable deformations in type IIA superstring theory uplift to normalizable deformations in M-theory \cite{Gaiotto:2009hg}. In type IIA limit of M-theory where the M-theory circle shrinks, $\Sigma$ must recover the original brane configuration. Two key insights are the following:
\begin{itemize}
    \item The singularities present in the type IIA brane configuration are generically smoothed out in $\Sigma$. This implies that knowledge of the low-energy effective approximation of M-theory is sufficient to determine the low-energy physics of the worldvolume theory on the D4-branes.
    \item The Riemann surface $\Sigma$ is the \emph{Seiberg--Witten curve}\index{Seiberg--Witten!curve} of the $4d$ $\mathcal N=2$ theory on the D4-branes. 
\end{itemize}

This second point can be argued as follows. The worldvolume theory on an M5-brane describes a self-dual two-form gauge field $\beta$, i.e., a two-form satisfying $T = \mathrm{d}\beta = \star_6 \mathrm{d}\beta$. Assume the M5-brane is compactified on a Riemann surface $\Sigma$, so that it takes the form $\mathbb{R}^{1,3} \times \Sigma$. In the scenario of interest, $\Sigma$ can be naturally compactified by adding a finite number of points, resulting in a compact Riemann surface $\overline{\Sigma}$ of genus $g$. The zero modes of $\beta$ on $\overline{\Sigma}$ then effectively give rise to $g$ abelian gauge fields on $\mathbb{R}^4$, with complexified coupling constants described by the Jacobian $J(\overline{\Sigma})$. This arises from the fact that each harmonic one-form $\Lambda$ on $\overline{\Sigma}$ provides a way to embed solutions to Maxwell's equations in $\mathbb{R}^{1,3}$ into solutions for $T$ in $\mathbb{R}^{1,3} \times \Sigma$ \cite{Verlinde:1995mz}.

In other words, the low-energy physics on the Coulomb branch of the D4-brane worldvolume theory is described by the Riemann surface $\Sigma$ and its Jacobian variety $J(\overline{\Sigma})$. The complex structure of $\Sigma$ is determined by the specific brane configuration, or equivalently, by the point on the Coulomb branch corresponding to that configuration. The Jacobian variety $J(\overline{\Sigma})$ encodes the low-energy effective action of the resulting abelian gauge fields in $\mathbb{R}^{1,3}$. The pair $(\Sigma, J(\overline{\Sigma}))$ represents the fiber of the Hitchin integrable system over a particular point on the Coulomb branch, as discussed in \Cref{Sec:SWgeneral}. This identifies $\Sigma$ with the Seiberg--Witten curve of the theory at hand, at the specific point on its Coulomb branch corresponding to the given brane configuration.

\begin{remark}
    The Seiberg--Witten differential $\lambda$ for linear $\mathcal N=2$ quivers can be identified with the Liouville 1-form on $Q = \mathbb{C}_v\times\mathbb{C}^\times_s$, seen as the holomorphic cotangent bundle of $\mathbb{C}^\times_s$. In coordinates:
    \begin{equation*}
        \lambda = v\frac{\mathrm{d}t}{t}\; ,
    \end{equation*}
    where $t=\exp(-s)$ is a globally-defined complex coordinate on $\mathbb{C}^\times_s$.
\end{remark}

\subsection{Computation of Seiberg--Witten curves}

Consider a type IIA brane configuration with two NS5-branes, labeled as $\mathrm{NS5}_0$ and $\mathrm{NS5}_1$, $N_0$ semi-infinite D4-branes to the left of $\mathrm{NS5}_0$, $N$ D4-branes stretching between $\mathrm{NS5}_0$ and $\mathrm{NS5}_1$, and $N_2$ semi-infinite D4-branes to the right of $\mathrm{NS5}_1$. This configuration is depicted in \Cref{fig:computationSWcurve}, where we schematize the logarithmic bending of the NS5-branes at large values of $v$ (at small values of $v$ and $x^6$, the type IIA brane description is not accurate: there is no known precise description of D4-endings within NS5-branes).

\begin{figure}[!ht]
    \centering
    \includegraphics[scale=0.83]{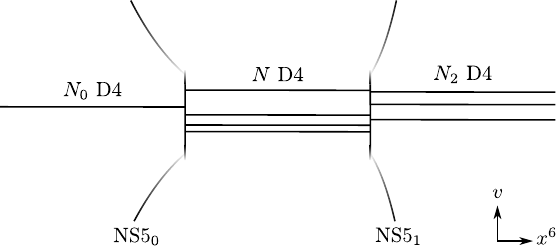}
    \caption{Type IIA brane configuration.}
    \labelx{fig:computationSWcurve}
\end{figure}

Since $s=(x^6 + \I x^{10})/R$ is a multivalued function, we use $t = \E^{-s}$ to parametrize $\mathbb{C}^\times_{6,10}$ instead of $s$. The M-theory uplift of the configuration depicted in \Cref{fig:computationSWcurve} consists of a single M5-brane extending along $x^{0,1,2,3}$ and wrapping a Riemann surface $\Sigma$ in $\mathbb{C}_v \times \mathbb{C}^\times_t$. This curve $\Sigma$---which is the Seiberg--Witten curve of the worldvolume theory on the D4-branes, in the present case $\mathrm{SU}(N)$ SYM with $N_0+N_2$ flavors---is described by a polynomial equation:
\begin{equation*}
    F(t,v)=0\; .
\end{equation*}

The precise form of $F$ can be derived as follows:
\begin{itemize}
    \item At generic fixed values $v_0$ of $v$, the roots of $F(t,v_0)$, viewed as a polynomial in $t$, correspond to the positions of the NS5-branes in the type IIA configuration. Therefore, for the configuration shown in \Cref{fig:computationSWcurve}, $F(t,v)$ must be a quadratic polynomial in $t$, i.e.:
    \begin{equation*}
        F(t,v) = A(v)t^2 + B(v)t + C(v)\; .
    \end{equation*}
    \item At fixed values $t_0$ of $t$, such that $x^6 = -R\ln\vert t \vert$ lies between the classical positions of $\mathrm{NS5}_0$ and $\mathrm{NS5}_1$ along the $x^6$ direction, we expect $N$ roots for $F(t_0,v)$, corresponding to the positions of the D4-branes stretching between $\mathrm{NS5}_0$ and $\mathrm{NS5}_1$. As a result, $A(v)$, $B(v)$, and $C(v)$ are polynomials of degree at most $N$.
    \item At a zero $v_0$ of $C(v)$, one of the roots of $F(t,v_0)$ goes to $t=0$, i.e. $x^6=\infty$, which is to be interpreted as translating the presence of a semi-infinite D4-brane to the right of $\mathrm{NS5}_1$ placed at $v=v_0$. Thus, the degree of $C(v)$ is $N_2$.
    \item At a zero $v_0$ of $A(v)$, one of the roots of $F(t,v_0)$ goes to $t=\infty$, i.e. $x^6=-\infty$, which is to be interpreted as translating the presence of a semi-infinite D4-brane to the left of $\mathrm{NS5}_0$ placed at $v=v_0$. Thus, the degree of $A(v)$ is $N_0$.
\end{itemize}

\begin{example}
    In the case of pure $4d$ $\mathcal{N}=2$ $\mathfrak{su}(N)$ SYM, we can rescale $t$ so that the constant polynomials $A$ and $C$ are equal to 1. This gives the Seiberg--Witten curve:
    \begin{equation}\labelx{eq:SWcurvepuresuN}
        t^2+B(v)t+1=0\; ,
    \end{equation}
    where $B(v)$ is a polynomial of degree $N$. In particular, for $N=2$, by letting $z=t$ and rescaling/shifting v into a new coordinate $x$, we obtain an equivalent equation of the form:
    \begin{equation*}
        \Lambda^2z + \frac{\Lambda^2}{z} = x^2-u\; ,
    \end{equation*}
    which is the Seiberg--Witten curve for pure $\mathfrak{su}(2)$ SYM as given in \Cref{eq:SWpureSU2}. 
    
    Now, letting $\widetilde{t} = t+ B(v)/2$, we rewrite the equation as:
    \begin{equation}\labelx{eq:newSWcurve}
        \widetilde{t}^2 = \frac{B(v)^2}{4}-1\; .
    \end{equation}
    By appropriately rescaling and shifting $v$, we can further assume that $B(v)$ has the following form:
    \begin{equation*}
        B(v) = v^N + u_2v^{N-2} + u_3v^{N-3} + \dots + u_N\; .
    \end{equation*}
    \Cref{eq:SWcurvepuresuN} with this $B(v)$ is a standard expression of the Seiberg--Witten curve for pure $\mathfrak{su}(N)$ SYM.
\end{example}

\begin{remark}
    Note that for $\vert t\vert$ large, the roots of \Cref{eq:SWcurvepuresuN} are approximately given by $t\propto v^N$. Similarly, for $\vert t\vert$ small, the roots of \Cref{eq:SWcurvepuresuN} are approximately given by $t\propto v^{-N}$. At fixed $t$, $F(t,v)$ viewed as a polynomial in $v$ has $N$ roots for any generic values of $t$, not just those between the classical positions of the NS5-branes along $x^6$: $t\propto v^{\pm N}$ matches the asymptotic form of \Cref{eq:bendingcomplexified}, which reflects the logarithmic bending of the NS5-branes.
\end{remark}

\begin{example}
    The case $N_0=0$, $N=2$, and $N_2=1$, corresponds to $\mathrm{SU}(2)$ SYM theory with one fundamental hypermultiplet. The Seiberg--Witten curve writes:
    \begin{equation*}
        t+B(v)+\frac{\alpha v+\beta}{t} = 0\; ,
    \end{equation*}
    where $B(v)$ is a polynomial of degree 2, and $\alpha$ and $\beta$ are constants. By appropriately rescaling and shifting $v$ and $t$, this equation can be written in the form of \Cref{eq:SWcurveF=1}:
    \begin{equation*}
        \Lambda_1^2z + \frac{2\Lambda_1(x-m_1)}{z} = x^2-u\; .
    \end{equation*}
    \end{example}

    \begin{example}
    More generally, if $N_0=0$, $N$ is arbitrary and $N_2=F$, \Cref{eq:newSWcurve} becomes:
    \begin{equation*}
        \widetilde{t}^2 = \frac{B(v)^2}{4}-f\prod_{j=1}^F(v-m_j)\; ,
    \end{equation*}
    with $f$ a complex constant and where, possibly after shifting $v$:
    \begin{equation*}
        B(v) = e(v^n+u_2v^{n-2}+\dots+u_{n-1}v+u_n)\; ,
    \end{equation*}
    with $e$ another complex constant. 
    
    For $F\neq 2N$, one can rescale $\widetilde{t}$ and $v$ to set $e=f=1$. In contrast, for $F=2N$, one can only fix a single combination of $e$ and $f$, leaving an unfixed degree of freedom. This additional parameter corresponds to the coupling constant of the theory, which is exactly marginal when $F=2N$. In the brane picture the two NS5-branes become parallel as $\vert v\vert \rightarrow\infty$. The distance between the NS5-branes at large $\vert v\vert $ approaches a constant interpreted as the dimensionless UV gauge coupling of the theory.
\end{example}

\begin{remark}
    When $N_0+N_2 > 2N$, the NS5-branes also remain parallel as $\vert v\vert \rightarrow \infty$. This is interpreted as the fact that the M-theory uplift of the corresponding type IIA brane configurations defines a UV completion of these infrared-free theories. 
\end{remark}

More general setups involving D4-branes and NS5-branes can be studied using the same approach, allowing for the derivation of the Seiberg--Witten curves for a wide class of linear $4d$ $\mathcal N=2$ quiver theories. These curves take the form of a Riemann surface in $Q=\mathbb{C}_v \times \mathbb{C}^\times_t$, with the Seiberg--Witten differential represented by the Liouville 1-form:
\begin{equation*}
    \lambda = x\frac{\mathrm{d}t}{t}\; .
\end{equation*}
The most general linear quivers are obtained by incorporating D6-branes into these configurations. The M-theory uplift of setups that include D6-branes results in more complex geometries, specifically, the complex space $Q$ is replaced by a \emph{Taub--NUT geometry}\index{Taub--NUT geometry}. The Seiberg--Witten curve is still a Riemann surface in $Q$. One can also compute the Seiberg--Witten curves for necklace quivers. In this scenario, $Q$ corresponds to the holomorphic cotangent bundle of an elliptic curve $E$, hence the name \emph{elliptic models}\index{elliptic model}.

\begin{example}
Four-dimensional $\mathcal N=2$ $\mathfrak{su}(N)$ SYM theory with a massless hypermultiplet in the adjoint representation—i.e., $4d$ $\mathcal N=4$ $\mathfrak{su}(N)$ SYM—is described by $N$ D4-branes extending along the $x^{0,1,2,3,6}$ directions, where the $x^6$ direction is compactified on a circle of length $2\pi L$. A single NS5-brane is positioned at a point along this circle. When this configuration is uplifted to M-theory, the $x^6$ and $x^{10}$ directions combine to form an elliptic curve $E$, defined such that when $x^6$ shifts by $2\pi L$, the $x^{10}$ coordinate also shifts by $\theta R$ for some $\theta \in [0,2\pi[$. The type IIA branes in this setup become a single M5-brane that wraps a degree-$N$ ramified covering $\Sigma$ of $E$ in the space $E \times \mathbb{C}_v \simeq T^*_{1,0} E$. This surface $\Sigma$ corresponds to the Seiberg–Witten curve of $4d$ $\mathcal N=4$ $\mathfrak{su}(N)$ SYM.

To introduce a bare mass $m$ for the adjoint hypermultiplet, one must consider non-trivial affine fibrations of $\mathbb{C}_v$ over the circle $S^1_6$, such that as $x^6$ shifts by $2\pi L$, the coordinate $v$ transforms as $v \mapsto v + m$, where $m \in \mathbb{C}$. Uplifting this modified brane configuration to M-theory results in the Seiberg–Witten curve for the $4d$ $\mathcal N=2^*$ $\mathfrak{su}(N)$ theory, matching the results of \cite{Donagi:1995cf}.
\end{example}

\subsection{BPS states}

In this geometric realization of Seiberg--Witten curves, BPS states are associated with M2-branes extending along $x^0$ (the worldline of \emph{BPS particles}\index{BPS!state} in $\mathbb{R}^{1,3}_{0,1,2,3}$) and wrapping a Riemann surface~$D$ in $Q$ with non-empty boundary lying on the Seiberg--Witten curve $\Sigma$. If $D$ minimizes its volume within its homology class relative to its boundary, then its quantization yields a BPS state \cite{Mikhailov:1997jv}. We will develop this geometric perspective in \Cref{Chap:BPSstatesclassS}.

\section{\texorpdfstring{$\mathrm{SU}(2)$}{SU(2)} SYM with \texorpdfstring{$F=4$}{F=4} flavors revisited}\labelx{Sec:genSU2quivers}

\subsection{Electric-magnetic duality}

Recall the discussion of $S$-duality of $4d$ $\mathcal N=2$ $\mathrm{SU}(2)$ SYM with $F=4$ massless fundamental flavors from \Cref{Sec:SdualitySU2flavors}. The duality group $\mathrm{SL}_2(\mathbb{Z})$ jointly acts on the special coordinates $a$ and $a_D$ on the Coulomb branch, which in this special case satisfy (cf. \Cref{rem:scaleinvariance}):
\begin{equation*}
    \begin{split}
        a(u) &= \sqrt{u}\; ,\\
        a_D(u) &= 2\widetilde{\tau}_\mathrm{cl} a(u)\; ,
    \end{split}
\end{equation*}
and shuffles the representations of the flavor group $\mathrm{SO}(8)$---or rather, $\mathrm{Spin}(8)$---by triality. 

Let us focus on the subgroup $\mathrm{Spin}(4)\times\mathrm{Spin}(4)$ of the flavor group, which arises as one splits the four fundamental hypermultiplets into two pairs. Exploiting the isomorphism $\mathfrak{so}(4)\simeq \mathfrak{su}(2)\times \mathfrak{su}(2)$, we rewrite the first $\mathrm{Spin}(4)$ factor as $\mathrm{SU}(2)_a\times\mathrm{SU}(2)_b$, and the second, as $\mathrm{SU}(2)_c\times\mathrm{SU}(2)_d$. Let $\mathbf{2}_a$, $\mathbf{2}_b$, $\mathbf{2}_c$, and $\mathbf{2}_d$, denote the fundamental representation of each copy of $\mathrm{SU}(2)$.
\begin{proposition}
    With respect to the subgroup $\mathrm{SU}(2)_a\times\mathrm{SU}(2)_b\times \mathrm{SU}(2)_c\times\mathrm{SU}(2)_d$, the three eight-dimensional representations of $\mathrm{Spin}(8)$ decompose as:
\begin{equation}\labelx{eq:decompositionSO8SU2}
    \begin{split}
        \mathbf{8}_v &\simeq (\mathbf{2}_a\otimes\mathbf{2}_b)\oplus(\mathbf{2}_c\otimes\mathbf{2}_d)\; ,\\
    \mathbf{8}_s &\simeq (\mathbf{2}_a\otimes\mathbf{2}_c)\oplus(\mathbf{2}_b\otimes\mathbf{2}_d)\; ,\\
    \mathbf{8}_c &\simeq (\mathbf{2}_a\otimes\mathbf{2}_d)\oplus(\mathbf{2}_b\otimes\mathbf{2}_c)\; .
    \end{split} 
\end{equation}
Here $\mathbf{8}_v$, $\mathbf{8}_s$, and $\mathbf{8}_c$, respectively denote the vector, spinor, and conjugate spinor representations of $\mathrm{Spin}(8)$. Elements of $\mathrm{SL}_2(\mathbb{Z})$ permute $\mathbf{8}_v$, $\mathbf{8}_s$, and $\mathbf{8}_c$ according to their reduction modulo 2---recall that $\mathrm{SL}_2(\mathbb{Z}_2)\simeq\mathfrak{S}_3$. Hence, the duality group $\mathrm{SL}_2(\mathbb{Z}_2)$ permutes the $\mathrm{SU}(2)$ factors in all possible ways. 
\end{proposition}

The action of the duality group on the $\mathrm{SU}(2)$ flavor factors is illustrated by the refined quiver notation in \Cref{fig:dualitySU2F=4}. As before, each round node labeled by $2$ represents an $\mathrm{SU}(2)$ gauge factor, while each square node labeled by $2$ denotes an $\mathrm{SU}(2)$ flavor factor\footnote{In contrast with the above, square nodes here correspond to $\mathrm{SU}$ rather than $\mathrm{U}$ flavor factors.}. Trivalent vertices correspond to pairs of fundamental hypermultiplets: for instance, the leftmost trivalent vertex in \Cref{fig:dualitySU2F=4} represents two hypermultiplets in the fundamental representation of the $\mathrm{SU}(2)$ gauge group. The flavor symmetry associated with these hypermultiplets is $\mathrm{SU}(2)_a \times \mathrm{SU}(2)_b$.

\begin{figure}[!ht]
    \centering
    \includegraphics[width=\textwidth]{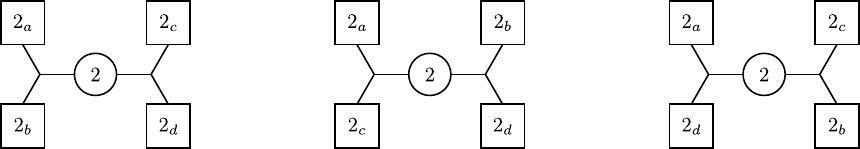}
    \caption{$S$-duality of $4d$ $\mathcal N=2$ $\mathrm{SU}(2)$ SYM with $4$ fundamental flavors.}
    \labelx{fig:dualitySU2F=4}
\end{figure}

The quivers in \Cref{fig:dualitySU2F=4} represent the weak-coupling limits of $4d$ $\mathcal{N}=2$ $\mathrm{SU}(2)$ SYM theory with $F=4$ fundamental hypermultiplets. Initially, the theory is depicted by the leftmost quiver in \Cref{fig:dualitySU2F=4}, where the fundamental hypermultiplets transform in the vector representation $\mathbf{8}_v$ of $\mathrm{Spin}(8)$. By increasing the gauge coupling of the $\mathrm{SU}(2)$ gauge group—in the limit $\tau \rightarrow 0$—applying the transformation $S$ yields a better description of the theory as the middle quiver in \Cref{fig:dualitySU2F=4}. In this dual description, the gauge coupling is given by $\tau_D = -1/\tau$, meaning the new gauge group is weakly coupled. 

The transition to the middle quiver arises because $S$ exchanges the $\mathbf{8}_v$ and $\mathbf{8}_s$ representations of $\mathrm{Spin}(8)$ while keeping the $\mathbf{8}_c$ representation unchanged. This follows\footnote{As discussed below \Cref{rem:scaleinvariance}, the representations of $\mathrm{Spin}(8)$ are classified by their charges under the center $\mathcal{Z}(\mathrm{Spin}(8)) = \mathbb{Z}_2 \times \mathbb{Z}_2$. Specifically, $\mathbf{8}_v$ corresponds to $(1,0)$, $\mathbf{8}_s$ to $(0,1)$, and $\mathbf{8}_c$ to $(1,1)$.} from the reduction of $S$ modulo $2$.

\subsubsection*{Massive hypermutliplets} 

The mass matrix $m$ encoding the masses of the fundamental hypermultiplets belongs to the Cartan subalgebra of $\mathfrak{so}_8(\mathbb{C})$. With respect to the subgroup $\mathrm{SU}(2)_a \times \mathrm{SU}(2)_b \times \mathrm{SU}(2)_c \times \mathrm{SU}(2)_d \subset \mathrm{Spin}(8)$, the matrix $m$ decomposes into four complex mass parameters $m_a$, $m_b$, $m_c$, and $m_d$, each associated with one of these $\mathrm{SU}(2)$ factors. If the original masses of the fundamental hypermultiplets are denoted by $m_1$, $m_2$, $m_3$, and $m_4$, then the relations between these sets of mass parameters are given by:
\begin{equation}\labelx{eq:changebasismasses}
    \begin{split}
        m_a &= \frac{1}{2}(m_1+m_2)\; , \\
        m_b &= \frac{1}{2}(m_1-m_2)\; , \\
        m_c &= \frac{1}{2}(m_3+m_4)\; , \\
        m_d &= \frac{1}{2}(m_3-m_4)\; .
    \end{split}
\end{equation}
The action of the duality group $\mathrm{SL}_2(\mathbb{Z})$ on the parameters $m_{a,b,c,d}$ is simpler than its action on $m_{1,2,3,4}$, as it permutes $m_{a,b,c,d}$ following the pattern shown in \Cref{eq:decompositionSO8SU2}. For instance, under $S$, $m_b$ and $m_c$ are exchanged, while $m_a$ and $m_d$ remain fixed. It can be verified that this duality action corresponds to the transformations in \Cref{eq:transfomasses}, when considering the relations of \Cref{eq:changebasismasses}.

\subsubsection*{Space of marginal deformations and dualities}

When all hypermultiplets are massless, the low-energy gauge coupling $\tau$ on the Coulomb branch reads $\tau=a_D(u)/a(u)=2\widetilde{\tau}_\mathrm{cl}$, where $\widetilde{\tau}_\mathrm{cl}$ is the renormalization-invariant UV complexified gauge coupling. In terms of the UV gauge coupling $g$ and theta angle $\Theta$, one has:
\begin{equation}\labelx{eq:UVcoupling}
    \tau = \frac{\Theta}{\pi} + \frac{8\pi \I}{g^2}\; .
\end{equation}
Therefore, $\tau$ takes values in the complex upper half-plane $\mathbb{H}$. The duality group $\mathrm{SL}_2(\mathbb{Z})$ acts on $\tau$ by fractional linear transformations, inducing the standard action of $\mathrm{PSL}_2(\mathbb{Z}) = \mathrm{SL}_2(\mathbb{Z})/\mathbb{Z}_2$ on $\mathbb{H}$. In other words, the space of gauge couplings for the theory is $\mathbb{H}$, and the space of gauge couplings modulo $\mathrm{SL}_2(\mathbb{Z})$ duality is the quotient $\mathbb{H}/\mathrm{PSL}_2(\mathbb{Z})$. Note that the quotient $\mathbb{H}/\mathrm{PSL}_2(\mathbb{Z})$ has a single cusp which corresponds to the weak-coupling limit $\Im\tau\rightarrow\infty$.

\subsection{Seiberg--Witten curve}

The Seiberg--Witten curve for $4d$ $\mathcal{N}=2$ $\mathrm{SU}(2)$ super-Yang--Mills (SYM) theory with four (massive) fundamental hypermultiplets follows from \Cref{Sec:Mtheoryuplifts}:
\begin{equation*} 
(\Sigma)\colon\quad f\frac{(v-m_1)(v-m_2)}{t} + f'(v-m_3)(v-m_4)t = v^2 - u\; .
\end{equation*} 
Here, $f$ and $f'$ are complex constants, $m_{1,2,3,4}$ encode the bare masses of the hypermultiplets, and $u$ is a coordinate on the Coulomb branch. The Seiberg--Witten differential is given by $v\mathrm{d}t/t$. We will rewrite $\Sigma$ to highlight the geometric nature of $\tau$, which will be useful for discussing generalizations. We follow the presentation of \cite[Chap. 9]{Tachikawa:2013kta}, where further details can be found.

One can rescale $t$ so that $f' = 1$, and collect the powers of $v$, which leads to:
\begin{equation*} 
(\Sigma)\colon\quad v^2 + \frac{f(m_1 + m_2)/t + (m_3 + m_4)t}{1 - t - f/t}v - \frac{u + f m_1 m_2/t + m_3 m_4 t}{1 - t - f/t} = 0\; . 
\end{equation*}
Letting:
\begin{equation*} 
x = v - \frac{1}{2}\frac{f(m_1 + m_2)/t + (m_3 + m_4)t}{1 - t - f/t}\; , 
\end{equation*} 
yields:
\begin{equation*} 
(\Sigma)\colon\quad x^2 - \left( \frac{u + f m_1 m_2/t + m_3 m_4 t}{1 - t - f/t} + \frac{1}{4}\left( \frac{f(m_1 + m_2)/t + (m_3 + m_4)t}{1 - t - f/t} \right)^2 \right) = 0\; . 
\end{equation*}
The relation between the differentials $\lambda = x \mathrm{d}t/t$ and $v \mathrm{d}t/t$ only involves the mass parameters, hence $\lambda = x \mathrm{d}t/t$ is another fine choice of Seiberg--Witten differential. Let $t_1$ and $t_2$ be the roots of $1 - t - f/t = -(t - t_1)(t - t_2)/t$, and let $z = t/t_1$ and $q = t_2/t_1$. Then:
\begin{equation}\labelx{eq:SWcurveSU2F=4} 
(\Sigma)\colon\quad \lambda^2 - \phi_2(z) = 0\; ,
\end{equation} 
where
\begin{equation*}
    \phi_2(z) = \frac{P(z)}{(z - 1)^2(z - q)^2}\frac{\mathrm{d}z^2}{z^2}
\end{equation*}
is a quadratic differential on $\mathbb{CP}^1_z$ with double poles at $z = 0$, $q$, $1$, and $\infty$, with $P(z)$ is a polynomial of degree 4. The action of $\mathrm{PSL}_2(\mathbb{C})$ on $\mathbb{CP}^1_z$ allows us to treat the four singular points of the quadratic differential $\phi_2(z)$ symmetrically. 

Recall that the mass of BPS particles are computed as the integral of $\vert\lambda\vert$ along cycles on $\Sigma$. Therefore, $\lambda$ has mass dimension $1$, and $P(z)$ has mass dimension 2. The bare masses of the hypermutliplets are the residues of $\lambda$ at poles on $\Sigma$; one finds that the residues of $\lambda$ at the poles above $z=0,q,1,\infty$ are $\pm m_b$, $\pm m_a$, $\pm m_c$, and $\pm m_d$, respectively\footnote{This holds up to a finite renormalization discussed e.g. in \cite{Tachikawa:2013kta}.}. Correspondingly, the coefficient of $\phi_2$ at the double pole at $0$, $q$, $1$, and $\infty$, is $m_b^2$, $m_a^2$, $m_c^2$, and $m_d^2$, respectively.  For instance:
\begin{equation}\labelx{eq:doublepole}
    \phi_2(z) = \frac{m_b^2}{z^2}\mathrm{d}z^2 + \text{less singular terms},
\end{equation}
for $z$ close to $0$. Therefore, there is a natural one-to-one correspondence between the flavor factors $\mathrm{SU}(2)_a$, $\mathrm{SU}(2)_b$, $\mathrm{SU}(2)_c$, and $\mathrm{SU}(2)_d$, and the punctures $0$, $q$, $1$, and $\infty$, on $\mathbb{CP}^1_z$. Note that $\phi_2(z)$ is holomorphic on $\mathbb{CP}^1$ away from the punctures.

\begin{remark}
    When all hypermultiplets are massless, i.e. $m_{1,2,3,4}=0$, the same computation leads to a Seiberg--Witten curve of the form of \Cref{eq:SWcurveSU2F=4}, with
    \begin{equation*}
        \phi_2(z) = \frac{P}{(z - 1)(z - q)}\frac{\mathrm{d}z^2}{z}\; ,
    \end{equation*}
    where $P$ is a constant. Note that $\phi_2(z)$ has now poles of order 1 at $0,q,1,\infty$.
\end{remark}

\begin{definition}
    Punctures where $\phi_2$ has order $1$ without mass deformation and order $2$ for a generic mass deformation, are called \emph{simple $\mathrm{SU}(2)$ punctures}\index{puncture!simple}.
\end{definition}

\begin{definition}
    The Riemann sphere with four simple punctures is the \emph{UV curve}\index{UV curve} of $\mathrm{SU}(2)$ SYM with four fundamental flavors; it is depicted in \Cref{fig:UVcurveSU2F4}.
    \begin{figure}[!ht]
        \centering
        \includegraphics[scale=0.83]{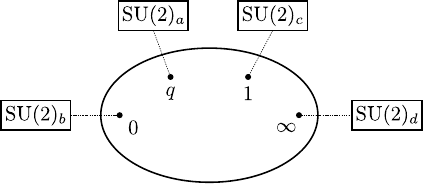}
        \caption{UV curve of $\mathrm{SU}(2)$ SYM with four fundamental flavors.}\labelx{fig:UVcurveSU2F4}
    \end{figure}
\end{definition}

\begin{remark}
    The complex number $q\in\mathbb{CP}^1$ is the cross ratio of the punctures $0,q,1,\infty\in\mathbb{CP}^1_z$. The cross ratio is invariant under the action of the Möbius group $\mathrm{PSL}_2(\mathbb{C})$, which is the group of complex automorphisms of the Riemann sphere.
\end{remark}

\begin{remark}
    The definition of $q$ above shows that $q$ is related to the constant $f$, itself related to the UV gauge coupling $\widetilde{\tau}$ which is an exactly marginal parameter of the theory---actually, the only exactly marginal parameter of the theory.
\end{remark}

The UV gauge coupling $\widetilde{\tau}$ naturally takes value in the complex upper-half plane $\mathbb{H}$. In fact, $\mathbb{H}$ can be identified with Teichmüller space of the Riemann sphere with four punctures (cf. \Cref{subsec_Teich_sp}).

\begin{proposition}
    The Teichmüller space of the UV curve of $\mathrm{SU}(2)$ SYM with four fundamental flavors is the space of exactly marginal parameters of the theory. The duality group acting on $\widetilde{\tau}$, i.e. $\mathrm{PSL}_2(\mathbb{Z})$, can be identified with the subgroup\footnote{The mapping class group of the sphere with four punctures $S_{0,4}$ is a semi-direct product $\mathrm{PSL}_2(\mathbb{Z})\ltimes(\mathbb{Z}_2\times\mathbb{Z}_2)$ \cite{Farb-Margalit}.} of the mapping class group of the UV curve acting non-trivially on the Teichmüller space $\mathbb{H}$. The quotient $\mathbb{H}/\mathrm{PSL}_2(\mathbb{Z})$ is the moduli space of complex structures on a sphere with four punctures $S_{0,4}$.
\end{proposition} 

\begin{remark}
    Here, by Teichmüller space of the smooth surface $S_{g,n}$ of genus $g$ and with $n$ punctures, we mean the space of all hyperbolic metrics that $S_{g,n}$ can be endowed with, where the punctures are hyperbolic cusps, modulo diffeomorphisms connected to the identity.
\end{remark}

In contrast to the massless case, when the hypermultiplets are massive, the correspondence between points in $\mathbb{H}$ and the low-energy gauge couplings becomes non-trivial. Nevertheless, the UV gauge coupling in \Cref{eq:UVcoupling} remains an exactly marginal parameter of the theory. Thus, in general, $\mathbb{H}=\mathcal{T}(S_{0,4})$ is to be understood as the space of exactly marginal deformations of the theory.

The \emph{pure mapping class group}\index{mapping class group} is the distinguished subgroup of the mapping class group $\mathrm{Mod}(S_{0,4})$ which preserves the punctures (cf. \Cref{def:pure_Mod_marked} and \cite{Farb-Margalit}). In other words, the pure mapping class group acting non-trivially on $\mathbb{H}$ is the subgroup of $\mathrm{PSL}_2(\mathbb{Z})$ which preserves the masses $m_{a,b,c,d}$, hence it can be characterized as the kernel $\Gamma(2)$ of the reduction modulo 2: $\mathrm{PSL}_2(\mathbb{Z})\rightarrow\mathrm{PSL}_2(\mathbb{Z}_2)$. Therefore, modulo dualities that preserve the masses $m_{a,b,c,d}$, the space of exactly marginal deformations of the theory is the quotient $\mathbb{H}/\Gamma(2)$.

\begin{remark}
    The quotient $\mathbb{H}/\Gamma(2)$ is the moduli space of a sphere with four marked punctures.
\end{remark}

The moduli space $\mathbb{H}/\Gamma(2)$ has three cusps. These correspond bijectively to the quivers shown in \Cref{fig:dualitySU2F=4}, hence to the three distinct weakly-coupled descriptions of $4d$ $\mathcal{N}=2$ $\mathrm{SU}(2)$ SYM with four fundamental flavors, when the mass parameters are labeled. Geometrically, the three cusps in the moduli space $\mathbb{H}/\Gamma(2)$ correspond to the different ways in which the four-punctured sphere $\Sigma_{0,4}$ can degenerate into two spheres, each containing two simple punctures, connected by a long thin tube, as illustrated in \Cref{fig:degenerationspherefourpunct}. For an explicit description of these weakly-coupled limits, we refer to \cite[Chap. 9]{Tachikawa:2013kta}.

\begin{figure}[!ht]
    \centering
    \includegraphics[width=\textwidth]{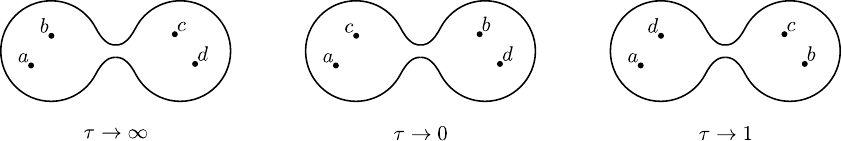}
    \caption{Possible degenerations of a sphere with four punctures.}
    \labelx{fig:degenerationspherefourpunct}
\end{figure}

At each of these limits, the Riemann sphere with four punctures completely decouples into two spheres, each featuring three simple punctures. For example, the limit $\tau = \infty$ corresponds to the total decoupling of the original $\mathrm{SU}(2)$ gauge group, resulting in two sets of four hypermultiplets. Consequently, the Riemann sphere with three punctures is identified as the UV curve of the theory that includes only the four hypermultiplets.

\subsection{Hitchin description}

\Cref{eq:SWcurveSU2F=4} can be recast in a way which makes the relationship with Hitchin systems very explicit. The \emph{Hitchin field}\index{Hitchin!field} corresponding to the quadratic differential $\phi_2(s)$ is a traceless $2\times 2$ matrix $\Phi(z)$ such that $\tr\Phi(z)^2=2\phi_2(z)$. In terms of $\Phi(z)$, \Cref{eq:SWcurveSU2F=4} rewrites:
\begin{equation}\labelx{eq:HitchinSU2F=4}
    \det(\lambda\otimes\mathbf{1}_2-\Phi(z)) = 0\; .
\end{equation}
Note that \Cref{eq:HitchinSU2F=4} is invariant under $\mathrm{SU}(2)$ ``gauge'' transformations acting on $\Phi(z)$ by conjugation, i.e. $\Phi(z)$ transforms in the adjoint representation of $\mathrm{SU}(2)$. At a simple puncture $z=z_*$ for which the mass parameter $m$ does not vanish, the quadratic differential $\phi_2(z)$ has a pole of order 2 with coefficient $m^2$ (cf. \Cref{eq:doublepole}). Equivalently, up to $\mathrm{SU}(2)$ gauge transformations, for $z$ close to $z_*$ one has:
\begin{equation*}
    \Phi(z) \sim \begin{pmatrix}
        m & 0 \\ 0 & -m
    \end{pmatrix}\frac{\mathrm{d}(z-z_*)}{z-z_*}\; .
\end{equation*}
When the residue $m$ of $\lambda$ vanishes, the order of the pole of the quadratic differential $\phi_2$ at $z_*$ reduces to one. Up to gauge transformations, for $z$ close to $z_*$ one can assume that:
\begin{equation*}
    \Phi(z) \sim \begin{pmatrix}
        0 & 1 \\ 0 & 0
    \end{pmatrix}\frac{\mathrm{d}(z-z_*)}{z-z_*}\; .
\end{equation*}
Note that the Hitchin field $\Phi(z)$ has a simple pole at simple punctures, independently of whether one is considering mass deformations. For more details on the relationship between Hitchin systems and Seiberg--Witten fibrations for theories of class $\mathcal S$, the reader can for example consult \cite{Gaiotto:2009hg}.

\subsection{\texorpdfstring{$\mathrm{SU}(2)$} {SU(2)} SYM with \texorpdfstring{$F<4$}{F<4} flavors}\labelx{Subsec:SU2F<4revisited}

It is instructive to apply the reasoning above to the Seiberg--Witten curves of $\mathrm{SU}(2)$ SYM with $F<4$ flavors. For $F=0$, the Seiberg--Witten curve reads
\begin{equation*}
    \frac{\Lambda^2}{t}+\Lambda^2t = v^2-u\; .
\end{equation*}
One can rewrite it as:
\begin{equation}\labelx{eq:SWcurveF=0}
    \lambda^2-\phi_2(z) = 0\; ,
\end{equation}
where $\lambda=x\mathrm{d}z/z$ and $\phi_2$ a quadratic differential on the sphere $\mathbb{CP}^1_z$ with order three poles at zero and infinity. Considering $F=1$ instead obtained from a type IIA configuration with one semi-infinite D4-brane to the left of the leftmost NS5-brane, the Seiberg--Witten curve reads:
\begin{equation*}
    \frac{2\Lambda_1(v-m)}{t}+\Lambda_1^2t = v^2-u\; .
\end{equation*}
One can rewrite it as in \Cref{eq:SWcurveF=0}, with a pole of order 4 at 0 and another of order 3 at $\infty$ for $\phi_2$. There is a similar description with the role of the punctures at $0$ and $\infty$ exchanged; it is obtained starting from a type IIA configuration with the semi-infinite D4-brane to the right of the rightmost NS5-brane.

The cases $F=2$ and $F=3$ also admit similar descriptions. For instance, when $F=2$, the differential $\phi_2$ can either be taken to have two poles of order 4, one at 0 and the other at $\infty$, or two simple poles at $0,1$ and one pole of order 3 at $\infty$. Even in the case of $\mathrm{SU}(2)$ SYM with four fundamental flavors, one can consider \emph{unbalanced} type IIA descriptions where the numbers of semi-infinite D4-branes to the left and to the right of the configuration do not match. Punctures where the differential $\phi_2(z)$ has a pole of order greater than 2 are termed \emph{irregular $\mathrm{SU}(2)$ punctures}\index{puncture!irregular}. A higher-order pole for $\phi_2$ encodes the bending of an NS5-brane caused by an imbalance in the number of D4-branes ending on each side.

In the Hitchin description of the Seiberg--Witten fibration, such punctures where $\phi_2$ has poles of order greater than two correspond to irregular, higher-order, poles of the Hitchin field. This leads to the following distinction, which will generalize to other gauge groups:

\begin{definition}\labelx{def:puncturetype}
    Punctures at which the Hitchin field has a pole of order (at most) one are said to be \emph{regular}\index{puncture!regular}, while those where it has a pole of higher order are called \emph{irregular}\index{puncture!irregular}.
\end{definition}

\section{Theories of class \texorpdfstring{$\mathcal S$}{S} of type \texorpdfstring{$A_1$}{A1}}

\subsection{Generalized \texorpdfstring{$\mathrm{SU}(2)$}{SU(2)} quivers}

Let us now turn to more general linear conformal quivers involving $\mathrm{SU}(2)$ gauge and flavor groups (\Cref{fig:moreSU2quivers}). The action of $\mathrm{SL}_2(\mathbb{Z})$-duality on the weakly-coupled descriptions of the associated $4d$ $\mathcal{N}=2$ theories can be understood by focusing on scenarios where all but one of the gauge groups are very weakly coupled. The remaining gauge group behaves like an $\mathrm{SU}(2)$ SYM theory with four fundamental flavors. Increasing the gauge coupling of this remaining group until it becomes strongly coupled, it can be then better described as a dual weakly-coupled gauge group.

\begin{figure}[!ht]
    \centering
    \includegraphics[width=\textwidth]{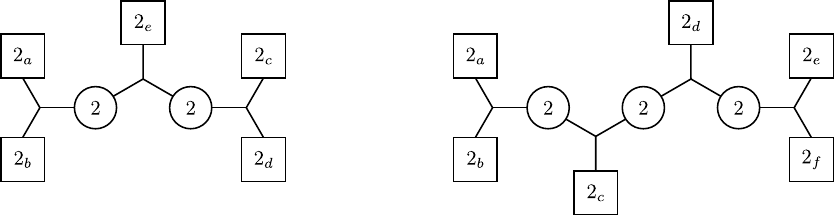}
    \caption{More involved $\mathrm{SU}(2)$ quivers.}
    \labelx{fig:moreSU2quivers}
\end{figure}

The quiver on the left of \Cref{fig:moreSU2quivers} represents a $4d$ $\mathcal{N}=2$ theory with gauge group $\mathrm{SU}(2)_1\times\mathrm{SU}(2)_2$, where each gauge factor has two fundamental flavors. Additionally, there is a hypermultiplet in the bifundamental representation $\mathbf{2}_1\otimes \overline{\mathbf{2}}_2$. This bifundamental hypermultiplet introduces an additional $\mathrm{SU}(2)$ flavor symmetry\footnote{The fact that this symmetry is $\mathrm{SU}(2)\simeq\mathrm{USp}(2)$ rather than $\mathrm{U}(1)$ stems from the fact that $\mathbf{2}_1\otimes\overline{\mathbf{2}}_2$ is a real representation.}, denoted $\mathrm{SU}(2)_e$ in the figure. Therefore, each $\mathrm{SU}(2)$ gauge factor effectively has four fundamental flavors, and the gauge couplings $\tau_1$ and $\tau_2$ are \emph{exactly marginal}\index{exactly marginal parameter} parameters of the theory. The $S$-duality properties of this theory were first explored in \cite{Argyres:1999fc}.

Let us first consider a \emph{duality frame}\index{duality!frame} in which both gauge groups are weakly coupled, and increase the gauge coupling $g_1$ of $\mathrm{SU}(2)_1$, keeping $g_2$ small. The gauge group $\mathrm{SU}(2)_1$ is then better described as another weakly coupled $\mathrm{SU}(2)$ gauge factor. As discussed previously, this change in description can generally involve all possible permutations of the ``flavor'' symmetry factors associated with $\mathrm{SU}(2)_1$, namely $\mathrm{SU}(2)_{a,b,e}$ and $\mathrm{SU}(2)_2$. The same reasoning applies when the roles of the gauge factors $\mathrm{SU}(2)_1$ and $\mathrm{SU}(2)_2$ are interchanged. Thus, by repeated actions of the duality group $\mathrm{SL}_2(\mathbb{Z})$ on either of the two gauge factors, the flavor symmetry factors $\mathrm{SU}(2)_{a,b,c,d,e}$ are exchanged in all possible ways. Importantly, one sees that the copies of $\mathrm{SL}_2(\mathbb{Z})$ associated with the two gauge factors do not commute.

The Seiberg--Witten curve for that theory can be computed with the methods discussed in \Cref{Sec:Mtheoryuplifts}, and one can again present it under the form:
\begin{equation}\labelx{eq:SWgenSU2}
    \lambda^2-\phi_2(z)=0\; ,
\end{equation}
where $\lambda = x\mathrm{d}z/z$ is the Seiberg--Witten differential, and $\phi_2(z)$ is a quadratic differential on the sphere $\mathbb{CP}^1_z$ with five simple punctures. Correspondingly:

\begin{proposition}
    The space of exactly marginal deformations of that theory can be identified with the Teichmüller space of $\mathcal{T}(S_{0,5})$ of a sphere with five punctures. The mapping class group $\mathrm{Mod}(S_{0,5})$ acts as the duality group. The moduli space $\mathcal{T}(S_{0,5})/\mathrm{Mod}(S_{0,5})$ has a single cusp that corresponds to the weakly-coupled description of the theory as a quiver of the same form as the one on the left of \Cref{fig:moreSU2quivers}. Restricting to the subgroup $\widetilde{\mathrm{Mod}}(S_{0,5})$ of $\mathrm{Mod}(S_{0,5})$ that fixes the punctures, the quotient $\mathcal{T}(S_{0,5})/\widetilde{\mathrm{Mod}}(S_{0,5})$ has 15 cusps corresponding to all possible configurations of the flavor factors $\mathrm{SU}(2)_{a,b,c,d,e}$. Geometrically, each of these cusps corresponds to a degeneration limit of $S_{0,5}$ into three spheres connected by long, thin tubes: two of the spheres carry two punctures and connect to one tube each, while the third carries one puncture and connects to two tubes. The \emph{UV curve} of the theory is $\mathbb{CP}^1$ with five punctures.
\end{proposition}

Let us now turn to the quiver on the right of \Cref{fig:moreSU2quivers}. Dualizing the quiver at the first and third gauge factors results in a similar-looking quiver, albeit with the flavor factors shuffled. Dualizing the quiver at its second gauge node yields all possible arrangements of the four ``flavor'' factors of $\mathrm{SU}(2)_2$, namely $\mathrm{SU}(2)_{1,c,d,3}$, into two pairs.

\begin{figure}[!ht]
    \centering
    \includegraphics[scale=0.83]{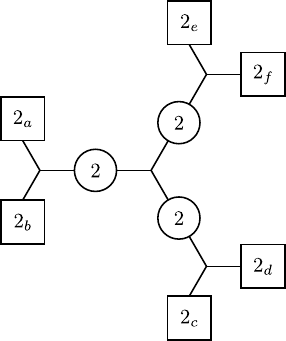}
    \caption{A generalized $\mathrm{SU}(2)$ quiver.}
    \labelx{fig:dualitiesSU23}
\end{figure}

One such arrangement yields the quiver depicted in \Cref{fig:dualitiesSU23}, which is an example of a \emph{generalized $\mathrm{SU}(2)$ quiver}\index{quiver!$\mathcal N=2$!generalized}. The weakly-coupled theory it describes can be thought of as obtained from the quiver on the left of \Cref{fig:moreSU2quivers}, gauging the flavor symmetry factor $\mathrm{SU}(2)_e$ and adding two fundamental flavors for this new gauge group to ensure scale invariance.

Again, the SW curve of this theory can be presented as in \Cref{eq:SWgenSU2}, with $\phi_2$ a quadratic differential on $\mathbb{CP}^1_z$ with six simple punctures. The weakly-coupled descriptions of this theory as $\mathrm{SU}(2)$ quivers correspond to the degenerations of $\mathbb{CP}^1_z$ with six simple punctures depicted in \Cref{fig:degenerationspheresixpunct}.

\begin{figure}[!ht]
        \centering
        \includegraphics[scale=0.83]{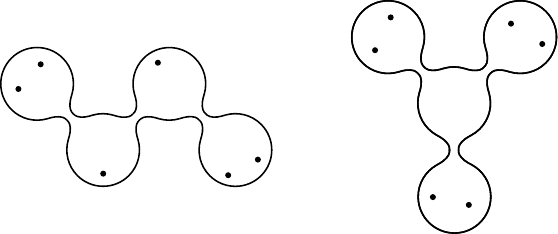}
        \caption{Possible degenerations of a sphere with six punctures.}
        \labelx{fig:degenerationspheresixpunct}
\end{figure}

The space of exactly marginal deformations can be identified with the Teichmüller space of a sphere with six punctures, on which the mapping class group acts as the duality group.

\vspace{0.3cm}

The quivers of \Cref{fig:moreSU2quivers} are specific cases of linear conformal quivers involving $\mathrm{SU}(2)$ gauge and flavor groups. By performing dualities on the gauge nodes in all possible configurations, one obtains quivers that are not necessarily linear but always correspond to trivalent trees with external edges. In this representation, each internal edge denotes an $\mathrm{SU}(2)$ gauge factor, each external edge represents an $\mathrm{SU}(2)$ flavor factor, and each trivalent vertex corresponds to a configuration of four hypermultiplets, specifically one of the following:
\begin{itemize}
\item A pair of fundamental hypermultiplets for an $\mathrm{SU}(2)$ gauge factor. The flavor symmetry is $\mathrm{SU}(2) \times \mathrm{SU}(2)$. 
\item A hypermultiplet in the bifundamental representation of two $\mathrm{SU}(2)$ gauge factors, with associated flavor symmetry $\mathrm{SU}(2)$. 
\item A hypermultiplet in a trifundamental representation of three $\mathrm{SU}(2)$ gauge factors, with trivial associated flavor symmetry. 
\end{itemize}
That these graphs are connected trivalent trees implies that for $E_i=n$ the number of internal edges, there must be $V=n+1$ trivalent vertices, and
\begin{equation*}
      E_e = 3V - 2E_i = 3(n+1) - 2n = n+3  
\end{equation*}
external edges. The quiver description makes an $\mathrm{SU}(2)^{n+3}$ subgroup of the flavor symmetry group explicit. By $S$-duality, one can reach all possible trivalent tree topologies with $n$ internal edges. For fixed $n$, each such topology corresponds to a specific weak-coupling description of a unique underlying theory. Equivalently, these descriptions map to different degeneration limits of the Riemann sphere with $n+3$ punctures. The Seiberg--Witten curve of such a theory can be presented as in \Cref{eq:SWgenSU2}, with $\phi_2(z)$ a quadratic differential on $\mathbb{CP}^1_z$ with $n+3$ simple punctures. The space of exactly marginal deformations of the theory is identified with the Teichmüller space $\mathcal T(S_{0,n+3})$ of a sphere with $n+3$ punctures, thus the unique underlying theory is naturally associated with $S_{0,n+3}$.

\begin{remark}
    Weakly-coupled descriptions are in one-to-one correspondence with the decompositions of $S_{0,n+3}$ into pairs of pants (cf. \Cref{sec: triangles_hexagons}).
\end{remark}

One can generalize even further the quivers under consideration by relaxing the constraint that they are trees. In particular, Witten's construction for $\mathrm{SU}(2)$ \emph{elliptic models}\index{elliptic model} yields necklace quivers. S-duality applied to such quivers yields all possible trivalent graphs with external edges and one loop. An example is shown on the left of \Cref{fig:generalizedSU2quivers}. In such cases, the Seiberg--Witten curve can be presented as 
\begin{equation*}
    \lambda^2-\phi_2(z) = 0\; ,
\end{equation*}
where $\lambda=x\mathrm{d}z$ is the Liouville 1-form on the holomorphic cotangent bundle to an elliptic curve $E_z$, identified with the Seiberg--Witten differential when restricted to the Seiberg--Witten curve, and $\phi_2(z)$ is a quadratic differential on $E_z$ with simple punctures. For instance, the quiver displayed on the left of \Cref{fig:generalizedSU2quivers} corresponds to a quadratic differential $\phi_2$ with 6 simple punctures on $E_\tau$. Here, the UV curve of the theory is the elliptic curve $E_\tau$ with 6 punctures.

The family of one-loop quivers with fixed number $n$ of internal edges (i.e. $\mathrm{SU}(2)$ gauge factors) encodes all possible weakly-coupled descriptions of an underlying theory associated with the (smooth) torus with $n$ punctures $S_{1,n}$. Equivalently, such quivers are in one-to-one correspondence with inequivalent decompositions of $S_{1,n}$ into pairs of pants.

\begin{figure}[!ht]
    \centering
    \includegraphics[scale=0.83]{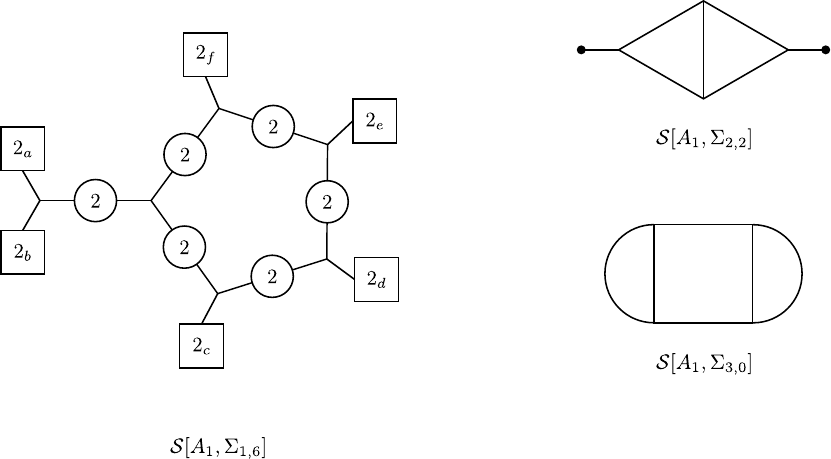}
    \caption{Generalized $\mathrm{SU}(2)$ quivers corresponding to weakly-coupled descriptions of $\mathcal S[A_1,S_{g,n}]$. The two quivers on the right are depicted as trivalent graphs with external edges.}
    \labelx{fig:generalizedSU2quivers}
\end{figure}

More generally, one can consider \emph{generalized $\mathrm{SU}(2)$ quivers}\index{quiver!$\mathcal N=2$!generalized} with $g$ loops. There is no direct way to construct the Seiberg--Witten curves for these theories via type IIA superstring theory and M-theory with the methods of \Cref{Sec:Mtheoryuplifts}, but the examples of linear and necklace quivers suggest the following. If the quiver under consideration has $g$ loops and $n$ external edges, then the Seiberg--Witten curve writes:
\begin{equation*}
    \lambda^2-\phi_2(z) = 0\; ,
\end{equation*}
where $\lambda=x\mathrm{d}z$ is the Liouville 1-form on the holomorphic cotangent bundle to a Riemann surface $C_{g,n}$ of genus $g$ and with $n$ punctures (the UV curve), identified with the Seiberg--Witten differential when restricted to the Seiberg--Witten curve, and $\phi_2(z)$ is a quadratic differential on $C_{g,n}$ with $n$ simple punctures. Two examples are shown on the right of \Cref{fig:generalizedSU2quivers}.

As before, one can show that by applying repeated $S$-dualities at the gauge nodes of a generalized $\mathrm{SU}(2)$ quiver with $g$ loops and $n$ external edges, all possible configurations of generalized $\mathrm{SU}(2)$ quivers with $g$ loops and $n$ external edges can be achieved. This process suggests the existence of an underlying $4d$ $\mathcal{N}=2$ SCFT associated with the smooth surface $S_{g,n}$ of genus $g$ and $n$ punctures. The space of exactly marginal deformations of this theory is identified with the Teichmüller space $\mathcal{T}(S_{g,n})$, on which the mapping class group $\mathrm{Mod}(S_{g,n})$ acts as the duality group. The weakly-coupled descriptions of the theory, or equivalently, the corresponding generalized $\mathrm{SU}(2)$ quivers, correspond to the decompositions of $S_{g,n}$ into pairs of pants.

\begin{remark}
    For every decomposition of $S_{g,n}$ into pairs of pants, there is a corresponding system of Fenchel--Nielsen coordinates on $\mathcal T(S_{g,n})$ (see \Cref{Sec:Teichcoordinates}). The weak-coupling limit associated to this decomposition corresponds to the region in $\mathcal T(S_{g,n})$ in which the length of the simple closed curves on $S_{g,n}$ appearing in the decomposition in pairs of pants is very small. In this limit, the Fenchel–Nielsen coordinates can be associated with the parameters of the UV theory: the length of each simple closed curve corresponds to the UV gauge coupling of the associated $\mathrm{SU}(2)$ gauge factor, while the twist parameter is identified with its theta angle $\Theta$ \cite{Gaiotto:2009gz}.
\end{remark}

\begin{remark} 
    Given a \emph{simple closed curve}\index{simple closed curve} $\gamma$ on $S_{g,n}$, one can consider the limit where its length $l(\gamma)$ becomes zero. In this limit, for any decomposition of $S_{g,n}$ into pairs of pants that includes $\gamma$, the $\mathrm{SU}(2)$ gauge group corresponding to $\gamma$ decouples entirely. Then:
    \begin{itemize}
    \item If $\gamma$ is \emph{non-separating}\index{simple closed curve!non-separating}\footnote{A simple closed curve $\gamma$ on a smooth surface $S$ is called separating if $S\setminus\gamma$ has two connected components, and non-separating otherwise.}, this leaves a generalized quiver corresponding to $S_{g-1,n+2}$.
    \item If $\gamma$ is \emph{separating}\index{simple closed curve!separating}, this results in two generalized quivers corresponding to $S_{g_1,n_1}$ and $S_{g_2,n_2}$, where $g_1+g_2=g$ and $n_1+n_2=n+2$. 
    \end{itemize} 
    This corresponds to infinities in the Teichmüller space $\mathcal T(S_{g,n})$ (and the corresponding moduli spaces) corresponding either to $\mathcal T(S_{g-1,n+2})$ if $\gamma$ is non-separating, or to products $\mathcal T(S_{g_1,n_1})\times\mathcal T(S_{g_2,n_2})$ if $\gamma$ is separating.

    By considering the $4d$ $\mathcal{N}=2$ theories associated with $S_{g,n}$, $S_{g_1,n_1}$, or with $S_{g_1,n_1}$ and $S_{g_2,n_2}$, and more specifically the branches of their moduli spaces (either Coulomb or Higgs), this hints at the existence of interesting $1+1$-dimensional \emph{topological quantum field theories}\index{topological!quantum field theory} valued in the category of holomorphic symplectic manifolds \cite{Moore:2011ee}.
\end{remark}

\subsection{Six-dimensional definition}

In Witten's construction of Seiberg--Witten curves \cite{Witten:1997sc}, the uplift of a Type IIA brane configuration to M-theory typically yields a single M5-brane wrapping a complex curve $\Sigma$ within $\mathbb{C} \times C_\mathrm{UV}$, where $C_\mathrm{UV}$ denotes the UV curve—either a punctured sphere for linear quivers or a punctured elliptic curve for necklace quivers. The Riemann surface $\Sigma$ is identified with the Seiberg--Witten curve of the worldvolume theory on the D4-branes, and is a ramified covering of degree $2$ (for $\mathrm{SU}(2)$ quivers) over $C_\mathrm{UV}$.

The M5-brane wrapping $\Sigma$ is a normalizable deformation of a configuration in which two M5-branes wrap $C_\mathrm{UV}$ and intersect transverse M5-branes extending along $\mathbb{C}$, located at fixed points $t_a\in C_\mathrm{UV}$. The worldvolume theory on a pair of M5-branes is the six-dimensional $\mathcal{N} = (2,0)$ superconformal theory of type $A_1$. At sufficiently low energies, the transverse M5-branes manifest as codimension-2 \emph{defects}\index{defect} at $t_a \in C_\mathrm{UV}$ within the $6d$ $\mathcal{N} = (2,0)$ theory. The deformation leading to a single M5-brane wrapping $\Sigma$ is determined by the vacuum expectation values of scalar fields in the $6d$ $\mathcal{N} = (2,0)$ $A_1$ theory, which describe the motion of the M5-branes along $\mathbb{C}$. Accordingly, $4d$ $\mathcal{N} = 2$ linear or necklace conformal quiver theories can be defined as the low-energy limit of the $6d$ $\mathcal{N} = (2,0)$ $A_1$ theory compactified on $C_\mathrm{UV}$, with codimension-2 simple defects at $t_a\in C_\mathrm{UV}$.

This framework extends to UV curves beyond $\mathbb{C}^*$ or $E_\tau$. Given a Riemannian surface $C_{g,n}$ of genus $g$ with $n$ punctures, one can compactify the $6d$ $\mathcal{N} = (2,0)$ $A_1$ theory on $C_{g,n}$ with codimension-2 simple defects at the punctures, and then take the low-energy limit. This compactification defines a $4d$ $\mathcal{N} = 2$ superconformal field theory \cite{Gaiotto:2009we}. To preserve $\mathcal{N}=2$ supersymmetry in four dimensions, a partial \emph{twist}\index{twist} of the $6d$ theory along $C_{g,n}$ is required. The effect of the twist is that the resulting $4d$ $\mathcal{N} = 2$ theory solely depends on the conformal class of the metric on $C_{g,n}$, i.e. on its complex structure. The Riemann surface $C_{g,n}$ is the UV curve of the theory. The $6d$ $\mathcal N=(2,0)$ theory of type $A_1$ has a protected operator of $R$-charge $2$ which becomes the quadratic differential $\phi_2(z)$ under the twist. 

The space of exactly marginal deformations of the resulting theory is identified with the Teichmüller space $\mathcal T(S_{g,n})$, where $S_{g,n}$ is the smooth surface underlying $C_{g,n}$. The duality group is the mapping class group $\mathrm{Mod}(S_{g,n})$. Thus, for each smooth surface $S_{g,n}$, one obtains a family of theories (modulo exactly marginal deformations). Each choice of complex structure on $S_{g,n}$---equivalently, each point in $\mathcal T(S_{g,n})$---corresponds to setting the exactly marginal parameters of the theory to specific values.

\begin{definition} 
The $4d$ $\mathcal{N} = 2$ theories obtained in this manner are referred to as \emph{theories of class $\mathcal S$ of type $A_1$}, and are denoted $\mathcal{S}[A_1, C_{g,n}]$, where $C_{g,n}$ represents the UV curve. The Coulomb branch of such a theory corresponds to the space of quadratic differentials $\phi_2(z)$ on $C_{g,n}$, which have order-1 poles at the punctures without masses and order-2 poles when masses are present. The Seiberg--Witten (SW) curve $\Sigma$ is a degree-2 ramified cover of $C_{g,n}$ in $T^*_{1,0}C_{g,n}$, specified by the equation: 
\begin{equation*} 
    \lambda^2 = \phi_2(z)\; , 
\end{equation*} 
where $\lambda$ denotes the holomorphic Liouville 1-form on $T^*_{1,0}C_{g,n}$. The space of exactly marginal couplings for the theory is $\mathcal{T}(S_{g,n})$, and its duality group is $\mathrm{Mod}(S_{g,n})$.
\end{definition}

\begin{remark}
    As noted above in the context of $\mathrm{SU}(2)$ SYM theories with $F \leq 3$ flavors, it is also possible to compactify the $6d$ $\mathcal{N}=(2,0)$ theory of type $A_1$ in the presence of irregular punctures. This approach allows the description of a wide class of $4d$ $\mathcal{N}=2$ asymptotically free theories as class $\mathcal{S}$ theories \cite{Gaiotto:2009hg}. In general, these theories are denoted $\mathcal{S}[A_1, C_{g,n},\mathcal{D}]$, where $\mathcal{D}$ encodes the nature of each puncture and is called the \emph{defect data}\index{defect data}.
\end{remark}

\section{Theories of class \texorpdfstring{$\mathcal S$}{S} of type \texorpdfstring{$A_2$}{A2}}

Four-dimensional $\mathcal N=2$ $\mathrm{SU}(N)$ SYM with $F=2N$ fundamental hypermultiplets is superconformal not only for $N=2$ but also for $N \geq 3$. The Seiberg–Witten curves for these theories were first computed in \cite{Argyres:1995wt}. Notably, these curves are not $\mathrm{SL}_2(\mathbb{Z})$-invariant\footnote{In $4d$ $\mathcal N=2$ $\mathrm{SU}(2)$ SYM with $F=4$ massless fundamental hypermultiplets, the low-energy gauge coupling $\tau$ and the UV coupling $\widetilde{\tau}$ are related by $\tau = 2\widetilde{\tau} = \theta/\pi + 2\pi i/g^2$. The duality group acts on $\tau$ via fractional linear transformations, so $T$ corresponds to $\theta \rightarrow \theta + \pi$, which is not \emph{a priori} a symmetry of the weakly-coupled description, unlike $T^2$. Note that $T$ permutes the three 8-dimensional representations of $\mathrm{SO}(8)$ in a non-trivial manner.}, but instead are invariant under the congruence subgroup:
\begin{equation*}
    \Gamma_0(2) = \left.\left\{\begin{pmatrix}
        a & b \\ c & d 
    \end{pmatrix}\in\mathrm{PSL}_2(\mathbb{Z})~\right\vert~c\equiv 0~[2] \right\} < \mathrm{PSL}_2(\mathbb{Z})\; .
\end{equation*}

Fundamental domains in $\mathbb{H}$ for the groups $\mathrm{PSL}_2(\mathbb{Z})$ and $\Gamma_0(2)$ are shown in \Cref{fig:gamma02}, together with their orbifold points of order 2 (purple rhombi) and 3 (blue triangles).

\begin{figure}[!ht]
    \centering
    \includegraphics[scale=0.83]{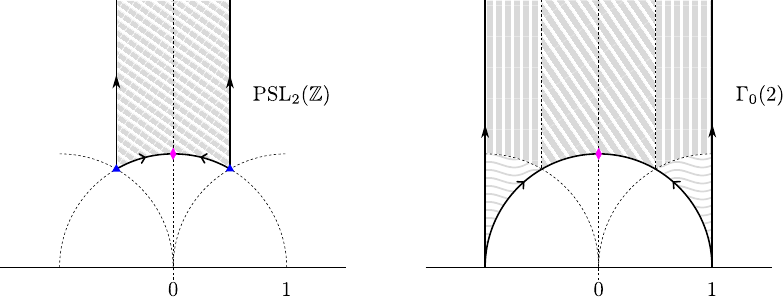}
    \caption{Fundamental cells for $\mathrm{PSL}_2(\mathbb{Z})$ and $\Gamma_0(2)$ in $\mathbb{H}$.}
    \labelx{fig:gamma02}
\end{figure}

The fundamental domain for $\mathrm{PSL}_2(\mathbb{Z})$ depicted on the left of \Cref{fig:gamma02} shows that, up to dualities, the $4d$ $\mathcal N=2$ SCFT can always be described as an $\mathrm{SU}(2)$ gauge theory with four fundamental flavors and gauge coupling $\vert \tau \vert \geq 1$. Correspondingly, $\mathbb{H}/ \mathrm{PSL}_2(\mathbb{Z})$ has a single cusp. In contrast, within the fundamental domain for $\Gamma_0(2)$ shown on the right of \Cref{fig:gamma02}, one can take the limit $\tau \rightarrow 1$. As $\mathbb{H}/\Gamma_0(2)$ has two cusps, this limit is \emph{a priori} not equivalent to $\tau\rightarrow\infty$. This raises the question of how $4d$ $\mathcal N=2$ $\mathrm{SU}(N)$ $F=2N$ SYM can be characterized as $\tau\rightarrow 1$.

\subsection{Argyres--Seiberg duality} 

The $S$-duality properties of $4d$ $\mathcal N=2$ $\mathrm{SU}(3)$ SYM with $F=6$ fundamental hypermultiplets were explored by Argyres and Seiberg in \cite{Argyres:2007cn}. The Seiberg–Witten curve for this theory has genus two; at the cusp $\Im \tau \to \infty$, three non-intersecting cycles shrink. In this limit, the gauge coupling $g$ goes to 0 and the full $\mathrm{SU}(3)$ gauge symmetry is restored. The three cycles that pinch correspond to the three $W$-bosons (along with their antiparticles) becoming massless for $g = 0$. In contrast, as $\tau \to 1$, only one cycle pinches on the Seiberg–Witten curve, indicating that a single BPS particle becomes massless. Thus, the physics at $\tau = 1$ cannot be that of a weakly-coupled $4d$ $\mathcal N=2$ $\mathrm{SU}(3)$ SYM with $F=6$ fundamental hypermultiplets.

The pinching of a single cycle, rather, suggests a description involving an $\mathrm{SU}(2)$ gauge factor that becomes weakly coupled as $\tau \to 1$. However, this description is necessarily incomplete, as the rank of a $4d$ $\mathcal N=2$ $\mathrm{SU}(2)$ gauge theory is one, whereas the rank of $4d$ $\mathcal N=2$ $\mathrm{SU}(3)$ SYM with $F=6$ fundamental hypermultiplets is two.

\begin{proposition}
Near the cusp $\tau \to 1$, $\mathrm{SU}(3)$ SYM with $F=6$ massless fundamental hypermultiplets is better described as a weakly-coupled $\mathrm{SU}(2)$ gauge theory coupled to a single massless hypermultiplet and to an isolated rank-1 $4d$ $\mathcal N=2$ SCFT, specifically the $E_6$ \emph{Minahan--Nemeschansky SCFT}\index{Minahan--Nemeschansky theory} \cite{Minahan:1996fg}. 
\end{proposition}

The coupling between the $\mathrm{SU}(2)$ gauge group and the $E_6$ SCFT arises from gauging the $\mathrm{SU}(2)$ factor in the maximal subgroup $\mathrm{SU}(2) \times \mathrm{SU}(6)$ of the global symmetry group $E_6$ in the Minahan–Nemeschansky theory. Gauging this $\mathrm{SU}(2)$ factor breaks the global symmetry $E_6$ down to $\mathrm{SU}(6)$, while the pseudo-real hypermultiplet in the fundamental of $\mathrm{SU}(2)$ contributes an $\mathrm{SO}(2)\simeq \mathrm{U}(1)$ factor. Hence, the total flavor symmetry is $\mathrm{SU}(6) \times \mathrm{U}(1) \simeq \mathrm{U}(6)$, which matches the flavor symmetry of $4d$ $\mathcal N=2$ $\mathrm{SU}(3)$ SYM with $F=6$ massless fundamental hypermultiplets\footnote{The fundamental of $\mathrm{SU}(3)$ is a genuine complex representation.}. The $S$-duality between $\mathrm{SU}(3)$ $F=6$ SYM and $\mathrm{SU}(2)$ SYM with a fundamental flavor and coupled to the rank-1 $E_6$ SCFT is called \emph{Argyres--Seiberg duality}\index{Argyres--Seiberg duality}.

Four-dimensional $\mathcal N=2$ $\mathrm{SU}(3)$ SYM with $F=6$ massless fundamental hypermultiplets can be represented as the quiver shown on the left in \Cref{fig:dualitiesSU3F=6}, where flavor nodes labeled by $1$ (resp. $3$) correspond to $\mathrm{U}(1)$ (resp. $\mathrm{SU}(3)$) factors. This representation follows from evenly splitting the set of six fundamental hypermultiplets into two, corresponding to the maximal subgroup $\mathrm{U}(3) \times \mathrm{U}(3)$ of the flavor symmetry group $\mathrm{U}(6)$. Each $\mathrm{U}(3)$ factor is then decomposed as $\mathrm{U}(3) \simeq \mathrm{SU}(3) \times \mathrm{U}(1)$.

The Argyres–Seiberg $S$-dual theory is represented by the quiver on the right in \Cref{fig:dualitiesSU3F=6}, focusing on the maximal subgroup $\mathrm{SU}(3) \times \mathrm{SU}(3) \times \mathrm{SU}(3)$ of the $E_6$ global symmetry from the Minahan–Nemeschansky theory, which appears as the rightmost three-pronged junction in the quiver. For further details, we refer to \cite[Sect. 3]{Gaiotto:2009we}. Note that the $E_6$ theory with the $\mathrm{SU}(2)$ factor gauged plays an analogous role in such \emph{generalized $\mathrm{SU}(3)$ quivers}\index{quiver!$\mathcal N=2$!generalized} as the blocks of four hypermultiplets in generalized $\mathrm{SU}(2)$ quivers.

\begin{figure}[!ht]
    \centering
    \includegraphics[scale=0.83]{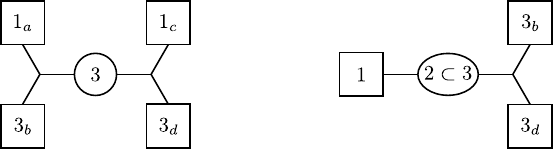}
    \caption{Quivers for $\mathrm{SU}(3)$ SYM with $F=6$ fundamental flavors and Argyres--Seiberg dual frame.}
    \labelx{fig:dualitiesSU3F=6}
\end{figure}

The quivers of \Cref{fig:dualitiesSU3F=6} suggest that $4d$ $\mathcal N=2$ $\mathrm{SU}(3)$ SYM with $F=6$ massless fundamental hypermultiplets can be associated with the Riemann sphere with four punctures of two different types (again called \emph{UV curve}\index{UV curve}), as shown in \Cref{fig:degenerationsSU3}, which illustrates the two degeneration limits of the surface. Each degeneration limit corresponds to one of the quivers shown in \Cref{fig:dualitiesSU3F=6}. In the Argyres--Seiberg dual, the $\mathrm{U}(1)$ flavor factor corresponding to the single fundamental hypermultiplet of $\mathrm{SU}(2)$ emerges as a combination of the flavor factors $\mathrm{U}(1)_a$ and $\mathrm{U}(1)_c$ in the original description, while the other combination is the commutant of $\mathrm{SU}(2)$ in $\mathrm{SU}(3)$.

\begin{figure}[!ht]
    \centering
    \includegraphics[scale=0.83]{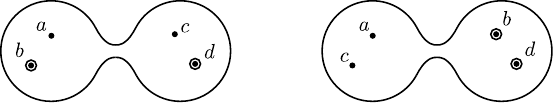}
    \caption{Degenerations of a sphere with $2+2$ punctures, where the simple dots correspond to $\mathrm{U}(1)$ flavor factors (simple punctures), and circled ones, to $\mathrm{SU}(3)$ flavor factors (full punctures).}
    \labelx{fig:degenerationsSU3}
\end{figure}

The justification for this geometric picture again comes from the computation of the Seiberg--Witten curve for $\mathrm{SU}(3)$ SYM with 6 fundamental hypermultiplets, following the method described in \Cref{Sec:Mtheoryuplifts}. One can present the Seiberg--Witten curve (without mass deformations) as:
\begin{equation*}
    (\Sigma)\colon\;\;\lambda^3 - \phi_2(z)\lambda - \phi_3(z) = x^3\frac{\mathrm{d}z^3}{z^3} - \frac{u^{(2)}\mathrm{d}z^2}{(z-1)(z-q)z}\left(x\frac{\mathrm{d}z}{z}\right) + \frac{u^{(3)}\mathrm{d}z^3}{(z-1)(z-q)z^2} = 0\; ,
\end{equation*}
where $\lambda = x\mathrm{d}z/z$ is the Liouville 1-form on the holomorphic cotangent bundle of $\mathbb{CP}^1_z$. Note that $\phi_2(z)$ is a quadratic differential on $\mathbb{CP}^1_z$ with order-1 poles at $0,q,1,\infty$, and $\phi_3(z)$ is a cubic differential on $\mathbb{CP}^1_z$ with order-1 poles at $z=q,1$ and order-2 poles at $z=0,\infty$, where $u^{(2)}$ and $u^{(3)}$ are complex parameters. 

\begin{definition}
    Without mass deformation, the points on $\mathbb{CP}^1_z$ where $\phi_2$ has order-1 poles and $\phi_3$ order-2 poles are called \emph{full punctures}\index{puncture!full}. The points where both $\phi_2$ and $\phi_3$ have order-1 poles are called \emph{simple punctures}\index{puncture!simple}. 
\end{definition}

In \Cref{fig:degenerationsSU3}, the simple punctures $a$ and $c$ are located at $z=q,1$ while the full punctures $b$ and $d$ are located at $z=0,\infty$. The limit $\tau \to 1$ corresponds to colliding the punctures $a$ and $c$, i.e. $q \to 1$. In this limit, the Seiberg–Witten curve simplifies to:
\begin{equation*}
    \lambda^3 = \frac{u^{(2)}\mathrm{d}z^2}{(z-1)^2z}\left(x\frac{\mathrm{d}z}{z}\right) + \frac{u^{(3)}\mathrm{d}z^3}{(z-1)^2z^2}\; .
\end{equation*}
In this limit, the parameter $u^{(2)}$ can be identified with the Coulomb branch parameter of the weakly-coupled $\mathrm{SU}(2)$ gauge group appearing in the Argyres–Seiberg dual frame. Setting $u^{(2)}=0$, the curve reduces to:
\begin{equation*}
    \lambda^3 = \frac{u^{(3)}}{(z-1)^2z^2}\; ,
\end{equation*}
which describes the Seiberg–Witten curve of the $E_6$ Minahan–Nemeschansky SCFT.

\begin{remark}
    In the presence of generic masses for the hypermultiplets, $\phi_2$ (resp. $\phi_3$) has order-2 (resp. order-3) poles at $z=0,q,1,\infty$. The residues of $\lambda_{\vert \Sigma}$ at the poles above $z=0,\infty$ are of the form $(m_1+m_2,-m_1,-m_2)$ for $m_1,m_2\in\mathbb{C}$, while above $z=q,1$, they are of the form $(2m,-m,-m)$ for $m\in\mathbb{C}$. This indicates that full punctures correspond to $\mathrm{SU}(3)$ flavor factors, with $(m_1+m_2,-m_1,-m_2)$ in their complexified Cartan subalgebra, and that simple punctures correspond to $\mathrm{U}(1)$ flavor factors. Both full and simple punctures are \emph{regular $\mathrm{SU}(3)$ punctures}\index{puncture!regular} in the sense of \Cref{def:puncturetype}.
\end{remark}

\subsection{Generalized \texorpdfstring{$\mathrm{SU}(3)$}{SU(3)} quivers}

The Seiberg--Witten curve of the linear quiver on the left in \Cref{fig:moreSU3quivers} can be computed with the methods of \Cref{Sec:Mtheoryuplifts}, from which one learns that this theory naturally corresponds to a specific degeneration limit of the UV curve, which in that case is the Riemann sphere with five punctures, two full and three simple. As in \Cref{Sec:genSU2quivers}, we first assume that the two $\mathrm{SU}(3)$ gauge factors are very weakly coupled. In the limit $\tau_2 \rightarrow 1$ where $\tau_2$ is the gauge coupling of the second node, the second $\mathrm{SU}(3)$ gauge factor is better described in the Argyres–Seiberg dual frame, yielding the quiver to the right of \Cref{fig:moreSU3quivers}, with the corresponding degeneration of $S_{0,5}$ depicted below.

\begin{figure}[!ht]
    \centering
    \includegraphics[scale=0.83]{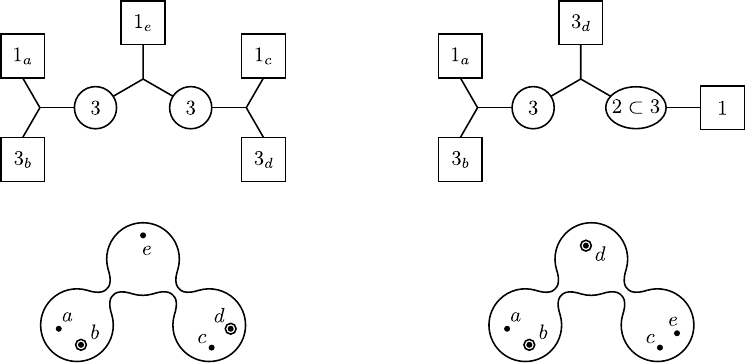}
    \caption{More general $\mathrm{SU}(3)$ quivers and the corresponding degenerations of $S_{0,5}$.}
    \labelx{fig:moreSU3quivers}
\end{figure}

At this stage, one cannot proceed as in \Cref{Sec:genSU2quivers}: the $S$-dual description of an $\mathrm{SU}(3)$ gauge factor coupled to the $E_6$ SCFT is not known. However, similarly to \Cref{Sec:genSU2quivers}, one can investigate all possible generalized $\mathrm{SU}(3)$ quivers. It can then be argued that these generalized $\mathrm{SU}(3)$ quivers encompass all possible weakly-coupled descriptions of the theories $\mathcal{S}[A_2, S_{g,n},D]$ introduced below. The elementary building blocks of generalized $\mathrm{SU}(3)$ quivers are the following:
\begin{itemize}
    \item \textbf{Blocks of nine hypermultiplets:}
    \begin{itemize}
        \item When these consist of three hypermultiplets in the fundamental representation of an $\mathrm{SU}(3)$ gauge factor, the flavor symmetry is $\mathrm{SU}(3) \times \mathrm{U}(1)$.
        \item When they consist of a single hypermultiplet in the bifundamental representation of a product $\mathrm{SU}(3) \times \mathrm{SU}(3)$, the flavor symmetry reduces to $\mathrm{U}(1)$.
    \end{itemize} 
    Such blocks are represented as trivalent vertices attached to two $\mathrm{SU}(3)$ and one $\mathrm{U}(1)$ factor.
    \item $\mathbf{E_6}$ \textbf{SCFTs}, with a choice of an $\mathrm{SU}(3)^3$ subgroup of the $E_6$ flavor symmetry group. Such an $E_6$ SCFT is represented as a trivalent vertex connected to three $\mathrm{SU}(3)$ factors.
\end{itemize}

Given a pair of $\mathrm{SU}(3)$ flavor factors, one can \emph{gauge the diagonal SU(3) subgroup} of their product, resulting in an $\mathrm{SU}(3)$ gauge group. Moreover, one can \emph{gauge an SU(2) subgroup} of an $\mathrm{SU}(3)$ flavor factor and couple it to a fundamental hypermultiplet. Further examples of generalized $\mathrm{SU}(3)$ quivers are shown in \Cref{fig:generalizedSU3quivers}. In general, if the quiver displays $m_2$ blocks of hypermultiplets, $m_3$ $E_6$ SCFTs, $n_3$ $\mathrm{SU}(3)$ gauge groups and $n_2$ $\mathrm{SU}(2)$ gauge groups, then there are $n_2+n_3$ (exactly marginal) gauge couplings. The explicit flavor symmetry of the quiver consists of $f_3=3m_3+2m_2-2n_3-n_2$ $\mathrm{SU}(3)$ factors and $f_1 = m_2+2n_2$ $\mathrm{U}(1)$ factors. The quiver has $g=n_3+1-m_3-m_2$ loops.

\begin{figure}[!ht]
    \centering
    \includegraphics[scale=0.83]{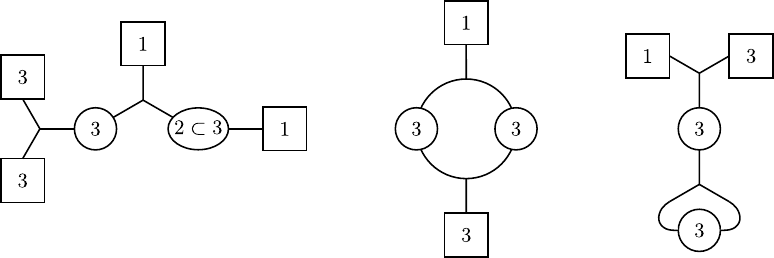}
    \caption{Examples of generalized $\mathrm{SU}(3)$ quivers.}
    \labelx{fig:generalizedSU3quivers}
\end{figure}

Full punctures can emerge from surface degenerations or from collisions of simple punctures. Consider, for instance, the generalized $\mathrm{SU}(3)$ quiver on the top row of \Cref{fig:decouplingsSU3}. The Seiberg–Witten curve associated with this linear quiver can be calculated using the methods of \Cref{Sec:Mtheoryuplifts}. In this case, one finds that the UV curve is a Riemann sphere with six simple punctures. For convenience, let us denote here the Riemann sphere with $f_1$ simple punctures and $f_3$ full punctures as $C_{0,(f_1,f_3)}$.

One possible degeneration limit of $C_{0,(6,0)}$ involves decoupling the $\mathrm{SU}(3)$ node, as depicted in the second row of \Cref{fig:decouplingsSU3}. In the limit, $C_{0,(6,0)}$ becomes the disjoint union of two copies of $C_{0,(3,1)}$, as one obtains two decoupled quivers with an additional $\mathrm{SU}(3)$ flavor factor each. Another degeneration occurs, for instance, when the left $\mathrm{SU}(2)$ gauge group decouples. This restores the full $\mathrm{SU}(3)$ flavor symmetry and, moreover, corresponds to the collision of two simple punctures. From this, we see that colliding two simple punctures yields a full puncture, as shown in the third row of \Cref{fig:decouplingsSU3}. The remaining simple punctures can then be paired off by decoupling the associated $\mathrm{SU}(2)$ gauge groups, one of which is implicit and only appears explicitly in the Argyres–Seiberg dual phase of the $\mathrm{SU}(3)$ gauge group. From this, we conclude that the UV curve of the Minahan–Nemeschansky $E_6$ SCFT is the Riemann sphere with three full punctures.

\begin{figure}[!ht]
	\centering
	\includegraphics[width=\textwidth]{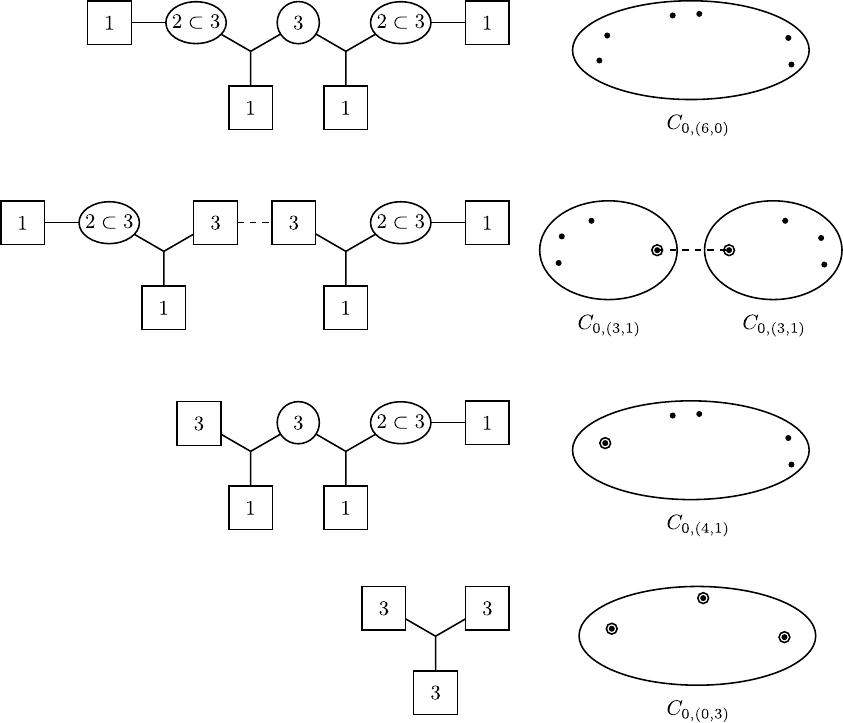}
	\caption{Different decoupling limits of a generalized $\mathrm{SU}(3)$ quiver.}
	\labelx{fig:decouplingsSU3}
\end{figure}

\subsection{Six-dimensional definition}

The M-theory uplift of the $\mathrm{SU}(3)$ $\mathcal N=2$ quiver theories studied in \cite{Witten:1997sc} involve a single M5-brane wrapping the Seiberg--Witten curve $\Sigma$, where $\Sigma$ is a ramified covering of degree $3$ of some UV curve $C_\mathrm{UV}$, which is either a punctured Riemann sphere (for linear quivers) or a punctured elliptic curve (for necklace quivers). This is a normalizable deformation of a configuration involving three M5-branes wrapping $C_\mathrm{UV}$ and intersecting M5-branes localized at some point of $C_\mathrm{UV}$. The worldvolume theory on a stack of three M5-branes is the $6d$ $\mathcal N=(2,0)$ SCFT of type $A_2$. At sufficiently low energy, the transverse M5-branes manifest as codimension-2 \emph{defects}\index{defect} in the 6d theory, localized at points in $C_\mathrm{UV}$ and extending along $x^{0,1,2,3}$. The $6d$ $\mathcal N=(2,0)$ SCFT of type $A_2$ admits simple and full defects, corresponding to the simple and full punctures described earlier. 

This framework extends to UV curves $C_\mathrm{UV}=C_{g,n}$ of arbitrary genus $g$ and number of punctures $n$. Every puncture on $C_\mathrm{UV}$ is either simple or full. The four-dimensional limit of the compactification of the $6d$ $\mathcal N=(2,0)$ theory on $C_\mathrm{UV}$ with prescribed defects at the punctures, together with a twist so as to preserve $\mathcal N=2$ supersymmetries in four-dimensions, yields a $4d$ $\mathcal N=2$ superconformal field theory. The $6d$ $\mathcal N=(2,0)$ SCFT admits protected operators of $R$-charge $2$ and $3$, which become quadratic and cubic differentials on $C_\mathrm{UV}$ after the twist, respectively. The space of exactly marginal deformations of the resulting theory is identified with the Teichmüller $\mathcal T(S_{g,n})$, where $S_{g,n}$ is the smooth surface underlying $C_{g,n}$. The duality group is identified with the mapping class group $\mathrm{Mod}(S_{g,n})$.
 
\begin{definition}\labelx{def:classSN=3}
The $4d$ $\mathcal{N} = 2$ theories obtained in this manner are referred to as \emph{theories of class $\mathcal S$ of type $A_2$}, and are denoted $\mathcal{S}[A_2, C_{g,n},\mathcal{D}]$, where $C_{g,n}$ is the UV curve and $\mathcal{D}$ is defect data, namely which punctures of $C_{g,n}$ are simple, and which are full. The Coulomb branch of such a theory corresponds to the space of quadratic and cubic differentials $\phi_2$ and $\phi_3$ on $C_{g,n}$. Without mass deformation, $\phi_2$ has order-1 poles at every puncture, while $\phi_3$ has order-1 poles at simple punctures and order-2 poles at full punctures. The Seiberg--Witten (SW) curve is a degree-3 ramified covering of $C_{g,n}$ in $T^*_{1,0}C_{g,n}$, specified by the equation: 
\begin{equation*} 
    \lambda^3 = \phi_2\lambda+\phi_3\; , 
\end{equation*} 
where $\lambda$ denotes the holomorphic Liouville 1-form on $T^*_{1,0}C_{g,n}$. The space of exactly marginal deformations is $\mathcal{T}(S_{g,n})$, and the duality group is $\mathrm{Mod}(S_{g,n})$.
\end{definition}

\begin{remark}
    One can also consider theories of class $\mathcal S$ with irregular punctures. Such theories are again denoted $\mathcal{S}[A_2, C_{g,n},\mathcal{D}]$, where $\mathcal{D}$ encodes the precise information on each puncture. This allows the description of a large class of asymptotically free theories. In contrast to type $A_1$, there is yet another kind of puncture that can be considered in type $A_2$, related to the fact that the compact Lie algebra $\mathfrak{su}(3)$ has a non-trivial outer automorphism group $\mathbb{Z}_2$ \cite{Tachikawa:2010vg}. Such punctures involve a twist by this outer automorphism, and are therefore called \emph{twisted punctures}\index{puncture!twisted}.
\end{remark}

\section{Theories of class \texorpdfstring{$\mathcal S$}{S} of type \texorpdfstring{$A_{N-1}$}{A(N-1)}}

The construction of theories of class $\mathcal S$ of type $A_{N-1}$ is very similar to type $A_2$, but for the fact that the classification of regular punctures includes new cases besides simple and full punctures. 

The Seiberg--Witten curve for $\mathrm{SU}(N)$ SYM with $2N$ massless fundamental hypermultiplets can be obtained with the methods of \Cref{Sec:Mtheoryuplifts}. One finds the equation:
\begin{equation*}
    \lambda^N  = \sum_{i=2}^N \frac{u^{(i)}\mathrm{d}z^i}{(z-1)(z-q)z^{i-1}}\lambda^{N-i} = \sum_{i=2}^N \phi_i(z)\lambda^{N-i}\; ,
\end{equation*}
where $\lambda$ is the Liouville 1-form on $T^*_{1,0}\mathbb{CP}^1_z$, and where each $\phi_i$ is an $i$-differential on $\mathbb{CP}^1_z$ holomorphic away from the punctures $z=0,q,1,\infty$. Each $\phi_i$ has a pole of order $1$ at $z=q,1$ and a pole of order $i-1$ at $z=0,\infty$.

\begin{definition}
    The punctures where all differentials $\phi_i$ have an order-1 pole are called \emph{simple}\index{puncture!simple}; those where each $\phi_i$ has an order $(i-1)$-pole are called \emph{full}\index{puncture!full}. Both simple and full punctures are \emph{regular}\index{puncture!regular}.
\end{definition}

As before, the \emph{UV curve} of $\mathrm{SU}(N)$ SYM with $2N$ fundamental flavors is the Riemann sphere $\mathbb{CP}^1$ with two simple punctures and two full punctures. When the fundamental hypermultiplets are assigned generic masses, one can verify explicitly that the residues of $\lambda$ at the poles above $0$ and $\infty$ take the form $(m_1+\dots+m_{N-1},-m_1,\dots,-m_{N-1})$. This vector belongs to the (complexified) Cartan subalgebra of $\mathfrak{su}(N)$. Meanwhile, the residues at the poles above $q$ and $1$ are of the form $((N-1)m,-m,\dots,-m)$. This vector rather belongs to the (complexified) Cartan subalgebra of $\mathfrak{u}(1)$, where $\mathfrak{u}(1)\simeq \mathbb{C}$. Therefore, one associates a $\mathrm{U}(1)$ flavor factor to each simple puncture, and an $\mathrm{SU}(N)$ flavor factor to each full puncture.

The linear conformal quiver shown in \Cref{fig:SUNquiver}, where round nodes correspond to $\mathrm{SU}(N)$ gauge factors, and square ones, to $\mathrm{U}(N)$ flavor factors, consists of $k$ $\mathrm{SU}(N)$ gauge groups with bifundamental hypermultiplets, together with $N$ fundamentals for the first and last $\mathrm{SU}(N)$ gauge factor. The manifest $\mathrm{SU}(N)\times\mathrm{U}(1)^{k+1}\times\mathrm{SU}(N)$ flavor symmetry of this quiver in encoded in the UV curve of the theory, which is the Riemann sphere with $k+1$ simple punctures and 2 full punctures.

\begin{figure}[!ht]
    \centering
    \includegraphics[scale=0.83]{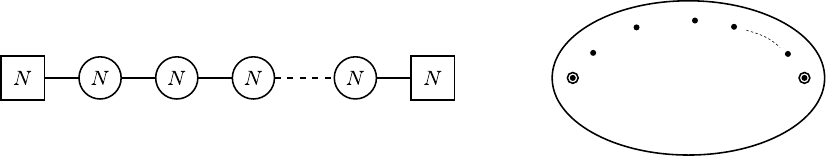}
    \caption{A linear conformal quiver of $\mathrm{SU}(N)$ gauge groups.}
    \labelx{fig:SUNquiver}
\end{figure}

Similarly to the previous sections, one defines class $\mathcal S$ theories of type $A_{N-1}$ in the following way:
\begin{definition}
    A \emph{class $\mathcal S$ theory of type $A_{N-1}$} is the four-dimensional limit of the compactification of the $6d$ $\mathcal N=(2,0)$ theory of type $A_{N-1}$ (which is the worldvolume theory of $N$ coincident NS5-branes) on a punctured Riemann surface $C_{g,n}$ called the UV curve, with a partial twist so as to preserve $4d$ $\mathcal N=2$ supersymmetry in four dimensions. The $6d$ $\mathcal N=(2,0)$ SCFT of type $A_{N-1}$ has protected operators of R-charge $2,3,\dots,N$ which become holomorphic differentials $\phi_2,\phi_3,\dots,\phi_N$ on $C_{g,n}$ of order $2,3,\dots,N$, respectively, after the twist. The behavior of the holomorphic differentials at the punctures is encoded in the \emph{defect data}\index{defect!data} $\mathcal{D}$, as prescribing it corresponds to inserting codimension-2 defects in the 6d theory, localized at the punctures of $C_{g,n}$ and extending along $x^{0,1,2,3}$. The Seiberg--Witten curve for this theory is a degree-$N$ ramified covering of $C_{g,n}$ in $T^*_{1,0}C_{g,n}$, specified by the equation:
    \begin{equation}\labelx{eq:SWclassSAN-1}
        \lambda^N = \sum \phi_i \lambda^{N-i}\; ,
    \end{equation}
    where $\lambda$ denotes the holomorphic Liouville 1-form on $T^*_{1,0}C_{g,n}$. Theories of class $\mathcal S$ of type $A_{N-1}$ are denoted $\mathcal S[A_{N-1},C_{g,n},\mathcal{D}]$. For superconformal theories, the space of exactly marginal deformations is $\mathcal{T}(S_{g,n})$ and the duality group is $\mathrm{Mod}(S_{g,n})$, where $S_{g,n}$ is the smooth surface underlying $C_{g,n}$.
\end{definition}

\begin{remark}
    A standard approach to constructing $4d$ $\mathcal{N}=2$ SCFTs within the class $\mathcal{S}$ of type $A_{N-1}$ involves considering only \emph{regular punctures}\index{puncture!regular}, as defined in \Cref{def:puncturetype}. However, as in type $A_2$, punctures may also be \emph{irregular}\index{puncture!irregular} or \emph{twisted}\index{puncture!twisted}. Irregular and twisted punctures for theories of class $\mathcal S$ of type $A_{N-1}$ have been studied in \cite{Chacaltana:2010ks,Chacaltana:2012ch}.
\end{remark}

Setting aside the cases of irregular and twisted punctures, we now review the classification of regular punctures in type $A_{N-1}$.

\subsection{Regular punctures}

\begin{proposition}\labelx{prop:regpuncture}
    Regular punctures can be characterized as punctures at which, for all $i$, the holomorphic differential $\phi_i$ has a pole of order at most $i-1$, in the absence of mass deformations. When mass deformations are present, the differential $\phi_i$ may exhibit an order-$i$ pole, ensuring that the differential $\lambda$ has simple poles at the preimages of the punctures under the covering map $\Sigma \rightarrow C_{g,n}$, with residues corresponding to the mass parameters, where $\Sigma$ is the Seiberg--Witten curve defined by \Cref{eq:SWclassSAN-1}.
\end{proposition}

Simple and full punctures represent extreme cases of regular punctures. When $N > 3$, however, intermediate cases also become possible. For each $i$, let $p_i < i$ denote the order of the pole of the differential $\phi_i$ at a given puncture, assuming no mass deformations are present. Near a puncture of $C_{g,n}$, the branches of the covering $\Sigma \rightarrow C_{g,n}$ group into blocks, each containing branches that behave identically. Consequently, each puncture is characterized by a sequence of integers $(q_1, \dots, q_s)$, where each $q_i$ represents the number of branches in a block. This sequence satisfies:
\begin{equation*} 
    \sum_s q_s = N\; , 
\end{equation*}
indicating that regular punctures correspond to \emph{partitions} of $N$: without loss of generality, one can assume $(q_1, \dots, q_s)$ is a non-increasing sequence.

\begin{definition}
    A \emph{partition}\index{partition} of $N$ is a non-increasing sequence of integers $q = (q_1, \dots, q_s)$ such that:
    \begin{equation*}
        \sum_{i=1}^s q_i = N\; .
    \end{equation*}
    Another common notation for partitions is the following. If $q_1 = \dots = q_{i_1} > q_{i_1+1} = \dots = q_{i_1 + i_2} > q_{i_1 + i_2 + 1} \dots$, then, letting $\chi_1 = q_1$, $\chi_2 = q_{i_1 + 1}$, etc., one writes $q = \chi_1^{i_1} \chi_2^{i_2} \dots$. 
    
    Every partition of $N$ can be represented by a \emph{Young diagram}\index{Young diagram} consisting of $N$ boxes arranged in left-justified rows of lengths $q_1, q_2, \dots, q_s$ from bottom to top. There are $i_1$ rows of length $\chi_1 = q_1$, $i_2$ rows of length $\chi_2 = q_{i_1 + 1}$, and so on.

    If the Young diagram contains $j_1$ rows of length $\psi_1$, $j_2$ rows of length $\psi_2$, etc., then the partition \emph{conjugate} to $q$ is the partition $q^T = \psi_1^{j_1} \psi_2^{j_2} \dots$, which is also a partition of $N$. Three examples of partitions of $N=15$ are shown in \Cref{fig:partition}, together with their conjugate.
\end{definition}

\begin{figure}[!ht]
    \centering
    \includegraphics[scale=0.83]{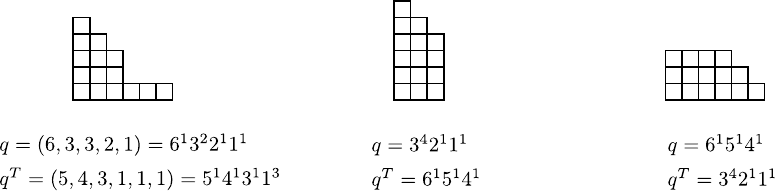}
    \caption{Three partitions of $15$ and their conjugate.}
    \labelx{fig:partition}
\end{figure}

In the presence of mass deformations, punctures are characterized by the form of the set of residues for $\lambda$ at the preimages of the puncture under the covering map $\Sigma \rightarrow C_{g,n}$. As seen in the examples of simple and full punctures above, not all residues are necessarily distinct. Residues can instead be grouped into blocks of $h_t$ residues with the sum of all residues vanishing, producing another partition $h$ of $N$.

The link between $q$, $h$, and the pole orders $p_i$ without mass deformation, is detailed in \cite[Sect. 4]{Gaiotto:2009we}:
\begin{proposition}
    Without mass deformation, let $q$ denote the partition characterizing a puncture. Then, in the presence of masses, one has $h = q^T$. Furthermore, if $h$ is expressed as $h = \psi_1^{j_1} \psi_2^{j_2} \dots$, then the flavor symmetry associated with this puncture is
    \begin{equation*} 
        G_q = \mathrm{S}\left(\prod_\alpha \mathrm{U}(j_\alpha)\right)\; . 
    \end{equation*}
    In fact, the set of residues corresponding to the partition $h$ can be naturally regarded as belonging to the complexified Cartan subalgebra of $G_q$. Lastly, to proceed with the construction, assign the numbers from $1$ to $N$ to the boxes of the Young diagram corresponding to $q$, beginning with $1$ in the bottom-left box and filling from left to right along rows and then moving up column by column. Let $h(i)$ represent the height of the box containing the number $i$. Then, for each $i$, $p_i = i - h_i$. Note that one always has $p_1 = 0$---note that there is no order-1 holomorphic differential in \Cref{eq:SWclassSAN-1}---and that $p_i \leq i-1$. This shows that the punctures under consideration are regular punctures, according to \Cref{prop:regpuncture}.
\end{proposition}

\begin{example}
    For $N=3$, the possible partitions are:
    \begin{itemize}
        \item $\mathbf{q=1^3}$, $h=3^1$, corresponding to three equal residues with vanishing sum, i.e. only zero residues, and $p_1 = p_2 = p_3 = 0$. This case corresponds to the absence of a puncture rather than to a puncture.
        \item $\mathbf{q=2^1 1^1}$, $h=2^11^1$, corresponding to sets of residues $\{2m,-m,-m\}$, and $p_2 = p_3 = 1$. This is the case of simple punctures, with flavor symmetry $\mathrm{S}(\mathrm{U}(1)\times\mathrm{U}(1))=\mathrm{U}(1)$.
        \item $\mathbf{q=3^1}$, $h=1^3$, corresponding to sets of residues $\{m_1+m_2,-m_1,-m_2\}$, $p_2 = 1$, and $p_3 = 2$. This is the case of full punctures, with corresponding flavor symmetry $\mathrm{SU}(3)$.
    \end{itemize}
\end{example}

\begin{example}
    For $N=4$, the possible partitions are:
    \begin{itemize}
        \item $\mathbf{q=1^4}$, corresponding no singularity.
        \item $\mathbf{q=2^1 1^2}$, $h=3^1 1^1$, corresponding to residues $\{3m,-m,-m,-m\}$, and $p_2 = p_3 = p_4 = 1$. This is the case of simple punctures, with flavor symmetry $\mathrm{S}(\mathrm{U}(1)\times\mathrm{U}(1)\simeq\mathrm{U}(1)$.
        \item $\mathbf{q = 2^2}$, $h = 2^2$, corresponding to residues $\{m,m,-m,-m\}$, $p_2 = p_3 = 1$, and $p_3 = 2$. The flavor symmetry is $\mathrm{SU}(2)$.
        \item $\mathbf{q=3^1 1^1}$, $h=2^1 1^2$, corresponding to residues $\{2 m_1+m_2,-m_1,-m_1,-m_2\}$, $p_2 = 1$, and $p_3 = p_4 = 2$. The flavor symmetry is $\mathrm{S}(\mathrm{U}(1)\times\mathrm{U}(2)\simeq\mathrm{U}(1)\times\mathrm{SU}(2)$.
        \item $\mathbf{q=4^1}$, $h=1^4$, corresponding to residues $\{m_1+m_2+m_3,-m_1,-m_2,-m_3\}$, $p_2 = 1$, $p_3 = 2$, $p_4=3$. This is the case of full punctures, with flavor symmetry $\mathrm{SU}(4)$.
    \end{itemize}
\end{example}

Therefore, the defect data $\mathcal{D}$ in a theory $\mathcal{S}[A_{N-1}, C_{g,n}, \mathcal{D}]$ with regular punctures only consists of assigning a Young diagram with $N$ boxes to each puncture on the UV curve $C_{g,n}$. Without mass deformations, this Young diagram encodes the pole orders of the holomorphic differentials $\phi_2, \dots, \phi_N$ at each puncture. The form of generic mass deformations is also prescribed by $\mathcal{D}$, as discussed above. Note that when $N=3$, regular punctures must be either simple or full, in alignment with \Cref{def:classSN=3}.

\subsection{Degeneration limits}

When the genus $g$ of the UV curve $C_{g,n}$ is $0$ or $1$, certain degeneration limits of $C_{g,n}$ with generic regular punctures correspond to weakly-coupled linear or necklace conformal quiver descriptions of the theory $\mathcal{S}[A_{N-1}, C_{g,n}, \mathcal{D}]$. The Seiberg–Witten curve for such quivers can be computed using the methods in \Cref{Sec:Mtheoryuplifts}.

Consider, for instance, a general linear conformal quiver with gauge group $\mathrm{SU}(N_1) \times \mathrm{SU}(N_2) \times \dots \times \mathrm{SU}(N_k)$, as depicted in \Cref{fig:N=2quiverD6}, where $\max(N_i) = N$. Let $N_0 = N_{k+1} = 0$, and let $F_i$ denote the number of fundamental hypermultiplets associated with each gauge factor $\mathrm{SU}(N_i)$, for $i = 1, \dots, k$. This quiver theory is conformal, provided that: 
\begin{equation*} 
	\text{for all } 1 \leq i \leq k, \quad 2N_i - N_{i-1} - N_{i+1} = F_i \geq 0\; . 
\end{equation*} 
This requirement implies that the sequence $(N_1, \dots, N_k)$ is concave and thus satisfies: 
\begin{equation*} 
    0 < N_1 < \dots < N_{k_1} = N = N_{k_1+1} = \dots = N_{k_2} > \dots > N_k > 0\; , 
\end{equation*} 
for some indices $1 \leq k_1 \leq k_2 \leq k$. For each $1 \leq i \leq k-1$, define $q^L_i := N_i - N_{i-1}$, and set: 
\begin{equation*} 
    q^L = (q^L_1, \dots, q^L_{k_1})\; . 
\end{equation*} 
The concavity of $(N_1, \dots, N_k)$ implies that $q^L$ is a partition of $N$. Similarly, there exists a partition $q^R$ corresponding to the right side of the quiver. The computation of the Seiberg--Witten curve for such a quiver, along the lines of \Cref{Sec:Mtheoryuplifts}, shows that the UV curve is a punctured Riemann sphere $C_{0,n}$, where $n-2$ punctures are simple, and the remaining two are characterized by the partitions $q^R$ and $q^L$ determined by the left and right ends of the quiver. Two examples for $N=4$ are shown in \Cref{fig:SUNquivergenend}.

\begin{figure}[!ht]
    \centering
    \includegraphics[scale=0.83]{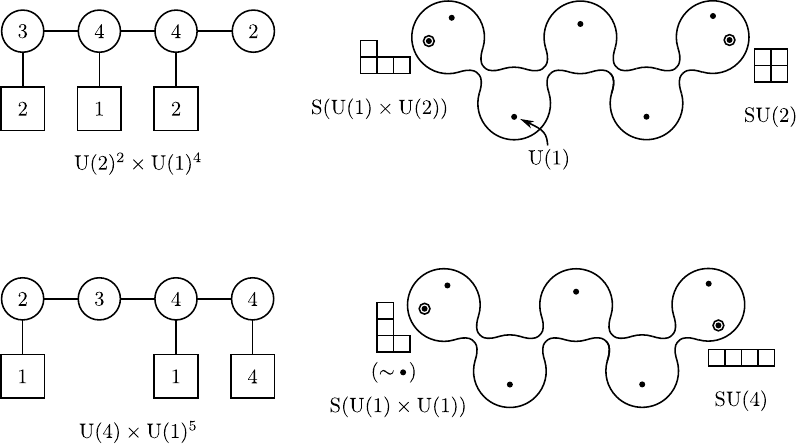}
    \caption{Weak coupling limits of class $\mathcal{S}$ theories of type $A_3$.}
    \labelx{fig:SUNquivergenend}
\end{figure}

In general, weakly-coupled descriptions of $\mathcal{S}[A_{N-1}, C_{g,n}, \mathcal{D}]$ can be formulated as \emph{generalized $\mathrm{SU}(N)$ quivers}\index{quiver!$\mathcal N=2$!generalized}. Among the building blocks of these generalized $\mathrm{SU}(N)$ quivers are intrinsically strongly coupled $4d$ $\mathcal{N}=2$ SCFTs, which extend the case of the $E_6$ theory found in generalized $\mathrm{SU}(3)$ quivers. The variety of regular punctures available in the $A_{N-1}$ case allows for numerous such generalizations.

For instance, the UV curve of $\mathrm{SU}(N)$ SYM with $2N$ fundamental hypermultiplets is the Riemann sphere with two simple punctures and two full punctures. For $N=3$, we have seen that the degeneration limit, in which the two simple punctures merge, corresponds to the Argyres–Seiberg dual phase of $\mathrm{SU}(3)$ SYM with six fundamental hypermultiplets. More generally, for any $N$, one can consider the degeneration limit in which the two simple punctures collide, resulting in a generalization of Argyres–Seiberg duality: as $\tau \rightarrow 1$, $\mathrm{SU}(N)$ SYM with $2N$ fundamental hypermultiplets is better described as an $\mathrm{SU}(2)$ gauge theory coupled to a single fundamental hypermultiplet and an interacting $4d$ $\mathcal{N}=2$ SCFT of rank $N-2$ with flavor symmetry $\mathrm{SU}(2) \times \mathrm{SU}(2N)$. This interacting theory is sometimes denoted $R_N$, and its UV curve is a sphere with two full punctures and one simple puncture (cf. \cite{Chacaltana:2010ks} and \cite[Sect. 12]{Tachikawa:2013kta}).

Another possible generalization of the $E_6$ SCFT is the class $\mathcal{S}$ theory of type $A_{N-1}$, whose UV curve is the Riemann sphere with three full punctures. This theory is an interacting $4d$ $\mathcal{N}=2$ SCFT with $\mathrm{SU}(N)^3$ flavor symmetry, sometimes referred to as $T_N$ (cf. \cite[Sect. 12]{Tachikawa:2013kta}).

Degeneration limits where generic regular punctures collide are examined in \cite{Gaiotto:2009we}. Generally, such collisions result in weakly coupled $\mathrm{SU}(M)$ gauge groups with $M \leq N$. However, a special case occurs when two punctures, each associated with Young diagrams having only two columns, collide. This does not produce a weakly-coupled special unitary gauge group but instead yields a weakly coupled $\mathrm{USp}$ gauge group. Consequently, quivers of special unitary groups that are terminated by $\mathrm{USp}$ gauge groups fall within the class of generalized $\mathrm{SU}(N)$ quivers. A specific instance of this leads to another example of Argyres–Seiberg duality: in the strong coupling limit, $\mathrm{USp}(4)$ with six flavors corresponds to a weakly coupled $\mathrm{SU}(2)$ gauge theory coupled to the Minahan–Nemeschansky $E_7$ SCFT \cite{Argyres:2007cn}.

\section{General class \texorpdfstring{$\mathcal S$}{S} theories}

The existence of six-dimensional superconformal field theories (SCFTs) with $\mathcal{N}=(2,0)$ supersymmetry was initially argued from superstring theory and M-theory considerations \cite{Witten:1995zh, Seiberg:1996vs, Seiberg:1997ax, Seiberg:1997zk}. These theories exhibit the spacetime symmetry algebra $\mathfrak{osp}(6,2\vert 4)$, which is the maximal possible superconformal algebra. Known $6d$ $\mathcal{N}=(2,0)$ SCFTs fall into two primary categories: abelian $6d$ $\mathcal{N}=(2,0)$ SCFTs and interacting $6d$ $\mathcal{N}=(2,0)$ SCFTs, each associated with simple compact Lie algebras of Dynkin types $A$, $D$, or $E$. The abelian theory, connected to the algebra $\mathfrak{u}(1)$, describes the worldvolume dynamics on a single M5-brane. In contrast, the interacting theory corresponding to $\mathfrak{su}(N)$ (of type $A_{N-1}$) describes the worldvolume theory of a stack of $N$ M5-branes. For a more extensive discussion of $6d$ $\mathcal{N}=(2,0)$ SCFTs, see \cite{Heckman:2018jxk} and references therein.

A stack of $N$ NS5-branes in type IIA superstring theory is $T$-dual to the $\mathbb{C}^2/\mathbb{Z}_N$ singularity in type IIB superstring theory. The preservation of $\mathcal{N}=(2,0)$ supersymmetry in six dimensions is related to the fact that $\mathbb{C}^2/\mathbb{Z}_N$ is an affine Calabi--Yau two-fold singularity, which preserves half of the 32 supercharges in type II superstring theory. More generally, singularities that preserve 16 of the 32 supercharges must have holonomy in $\mathrm{SU}(2)$, indicating that they are of the form $\mathbb{C}^2/\Gamma$, where $\Gamma$ is a discrete subgroup of $\mathrm{SU}(2)$. The classification of discrete $\mathrm{SU}(2)$ subgroups follows an $ADE$ pattern, which reflects the types of these singularities. Through $T$-duality, these singularities realize the $6d$ $\mathcal{N}=(2,0)$ SCFTs corresponding to each $ADE$ type.

The $A_{N-1}$ theory, tied to the worldvolume dynamics on stacks of M5-branes, has been central in our discussions because its compactification on a punctured Riemann surface $C_{g,n}$ with defect data $\mathcal{D}$ and a partial twist leads, in the four-dimensional limit, to class $\mathcal{S}$ theories of type $A_{N-1}$. In these theories, the Seiberg–Witten curves $\Sigma$ are holomorphic ramified coverings of $C_{g,n}$ within the cotangent bundle $T^*_{(1,0)}C_{g,n}$, where the restricted holomorphic Liouville 1-form defines a Seiberg–Witten differential on $\Sigma$. Class $\mathcal{S}$ theories of type $A_{N-1}$ encompass many known four-dimensional $\mathcal{N}=2$ theories, though there exist $4d$ $\mathcal{N}=2$ theories whose realization within class $\mathcal{S}$ remains elusive or even possibly unattainable.

An important feature of the class $\mathcal{S}$ construction, as previously mentioned, is its explicit link between Seiberg–Witten theory and Hitchin systems \cite{Donagi:1995cf}. Specifically, the Coulomb branch of a four-dimensional $\mathcal{N}=2$ theory serves as the base of a holomorphic integrable system, with Hitchin systems providing key examples. Although the entire integrable system structure—namely, the torus fibration over the Coulomb branch—might appear auxiliary for $4d$ theories, it gains physical relevance when the theory is further compactified on a circle. In this case, the integrable system directly corresponds to the Coulomb branch of the resulting three-dimensional theory (see, e.g., \cite{Seiberg:1996nz, Gaiotto:2010okc}).

In particular, for class $\mathcal{S}$ theories of type $A_{N-1}$, the integrable system encoding their low-energy Coulomb branch physics is a Hitchin system corresponding to $G_\mathbb{C} = \mathrm{SL}_n(\mathbb{C})$. This structure fully manifests only after a compactification on a circle. Starting from the six-dimensional $\mathcal{N}=(2,0)$ theory of type $A_{N-1}$, this corresponds to first compactifying on $C_{g,n}$ with defect data $D$, and then on $S^1$. Alternatively, if one first compactifies the $6d$ theory on $S^1$, it yields a five-dimensional $\mathcal{N}=2$ SYM theory with $A_{N-1}$ gauge group. Under a further compactification on $C_{g,n}$, the BPS equations for this $5d$ SYM theory reduce precisely to Hitchin’s equations. This explains why Hitchin systems appear as the integrable systems corresponding to class $\mathcal{S}$ theories of type $A_{N-1}$ \cite{Gaiotto:2009hg}. More generally, this demonstrates that theories of class~$\mathcal S$ are exactly the $4d$ $\mathcal N=2$ theories whose Seiberg--Witten integrable systems are Hitchin systems. 

The same construction applies starting from a general interacting $6d$ $\mathcal{N}=(2,0)$ theory of type $D$ or $E$, resulting in class $\mathcal{S}$ theories of type $D$ or $E$, also denoted $\mathcal{S}[\mathfrak{g}, C_{g,n}, \mathcal{D}]$ (with $\mathfrak{g}$ of type $D$ or $E$), which correspond to Hitchin systems of type $D$ or $E$. These theories allow for weakly-coupled descriptions as gauge theories with gauge factors of type $D$ or $E$. Punctures in these types of class $\mathcal S$ theories can also be regular, irregular and twisted \cite{Chacaltana:2011ze,Chacaltana:2012zy,Chacaltana:2013oka,Chacaltana:2014jba,Chacaltana:2015bna,Chacaltana:2016shw,Chacaltana:2017boe,Chacaltana:2018vhp}.

Hitchin systems associated with non-simply laced groups (i.e., the series $B$, $C$, $F$, and $G$) arise from compactifying $6d$ $\mathcal{N}=(2,0)$ SCFTs related to an algebra $\mathfrak{g}$ of type $A$, $D$, or $E$ on a circle with a twist by an outer automorphism of $\mathfrak{g}$, yielding $\mathcal{N}=2$ $5d$ SYM theories of types $B$, $C$, $F$, and $G$ \cite{HKS1}, \cite{HKS2}, \cite{Keller:2011ek}.

For an overview of recent advancements in theories of class $\mathcal{S}$, readers are referred to the review \cite{Akhond:2021xio} and the references cited therein.
\chapter{BPS states in class \texorpdfstring{$\mathcal S$}{S} theories}\labelx{Chap:BPSstatesclassS}




\abstract{We review the geometric approach to BPS states in four-dimensional $\mathcal N=2$ theories, with a focus on class $\mathcal S$ theories. Owing to the advanced nature of this material, our treatment is intended as a guided survey of the literature rather than a comprehensive introduction. We begin with the description of BPS states as finite string webs of variable tension on the ultraviolet curve $\mathrm{C}_\mathrm{UV}$ associated with class $\mathcal S$ theories, together with the notion of WKB triangulations in the case of type $A_1$ theories, following \cite{Gaiotto:2010okc,Gaiotto:2009hg}. We then turn to the framework of spectral networks, introduced in \cite{Gaiotto:2012rg}, which builds on the theory of $2d$–$4d$ framed BPS states developed in \cite{Gaiotto:2010be,Gaiotto:2011tf}.}

\section{Strings with variable tension}

\subsection{Strings from M-theory}

In what follows, we will focus specifically on class $\mathcal S$ theories of type $A_{N-1}$, which are constructed by compactifying the interacting $6d$ $\mathcal N=(2,0)$ SCFTs of type $A_{N-1}$ on a punctured Riemann surface. We will frequently rely on the M-theory perspective, where the $6d$ theory emerges as the worldvolume theory on a stack of $N$ M5-branes in flat space $\mathbb{R}^{1,10}$. We follow the presentation of \cite{Gaiotto:2009hg}. 

Similar to the case of D-branes, these $6d$ $\mathcal N=(2,0)$ SCFTs exhibit a kind of \emph{Coulomb branch}\index{branch of moduli space!Coulomb} on which the dynamics generically reduces to that of an abelian $6d$ $\mathcal N=(2,0)$ SCFT of rank $N-1$. In M-theory, this Coulomb branch corresponds to separating the M5-branes in the five directions transverse to their worldvolume. This separation is achieved by assigning vacuum expectation values (VEVs) to specific chiral operators in the $6d$ theory of type $A_{N-1}$.

On its Coulomb branch, the $6d$ $\mathcal N=(2,0)$ SCFT of type $A_{N-1}$ supports BPS strings labeled by pairs of integers $(ij)$, where $1 \leq i \neq j \leq N$. In the M-theory picture, these BPS strings manifest as M2-branes stretched between the $i$-th and $j$-th M5-branes (\Cref{fig:stringsCoulomb6d}). The tension of an $(ij)$ BPS string is proportional to the separation between the $i$-th and $j$-th M5-branes.

\begin{figure}[!ht]
    \centering
    \includegraphics[scale=0.83]{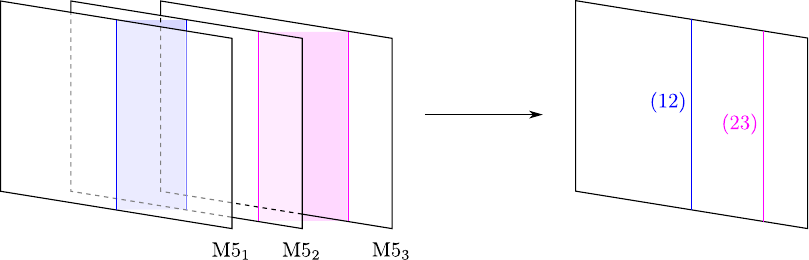}
    \caption{Strings in the $6d$ $\mathcal N=(2,0)$ SCFT of type $A_2$. The picture on the left is an M-theory picture showing M2-branes stretching between M5-branes. The picture on the right shows the corresponding strings in the $6d$ $\mathcal N=(2,0)$ theory.}
    \labelx{fig:stringsCoulomb6d}
\end{figure}

Strings of types $(ij)$, $(jk)$, and $(ik)$ can meet at junctions, provided that a specific condition is met to preserve supersymmetry: the corresponding junction of M2-branes must satisfy ``mechanical equilibrium''.

Let us now turn to theories of class $\mathcal S$ of type $A_{N-1}$. In the previous section, we observed that the Coulomb branch of these theories is described in M-theory as the space of normalizable deformations of a stack of $N$ M5-branes wrapping a UV curve $C_{g,n}$. More precisely, the setup under consideration is as follows: we consider M-theory on $\mathbb{R}^{1,6} \times Q$, where $Q$ is a hyperkähler manifold, with $N$ M5-branes wrapping a complex curve $C_{g,n} \subset Q$. The Coulomb branch of the theory $\mathcal{S}[A_{N-1}, C_{g,n}, \mathcal D]$ corresponds to normalizable connected deformations $\Sigma$ of the M5-brane stack in $Q$ that form degree-$N$ ramified coverings of $C_{g,n}$. The defect data $\mathcal D$ specifies the boundary condition for $\Sigma$ at the punctures of $C_{g,n}$.

The neighborhood of $C_{g,n}$ in $Q$ can be identified with the holomorphic cotangent bundle $T^*_{1,0}(C_{g,n})$ of $C_{g,n}$. Excitations close to $C_{g,n}$ correspond to the field theory limit of the M-theory configuration. Therefore, the Coulomb branch of the theory of class $\mathcal S$ can be identified with a specific set of holomorphic ramified coverings $\Sigma$ of $C_{g,n}$ in $T^*_{1,0}(C_{g,n})$, of degree $N$ and with prescribed behavior near the punctures. The ramified covering $\Sigma$ of $C_{g,n}$ is identified with the Seiberg--Witten curve corresponding to the point on the Coulomb branch under consideration, and since it lives in $T^*_{1,0}(C_{g,n})$ it can be defined by an equation of the form:
\begin{equation}\labelx{eq:SWcurvegeneral}
    \lambda^N + \sum_{i=2}^N \phi_i\lambda^{N-i} = 0\; ,
\end{equation}
where $\lambda$ is the holomorphic Liouville 1-form on $T^*_{1,0}(C_{g,n})$, and for each $i=2,\dots,N$, $\phi_i$ is a holomorphic $i$-differential on $C_{g,n}$ with poles at the punctures of $C_{g,n}$ prescribed by $\mathcal D$.

Let us assume that one is considering a generic point on the Coulomb branch of the theory, where the covering $\Sigma \rightarrow C_{g,n}$ has only simple branch points. Upon choosing branch cuts on $C_{g,n}$ terminating at the branch points, one can label the sheets of the covering by indices $i = 1, \dots, N$, while the branch points and branch cuts are indexed by pairs $\{ij\}$.

As in the flat space case, M2-branes can extend between the sheets of the covering $\Sigma \rightarrow C_{g,n}$. Such configurations are BPS if and only if the M2-branes wrap \emph{special Lagrangian cycles}\index{special!Lagrangian cycle}\index{Lagrangian!special cycle} within the hyperkähler space $Q$. In the field theory limit, this condition requires, in particular, that the M2-branes be ``vertical'', meaning they project to strings on $C_{g,n}$ under the covering map $\Sigma \rightarrow C_{g,n}$. However, not all such configurations are necessarily BPS. 

\subsection{Local BPS condition}\labelx{Subsec:localBPS}

Let $(z,x)$ be local holomorphic coordinates on $T^*_{1,0}(C_{g,n})$, where $z$ is a coordinate on an open patch $U$ of $C_{g,n}$ and $x$ is a coordinate on the fibers of $T^*_{1,0}(C_{g,n})_{\vert U} \rightarrow U$. Assume that $U$ contains no branch point nor branch cut. In terms of $z$ and $x$, the 1-form $\lambda$ reads $\lambda = x\mathrm{d}z$, and the $N$ sheets of the covering $\Sigma_{\vert U} \rightarrow U$ can be described by $N$ functions $x_1(z), \dots, x_N(z)$ on $U$. For each $i = 1, \dots, N$, let
\begin{equation*} 
    \lambda_i = x_i \mathrm{d}z 
\end{equation*}
be the restriction of the Liouville form $\lambda$ to the $i$-th sheet of the covering, and for any $j \neq i$, let $\lambda_{ij} = \lambda_i - \lambda_j$. Note that $\lambda_{ij}$ is a one-form on $U$ which encodes the separation between the $i$-th and $j$-th sheets of $\Sigma_{\vert U} \rightarrow U$.

\begin{figure}[!ht]
    \centering
    \includegraphics[scale=0.83]{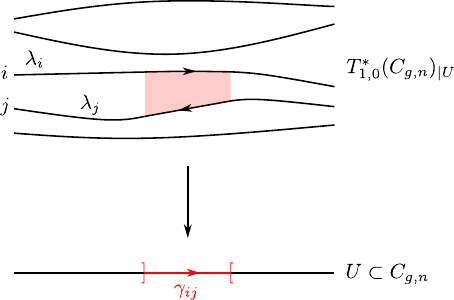}
    \caption{Strings on $C_{g,n}$ from M2-branes.}
    \labelx{fig:BPSconditionstrings}
\end{figure}

Let $\gamma_{ij}$ be a small segment of an oriented $(ij)$-string in $U$, defined by a smooth map
\begin{equation*} 
    \gamma_{ij} \colon t \in ]-1,1[ \; \longmapsto \; \gamma_{ij}(t) \in U\; . 
\end{equation*}
This setup is illustrated in \Cref{fig:BPSconditionstrings}. Similar to the flat space case, the tension of $\gamma_{ij}$ is proportional to the separation between the $i$-th and $j$-th sheets. However, the string tension now varies along $\gamma_{ij}$ as $\vert x_i(\gamma_{ij}(t)) - x_j(\gamma_{ij}(t)) \vert$. Such strings are thus called \emph{strings with variable tension}\index{string!with variable tension}. Integrating the tension along $\gamma_{ij}$ yields the mass of this $(ij)$-string segment:
\begin{equation}\labelx{eq:massstring}
    M(\gamma_{ij}) = \frac{1}{\pi}\int_{\gamma_{ij}} \vert \lambda_{ij}\vert\; .
\end{equation}
The mass $M(\gamma_{ij})$ is independent of the orientation of $\gamma_{ij}$.

Recall that $\lambda_{\vert \Sigma}$ serves as Seiberg--Witten differential for the class $\mathcal S$ theory, and that integrating $\lambda_{\vert \Sigma}$ along an oriented cycle $\gamma$ on the Seiberg--Witten curve $\Sigma$ computes the $4d$ $\mathcal N=2$ central charge of the BPS state with low-energy charges given by the homology class of $\gamma$ in some symplectic basis. Let $\gamma$ be an oriented cycle on $\Sigma$ such that $\gamma_{ij}^{(i)} - \gamma_{ij}^{(j)} \subset \gamma$, where for each $k$, $\gamma_{ij}^{(k)}$ denotes the uplift of $\gamma_{ij}$ to the $k$-th sheet of the covering $\Sigma_{\vert U} \rightarrow U$, and where the minus sign indicating orientation reversal. The contribution of $\gamma_{ij}^{(i)} - \gamma_{ij}^{(j)}$ to the central charge reads:
\begin{equation}\labelx{eq:centralchargestring}
    Z(\gamma_{ij}) := \frac{1}{\pi} \left( \int_{\gamma_{ij}^{(i)}} \lambda_{\vert \Sigma} - \int_{\gamma_{ij}^{(j)}} \lambda_{\vert \Sigma} \right) = \frac{1}{\pi}\int_{\gamma_{ij}} \lambda_{ij}\; . 
\end{equation}
Slightly abusing notation, we shall refer to $Z(\gamma_{ij})$ as the \emph{central charge}\index{central charge} of the string segment $\gamma_{ij}$. Note that if $\gamma_{ij}^\mathrm{opp}(t) := \gamma_{ij}(-t)$ denotes the orientation reversed string and if $\gamma_{ji}(t)=\gamma_{ij}(t)$, then $Z(\gamma_{ij}^\mathrm{opp})=Z(\gamma_{ji})$.

BPS states correspond to 1-cycles in $\Sigma$ where M2-branes attach to the M5-brane wrapping $\Sigma$, which project to strings or, more generally, webs of strings on $C_{g,n}$. The analysis above shows that the mass and central charge of such string webs can be computed locally on $C_{g,n}$. A state in the class $\mathcal{S}$ theory, represented by such a string web, is BPS if and only if it satisfies the BPS condition $M = \vert Z \vert$. Equivalently, each segment $\gamma_{ij}$ in the web must satisfy the local BPS condition\footnote{In this chapter we adopt the conventions of \cite{Gaiotto:2009hg}, which differ from those of \Cref{Chap:SWtheory} by a factor $\sqrt{2}$.} $M_{\gamma_{ij}} = \vert Z_{\gamma_{ij}} \vert$, where $M_{\gamma_{ij}}$ and $Z_{\gamma_{ij}}$ are defined in \Cref{eq:massstring,eq:centralchargestring}. More precisely:
\begin{definition}
    A small segment of an $(ij)$-string $\gamma_{ij}$ on $C_{g,n}$ satisfies the local BPS condition if and only if there exists some $\vartheta \in [0, 2\pi[$ such that, for any $t \in ]0,1[$, denoting by $\partial_t$ the vector tangent to $\gamma_{ij}$ at $\gamma_{ij}(t)\in U$, the following equation holds:
    \begin{equation}\labelx{eq:localBPScondition}
        \lambda_{ij}\cdot\partial_t \in \E^{\I\vartheta}\mathbb{R}_+\; .
    \end{equation}
    When this condition is satisfied, we have $Z(\gamma_{ij}) = \E^{\I \vartheta} M(\gamma_{ij})$.
\end{definition}

\begin{remark}
    For a string web to represent a BPS state, the angle $\vartheta$ in \Cref{eq:localBPScondition} must be identical across all string segments that make up the web.
\end{remark}

\begin{remark}
    Note that condition (\ref{eq:localBPScondition}) is (a mild generalization of) the one used in the analytic definition of spectral networks in \Cref{sec:analytic_con}. In practice, most of the spectral networks considered in this chapter will be of \emph{WKB type}, as this is the most natural framework from the perspective of class~$\mathcal S$ theories. The construction in \cite{Gaiotto:2012rg} begins with such WKB spectral networks, and subsequently abstracts the notion to a topological one, introducing the processes of abelianization and non-abelianization (cf. \Cref{sec:non-abel}).
\end{remark}

Letting $\gamma_{ij}(t)=:z(t)$ for clarity, \Cref{eq:localBPScondition} can be reformulated as follows: 
\begin{equation} 
    \left[x_i(z(t)) - x_j(z(t))\right] \frac{\mathrm{d}z}{\mathrm{d}t}(t) \in \E^{\I \vartheta}\mathbb{R}_{>0}\; . 
\end{equation} 

\subsection{Finite string webs}\labelx{Subsec:Finitestringwebs} 

In order for a string web satisfying the local BPS condition with a given angle $\vartheta$ to correspond to a genuine BPS state, it must also be finite. This requirement stems from \Cref{eq:massstring}, which shows that non-finite webs would have infinite mass. We thus arrive at the following characterization of BPS states in theories of class $\mathcal{S}$:

\begin{proposition}
    BPS states in theories of class $\mathcal{S}$ correspond to string webs on the UV curve $C_\mathrm{UV}$ that satisfy the local BPS condition of \Cref{eq:localBPScondition} for some fixed angle $\vartheta$. The mass and central charge of the state are obtained by summing the contributions of each string segment constituting the web (cf. \Cref{eq:massstring,eq:centralchargestring}). In particular, the argument of the central charge is $\vartheta$.
\end{proposition}

Examples of finite BPS strings together with their M-theory origin are shown in \Cref{fig:BPSstatesM2}. 
\begin{figure}[!ht]
        \centering
        \includegraphics[width=\textwidth]{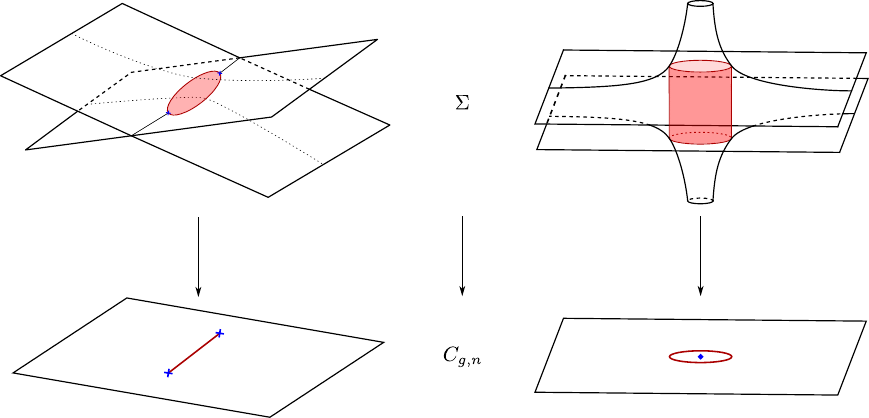}
        \caption{BPS states as finite M2-branes and as finite strings.}
        \labelx{fig:BPSstatesM2}
    \end{figure}
    
Theoretically, this provides a framework to compute and analyze the BPS states of theories of class $\mathcal{S}$, even though there is no straightforward algorithm for implementing this method in practice. Explaining how this can be done will be the focus of the next sections. 

\begin{remark}
    This approach to BPS states in $4d$ $\mathcal{N}=2$ theories was first pioneered in the context of gauge theories engineered from string theory \cite{Klemm:1996bj, Brandhuber:1996xk, Shapere:1999xr}.
\end{remark}

\section{BPS foliation in type \texorpdfstring{$A_1$}{A1}}

\subsection{Local description}\labelx{Subsec:localdesc}

In this section, we focus on theories of class $\mathcal{S}$ of type $A_1$. In these theories, the Seiberg–Witten curve $\Sigma$ is a ramified covering of degree 2 of the UV curve $C_{g,n}$, determined by the equation
\begin{equation}\labelx{eq:SWtypeA1}
    \lambda^2 - \phi_2 = 0\; ,
\end{equation}
where $\lambda$ is the holomorphic Liouville 1-form on $T^*C_{g,n}$ and $\phi_2$ is a quadratic differential on $C_{g,n}$. For any given angle $\vartheta \in [0, 2\pi[$, the local BPS condition in \Cref{eq:localBPScondition} defines a foliation of $C_{g,n}$, which we now proceed to describe. All the leaves of the foliation are $(12)$-strings (or $(21)$-strings), which will often be referred to simply as \emph{BPS strings} in this section. 

\begin{remark}
    These foliations of Riemann surfaces defined by quadratic differentials coincide with those introduced and studied in \Cref{Sec:Quaddifftrajectories} (see also \Cref{Sec:StockescurvesSchroding}), which we now revisit from the perspective of class~$\mathcal S$ theories.
\end{remark}

\subsubsection*{Regular point}

In a sufficiently small open neighborhood $U$ of a regular point in $C_{g,n}$, the functions $x_{1,2}(z)$ can be approximated by constants. Then, the BPS differential equation \ref{eq:localBPScondition} reduces to 
\begin{equation*} 
    \frac{\mathrm{d}z}{\mathrm{d}t}(t) = \alpha\; , 
\end{equation*} 
where $\alpha$ is a complex constant whose argument is fixed. The integral curves of this differential equation within $U$ are parallel lines with slope $\arg(\alpha)$.

\subsubsection*{Branch point} 

Consider a (simple) branch point $B = B_{12} \in C_{g,n}$. Let $z$ be a holomorphic coordinate near $B$ such that $B$ is at $z=0$, and such that $x_1(z) \sim \sqrt{z}$ and $x_2(z) \sim -\sqrt{z}$ close to $B$, for some choice of branch for the square root $\sqrt{z}$. Let $U = {\vert z \vert < \epsilon }$ for some $\epsilon$. We choose ${z \in \mathbb{R}_{<0}} \cap U$ as branch cut. With this notation, \Cref{eq:localBPScondition} becomes:
\begin{equation*} 
    2\sqrt{z}\frac{\mathrm{d}z}{\mathrm{d}t}(t) = \frac{4}{3} \frac{\mathrm{d}}{\mathrm{d}t}z^{3/2}(t) \in \E^{\I\vartheta}\mathbb{R}_{>0}\; . 
\end{equation*} 
Among the integral curves of this differential equation in $U$ are three BPS strings, each either beginning or ending at the branch point. These strings can be described by $z_k(t) \propto t^{2/3} \E^{\I \phi}$ with $\phi = (2\vartheta + 4k\pi)/3$, where $k = 0,1,2$. Due to the branch cut, $z_0$ represents a $(12)$-curve that emanates from $B$, whereas $z_1$ and $z_2$ correspond to $(12)$-curves terminating at $B$. Forgetting about orientation, one sees that at each branch point $B$, there are three strings emanating from $B$. The remaining integral curves in $U$ approach $z_1$, $z_2$, or $z_3$ away from $B$. This local configuration is illustrated in \Cref{fig:branchpointsfoliation}.

\begin{figure}[!ht]
    \centering
    \includegraphics[scale=0.83]{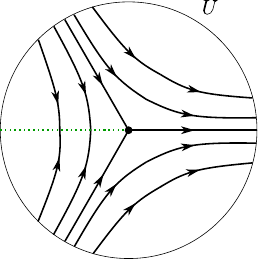}
    \caption{Integral curves of the local BPS equation \Cref{eq:localBPScondition} close to a branch point $B$. All oriented plain curves are $(12)$-strings, while the dotted line is the branch cut.}
    \labelx{fig:branchpointsfoliation}
\end{figure}

\subsubsection*{Regular puncture} 

We now turn to the description of the integral curves of \Cref{eq:localBPScondition} near a regular puncture of $C_{g,n}$. Let $z$ be a holomorphic coordinate in a neighborhood of the puncture, such that the puncture is located at $z=0$. Let $m\in\mathbb{C}$ be the mass parameter associated with the puncture, which in type $A_1$ is necessarily simple (or, equivalently, full). The quadratic differential $\phi_2$ has a double pole with coefficient $m^2$, and the solutions of \Cref{eq:SWtypeA1} behave as $x_\pm(z)\mathrm{d}z \sim \pm m\mathrm{dz}/z$. Thus, the BPS differential equation from \Cref{eq:localBPScondition} reads:
\begin{equation*} 
    \frac{2m}{z} \frac{\mathrm{d}z}{\mathrm{d}t} \in \E^{\I\vartheta}\mathbb{R}_{>0}\; . 
\end{equation*} 
The integral curves near the puncture can be parameterized as: 
\begin{equation*} 
    z(t) = C \exp(\alpha t)\; , 
\end{equation*} 
where $C$ and $\alpha$ are complex constants, with $\alpha = \E^{\I\vartheta}/m$. The value of $\vartheta$ that makes $\Re(\alpha) = 0$ is special, as in this case, the integral curves circle around the puncture. For other values of $\vartheta$, the integral curves either emerge from the puncture as $t\rightarrow -\infty$ or converge to it as $t\rightarrow \infty$, tracing out logarithmic spirals. Specifically, let $\alpha = \alpha_1 + \I\alpha_2$. Then: 
\begin{equation*} 
    z(t) = C \E^{\alpha_1 t} \left( \cos(\alpha_2 t) + \I \sin(\alpha_2 t) \right)\; . 
\end{equation*} 
Such integral curves are illustrated in \Cref{fig:regularpuncture}.

\begin{figure}[!ht]
    \centering
    \includegraphics[scale=0.83]{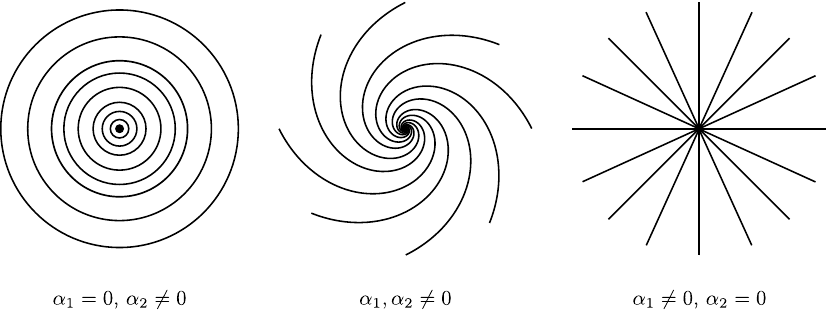}
    \caption{Integral curves of the local BPS equation in the neighborhood of a regular puncture.}
    \labelx{fig:regularpuncture}
\end{figure}

\subsubsection*{Irregular puncture}

In \Cref{Subsec:SU2F<4revisited}, we observed that irregular punctures are characterized by the poles of the holomorphic differentials defining the ramified covering $\Sigma\rightarrow C_{g,n}$ being of higher order. Such punctures typically arise in the context of $\mathrm{SU}(2)$ SYM theories with $F \leq 3$ flavors, where poles of order $3$ and $4$ appear in the quadratic differential $\phi_2$. Let $z$ be a holomorphic coordinate in a neighborhood of an irregular puncture situated at $z=0$, and let $k+2$ denote the order of the quadratic differential $\phi_2$, where $k$ is a positive integer:
\begin{equation*} 
    \phi_2(z) \sim \frac{c}{z^{k+2}}\; . 
\end{equation*} 
Near the puncture, the branches of $\Sigma$ behave as $x_\pm(z) \sim \pm \sqrt{c}z^{-k/2-1}$, hence \Cref{eq:localBPScondition} becomes: 
\begin{equation*} 
    \frac{2\sqrt{c}}{z^{k/2+1}} \frac{\mathrm{d}z}{\mathrm{d}t} \in \E^{\I\vartheta}\mathbb{R}_{>0}\; . 
\end{equation*} 
The integral curves for this equation can be parameterized by: 
\begin{equation*} 
    z(t) = \left(\frac{1}{\alpha t + C}\right)^{2/k}\; , 
\end{equation*} 
where $C$ is a complex constant and $\alpha = -k \E^{\I\vartheta} / 4\sqrt{c}$. As $t \rightarrow \infty$, these integral curves approach the puncture along $k$ distinct asymptotic directions. Examples for $k=1$ and $k=4$ are illustrated in \Cref{fig:irregpuncture}.

\begin{remark}\labelx{rem:irreg=cilia}
    The behavior of integral curves of the local BPS equation near irregular punctures connects with the notion of ciliated surfaces introduced in \Cref{Sec:ciliatedsurfaces}. For an irregular puncture where the quadratic differential $\phi_2$ has order $k=2$, we can excise a small disk around the puncture and mark the $k$ intersection points between the boundary of this disk and the asymptotic directions along which the integral curves converge toward the puncture, thereby introducing a boundary component with $k$ cilia. We will denote this ciliated surface by $C_\mathrm{UV}$.
\end{remark}

\begin{figure}[!ht]
    \centering
    \includegraphics[width=\textwidth]{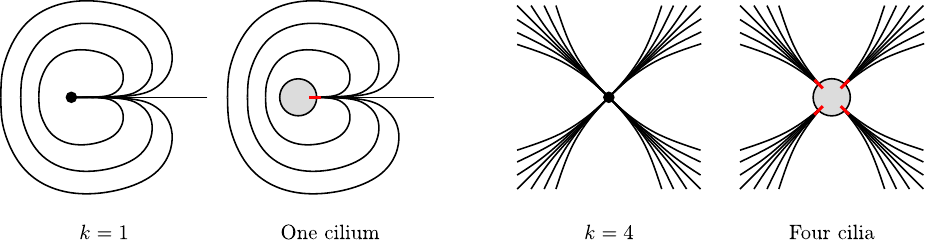}
    \caption{Integral curves of the local BPS equation in the neighborhood of an irregular puncture.}
    \labelx{fig:irregpuncture}
\end{figure}

This concludes our local description of the integral curves of the local BPS equation \ref{eq:localBPScondition}.

\subsection{WKB triangulation}\labelx{Sec:WKBtriangulation}

A given string within the foliation defined by \Cref{eq:localBPScondition} must be either: 
\begin{itemize}
    \item A \emph{generic} string, which begins at a puncture and ends at a puncture (potentially the same). 
    \item A \emph{separating} string, which either begins at a puncture and ends at a branch point, or vice versa. 
    \item A \emph{finite} string, which begins at a branch point and ends at a branch point, or forms a closed loop. Such finite strings and their M-theory origin are depicted in \Cref{fig:BPSstatesM2}.
    \item A \emph{divergent} string, which is not closed and has no limit in at least one direction. 
\end{itemize}

\begin{remark} 
    When $C_\mathrm{UV}$ has at least one puncture or cilium, in the absence of finite strings, divergent strings cannot exist \cite{Strebel}. Values of $\vartheta$ for which there exist no finite strings are dubbed generic. 
\end{remark}

Hence, for generic values of $\vartheta$, the BPS strings are either generic or separating, and the generic BPS strings form 1-parameter families bounded by a union of separating BPS strings. There are two possible topologies for these 1-parameter families, depending on whether the separating strings are associated with two branch points or a single one. The two topologies are illustrated in \Cref{fig:top1paramgenstrings}, where generic BPS strings are represented by solid black curves, separating BPS strings, by dashed red curves, punctures, by black dots and branch points, by red stars.

\begin{figure}[!ht]
    \centering
    \includegraphics[scale=0.83]{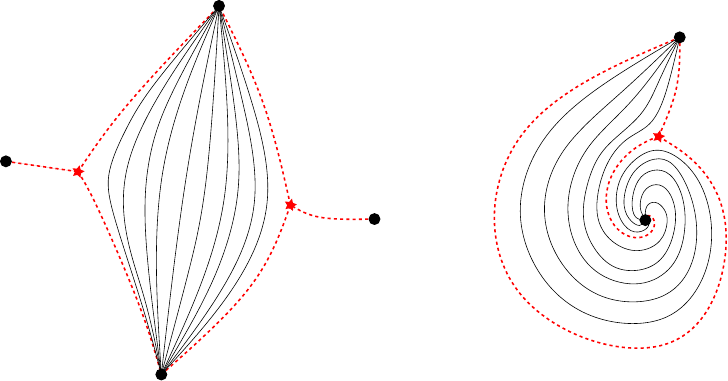}
    \caption{1-parameter families of generic BPS strings.}
    \labelx{fig:top1paramgenstrings}
\end{figure}

Selecting a single representative from each one-parameter family of generic BPS strings yields a triangulation of $C_\mathrm{UV}$ in the sense of \Cref{Def:triang}, called the \emph{WKB triangulation}\index{WKB!triangulation} of $C_\mathrm{UV}$ corresponding to the angle~$\vartheta$. We will now briefly explain the reasoning behind this terminology, following the original articles \cite{Gaiotto:2010okc,Gaiotto:2009hg} to which we refer for details.

\begin{remark}\labelx{Rem:decortriang}
    The WKB triangulation defined here coincides with the triangulation introduced in \Cref{sec:Crit_graph_quadratic} from the critical graph of a quadratic differential, up to the fact that the WKB triangulation encodes slightly more data than the bare triangulation as we will see shortly: it is a \emph{decorated triangulation}, a notion introduced in \cite{Gaiotto:2009hg} in reference to the \emph{decorated local systems}\footnote{Strictly speaking, what plays a role in \cite{Gaiotto:2009hg} is not the decoration itself, but rather the \emph{framing}.} of \cite{FG}, cf. \Cref{Sec:FockGoncharov}.
\end{remark}

\subsection{BPS state counting}

\subsubsection*{BPS index}

To clarify what is meant by \emph{BPS counting}, note that in general one cannot expect the number of BPS states at a given point $u \in \mathcal{C}$ in the Coulomb branch $\mathcal{C}$ of a $4d$ $\mathcal{N}=2$ theory to remain constant as $u$ varies, even if no wall of marginal stability is crossed.

As discussed in \Cref{Chap:SWtheory}, the charges of low-energy excitations at generic points in $\mathcal{C}$ form a local system of lattices $\Gamma=(\Gamma_u)_{u\in\mathcal C}$ over $\mathcal{C}$, with rank $2r+f$, where $r$ denotes the rank of the theory (i.e., the dimension of the Coulomb branch) and $f$ is the number of $\mathrm{U}(1)$ flavor symmetry factors. This local system is equipped with the antisymmetric \emph{Dirac--Schwinger--Zwanziger pairing}\index{DSZ!pairing} $\langle \cdot,\cdot \rangle$, which is trivial on the flavor charges. The local system $\Gamma$ exhibits monodromy around the complex codimension-1 locus in $\mathcal{C}$ where some BPS particles become massless. Additionally, there is a complex-valued linear function $Z(u)$ on $\Gamma$, encoding the central charge, i.e., $Z(u)\in\Gamma_u^*\otimes_{\mathbb{Z}}\mathbb{C}$, which varies holomorphically with $u$. The function $Z(u)$ is calculated by integrating the Seiberg--Witten differential $\lambda$ over 1-cycles on the Seiberg--Witten curve $\Sigma_u$ at $u$. Specifically, for each $\gamma\in\Gamma_u$, we have: 
\begin{equation*} 
    Z_\gamma(u):=Z(u)(\gamma) = \oint_{\widehat{\gamma}} \lambda_u\; , 
\end{equation*} 
where $\lambda_u$ is the Seiberg--Witten differential at $u\in\mathcal C$, and $\widehat{\gamma}$ is a 1-cycle on $\Sigma_u$ whose homology class is given by $\gamma\in\Gamma_u$. Furthermore, each element $\gamma \in \Gamma_u$ determines a subspace $\mathcal{H}_{\gamma,\mathrm{BPS}}(u)$ of the Hilbert space of 1-particle states at $u \in \mathcal{C}$, comprising all one-particle states in the theory with electric, magnetic, and flavor charges $\gamma$, which also satisfy the BPS bound $M \geq \vert Z_\gamma(u)\vert$.

BPS states cannot be counted in a straightforward way, meaning that one cannot hope for a well-defined integer for each connected component of the complement of the walls of marginal stability in $\mathcal{C}$, is that (half) BPS states can recombine in pairs as $u$ is varied. Conversely, non-BPS representations of the supersymmetry algebra may split into pairs of BPS states. To address this, a supersymmetric ``protected'' index is used instead, similar to the protected indices defined in \cite{Witten:1982df,Cecotti:1992qh}. 

\begin{proposition}
    For BPS states in four-dimensional theories with $\mathcal{N}=2$ supersymmetry, the relevant index is called \emph{second helicity supertrace}\index{second helicity supertrace}, and writes: 
    \begin{equation}\labelx{eq:secondhelsup} 
        \Omega(\gamma;u) = -\frac{1}{2}\Tr_{\mathcal{H}^{1,\mathrm{BPS}}_{\gamma}(u)} (-1)^{2J_3}(2J_3)^2\; , 
    \end{equation} 
    where $J_3$ is a generator of the rotation subgroup of the massive little group.
\end{proposition}

This index ``counts'' BPS particles while accounting for the above ambiguity, meaning it remains constant on the connected components of the complement of the walls of marginal stability in $\mathcal{C}$. As discussed in \Cref{Chap:SWtheory}, the BPS spectrum may change when crossing walls of marginal stability; correspondingly, $\Omega(\gamma;u)$ can jump across these walls. One can show that every BPS hypermultiplet contributes a factor of $1$ to $\Omega(\gamma;u)$, whereas every BPS vector multiplet contributes a factor of $-2$.

\subsubsection*{Kontsevich--Soibelman wall-crossing formula}

In \cite{Kontsevich-Soibelman}, Kontsevich and Soibelman derived a \emph{wall-crossing formula}\index{wall-crossing!formula} for the \emph{Donaldson--Thomas invariants}\index{Donaldson--Thomas invariants} of three-dimensional Calabi--Yau manifolds $\widehat{\Omega}(\gamma;u)$. Although this context appears unrelated to BPS counting in $4d$ $\mathcal{N}=2$ theories, their wall-crossing formula is applicable to the second helicity supertrace $\Omega(\gamma;u)$. This connection was demonstrated in \cite{Gaiotto:2010okc} by examining $4d$ $\mathcal{N}=2$ theories compactified on a circle, i.e., on $\mathbb{R}^3 \times S^1_R$, with $R$ denoting the radius of $S^1$. At large radius and low energies, the effective theory is a three-dimensional $\mathcal{N}=4$ sigma model into the integrable system $\mathcal M\rightarrow \mathcal C$ capturing the low-energy dynamics of the original $4d$ $\mathcal{N}=2$ theory \cite{Seiberg:1996nz}. Given that the base of the Seiberg--Witten integrable system corresponds to the Coulomb branch $\mathcal{C}$ of the $4d$ $\mathcal{N}=2$ theory, the fibers, which are generically $2r$-dimensional tori, are parameterized by the expectation values of ``electric'' ($\theta_e^I$) and ``magnetic'' ($\theta_m^I$) Wilson loops: 
\begin{equation*} 
    \theta_e^I = \oint_{S^1} A_4^I \mathrm{d}x^4\; , \quad \theta_{m,I} = \oint_{S^1} (A_{D,4})_I \mathrm{d}x^4\; , 
\end{equation*} 
where $A_4^I$ is the component of the $I$-th low-energy $4d$ abelian gauge field along $x^4$, and $A_{D,4}$ is the dual photon in three-dimensions. $\mathcal{N}=4$ supersymmetries in three dimensions requires that the sigma model's target manifold be \emph{hyperkähler}\index{hyperkähler!manifold}.

The wall-crossing formula of Kontsevich and Soibelman is (roughly) derived as follows. The lattice $\Gamma_u$ defines the complex split torus $\widetilde{T}_u = \Gamma_u^* \otimes_\mathbb{Z} \mathbb{C}^*$, and each $\gamma \in \Gamma_u$ naturally defines a function $X_\gamma$ on $\widetilde{T}$. These functions satisfy the relation $X_{\gamma_1} X_{\gamma_2} = X_{\gamma_1+\gamma_2}$, and any basis of $\Gamma_u$ defines a coordinate system on $\widetilde{T}$. Furthermore, the DSZ pairing $\langle \cdot, \cdot \rangle$ induces a holomorphic symplectic form on~$\widetilde{T}_u$ as in \Cref{Sec:clustervarieties}. Let $\sigma$ be a quadratic refinement of $(-1)^{\langle \gamma_1, \gamma_2 \rangle}$, i.e.
\begin{equation*} 
    \sigma(\gamma_1) \sigma(\gamma_2) = (-1)^{\langle \gamma_1, \gamma_2 \rangle} \sigma(\gamma_1 + \gamma_2)\; , 
\end{equation*} 
and let $e_\gamma$ be the symplectomorphism of $\widetilde{T}_u$ generated by the ``Hamiltonian'' $\sigma(\gamma)X_\gamma$. The collection of all $e_\gamma$ for $\gamma \in \Gamma_u$ generates a Lie algebra with the relations: 
\begin{equation*} 
    [e_{\gamma_1}, e_{\gamma_2}] = (-1)^{\langle \gamma_1, \gamma_2 \rangle} \langle \gamma_1, \gamma_2 \rangle e_{\gamma_1 + \gamma_2}\; . 
\end{equation*}
Let now
\begin{equation*}
    K_\gamma = \exp \sum_{n=1}^\infty \frac{1}{n^2}e_{n\gamma}\; .
\end{equation*}
This transformation acts on the functions $X_{\gamma'}$ as:
\begin{equation}\labelx{eq:KSactionontheX}
    K_\gamma \colon X_{\gamma'} \longrightarrow X_{\gamma'}(1-\sigma(\gamma)X_\gamma)^{\langle \gamma',\gamma\rangle}\; .
\end{equation}

\begin{remark}
    These are cluster transformations; see \Cref{Sec:clustervarieties} and \cite{Kontsevich-Soibelman}.
\end{remark}

Let $u_0 \in \mathcal{C}$ be a generic point on a wall of marginal stability in the four-dimensional Coulomb branch $\mathcal{C}$, and let $u_+$ and $u_-$ be points located close to $u_0$ on opposite sides of the wall. For each $u\in\mathcal C$ and each BPS state with charge $\gamma$, let $l_{\gamma,u} \subset \mathbb{C}$ denote the half-line: 
\begin{equation}\labelx{eq:halflinetwistor} 
    l_{\gamma,u} = \{ \zeta \cdot Z_\gamma(u) \mid \zeta \in \mathbb{R}_- \}\; . 
\end{equation} 
Let $Z_{\gamma_1}$ and $Z_{\gamma_2}$ be the central charges that align on the wall of marginal stability on which $u_0$ lies. Under the assumption of the \emph{support property}, the charges of particles (as opposed to anti-particles) that become massless on the wall belong to the set $\mathcal{ML} = \{ n\gamma_1 + m\gamma_2 \mid n, m > 0 \}$. Note that the ordering of the rays $l_{\gamma,u}$ in $\mathbb{C}$, for $\gamma \in \mathcal{ML}$, changes as one crosses the wall. Lastly, for $u$ away from the wall, let 
\begin{equation*} 
    \prod_{\gamma \in \mathcal{ML}}^\curvearrowright K_\gamma^{\Omega(\gamma; u_+)} 
\end{equation*} 
denote the product of the $K_\gamma^{\Omega(\gamma; u_+)}$'s ordered according to the clockwise orientation of the $l_{\gamma,u}$ in $\mathbb{C}$.

\begin{proposition}[Kontsevich--Soibelman wall-crossing formula] 
\begin{equation}\labelx{eq:KSWCF} 
    \prod_{\gamma \in \mathcal{ML}}^\curvearrowright K_\gamma^{\Omega(\gamma; u_+)} = \prod_{\gamma \in \mathcal{ML}}^\curvearrowright K_\gamma^{\Omega(\gamma; u_-)}\; . 
\end{equation} 
\end{proposition}

In the original wall-crossing formula, Donaldson--Thomas invariants $\widehat{\Omega}(\gamma;u)$ replace the second helicity supertraces $\Omega(\gamma;u)$ in \Cref{eq:KSWCF}. However, Gaiotto, Moore, and Neitzke proved that this formula also applies to BPS states in $4d$ $\mathcal{N}=2$ theories, and that it can actually be derived from the smoothness properties of the hyperkähler metric on the target space $\mathcal{M}$ of the effective $3d$ $\mathcal{N}=4$ sigma model, obtained by compactifying the four-dimensional theory on a circle $S^1_R$. Naive dimensional reduction yields a hyperkähler metric $g^{\mathrm{sf}}_{\mathcal{M}}$ on $\mathcal{M}$ with singularities along the singular loci in $\mathcal{C}$. The actual hyperkähler metric $g_{\mathcal{M}}(R)$ on $\mathcal{M}$ receives contributions from $4d$ BPS particles only, which smooth out these singularities. The relation between the two metrics is that $g_{\mathcal{M}}(R)\sim g^{\mathrm{sf}}_{\mathcal{M}}$ as $R\rightarrow\infty$. As one crosses a wall of marginal stability in $\mathcal{C}$, the BPS spectrum changes; nevertheless, the metric $g_{\mathcal{M}}$ is expected to vary smoothly. This imposes stringent constraints on wall-crossing, specifically that it must follow the Kontsevich--Soibelman wall-crossing formula (\Cref{eq:KSWCF}).

\begin{example}\labelx{ex:WCFpureSU2}
    A special case of the Konstevich--Soibelman wall-crossing formula is:
\begin{equation*}
    K_{2,-1}K_{0,1} = (K_{0,1}K_{2,1}K_{4,1}\dots) K_{2,0}^{-2} (\dots K_{6,-1}K_{4,-1}K_{2,-1})\; .
\end{equation*}
Comparing with the discussion in \Cref{Sec:BPSwallcrossSU2}, one can see that this equation encodes the wall-crossing observed in pure $\mathrm{SU}(2)$ SYM.
\end{example}

\begin{example}\labelx{ex:simpleWCF}
    Some instances of the wall-crossing formula are simpler than general ones, in that they involve only finitely many terms on the left and right hand sides of \Cref{eq:KSWCF}, e.g.:
    \begin{equation*}
        K_{1,0}K_{0,1} = K_{0,1}K_{1,1}K_{1,0}\; .
    \end{equation*}
    This formula encodes the wall-crossing in a simple Argyres--Douglas theory (see e.g. \cite{Shapere:1999xr}).
\end{example}

How $4d$ BPS states provide corrections to the ``naive'' hyperkähler metric $g_{\mathcal M}^\mathrm{sf}$ is understood from the perspective of the \emph{twistorial} construction of hyperkähler metrics \cite{Hitchin:1986ea,Hitchin:1992bo}. As a hyperkähler manifold, the target space $\mathcal M$ of the effective $3d$ $\mathcal N=4$ sigma model admits a $\mathbb{CP}^1$-worth of complex structures. The space $\mathbb{CP}^1$ encoding the complex structures on $\mathcal M$ is called the \emph{twistor sphere}\index{twistor!sphere} associated with $\mathcal M$, whereas the \emph{twistor space}\index{twistor!space} associated with the hyperkähler manifold $\mathcal M$ is the product $\mathcal M\times\mathbb{CP}^1$, which is naturally endowed with a structure of complex manifold. Let $\zeta$ be the standard coordinate on the twistor sphere $\mathbb{CP}^1_\zeta$.

Topologically, $\mathcal{M}$ is a torus fibration over $\mathcal{C}$, i.e. it can locally be parameterized by $(u, \theta) \in \mathcal{M}$. The hyperkähler metric on $\mathcal{M}$ is constructed from a map 
\begin{equation*}
    \mathcal{X}^\mathrm{RH}\colon \mathcal M \times\mathbb{C}^\times_\zeta \longrightarrow \Gamma^*\times\mathbb{C}^\times\; .
\end{equation*}
The contraction of $\mathcal{X}^\mathrm{RH}$ with some element $\gamma\in\Gamma$ of the charge lattice is denoted $\mathcal{X}^\mathrm{RH}_\gamma(u, \theta, \zeta)$. One has $\mathcal{X}^\mathrm{RH}_\gamma \mathcal{X}^\mathrm{RH}_{\gamma'} = \mathcal{X}^\mathrm{RH}_{\gamma+\gamma'}$. The DSZ pairing on $\Gamma$ defines a Poisson structure:
\begin{equation*}
    \left\{\mathcal{X}^\mathrm{RH}_\gamma,\mathcal{X}^\mathrm{RH}_{\gamma'}\right\} = \langle \gamma,\gamma'\rangle \mathcal{X}^\mathrm{RH}_{\gamma+\gamma'}\; .
\end{equation*}
Moreover, the $\mathcal{X}^\mathrm{RH}_\gamma$ are required to be holomorphic on the twistor sphere $\mathbb{CP}^1_\zeta$ away from the half-lines $l_{\gamma,u}$ defined in \Cref{eq:halflinetwistor}, for each $\gamma\in\Gamma_u$ such that $\Omega(\gamma;u)\neq 0$. As $\zeta$ crosses a line $l_{\gamma,u}$ in $\mathbb{CP}^1_\zeta$, the functions $\mathcal{X}^\mathrm{RH}_\gamma$ are discontinuous, with discontinuity given by the Kontsevich--Soibelman transformations as in \Cref{eq:KSactionontheX}. One can then construct a holomorphic symplectic form and a hyperkähler metric on $\mathcal M$ such that the $\mathcal{X}_\gamma(u, \theta, \zeta)$ serve as holomorphic Darboux coordinates.

The functions $\mathcal{X}_\gamma^{\mathrm{sf}}(u, \theta, \zeta)$ corresponding to the semi-flat metric $g^{\mathrm{sf}}$ can be expressed explicitly as
\begin{equation*}
    \mathcal{X}_\gamma^{\mathrm{sf}}(u, \theta, \zeta) = \exp(\pi R\zeta^{-1}Z_\gamma+\I\theta_\gamma+\pi R\zeta\overline{Z}_\gamma)\; ,
\end{equation*}
where $\theta_\gamma = \Gamma^*\otimes\mathbb{R}/2\pi\mathbb{Z} \rightarrow \mathbb{R}/2\pi\mathbb{Z}$ are dual to the coordinates on $\Gamma$.
The functions $\mathcal{X}^\mathrm{RH}_\gamma$ corresponding to the genuine hyperkähler metric on $\mathcal M$ are then obtained by factoring in corrections arising from $4d$ BPS particles. As a result, the $\mathcal{X}^\mathrm{RH}_\gamma$ are equivalent to the $\mathcal{X}_\gamma^{\mathrm{sf}}$ both as $\zeta\rightarrow 0,\infty$ and as $R\rightarrow \infty$. The superscript ``RH'' in $\mathcal{X}^\mathrm{RH}_\gamma$ indicates that these functions are solutions to an infinite-dimensional Riemann--Hilbert problem.

\subsection{Theories of class \texorpdfstring{$\mathcal S$}{S} of type \texorpdfstring{$A_1$}{A1}}\labelx{Subsec:BPSclassSA1}

In the special case of theories of class $\mathcal{S}$ of type $A_1$, the moduli space $\mathcal{M}$ corresponds to the Hitchin moduli space of $\mathrm{SU}(2)$ Higgs bundles on the UV curve $C_\mathrm{UV}$. Given a solution $(A, \phi)$ to Hitchin's equations, one can construct an $\mathrm{SL}_2(\mathbb{C})$-connection $\mathcal{A}$ on $C_\mathrm{UV}$ for any $\zeta \in \mathbb{C}^\times$ as follows:
\begin{equation*} \mathcal{A} = \frac{R}{\zeta} \phi + A + R \zeta \overline{\phi}\; . 
\end{equation*}
The condition that $(A, \phi)$ solves Hitchin's equations ensures that $\mathcal{A}$ is a flat $\mathrm{SL}_2(\mathbb{C})$-connection. Conversely, if $\mathcal{A}$ is flat for all $\zeta \in \mathbb{C}^\times$, then the pair $(A, \phi)$ satisfies Hitchin's equations. This correspondence motivates studying the moduli space of flat $\mathrm{SL}_2(\mathbb{C})$-connections on $C_{g,n}$ with prescribed monodromy at the punctures.

A remarkable set of coordinates can be introduced on this moduli space: \emph{Fock--Goncharov coordinates}\index{Fock--Goncharov coordinates} (cf. \Cref{Sec:FockGoncharov} and the discussion at the end of \Cref{Sec:connectedComponents}). The coordinates employed in \cite{Gaiotto:2009hg} are closely related to those defined in \cite{FG}, with a few minor adjustments. We have explained in \Cref{Sec:WKBtriangulation} how the local BPS condition for strings with variable tension on $C_\mathrm{UV}$ defines a triangulation on $C_\mathrm{UV}$. In fact, as alluded in \Cref{Rem:decortriang}, the WKB triangulation encodes slightly more data than the bare triangulation: it is a \emph{decorated triangulation}\index{triangulation!decorated}, meaning a triangulation of $C_\mathrm{UV}$ together with a solution $s$ to the flatness equation $(\mathrm{d} + \mathcal{A})s = 0$ near the punctures or cilia on $C_\mathrm{UV}$, determined up to scale. The WKB triangulation on $C_\mathrm{UV}$ is naturally equipped with a canonical decoration at each puncture or cilium. This decoration arises from the WKB analysis (cf. \Cref{sec:WKB_Stokes}) of the differential equation $(\mathrm{d} + \mathcal{A})s = 0$ as $\zeta\rightarrow 0$, thereby justifying the term ``WKB triangulation''.

Each WKB triangulation of $C_\mathrm{UV}$ defines Fock--Goncharov coordinates 
\begin{equation*}
    \mathcal X_\gamma^\vartheta \colon \mathcal M \times\mathbb{C}^\times \longrightarrow \mathbb{C}\; ,
\end{equation*}
where $\vartheta$ is the angle that determines the WKB triangulation. The functions $\mathcal{X}_\gamma^\mathrm{RH}$ are then identified with the coordinates $\mathcal{X}_\gamma^\vartheta$ by specializing $\vartheta$ to $\arg(\zeta)$.

As $\vartheta$ varies, the WKB triangulation of $C_\mathrm{UV}$ generically remains constant but undergoes discrete jumps when $\vartheta$ crosses specific values corresponding to angles $\vartheta = \arg(Z_\gamma)$ of BPS states. The transformations that relate distinct triangulations can take the following forms:
\begin{itemize}
    \item \emph{flips}\index{flip}, discussed in \Cref{partIII,partIV} of the monograph, which arise because of BPS hypermultiplets, 
    \item \emph{juggles}\index{juggle}, related to the existence of \emph{limit triangulations}\index{triangulation!limit}, and physically, to BPS vector multiplets,
    \item \emph{pops}\index{pop}, corresponding to a change of flat section at a puncture of cilium of $C_\mathrm{UV}$, and do not correspond to BPS states.
\end{itemize}

\begin{remark}
    Flips and pops are precisely those operations discussed in \Cref{Sec:connformulaVorossymb} in connection with Voros symbols.
\end{remark}

Fock--Goncharov coordinates undergo cluster transformations when the triangulation is flipped. These cluster transformations are specific instances of Kontsevich--Soibelman transformations, as described in \Cref{eq:KSactionontheX}. The transformations of the functions $\mathcal X_\gamma^\mathrm{RH}$ under juggles and pops can also be computed explicitly and correspond to the action of Kontsevich--Soibelman symplectomorphisms on the $\mathcal X_\gamma^\mathrm{RH}$. 

Varying $\vartheta$ from a generic value $\vartheta$ to $\vartheta+\pi$, one finds that the WKB triangulation of $C_\mathrm{UV}$ at $\vartheta+\pi$ is topologically identical to the WKB triangulation at $\vartheta$. The two decorated triangulations are related by a pop at every puncture or cilium. This transformation is called an \emph{omnipop}\index{omnipop}. The corresponding transformation is given by:
\begin{equation*}
    \mathbf{S} = \prod^{\curvearrowleft}_{\left\{\gamma \mid \vartheta < \arg -Z_\gamma(u) < \vartheta+\pi\right\}} \mathcal K_\gamma^{\Omega(\gamma;u)}\; ,
\end{equation*} 
which relates the functions $\mathcal X^\vartheta$ and $\mathcal X^{\vartheta+\pi}$. This is known as the \emph{spectrum generator}\index{spectrum generator}, as knowing $\mathbf{S}$ fully determines the BPS multiplicities $\Omega(\gamma;u)$. The computation of the spectrum generator $\mathbf{S}$ is carried out in \cite[Sect. 11]{Gaiotto:2009hg}, along with numerous examples.

One of the simplest cases is that of the rank-1 Argyres--Douglas theories, where the UV curve is a sphere with a single irregular puncture. At this puncture, the quadratic differential $\phi_2$ has a pole of order $k+2$, with $k\geq 3$. Recalling \Cref{rem:irreg=cilia}, one obtains WKB triangulations of polygons with $k+2$ vertices. The cases $k=3,4$ are particularly simple. For $k=3$ there are no BPS states, while for $k=4$, there is a single BPS state, with no wall-crossing. The case $k=5$ corresponds to triangulations of a pentagon, and the pentagon identity discussed in \Cref{Sec:clustervarieties} corresponds to the simple wall-crossing formula in \Cref{ex:simpleWCF}:
\begin{equation*}
    K_{1,0}K_{0,1} = K_{0,1}K_{1,1}K_{1,0}\; .
\end{equation*}

The cases of $\mathrm{SU}(2)$ super-Yang--Mills theories with $F\leq 4$ flavors are discussed extensively in \cite[Sect. 10]{Gaiotto:2009hg}. For $F=0$, the UV curve is recast as an annulus with one cilium on each boundary component, and one finds the wall-crossing formula of \Cref{ex:WCFpureSU2}. For $F=1$, the UV curve is an annulus with one cilium on one boundary component and two cilia on the other. For $F=2$, there are two standard realizations: one where the UV curve is an annulus with two cilia on each boundary component, and the other in which the UV curve is a disk with one cilium on the boundary component and two punctures in the interior. The cases $F=3$ and $F=4$ are treated similarly; for instance, the ``balanced'' realization of the $F=4$ case corresponds to a sphere with four punctures as UV curve.

\section{Spectral networks}\labelx{Sec:SpectralNetworksPhysics}

Spectral networks are a far-reaching generalization of the methods described in the previous two sections, allowing the study of BPS spectra in general theories of class $\mathcal{S}$. Spectral networks are very similar to the networks of strings with variable tension introduced in the case of theories of type $A_1$, with the main difference being the interpretation: spectral networks are defined with respect to certain surface defects in the theory of class $\mathcal{S}$ corresponding to $C_\mathrm{UV}$. Correspondingly, each edge of the network carries an index $\mu$ called the \emph{soliton index}\index{soliton!index}.

In the presence of such surface defects, one can define the notion of \emph{framed BPS states}\index{BPS!framed state}, which are BPS states associated with certain supersymmetric interfaces. The spectrum of framed BPS states turns out to be overdetermined, which in turn allows for the computation of the soliton indices for each edge of the network. Moreover, the extension of the Kontsevich--Soibelman wall-crossing formula, which accounts for the coupling between $2d$ and $4d$ BPS states derived in \cite{Gaiotto:2011tf}, enables the computation of the $4d$ BPS degeneracies directly from the topology of finite BPS webs on $C_\mathrm{UV}$.

We follow the seminal article \cite{Gaiotto:2012rg}; in particular, we restrict to theories of class $\mathcal{S}$ of type $A_{N-1}$ for $N \geq 2$. Spectral networks have been examined in the context of general theories of class $\mathcal{S}$ in \cite{LoPa}.

\subsection{BPS states, solitons, and framed BPS states}

\subsubsection*{BPS states}

Recall the geometric description of $4d$ BPS states provided in \Cref{Subsec:Finitestringwebs}: given an angle $\vartheta$, a BPS state with charge $\gamma$ and central charge $Z_\gamma$ such that $\arg (Z_\gamma) = \vartheta$ corresponds to a finite web of strings on the UV curve $C_\mathrm{UV}$. Each edge\footnote{Edges of the web are defined as the connected components of the complement of branch points and junctions.} of the web satisfies the local BPS condition from \Cref{eq:localBPScondition} with angle $\vartheta$. 

There exist junctions where $ij$-, $jk$-, and $ik$-strings meet, where $i \neq j \neq k \in \{ 1, \dots, N\}$ (other types of junctions can occur for theories of class $\mathcal S$ of type $D$ or $E$). A general junction is shown in \Cref{fig:junction}. 

\begin{figure}[!ht]
    \centering
    \includegraphics[scale=0.83]{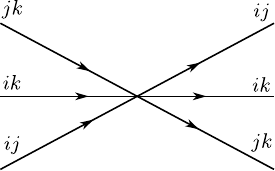}
    \caption{A general string junction (type $A$).}
    \labelx{fig:junction}
\end{figure}

Both branch points and branch cuts are labeled by an unordered pair of integers $(ij)$. For a web to be finite, all its edges must either close, or start and end at branch points or junctions. Three examples of finite BPS webs are shown in \Cref{fig:finiteBPSwebs}. 

Any such finite web corresponds to a cycle $C$ on the Seiberg--Witten curve $\Sigma$, and the charge of the BPS state is the homology of $C$: the charge lattice is $\Gamma = \mathrm{H}_1(\Sigma,\mathbb{Z})$. The degeneracies of $4d$ BPS states are encoded in the second helicity supertrace $\Omega(\gamma; u)$, as defined in \Cref{eq:secondhelsup}, where $\gamma\in \Gamma$, and where $u$ is a point on the Coulomb branch $\mathcal C$.

\begin{figure}[!ht]
    \centering
    \includegraphics[scale=0.83]{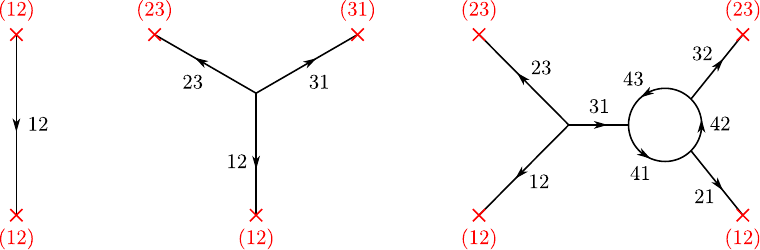}
    \caption{Three finite BPS webs (branch points are shown as red crosses).}
    \labelx{fig:finiteBPSwebs}
\end{figure}

\subsubsection*{Surface defects and solitons} 

\begin{proposition}
    Theories in class $\mathcal{S}$ corresponding to a Lie algebra $\mathfrak{g}$ of type $A$, $D$, or $E$ admit canonical half-BPS surface defects $\mathbb{S}_z$, depending on a point $z\in C_{g,n}$ and a representation of $\mathfrak{g}$ \cite{Gaiotto:2012}.
\end{proposition} 

We focus on cases where the representation is the fundamental representation of $\mathfrak{su}(N)$. In the M-theoretic realization of type $A_{N-1}$ class $\mathcal{S}$ theories, these surface defects correspond to open M2-branes whose boundaries extend along $\mathbb{R}^{1,3}$ and are localized at a point $z$ in $C_\mathrm{UV}$, thereby defining a surface defect in the low-energy four-dimensional theory within class $\mathcal{S}$.

At a point $u \in \mathcal{C}$ (specifying the Seiberg--Witten covering $\Sigma \rightarrow C_\mathrm{UV}$), the preimages of $z$ in $\Sigma$ correspond to distinct vacua for the surface defect $\mathbb{S}_z$: if $z$ is a generic point of the fibration $\Sigma \rightarrow C_\mathrm{UV}$, the surface defect $\mathbb{S}_z$ admits $N$ distinct vacua. In the presence of a surface defect $\mathbb{S}_z$, the theory supports BPS particles that are bound to the surface defect and interpolate between different vacua of $\mathbb{S}_z$. 

\begin{definition}
    These BPS particles are referred to as \emph{solitons}\index{soliton}, or \emph{$2d$-$4d$ BPS states}\index{BPS!$2d$-$4d$ state}.
\end{definition} 
If $z$ is generic, a soliton interpolating between the $i$-th vacuum $z^{(i)}$ and the $j$-th vacuum $z^{(j)}$ (where $i \neq j$) of $\mathbb{S}_z$ is called an $ij$-soliton. Solitons can also be described as finite string webs on $C_\mathrm{UV}$ terminating at $z \in C_\mathrm{UV}$. Three examples of $12$-solitons for the surface defect $\mathbb{S}_z$ are shown in \Cref{fig:solitonBPSwebs}.

\begin{figure}[!ht]
    \centering
    \includegraphics[scale=0.83]{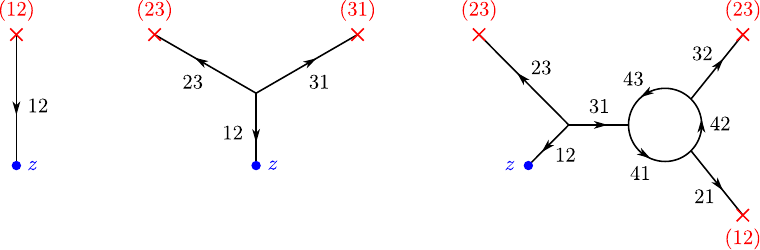}
    \caption{Three 12-solitons for the surface defect $\mathbb{S}_z$.}
    \labelx{fig:solitonBPSwebs}
\end{figure}

The uplift to $\Sigma$ of a string web on $C_\mathrm{UV}$ corresponding to an $ij$-soliton for $\mathbb{S}_z$ is a 1-chain with boundary $z^{(j)} - z^{(i)}$. Denoting by $\Gamma_{ij}(z,z)$ the set of relative homology classes on $\Sigma$ whose boundary is $z^{(j)} - z^{(i)}$, the charges of solitons for $\mathbb{S}_z$ belong to the set
\begin{equation*} 
    \Gamma(z,z) = \bigcup_{i \neq j} \Gamma_{ij}(z,z)\; . 
\end{equation*} 
The central charge $Z_{\overline{a}}(z)$ of a soliton with charge $\overline{a} \in \Gamma(z,z)$ can be computed by integrating the Liouville 1-form $\lambda$ along any 1-chain with boundary representing $\overline{a}$, similarly to the case of $4d$ BPS states. Furthermore, away from the walls of marginal stability, there exists a protected index that counts solitons of charge $\overline{a}$ (subject to a ``modulo 2 caveat'' to be addressed shortly): 
\begin{equation}\labelx{eq:solmultsignissue}
\mu(\overline{a}) = \Tr_{\mathcal{H}^{1,\mathrm{BPS}}_{\mathbb{S}_z,\overline{a}}(u)} F \E^{\I\pi F}\; , 
\end{equation} 
where $F$ is a fermion number operator.

\subsubsection*{Framed $2d$-$4d$ BPS states}

In the presence of two surface defects $\mathbb{S}_{z_1}$ and $\mathbb{S}_{z_2}$, where $z_1, z_2 \in C_\mathrm{UV}$, additional BPS states can be defined as follows. One may consider supersymmetric interfaces between the surface defects $\mathbb{S}_{z_1}$ and $\mathbb{S}_{z_2}$. If $\mathbb{R}^{1,3}$ is coordinatized by $x^{0,1,2,3}$ as usual, let $\mathbb{S}_{z_1}$ extend along $x^0$ and $x^1 < 0$, while being localized at $x^{2,3} = 0$, and let $\mathbb{S}_{z_2}$ extend along $x^0$ and $x^1 > 0$, also localized at $x^{2,3} = 0$. The interface between these defects extends along $x^0$ and is localized at $x^{1,2,3} = 0$.

Distinct interfaces between $\mathbb{S}_{z_1}$ and $\mathbb{S}_{z_2}$ are classified by the homotopy class of a path $\wp$ on $C_\mathrm{UV}$ connecting $z_1$ and $z_2$, along with a real parameter $\vartheta$. These interfaces are denoted $L_{\wp,\vartheta}$ and are half-BPS, meaning they preserve two out of the four supersymmetries preserved by $\mathbb{S}_{z_1}$ and $\mathbb{S}_{z_2}$ \cite{Gaiotto:2011tf}.

\begin{definition} 
A \emph{framed $2d$-$4d$ BPS state}\index{BPS!framed $2d$-$4d$ state} corresponding to a supersymmetric interface $L_{\wp,\vartheta}$ is a state in the Hilbert space $\mathcal{H}_{L_{\wp,\vartheta}}^1$ that preserves the two supercharges of the combined system. 
\end{definition}

\begin{remark} 
The parameter $\vartheta$ is said to be generic if there is no $\gamma \in \Gamma$ such that $\E^{-\I\vartheta} Z_\gamma \in \mathbb{R}_-$. Similarly, $(\wp, \vartheta)$ is generic if there exists no $\overline{a} \in \Gamma(z_1, z_2) \cup \Gamma(z_2, z_2)$ such that $\E^{-\I\vartheta} Z_{\overline{a}} \in \mathbb{R}_-$. When $(\wp, \vartheta)$ is generic, framed $2d$-$4d$ BPS states corresponding to $(\wp, \vartheta)$ are localized near $L_{\wp,\vartheta}$. 
\end{remark}

The charges of framed $2d$-$4d$ BPS states are classified by the relative homology classes of 1-chains with boundary $z_2^{(j)} - z_1^{(i)}$, where $z_2^{(j)}$ (respectively, $z_1^{(i)}$) is the $j$-th (respectively, $i$-th) lift of $z_2$ (respectively, $z_1$). Let $\Gamma_{ij}(z_1, z_2)$ denote the set of such homology classes, and define: 
\begin{equation*} 
\Gamma(z_1, z_2) = \bigcup_{i,j} \Gamma_{ij}(z_1, z_2)\; . 
\end{equation*}

There exists another protected index, referred to as the \emph{framed BPS index}\index{BPS!framed index}, which counts framed $2d$-$4d$ BPS states \cite{Gaiotto:2011tf}:
\begin{equation}\labelx{eq:framedBPSindex}
    \overline{\underline{\Omega}}' (L_{\wp,\vartheta}, \overline{a}) = \Tr_{\mathcal{H}^{1,\mathrm{BPS}}_{L_{\wp,\vartheta}},\overline{a}} \E^{\I\pi F}\; , 
\end{equation}
where $\overline{a} \in \Gamma(z_1, z_2)$.

\subsubsection*{$\mathbb{Z}_2$-twist}

The operator $F$ in \Cref{eq:framedBPSindex} is defined only up to shifts by complex numbers. Requiring that $\overline{\underline{\Omega}}'(L_{\wp,\vartheta}, \overline{a})$ is real constrains $F$, but still permits integer shifts. Consequently, $\overline{\underline{\Omega}}'(L_{\wp,\vartheta}, \overline{a})$ remains defined only up to a sign. Similarly, the index in \Cref{eq:solmultsignissue} depends on the choice of fermion number operator $F$, which also leaves it determined only up to sign.

As proposed in \cite{Gaiotto:2012rg}, there is a systematic way to resolve these ambiguities globally, analogous to the twist used in defining the moduli space $\mathcal{A}_{G^L, S}$ of \emph{twisted local systems}\index{local system!twisted} in \cite{FG} (see also \Cref{rem:Amodulispace}). For any smooth surface $S$, let $\widetilde{S}$ denote its unit tangent bundle, and let $H_S$ denote the class of the fiber in the circle bundle $\widetilde{S} \to S$. Note that any smooth oriented path on $S$ canonically lifts to $\widetilde{S}$, while a smooth unoriented path on $S$ admits two canonical lifts to $\widetilde{S}$.

Now consider a 1-chain on $\Sigma$ with boundary $z^{(j)} - z^{(i)}$. Recall that $\Gamma_{ij}(z,z)$ denotes the set of homology classes of such 1-chains, corresponding to the charges of $ij$-solitons for the defect $\mathbb{S}_z$. Let $z_+^{(i)},z_+^{(j)} \in \widetilde{\Sigma}$ represent the lifts of $z^{(i)}$ and $z^{(j)}$ to $\widetilde{\Sigma}$ with orientation consistent with that of the 1-chain. By considering the charges of $ij$-solitons in the set $\widetilde{\Gamma}(z_+^{(i)}, z_+^{(j)})$, consisting of relative homology classes in $\widetilde{\Sigma}$ with boundary $z_+^{(j)} - z_+^{(i)}$, modulo $2H_\Sigma$, rather than in $\Gamma_{ij}(z,z)$, the soliton degeneracy $\mu$ can be made completely well-defined. This is achieved by imposing the condition that for two lifts $a$ and $a'$ of $\overline{a} \in \Gamma_{ij}(z,z)$ within $\widetilde{\Gamma}(z_+^{(i)}, z_+^{(j)})$, the relation
\begin{equation*}
    \mu(a)/\mu(a') = (-1)^{w(a,a')}
\end{equation*}
holds, $w(a, a')$ the difference in winding between $a$ and $a'$ in units of $H_\Sigma$, modulo $2H_\Sigma$. Let:
\begin{equation*}
    \widetilde{\Gamma}(\widetilde{z},\widetilde{z}) = \bigcup_{i\neq j} \widetilde{\Gamma}(z_+^{(i)}, z_+^{(j)})\; .
\end{equation*}

\begin{definition} 
The integer $\mu(a)$, for $a \in \widetilde{\Gamma}(z_+^{(i)}, z_+^{(j)})$, is called the \emph{soliton degeneracy}. 
\end{definition}

Similarly, the framed BPS index can also be made well-defined by considering lifts $a$ of $\overline{a}$ to $\widetilde{\Sigma}$ modulo $2H_\Sigma$. Precisely, instead of a class $\overline{a}\in\Gamma(z_1,z_2)$, one considers lifts $a$ of $\overline{a}$ to $\widetilde{\Sigma}$ which project to $a$, i.e. relative homology classes with boundary $z_{2,+}^{(j)}-z_{1,+}^{(i)}$ for some $i$ and $j$, where the $+$ denotes the lifts of $z_2^{(j)}$ and $z_1^{(i)}$ to $\widetilde{\Sigma}$ prescribed by $\wp$, modulo $2H_\Sigma$. These form the set $\widetilde{\Gamma}(\widetilde{z_1},\widetilde{z_2})$. The relative homology of two such lifts $a$ and $a'$ differ by a winding number $w(a,a')$ which is either $0$ or $1$. One sets:
\begin{equation*}
    \overline{\underline{\Omega}}' (L_{\wp,\vartheta}, a)/\overline{\underline{\Omega}}' (L_{\wp,\vartheta}, a') = (-1)^{w(a,a')}\; .
\end{equation*}
Likewise, given two paths $\wp$ and $\wp'$ whose lifts to $\widetilde{C_\mathrm{UV}}$ have the same starting and final endpoints in $\widetilde{C_\mathrm{UV}}$, one can define a winding number $w(\wp,\wp')$, as illustrated in \Cref{fig:liftspathwinding}, and declare:
\begin{equation*}
    \overline{\underline{\Omega}}' (L_{\wp,\vartheta}, a)/\overline{\underline{\Omega}}' (L_{\wp',\vartheta}, a) = (-1)^{w(\wp,\wp')}\; .
\end{equation*}
This means that the framed BPS indices $\overline{\underline{\Omega}}'$ of the interface $L_{\wp,\vartheta}$ are \emph{twisted homotopy invariants}\index{twisted homotopy invariant} of the path $\wp$, rather than genuine homotopy invariants.

\begin{figure}[!ht]
    \centering
    \includegraphics[scale=0.83]{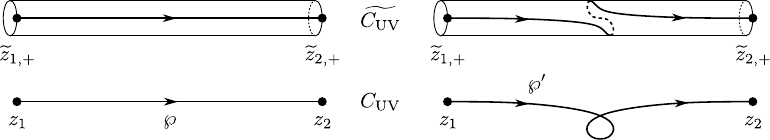}
    \caption{Two paths $\wp$ and $\wp'$ such that $w(\wp,\wp')=1$.}
    \labelx{fig:liftspathwinding}
\end{figure}

\subsubsection*{Generating functions of framed BPS states}

Given a path $\wp$ between two points $z_1$ and $z_2$ in $C_\mathrm{UV}$, the framed BPS indices $\overline{\underline{\Omega}}'(L_{\wp,\vartheta},a)$ can be encoded in a formal generating function $F(\wp,\vartheta)$ as follows. For each $a \in \widetilde{\Gamma}(\widetilde{z_1},\widetilde{z_2})$, let $X_a$ denote a formal variable, subject to the condition that if $a$ and $a'$ project to the same class $\overline{a} \in \Gamma(z_1,z_2)$, then: 
\begin{equation*} X_a/X_{a'} = (-1)^{w(a,a')}\; . 
\end{equation*}
This allows to define: 
\begin{equation} 
F(\wp,\vartheta) = \sum_{\overline{a}\in\Gamma(z_1,z_2)} \overline{\underline{\Omega}}'(L_{\wp,\vartheta},a) X_a\; , 
\end{equation} 
where $a$ is any lift of $\overline{a}$ to $\widetilde{\Gamma}(\widetilde{z_1},\widetilde{z_2})$, since the product $\overline{\underline{\Omega}}' (L_{\wp,\vartheta},a) X_a$ is independent of the choice of representative $a$ for each $\overline{a}$. The properties of $\widetilde{\Gamma}(\widetilde{z_1},\widetilde{z_2})$ imply that $F(\wp,\vartheta)$ are (twisted) homotopy invariants of the path $\wp$.

As discussed in \Cref{Sec:pathalgebra}, one can concatenate paths on $\widetilde{\Sigma}$ whenever the end of the first path coincides with the start of the second path. This gives rise to a (homological) \emph{path algebra}\index{path!algebra} on $\widetilde{\Sigma}$, with the following multiplication rule: 
\begin{equation*} 
X_a X_b = \begin{cases} X_{a+b} & \text{if}\; \mathrm{end}(a) = \mathrm{start}(b)\; , \\ 
0 & \text{otherwise}\; . 
\end{cases} \end{equation*}

\begin{proposition}\labelx{Prop:concatenation}
If $\wp$ and $\wp'$ are two paths on $C_\mathrm{UV}$ with $\mathrm{end}(\wp) = \mathrm{start}(\wp')$ in $\widetilde{C_\mathrm{UV}}$, i.e., if $\wp$ and $\wp'$ can be concatenated into a smooth path $\wp\wp'$, and if both $(\wp,\vartheta)$ and $(\wp',\vartheta)$ are generic, then: 
\begin{equation*} 
    F(\wp,\vartheta)F(\wp',\vartheta) = F(\wp\wp',\vartheta)\; . 
\end{equation*} 
\end{proposition}

\begin{proof} 
This result follows from the genericity assumption on $(\wp,\vartheta)$ and $(\wp',\vartheta)$. Under this assumption, framed $2d$-$4d$ BPS states are localized near the interfaces $L_{\wp,\vartheta}$ and $L_{\wp',\vartheta}$, respectively. Consequently, the product $F(\wp,\vartheta)F(\wp',\vartheta)$ does not depend on the distance between $L_{\wp,\vartheta}$ and $L_{\wp',\vartheta}$. In the limiting case where this distance vanishes, the framed $2d$-$4d$ BPS states are those of the interface $L_{\wp\wp',\vartheta}$. 
\end{proof}

We are interested in computing the generating function $F(\wp,\vartheta)$. This computation crucially depends on the way $\wp$ intersects the \emph{spectral network} $\mathcal{W}_\vartheta$ on $C_\mathrm{UV}$, as will be detailed in the next section.

\subsection{Computation of framed \texorpdfstring{$2d$-$4d$}{2d-4d} BPS indices}

\begin{definition}\labelx{def:spectralnetwork}
    Let $\vartheta$ be an angle. One says that a point $z\in C_\mathrm{UV}$ \emph{supports} the soliton charge $\overline{a}\in\Gamma(z,z)$ if $Z_{\overline{a}}(z)/\E^{\I\vartheta}\in\mathbb{R}_-$. The \emph{spectral network}\index{spectral network} $\mathcal W_\vartheta$ is the set of points $z\in C_\mathrm{UV}$ that support some $\overline{a}\in\Gamma(z,z)$ such that $\mu(a)\neq 0$, where $a$ is any lift of $\overline{a}$ to $\widetilde{\Gamma}(\widetilde{z},\widetilde{z})$.
\end{definition}

\begin{proposition}
    The spectral network $\mathcal W_\vartheta$ is of codimension 1 in $C_\mathrm{UV}$, and consists of a union of segments called \emph{$\mathcal{S}$-walls}\index{spectral network!$\mathcal S$-wall} ending either at branch points or at special points called \emph{joints}\index{spectral network!joint} where different $\mathcal S$-walls intersect.
\end{proposition}

We assume that when $\vartheta$ is generic, any generic point of $\mathcal W_\vartheta$ supports exactly one $\overline{a}\in\Gamma_{ij}(z,z)$ for some $i\neq j$, where generic points of $\mathcal W_\vartheta$ are all the points of $\mathcal W_\vartheta$ which are neither branch points nor junctions. In this case, all the points of a given $\mathcal S$-wall $p$ support the same charge $\overline{a}_{ij}\in\Gamma(z,z)$. One says that the wall is of type $ij$, or that it is an $ij$-trajectory, since by definition, this means that for all $z\in p$, one has $Z_{\overline{a}}/\E^{\I\vartheta}\in\mathbb{R}_-$, and, in turn, that $p$ satisfies the local BPS differential equation \Cref{eq:localBPScondition} with angle $\vartheta$, since
\begin{equation*}
    \mathrm{d}Z_{\overline{a}}(z) = \frac{1}{\pi}(\lambda^{(j)}-\lambda^{(i)})\; .
\end{equation*}
In other words, the $\mathcal S$-walls satisfy the same differential equation as the $4d$ BPS strings described in \Cref{Subsec:localBPS}.

\subsubsection*{The mass filtration} 

Given a parameter $\Lambda$ with the dimensions of mass, one can define the truncation $\mathcal{W}_\vartheta[\Lambda]$ as the subset of $\mathcal{W}_\vartheta$ containing only those points $z \in \mathcal{W}_\vartheta$ for which $z$ supports a charge $\overline{a} \in \Gamma(z, z)$ satisfying $\vert Z_{\overline{a}}(z)\vert < \Lambda$. For any two mass parameters $\Lambda < \Lambda'$, the inclusion $\mathcal{W}_\vartheta[\Lambda] \subset \mathcal{W}_\vartheta[\Lambda']$ holds, justifying the term \emph{mass filtration}\index{mass filtration}.

Similarly, one can define truncated soliton degeneracies as follows:
\begin{equation*}
    \mu[\Lambda]\colon \left\{ \left. a\in\widetilde{\Gamma}(\widetilde{z},\widetilde{z})\, \right\vert\, \vert Z_{\overline{a}}(z)\vert < \Lambda \right\} \longrightarrow \mu[\Lambda](a) := \mu(a)\; .
\end{equation*}

The truncated network $\mathcal{W}_\vartheta[\Lambda]$ for $\Lambda = 0$ comprises only the branch points of the covering $\Sigma \to C_\mathrm{UV}$. As $\Lambda$ is gradually increased from $0$ to $\infty$, the network grows incrementally, eventually reconstructing the entire spectral network in the limit $\Lambda\rightarrow\infty$:
\begin{equation*}
    \mathcal W_\vartheta = \lim_{\Lambda\rightarrow\infty} \mathcal W_\vartheta[\Lambda]\; .
\end{equation*}

For sufficiently small $\Lambda$, the truncated network $\mathcal{W}_\vartheta[\Lambda]$ consists only of small $\mathcal{S}$-walls emanating from the branch points of the covering $\Sigma \to C_\mathrm{UV}$ (cf. \Cref{Subsec:localBPS}). As $\Lambda$ increases, some $\mathcal{S}$-walls may intersect, forming junctions. When $\vartheta$ is generic, such junctions fall into one of the following categories:
\begin{itemize}
    \item When an $\mathcal{S}$-wall of type $ij$ intersects another $\mathcal{S}$-wall of type $kl$, where the indices $i,j,k,l$ are all distinct, the interaction is trivial: the two $\mathcal{S}$-walls simply cross each other and continue along their respective trajectories.
    \item When an $\mathcal{S}$-wall of type $ij$ intersects another $\mathcal{S}$-wall of type $jk$, where $k\neq i$, an $\mathcal{S}$-wall of type $ik$ may be created at the junction.
    \item Conversely, an $\mathcal{S}$-wall of type $ik$ can disappear at a junction where it meets two other $\mathcal{S}$-walls of respective types $ij$ and $jk$.
    \item There can also be junctions where $\mathcal{S}$-walls of types $ij$, $jk$, and $ik$ meet and cross (cf. \Cref{fig:junction}).
\end{itemize}

\begin{remark}
    The possibility of walls being created or disappearing at junctions is tied to \Cref{def:spectralnetwork} of spectral networks. According to this definition, $\mathcal{S}$-walls are not composed of all potential trajectories supporting a given soliton charge, but only those for which the soliton degeneracy $\mu(a)$ is nonzero. This distinction will be further explored in subsequent sections.
\end{remark}

\subsubsection*{Computation of $F(\wp,\vartheta)$}

The computation of $F(\wp,\vartheta)$ follows from the $2d$-$4d$ wall-crossing formula derived in \cite{Gaiotto:2011tf}. Given a path $\wp\subset C_\mathrm{UV}$, let us first define the element
\begin{equation}\labelx{eq:liftnocrossing}
    \mathcal{D}(\wp) := \sum_{i=1}^N X_{\wp^{(i)}}
\end{equation}
of the homological path algebra on $\Sigma$, where $\wp^{(i)}$, $i=1,\dots,N$, are the lifts of $\wp$ to $\Sigma$.

By \Cref{Prop:concatenation}, the computation of $F(\wp,\vartheta)$, assuming that $\wp$ intersects the spectral network $\mathcal{W}_\vartheta$ only at generic points, is fully determined by two cases:
\begin{enumerate}
    \item For small paths $\wp$ that do not intersect $\mathcal{W}_\vartheta$:
    \begin{equation}\labelx{eq:genfuncNOcrossing}
        F(\wp,\vartheta) = \mathcal{D}(\wp)\; ,
    \end{equation}
    where $\mathcal{D}(\wp)$ is defined in \Cref{eq:liftnocrossing}.
    \item For small paths $\wp$ that cross $\mathcal{W}_\vartheta$ exactly once at a generic point, let us assume that $\wp$ intersects $\mathcal{W}_\vartheta$ at a point $z$ on an $\mathcal{S}$-wall carrying the soliton charge $\overline{a} \in \Gamma_{ij}(z,z)$. Denote the soliton multiplicity as $\mu(a)$, where $a$ is any lift of $\overline{a}$ to $\widetilde{\Gamma}(\widetilde{z},\widetilde{z})$. In this situation, split $\wp$ into two paths $\wp_+$ and $\wp_-$, as illustrated in \Cref{fig:computationFwptheta}.
    
    \begin{figure}[!ht]
        \centering
        \includegraphics[scale=0.83]{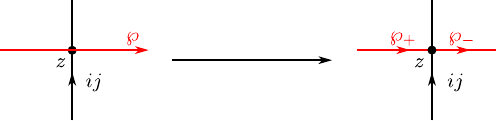}
        \caption{Computation of $F(\wp,\vartheta)$ when $\wp$ crosses $\mathcal W_\vartheta$ once.}
        \labelx{fig:computationFwptheta}
    \end{figure}
    
    Then, the generating function $F(\wp,\vartheta)$ is expressed as:
    \begin{equation*}
        F(\wp,\vartheta) = \mathcal{D}(\wp) + \mathcal{D}(\wp_+)(\mu(a) X_{a_{\widetilde{z}}})\mathcal{D}(\wp_-)\; ,
    \end{equation*}
    where $a_{\widetilde{z}}$ is the path in $\widetilde{\Sigma}$ obtained by concatenating the shortest path from $\widetilde{z}^{(i)}_+$ to the starting point of $a$, the path $a$, and the shortest path from the final point of $a$ to  $\widetilde{z}^{(j)}_-$. Here, $\widetilde{z}^{(i)}_+$ (resp. $\widetilde{z}^{(j)}_-$) denote the lift of $z^{(i)}$ (resp. $z^{(j)}$) to $\widetilde{\Sigma}$ prescribed by $\wp_+$ (resp. $\wp_-$). Note that the only non-zero contribution from the second term in the equation is 
    \begin{equation*}
        \mu(a) X_{\wp_+^{(i)}}X_{a_{\widetilde{z}}}X_{\wp_-^{(j)}}\; ,
    \end{equation*} 
    since the other paths cannot be concatenated. Thus, crossing an $\mathcal S$-wall introduces an additional contribution to $F(\wp,\vartheta)$, which corresponds to a detour of $\wp$ along the path associated with the BPS soliton supported on the wall.
    
    Finally, $F(\wp,\vartheta)$ can be rewritten as:
    \begin{equation}\labelx{eq:genfunccrossing}
        F(\wp,\vartheta) = \mathcal{D}(\wp_+)(1+\mu(a) X_{a_{\widetilde{z}}})\mathcal{D}(\wp_-)\; .
    \end{equation}
\end{enumerate}

\subsection{Homotopy invariance and soliton degeneracies}

An important constraint on the framed BPS indices $\overline{\underline{\Omega}}' (L_{\wp,\vartheta},a)$, and thus on the generating functions $F(\wp,\vartheta)$, is that they are expected to be (twisted) homotopy invariants of the path $\wp$. This, in turn, allows for the determination of the soliton degeneracies $\mu(a)$ for each $\mathcal S$-wall in the spectral network $\mathcal W_\vartheta$. In this section, we again assume that $\vartheta$ is generic.

\begin{proposition}
    The soliton degeneracy $\mu(a)$ for any lift $a$ of $\overline{a}$ is constant along each $\mathcal S$-wall. Therefore, if $p$ is an $\mathcal S$-wall carrying the charge $\overline{a}\in\Gamma(z,z)$, one denotes $\mu(a,p)$ the soliton degeneracy corresponding to the lift $a$ of $\overline{a}$, along $p$.
\end{proposition}

\begin{proof}
    Consider the two setups of \Cref{fig:solitondegconstant}. Since $\wp' = \wp'_+\wp'_0\wp'_-$ is in the same twisted homotopy class as $\wp$, it must be the case that:
    \begin{equation*}
        F(\wp,\vartheta) = F(\wp',\vartheta)\; .
    \end{equation*}
    \begin{figure}[!ht]
        \centering
        \includegraphics[scale=0.83]{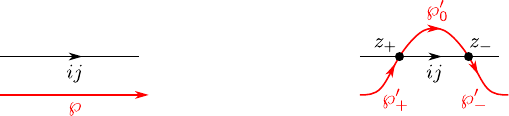}
        \caption{Two (regular) homotopy-equivalent setups.}
        \labelx{fig:solitondegconstant}
    \end{figure}
    From \Cref{eq:genfuncNOcrossing}, one has
    \begin{equation*}
        F(\wp,\vartheta) = \mathcal{D}(\wp)\; ,
    \end{equation*}
    whereas \Cref{eq:genfunccrossing} implies that $F(\wp',\vartheta)$ decomposes as:
    \begin{align*}
        &\mathcal{D}(\wp'_+) (1+\mu(a;z_+)X_{a_{\widetilde{z}_+}}) \mathcal{D}(\wp'_0) (1+\mu(a;z_-)X_{a_{\widetilde{z}_-}}) \mathcal{D}(\wp'_-) \\
        = &\mathcal{D}(\wp') + \mu(a;z_+)X_{(\wp'_+)^{(i)} a_{\widetilde{z}_+} (\wp'_0)^{(j)} (\wp'_-)^{(j)}} + \mu(a;z_-)X_{(\wp'_+)^{(i)} (\wp'_0)^{(i)} a_{\widetilde{z}_-} (\wp'_-)^{(j)}}\\
        = &\mathcal{D}(\wp') + \left(\mu(a;z_+)-\mu(a;z_-)\right)X_{(\wp'_+)^{(i)} a_{\widetilde{z}_+} (\wp'_0)^{(j)} (\wp'_-)^{(j)}}\; ,
    \end{align*}
    where the last equality follows from the fact that the paths $(\wp'_+)^{(i)} a_{\widetilde{z}_+} (\wp'_0)^{(j)} (\wp'_-)^{(j)}$ and $(\wp'_+)^{(i)} (\wp'_0)^{(i)} a_{\widetilde{z}_-} (\wp'_-)^{(j)}$ only differ by a unit of winding around the circle fiber of $\widetilde{\Sigma} \rightarrow \Sigma$. Therefore, homotopy invariance imposes that $\mu(a;z_+)=\mu(a;z_-)$, and hence that $\mu(a)$ is constant along the $\mathcal S$-wall.
\end{proof}

\begin{proposition}
    The soliton degeneracy of any $\mathcal S$-wall emerging from a branch point is 1.
\end{proposition}

\begin{proof}
    This again follows from homotopy invariance in the setup depicted in \Cref{fig:hominvbranchpoint}, where a choice of branch cut is shown in purple.
    \begin{figure}[!ht]
        \centering
        \includegraphics[scale=0.83]{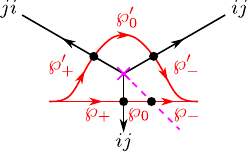}
        \caption{Homotopy invariance close to a branch point.}
        \labelx{fig:hominvbranchpoint}
    \end{figure}
    The generating function $F(\wp,\vartheta)$ can be expressed directly from \Cref{eq:genfunccrossing}, except for the fact that the branch cut connects $\wp_0^{(i)}$ to $\wp_-^{(j)}$ and vice-versa. On the other hand, the generating function $F(\wp',\vartheta)$ is derived by applying \Cref{eq:genfunccrossing} twice, as in the proof of the previous proposition.

    Note that close to the branch point, the soliton supported on each $\mathcal{S}$-wall is represented by the segment of the $\mathcal{S}$-wall linking the point under consideration to the branch point. This implies that $\mu(a) = \pm 1$ for each $\mathcal{S}$-wall. A direct verification shows that homotopy invariance imposes $\mu(a) = 1$.
\end{proof}

\begin{figure}[!ht]
    \centering
    \includegraphics[scale=0.83]{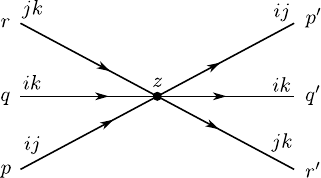}
    \caption{A general junction.}
    \labelx{fig:junction2}
\end{figure}

\begin{proposition}\labelx{prop:junctions}
    Consider a general junction as depicted in \Cref{fig:junction2}. Then:
    \begin{align*}
        \mu(a,p') &= \mu(a,p)\; , \\
        \mu(b,r') &= \mu(b,r)\; ,\\
        \mu(c,q') &= \mu(c,q)+ (-1)^{w(a_{\widetilde{z}}+b_{\widetilde{z}},c_{\widetilde{z}})}\mu(a,p)\mu(b,r)\; .
    \end{align*} 
\end{proposition}

\begin{proof}
    This again follows from homotopy invariance. To establish the proposition, one applies \Cref{eq:genfunccrossing} three times: first to a path $\wp$ passing vertically to the left of $z$, and then to another path $\wp'$ passing vertically to the right of $z$. These paths can then be deformed such that their only intersections with the $\mathcal{S}$-walls occur at the junction $z$.
\end{proof}

Important special cases of the general formula of \Cref{prop:junctions} are the cases where either $\mu(c,q) = 0$ and $\mu(c,q') \neq 0$, which correspond to $ik$-walls being created at an $ij$-$jk$ junction, or $\mu(c,q') = 0$ and $\mu(c,q) \neq 0$, which corresponds to $ik$-walls being annihilated at an $ij$-$jk$-$ik$ junction.

\subsection{\texorpdfstring{$4d$}{4d} BPS degeneracies from spectral networks}

BPS spectra are determined by examining how the generating functions $F(\wp,\vartheta)$ of framed $2d$-$4d$ BPS states change as $\vartheta$ varies. In general, $F(\wp,\vartheta)$ remains constant under small variations of $\vartheta$. However, there are specific values of $\vartheta$ at which $F(\wp,\vartheta)$ undergoes a discontinuous change. This can occur for two distinct reasons.

First, as $\vartheta$ varies, the $\mathcal{S}$-walls of the spectral network $\mathcal{W}_\vartheta$ move on $C_\mathrm{UV}$. Due to this motion, $\mathcal{S}$-walls may cross the starting or endpoint of the path $\wp$ under consideration, resulting in the addition or removal of a contribution as described in \Cref{eq:genfunccrossing}. This provides one reason for the jump in $F(\wp,\vartheta)$, which depends both on $\vartheta$ and on the endpoints of $\wp$.

The second reason for the discontinuity in $F(\wp,\vartheta)$ at critical values $\vartheta_c$ of $\vartheta$ lies in changes in the topology of $\mathcal{W}_\vartheta$. As discussed in \Cref{Subsec:BPSclassSA1}, such changes occur at specific values of $\vartheta$ referred to as \emph{$\mathcal{K}$-walls}\index{spectral network!$\mathcal K$-wall}, where an $ij$-trajectory encounters an $(ij)$-branch point. The simplest example of such a topology change, known as a \emph{saddle connection}\index{saddle!connection}, is depicted in \Cref{fig:saddlepoint}. 

\begin{figure}[!ht]
    \centering
    \includegraphics[width=\textwidth]{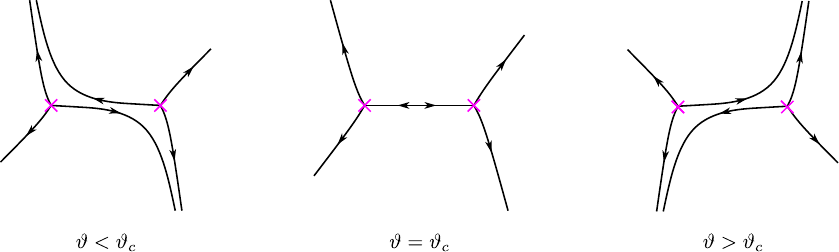}
    \caption{The topology change of the spectral network $\mathcal W_\vartheta$ at a saddle connection.}
    \labelx{fig:saddlepoint}
\end{figure}

An \textit{in-situ} example of such connections is provided in \Cref{fig:KwallsSU2}, which shows the spectral network for pure $\mathrm{SU}(2)$ SYM theory at the origin of the Coulomb branch for different values of $\vartheta$. In this example, $\mathcal K$-walls occur at $\theta_c=\pi\sim 0$ and $\theta_c=\pi/2$.

\begin{figure}[!ht]
    \centering
    \includegraphics[scale=0.83]{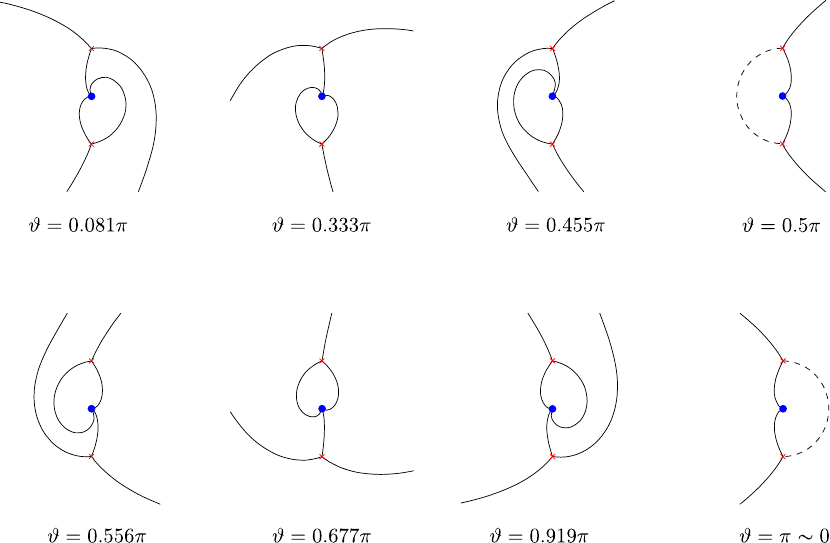}
    \caption{Evolution of the spectral network for pure $\mathrm{SU}(2)$ SYM theory at the origin of the Coulomb branch, as $\vartheta$ varies from $0$ to $\pi$.}
    \labelx{fig:KwallsSU2}
\end{figure}

\subsubsection*{$\mathcal K$-wall formula}

In general, at $\vartheta = \vartheta_c$, the spectral network $\mathcal{W}_{\vartheta_c}$ exhibits a non-generic $\mathcal{S}$-wall, formed from the collision of two or more generic $\mathcal{S}$-walls. When these $\mathcal{S}$-walls are traveling in opposite directions---as is the case for saddle connections---the resulting non-generic $\mathcal{S}$-wall is referred to as a \emph{two-way street}\index{spectral network!two-way street} or \emph{double wall}.

Every two-way street admits two distinct resolutions depending on whether $\vartheta = \vartheta_c + \epsilon$ or $\vartheta = \vartheta_c - \epsilon$, where $\epsilon > 0$ is infinitesimally small. If $\wp$ is a path crossing a two-way street, one can compute $F(\wp,\vartheta_c + \epsilon)$ and $F(\wp,\vartheta_c - \epsilon)$ using \Cref{eq:genfunccrossing}, finding that $F(\wp,\vartheta_c + \epsilon) \neq F(\wp,\vartheta_c - \epsilon)$: the generating function $F(\wp,\vartheta)$ exhibits a jump at $\vartheta = \vartheta_c$.

The jump of the generating function $F(\wp,\vartheta)$ as $\vartheta$ crosses a $\mathcal K$-wall can be expressed in terms of a universal substitution $\mathcal K$ acting on the formal path variables $X_a$:
\begin{equation}\labelx{eq:Kwallformula}
    F(\wp,\vartheta_c+\epsilon) = \mathcal{K}\left[F(\wp,\vartheta_c-\epsilon)\right]\; .
\end{equation}
\Cref{eq:Kwallformula} is called the \emph{$\mathcal K$-wall formula}\index{spectral network!$\mathcal K$-wall!formula}.

The soliton degeneracies on a two-way street $p$ are collected into a generating function $Q(p)$, for which one must introduce new variables $X_{\widetilde{\gamma}}$ corresponding to each homology class $\widetilde{\gamma}$ of closed paths on $\widetilde{\Sigma}$, i.e., $\widetilde{\gamma} \in \widetilde{\Gamma} = \mathrm{H}^1(\widetilde{\Sigma},\mathbb{Z})$, modulo shifts by $2H_\Sigma$. The homological path algebra is then extended to include the generators $X_{\widetilde{\gamma}}$, with the multiplication rules: \begin{equation*} X_{\widetilde{\gamma}} X_{\widetilde{\gamma}'} = X_{\widetilde{\gamma}+\widetilde{\gamma}'}\; ,\quad X_{\widetilde{\gamma}} X_{a} = X_{\widetilde{\gamma}+a}\; , \end{equation*} where $\widetilde{\gamma}+a$ denotes the natural action of $\widetilde{\Gamma}$ on $\widetilde{\Gamma}(\widetilde{z},\widetilde{z})$ for any $z \in C_\mathrm{UV}$. There is also the additional relation $X_{H_\Sigma} = -1$.

Moreover, for any open path $a \in \widetilde{\Gamma}_{ij}(\widetilde{z},\widetilde{z})$, one denotes $\mathrm{cl}(a) \in \widetilde{\Gamma}$ as the corresponding closed homology class. With this notation, the generating function $Q(p)$ is defined as: 
\begin{equation*} 
Q(p) = 1 + \sum_{\overline{a} \in \Gamma_{ij}(p),\; \overline{b} \in \Gamma_{ji}(p)} \mu^-(a,p)\mu^-(b,p) X_{\mathrm{cl}(a+b)}\; , 
\end{equation*} 
where $\Gamma_{ij}(p)$ and $\Gamma_{ji}(p)$ denote the sets of charges supported by the two-way street $p$, and where $\mu^-(a,p)$ and $\mu^-(b,p)$ represent the soliton degeneracies on $p$ in the limit $\vartheta \to \vartheta_c^-$. Replacing $\mu^-$ with $\mu^+$, where $\mu^+$ is defined in the limit $\vartheta \to \vartheta_c^+$, yields the same generating function $Q(p)$. 

Note that all the $X_{\widetilde{\gamma}}$ appearing in $Q(p)$ satisfy $\E^{\I\vartheta}Z_{\widetilde{\gamma}} \in \mathbb{R}_-$. Generically, and typically when considering a point on the Coulomb branch away from the walls of marginal stability, one can assume that the set $\Gamma_c$ of charges $\gamma \in \Gamma$ such that $\E^{\I\vartheta}Z_\gamma \in \mathbb{R}_-$ is generated by a single element $\gamma_0$. In this case, $Q(p)$ becomes a power series in a single variable $X_{\widetilde{\gamma}_0}$, where $\widetilde{\gamma}_0$ is a lift of $\gamma_0$ to $\widetilde{\Gamma}$. In fact, for each $\gamma \in \Gamma$, there exists a ``preferred'' lift $\widetilde{\gamma} \in \widetilde{\Gamma}$ such that $Q(p)$ factorizes as: 
\begin{equation*} 
    Q(p) = \prod_{\gamma \in \Gamma_c} \left(1 - X_{\widetilde{\gamma}}\right)^{\alpha_\gamma(p)}\; , 
\end{equation*} 
where the exponents satisfy $\alpha_\gamma(p) \in \mathbb{Z}$. Then, for each $\gamma \in \Gamma_c$, one defines: 
\begin{equation*} 
    L(\gamma) = \sum_{p} \alpha_\gamma(p) p_\Sigma\; , 
\end{equation*} 
where the sum runs over two-way streets in $\mathcal{W}_{\vartheta_c}$, and $p_\Sigma$ denotes the 1-chain lift of $p$ in $\Sigma$. One can show that $L(\gamma)$ is, in fact, a 1-cycle. This, in turn, allows for the definition of the intersection number $\langle \overline{c}, L(\gamma)\rangle$ of $L(\gamma)$ with any $\overline{c} \in \Gamma(z_1, z_2)$. This leads to the following result \cite{Gaiotto:2012rg}:

\begin{proposition} 
    The universal substitution $\mathcal{K}$ is given by: 
    \begin{equation*} 
        \mathcal{K}\colon X_a \longmapsto \prod_{\gamma \in \Gamma_c} \left(1 - X_{\widetilde{\gamma}}\right)^{\langle \overline{a}, L(\gamma)\rangle} X_a\; . 
    \end{equation*} 
\end{proposition}

\subsubsection*{BPS degeneracies}

We observe that 
\begin{equation*} 
    \E^{-\I\vartheta_c} \int_{p_\Sigma} \lambda \in \mathbb{R}_-\; . 
\end{equation*} 
Under the assumption that $\Gamma_c$ is generated by a single element, this implies that the homology class $[L(\gamma)]$ of $L(\gamma)$ is proportional to $\gamma$.

\begin{proposition} 
    The ratio $[L(\gamma)] / \gamma$ is the second helicity supertrace $\Omega(\gamma)$. 
\end{proposition}

\begin{remark}
    In fact, spectral networks provide a way to compute not only $4d$ BPS degeneracies $\Omega(\gamma)$ but also \emph{enhanced BPS degeneracies}, denoted $\omega(\gamma,\overline{a})$, for $\gamma \in \Gamma$ and $\overline{a} \in \Gamma(z_1, z_2)$ \cite{Gaiotto:2011tf}, as follows: 
    \begin{equation*} 
        \omega(\gamma,\overline{a}) = \langle L(\gamma), \overline{a} \rangle\; . 
    \end{equation*} 
    The enhanced BPS degeneracies $\omega(\gamma,\overline{a})$ satisfy, in particular: 
    \begin{equation*}  
        \omega(\gamma,\overline{a}+\gamma') - \omega(\gamma,\overline{a}) = \Omega(\gamma) \langle \gamma, \gamma' \rangle\; , 
    \end{equation*} 
    i.e., they encode the information of the $4d$ BPS degeneracies $\Omega(\gamma)$. Moreover, unlike $\Omega(\gamma)$, the enhanced BPS degeneracies $\omega(\gamma,\overline{a})$ provide details about the location of $4d$ BPS states along $C_\mathrm{UV}$ and their interactions with surface defects.
\end{remark}

\begin{example} 
In the case of a saddle connection, the two $\mathcal{S}$-walls forming the two-way street $p$ carry charges $\overline{a}, \overline{b}$ such that $\mu^-(a) = \mu^-(b) = 1$, where $a$ and $b$ are the natural lifts of $\overline{a}$ and $\overline{b}$. Denoting $\widetilde{\gamma}$ the preferred lift of $\gamma = \mathrm{cl}(\overline{a} + \overline{b})$, one has: 
\begin{equation*} 
    Q(p) = 1 - X_{\widetilde{\gamma}}\; . 
\end{equation*} 
This implies $L(\gamma) = p_\Sigma$, where $[L(\gamma)] = \gamma$, and $L(n\gamma) = 0$ for any $n > 1$. Thus: 
\begin{equation*} 
    \Omega(\gamma) = 1\; , 
\end{equation*} 
showing that saddle connections correspond to $4d$ BPS hypermultiplets. 
\end{example}

\begin{example} 
Another fundamental example of a $\mathcal{K}$-wall occurs when $\mathcal{S}$-walls winding around a cylinder collapse into two two-way streets, $p^1$ and $p^2$. Each of these two-way streets begins and ends at the same branch point. This case is described in \cite[Sect. 7.2]{Gaiotto:2012rg}. One finds: 
\begin{equation*} 
    L(\gamma) = -p^1_\Sigma - p^2_\Sigma\; , 
\end{equation*} 
where $p^1_\Sigma$ and $p^2_\Sigma$ are both in the homology class $[\gamma]$. Therefore: 
\begin{equation*} 
    \Omega(\gamma) = -2\; , 
\end{equation*} 
showing that such closed loops in the spectral networks correspond to BPS vector multiplets. 
\end{example}

One can analyze more complicated finite string webs appearing at $\mathcal{K}$-walls in a similar manner. In particular, it is possible to compute the BPS degeneracies $\Omega(\gamma)$ and the enhanced BPS degeneracies $\omega(\gamma, \overline{a})$ of the corresponding $4d$ BPS states.

These techniques enable the computation of the $4d$ BPS spectra in higher-rank theories in class $\mathcal{S}$. Examples include the ``pentagon'' Argyres–Douglas theory mentioned at the end of \Cref{Subsec:BPSclassSA1}, strongly-coupled pure $\mathrm{SU}(3)$ SYM, and the theory of nine free hypermultiplets, as discussed in \cite[Sect. 8]{Gaiotto:2012rg}. Many other significant examples were subsequently addressed in later works, including general Argyres–Douglas theories \cite{Gaiotto:2012db, Maruyoshi:2013fwa}, class $\mathcal{S}$ theories of type\footnote{The figure on the front page of this monograph depicts a spectral network for pure $\mathrm{SO}(8)$ SYM theory at the origin of its Coulomb branch.} $D$ and $E$ \cite{LoPa}, and Minahan–Nemeschansky SCFTs \cite{HN:2017,HHN:2020}.

For instance, the BPS spectrum of the $E_7$ Minahan–Nemeschansky theory is studied in \cite{HHN:2020} by considering a deformation of the theory by a small mass parameter $m$, taking the meromorphic cubic differentials
\begin{equation*}
    \phi_2=\frac{mz\mathrm{d}z^2}{(z^3-1)^2}\; , \quad \phi_3=-\frac{\mathrm{d}z^3}{(z^3-1)^2}\; ,
\end{equation*}
to parameterize the Coulomb branch. One can define and study the spectral network $\mathcal{W}_\vartheta$ for the mass-deformed theory using the methods outlined in the previous sections, and derive spectral networks $\mathcal{W}_\vartheta$ for the massless theory by considering the limit $m \to 0$. For example, one can show that
\begin{equation*}
    \vert \Omega(n \gamma_1)\vert \sim cn^{-\frac{5}{2}}(17+12 \sqrt{2})^n\; , 
\end{equation*}
asymptotically at large $n$, where $c$ is a constant, and deduce that, similarly to the $E_6$ Minahan–Nemeschansky theory, the BPS states in the $E_7$ theory carrying electromagnetic charges that are $n$ times a primitive charge always come with index a positive integer multiple of $(-1)^{n+1}n$.

\begin{remark}
    Spectral networks, as described in this section, are subordinate to a Seiberg--Witten holomorphic ramified covering $\Sigma \to C_\mathrm{UV}$ and are thus sometimes referred to as \emph{WKB spectral networks}\index{spectral network!WKB}. However, the notion of a spectral network can be generalized to purely topological setups. In this case, they are subordinate to ramified coverings $\Sigma \to C$ without analytic structures and are referred to as \emph{topological spectral networks} or \emph{general spectral networks}\index{spectral network!general}.

    Although topological spectral networks do not directly connect to BPS counting questions in $4d$ $\mathcal{N}=2$ theories, they preserve their path-lifting properties. Specifically, the twisted homotopy invariance properties of the generating function $F(\wp,\vartheta)$ enable a process known as non-abelianization. Given an abelian flat connection on $\Sigma$ and a spectral network on $C$, one can construct a non-abelian flat connection on $C$ by defining its parallel transport. This construction is developed for \emph{twisted $\mathrm{GL}_n(\mathbb{C})$-local systems} on $C$ in \cite[Sects. 9 and 10]{Gaiotto:2012rg}. The inverse process is known as abelianization.

    Ultimately, spectral networks provide charts on moduli spaces of twisted, decorated non-abelian flat connections on smooth Riemann surfaces. The spectral networks and path-lifting rules discussed here are adapted to twisted decorated local systems, whose moduli spaces are described by Fock–Goncharov $\mathcal A$-coordinates. In contrast, the slightly modified notion which is the focus of \Cref{partIV}, particularly \Cref{sec:non-abel}, is adapted to framed local systems and $\mathcal X$-coordinates. This ``framed'' version of (non-)abelianization furnishes charts on moduli spaces of framed non-abelian flat connections, as detailed in \Cref{partIV}.
\end{remark}

\begin{remark}
    Another fundamental approach to the study of BPS spectra in $4d$ $\mathcal{N}=2$ theories is the theory of \emph{BPS quivers}\index{BPS!quiver}\index{quiver!BPS}, first introduced and explored in \cite{Alim:2011ae,Alim:2011kw} (see also \cite{Cecotti:2012se} for a review). This approach applies to theories possessing the \emph{quiver property}, which requires the existence of a set of generators ${e_i}$ for the charge lattice $\Gamma$, such that for every BPS particle with charge $\gamma \in \Gamma$, either $\gamma$ or $-\gamma$ lies within the positive cone generated by ${e_i}$.

    The BPS spectrum of such theories can be computed through representation-theoretic methods applied to the BPS quivers associated with the theory. A key concept in the theory of BPS quivers is the notion of \emph{mutation}\index{mutation}, introduced in the context of cluster varieties in \Cref{Sec:clustervarieties}. In particular, the so-called \emph{mutation method}\index{mutation!method} provides an efficient framework for determining the BPS spectrum.

    There have been attempts to establish connections between the theories of spectral networks and BPS quivers. For example, see \cite{Gabella:2017hpz,Gabella:2017hxq} for explorations of such relations.
\end{remark}

\newpage
\part{Generalizations}\labelx{partVI}

\vspace{1.5cm}

In this final part, we briefly survey several important generalizations of spectral networks that have recently emerged in mathematical physics, directing the reader to relevant references for further study. Among the many significant directions, we highlight \emph{exponential networks}, \emph{three-dimensional networks}, \emph{WKB cameral networks}, and the \emph{non-abelianization method for conformal Virasoro blocks}. Other developments, which require additional background and are therefore omitted here, are nonetheless noteworthy.

A prominent example is the geometric construction of a $q$-nonabelianization procedure. Its first manifestation is the quantum trace map \cite{Bonahon-Wong}, which defines a morphism between two distinct quantizations of the $\mathrm{SL}_2(\mathbb{C})$ character variety of a surface, later interpreted as a non-abelianization map in \cite{Korinman-Quesney}. Related advances include the study of quantum holonomies in the physics literature \cite{Gabella:2017qh,KLS}. In three dimensions, $q$-nonabelianization induces a map from links in a 3-manifold $M$ to links in a branched $N$-fold cover $\widetilde{M}$ of $M$, realized as a morphism between skein modules associated to $M$ and $\widetilde{M}$ \cite{NeFe1,NeFe2,GMNY}.

Another striking direction is the relation between the Kontsevich–Soibelman operator and the superconformal index explored in \cite{CoSh}. Moreover, a generalized notion of spectral networks associated with families of Bridgeland stability conditions on surfaces has been introduced in \cite{HKSim}, formulated within the Fukaya category with coefficients in a triangulated dg-category. Finally, the abelianization procedure has further applications to the study of monodromy and Stokes data of families of flat connections \cite[Rmk. 3.9.6]{Neitzke_BPS-lec}. According to a conjecture of Gaiotto \cite{Gaiotto_Opers}, knowledge of central charges and BPS indices suffices to formulate an integral equation for these data, a conjecture that can be viewed as a generalization of the ODE/IM correspondence \cite{DoDuTa,DoTa}.

\vspace{0cm}

\parttoc


\chapter{Exponential networks}

\abstract{Exponential networks were introduced in \cite{ESW2017} by Eager, Selmani, and Walcher as a tool to study (framed) BPS B-branes on toric affine Calabi–Yau threefolds $X$, extending both the construction of spectral networks in $4d$ $\mathcal N=2$ gauge theories and the geometric approach to BPS states pioneered in \cite{Klemm:1996bj}. By mirror symmetry, BPS B-branes on $X$, i.e. \emph{coherent sheaves} on $X$, correspond to \emph{special Lagrangian submanifolds} in the mirror $Y$ of $X$. In the framework of local mirror symmetry, this correspondence reduces further to the study of calibrated Lagrangian cycles in the \emph{mirror curve} $\Sigma \subset (\mathbb{C}^*)^2$ of $X$. Upon choosing a projection $\pi \colon \Sigma \to \mathbb{C}^*$, one then studies saddle trajectories on $\mathbb{C}^*$, arising as the images under $\pi$ of calibrated cycles in $\Sigma$. It is precisely in this setting that exponential networks emerge.}

\vspace{0.5cm}

Exponential networks have been further developed by Banerjee, Longhi, and Romo in a series of papers \cite{BLR1, BLR3, BLR2, BLR4}; further recent developments in this direction include \cite{AHT, DeLo1, DeLo2, GHN}. We also refer to \cite{BRSW} for a most recent bibliographical overview on the topic.

\section{Definition and properties}

Similar to spectral networks, the formal definition of exponential networks relies on two kinds of data:
\begin{itemize}
    \item A piece of \emph{geometric data}, depending on a covering $\Sigma \to C$ of Riemann surfaces induced by an embedding $\Sigma \hookrightarrow (T^*C)^{\times}$, where $(T^*C)^{\times}$ denotes the cotangent bundle of $C$ with the zero section removed. For each choice of phase $\vartheta \in \mathbb{R}/2\pi\mathbb{Z}$, this data specifies a family of calibrations for one-dimensional submanifolds of $C$. In most applications, one takes $C = \mathbb{C}^\times$, which is the natural case arising in local mirror symmetry.
    \item A piece of \emph{combinatorial data} capturing the soliton content of a certain effective 3-dimensional quantum field theory engineered from the geometric data.
\end{itemize}

For $C=\mathbb{C}^\times_x$, one has $(T^*C)^\times\equiv\mathbb{C}^\times_x\times\mathbb{C}^\times_y$. This space is endowed with the symplectic form:
\begin{equation*}
    \omega = \frac{\D x}{x}\wedge\frac{\D y}{y}\; .
\end{equation*}
One considers the Liouville-like 1-form:
\begin{equation*}
    \lambda = \log{y}\,\frac{\D x}{x} = \log{y}\;\D\log{x}\; ;
\end{equation*}    
importantly, it is multi-valued on $(T^*C)^\times$. One can define an infinite cyclic cover $\tilde{\pi}\colon \widetilde{\Sigma}\to \Sigma$ given by the exponential map. Hence we have the following sequence of coverings:
\[\widetilde{\Sigma} \xlongrightarrow{\tilde{\pi}} \Sigma\xlongrightarrow{\pi} \mathbb{C}^\times\; .\]

For a ramified covering $\Sigma \to \mathbb{C}^\times_x$ where $\Sigma \subset (T^*C)^\times$, and a phase $\vartheta \in \mathbb{R}/2 \pi \mathbb{Z}$, an \emph{exponential network}\index{exponential networks} $\mathcal{W}(\vartheta)$ on $\mathbb{C}_{x}^{\times}$ is a web of local trajectories, called $\mathcal{E}$-walls, that are integral curves of the ordinary differential equation: 
\[(\mathrm{log}(y_j) -\mathrm{log}(y_k) +2 \pi \I n) \frac{\mathrm{d}\mathrm{log}x(t)}{\mathrm{d}t} \in \E^{\I \vartheta} \mathbb{R}_{+}\; ,\]
parameterized by $t \in \mathbb{R}$. These integral curves are labeled by all possible $(jk)_{n}$ (including $j=k$) and obey certain local crossing rules which generalize those of spectral networks. From the point of view of the logarithmic cover, these rules are obtained by lifting the multiple open paths to successive logarithmic sheets over the junction and matching up the ends in a way which respects orientation, so as to form a closed circle. Precisely, the \emph{joints} or \emph{junctions} appearing in exponential networks consist of joints similar to those of spectral networks of type $A$ (involving $(ij)_n$, $(ik)_n$ and $(jk)_n$ $\mathcal{E}$-walls for some $i,j,k,n$), as well as a new type of joint where $(ij)_n$ and $(ji)_m$ $\mathcal{E}$-walls intersect and give rise to an infinite family of new $\mathcal{E}$-walls, shown in \Cref{fig:newjointexpnet}. Because of this new kind of joints, exponential networks do not satisfy the local finiteness condition of spectral networks.

\begin{figure}
    \centering
    \includegraphics[scale=0.83]{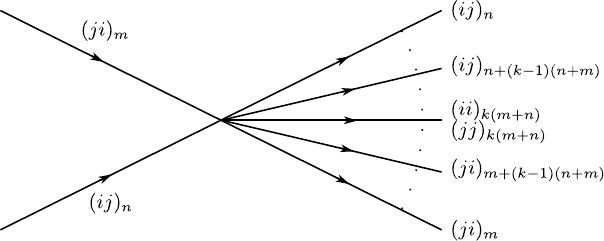}
    \caption{A type of joint occurring in exponential networks.}
    \labelx{fig:newjointexpnet}
\end{figure}

As for spectral networks, for generic $\vartheta$, exponential networks do not have any subwebs that would lift to finite length closed and calibrated cycles on the mirror curve (and hence to compact Lagrangians in $Y$). However, finite webs occur at non-generic values of $\vartheta$; lifting the multiple open paths and matching up the ends of the lifts in a way which respects orientation, one obtains a closed circle in $\Sigma$ which, in turn, corresponds to a compact special Lagrangian cycle in $Y$.

Another specificity of exponential networks, compared to spectral networks, in the presence of a new kind of asymptotic behavior for $\mathcal E$-walls near punctures: logarithmic spirals. This is of physical relevance: it accounts for \emph{non-compact D4-branes}.

\subsubsection*{Mirror symmetric interpretation and BPS counting}

For a toric affine Calabi--Yau threefold $X$, exponential networks capture the BPS spectrum of M-theory on $S^1 \times X$. The mirror $Y$ of an affine toric Calabi--Yau threefold $X$ is defined by a polynomial equation:
\[Y=\{H(x,y)=uv\} \subset \mathbb{C}_{x}^{\times} \times \mathbb{C}_{y}^{\times} \times \mathbb{C}_{u} \times \mathbb{C}_{v}\; ,\]
with $H(x,y)$ a Laurent polynomial whose Newton polygon is the toric diagram of $X$ (subject to a framing ambiguity). The Calabi--Yau $Y$ is equipped with the holomorphic 3-form: 
\[\Omega = \frac{\mathrm{d}x}{x} \wedge \frac{\mathrm{d}y}{y} \wedge\frac{\mathrm{d}u}{u}\; .\]
One fixes a local trivialization of the covering $\pi\colon \Sigma \to \mathbb{C}_{x}^{\times}$, so that the branches can be labeled as $y_k$ for $k=1,...,d$, where $d$ is the degree of the covering. The logarithmic one-form $\lambda$ of above is obtained by integrating $\Omega$ along the fibers of the projection $Y\rightarrow (\mathbb{C}^\times)^2_{x,y}$. By construction, exponential networks allow for a counting of BPS A-branes on $Y$, hence of BPS B-branes on $X$, which can be recast in terms of Donaldson--Thomas invariants of $X$.

The counting of exponential networks also admits a gauge-theoretic interpretation, though this occurs in the context of $5d$ $\mathcal{N}=1$ theories, and not of $4d$ $\mathcal{N}=2$ theories where spectral networks arise. Specifically, a $5d$ $\mathcal N=1$ theory on $\mathbb{S}^1\times \mathbb{R}^4$ corresponds to exponential networks on $\mathbb{C}^\times$. The Seiberg--Witten curve $\Sigma$, which is sometimes referred to as the Seiberg--Witten--Nekrasov curve in this context \cite{Nekrasov:1996cz}, is defined by a polynomial equation: 
\[\left\{F(x,y)=0\right\} \subset (\mathbb{C}^\times)^2=(T^*\mathbb{C}^\times)^\times\; ,\] 
where $F$ is a Laurent polynomial, and can be identified with the mirror curve of some toric affine Calabi--Yau threefold.

\vspace{0.5cm}

In practice, exponential networks have been explicitly constructed for some simple toric Calabi–Yau threefolds, including the conifold \cite{ESW2017}, local $\mathbb{P}^2$ \cite{ESW2017,BLR2}, and the local Hirzebruch surface $\mathbb{F}_0$ \cite{BLR3}.

\section{Path lifting and (non-)abelianization}

Similarly to spectral networks, exponential networks provide a framework for homotopy-invariant \emph{path lifting}. Concretely, any exponential network on $\mathbb{C}^\times$ subordinate to a ramified covering $\Sigma \to \mathbb{C}^\times$ induces a map that assigns to each homotopy class on $\mathbb{C}^\times$ a collection of homotopy classes on $\Sigma$ (or on its covering $\widetilde{\Sigma}$, subject to certain constraints) \cite{BLR1}.

This path-lifting map underlies a formal (non-)abelianization procedure, closely paralleling the case of spectral networks. Let $\mathcal{L}$ be a line bundle over $\widetilde{\Sigma}$ equipped with a flat abelian connection. Its push-forward to $\mathbb{C}^\times$ produces an infinite-rank bundle endowed with a formal non-abelian flat connection, i.e. a notion of parallel transport. The fact that the fiber of the projection $(T^*\mathbb{C}^\times)^\times \to \mathbb{C}^\times$ is $\mathbb{C}^\times$ itself suggests that the (non-)abelianization defined by exponential networks is naturally connected to character varieties of loop groups, rather than to those of finite-dimensional Lie groups. 

\chapter{Three-dimensional networks}

\abstract{In \cite{Freed-Neitzke}, Freed and Neitzke introduce an abstract framework for spectral networks on manifolds of dimension less than three and describe a stratified abelianization process for three-dimensional Chern--Simons invariants.
Specifically, they demonstrate that the $\SL_2(\mathbb{C})$-Chern--Simons invariant of a three-manifold $M$ equipped with a principal $\SL_2(\mathbb{C})$-bundle and a flat connection, can be computed in terms of the spin $\mathbb{C}^*$-Chern--Simons invariant of a ramified double cover of $M$. Moreover, the authors extend the construction of three-dimensional ($3d$) spectral networks to the groups $\GL_2(\mathbb{C})$ and $\PSL_2(\mathbb{C})$.}

\section{Chern--Simons field theory}

Chern--Simons theory\index{Chern--Simons!theory} is a three-dimensional topological gauge theory, providing topological invariants of $3d$ manifolds. It can be expressed as a $3d$ topological quantum field theory\index{topological!quantum field theory} (TQFT) as described in \cite{Atiyah:1988TQFT}. Specifically, the \textit{exponentiated Chern--Simons invariant} defines a symmetric monoidal functor from a specific category of 3-bordisms to the groupoid $\mathrm{Line}_\mathbb{C}$ of 1-dimensional complex vector spaces.

\begin{remark}
    The Yang--Mills functional
    \[S_\mathrm{YM}(A) = \int_M \tr F(A)\wedge \star F(A)\; ,\]
    where $M$ is a pseudo-Riemannian manifold, $A$ is a connection on a fixed principal $G$-bundle $P$ over $M$, $F(A)$ is the curvature of $A$, and $\tr$ denotes a Killing form on the Lie algebra of $G$, depends explicitly on the metric of $M$ via the Hodge star operator. In contrast, for $M$ a four-manifold, the functional
    \[S_\Theta = \frac{1}{8\pi^2}\int_M \tr F(A)\wedge F(A)\] 
    is topological. This functional computes a characteristic class of the principal bundle $P \rightarrow M$ and, notably, is independent of the connection $A$. The differential form $\tr F(A)\wedge F(A)$ admits the \emph{Chern--Simons 3-form}
    \[\tr (A\wedge \D A+\tfrac{2}{3}A\wedge A\wedge A)\]
    as local primitive.
\end{remark}

Let $X$ be a three-manifold and let $P \to X$ be a principal $G$-bundle, where $G$ is a Lie group with finitely many components, equipped with a $G$-invariant symmetric bilinear form $\mathrm{tr}$ on its Lie algebra $\mathfrak{g}$ (e.g., the Killing form if $G$ is semisimple). For any $1$-form $\theta \in \Omega^1(P, \mathfrak{g})$ on $P$, define:
\[\eta(\theta) = \mathrm{tr}(\theta\wedge F(\theta)-\tfrac{1}{6}\theta\wedge[\theta\wedge\theta])\in \Omega^3(P,\C)\; ,\]
where $F(\theta)$ is the curvature of $\theta$. The $3$-form $\eta(\theta)$ is closed and can be used to define an invariant of $X$ as follows. Let $s \colon X \to P$ be a section of $P$, and consider the integral $\int_X s^* \eta(\theta)$. This complex number depends only on the homotopy class of the section, because $\eta(\theta)$ is closed. However, since two sections are not necessarily homotopic, an additional modification is required to make the resulting quantity independent of the choice of $s$: the symmetric bilinear form $\mathrm{tr}$ must be integral, i.e., an element of $\mathrm{H}^4(BG, \mathbb{Z})$, where $BG$ denotes the classifying space of $G$. Moreover, the invariant is only well-defined modulo $2\pi \I$.

\begin{definition}
The \emph{exponentiated Chern--Simons invariant}\index{Chern--Simons!invariant} of the three-manifold $X$ endowed with a principal $G$-bundle $P \to X$ and $\theta \in \Omega^1(P, \mathfrak{g})$ is defined as:
\[\mathcal{F}_G(X,\theta)=\exp\left(2\pi \I \int_X s^*\eta(\theta)\right)\in\C^*\; ,\]
where $s$ is a section of $P$. The exponentiated Chern--Simons invariant is independent of the choice of $s$.
\end{definition}

Let us now describe the Chern--Simons TQFT in dimension 3. The basic idea of a TQFT is to provide a ``representation theory for manifolds with boundary''. In more precise terms, a TQFT in dimension $d$ assigns vector spaces to $(d-1)$-dimensional closed and oriented manifolds, and associates linear maps to $d$-dimensional oriented manifolds. The linear map goes from the vector space associated with the incoming boundary (i.e., boundary components whose orientation matches the induced orientation) to the vector space associated with the outgoing boundary. These $d$-dimensional manifolds are referred to as \textit{bordisms}.

\begin{figure}[!ht]
    \centering
    \includegraphics[scale=0.8]{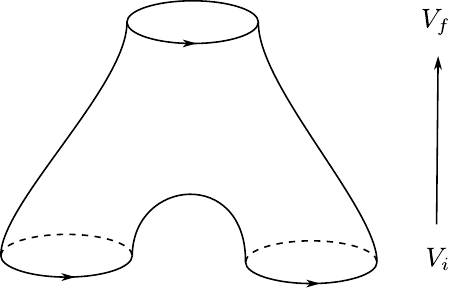}
    \caption{Toy picture of a TQFT.}
    \labelx{fig:TQFT-toy}
\end{figure}    

This assignment should depend solely on the topology of the manifolds, ensuring the composition of linear maps when manifolds are glued together, and must be multiplicative. This means that the vector space associated with the disjoint union of two $(d-1)$-dimensional manifolds is the tensor product of the vector spaces for each individual manifold. In particular, a TQFT assigns a scalar topological invariant to any closed $d$-dimensional manifold.

The Chern--Simons $3d$ TQFT assigns complex lines (i.e., complex vector spaces of dimension 1) to closed 2-dimensional manifolds equipped with a bundle and a flat connection, along with isomorphisms between these complex lines corresponding to bordisms. For a closed 3-dimensional manifold $X$ with connection $\theta$, the invariant is given by the exponentiated Chern--Simons invariant $\mathcal{F}_G(X,\theta)$.

The precise definition of this $3d$ TQFT is somewhat involved. Using $3d$ spectral networks, Freed and Neitzke are able to compute the Chern--Simons invariant for $G=\SL_2(\C)$ by utilizing the spin $\C^*$-Chern--Simons invariant of a double cover of $X$.

\section{Three-dimensional spectral networks and applications}

The basic concept that Freed and Neitzke use is the notion of an \textit{SN-stratification} of the underlying manifold.

\begin{definition}[{\cite[Def. 4.15]{Freed-Neitzke}}] Let $X$ be a manifold with corners, of dimension at most three. An \emph{SN-stratification}\index{spectral network!stratification} of $X$ consists of the following data: 
\begin{enumerate}
\item a smooth manifold of codimension 0, 
\item codimension 1 submanifolds, 
\item two types of codimension 2 submanifolds: \emph{type b}, where $b$ stands for "branch", with the linking manifold being a circle with exactly 3 marked points, and \emph{type a}, with the linking manifold being a circle with any number of marked points, 
\item two types of codimension 3 submanifolds: \emph{type b}, where the linking manifold is a 2-sphere with a stratification induced by a tetrahedron, and \emph{type a}, where the linking manifold is a 2-sphere stratified by trivalent graphs with vertices of type a. 
\end{enumerate}
\end{definition}

Note that the classification into type a or type b is part of the data: for example, in codimension 2, there can be strata of type a with a linking circle that has exactly 3 marked points. The different strata are denoted by $M_{-1}$, $M_{-2a}$, and so on. One defines $M_{\geq -3a} := M \setminus (M_{-2b} \sqcup M_{-3b})$, which represents the part of $M$ excluding the branch locus.

\begin{definition}[{\cite[Def. 4.24 (i)]{Freed-Neitzke}}]
Let $M$ be a compact manifold of dimension at most 3 with boundary, equipped with an SN-stratification such that $M \setminus \partial M = M_0 \sqcup M_{-1} \sqcup M_{-2a} \sqcup M_{-3a} \sqcup M_{-2b} \sqcup M_{-3b}$. A \emph{spectral network $(\pi, s)$ subordinate to the stratification on $M$} consists of the following:
\begin{itemize}
\item A double cover $\pi \colon \widetilde{M}_{\geq 3a} \to M_{\geq 3a}$, which restricts non-trivially to the link of each point in $M_{-2b}$.
\item A section $s$ of $\pi$ over $M_{-1} \sqcup M_{-2a} \sqcup M_{-3a}$. \end{itemize}
\end{definition}

Finally, we can define the data required to abelianize the Chern--Simons invariant. Let $T$ denote a maximal torus in $\SL_2(\C)$, and let $H$ be its normalizer.

\begin{definition}[{\cite[Def. 4.24 (ii)]{Freed-Neitzke}}]
Let $M$ be as in the previous definition. We define the following as \emph{stratified abelianization data}\index{abelianization!stratified data}: 
\begin{itemize} 
\item A principal $\SL_2(\C)$-bundle $P \to M$ with a flat connection, 
\item A principal $H$-bundle $Q \to M_{\geq 3a}$ with a flat connection, and a flat isomorphism $Q \cong P$ over $M_0$, 
\item An isomorphism of double covers $\widetilde{M}_{-3a} \cong Q/T$ over $M_{\geq -3a}$. 
\end{itemize}
\end{definition}

There is a standard SN-stratification associated with a (truncated) tetrahedron. As a result, each triangulated 3-manifold is assigned a standard SN-stratification, which generalizes (some of) the spectral networks that appear in dimension 2 on Riemann surfaces equipped with a holomorphic quadratic differential. The main result of Freed and Neitzke is a global abelianization procedure for the Chern--Simons invariant, expressed as an isomorphism of invertible field theories:

\begin{theorem}[{\cite[Thm. 5.106]{Freed-Neitzke}}] 
Let $M$ be a closed, oriented 3-manifold equipped with stratified abelianization data. Then the $\SL_2(\C)$-Chern--Simons field theory on $M$ is isomorphic to the tensor product of the $\C^*$-spin Chern--Simons theory on $\widetilde{M}$ and a spin $\Z/2\Z$-Chern--Simons theory, which does not depend on the bundle on $M$. 
\end{theorem}

As an application, Freed and Neitzke provide a detailed description of the Chern--Simons TQFT. In particular, they determine the canonical isomorphisms of the complex lines associated with the boundary surfaces and compute the exponentiated Chern--Simons invariant for closed 3-manifolds.

\medskip
In conclusion, 3-dimensional spectral networks provide a method to abelianize the Chern--Simons invariant and may lead to many exciting developments. Several open problems arise from the paper, particularly the following:
\begin{itemize}
\item Generalize 3d spectral networks to encompass the full scope of 2d spectral networks. 
\item Extend the framework to higher-rank Lie groups. 
\item Connect 3d spectral networks to BPS states or the WKB approximation. 
\end{itemize}
\chapter{WKB cameral networks}

\abstract{In this chapter we are interested in generalizing the study of the non-abelianization map for $G$-local systems where $G$ is a general complex reductive Lie group. The first tool to use for this important generalization is that of a cameral cover which effectively generalizes the notion of a spectral cover for our purposes. This leads to a new construction of networks, called WKB cameral networks.}

\vspace{0.5cm}

The non-abelianization map of Gaiotto, Moore and Neitzke \cite{Gaiotto:2012rg} that we have explored in \Cref{sec:non-abel} was generalized from the case of $\mathrm{SL}_n(\mathbb{C})$ and $\mathrm{GL}_n(\mathbb{C})$ to arbitrary complex reductive Lie groups $G$ in the dissertations of Ionita \cite{Ionita} and Morrissey \cite{Morrissey} (cf. their joint article \cite{IM} as well). The notion of a cameral cover, first introduced by Donagi \cite{Donagi93}, provides a generalization of spectral covers and subsequently the Hitchin system for the case of an arbitrary reductive group (\cite{Faltings} and \cite{Scognamillo}). 
The non-abelianization map for a general group $G$ is an algebraic map produced via a ``cutting and re-gluing'' process along a basic spectral network, thus determining a morphism of algebraic stacks from a moduli space of certain local systems with additional structure over an open subset of a cameral cover $\widetilde{X} \to X$ of a non-compact (punctured) Riemann surface $X$ to the moduli space of $G$-local systems over $X$. This map agrees with the map of \cite{Gaiotto:2012rg} when $G=\mathrm{GL}_n(\mathbb{C})$ and $\mathrm{SL}_n(\mathbb{C})$, yet, unlike the construction in \cite{Gaiotto:2012rg}, it does not rely on choices of trivializations for the local systems or for the branched cover $\widetilde{X} \to X$. Furthermore, the non-abelianization map for general $G$-local systems is conjectured to be a local isomorphism; evidence in the case when $G=\mathrm{GL}_n(\mathbb{C})$ was provided in \cite{Gaiotto:2012rg} at the physical level of rigor. This is known to be true for $G=\mathrm{SL}_2(\mathbb{C})$ at the level of coarse moduli spaces due to work from \cite{FG}, \cite{Gaiotto:2009hg}, \cite{HoNe}. Whenever the conjecture holds, the map provides an \'{e}tale coordinate chart on the space of $G$-local systems. 
Our purpose in this short Chapter is to introduce the notion of a cameral cover $\widetilde{X} \to X$ and explain the tool of a WKB cameral network consisting of Stokes curves which are oriented curve segments on $\widetilde{X}$ each labeled by a root of $\mathfrak{g}$. Moreover, we describe how these Stokes curves descend to a set of oriented curves on $X$ resulting to a spectral network on $X$. For the construction of the non-abelianization map for general $G$-local systems, and for evidence supporting the conjecture that this map is a local isomorphism, we direct the reader to the original work of Ionita and Morrissey \cite{Ionita}, \cite{Morrissey} and \cite{IM}.

\section{Cameral covers}

Spectral networks were originally introduced for groups of type A and extended for groups of type D and E using a spectral cover $\widetilde{X} \to X$ over a Riemann surface $X$. In order to expand the theory for more general reductive Lie groups $G$, the primary step is to introduce a branched cover over $X$ that would not depend on the choice of a representation $\rho\colon G \to \mathrm{GL}_n(\mathbb{C})$. This cover is  equipped with a fiberwise action of the Weyl group of $G$ which is free and transitive away from the ramification points. We make this more precise in the sequel.

\vspace{2mm}

Let $G$ be a complex reductive Lie group with corresponding complex Lie algebra $\mathfrak{g}$ and let $\mathcal{E}$ be a principal $G$-bundle over a closed connected and oriented  Riemann surface $X$ of genus $g \ge 2$. Let also $\mathrm{ad}(\mathcal{E}):= \mathcal{E} \times _{G} \mathfrak{g}$ denote the adjoint bundle of $\mathcal{E}$, which is a vector bundle over $X$. Here, the twisted product $\mathcal{E} \times _{G} \mathfrak{g}$ is the quotient of the product bundle $\mathcal{E} \times \mathfrak{g}$ by the equivalence relation given by $(e \cdot g, x) \sim (e, \mathrm{ad}_g(x))$, for every section $e$ of $\mathcal{E}$ and any $g \in G$, $x \in \mathfrak{g}$. 

\begin{definition}
 For $X$ and $G$ as above, fix a line bundle $L$ over $X$. An \emph{$L$-twisted $G$-Higgs bundle}\index{Higgs bundle!L-twisted} over $X$ is a pair $(\mathcal{E}, \varphi)$, where $\mathcal{E}$ is a principal $G$-bundle over $X$ and $\varphi$ is a section in $\Gamma \left( X, \mathrm{ad}(\mathcal{E}) \otimes L\right)$; the section $\varphi$ is called the \emph{Higgs field}.  
\end{definition}

\begin{remark}
For a group representation $\rho\colon G \to \mathrm{GL}(V)$, one may consider the weight space decomposition of $V$ with respect to the maximal torus $T \subset G$ as $V= \bigoplus _{\lambda \in \Lambda} V_{\lambda}$, where $\Lambda$ is a collection of weights of $\rho$ from the lattice of weights of $G$. Now one may consider the spectral covers $\widetilde{X}_{\lambda}$ indexed by $W$-orbits of weights $\lambda$ of the representation $\rho$ in the light of \Cref{def:spectral_curve}, and subsequently consider the resulting spectral cover $\widetilde{X}_{\rho} \to X$ as the union of the sub-covers $\widetilde{X}_{\lambda}$. A natural question would be though how much would such a cover depend on the representation $\rho$. In general, there are infinitely many non-isomorphic such sub-covers $X_{\lambda}$, which lie into a finite number of birational equivalence classes. Donagi has yet constructed a cover $\widetilde{X}$ which dominates all of them. We review this construction next and refer the reader to \cite[Sect. 2]{Donagi93} for the original version and a complete discussion.
\end{remark}

Fix $T \subset G$ a maximal torus of $G$ and let $N$ be the normalizer of $T$ in $G$. Denote by $W \cong N/T$ the Weyl group, and by $\mathfrak{t}$ the corresponding Lie algebra of $T$.  Chevalley's theorem provides that the restriction map 
\[ \mathbb{C}[\mathfrak{g}]^{G} \to \mathbb{C}[\mathfrak{t}]^{W}\]
is an isomorphism from $\mathrm{Ad}$-invariant polynomial functions on the Lie algebra $\mathfrak{g}$ to $W$-invariant polynomial functions on the Cartan subalgebra $\mathfrak{t} \subset \mathfrak{g}$. We then consider the injective ring homomorphism  
\[\mathbb{C}[\mathfrak{t}]^{W} \cong \mathbb{C}[\mathfrak{g}]^{G} \hookrightarrow \mathbb{C}[\mathfrak{g}]\; ,\]
and take the prime spectrum of the rings to define a surjective $G$-invariant morphism of affine varieties
\begin{equation}\labelx{G_inv_morph}
\mathfrak{g} \twoheadrightarrow  \mathfrak{t}/{W}\; .   
\end{equation}
Thus, taking fiber product with the quotient map $\mathfrak{t} \to \mathfrak{t}/W$, we get 
\begin{equation}\labelx{cam_cover_Lie_alg}
\tilde{\mathfrak{g}}:=\mathfrak{g} \times _{\mathfrak{t}/W} \mathfrak{t}   \end{equation}
and now the projection $\pi\colon\tilde{\mathfrak{g}} \to \mathfrak{g}$ is a finite $W$-Galois morphism. This is called the \emph{cameral cover} of the Lie algebra $\mathfrak{g}$. The fiber $\pi^{-1}(g)$ of a regular semisimple element $g \in \mathfrak{g}$ is identified with the set of chambers in $\mathfrak{t}^{*}$. 

Now, let $K_X=T^*X$ denote the canonical line bundle over $X$ and $D$ a reduced effective divisor of finitely many points in $X$. We will also denote by $K_X(D)$  the line bundle $K_X \otimes \mathcal{O}(D)$. The morphism (\ref{G_inv_morph}) above is $G$-invariant and $\mathbb{C}^*$-equivariant, thus, for a given $K_X(D)$-twisted $G$-Higgs bundle $(\mathcal{E}, \varphi)$ over $X$, this morphism can be extended to a morphism 
\[\vert \mathrm{ad}(\mathcal{E}) \otimes K_X(D) \vert \to \vert \mathfrak{t} \otimes K_X(D) \vert /W\; ,\]
where $\vert \cdot \vert $ denotes here the total space of the bundle. Now one may form the fiber product with $\mathfrak{t} \otimes K_X(D)$ as in (\ref{cam_cover_Lie_alg}) and pull-back via the Higgs field $\varphi$ to get the following:

\begin{definition}\labelx{def:cameral_cover}

The \emph{cameral cover}\index{covering!cameral} determined by the $K_X(D)$-twisted $G$-Higgs bundle $(\mathcal{E}, \varphi)$ is given by the map $\pi\colon \widetilde{X} \to X$, for 
\[\widetilde{X}:= \varphi^*\left( \vert \mathrm{ad}(\mathcal{E}) \otimes K_X(D) \vert \times _{\vert \mathfrak{t} \otimes K_X(D) \vert /W} \vert \mathfrak{t} \otimes K_X(D) \vert\right)\; ,\]
and $\pi$ is the projection onto the first factor.
\end{definition}

\noindent More concretely, a point $b$ in the space of sections 
$$\mathcal{B}\left(X,G,K_X(D) \right):=\Gamma\left(X, \mathfrak{t}\otimes K_X(D)/W \right)\; ,$$ called the \emph{Hitchin base}\index{Hitchin!base}, determines a morphism $X \xrightarrow{b} \mathfrak{t}\otimes K_X(D)/W$. Then, the cameral cover $\widetilde{X}_b$ associated to the point $b$ is the fiber product in the diagram 
\vspace{2mm}
\begin{center}
\begin{tikzcd}
\tilde{X}_b \arrow[r, "\tilde{b}"] \arrow[d, "\pi"] & \mathfrak{t}\otimes K_X(D) \arrow[d]\\
X \arrow[r, "b"]  & \mathfrak{t}\otimes K_X(D)/W \; .
\end{tikzcd}
\end{center}
\vspace{2mm}

Whenever the divisor $D$ has at least 3 distinct points, the Hitchin base $\mathcal{B}\left(X,G,K_X(D) \right)$ has a non-empty Zariski open subset $\mathcal{B}^{\diamondsuit}\left(X,G,K_X(D) \right)$, for which $b \in \mathcal{B}^{\diamondsuit}\left(X,G,K_X(D) \right)$ if and only if the cameral cover $\tilde{X}_b$ is smooth. Moreover, for all such points $b \in \mathcal{B}^{\diamondsuit}\left(X,G,K_X(D) \right)$, $\widetilde{X}_b$ is connected. The cameral cover $\pi\colon\widetilde{X}_b \to X$ is a principal $W$-bundle over $X$ away from the ramification locus. For all points $b \in \mathcal{B}^{\diamondsuit}\left(X,G,K_X(D) \right)$, every ramification point of the cameral cover $\pi\colon\widetilde{X}_b \to X$ has order 2; see \cite[Sects. 4.6 and 4.7]{Ngo} for more details on these assertions. Let us denote at this point by $P \subset X$ the branch points and by $R \subset \widetilde{X}_b$ the ramification points of the map $\pi$.

\begin{example}
In the case when $G=\mathrm{GL}_n(\mathbb{C})$, the Weyl group $W$ is the symmetric group $S_{n}$. The cameral cover $\widetilde{X}_b \to X$ is then a degree $n!$-cover of $X$, while the spectral cover $\widetilde{Y}_b$ in this case has degree $n$. The difference is that $\widetilde{Y}_b$ here parameterizes eigenvalues of the characteristic polynomial, whereas $\widetilde{X}_b$ parameterizes orderings of the eigenvalues. When the point $b$ in the Hitchin base is chosen to be generic enough, and letting $S_{n-1}$ be the stabilizer of one of the eigenvalues, then the two kinds of covers above, spectral and cameral, are related by 
\[\widetilde{X}_b/S_{n-1} \cong \widetilde{Y}_b\; .\]  
Note that when $n=2$, then the spectral and cameral covers are isomorphic.
\end{example}

\begin{remark}
Building on the previous example, we note that in general the cameral cover dominates the root and the spectral coverings successively over a base surface (see \cite[Lem. 3.3]{Donagi93} and \Cref{fig:0} in \Cref{sec:analytic_con}). In the cases when $n=2$ or $n=3$ for the group $\mathrm{GL}_n(\mathbb{C})$ though, the root covering agrees with the cameral cover.
\end{remark}

\section{WKB cameral networks} 

In \Cref{sec:analytic_con} we have seen the construction of WKB spectral networks, originally introduced in the case when the group $G$ is $\mathrm{SL}_n(\mathbb{C})$, $\mathrm{GL}_n(\mathbb{C})$ or of type ADE (following \cite{Gaiotto:2012rg} and \cite{LoPa}). Below we summarize the generalization of this construction for the case of an arbitrary reductive group $G$ following the recent work of Ionita and Morrissey from \cite{Ionita}, \cite{Morrissey}, \cite{IM}. 

Let $G$ be a complex reductive Lie group and let $\Delta$ denote the set of roots of the associated Lie algebra~$\mathfrak{g}$. Fixing a Chevalley basis for $\mathfrak{g}$, then each root $\alpha \in \Delta$ determines an $\mathfrak{sl}_2$-triple. Now, let $s_{\alpha} \in W$ denote the reflection with respect to the root hyperplane $H_{\alpha}$ and denote by $p$ the quotient map $N \to T$, for the normalizer $N$ in $G$, a maximal torus $T\subset G$, and the Weyl group $W \cong N/T$. 

The Chevalley basis determines for each root $\alpha \in \Delta$ an element $n_{\alpha} \in N$ and, in fact, $n_{\alpha} \in p^{-1}(s_{\alpha})$. We may now introduce the following: 

\begin{definition}
Let $b \in \mathcal{B}^{\diamondsuit}\left(X,G,K_X(D) \right)$. For any point $d \in D$ we may write
\[b(x)=\left( \sum_{i=1}^{\infty} a_i x^i\right)\D x\; ,\]
where $a_i \in \mathfrak{t}/W$ and $x$ is a local coordinate around $d$. We call a \emph{residue} of $b$ at $d$ to be any lift $\tilde{a}_{-1} \in \mathfrak{t}$ of $a_{-1}$; these do not depend on the local coordinate $x$ chosen around the point $d$. We also say that $b$ \emph{satisfies condition $\textbf{R}$}, if for every point $d \in D$ it is 
\[\tilde{a}_{-1} \in \mathfrak{t} \setminus \bigcup_{a \in \Delta} H_a\; .\]
\end{definition}
Note that condition \textbf{R} above is $W$-invariant, thus if one residue satisfies it, then all residues do. Moreover, condition $R$ suggests that the cameral cover $\pi$ is unramified at the divisor $D$. 

Denote by  $\mathcal{B}_{R}^{\diamondsuit}\left(X,G,K_X(D) \right)$ the subset of $\mathcal{B}^{\diamondsuit}\left(X,G,K_X(D) \right)$ consisted of the points $b$ which satisfy condition \textbf{R}; this is a Zariski open subset. One may now associate certain meromorphic quadratic differentials to such points $b$:

\begin{proposition}[\cite{Ionita}, \cite{Morrissey}, \cite{IM}]
Let $b \in \mathcal{B}_{R}^{\diamondsuit}\left(X,G,K_X(D) \right)$ and let $\pi\colon \widetilde{X}_b \to X$ be the associated smooth cameral cover of $X$. For each root $\alpha \in \Delta$, let $\widetilde{X}_{b, \alpha} := \widetilde{X}_b / \langle s_{\alpha} \rangle$, and consider the decomposition of the map $\pi$ as 
\[\widetilde{X}_b \xrightarrow{\pi_{\alpha}} \widetilde{X}_{b, \alpha} \xrightarrow{p_{\alpha}} X\; .\]
Then $b$ determines a meromorphic quadratic differential on $\widetilde{X}_{b, \alpha}$ having simple zeroes at the branch points of $\pi_{\alpha}$ and double poles at pre-images of $D$. 
\end{proposition}

Denote the differentials obtained from this result by 
\[\omega_{b,\alpha} \in \Gamma \left( \widetilde{X}_{b, \alpha}, \left(K_{\widetilde{X}_{b, \alpha}} (p_{\alpha}^{*}D) \right)^{\otimes 2} \right)\]
and denote by 
\[\chi_{b, \alpha} \in \Gamma \left( \widetilde{X}_{b}, \left(K_{\widetilde{X}_{b, \alpha}} (p_{\alpha}^{*}D) \right)\right)\] 
the linear differentials on $\widetilde{X}_b$ obtained by composition of the following maps
\[\widetilde{X}_b \xrightarrow{\widetilde{b}} \mathfrak{t}_{\pi^*K_X(D)} \xrightarrow{} \mathfrak{t}_{\pi^*_{\alpha}(K_{\widetilde{X}_{b, \alpha}}(p_{\alpha}^*D))} \xrightarrow{\alpha} \pi^*_{\alpha}(K_{\widetilde{X}_{b, \alpha}}(p_{\alpha}^*D))\; .\]
These linear differentials each determines an oriented foliation on $\widetilde{X}_b$ by curves $\gamma\colon \mathbb{R} \to \widetilde{X}_b$ that satisfy
\[\int_{t_0}^{t_1} \chi_{b, \alpha} (\gamma(s)) \in \mathbb{R}^+\; .\]
We call \emph{oriented trajectories} of the differential $\chi_{b, \alpha}$, the maximal leaves of this foliation. We are finally ready to give the construction of Stokes curves in this general reductive group $G$ setting:

\begin{definition}[\cite{Ionita}, \cite{Morrissey}, \cite{IM}]
For a point $b \in \mathcal{B}_{R}^{\diamondsuit}\left(X,G,K_X(D) \right)$, we define the \emph{WKB construction} $\widetilde{\mathcal{W}}_b$ consisted of oriented curve segments on the cameral cover $\widetilde{X}_b$ called \emph{Stokes curves}\index{Stokes!curves}. The \emph{Stokes curves} are labeled by roots in $\Delta$ and are constructed algorithmically by the following procedure:
\begin{enumerate}
\item A \emph{primary Stokes curve} is a critical oriented trajectory of one of the linear differentials $\chi_{b, \alpha}$ on $\widetilde{X}_b$. There are exactly six primary Stokes curves starting from each ramification point of the map $\pi$. 
\item Let $x \in \widetilde{X}_b$ be an isolated intersection point of Stokes curves labeled by distinct roots in some convex set of roots $C_{in} \subset \Delta$. Then, for each root $\beta \in \mathrm{Conv}^{\mathbb{N}}_{C_{in}}$ in the restricted convex hull of the subset $C_{in}$, a secondary Stokes curve $g_{\beta}$ starts at the intersection point $x$. This secondary Stokes curve is the unique leaf from $x$ of the oriented foliation determined by $\chi_{b, \beta}$. 
\item The secondary Stokes curves are recursively created each time two or more of the existing Stokes curves intersect. See \Cref{G-Stokes_curves}. 
\end{enumerate}
We denote by $J$ the collection of intersection points of Stokes curves called the \emph{joints} of the WKB construction. 
\end{definition}

\begin{figure}[!ht]
    \centering    \includegraphics{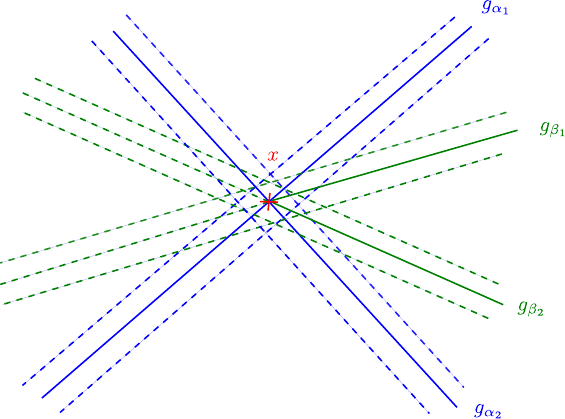}
    \caption{Stokes curves (solid lines) are segments of certain oriented trajectories (transparent lines). Primary Stokes curves $g_{\alpha_1}$ and $g_{\alpha_2}$ intersect at the point $x$. Then, secondary Stokes curves $g_{\beta_1}$ and $g_{\beta_2}$ are emanating from $x$.}
    \labelx{G-Stokes_curves}
\end{figure}

The WKB construction $\widetilde{\mathcal{W}}_b$ can be thought of as a directed graph with vertex set $R \cup J$ and a directed edge for each connected component of $\widetilde{\mathcal{W}}_b \setminus \{R \cup J\}$. In that sense we set the following:
\begin{definition}
We call $\widetilde{\mathcal{W}}_b$ \emph{admissible} if the following conditions are satisfied:
\begin{itemize}
    \item No point of the cameral cover $\widetilde{X}_b \setminus \pi^{-1}(D)$ is an accumulation point of the set of joints $J$.
    \item Stokes curves do not intersect at ramification points of $\widetilde{X}_b$, in other words, it is $R \cap J = \varnothing$.
    \item The directed graph corresponding to $\widetilde{\mathcal{W}}_b$ is acyclic.
    \item For every joint, the set of roots $C_{in} \subset \Delta$ of labels of the incoming Stokes curves is convex. 
\end{itemize}
\end{definition}
Next we give the definition of a WKB cameral network due to Ionita and Morrissey:

\begin{definition}[\cite{Ionita}, \cite{Morrissey}, \cite{IM}]
An admissible WKB construction $\widetilde{\mathcal{W}}_b$ is called a \emph{WKB cameral network}\index{WKB!cameral network} if there exists a subset $X' \subset X \setminus D$ such that the inclusion map $X' \hookrightarrow X \setminus D$ is a homotopy equivalence, and the restriction of $\widetilde{\mathcal{W}}_b$ to  $\widetilde{X}'_{b}:=\pi^{-1}(X') \subset \widetilde{X}_b$ consists of finitely many Stokes curves. 
\end{definition}

An important conjecture by Ionita and Morrissey asserts that the set of points $b \in \mathcal{B}_{R}^{\diamondsuit}\left(X,G,K_X(D) \right)$ for which the WKB construction  $\widetilde{\mathcal{W}}_b$ is admissible is open and dense in the classical topology. In fact, it is true that any admissible $\widetilde{\mathcal{W}}_b$ is a WKB cameral network, that is, the admissibility condition provides that a WKB construction as defined above always has finitely many Stokes curves away from a neighborhood of $\pi^{-1}(D)$.

Lastly, from a WKB cameral network $\widetilde{\mathcal{W}}_b$ on a cameral cover $\pi\colon \widetilde{X}_b \to X$ there is an associated \emph{WKB spectral network} $\mathcal{W}_b$ on $X$ defined as a union of oriented labeled curves on $X$ by $\mathcal{W}_b=\pi_{*}\widetilde{\mathcal{W}}_b$; the label of the Stokes curve $g \in \mathcal{W}_b$ is a locally constant section $\psi_{g}$ of $\Hom_W(\widetilde{X}_b \vert _{g}, \Delta)$ mapping any preimage $\widetilde{g}$ of the curve $g$ to the root labeling $\widetilde{g}$.  

For a WKB cameral network, there are six Stokes curves starting from a ramification point $r \in R$ of the covering map $\pi\colon \widetilde{X}_b \to X$. The restriction of the map $\pi$ to a small neighborhood of the point $r$ has degree 2, and, since the WKB cameral networks are $W$-equivariant, it follows that the corresponding spectral network has three Stokes curves starting from each branch point. This phenomenon agrees with the constructions of spectral networks from \cite{Gaiotto:2012rg} and \cite{LoPa}.
\chapter{Non-abelianization for conformal Virasoro blocks}

\abstract{We briefly include here a recent application of spectral networks by Hao and Neitzke \cite{HN_Virasoro}, namely the non-abelianization of conformal blocks. Hao and Neitzke produce a passage between Virasoro conformal blocks at $c=1$ on a Riemann surface $C$ and Heisenberg conformal blocks over a double cover of $C$. The construction builds upon Zamolodchikov's branched free-field construction \cite{Zam1987}, yet the normalized correlation functions involved now are everywhere regular, thus no additional insertions of Virasoro primary fields have to be considered.}

\section{Virasoro and Heisenberg conformal blocks}

One way to approach conformal blocks abstractly is by considering them as systems of correlation functions satisfying chiral Ward identities (see for instance  \cite{FrBZ}, \cite{Teschner}). This approach allows one to introduce the space of conformal blocks $\mathrm{Conf}(C, \mathcal{V})$ as a canonically defined vector space that involves no arbitrary choices such as decompositions in pairs of pants: it only depends on the data of a vertex algebra $\mathcal{V}$ and a Riemann surface $C$, with the specification of primary fields inserted at punctures (if any). We will only consider the cases of Virasoro and the Heisenberg vertex algebras.

For a fixed constant $c \in \mathbb{C}$, the \emph{Virasoro vertex algebra}\index{Virasoro vertex algebra} $\mathrm{Vir}_c$ is generated by a single field $T$ satisfying the operator product expansion (OPE): \index{Operator product expansion}
\begin{equation}\labelx{Vir_op_prod}
T(p)T(q)=\frac{c/2}{(z(p)-z(q))^4}+\frac{2T(q)}{(z(p)-z(q))^2}+\frac{\partial_{z(q)}T(q)}{z(p)-z(q)}+\text{ regular terms}\; .
\end{equation}
The primary fields $W_h$, for $h \in \mathbb{C}$, are defined by the condition:
\[T(p)W_h(q)= \frac{hW_h(q)}{(z(p)-z(q))^2}+\frac{\mathrm{reg}}{z(p)-z(q)}\; .\]

The \emph{Heisenberg vertex algebra}\index{Heisenberg vertex algebra} $\mathrm{Heis}$ is generated by a single field $J$ with OPE:
\begin{equation}\labelx{Heis_OPE_rel}J(p)J(q)= \frac{1}{(z(p)-z(q))^2}+\text{ regular terms}\; .
 \end{equation}
The primary fields $V_{\alpha}$, for $\alpha \in \mathbb{C}$, are defined by the condition:
\begin{equation}\labelx{Heis_condJV}
J(p)V_{\alpha}(q)=\frac{\alpha V_{\alpha}(q)}{z(p)-z(q)}+\text{ regular terms}\; ,
\end{equation}
and one can check that the Heisenberg vertex algebra contains a Virasoro vertex algebra of central charge $c=1$ (see \cite[Ex. 2.5.9]{FrBZ}). 

Let now $C$ be a Riemann surface covered by a complex atlas $(U_i,z_i)$, and a set of points $p_1,...,p_n \in C$, for some fixed $n \in \mathbb{N}$, with $p_i \in U_i$. A \emph{Heisenberg conformal block}\index{conformal!block} $\Psi \in \mathrm{Conf}(C, \mathrm{Heis})$ is defined as a system of correlation functions given for every $n$ by: 
\[\langle J(p_1)^{z_1} \cdots J(p_n)^{z_n} \rangle _{\Psi}\colon U_1 \times U_2 \times \cdots \times U_n \to \mathbb{C}\; ,\]
which are holomorphic away from the diagonals $p_i=p_j$ with the singularities at the diagonal governed by (\ref{Heis_OPE_rel}). Now take the vertex algebra $\mathrm{Vir}_{c=1} \otimes \mathrm{Heis}$ and define similarly a \emph{$c=1$ Virasoro conformal block} $\Psi \in \mathrm{Conf}(C, \mathrm{Vir}_{c=1} \otimes \mathrm{Heis})$, as a system of correlation functions:
\[\langle T(p_1)^{z_1} \cdots T(p_n)^{z_n} J(q_1)^{w_1} \cdots J(q_m)^{w_m} \rangle _{\Psi}\colon U_1 \times U_2 \times \cdots \times U_n \to \mathbb{C}\; ,\]
which are holomorphic away from the points $p_i = p_j$ or $q_i = q_j$, obey the OPEs (\ref{Vir_op_prod}) and (\ref{Heis_OPE_rel}), as well as certain coordinate transformation rules.

The space of conformal blocks with primary fields inserted $\mathrm{Conf}(C, \mathrm{Heis}; V_{\alpha_1}(q_1) \cdots V_{\alpha_k}(q_k))$ is defined as the space of systems of correlated functions: 
\[\langle J(p_1)^{z_1} \cdots J(p_n)^{z_n}V_{\alpha_1}(q_1) \cdots V_{\alpha_k}(q_k) \rangle _{\Psi}\colon U_1 \times U_2 \times \cdots \times U_n \to \mathbb{C}\; ,\]
with the same OPE and coordinate transformations for the insertions $J$ as before, but now with extra first-order poles when any $p_i$ meets any $q_j$, as dictated by (\ref{Heis_condJV}).

\section{The non-abelianization map}

Virasoro blocks are generally difficult to compute; however, one can use specific recursion relations (see \cite{Zam}, \cite{CCY2019}) or the representations provided by the AGT correspondence \cite{AGT2010}. Another possible approach for computing them is to reduce the problem to Heisenberg conformal blocks, which are easier to obtain. This pertains to the so-called \emph{free-field construction} producing Virasoro blocks of a very special kind. Namely, for a constant $Q \in \mathbb{C}$, substituting the expression for $T= \frac{1}{2}J^2+Q\partial J$ in the Heisenberg correlation functions provides a map
\[\mathrm{Conf}(C, \mathrm{Heis}) \to \mathrm{Conf}(C, \mathrm{Vir}_{c=1+6Q^2})\]
between conformal blocks. A variation of this technique when $c=1$, called the \emph{branched free-field construction}, was given by Zamolodchikov in \cite{Zam1987}. For this construction, one considers a branched double cover $\pi\colon \tilde{C} \to C$, and Heisenberg fields $\tilde{J}$ over $\tilde{C}$. Let
\[\tilde{J}^{(-)}:=\frac{1}{\sqrt{2}}(\tilde{J}^{(1)}-\tilde{J}^{(2)})\]
denote the anti-invariant combination of insertions on the two sheets of $\tilde{C}$, which gives a well-defined operator on $C$ up to the $\mathbb{Z}_2$-action $\tilde{J}^{(-)} \to -\tilde{J}^{(-)}$. Then, substituting $T:=\frac{1}{2}(\tilde{J}^{(-)})^2$ in the Heisenberg correlation functions, provides Virasoro correlators on $C$, and the induced Virasoro blocks have additional singularities at the branch points $b_1, ..., b_k$ of the covering. These singularities can be viewed as insertions of Virasoro primary fields $W_h(b_i)$ with weight $h=\frac{1}{16}$, and so there is a well-defined map
\[\mathrm{Conf}(\tilde{C}, \mathrm{Heis}) \to \mathrm{Conf}(C, \mathrm{Vir}_{c=1}; W_{\frac{1}{16}}(b_1) \cdots W_{\frac{1}{16}}(b_k))\; .\]

Spectral networks can be used in order to modify the branched free-field construction outlined above, in such a way that the $W_{\frac{1}{16}}$ insertions at the branch points do not appear, and moreover no extra singularities are created elsewhere. This can be indeed achieved by inserting an extra operator in the correlation functions on $\tilde{C}$. We summarize below this construction of Hao and Neitzke from \cite{HN_Virasoro}.

For a smooth double cover $\pi\colon \tilde{C} \to C$, fix a generic choice of a $\mathfrak{gl}_2$-spectral network $\mathcal{W}$ subordinate to $\pi$ with three arcs ending on each branch point of the covering $\pi$, and a spin structure on $C$. Then, consider the spectral network operator
\[E(\mathcal{W})= \mathrm{exp} \left[ \frac{1}{2\pi \I } \int_{\mathcal{W}}\psi_{+}(z^{(+)})\psi_{-}(z^{(-)})\D z\right]\; ,\]
where $\psi_{\pm}$ are the free fermions built out of the Heisenberg field $\tilde{J}$ via fermionization, and $z^{(+)}, z^{(-)} \in \tilde{C}$ are the two pre-images of a point $z \in C$ on the double cover $\tilde{C}$. 

It is then computed in \cite{HN_Virasoro} that with $E(\mathcal{W})$ inserted in the correlation functions on $\tilde{C}$, the normalized correlation functions on $T(z)$ are everywhere regular, even at the branch points. Therefore, the insertion of $E(\mathcal{W})$ removes the unwanted insertions $W_{\frac{1}{16}}$ at the branch points and does not create any further singularities. Moreover, because the 0-point function with $E(\mathcal{W})$ inserted is divergent, the operator $E(\mathcal{W})$ is replaced by a renormalized one $E_{\text{ren}}(\mathcal{W})$ to render finite the 0-point function. This renormalization introduces an anomalous dependence on a local coordinate near each branch point with weight $-\frac{1}{16}$ so as to cancel the insertion of weight $\frac{1}{16}$ at these points. 

In conclusion, there is a well-defined linear map between spaces of conformal blocks
\begin{equation}\labelx{map_conf_bl}
\mathcal{F}_{\mathcal{W}}\colon \mathrm{Conf}(\tilde{C}, \mathrm{Heis}) \to \mathrm{Conf}(C, \mathrm{Vir}_{c=1} \otimes \mathrm{Heis})\; ,
\end{equation}
called the \emph{non-abelianization map for conformal blocks}\index{non-abelianization!conformal blocks}, and, for a Heisenberg block $\tilde{\Psi}$ over $\tilde{C}$, the 1-point function of the Virasoro generator in the corresponding block $\mathcal{F}_{\mathcal{W}}(\tilde{\Psi})$ is given by 
\[ \langle T(z) \rangle _{\mathcal{F}_{\mathcal{W}}(\tilde{\Psi})} = \frac{1}{4}  \langle \tilde{J}(z^{(1)}) -\tilde{J}(z^{(2)})^2 E_{\mathrm{ren}}(\mathcal{W})\rangle _{\tilde{\Psi}}\; .\]

There is an action on the space of conformal blocks 
$\mathrm{Conf}(\tilde{C}, \mathrm{Heis})$ by loop operators $L_{\gamma}$, labeled by loops $\gamma$ on $\tilde{C}$. These loop operators generate the commutative skein algebra $\mathrm{Sk}_{-1}(\tilde{C}, \mathrm{GL}(1))$, and the algebra of loop operators is dually the algebra of functions on the moduli space of (twisted) flat connections $\mathcal{M}(\tilde{C}, \mathrm{GL}(1))$. Now, for a point $X$ of $\mathcal{M}(\tilde{C}, \mathrm{GL}(1))$, take its decomposition with respect to a fixed basis of $A$ and $B$ cycles on $\tilde{C}$ as $X=(\E^x,\E^y)$. In general, we can construct a collection of linearly independent Heisenberg blocks parameterized by $\tilde{g}$ continuous parameters 
\[\tilde{\Psi}_a \in \mathrm{Conf}(\tilde{C}, \mathrm{Heis})\; ,\]
for $a=(a_1,..., a_{\tilde{g}}) \in \mathbb{C}^{\tilde{g}}$ (see \cite[Sect. 6]{HN_Virasoro}). In particular, we write an eigenblock $\tilde{\Psi}_{x,y} \in \mathrm{Conf}(\tilde{C}, \mathrm{Heis})$ as a linear combination of the $\tilde{\Psi}_a$:
\[\tilde{\Psi}_{x,y}= \sum_{n\in \mathbb{Z}^{\tilde{g}}}\mathrm{exp}\left( -\frac{(x+2\pi \I n)\cdot y}{2 \pi \I}\right)\tilde{\Psi}_{a=x+2 \pi \I n}\; .\]
In order to construct eigenblocks $\Psi \in \mathrm{Conf}(C, \mathrm{Vir}_{c=1} \otimes \mathrm{Heis})$ one uses the covariance of the map $\mathcal{F}_{\mathcal{W}}$ from (\ref{map_conf_bl}) with respect to the loop operators. In this way, we get a family of eigenblocks
\[\Psi_{x,y}^{\mathcal{W}}:= \mathcal{F}_{\mathcal{W}}(\tilde{\Psi}_{x,y})\; .\]

Translating the abstract language of conformal blocks to free fermionic field theory, correlation functions on $\tilde{C}$ in the eigenblocks $\tilde{\Psi}_{x,y}$ should be understood as related to the theory of a chiral free fermion on $\tilde{C}$, twisted by a complex background gauge field on $\tilde{C}$ with holonomies $(\E^x, \E^y)$. On the other hand, correlation functions on $C$ in the eigenblocks $\Psi_{x,y}^{\mathcal{W}}$ should likewise have to do with the theory of 2 chiral free fermions on $C$, twisted by a complex background gauge field on $C$ with holonomies $\lambda = \mathcal{F}_{\mathcal{W}}^{\flat}((\E^x, \E^y))$. From this point of view, the non-abelianization map $\mathcal{F}_{\mathcal{W}}$ from (\ref{map_conf_bl}) outlines a passage between these two field theories, namely, the two fermion determinants are not exactly equal but they do become equal up to a constant after inserting the operator $E_{\mathrm{ren}}(\mathcal{W})$ in the theory on $C$.

\appendix
\newpage
\part*{Appendix}\markboth{}{}\labelx{partVII}
\addcontentsline{toc}{part}{Appendix}%


\chapter{Lie theory essentials}\labelx{app:Lietheory}

\abstract{In this Appendix we collect for the reader's convenience essential notions and results from Lie theory supporting deeper Lie theoretic aspects that appear in the main body of the manuscript. We focus in particular on the classification of complex reductive Lie groups, root systems, real Lie groups, Levi decomposition and parabolic subalgebras. For instance, the classification by Dynkin diagrams and the description of the roots in each case is useful in \Cref{sec:positivity} for introducing the notions of a positive structure on a Lie algebra and of a positive representation. Standard comprehensive references for further reading include \cite{Helgason} or \cite{Knapp}.}

\section{Complex Lie groups and their classification}

\subsection{Complex reductive Lie algebras}

\begin{definition}
Let $\mathfrak{g}$ be a complex finite-dimensional Lie algebra. We call $\mathfrak{g}$ \emph{reductive}, if for each ideal $\mathfrak{a}$ in $\mathfrak{g}$, there is an ideal $\mathfrak{b}$ in $\mathfrak{g}$ such that $\mathfrak{g}=\mathfrak{a} \oplus \mathfrak{b}$. 

\noindent The \emph{radical} of $\mathfrak{g}$, denoted by $\mathrm{rad}\,\mathfrak{g}$, is defined as the maximal solvable ideal in $\mathfrak{g}$.
\end{definition}

The center of a complex reductive Lie algebra $\mathfrak{g}$ is isomorphic to its radical $\mathrm{rad}\,\mathfrak{g}$. Moreover, one has the decomposition 
\[\mathfrak{g}=\mathrm{rad}\,\mathfrak{g} \oplus [\mathfrak{g},\mathfrak{g}]\; ,\]
where $[\mathfrak{g},\mathfrak{g}]$ is the semisimple part of $\mathfrak{g}$.

The adjoint representation of a complex semi-simple Lie algebra $\mathfrak{g}$ is the Lie algebra representation $\mathrm{ad}\colon \mathfrak{g} \to \mathrm{End}(\mathfrak{g})$ defined by the Lie bracket 
\[\mathrm{ad}_x(y)=[x,y]\; .\] 
There exists a maximal abelian subalgebra $\mathfrak{t} \subset \mathfrak{g}$ such that the adjoint action on $\mathfrak{g}$ is diagonal. The latter means that the adjoint representation $\mathrm{ad}\vert_{\mathfrak{t}}$ decomposes the Lie algebra $\mathfrak{g}$ into a direct sum of irreducible representations called \textit{weight spaces}. The subalgebra $\mathfrak{t}=\mathfrak{g}_{0}$ corresponding to weight zero is called the \textit{Cartan subalgebra}  of $\mathfrak{g}$  and any two Cartan subalgebras are conjugate by an automorphism of $\mathfrak{g}$. Thus all Cartan subalgebras have the same dimension; we define the rank of the semisimple complex Lie algebra $\mathfrak{g}$ to be the common value of this dimension:
\[\mathrm{rk}_\mathbb{C}(\mathfrak{g})=\mathrm{dim}(\mathfrak{t})\; .\]

Now, let $\mathfrak{t} \subset \mathfrak{g}$ be a choice of a maximal abelian subalgebra of $\mathfrak{g}$, up to an inner automorphism of $\mathfrak{g}$. Note that $\mathfrak{t}\supset \mathrm{rad}\,\mathfrak{g}$. The adjoint representation
\[\mathrm{ad}\colon  \mathfrak{g} \to \mathrm{End}(\mathfrak{g})\]
restricted to $\mathfrak{t}$ is diagonalizable and one gets the decomposition 
\[\mathfrak{g} = \mathfrak{t} \oplus \bigoplus\limits_{\alpha \in R}^{}{}\mathfrak{g}_{\alpha} \; ,\]
for $R \subset \mathfrak{t}^*$, the \emph{set of roots} of $\mathfrak{g}$. The space $\mathfrak{g}_{\alpha}$, for each $\alpha$, is the eigenspace for the root $\alpha \in R$, that is, $X \in \mathfrak{g}_{\alpha}$ if and only if 
\[[h, X]=\alpha(h)X \; ,\] 
for every $h \in \mathfrak{t}$. 

The \emph{Killing form} $K\colon \mathfrak{g} \times \mathfrak{g} \to \mathbb{C}$ is defined as the complex bilinear form on $\mathfrak{g}$ given by 
\[K(X,Y)= \mathrm{Tr}(\mathrm{ad}X, \mathrm{ad}Y) \; .\]
When restricted to the semisimple part $[\mathfrak{g}, \mathfrak{g}]$ of $\mathfrak{g}$, the Killing form is non-degenerate, thus allowing to introduce the following:

\begin{definition}
Let $\mathfrak{g}$ be a complex reductive Lie group and let $\mathfrak{t} \subset \mathfrak{g}$ be a maximal abelian subalgebra. For every root $\alpha \in \mathfrak{t}^*$, a \emph{coroot} $h_{\alpha} \in \mathfrak{t}$ is the unique element in $[\mathfrak{g}, \mathfrak{g}] \cap \mathfrak{t}$ satisfying
\[\alpha = 2 \frac{K(-,h_{\alpha})}{K(h_{\alpha}, h_{\alpha})} \; .\]
We denote by $C \subset \mathfrak{t}$, the set of all coroots of $\mathfrak{g}$. 
\end{definition}

\begin{remark}
The coroots $C$ span $\mathfrak{t}$ as a complex vector space if and only if $\mathfrak{g}=[\mathfrak{g},\mathfrak{g}]$, that is, if $\mathfrak{g}$ is a semisimple complex Lie algebra. 
\end{remark}

An implication of the decomposition into weight spaces is that for a choice of a linear function $l\colon  \mathfrak{t}^{*} \to \mathbb{R}$ so that $l(\Delta)\neq 0$, there is a splitting of the root space $\Delta$ into positive and negative roots, namely, 
\[\Delta = \Delta_{+} \cup \Delta_{-} \; ,\]
where $\Delta_{+}:=\{\alpha \in \Delta \mid l(\alpha)>0\}$ and $\Delta_{-}$ is defined analogously. We call a positive root $\alpha \in \Delta_{+}$ \textit{simple} if it can not be written as a linear combination with positive coefficients of elements of $\Delta_{+}$. The set of simple roots will be denoted by $\Pi=\{\alpha_{1}, ..., \alpha_{n}\}$, for $n$ the rank of $\mathfrak{g}$  and it forms a basis for the Lie algebra $\mathfrak{t}^{*}$. 

An element $e\in \mathfrak{g}$ is called \textit{nilpotent} if $\mathrm{ad}_e$ is nilpotent as a linear map on $\mathfrak{g}$. By the Morozov--Jacobson theorem, any nilpotent element embeds in a 3-dimensional simple subalgebra of $\mathfrak{g}$, that is, a subalgebra isomorphic to the Lie algebra $\mathfrak{sl}_2$ (see \cite[Chap. VIII, Sect. 11, Prop. 2]{Bourbaki_Lie} for a proof). Indeed there exists another nilpotent element $\tilde{e} \in \mathfrak{g}$ such that   
\begin{equation}\labelx{Lie_algebra_sl2}
\left<e,\tilde{e},h\right>\cong \mathfrak{sl}_2(\mathbb{C}) \; ,
\end{equation}
subject to the relations $[h,e]=e$, $[h,\tilde{e}]=-\tilde{e}$, $[e,\tilde{e}]=h$.
A nilpotent element is, in particular, called a \textit{regular nilpotent} or \textit{principal nilpotent} whenever
\[e=\sum_{\alpha\in \Delta^{+}}c_{\alpha}e_{\alpha} \; ,\]
for positive integer coefficients $c_{\alpha}$ and for the generators $e_{\alpha}\in \mathfrak{g}_{\alpha}$, $\tilde{e}_{\alpha}\in \mathfrak{g}_{-\alpha}$, with $\alpha \in \Delta_{+}$. 

\begin{definition}\labelx{principal_sl2}
A subalgebra of a complex semisimple Lie algebra $\mathfrak{g}$ is called a \textit{principal $\mathfrak{sl}_2$-subalgebra} if it is isomorphic to $\mathfrak{sl}_2$ as in (\ref{Lie_algebra_sl2}) containing a regular nilpotent element $e$ of $\mathfrak{g}$. There is a unique conjugacy class of these subalgebras, thus we sometimes refer to \emph{the principal $\mathfrak{sl}_2$-subalgebra of $\mathfrak{g}$}.
\end{definition}
We next define:

\begin{definition}  
Let $\mathfrak{g}$ be a complex semisimple Lie algebra, $\mathfrak{t}$ a Cartan subalgebra of $\mathfrak{g}$ and $\Delta_{+}$ the set of positive roots. The standard Borel subalgebra with respect to the choice of a Cartan subalgebra and a set positive roots is defined as the subalgebra
\[\mathfrak{b}=\mathfrak{t}\oplus \bigoplus_{\alpha \in \Delta_{+}} \mathfrak{g}_{\alpha} \; .\]
The standard Borel subalgebra is also called a minimal parabolic subalgebra, and we call a  standard parabolic subalgebra any subalgebra containing $\mathfrak{b}$.   
\end{definition}

\subsection{Characters and cocharacters}

Analogously to the definitions we gave for the complex Lie algebras, we have the following:

\begin{definition}
A complex Lie group $G$ is called \emph{reductive} if it is connected and its radical is a torus, that is, the radical is a complex Lie group isomorphic to a direct product of copies of the multiplicative group $\mathbb{C}^{\times}=\mathbb{C}\setminus\{0\}$. 
\end{definition}

For a complex reductive Lie group $G$ and for a maximal torus $T \subset G$ in $G$, the corresponding Lie algebra $\mathfrak{t}$ of $T$ is maximal abelian in $\mathfrak{g}$. Thus, there is  a corresponding set of roots $R \subset \mathfrak{t}^*$ and a set of coroots $C \subset \mathfrak{t}$ relative to $T$. We have the following:

\begin{definition}
For $G$ and $T$ as above, the \emph{character lattice} of $T \subset G$ is defined as the free abelian group 
\[\Lambda^{\perp}= \mathrm{Hom}(T, \mathbb{C}^{\times})\]
of morphisms between complex Lie groups. 
\end{definition}

The exponential map $\mathrm{exp}\colon  \mathbb{C} \to \mathbb{C}^{\times}$, $z \mapsto \E^z$ provides the standard identification between the Lie algebra of $\mathbb{C}^{\times}$ and $\mathbb{C}$. Thus, there is an inclusion
\[\Lambda^{\perp}:=\mathrm{Hom}(T, \mathbb{C}^{\times}) \subset \mathrm{Hom}(\mathfrak{t}, \mathbb{C})= \mathfrak{t}^* \; .\]
Then, by definition, we have $R \subset \Lambda^{\perp}$, and so the following:

\begin{definition}
For $G$ and $T$ as above, the \emph{cocharacter lattice} of $T \subset G$ is defined as the free abelian group 
\[\Lambda = \mathrm{Hom}(\mathbb{C}^{\times}, T) \subset \mathrm{Hom}(\mathbb{C}, \mathfrak{t})= \mathfrak{t} \; .\]
Note that all coroots of $\mathfrak{g}$ lie in the cocharacter lattice, i.e. $C \subset \Lambda$. 

\end{definition}

Remember the natural pairing between a vector space $\mathfrak{t}$ and its dual $\mathfrak{t}^*$. There is a canonical pairing between the lattices $\Lambda$ and $\Lambda^{\perp}$ 
\begin{equation}\labelx{pairing_lattices}
\langle \cdot, \cdot \rangle\colon  \Lambda \times \Lambda^{\perp} \to \mathbb{Z}
\end{equation}
defined by composition
\[\mathrm{Hom}(\mathbb{C}^{\times}, T) \times \mathrm{Hom}(T, \mathbb{C}^{\times}) \to \mathrm{Hom}(\mathbb{C}^{\times}, \mathbb{C}^{\times})=\mathbb{Z} \; ,\]
which agrees with the pairing between a vector space and its dual. The pairing (\ref{pairing_lattices}) is perfect, thus $\Lambda^{\perp}$ is the dual lattice of $\Lambda$. 

Note also that the pairing (\ref{pairing_lattices}) extends to $\mathbb{Q}$ by tensoring $\Lambda$ with $\mathbb{Q}$. We will call \emph{(rational) weight}, an element $\theta \in \Lambda \otimes_{\mathbb{Z}} \mathbb{Q}$. Under differentiation, $\theta$ can be regarded as a point in the Lie algebra $\mathfrak{t}$ of the maximal torus $T$. Similarly, we call \emph{(rational) coweight}, an element in $ \Lambda^{\perp} \otimes_{\mathbb{Z}} \mathbb{Q}$.

\begin{remark}
Let $\{t_i\}$ be the generators of the lattice $\Lambda$. By an abuse of notation, we may regard the collection $\{t_i\}$ as a basis for the Lie algebra $\mathfrak{t}$. Then, a rational weight $\theta$ (regarded as the corresponding element in $\mathfrak{t}$) may be written as $\theta = \sum \frac{a_i}{d_i}t_i$.
\end{remark}
 
\subsection{Root datum and its Langlands dual}

Following the introduction of the character and the cocharacter lattice, we now have the following:

\begin{definition}
Let $G$ be a connected complex Lie group with a choice of a maximal torus $T \subset G$. The \emph{root datum} of $G$ is the data:
\[\mathcal{R}(G,T):=\{C \subset \Lambda, R \subset \Lambda^{\perp}\} \; ,\]
which implicitly also induces the pairing (\ref{pairing_lattices}) and the bijection $R \to C$, $\alpha \mapsto h_{\alpha}$ between roots and coroots of $\mathfrak{g}$. 
\end{definition}
The notion of root datum was introduced by Demazure \cite{Demazure} as a tool to classify the complex reductive Lie groups. Indeed, a root datum determines the reductive Lie group $G$ up to isomorphism, as shown by Steinberg \cite{Steinberg}. 

For a complex reductive Lie group $G$ with a choice of a maximal torus $T \subset G$, we may interchange the roles of roots and coroots as well as of the character and cocharacter 
lattices, thus obtaining a new root datum

\[\mathcal{R}(G,T)^{\vee}:=\{R \subset \Lambda^{\perp}, C \subset \Lambda\} \; ,\]
called the \emph{dual root datum}.
This introduces the following:

\begin{definition}
For $G$ and $T$ as above, the \emph{Langlands dual group} of $G$  is the complex reductive group (unique up to isomorphism) determined by the dual root datum $\mathcal{R}(G,T)^{\vee}$. We denote the Langlands dual group of $G$ by $G^{\vee}$.
\end{definition}

\begin{example}
    The Langland dual of $\PSL_2(\mathbb C)$ is $\PSL_2(\mathbb C)^\vee = \mathrm{SL}_2(\mathbb C)$. More generally the Langland dual of a centerless reductive group is simply connected.
\end{example}

\subsection{Root systems}

The concept of a \emph{root system}\index{root!system} of a finite-dimensional vector space is important for many fields of mathematics, but becomes particularly essential in the study of simple Lie algebras, leading to their classification. We review some of its basic features here and refer to \cite[Chap. 10, Sect. 3]{Helgason} for a very thorough study.

\begin{definition}
 Let $V$ be a real vector space of finite dimension and let $\alpha$ be a nonzero element in $V$. A \emph{reflection} along $\alpha$ is a linear transformation $s$ on $V$ satisfying the following:
 \begin{enumerate}
     \item It holds that $s(\alpha)=-\alpha$.
     \item The fixed points of $s$ constitute a hyperplane in $V$. 
 \end{enumerate}
\end{definition}

Note that any reflection has order 2 and can be determined by its fixed points. An important property of reflections is that given a nonzero element $\alpha \in V$, there is at most one reflection along $\alpha$ leaving a finite generating subset $R\subset V$ invariant. We now introduce the following:

\begin{definition}\labelx{def_reduced_root_sys}
Let $V$ be a real vector space of finite dimension and let $R \subset V$ be a finite set of nonzero vectors. The set $R$ is called a \emph{root system} in $V$, and its points are called \emph{roots}, if the following are satisfied:
\begin{enumerate}
    \item The set $R$ generates the vector space $V$.
    \item For each $\alpha \in R$ there exists a reflection $s_{\alpha}$ along $\alpha$ leaving $R$ invariant; this reflection has to be then unique.
    \item For all $\alpha, \beta \in R$, the number $a_{\alpha, \beta}$ determined by the equation $s_{\alpha}(\beta)= \beta - a_{\alpha, \beta}\alpha$ is an integer. 
\end{enumerate}
A root system is called \emph{reduced} if for any pair of elements $\alpha, \beta \in R$ with $\beta = m\alpha$ for a real number $m$, it is then $m=\pm 1$.
\end{definition}

Let now $\mathfrak{g}$ be a complex simple Lie algebra and let $\mathfrak{t}$ be a \emph{Cartan subalgebra}, that is, a self-normalizing nilpotent subalgebra of $\mathfrak{g}$. A \emph{root} of $\mathfrak{g}$ relative to $\mathfrak{t}$ is a nonzero element $\alpha \in \mathfrak{t}^{*}$ for which there exists a nonzero $X\in \mathfrak{g}$ such that
\[[H,X]=\alpha(H)X, \text{ for every } H\in \mathfrak{t} \; .\]
The set $\Delta =\Delta(\mathfrak{g}, \mathfrak{t}) \subset \mathfrak{t}^{*}$ of roots of $\mathfrak{g}$ with respect to a Cartan subalgebra $\mathfrak{t}$ is finite and it is a reduced root system in the sense of \Cref{def_reduced_root_sys}.

For a root system $R$ in $\mathfrak{t}^{*}$, let $\mathrm{Aut}(R)$ denote the finite group of linear transformations of $\mathfrak{t}^{*}$ leaving $R$ invariant. The subgroup $W(R)$ of $\mathrm{Aut}(R)$ generated by the reflections $\{s_{\alpha} \mid \alpha \in R \}$ is called the \emph{Weyl group} of $R$. We moreover define a subset $B\subset R$ to be a \emph{basis} of a root system $R$ if
\begin{enumerate}
    \item The set $B$ is a basis of $\mathfrak{t}^{*}$.
    \item Each element $\beta \in R$ can be written as $\beta = \sum_{\alpha \in B} n_{\alpha}\alpha$, where the factors $n_{\alpha}$ are integers of the same sign.
\end{enumerate}
Every root system of a complex simple Lie algebra has a basis, and any two bases are conjugate under a unique Weyl group element. 

There exists a unique positive definite inner product $\langle \;,\, \rangle$
 on $\mathfrak{t}^{*}$ invariant under $W(R)$ satisfying 
 \[\langle \lambda, \mu \rangle = \sum_{\alpha \in R} \langle \lambda, \alpha \rangle \langle \alpha, \mu \rangle, \text{ for } \lambda, \mu \in \mathfrak{t}^{*} \; .\]
For $a_{\alpha, \beta}$ as in \Cref{def_reduced_root_sys}, it is then 
\[a_{\alpha, \beta} = 2 \frac{\langle \beta, \alpha \rangle}{\langle \alpha, \alpha \rangle}, \text{ for }\alpha, \beta \in R \; .\]
This is called the \emph{Killing form} of the reduced root system $R$ of $\mathfrak{g}$.

An element $\gamma \in \mathfrak{t}^{*}$ is called \emph{regular} if $\langle \alpha, \gamma \rangle \neq 0$ for all $\alpha \in R$. For a regular element $\gamma$, let 
\[R^{+}(\gamma):= \{\alpha \in R \mid \langle \alpha, \gamma \rangle > 0\} \; .\]
A root $\alpha \in R^{+}(\gamma)$ is called \emph{simple} if it cannot be written as a sum of two elements in $R^{+}(\gamma)$. Simple roots do form a basis of $R$.  A root system is called \emph{irreducible} if it cannot be decomposed into a pair of disjoint nonempty orthogonal subsets; the orthogonality considered with respect to the Killing form.

\subsection{Classification of reduced root systems}
We now see how from a type of weighted graphs called \emph{Dynkin diagrams} we can determine the root system of a complex simple Lie algebra and, in fact, fully classify all such Lie algebras up to isomorphism. These diagrams determine a matrix of integer entries, called the \emph{Cartan matrix}, which in turn determines the root system. 

\begin{definition}
Let $\mathfrak{g}$ be a complex simple Lie algebra of rank $l$. Let $R$ be a reduced root system of $\mathfrak{t}^{*}$ and let $B=\{\alpha_{1},...,\alpha_{n}\}$ be a basis of $R$. The matrix $(A_{ij})_{1\leq i,j \leq l}$ with $A_{ij} = a_{\alpha_i , \alpha_j}$ is called the \emph{Cartan matrix} of $\mathfrak{g}$; remember that $a_{\alpha_{i}, \alpha_{j}} = 2 \frac{\langle \alpha_{i}, \alpha_{j} \rangle}{\langle \alpha_{i}, \alpha_{i} \rangle}, \text{ for }\alpha_{i}, \alpha_{j} \in R$.    
\end{definition}

Let now $R$ be an irreducible reduced root system of a Cartan subalgebra $\mathfrak{t}$ of a rank $l$ complex simple Lie algebra $\mathfrak{g}$ with a basis of $R$ given by $B=\{\alpha_1,...,\alpha_l\}$, and let $R^{+}$ be the corresponding set of positive roots. The \emph{Coxeter graph} of $R$ is a graph with $l$-many vertices, with the $i$th vertex joined to the $j$th vertex by $a_{ij}a_{ji}$-many non-intersecting lines, where $a_{ij}=a_{\alpha_{i}, \alpha_{j}}$.  The irreducibility of $R$ provides that the Weyl group $W(R)$ acts irreducibly on $\mathfrak{t}^{*}$, hence the invariant inner product $\langle \;,\; \rangle$ is unique up to a constant factor. Assigning a weight proportional to  $\langle \alpha_{i}, \alpha_{i} \rangle$ on the $i$th vertex in the Coxeter graph, we get a graph called the \emph{Dynkin diagram}. In the seminal works of Killing and Cartan, a full classification of the complex simple Lie algebras was provided in terms of their Dynkin diagrams. We list these diagrams below and include the roots in a basis of the irreducible reduced system as this will be useful later on in \Cref{sec:positivity}; for a detailed account we refer the reader to \cite[Chap. 10, Sect. 3.3]{Helgason}.

\begin{theorem}\labelx{classification_Dynkin}
The classification of Dynkin diagrams of reduced root systems of a Cartan subalgebra of a rank $l$ complex simple Lie algebra is as follows:
\begin{itemize}
    \item For the Lie algebra $\mathfrak{a}_l\cong \mathfrak{sl}_{l+1}(\mathbb C)$ with $l\ge 1$ the Dynkin diagram is 
   
\begin{dynkinDiagram}[text style/.style={scale=1},
edge length=1cm,
labels*={1,1,1,1},
labels={1,2,l-1,l},
label macro/.code={\alpha_{\drlap{#1}}}
]A{oo...oo}
\end{dynkinDiagram}

and a basis of the reduced root system is given by the roots 

$\alpha_i = e_i -e_{i+1}$, for $1\leq i \leq l$. 

\item For the Lie algebra $\mathfrak{b}_l\cong\mathfrak{so}_{2l+1}(\mathbb C)$ with $l\ge 2$ the Dynkin diagram is 
   
\begin{dynkinDiagram}[text style/.style={scale=1},
edge length=1cm,
labels*={2,2,2,1},
labels={1,2,l-1,l},
label macro/.code={\alpha_{\drlap{#1}}}
]B{oo...oo}
\end{dynkinDiagram}

and a basis of the reduced root system is given by the roots 

$\alpha_i = e_i -e_{i+1}$, for $1\leq i \leq l-1$, and $\alpha_l = e_l$. 

\item For the Lie algebra $\mathfrak{c}_l\cong\mathfrak{sp}_{2l}(\mathbb C)$ with $l\ge 3$ the Dynkin diagram is 
   
\begin{dynkinDiagram}[text style/.style={scale=1},
edge length=1cm,
labels*={1,1,1,2},
labels={1,2,l-1,l},
label macro/.code={\alpha_{\drlap{#1}}}
]C{oo...oo}
\end{dynkinDiagram}

and a basis of the reduced root system is given by the roots 

$\alpha_i = e_i -e_{i+1}$, for $1\leq i \leq l-1$, and $\alpha_l = 2e_l$.

\item For the Lie algebra $\mathfrak{d}_l\cong\mathfrak{so}_{2l}(\mathbb C)$ with $l\ge 4$ the Dynkin diagram is 
   
\begin{dynkinDiagram}[text style/.style={scale=1},
edge length=1cm,
labels*={1,1,1,1,1},
labels={1,2,l-2,l-1,l},
label macro/.code={\alpha_{\drlap{#1}}}
]D{oo...ooo}
\end{dynkinDiagram}

and a basis of the reduced root system is given by the roots 

$\alpha_i = e_i -e_{i+1}$, for $1\leq i \leq l-1$, and $\alpha_l = e_{l-1}+e_l$.

\item For the Lie algebra $\mathfrak{e}_6$ the Dynkin diagram is 
   
\begin{dynkinDiagram}[text style/.style={scale=1},
edge length=1cm,
labels*={1,1,1,1,1,1},
labels={6,2,5,4,3,1},
label macro/.code={\alpha_{\drlap{#1}}}
]E{oooooo}
\end{dynkinDiagram}

and a basis of the reduced root system is given by the roots 

$\alpha_1=\frac{1}{2}(e_1+e_6)-\frac{1}{2}(e_2+e_3+e_4+e_5),\; \alpha_2=e_1+e_2,\;\alpha_3=e_2-e_1,\;\alpha_4=e_3-e_2,\;\alpha_5=e_4-e_3,\;\alpha_6=e_5-e_4.$

\item For the Lie algebra $\mathfrak{e}_7$ the Dynkin diagram is 
   
\begin{dynkinDiagram}[text style/.style={scale=1},
edge length=1cm,
labels*={1,1,1,1,1,1,1},
labels={7,2,6,5,4,3,1},
label macro/.code={\alpha_{\drlap{#1}}}
]E{ooooooo}
\end{dynkinDiagram}

and a basis of the reduced root system is given by the roots 

$\alpha_1=\frac{1}{2}(e_1+e_7)-\frac{1}{2}(e_2+e_3+e_4+e_5+e_6),\; \alpha_2=e_1+e_2,\;\alpha_3=e_2-e_1,\;\alpha_4=e_3-e_2,\;\alpha_5=e_4-e_3,\;\alpha_6=e_5-e_4,\;\alpha_7=e_6-e_5.$

\item For the Lie algebra $\mathfrak{e}_8$ the Dynkin diagram is 
   
\begin{dynkinDiagram}[text style/.style={scale=1},
edge length=1cm,
labels*={1,1,1,1,1,1,1,1},
labels={8,2,7,6,5,4,3,1},
label macro/.code={\alpha_{\drlap{#1}}}
]E{oooooooo}
\end{dynkinDiagram}

and a basis of the reduced root system is given by the roots 

$\alpha_1=\frac{1}{2}(e_1+e_8)-\frac{1}{2}(e_2+e_3+e_4+e_5+e_6+e_7),\; \alpha_2=e_1+e_2,\;\alpha_3=e_2-e_1,\;\alpha_4=e_3-e_2,\;\alpha_5=e_4-e_3,\;\alpha_6=e_5-e_4,\;\alpha_7=e_6-e_5,\;\alpha_8=e_7-e_6.$

\item For the Lie algebra $\mathfrak{f}_4$ the Dynkin diagram is 
   
\begin{dynkinDiagram}[text style/.style={scale=1},
edge length=1cm,
labels*={2,2,1,1},
labels={1,2,3,4},
label macro/.code={\alpha_{\drlap{#1}}}
]F{oooo}
\end{dynkinDiagram}

and a basis of the reduced root system is given by the roots 

$\alpha_1=e_2-e_3,\; \alpha_2=e_3-e_4,\;\alpha_3=e_4,\;\alpha_4=\frac{1}{2}(e_1-e_2-e_3-e_4).$

\item For the Lie algebra $\mathfrak{g}_2$ the Dynkin diagram is 
   
\begin{dynkinDiagram}[text style/.style={scale=1},
edge length=1cm,
labels*={3,1},
labels={2,1},
label macro/.code={\alpha_{\drlap{#1}}}
]G{oo}
\end{dynkinDiagram}

and a basis of the reduced root system is given by the roots 

$\alpha_1=e_1-e_2,\; \alpha_2=-2e_1+e_2+e_3.$

\end{itemize}
\end{theorem}

\section{Real Lie groups}\labelx{sec:real_Lie_groups}

The structure theory of complex semisimple Lie algebras enjoys several nice properties and is, in general, simpler than that of real Lie algebras. For our purposes though, we will need to consider real Lie algebras and these can be introduced via real forms of complex Lie algebras. Standard references for an extensive treatment of the structure theory of semisimple Lie algebras include \cite{Helgason, Knapp, Onishchik}.

\begin{definition}
    Let $\mathfrak{g}_\mathbb{R}$ be a real Lie algebra. The complex Lie algebra $\mathfrak{g}_\mathbb{C} = \mathfrak{g}_\mathbb{R}\otimes \mathbb C$ is called the \emph{complex form} of $\mathfrak{g}_\mathbb{R}$, and $\mathfrak{g}_\mathbb{R}$ is called a \emph{real form} of $\mathfrak{g}_\mathbb{C}$.
\end{definition}

\begin{remark}
    The complex form $\mathfrak{g}_\mathbb{C}$ is simple (resp. semisimple) if and only if $\mathfrak{g}_\mathbb{R}$ is.
\end{remark}

\begin{example}
    The complex form of both $\mathfrak{sl}(n,\mathbb R)$ and $\mathfrak{su}(n)$ is $\mathfrak{sl}(n,\mathbb C)$, showing that two non-isomorphic real Lie algebra can have the same complex form.
\end{example}

\begin{proposition}
    Let $\mathfrak{g}_\mathbb{C}$ be a complex Lie algebra and let $\sigma\colon  \mathfrak{g}_{\mathbb{C}} \to \mathfrak{g}_{\mathbb{C}}$ be a complex antilinear involution. The involution $\sigma$ determines a decomposition of $\mathfrak{g}_\mathbb{C}$ into its $(\pm1)$-eigenspaces, as 
 $$\mathfrak{g}_\mathbb{C}=\mathfrak{g}_{+} \oplus \mathfrak{g}_{-} \; .$$ 
 The subalgebra $\mathfrak{g}_{+}$ is a real form of $\mathfrak{g}_\mathbb{C}$. Conversely, any real form $\mathfrak{g}_\mathbb{R}$ of $\mathfrak{g}_{\mathbb{C}}$ is the $(+1)$-eigenspace of the complex conjugation induced on $\mathfrak{g}_\mathbb{C}$ by the isomorphism $\mathfrak{g}_\mathbb{C} = \mathfrak{g}_\mathbb{R}\otimes \mathbb C$.
\end{proposition}

When $\mathfrak{g}_\mathbb{C}$ is semisimple, its Killing form $K_\mathbb{C}$ is non-degenerate and restricts to a non-degenerate Killing form $K_\mathbb{R}$ on any real form $\mathfrak{g}_\mathbb{R}$. By Sylvester's law of inertia, the real Killing form $K_\mathbb{R}$ can be diagonalized in a suitable basis with the diagonal entries $+1$ or $-1$.

\begin{definition}
    Let $\mathfrak{g}_\mathbb{R}$ be a real semisimple Lie algebra. A \emph{Cartan involution} is a Lie algebra involution $\theta \colon  \mathfrak{g}_\mathbb{R}\to \mathfrak{g}_\mathbb{R}$ such that the bilinear form $$ \begin{array}{rcl}
         K_\theta\colon \mathfrak{g}_\mathbb{R}\times \mathfrak{g}_\mathbb{R} &\to &\mathbb R\\
         (X,Y) & \mapsto &K_\mathbb{R}(X, -\theta(Y))
    \end{array}$$
    is positive definite.
    A Cartan involution $\theta$ determines a decomposition of $\mathfrak{g}_\mathbb{R}$ as a vector space into its $(-1)$-eigenspace $\mathfrak{p}$ and its $(+1)$-eigenspace $\mathfrak{k}$, called a \emph{Cartan decomposition}
    $$\mathfrak{g}_\mathbb{R} = \mathfrak{p}\oplus\mathfrak{k} \; .$$
\end{definition}

A semisimple real Lie algebra $\mathfrak{g}_\mathbb{R}$ may have several Cartan involutions, but they are all conjugate by an automorphism of $\mathfrak{g}_\mathbb{R}$. Note that the $(+1)$-eigenspace $\mathfrak{k}$ is a subalgebra of $\mathfrak{g}_\mathbb{R}$ while the $(-1)$-eigenspace $\mathfrak{p}$ is only a vector space as $[\mathfrak{p},\mathfrak{p}]\subset \mathfrak{k}$. As a result, if a real Lie group $G$ has its Lie algebra equal to $\mathfrak{g}_\mathbb{R}$, there is a subgroup $K\subset G$ whose Lie algebra is $\mathfrak{k}$. This subgroup is a maximal compact subgroup of $G$. Also note that the Killing form $K_\mathbb{R}$ is positive definite on $\mathfrak{p}$ and negative definite on $\mathfrak{k}$. 

A real semisimple Lie algebra also admits a decomposition into weight spaces. We call a maximal abelian subspace $\mathfrak{a}$ of $\mathfrak{p}$ a \emph{Cartan subspace}. The Lie algebra $\mathfrak{g}_\mathbb{R}$ then decomposes as 
\[\mathfrak{g}_\mathbb{R} = \mathfrak{g}_0 \oplus \bigoplus\limits_{\alpha \in R}^{}{}\mathfrak{g}_{\alpha} \; ,\]
for $R \subset \mathfrak{a}^*$, the \emph{set of restricted roots} of $\mathfrak{g}_\mathbb{R}$. The space $\mathfrak{g}_{\alpha}$, for each $\alpha$, is the eigenspace for the \emph{restricted} root $\alpha \in \mathfrak{a}^*$. Unlike in the complex case, $\mathfrak{a}\neq \mathfrak{g}_0$ in general. The dimension of $\mathfrak{a}$ is called the \emph{real rank} of $\mathfrak{g}_\mathbb{R}$, denoted by $\mathrm{rk}_\mathbb{R}(\mathfrak{g}_\mathbb{R})$.

Among all real forms of a complex semisimple Lie algebra, there are two specific ones we will be interested in: the \textit{split real form} and the \emph{compact form}. We start by describing the compact form.

\begin{definition}
    A real form $\mathfrak{g}_\mathbb{R}$ of $\mathfrak{g}_\mathbb{C}$ is called a \emph{compact form} if $K_\mathbb{R}$ is negative definite on $\mathfrak{g}_\mathbb{R}$.
\end{definition}

The name \emph{compact form} is justified by the following proposition:
\begin{proposition}
    A real simple Lie group is compact if and only if its Lie algebra is of compact form.
\end{proposition}

\begin{remark}
    Every complex semisimple Lie algebra has a unique (up to inner automorphism) compact form. In that case, there is a unique Cartan involution which is the identity.
\end{remark}

We now describe the split real form.

\begin{definition}
For a complex semisimple Lie algebra $\mathfrak{g}_\mathbb{C}$ with Cartan subalgebra $\mathfrak{t}$, a real form $\mathfrak{g}_\mathbb{R}$ whose Cartan subspace $\mathfrak{a}$ satisfies $\mathfrak{a}\otimes \mathbb{C} = \mathfrak{t}$, or equivalently if $\mathrm{rk}_\mathbb{C}(\mathfrak{g}_\mathbb{C}) =\mathrm{rk}_\mathbb{R}(\mathfrak{g}_\mathbb{R})$, is called a \textit{split real form} of $\mathfrak{g}_\mathbb{C}$.
\end{definition}
Any complex semisimple Lie algebra contains a split real form, and from the classification of simple Lie algebras, we may identify the list of split simple real Lie algebras up to isomorphism.

We now consider the exponentials of these Lie algebras to pass to Lie groups, and we thus define:
\begin{definition}
A simple real Lie group is called a \textit{split real} Lie group whenever $\mathrm{rk}_{\mathbb{R}}G=\mathrm{rk}_{\mathbb{C}}G^{\mathbb{C}}$, where $G^{\mathbb{C}}:=G\otimes \mathbb{C}$ denotes the complexification of $G$.
\end{definition}

\begin{fact}
Up to a covering, a simple split real Lie group is one of the following:
$$\mathrm{SL}_{n}(\mathbb{R}),\;\;\mathrm{SO}(n,n+1),\;\;\mathrm{Sp}_{2n}(\mathbb{R}),\;\; \mathrm{SO}(n,n),\;\;  E_{6(6)},\;\;E_{7(7)},\;\;E_{8(8)},\;\;F_{4(4)},\;\;G_2$$
for $n\in \mathbb N$. The first four groups are called \emph{classical} whereas the last five are called \emph{exceptional}. 
\end{fact}

A similar classification result holds for compact simple Lie groups:

\begin{fact}
Up to a covering, a compact simple Lie group is one of the following:
$$\mathrm{SU}(n),\;\;\mathrm{SO}(2n+1),\;\;\mathrm{USp}(2n),\;\;\mathrm{SO}(2n),\;\;E_{6}^c,\;\; E_{7}^c,\;\;E_{8}^c,\;\;F_{4}^c,\;\;G_2^c$$
for $n\in \mathbb N$. The first four groups are called \emph{classical} whereas the last five are called \emph{exceptional}. Note that $\mathrm{USp}(2n)=\mathrm{Sp}_{2n}(\mathbb C)\cap\mathrm{SU}(2n)$.
\end{fact}

\section{Levi decomposition, parabolics and flag varieties}\labelx{Sec:Levi-flags}

In some contexts, we need to consider Lie algebras which are not reductive, for instance the subalgebra of $\mathfrak{gl}_n$ consisting of upper triangular matrices. 
We briefly review the concepts of Levi decomposition, Borel subalgebra, parabolic subalgebra and partial flag variety here; a standard reference for these topics is the book of Helgason \cite{Helgason}.

In this section, we denote by $\mathfrak{g}$ a finite-dimensional Lie algebra over a field of characteristic zero (typically $\mathbb{R}$ or $\mathbb{C}$).
Recall that its radical $\mathrm{rad}\,\mathfrak{g}$ is the maximal solvable ideal.

\begin{theorem}[Levi decomposition]
    Any finite-dimensional Lie algebra over a field of characteristic zero is a semidirect product of its radical and a semisimple Lie subalgebra:
    $$\mathfrak{g} \cong \mathrm{rad}\,\mathfrak{g}\rtimes \mathfrak{l}\; .$$
    The semisimple subalgebra $\mathfrak{l}$ is called a \emph{Levi subalgebra}.
\end{theorem}

\begin{theorem}[Levi--Malcev]
    Any two Levi subalgebras are conjugated to each other by an inner automorphism.
\end{theorem}

This decomposition allows one to separate semisimple and solvable Lie algebras. We will typically apply it to generalizations of upper triangular matrices in $\mathfrak{gl}_n$.

\begin{definition}
    A \emph{Borel subalgebra} $\mathfrak{b}$ is a maximal solvable subalgebra of $\mathfrak{g}$. Its commutator $\mathfrak{u}=[\mathfrak{b},\mathfrak{b}]$ is called \emph{unipotent subalgebra}. 
\end{definition}

\begin{example}
    For $\mathfrak{g}=\mathfrak{gl}_n$, the space of upper triangular matrices forms a Borel subalgebra, while the space of strictly upper triangular matrices forms a unipotent subalgebra.
\end{example}
\begin{example}
    More generally, for $\mathfrak{g}=\mathfrak{gl}(V)$ where $V$ is a finite-dimensional vector space of dimension $n$, a Borel subalgebra is the stabilizer of a complete flag. Given $V_0=\{0\}\subset V_1\subset V_2\subset...\subset V_n=V$, with $\mathrm{dim}(V_i)=i$, the set $\{M\in\mathfrak{gl}(V)\mid M(V_i)\subset V_i \;\forall\,i\}$ is a Borel subalgebra. Any Borel subalgebra of $\mathfrak{gl}(V)$ is of this form.
\end{example}

\begin{proposition}
    Any two Borel subalgebras are conjugated by an inner automorphism.
\end{proposition}
    
A more general class of subalgebras of a Lie algebra are those containing a Borel:
\begin{definition}
    A strict subalgebra $\mathfrak{p}\subset \mathfrak{g}$ containing a Borel subalgebra is called \emph{parabolic}.
\end{definition}

\begin{example}
    For $\mathfrak{g}=\mathfrak{gl}(V)$ as before, a parabolic is the stabilizer of a partial flag. Given a partial flag $V_{i_0}\subset V_{i_1}\subset...\subset V_{i_m}$ (where $0\leq i_1<i_2<...i_m\leq n$), the space $\{M\in \mathfrak{gl}(V)\mid M(V_{i_k)}\subset V_{i_k}\}$ is a parabolic subalgebra. It contains the Borel subalgebra associated with any completion of the partial flag. Any parabolic subalgebra in $\mathfrak{gl}(V)$ is of this form.
\end{example}

In general, parabolic subalgebras are in correspondence with non-empty subsets of the set of simple roots $\Delta$:
\begin{proposition}
    Up to inner automorphisms, there is a bijection between parabolic subalgebras and non-empty subsets $\Theta\subset \Delta$.
\end{proposition}
In particular there are $2^{\mathrm{rk}(\mathfrak{g})}-1$ different kinds of parabolic subalgebras. If $\lvert \Theta\rvert =1$, the parabolic is maximal. For $\Theta=\Delta$, the parabolic is minimal and equals a Borel subalgebra.

For a given subset $\Theta\subset \Delta$, the parabolic $\mathfrak{p}_{\Theta}$ is the Lie algebra generated by all $f_i, h_i$ and those $e_i$ such that $\beta(e_i)\neq 0$ for all $\beta\in\Theta$, where $(f_i, h_i, e_i)$ denotes a Chevalley basis of $\mathfrak{g}$. The opposite parabolic $\mathfrak{p}_{\Theta}^{\mathrm{opp}}$ is generated by all $e_i, h_i$ and those $f_i$ such that $\beta(f_i)\neq 0$ for all $\beta\in \Theta$.

The Levi decomposition applied to a parabolic subalgebra gives the following:
\begin{proposition}
    A parabolic subalgebra is a semidirect product between the unipotent radical and a semisimple subalgebra.
\end{proposition}

\begin{example}
    Consider a vector space $V$ with an inner product. For $\mathfrak{g}=\mathfrak{gl}(V)$ and the parabolic $\mathfrak{p}$ associated to a partial flag $(V_{i_k})_{1\leq k\leq m}$, the opposite flag in $V^*$ can be identified with the orthogonals $(V^\perp_{i_k})$. The Levi subalgebra is then the intersection $\mathfrak{p}\cap \mathfrak{p}^{\mathrm{opp}}$, and the unipotent radical is the space of transformations preserving both $\mathfrak{p}$ and $\mathfrak{p}^{\mathrm{opp}}$.
\end{example}

\begin{figure}
    \centering
    \includegraphics[width=0.65\linewidth]{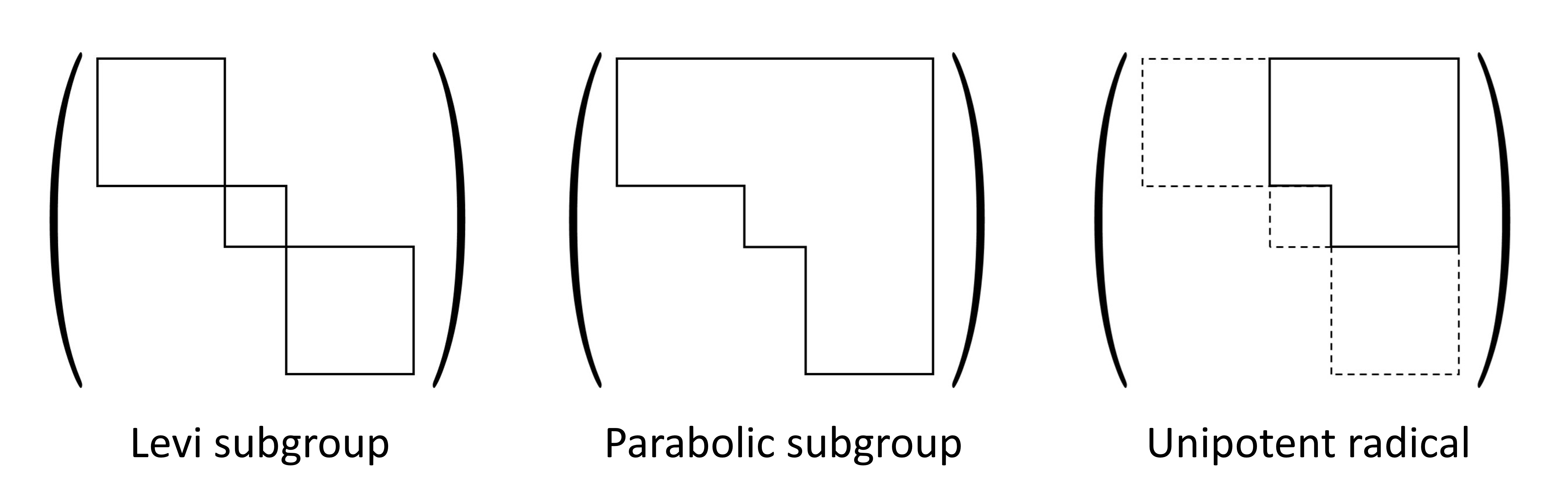}
    \caption{Cartoon for a Levi and parabolic subgroup, and its unipotent radical for $\mathrm{GL}_n$.}
    \labelx{fig:Levi-decompo-cartoon}
\end{figure}

\medskip
\noindent On the level of Lie groups, we have similar notions. Let $G$ be a Lie group with Lie algebra $\mathfrak{g}$.

\begin{definition}
    A \emph{Borel subgroup} $B$ is a maximal solvable subgroup of $G$. Its commutator subgroup $U=[B,B]$ is a \emph{unipotent subgroup}. A \emph{parabolic subgroup} $P$ is a strict subgroup of $G$ that contains a Borel subgroup.
\end{definition}

The general correspondence between a Lie group and its Lie algebras gives a bijection between Borel subgroups and Borel subalgebras, and similarly for parabolic and unipotent subgroups and subalgebras.
As before, there is a close relationship between parabolic subgroups and partial flags, the former being the stabilizer of the latter.

\begin{definition}
    The \emph{flag variety} of a Lie group $G$ is the coset space $G/B$, where $B$ is a Borel subgroup. A \emph{partial flag variety} is the coset space $G/P$, where $P$ is a parabolic subalgebra of $G$.
\end{definition}

\begin{example}\labelx{Ex:flag-sl2-C}
    For $G=\mathrm{SL}_2(\mathbb{C})$, the space of flags can be identified with $\mathbb{CP}^1$. The action of $G$ on complete flags being transitive, and the stabilizer being a Borel, we have that $G/B\cong \mathbb{CP}^1$.
\end{example}

More generally, for $G=\mathrm{GL}(V)$, the action of $G$ on the space of all complete flags is transitive. The stabilizer of a fixed flag is a Borel subgroup. Hence, we can identify $G/B$ with the space of all complete flags.
Similarly, the space $G/P$ can be identified with the space of all partial flags, where the type is specified by the stabilizer of $P$.

In the presence of an inner product, the orthogonalization process (or Gram--Schmidt algorithm) allows to associate an orthonormal basis to a complete flag. The following proposition can be seen as a generalization of this.

\begin{proposition}
    The flag variety $G/B$ is diffeomorphic to the coset space $K/T$, where $K$ is a maximal compact subgroup of $G$, and $T\subset K$ a maximal torus in $K$. More generally, for the partial flag variety, we have $G/P\cong K/L^{\mathbb{R}}$, where $L^{\mathbb{R}}=L\cap K$, and $L$ is the Levi subgroup corresponding to $P$.
\end{proposition}

\begin{example}
    The proposition applied to  \Cref{Ex:flag-sl2-C} gives
    $$\mathbb{CP}^1\cong\mathrm{SL}_2(\mathbb{C})/B\cong \mathrm{SU(2)}/T\cong \mathrm{SO}(3)/\mathrm{SO}(2) \; .$$
\end{example}


\backmatter


\newpage

\markboth{Glossary}{Glossary}

\glsaddall
\printglossaries

\newpage

\markboth{Bibliography}{Bibliography}

\printbibliography[title={Bibliography}]

\newpage

\printindex

\end{document}